\documentclass{siamart1116}
\usepackage{indentfirst}
\usepackage{graphicx}
\usepackage{epstopdf}
\usepackage{epsfig}
\usepackage{overpic}
\usepackage{array}
\usepackage{color}
\usepackage{multirow}
\usepackage{float}
\usepackage{subfigure} 
\usepackage{amsmath,amssymb}
\usepackage[mathscr]{euscript}
\usepackage{latexsym,bm}
\usepackage{booktabs}
\usepackage{bbm}
 \usepackage[all]{xy}
 \usepackage{color}
\usepackage{diagbox}
\usepackage{algorithm}  
\usepackage{algorithmicx} 
\usepackage{algpseudocode}  
\usepackage{hyperref}
\usepackage{makecell}
\usepackage{mathtools}
\usepackage{multirow}

\newtheorem{example}{\textup{\textbf{Example}}}

\newcommand\cp{\color{black}}
\newcommand\comment[1]{}

{\theoremstyle{remark} \newtheorem{remark}{\textup{\textbf{Remark}}}}
\def\mi{\mathbbm{i}}
\def\me{\mathbbm{e}}
\def\mone{\mathbbm{1}}
\newcommand\D{\textup{d}}
\def\bx{\bm{x}}

\def\bR{\bm{R}}

\def\bk{\bm{k}}
\def\by{\bm{y}}
\def\bu{\bm{u}}
\def\bv{\bm{v}}
\def\tD{\textup{D}}

\def\bxi{\bm{\xi}}

\def\pdo{\Psi \textup{DO}}

\makeatletter
\@addtoreset{equation}{section}
\makeatother
\renewcommand{\theequation}{\arabic{section}.\arabic{equation}}

\graphicspath{ {./} }

\newsavebox{\tablebox}

\begin{document}
\bibliographystyle{unsrt}

\title{Overcoming the numerical sign problem in the Wigner dynamics via adaptive particle annihilation}
\author{Yunfeng Xiong\footnotemark[1] \footnotemark[2], \and Sihong Shao\footnotemark[2] \footnotemark[3]
}
\renewcommand{\thefootnote}{\fnsymbol{footnote}}
\footnotetext[1]{{\cp School of Mathematical Sciences, Beijing Normal University, Beijing 100091, China. Email address:
{\tt yfxiong@bnu.edu.cn} (Y. Xiong).}}
\footnotetext[2]{CAPT, LMAM and School of Mathematical Sciences, Peking University, Beijing 100871, China. Email address:{\tt sihong@math.pku.edu.cn} (S. Shao).}
\footnotetext[3]{To
whom correspondence should be addressed.}
\maketitle


\begin{abstract}
The infamous numerical sign problem poses a fundamental obstacle to particle-based stochastic Wigner simulations in high dimensional phase space. Although the existing particle annihilation via uniform mesh significantly alleviates the sign problem when dimensionality D $\le$ 4, the mesh size grows dramatically when D $\ge$ 6 due to the curse of dimensionality and consequently makes the annihilation very inefficient.
 In this paper, we propose an adaptive particle annihilation algorithm, termed Sequential-clustering Particle Annihilation via Discrepancy Estimation (SPADE), to overcome the sign problem. SPADE follows a divide-and-conquer strategy: Adaptive clustering of particles via controlling their number-theoretic discrepancies and independent random matching in each cluster.
The target is to alleviate the oversampling problem induced by the over-partitioning of phase space and capture the non-classicality of the Wigner function simultaneously. Combining SPADE with the variance reduction technique based on the stationary phase approximation, we attempt to simulate the proton-electron couplings in 6-D and 12-D phase space. A thorough performance benchmark of SPADE is provided with the reference solutions in 6-D phase space produced by a characteristic-spectral-mixed scheme under a $73^3 \times 80^3$ uniform grid, which fully explores the limit of grid-based deterministic Wigner solvers.

\end{abstract}

 \vspace*{4mm}



\section{Introduction}


During the past few decades, it has  burgeoned with a wide spectrum of applications of the Wigner quantum dynamics \cite{Wigner1932} in the fields of semiconductor devices \cite{Frensley1987,KosinaSverdlovGrasser2006,SellierNedjalkovDimov2014,bk:Jacoboni2010}, nano-materials \cite{bk:NedjalkovQuerliozDollfusKosina2011,LucaRomano2019,BenamBallicchiaWeinbubSelberherrNedjalkov2021}, high energy physics \cite{GrazianiBauerMurillo2014} and quantum tomography \cite{KurtsieferPfauMlynek1997,DaviesRundleDwyerToddEveritt2019} for its huge theoretical advantage in resolving the classical-quantum dichotomy \cite{Wigner1932,KosinaNedjalkovSelberherr2003}, as well as its experimental observability \cite{DaviesRundleDwyerToddEveritt2019}. However, in contrast to the prosperity of both theoretical and experimental advances, there remains a huge gap in numerically solving the 6-D or higher dimensional Wigner quantum dynamics because of the well-known curse of dimensionality (CoD).


Grid-based deterministic Wigner solvers are able to produce highly accurate results owing to their solid mathematical theory and concise guiding principle \cite{XiongZhangShao2022,FurtmaierSucciMendoza2015,VandePutSoreeMagnus2017}, but both the computational cost and {\cp data} storage become extremely demanding when the dimensionality D $\ge$ 6 due to their unfavorable scaling. Alternatively, one can recourse to particle-based stochastic methods including particle affinity method \cite{bk:QuerliozDollfus2010}, signed-particle Wigner Monte Carlo \cite{KosinaNedjalkovSelberherr2003,KosinaSverdlovGrasser2006,bk:NedjalkovQuerliozDollfusKosina2011,SellierNedjalkovDimov2014,ShaoSellier2015},  random cloud model \cite{Wagner2016,MuscatoWagner2016} and Wigner branching random walk (WBRW) \cite{ShaoXiong2019,ShaoXiong2020}, {\cp in virtue of their convergence rate $N_0^{-1/2}$, where $N_0$ is the effective particle number (sample size), regardless of D}. However, even the state-of-the-art stochastic algorithms are still restricted in 4-D phase space  \cite{BenamBallicchiaWeinbubSelberherrNedjalkov2021,ShaoXiong2019} and few results have been reported for 6-D problems. The formidable obstacle there turns out to be the notorious numerical sign problem
\cite{SchmidtMoehring1993,NedjalkovKosinaSelberherr2003,ShaoXiong2020}, say, the exponential growth of both particle number and stochastic variance {\cp induced by increments of negative weights}, which is generally believed to be NP-hard \cite{TroyerWiese2005,IazziSoluyanovTroyer2016}.
This work follows the latter. Specifically, we propose an adaptive particle annihilation within the framework of WBRW, termed Sequential-clustering Particle Annihilation via Discrepancy Estimation (SPADE), to overcome the sign problem.

In our preceding work \cite{ShaoXiong2020}, we have pointed out that the sign problem is inherited in the widely used {\cp particle splitting technique \cite{KosinaNedjalkovSelberherr2003}} for the pseudodifferential operator ($\pdo$), because it ignores the cancelation of particle trajectories with opposite signs and leads to a rapid growth of variances. The remedies are to fully utilize the near-cancellation of positive and negative weights.
One approach directly aims at reducing stochastic variances, such as the semiclassical approximation \cite{GehringKosina2005}, the fractional particle weights \cite{ShaoXiong2019} and the stationary phase approximation (SPA) \cite{ShaoXiong2020}. These methods are able to  suppress the exponential growth of variances efficiently, albeit not completely eliminating it. The other approach is particle annihilation (PA), including PA via uniform mesh (PAUM for brevity) \cite{KosinaNedjalkovSelberherr2003,KosinaSverdlovGrasser2006,bk:NedjalkovQuerliozDollfusKosina2011,SellierNedjalkovDimov2014,ShaoXiong2019} and the particle resampling by filtering out the high-frequency components \cite{YanCaflisch2015}.  But the usage of  existing PA methods to 6-D problems is highly non-trivial. For instance, the most popular PAUM is bothered by CoD as the mesh size grows exponentially in D, so that many particles are left uncancelled  when the bin size largely exceeds the particle number \cite{YanCaflisch2015,bk:Silverman2018}.



The proposed SPADE tries to ameliorate CoD by a two-step strategy: Adaptive clustering of particles via controlling their number-theoretic discrepancies, partially borrowing the pioneering idea in the non-parametric high-dimensional density estimation \cite{LiYangWong2016}, and independent random matching among positive and negative particles in each cluster.
In other words, SPADE might potentially get rid of the over-partitioning problem in a uniform grid mesh and becomes less affected by CoD, thereby 
greatly facilitating realistic simulations, e.g., many-body problems in high-dimensional phase space. Moreover, SPADE can still recover the ``bottom line structure'' like in PAUM \cite{ShaoSellier2015}, an indicator that describes the minimal amount of particles that can accurately capture the non-classicality and oscillation of the Wigner function. It deserves to mention that the calculation of the discrepancy of a sequence, as a pivotal step in adaptive clustering, is NP-hard in nature \cite{GnewuchWahlstromWinzen2012}. In a sense, SPADE resolves the numerical sign problem inherited from CoD by seeking efficient heuristic approximations to another NP-hard problem.  

Combining SPADE and SPA together in WBRW, we succeed in simulating the proton-electron coupling, which is a typical non-equilibrium quantum dynamics under the Coulomb interaction \cite{PakHammesSchiffer2004,BenamBallicchiaWeinbubSelberherrNedjalkov2021} and serves as the prototype for the Coulomb collisions \cite{CarruthersZachariasen1983,GrazianiBauerMurillo2014}. A thorough  benchmark on 6-D simulations has been made to evaluate the performance of SPADE. For the sake of comparison, we endeavor to produce reference solutions by a massively parallel characteristic-spectral-mixed scheme  \cite{XiongZhangShao2022}, in which the Wigner function is represented as a tensor product of $75^3$ cubic spline basis in $\bx$-space and $80^3$ Fourier basis in $\bk$-space (with mesh size $73^3\times 80^3 \approx 2\times10^{11}$) to attain high accuracy. Numerical results manifest that SPADE may potentially avoid the oversampling problem by increasing the sample size. This constitutes the solid preparation for our attempt to obtain the first-principle solution to proton-electron coupling in 12-D phase space, where both proton and electron are treated quantum mechanically, and may potentially pave the way for the interlacement of kinetic theory and  molecular dynamics in high energy density physics \cite{GrazianiBauerMurillo2014}.

The rest is organized as follows. Section~\ref{sec.wigner} briefly reviews the Wigner function formalism for quantum mechanics in the phase space and illustrates the basic idea behind SPA. Section~\ref{sec.wbrw_spa} details the WBRW-SPA model (i.e.~using SPA in WBRW) for the Coulomb system. Section~\ref{sec.spade} focuses on the intuition and design of SPADE.  Numerical simulations on 6-D and 12-D proton-electron couplings are reported in Sections~\ref{sec.num} and \ref{sec.num_12d}, respectively. Finally, conclusion and discussion are drawn in Section~\ref{sec.conclusion}.


\section{Background}
\label{sec.wigner}

As a preliminary, we give a brief review of the Wigner dynamics and the physical intuition behind SPA. The $N$-body Wigner function is defined by the Weyl-Wigner transform of normalized density matrix $\rho(\bx, \by, t)$, 
\begin{equation}\label{def.Wigner_function}
f(\bx, \bk, t) = \int_{\mathbb{R}^{Nd}} \rho(\bx - \frac{\by}{2}, \bx + \frac{\by}{2}, t) \me^{-\mi \bk \cdot \by} \D \by
\end{equation}
with the spatial dimension $d$ and the dimensionality of phase space $\tD =2 Nd$. The quantum dynamics of the Wigner function is governed by the Wigner equation, 
\begin{equation}\label{eq.Wigner}
\begin{split}
\frac{\partial }{\partial t}f(\bm{x}, \bm{k}, t)+\frac{\hbar \bm{k}}{\bm{m}} \cdot \nabla_{\bm{x}} f(\bm{x},\bm{k}, t)  = \Theta_V[f](\bx, \bk, t),
\end{split}
\end{equation}
where $\bk/\bm{m}$ denotes $(\bk_1/m_1, \cdots, \bk_N/m_N)$ with $\bk_i$ and $m_i$ the wave vector and mass for the $i$-th body, respectively,
 $\hbar$ is the reduced Planck constant. For (many-body) particle interaction potential $V(\bx)
$, $\pdo$ reads as
\begin{equation}\label{pdo}
\Theta_V[f](\bx, \bk, t) = \frac{1}{\mi \hbar (2\pi)^{Nd}} \iint_{\mathbb{R}^{2Nd}} \me^{-\mi (\bk - \bk^{\prime}) \cdot \by }(V(\bx + \frac{\by}{2}) - V(\bx - \frac{\by}{2}))f(\bx, \bk^{\prime}, t) \D \by \D \bk^{\prime}.
\end{equation}

A profound advantage of the Wigner function is its simplicity in visualization in both position and momentum by lower dimensional projections \cite{Wigner1932}. For instance, the reduced Wigner function along the $j$-th phase space coordinate reads
\begin{equation}\label{reduced_Wigner_function}
W_j(x_j, k_j, t) = \iint_{\mathbb{R}^{Nd-1} \times \mathbb{R}^{Nd-1}} f(\bx, \bk, t) \D \bx_{\{j\}}  \D \bk_{\{j\}}
\end{equation}
with $\bx_{\{j\}} = (x_1, \dots, x_{j-1}, x_{j+1}, \dots, x_{Nd})$, $ \bk_{\{j\}}  = (k_1, \dots, k_{j-1}, k_{j+1}, \dots, k_{Nd})$. The spatial marginal distributions $P_{xy}$ and $P_x$ are obtained as follows, 
\begin{equation}\label{spatial_distribution}
P_{xy}(x_1, x_2, t) = \iint_{\mathbb{R}^{2Nd-2}} f(\bx, \bk, t) \D x_{3} \cdots \D x_{Nd} \D \bk, \quad P_x(x_1, t) = \int_{\mathbb{R}} P_{xy}(x_1, x_2, t) \D x_2.
\end{equation}

\subsection{Quantum two-body Coulomb collision and stationary phase approximation}
The Coulomb interaction is of great interest in quantum science \cite{BenamBallicchiaWeinbubSelberherrNedjalkov2021,GrazianiBauerMurillo2014}. Although the two-body Hydrogen wave functions are exactly solvable, the phase space solution from the first principle, especially the non-equilibrium dynamics,  is less than straightforward or complete due to the presence of proton-electron correlation  \cite{bk:CurtrightFairlieZachos2013}.

Consider a two-body system composed of one electron and one proton and treat both quantum mechanically, with  their coordinates in phase space denoted by  $(\bx, \bk)= (\bx_e, \bx_p, \bk_e, \bk_p)$, $\bx_e, \bx_p, \bk_e, \bk_p \in \mathbb{R}^3$. Under the attractive Coulomb potential $V(\bx) = -\gamma/|\bx_e - \bx_p|$, where $\gamma = e^2/(4\pi \epsilon_0)$ with the point charge $e$ and the dielectric constant $\epsilon_0$, $\pdo$ reads
\begin{equation}\label{pdo_scatter_manybody}
\begin{split}
\Theta_V[f](\bx, \bk,t) = &\frac{\gamma}{\mi \hbar c_{3,1}} \int_{\mathbb{R}^3}  \frac{ \me^{\mi \bk^{\prime} \cdot (\bx_e - \bx_p)}}{|\bk^{\prime}|^2} f(\bx_e, \bx_p, \bk_e - \frac{\bk^{\prime}}{2}, \bk_p + \frac{\bk^{\prime}}{2}, t) \D \bk^{\prime} \\
& - \frac{ \gamma}{\mi \hbar c_{3,1}} \int_{\mathbb{R}^3} \frac{\me^{\mi \bk^{\prime} \cdot (\bx_e - \bx_p)}}{|\bk^{\prime}|^2}  f(\bx_e, \bx_p, \bk_e + \frac{\bk^{\prime}}{2}, \bk_p - \frac{\bk^{\prime}}{2}, t) \D \bk^{\prime}
\end{split}
\end{equation}
with $c_{n, \alpha} = \pi^{n/2} 2^\alpha {\Gamma(\frac{\alpha}{2})}/{\Gamma(\frac{n-\alpha}{2})}$. $\pdo$ \eqref{pdo_scatter_manybody} has an intuitive scattering interpretation as it takes the average of the inner-scattering states $(\bx_e, \bx_p, \bk_e \mp \frac{\bk^{\prime}}{2}, \bk_p \pm \frac{\bk^{\prime}}{2})$ weighted by the phase factor $\me^{\mi \bk^{\prime} \cdot (\bx_e - \bx_p)}$, and the Riesz potential $|\bk|^{-2}$ plays the role as the kernel \cite{KosinaNedjalkovSelberherr2003,CarruthersZachariasen1983}. The quantum Coulomb interaction decays as the two-body displacement $|\bx_e - \bx_p|$ increases since the phase factor $\me^{\mi \bk^{\prime} \cdot (\bx_e - \bx_p)}$  becomes more and more oscillating. To  characterize the decay property more precisely, we need to introduce a filter $\lambda_0$ and a decomposition of $\pdo$ \eqref{pdo_scatter_manybody},
\begin{equation}
\Theta_V[f](\bx, \bk, t) = \Lambda^{\le \lambda_0}[f] (\bx, \bk,t) + \Lambda^{> \lambda_0}[f] (\bx, \bk,t),
\end{equation}
where the low-frequency component $\Lambda^{\le \lambda_0}[f](\bx, \bk, t)$ reads that 
\begin{equation*}
\begin{split}
\Lambda^{\le \lambda_0}[f](\bx, \bk,t) &= \int_{B(\frac{\lambda_0}{|\bx_e - \bx_p|})} \frac{ \gamma \sin((\bx_e - \bx_p) \cdot \bk^{\prime})}{\hbar c_{3,1} |\bk^{\prime}|^2} \underbracket{f(\bx_e, \bx_p, \bk_e - \frac{\bk^{\prime}}{2}, \bk_p + \frac{\bk^{\prime}}{2}, t)}_{\textup{e}-\textup{p}+ ~\textup{scattering}} \D \bk^{\prime} \\
& \hspace{3mm}- \int_{B(\frac{\lambda_0}{|\bx_e - \bx_p|})} \frac{\gamma\sin((\bx_e - \bx_p) \cdot \bk^{\prime})}{ \hbar c_{3,1} |\bk^{\prime}|^2}  \underbracket{f(\bx_e, \bx_p, \bk_e + \frac{\bk^{\prime}}{2}, \bk_p - \frac{\bk^{\prime}}{2}, t)}_{\textup{e}+\textup{p}-~\textup{scattering}} \D \bk^{\prime}
\end{split}
\end{equation*}
and $B(r)$ is a ball centered at $0$ with radius $r$. When $\lambda_0 \ge 1$, the high-frequency component allows an asymptotic expansion \cite{ShaoXiong2020}, in the light of SPA, 
\begin{equation}\label{truncation_high_frequency_component}
\Lambda^{> \lambda_0}[f] (\bx, \bk,t) =  \underbracket{\Lambda_+^{>\lambda_0}[f] (\bx, \bk,t)}_{\textup{e}-\textup{p}+ ~\textup{scattering}} + \underbracket{\Lambda_-^{>\lambda_0}[f] (\bx, \bk,t)}_{\textup{e}+\textup{p}- ~\textup{scattering}} + \mathcal{O}(\lambda_0^{-3/2}),
\end{equation}
where two principal terms in the asymptotic expansion are
\begin{equation*}
\begin{split}
 \Lambda_\pm^{>\lambda_0}[f] (\bx, \bk, t) = &\frac{\pm 4\pi \gamma}{\hbar c_{3,1}}\int_{\frac{\lambda_0}{|\bx_e - \bx_p|}}^{+\infty} \frac{\sin( r |\bx_e - \bx_p|)}{ r |\bx_e - \bx_p|}   f(\bx, \bk_e \mp \frac{r\sigma_\ast(\bx)}{2},  \bk_p \pm \frac{r\sigma_\ast(\bx)}{2}, t)  \D r, 
\end{split}
\end{equation*} 
and the critical point is parameterized by $\sigma_\ast(\bx) =(\cos\theta^\ast, \sin\theta^\ast \cos \phi^\ast, \sin\theta^\ast \sin\phi^\ast)$, 
\begin{equation}\label{critical_points}
\begin{split}
&\theta^\ast = \textup{atan2}(\sqrt{(x_{e, 2} - x_{p,2})^2 + (x_{e,3} - x_{p,3})^2}, x_{e,1} - x_{p, 1}), \\
& \phi^\ast = \textup{atan2}(x_{e,3} - x_{p,3}, x_{e,2} - x_{p,2}),
\end{split}
\end{equation}
with $\bx_e = (x_{e, 1}, x_{e, 2}, x_{e, 3})$ and $\bx_p = (x_{p, 1}, x_{p, 2}, x_{p, 3})$. 

Now $\pdo$  \eqref{pdo_scatter_manybody} decays asymptotically as $|\bx_e - \bx_p|$ increases. For the low-frequency component, by a scaling $\bk^{\prime} \to \bk^{\prime}/|\bx_e - \bx_p|$, it yields $\Lambda^{\le \lambda_0}[f](\bx, \bk,t) = \mathcal{O}(|\bx_e - \bx_p|^{-1})$ when $|\bx_e - \bx_p|$ is sufficiently large. At the same time, the major contribution of the high-frequency component is determined by the scattering event along or opposite to the direction ${(\bx_e - \bx_p)}/{|\bx_e - \bx_p|}$, while  the contributions deviated from that line almost cancel out. By the integration by parts, the principal asymptotic terms behave like $ \Lambda_\pm^{>\lambda_0}[f] (\bx, \bk, t) = \mathcal{O}(|\bx_e - \bx_p|^{-1})$ for large $|\bx_e - \bx_p|$.

\subsection{Quantum Coulomb collision with a fixed proton}
In many applications, as the proton moves much slower than the electron ($m_p \gg m_e$), it may pretend the nucleus has infinite mass   and investigate the single-body dynamics of the electron Wigner function $f_e(\bx_e, \bk_e, t)$ \cite{GrazianiBauerMurillo2014}, 
\begin{equation}\label{single_body_Wigner}
\begin{split}
\frac{\partial }{\partial t}f_e(\bm{x}_e, \bm{k}_e, t) &+\frac{\hbar \bm{k}_e}{m_e} \cdot \nabla_{\bm{x}_e} f_e(\bm{x}_e,\bm{k}_e, t)  = \Theta_V[f_e](\bx_e, \bk_e, t).
\end{split}
\end{equation}
Under the interacting potential $V(\bx) =  -\gamma/|\bx_e - \bx_A|$ with fixed $\bx_A$, $\pdo$ reads
\begin{equation}\label{pdo_scatter}
\Theta_V[f_e](\bx_e, \bk_e, t) =\frac{\gamma}{\mi \hbar c_{3,1}}  \int_{\mathbb{R}^3}\frac{ \me^{\mi \bk^{\prime} \cdot (\bx_e - \bx_A)}}{|\bk^{\prime}|^2}(f_e(\bx_e, \bk_e - \frac{\bk^{\prime}}{2}, t) - f_e(\bx_e, \bk_e + \frac{\bk^{\prime}}{2}, t)) \D \bk^{\prime},
\end{equation} 
which behaves like a scattering operator with outer-scattering states $(\bx_e, \bk_e \pm \frac{\bk^{\prime}}{2})$.
Similarly, the low-frequency component of single-body $\pdo$ \eqref{pdo_scatter} reads that
\begin{equation*}
\begin{split}
\Lambda^{\le \lambda_0}[f_e] (\bx_e, \bk_e,t) = & \frac{\gamma}{\hbar c_{3,1}}\int_{B(\frac{\lambda_0}{|\bx_e - \bx_A|})} \frac{\sin((\bx_e - \bx_A) \cdot \bk^{\prime})}{ |\bk^{\prime}|^2} \underbracket{f_e(\bx_e, \bk_e-\frac{\bk^{\prime}}{2}, t)}_{\textup{e}-~\textup{scattering}} \D \bk^{\prime} \\
&- \frac{\gamma}{\hbar c_{3,1}} \int_{B(\frac{\lambda_0}{|\bx_e - \bx_A|})} \frac{\sin((\bx_e - \bx_A) \cdot \bk^{\prime})}{ |\bk^{\prime}|^2}  \underbracket{f_e(\bx_e, \bk_e+\frac{\bk^{\prime}}{2}, t)}_{\textup{e}+~\textup{scattering}} \D \bk^{\prime}, 
\end{split}
\end{equation*}
and SPA to the high-frequency component reads that
\begin{equation*}
\begin{split}
 \Lambda_\pm^{>\lambda_0}[f_e] (\bx_e, \bk_e, t) = &\pm \frac{4\pi \gamma}{\hbar c_{3,1}}\int_{\frac{\lambda_0}{|\bx_e - \bx_A|}}^{+\infty} \frac{\sin( r |\bx_e - \bx_A|)}{ r |\bx_e - \bx_A|}   f(\bx_e, \bk_e \mp \frac{r\sigma_\ast(\bx_e)}{2}, t) \D r,
\end{split}
\end{equation*} 
where the critical point $\sigma_\ast(\bx_e)$ is given in Eq.~\eqref{critical_points} by replacing $\bx_p \to \bx_A$.

\section{Numerical sign problem:  Fundamental obstacle in negative particle method}
\label{sec.wbrw_spa}

The stochastic particle method for the deterministic Wigner equation \eqref{eq.Wigner} is based on its stochastic representation, which interprets the Neumann series expansion as the expectation of stochastic trajectories over the Poisson jumps \cite{KosinaNedjalkovSelberherr2003,Wagner2016,ShaoXiong2019}. The remarkable conceptual advance of the Wigner Monte Carlo is the particle splitting \cite{KosinaNedjalkovSelberherr2003}, making quantum algorithm distinct from the Direct Simulation Monte Carlo \cite{bk:Jacoboni2010}, 
  \begin{equation}
\Theta_V[f](\bx, \bk, t) = \underbracket{\Theta_V^+[f](\bx, \bk, t)}_{\textup{e}-\textup{p}+~\textup{scattering}}- \underbracket{\Theta_V^-[f](\bx, \bk, t)}_{\textup{e}+\textup{p}-~\textup{scattering}},
\end{equation}
so that two particles carrying opposite weights  are generated simultaneously \cite{KosinaNedjalkovSelberherr2003}. Despite its vivid physical intuition and convenience in implementation, the direct splitting of $\pdo$ ignores the cancelations of an oscillatory integral and leads to a rapid growth of random noises. One approach to alleviating such problem is to cancel out the stochastic trajectories
via SPA \cite{ShaoXiong2020}. In the subsequent part, we will discuss WBRW-SPA for the quantum Coulomb interaction, along with an illustrative description of numerical sign problem, and show how SPA can help alleviate such problem.

\subsection{Particle generation and variance reduction}

From the mathematical perspective, the Wigner Monte Carlo utilizes the fact that,  given an inner product $\langle f, g\rangle = \iint_{\mathbb{R}^{Nd}\times \mathbb{R}^{Nd}} f(\bx, \bk) g(\bx, \bk) \D \bx \D \bk$, it has 
\begin{equation*}
\begin{split}
&\langle \varphi(\bx, \bk), f(\bx, \bk, t) \rangle = \underbracket{\me^{-\gamma_0 t} \langle \varphi(\bx(t), \bk), f_0(\bx, \bk) \rangle}_{\textup{frozen state}} \\
&- \int_0^t \underbracket{\gamma_0 \me^{-\gamma_0 (t-t^{\prime})}}_{\textup{particle life}} \Big \langle (\underbracket{{\gamma_0^{-1}} \Theta_V^+[\varphi]}_{\textup{e}-\textup{p}+ \textup{scattering}} - \underbracket{{\gamma_0^{-1}} \Theta_V^-[\varphi]}_{\textup{e}+\textup{p}-\textup{scattering}}- \varphi)(\bx(t-t^{\prime}), \bk), f(\bx, \bk, t^{\prime}) \Big \rangle  \D t^{\prime}
\end{split}
\end{equation*}
for any test function $\varphi(\bx, \bk) \in L_{\textup{loc}}^2(\mathbb{R}^{Nd}\times \mathbb{R}^{Nd})$ and the initial data $f_0 \in L^2(\mathbb{R}^{Nd}\times \mathbb{R}^{Nd})$, where the exponential distribution is introduced by adding $\gamma_0 f$ on both sides of Eq.~\eqref{eq.Wigner} and $(\bx(\tau), \bk) = (\bx + {\hbar \bk \tau}/{\bm{m}}, \bk)$. One can expand $\langle \varphi, f(t^{\prime})\rangle$ and obtain an iterative integral related to a stochastic process. When $f$ has a compact $\bk$-support, the split $\pdo$ $\Theta_V^{\pm}$ can be normalized and there exist a stochastic process $X_t$ and constants $C_1, C_2 > 0$ such that  \cite{ShaoXiong2019,ShaoXiong2020}
\begin{equation}\label{variance}
 \mathbb{E} \langle {X}_t, f_0 \rangle = \langle \varphi, f(t) \rangle, \quad  \mathbb{E} |\langle {X}_t, f_0 \rangle - \langle \varphi, f(t) \rangle|^2 \le C_1 \exp(C_2 t).
\end{equation}
By taking average of realizations of WBRW-SPA, it yields the particle estimator
\begin{equation}\label{def.estimator}
\langle \varphi(\bx, \bk), f(\bx, \bk, t) \rangle \sim \langle \varphi(\bx, \bk), \nu_t \rangle, \quad \nu_t = \frac{1}{N_0} \sum_{i=1}^{P(t)} \delta_{(\bx_i^+, \bk_i^+)} - \frac{1}{N_0}\sum_{i=1}^{M(t)} \delta_{(\bx_i^-, \bk_i^-)},
\end{equation} 
where $\mathcal{S}^+ = \{(\bx_i^+, \bk_i^+)\}_{i=1}^{P(t)}$ and $\mathcal{S}^- = \{(\bx_i^-, \bk_i^-)\}_{i=1}^{M(t)}$  are positive and negative particles, carrying opposite particle weight $\pm 1$, respectively.  The normalizing constant  is $N_0 = P(t) - M(t)$. Namely, the particle method approximates  the Wigner function by an empirical signed measure $\nu_t$ in the weak sense. In particular, given a uniform partition $\mathbb{R}^{2} = \cup_{\mu=1}^{N_x} \cup_{\nu=1}^{N_k} \mathcal{X}_\mu \times \mathcal{K}_\nu$, where $|\mathcal{X}_\mu | |\mathcal{K}_\nu|$ denotes the volume of $\mathcal{X}_\mu \times \mathcal{K}_\nu$, the reduced Wigner function $W_j(x_j, k_j, t)$ along the $j$-th phase-space coordinate can be reconstructed by a piecewise constant histogram (let $\varphi(\bx, \bk) = \mone_{\mathcal{X}_\mu \times \mathcal{K}_\nu}(\bx, \bk)$)
\begin{equation}\label{histogram}
W_j(x_j, k_j, t) \approx \sum_{\mu=1}^{N_x} \sum_{\nu=1}^{N_k} (\sum_{i=1}^{P(t)}\mone_{\mathcal{X}_\mu \times \mathcal{K}_\nu}(x_{i,j}^+, k_{i,j}^+) - {\sum_{i=1}^{M(t)}}\mone_{\mathcal{X}_\mu \times \mathcal{K}_\nu}(x_{i,j}^-, k_{i,j}^-)) \frac{ \mone_{\mathcal{X}_\mu \times \mathcal{K}_\nu}(x_j, k_j)}{N_0 |\mathcal{X}_\mu | |\mathcal{K}_\nu|}. 
\end{equation}

However, Eq.~\eqref{variance} also states that both the stochastic variance and particle number (let $\varphi = 1$) grow exponentially, thereby posing a formidable limitation to the particle method especially for long-time simulations. Such phenomenon is well known as the numerical sign problem for the negative particle method \cite{SchmidtMoehring1993,TroyerWiese2005,YanCaflisch2015}, stemming from the near-cancellation of positive and negative weights in sampling oscillatory functions. Because of a rapid growth of stochastic variances, sample size must be large enough to obtain reliable results within a small relative uncertainty.

To alleviate the sign problem, we have suggested to replace the high-frequency component of $\pdo$ by its principal asymptotic terms in Eq.~\eqref{truncation_high_frequency_component}, yielding another stochastic model with lower variance, termed WBRW-SPA \cite{ShaoXiong2020}. That is, there exists a positive constant $\alpha^\ast < 1$ such that
\begin{equation}\label{SPA_theory}
 \mathbb{E} \langle \mathrm{X}_t, f_0 \rangle = \langle \varphi, f(t) \rangle + \mathcal{O}(\lambda_0^{-3/2}), \quad  \mathbb{E} |\langle \mathrm{X}_t, f_0 \rangle - \langle \varphi, f(t) \rangle|^2 \lesssim \exp(\alpha^\ast C_2 t),
\end{equation}
which implies that SPA suppresses the exponential growth of both particle number and stochastic variances compared with all existing stochastic algorithms, at the cost of introducing a small asymptotic error term $\mathcal{O}(\lambda_0^{-3/2})$. The implementation of WBRW-SPA for the Coulomb potential is illustrated in Algorithm~\ref{WBRW_SPA}, where single-body and two-body interactions are treated in a unified framework due to their strong resemblance. For more details, one can refer to \cite{ShaoXiong2020}.

 \begin{algorithm}
\caption{WBRW-SPA for two-body and single-body Coulomb systems}
\label{WBRW_SPA}
\vspace{1mm} 

{\bf Input parameters}: The time interval $[t_l, t_{l+1}]$, the constant rate $\gamma_0$, the filter $\lambda_0$, $\bk$-domain $\mathcal{K}$ and the upper band $r_{\max} > 4 |\mathcal{K}|$.

{\bf Sampling processes}: Suppose each particle in the branching particle system, carrying an initial weight $w$ either $1$ or $-1$, starts  at time $t_l$ at state $(\bx, \bk) = (\bx_e, \bx_p, \bk_e, \bk_p)$ for two-body system or $(\bx, \bk) = (\bx_e, \bk_e)$ for single-body system, and moves until $t_{l+1} = t_l +\Delta t$ according to the following rules.

1. ({\bf Frozen}) Generate a random $\tau \propto \gamma_0 \me^{-\gamma_0 t}$. For a particle at $(\bx, \bk)$ at instant $t \in [t_{l}, t_{l+1}]$, if $t+ \tau \ge t_{l+1}$, it becomes frozen at $(\bx+\frac{\hbar \bk(t_{l+1} - t)}{\bm{m}}, \bk, t_{l+1})$.

2. ({\bf Death}) If $\tau < \Delta t$, the particle is killed at shifted state $(\bx + \frac{\hbar \bk \tau}{\bm{m}}, \bk, t+\tau)$. 

3. ({\bf Branching}) When the particle is killed at $(\bx +  \frac{\hbar \bk \tau}{\bm{m}}, \bk, t+\tau)$, it produces at most three offsprings at states $(\bx^{(1)}, \bk^{(1)}, t+\tau)$, $(\bx^{(2)}, \bk^{(2)}, t+\tau)$ and $(\bx^{(3)}, \bk^{(3)}, t+\tau)$. The third offspring is produced at state $(\bx^{(3)}, \bk^{(3)}) = (\widetilde{\bx}, \bk)$ with probability $1$, carrying the weight $w$.

\begin{itemize}

\item[$\ast$] Two-body system: $\widetilde{\bx} = (\widetilde{\bx}_e, \widetilde{\bx}_p) = (\bx_e +  \frac{\hbar \bk_e \tau}{m_e}, \bx_p+  \frac{\hbar \bk_p \tau}{m_p})$.

\item[$\circ$] Single-body system: $\widetilde{\bx} = \widetilde{\bx}_e  = \bx_e +  \frac{\hbar \bk_e \tau}{m_e}$ and $ \widetilde{\bx}_p = \bx_A$.

\end{itemize}

4. ({\bf Scattering}) Generate a random number $r$ uniformly in $[0, r_{\max}]$.

(1) If $r < \lambda_0/|\widetilde{\bx}_e -\widetilde{\bx}_p|$, generate random $\theta$ uniformly in $[0, \pi]$ and $\phi$ uniformly in $[0, 2\pi]$, yielding a random vector $\bk^{\prime} = (\cos\theta, \sin\theta \cos\phi, \sin\theta\sin\phi)$. Two offsprings are produced at states $(\bx^{(1)}, \bk^{(1)})$ and $(\bx^{(2)}, \bk^{(2)})$, $\bx^{(1)} = \bx^{(2)} = \widetilde{\bx}$, with probability $\Pr(1)$, $\Pr(2)$, endowed with weights $w_1$ and $w_2$, respectively.
 
  \hspace{1cm} $\textbf{Probability:}$~~$\Pr(1) = \Pr(2) =  \frac{2\pi^2  \gamma}{ \hbar c_{3,1} \gamma_0} r_{\max} |\sin((\widetilde{\bx}_e - \widetilde{\bx}_p) \cdot \bk^{\prime}) \sin \theta |$. \

  \hspace{1cm} $\textbf{Random jump:}$

\begin{itemize}

\item[$\ast$] Two-body system: $\bk^{(1)} = (\bk_e - \frac{\bk^{\prime}}{2}, \bk_p + \frac{\bk^{\prime}}{2}), ~  \bk^{(2)} = (\bk_e + \frac{\bk^{\prime}}{2}, \bk_p - \frac{\bk^{\prime}}{2})$.

\item[$\circ$] Single-body system: $\bk^{(1)} = \bk_e - \frac{\bk^{\prime}}{2}, ~  \bk^{(2)} = \bk_e + \frac{\bk^{\prime}}{2}$.


\end{itemize}

  \hspace{1cm}  $\textbf{Update weight:}~~w_i = (-1)^{i-1} w \cdot \frac{ \sin((\widetilde{\bx}_e - \widetilde{\bx}_p) \cdot \bk^{\prime}) \sin \theta}{|\sin((\widetilde{\bx}_e - \widetilde{\bx}_p) \cdot \bk^{\prime}) \sin \theta |} \cdot \mone_{\{\bk^{(i)} \in \mathcal{K}\}}, ~~ i = 1, 2.$

(2) If $r \ge \lambda_0/|\widetilde{\bx}_e - \widetilde{\bx}_p|$, two offsprings are produced with the probability $\Pr(1)$, $\Pr(2)$ at states $(\bx_1, \bk_1)$ and $(\bx_2, \bk_2)$, $\bx^{(1)} = \bx^{(2)}  = \widetilde{\bx}$, $\sigma_\ast = \sigma_\ast(\widetilde{\bx})$, endowed with updated weights $w_1$ and $w_2$, respectively.

  \hspace{1cm} $\textbf{Probability:}~~\Pr(1) = \Pr(2) =  \frac{4\pi \gamma }{\hbar c_{3,1} \gamma_0} r_{\max}  \frac{ |\sin(r |\widetilde{\bx}_e - \widetilde{\bx}_p|)|}{r  |\widetilde{\bx}_e - \widetilde{\bx}_p|}$.

  \hspace{1cm} $\textbf{Random jump:}$

\begin{itemize}
\item[$\ast$] Two-body system: $\bk^{(1)} = (\bk_e - \frac{r\sigma_\ast}{2}, \bk_p + \frac{r\sigma_\ast}{2}), ~ \bk^{(2)} = (\bk_e + \frac{r\sigma_\ast}{2}, \bk_p - \frac{r\sigma_\ast}{2})$.

\item[$\circ$] Single-body system: $\bk^{(1)} = \bk_e - \frac{r\sigma_\ast}{2}, ~ \bk^{(2)} = \bk_e + \frac{r\sigma_\ast}{2}$.



\end{itemize}

  \hspace{1cm}  $\textbf{Update weight:}~~w_i = (-1)^{i-1} w \cdot \frac{\sin(r |\widetilde{\bx}_e - \widetilde{\bx}_p|)}{|\sin(r |\widetilde{\bx}_e - \widetilde{\bx}_p|)|} \cdot \mone_{\{\bk^{(i)} \in \mathcal{K}\}}, ~~ i = 1, 2$.

5. ({\bf Independence}) The offsprings continue to move independently.

{\bf Termination condition}: All particles in the branching particle system are frozen. 

\end{algorithm}

\subsection{Demonstration of the numerical sign problem}
\label{sec.nsp}

We provide an illustration of numerical sign problem by simulating the single-body 6-D Wigner equation \eqref{single_body_Wigner} (see Example \ref{example1} in Section \ref{sec.num}). In order to measure the empirical variances of the particle simulations, we calculate the $l^2$-error $\mathcal{E}_2[W_1](t)$ (see Eq.~\eqref{def.L2error}) by comparing the histogram \eqref{histogram} of $W_1(x_1, k_1, t)$ and reference solutions produced by a deterministic characteristic-spectral-mixed scheme \cite{XiongZhangShao2022}, as well as the deviation in total Hamiltonian  $\mathcal{E}_H(t)$ 
\begin{equation}
\mathcal{E}_{H}(t) = |H(t) - H(0)|, ~~ H(t) = \iint_{\mathbb{R}^{3N} \times \mathbb{R}^{3N}} \left(\frac{\hbar^2 |\bk|^2}{2\bm{m}} + V(\bx) \right) f(\bx, \bk, t)\D \bx \D \bk.
\end{equation}
Besides, we record the growth ratio of particle number, that is, the total particle number $\mathcal{N}(t) = P(t) + M(t)$ divided by $N_0$. 

\begin{figure}[!h]
\centering
\subfigure[The reduced Wigner function $W_1(x_1, k_1, t)$ at $t = 4$a.u. (left: deterministic, middle: direct particle splitting, right:WBRW-SPA). Here the effective particle number is $N_0 = 10^8$. \label{visualization_mc}]{{\includegraphics[width=0.32\textwidth,height=0.22\textwidth]{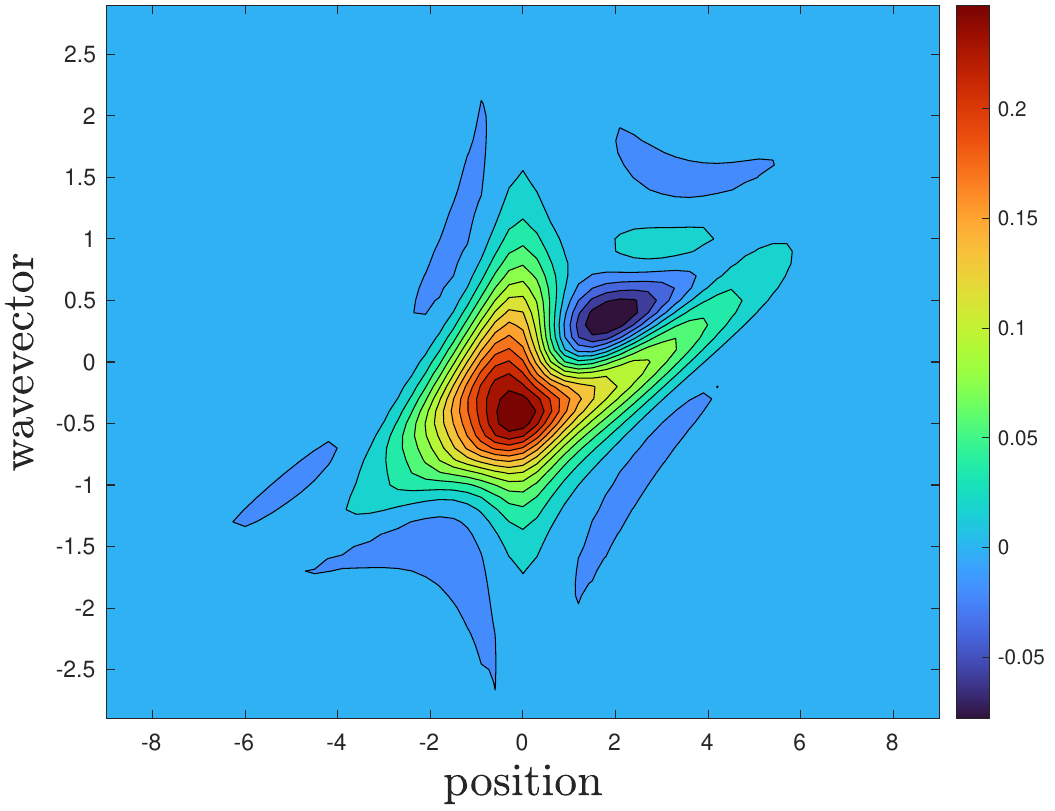}}
{\includegraphics[width=0.32\textwidth,height=0.22\textwidth]{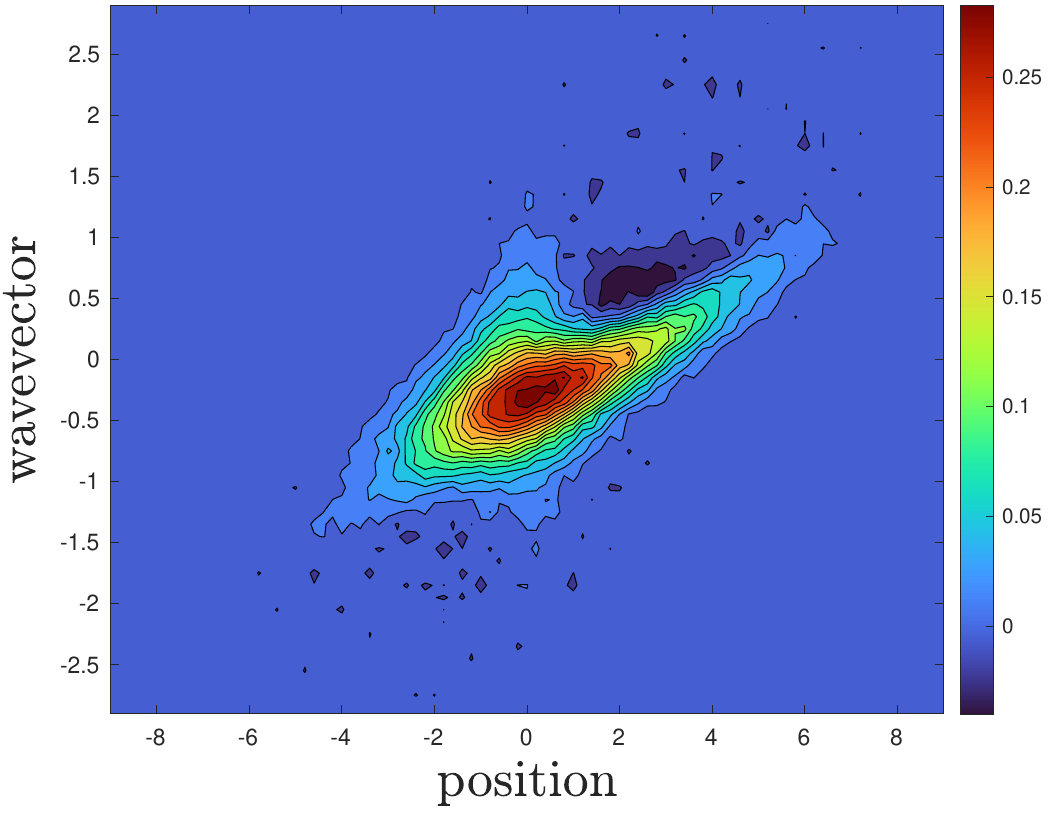}}
{\includegraphics[width=0.32\textwidth,height=0.22\textwidth]{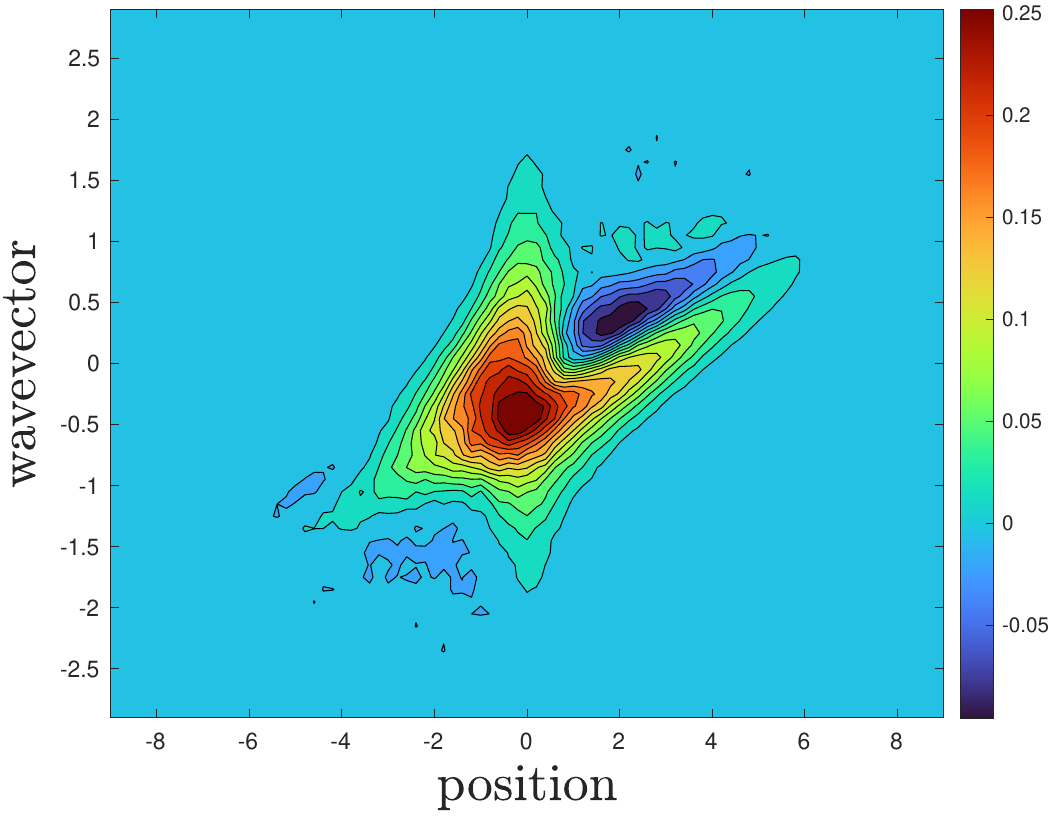}}}
\\
\centering
\subfigure[Errors without SPA (left) or with SPA (middle), and the convergence rate (right).\label{error_mc}]{\includegraphics[width=0.32\textwidth,height=0.22\textwidth]{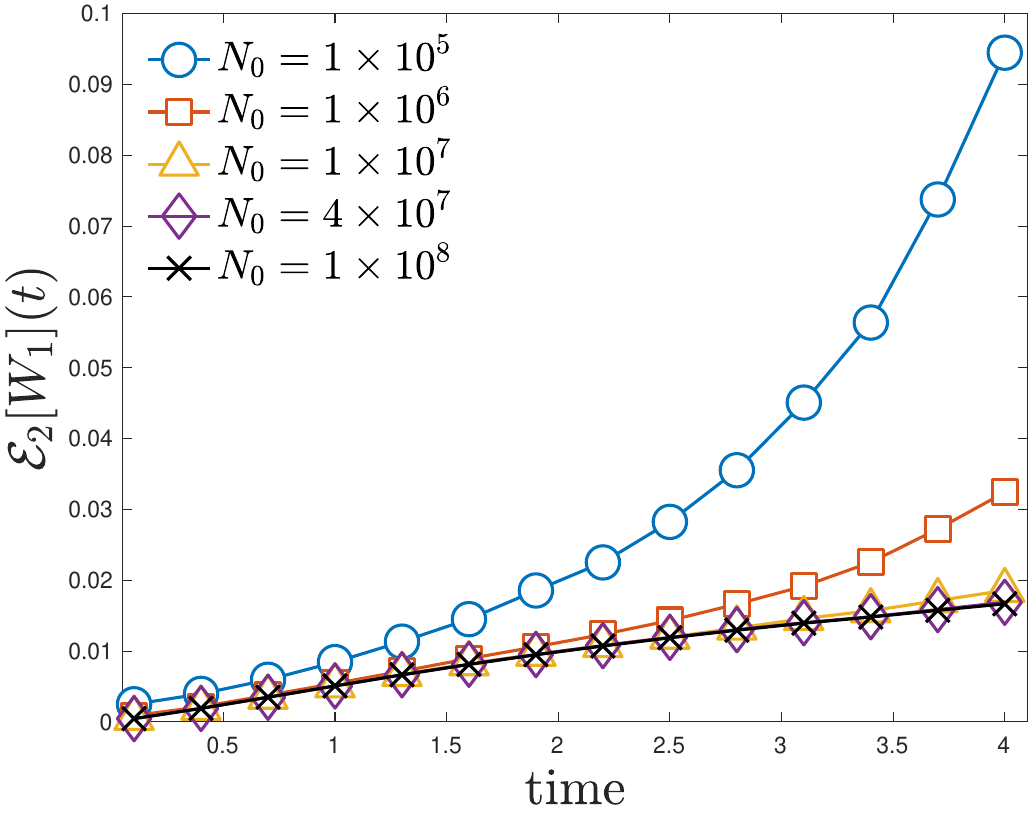}
{\includegraphics[width=0.32\textwidth,height=0.22\textwidth]{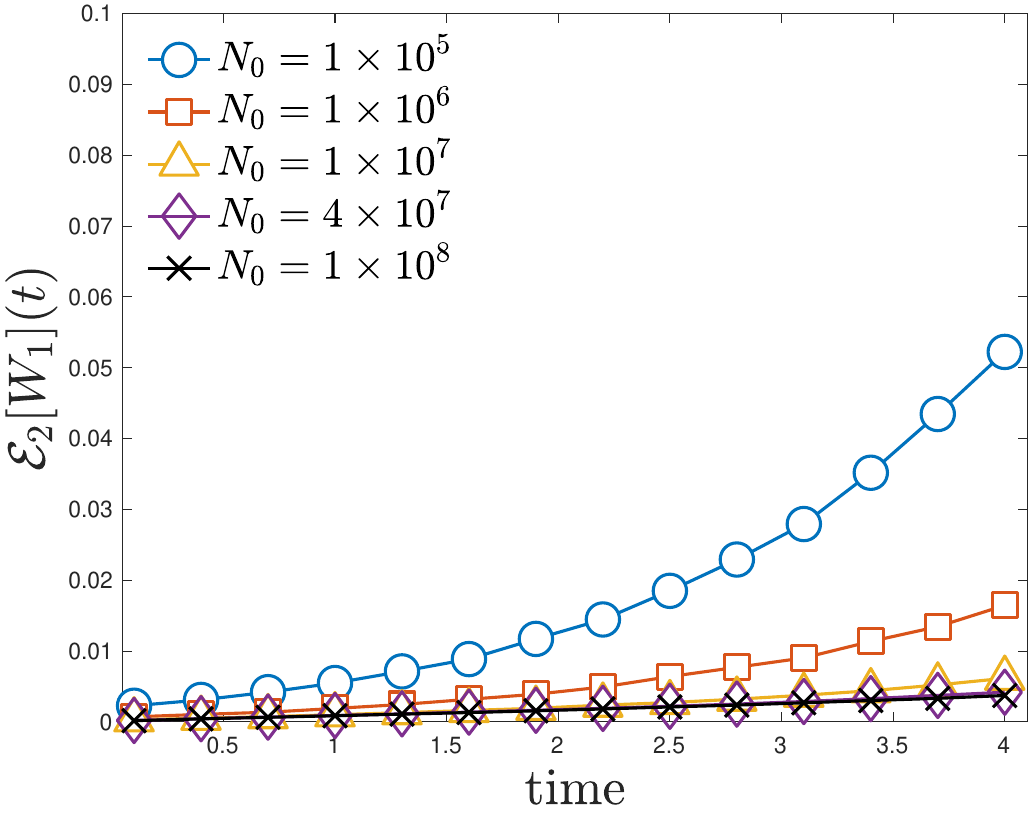}}
{\includegraphics[width=0.32\textwidth,height=0.22\textwidth]{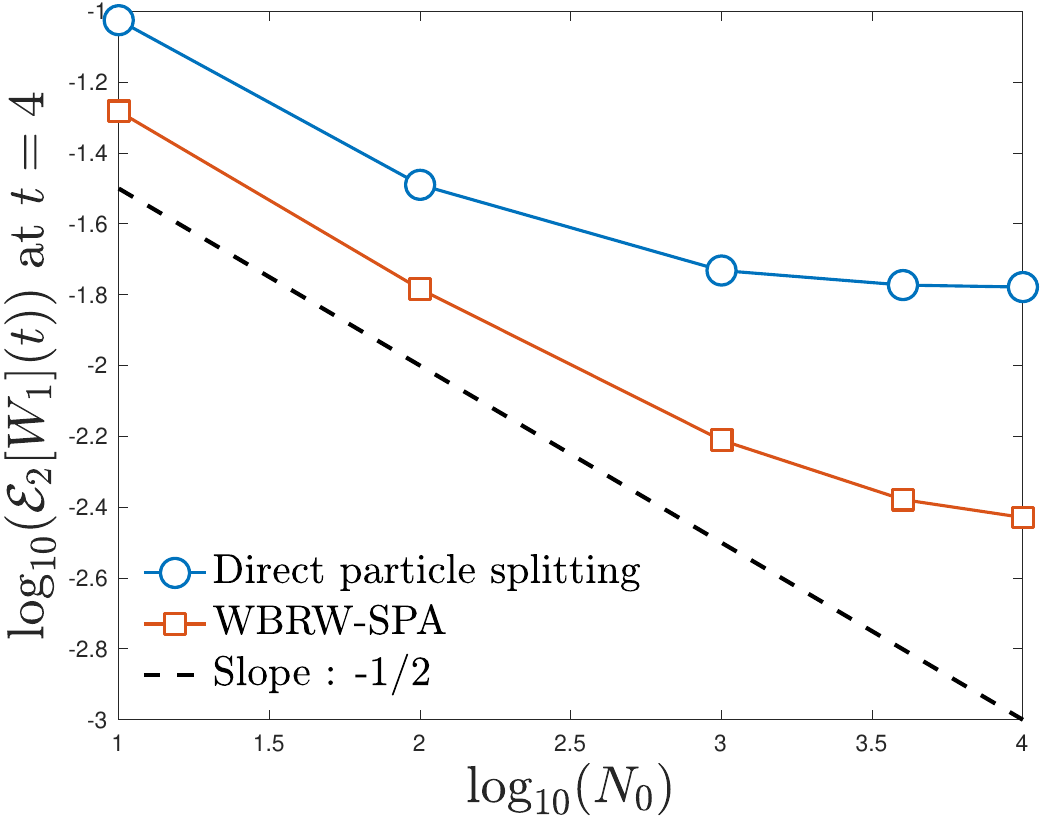}}}
\\
\centering
\subfigure[Errors (left), deviation in energy (middle) and particle growth (right) under $N_0 = 1\times10^7$. \label{error_N1000}]
{\includegraphics[width=0.32\textwidth,height=0.22\textwidth]{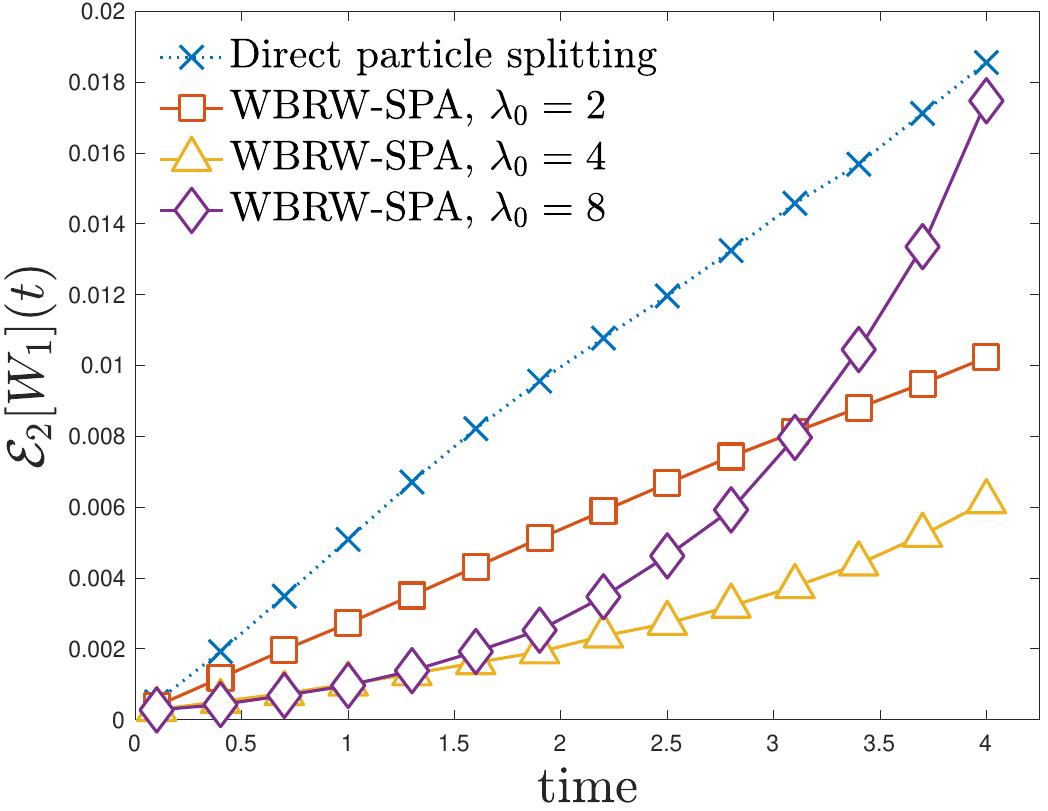}
{\includegraphics[width=0.32\textwidth,height=0.22\textwidth]{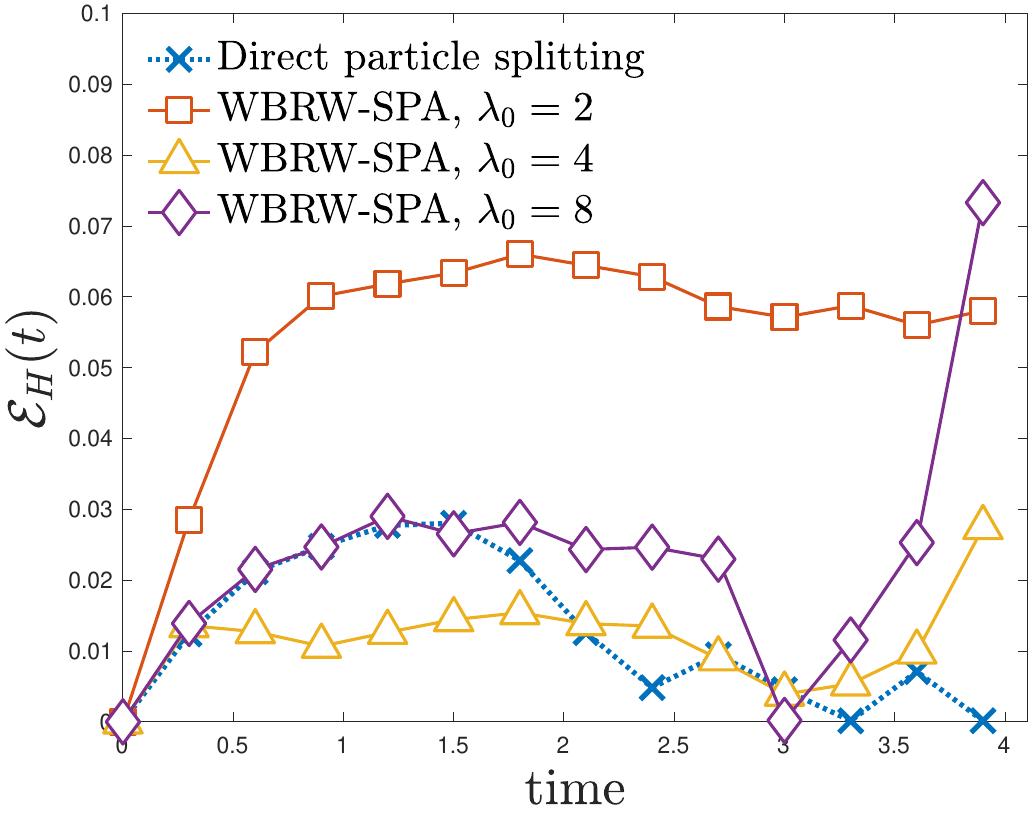}}
{\includegraphics[width=0.32\textwidth,height=0.22\textwidth]{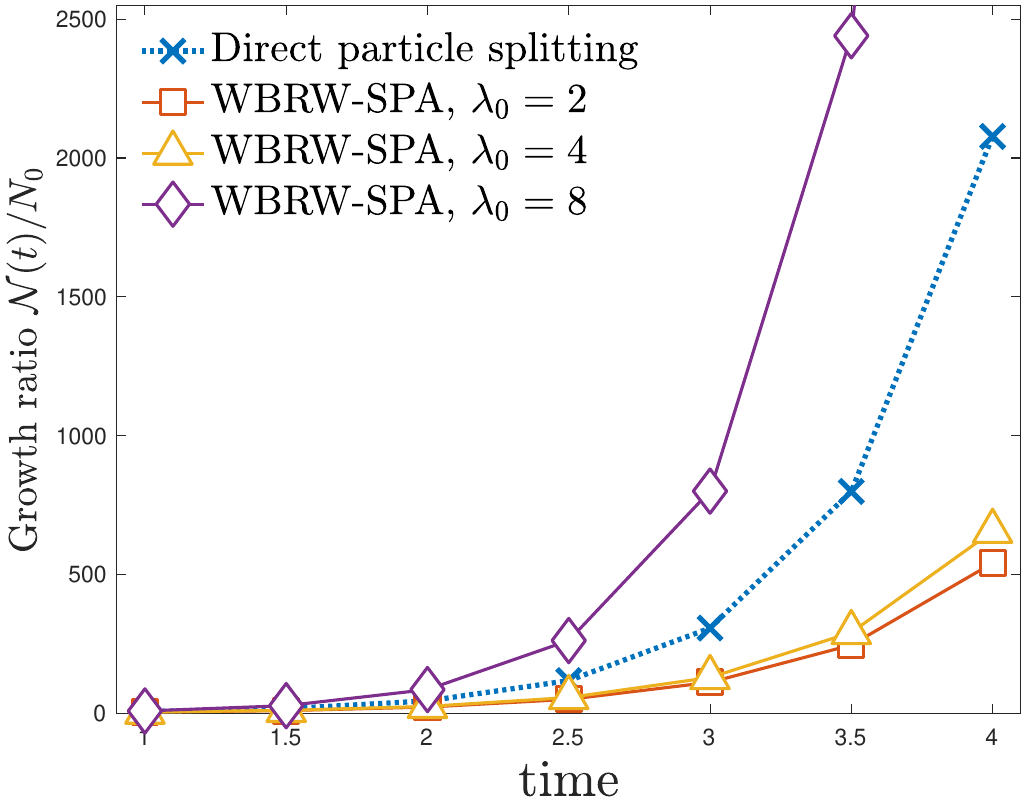}}}
\\
\subfigure[Errors (left), deviation in energy (middle) and particle growth (right) under $N_0 = 4\times10^7$. \label{error_N4000}]
{\includegraphics[width=0.32\textwidth,height=0.22\textwidth]{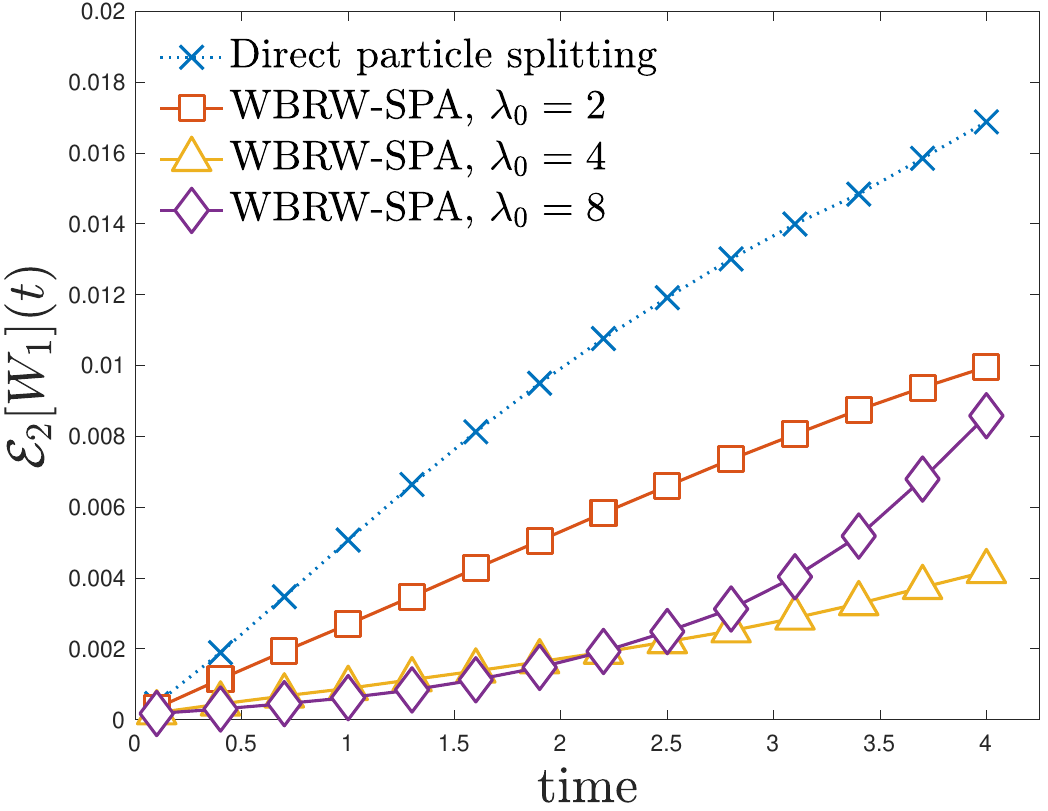}
{\includegraphics[width=0.32\textwidth,height=0.22\textwidth]{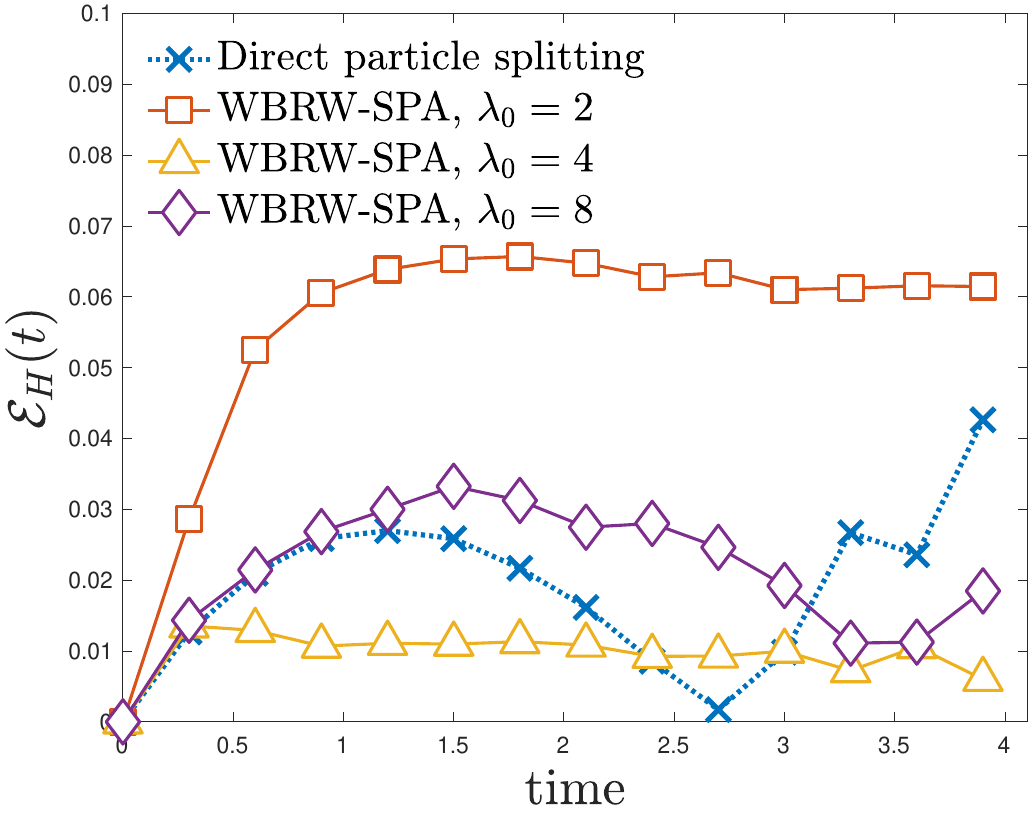}}
{\includegraphics[width=0.32\textwidth,height=0.22\textwidth]{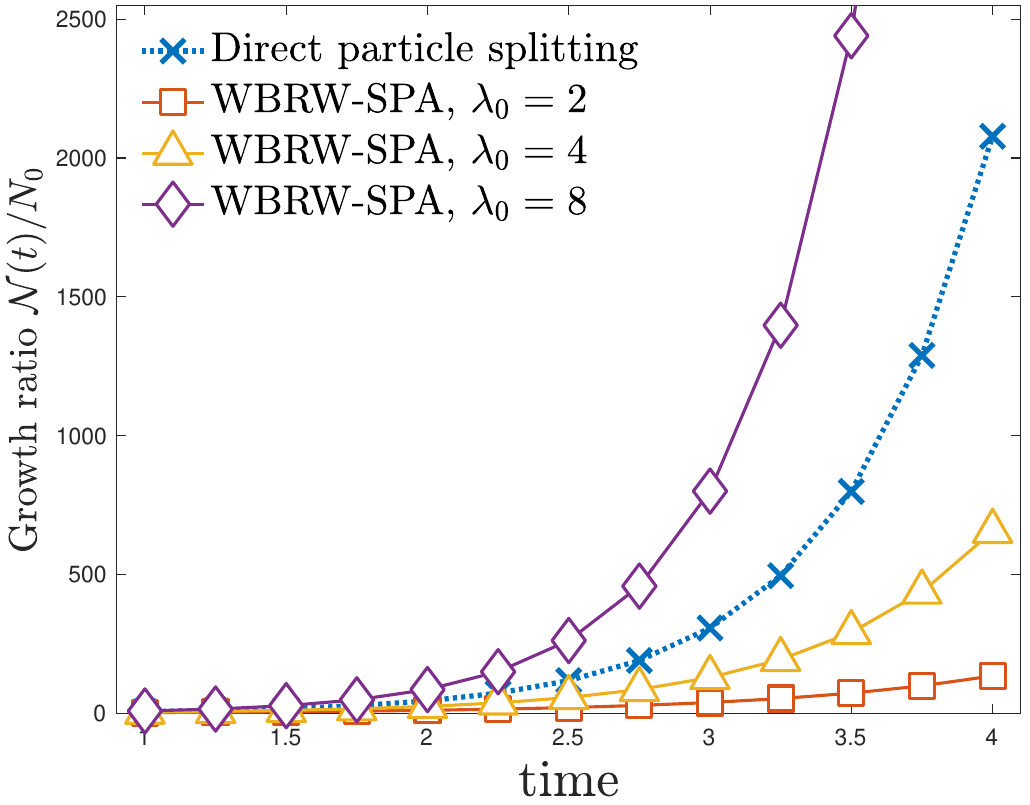}}}
\caption{\small  Numerical sign problem in stochastic Wigner simulations: Stochastic errors grow in time due to the accumulation of negative particle weights. WBRW-SPA is able to suppress the growth of errors as it properly accounts for the decay of $\pdo$ for large $\bk$. $\lambda_0 \le 2$ may underestimate the contribution of low-frequency components and amplify the asymptotic errors, while too large $\lambda_0$ may fail to cancel out the stochastic trajectories efficiently.}\label{comparsion_HJD_and_SPA}
\end{figure}

In Fig.~\ref{error_mc}, we first make a comparison between the Monte Carlo simulations with SPA ($\lambda_0=4$) and without SPA (direct splitting). The exponential growth of stochastic errors is clearly observed regardless of sample size $N_0$, and the convergence rate for the particle splitting deviates from the theoretical order $N_0^{-1/2}$ as  decay property of $\pdo$ is ignored. By contrast, when SPA is adopted, the exponential growth of variances can be suppressed to the large extent and the convergence rate becomes closer to $N_0^{-1/2}$ since it properly accounts for the decay for high-frequency componets. As visualized in Fig.~\ref{visualization_mc}, the reduced Wigner function $W_1(x_1, k_1, t)$  produced by the direct particle splitting is very noisy, while the noises are evidently suppressed by SPA. This indicates the sign problem is alleviated, albeit not eliminated.   

\subsection{Searching for an appropriate filter $\lambda_0$ in SPA}

To further  investigate the influence of $\lambda_0$ on  the accuracy and the growth of particle number, we make a comparison of WBRW-SPA under $\lambda_0 = 2, 4, 8$.

\begin{itemize}

\item[(1)] From Figs.~\ref{error_N1000} and  \ref{error_N4000}, WBRW-SPA under $\lambda_0 = 4$ achieves the smallest errors and deviation  in energy. Too small $\lambda_0$ underestimates the contribution from the low-frequency part,  while too large $\lambda_0$ fails to control the variances efficiently. This also coincides with our early observations made in \cite{ShaoXiong2020}.

\item[(2)] For the direct particle splitting, the number of particle grows two thousandfoldly up to $t = 4$ and the growth ratio $\mathcal{N}(t)/N_0$ is about $\me^{1.910t}$. This provides another evidence of the sign problem. By contrast, when SPA is adopted with $\lambda_0 = 4$, the growth ratio $\mathcal{N}(t)/N_0 $ reduces to $\me^{1.622t}$.  


\end{itemize}

Therefore, the parameter $\lambda_0$ should be adjusted dynamically to strike a balance between the asymptotic errors in SPA and the stochastic variances. For an energy-conserving system, we suggest to determine $\lambda_0$ by monitoring $\mathcal{E}_H(t)$, as given in  Algorithm \ref{optimal_lambda}. For a general case, it can be done by monitoring the growth of particle from low-frequency and high-frequency components of $\pdo$ \cite{ShaoXiong2020}. But it needs to emphasize that such $\lambda_0$ only achieves a balanced accuracy, while the optimal value might  not be attainable due to a subtle competition among various error sources.

\begin{algorithm}[!h]
\caption{Searching for the fliter $\lambda_0$ in SPA}\label{optimal_lambda}

{\bf Input parameters}: The positive particle set $\mathcal{S}^+ = \{(\bx_i^+, \bk_i^+)\}_{i=1}^{P(t_l)}$ and negative particle set  $\mathcal{S}^- = \{(\bx_i^-, \bk_i^-)\}_{i=1}^{M(t_l)}$ at the instant $t_l$ and a testing time $T$.
\vspace{1mm} 

{\bf Step 1:} Choose an interval $[\lambda_{\min}, \lambda_{\max}]$ and a uniform sequence $\lambda_{\min} = \lambda_1 < \dots < \lambda_i < \cdots < \lambda_M = \lambda_{\max}$, $\lambda_i = \lambda_{\min} + (i-1) \Delta \lambda$ with a fixed $\Delta \lambda$.
\vspace{1mm}

{\bf Step 2:} For each $\lambda_i$ ($1 \le i \le M$), start from $\mathcal{S}^+ \cup \mathcal{S}^-$ and simulate WBRW-SPA under $\lambda_i$ up to $t_l+T$, then record the maximal deviation of energy $\max_{t_l\le t \le t_l+T} \mathcal{E}_H(t)$.

{\bf Step 3:} Choose $\lambda_i$ to minimize the deviation $\max_{t_l\le t \le t_l+T} \mathcal{E}_H(t)$.
\end{algorithm}

\section{Particle annihilation: A remedy for the numerical sign problem}
\label{sec.spade}

Unfortunately, the numerical sign problem cannot be completely surmounted by SPA as demonstrated by Eq.~\eqref{SPA_theory} and Fig.~\ref{comparsion_HJD_and_SPA},  because it is rooted in the Monte Carlo evaluation of  the low-frequency component of $\pdo$. Moreover, it is more probably to be aggravated as the dimensionality (system size) increases due to the enrichment of fine structures, such as alternating local maxima and minima in phase space.

To further alleviate the sign problem, particle annihilation (PA) turns out to be indispensable. For a given empirical signed measure of the form \eqref{def.estimator} (the dependence on time is omitted), PA intends to remove $N_A$ positive particles from $\mathcal{S}^+$ and $N_A$ negative ones from $\mathcal{S}^-$,  and obtains another empirical signed measure $\widetilde \nu$ ,
\begin{equation}\label{empirical_signed_measure_after_annihilation}
\widetilde \nu = \frac{1}{N_0} \sum_{i=1}^{P-N_A} \delta_{(\tilde{\bx}_i^+, \tilde{\bk}_i^+)} - \frac{1}{N_0}\sum_{i=1}^{M-N_A} \delta_{(\tilde{\bx}_i^-, \tilde{\bk}_i^-)},
\end{equation}
where $(\tilde{\bx}_i^\pm, \tilde{\bk}_i^\pm)$ can be either chosen as a subset of $\mathcal{S}^\pm$, or be generated by certain operations of particles in $\mathcal{S}^\pm$ like the bootstrap filtering. The target of PA is to control the error function $\mathcal{E}(\varphi) = | \langle \varphi, \nu_t \rangle - \langle \varphi, \widetilde \nu \rangle |$ for suitable test functions $\varphi$. It is expected to annihilate two kinds of particles carrying opposite weights and to cancel out their contributions within a reasonable numerical accuracy. For this reason, PA is also named particle cancellation or particle resampling \cite{YanCaflisch2015}.

The prototype PAUM \cite{KosinaNedjalkovSelberherr2003,KosinaSverdlovGrasser2006,bk:NedjalkovQuerliozDollfusKosina2011} borrows the idea from the histogram statistics \cite{bk:Silverman2018}, that is, using a uniform grid to divide particles into several clusters and annihilate the particles in the same bin. But the cancellation might be very inefficient in high dimension since  a large amount of particles are left uncanceled \cite{YanCaflisch2015}. SPADE, the proposed adaptive PA method, intends to get rid of the severe limitation of regular mesh. The intuition, design and implementation of SPADE are detailed below, and a thorough comparison between PAUM and SPADE is left in Section \ref{sec.num}.

\subsection{PAUM: Particle annihilation via uniform mesh}  A more general setting is considered here. The  domain is a rectangular bin $\Omega = \prod_{i=1}^{s} [x_{\min}^{(i)}, x_{\max}^{(i)}] \times   \prod_{i=1}^{s} [k_{\min}^{(i)}, k_{\max}^{(i)}]$  with dimensionality $\tD = 2s$. Our target is to annihilate the positive particles $\mathcal{S}^+$ and negative particles $\mathcal{S}^-$ located in $\Omega$.

A straightforward idea is to utilize a uniform mesh for dividing $\Omega$: $\Omega  = \bigcup_{k=1}^K \mathrm{Q}_k$, where $\mathrm{Q}_k = \mathcal{X}_{i_1} \times \dots \times \mathcal{X}_{i_s} \times \mathcal{K}_{j_1} \times \dots \times \mathcal{K}_{j_s}$ is the tensor product of disjoint rectangular bins, $\mathcal{X}_{i_l} = [x_{\min}^{(l)} + (i_l-1)\Delta x_l, x_{\min}^{(l)} + i_l \Delta x_l]$,  $\mathcal{K}_{j_l} = [k_{\min}^{(l)} + (j_l-1)\Delta k_l, k_{\min}^{(l)} + j_l \Delta k_l]$, $ l =1, \dots, s$, 
and $\Delta x_l$ and $\Delta k_l$ are spatial and momental spacings in the $l$-th dimension, respectively. PAUM suggests to use a piecewise constant function $p(\bx, \bk)$ to approximate the Wigner function 
\begin{equation}
f(\bx, \bk) \approx p(\bx, \bk) = \sum_{k=1}^K \frac{P_k - M_k}{N_0} \cdot \frac{\mone_{\mathrm{Q}_k}(\bx, \bk)}{\textup{vol}(\mathrm{Q}_k)},
\end{equation}
where $K$ is total partition level,  $P_k$ and $M_k$ count the positive and negative particles in $\mathrm{Q}_k$, respectively, $\mone_{\mathrm{Q}_k}$ denotes the indicator function and $\textup{vol}(\mathrm{Q}_k)$ is the Lesbegue measure of $\mathrm{Q}_k$. In this way, the particles carrying opposite signs are eliminated directly and the error of PAUM scales as $\mathcal{O}(\prod_{l=1}^{s}\Delta x_l \Delta k_l)$ \cite{KosinaSverdlovGrasser2006,YanCaflisch2015}.

Despite its simplicity and easy implementation, PAUM might be very inefficient when either $K \ll N_0$ or $K \gg N_0$. For the former, the smoothing effect induced by the averaging is dominated \cite{Raviart1985}. For the latter, particles are divided into too many clusters and only a few are canceled \cite{YanCaflisch2015}. The sharp deterioration seems to be inevitable when the dimensionality increases, which is well known as the overfitting or over-partitioning problem in statistics when the partition level largely exceeds the effective sample size $N_0$ \cite{bk:Silverman2018}. In practice, we find that PAUM is still useful in 6-D simulations, albeit it requires a very strict balance between partition level $K$ and $N_0$. In Section \ref{sec.num}, we endeavor to test PAUM with a $61^3 \times 60^3$  uniform grid with $K = 4.9\times 10^{10}$ and find that particles might not be efficiently annihilated even under $N_0 = 1\times 10^{8}$. In addition, the storage of a huge uniform grid is very expensive and has to be distributed evenly in multiple nodes. This causes some difficulties in striking a load balance, especially when there is no symmetry inside underlying physical problems.

\subsection{SPADE: An adaptive particle annihilation}

Essentially, the uniform partition in PAUM divides particles into $K$ clusters, namely, $S^+ = \bigcup_{k=1}^K S_k^+$, $S^- = \bigcup_{k=1}^K S_k^-$, and the particles in the same bin are assumed to contribute to evaluating an integral \eqref{def.estimator} almost equally.   The motivation of SPADE to alleviate CoD is to replace the uniform mesh with an adaptive one, partially borrowing the idea from the discrepancy-based density estimation \cite{LiYangWong2016}. 
 
 First, it seeks an adaptive partition of $\Omega$ via the sequential binary splitting and controls the number-theoretic discrepancies of points in each group, so that  the particles located in the same bin contribute to the estimator \eqref{def.estimator} almost uniformly.  Once an adaptive partition $\Omega = \bigcup_{k=1}^K \mathrm{Q}_k$ is obtained, it also divides positive and negative particles into $K$ groups. Second, it seeks a random matching between the positive and negative particles in the same group independently, and the annihilation can be realized by removing the matched pairs.

\begin{figure}[h]
\centering
\includegraphics[width=0.9\textwidth,height = 0.65\textwidth]{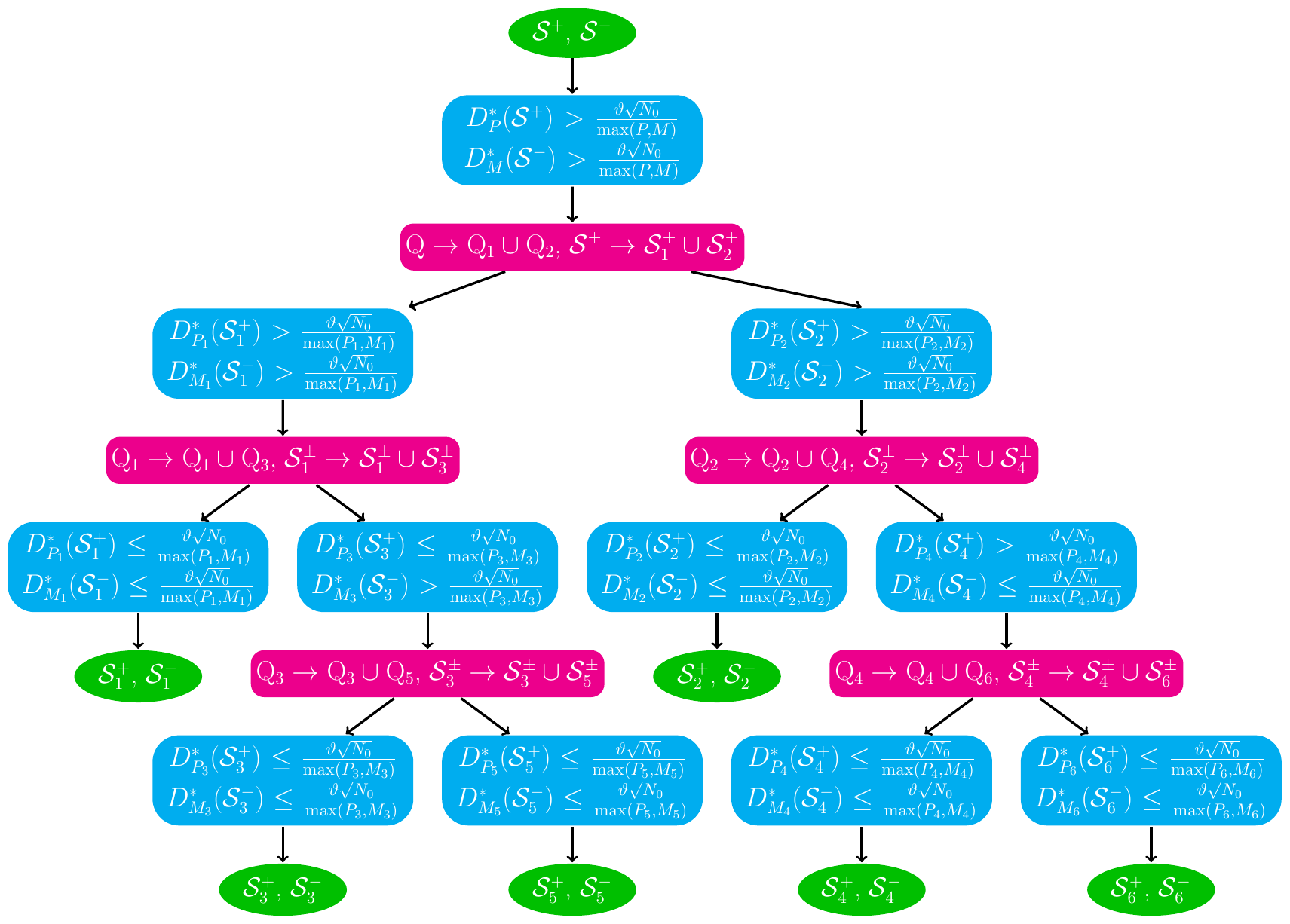}
\caption{\small The adaptive clustering via binary splitting and a decision tree. The binary partition of $\mathrm{Q}$ is $\mathcal{P}_6 = (\mathrm{Q}_1, \ldots, \mathrm{Q}_6)$, and each bin ceases to be split when both positive and negative particles satisfy the discrepancy bounds \eqref{def.discrepancy_bound} simultaneously; otherwise it will be split further into two parts. In the mean time, the particles are divided into 6 groups. }
\label{decision_tree}
\end{figure}

SPADE can be implemented via a recursive binary splitting (see Algorithm \ref{algorithm_SPADE}).  A binary partition $\mathcal{P}$ on domain $\mathrm{Q} = \Omega$ is the collection of sub-rectangles whose union is $\mathrm{Q}$. Starting with $\mathcal{P}_1 = \{ \mathrm{Q} \}$ at level $1$, for $\mathcal{P}_K = \{ \mathrm{Q}_1, \dots, \mathrm{Q}_K\}$ at level $K$, $\mathcal{P}_{K+1}$ is produced by dividing one of the regions in $\mathcal{P}_K$ along one coordinate and merging both sub-rectangles with the rest of regions in $\mathcal{P}_K$. This procedure corresponds to a decision tree as presented in Fig.~\ref{decision_tree}. Two key points shall be specified for a binary partition. The one is whether to split and the other is where to split. 
\begin{algorithm}[!h]
\caption{SPADE: An adaptive particle annihilation}\label{algorithm_SPADE}

{\bf Input parameters}: The positive particles $\mathcal{S}^+$, the negative particles $\mathcal{S}^-$, the domain $\Omega$, the normalizing constant $N_0$ and the parameter $\vartheta$.
\vspace{1mm} 

{\bf Clustering}: Start from  $K=1$, $\mathrm{Q}_1 = \Omega$, $\mathcal{P} = \{\mathrm{Q}_1\}$, $\mathcal{P}^{\prime}  =  \varnothing$, , $\mathcal{S}^\pm_1 = \mathcal{S}^\pm$. 
\begin{algorithmic}
  \While{ $\mathcal{P} \ne  \mathcal{P}^{\prime}$ } 
   \State $\mathcal{P}^{\prime} = \mathcal{P}$
   \ForAll{$\mathrm{Q}_k$ in $\mathcal{P}^{\prime} $} 
   \State $\mathcal{P} \leftarrow \mathcal{P} \setminus \mathrm{Q}_k$, $\mathcal{S}^+ \leftarrow \mathcal{S}^+ \setminus \mathcal{S}^+_k$, $\mathcal{S}^- \leftarrow \mathcal{S}^- \setminus \mathcal{S}^-_k$ 
   \State Calculate the star discrepancies $D_{P_k}^{\ast}(\mathcal{S}_k^+)$ and  $D_{M_k}^{\ast}(\mathcal{S}_k^-)$
   \If {$D_{P_k}^{\ast}(\mathcal{S}_k^+) >  \frac{\vartheta \sqrt{N_0}}{\max(P_k, M_k)} ~ \textup{or} ~ D_{M_k}^{\ast}(\mathcal{S}_k^-) >  \frac{\vartheta \sqrt{N_0}}{\max(P_k, M_k)}$}  
   \State $K \leftarrow K + 1$
   \State Choose a node $c_j^{(k)}$ to maximize the gap function \eqref{difference_gap} 
   \State Divide $\mathrm{Q}_k$ into $ \mathrm{Q}_k^{(1)}\bigcup \mathrm{Q}_k^{(2)}$ as given in Eq.~\eqref{split_subrectangle}:   $\mathrm{Q}_k^{(1)} \rightarrow \mathrm{Q}_k$, $\mathrm{Q}_k^{(2)} \rightarrow \mathrm{Q}_{K}$
   \State Divide the pointsets: $\mathcal{S}_k^{\pm} \to \mathcal{S}_k^{\pm} \cup \mathcal{S}_K^{\pm}$ 
   \State Update the partition and particles:  $\mathcal{P} \leftarrow \mathcal{P} \cup \mathrm{Q}_k \cup \mathrm{Q}_K$, $\mathcal{S}^\pm \leftarrow \mathcal{S}^\pm \cup \mathcal{S}_k^{\pm} \cup \mathcal{S}_K^{\pm}$
   \Else
   \State $\mathcal{P} \leftarrow \mathcal{P} \cup \mathrm{Q}_k$, $\mathcal{S}^\pm \leftarrow \mathcal{S}^\pm \cup \mathcal{S}^\pm_k$
   \EndIf
   \EndFor
   \EndWhile
   \State \Return{$\mathrm{Q} = \bigcup_{k=1}^K \mathrm{Q}_k$, $S^+ = \bigcup_{k=1}^K S_k^+$, $S^- = \bigcup_{k=1}^K S_k^-$}  

\end{algorithmic}

\vspace{1mm} 
{\bf Matching}: For the $k$-th group, when $P_k \ge M_k$, seeking a random matching from $\mathcal{S}_k^-$ to $\mathcal{S}_k^+$. Otherwise, seeking a random matching from $\mathcal{S}_k^+$ to $\mathcal{S}_k^-$. The random matchings in different bins are mutually independent. 

\vspace{2mm}
{\bf Annihilation}: Remove the paired particles in each group. 

\end{algorithm}

{\bf Whether to split}: For the stopping criterion, we try to control the irregularity of points distribution in each bin, measured by the star discrepancy.  The definition of the star discrepancy for a sequence $(\bx_1, \dots, \bx_P) \subset [0, 1]^{2s \times P}$ reads that  
\begin{equation}
D_P^\ast(\bx_1, \dots, \bx_P) = \sup_{\bu \in [0,1]^{2s}}  \Big | \frac{1}{P}\sum_{i=1}^P \mone_{[\bm{0}, \bu)}(\bx_i) - \textup{vol}([\bm{0}, \bu)) \Big |.
\end{equation}
For two sequences $\mathcal{S}_k^+$ and $\mathcal{S}_k^-$ in $\mathrm{Q}_k = [a^{(k)}_{1}, b^{(k)}_{1}] \times \dots \times [a^{(k)}_{2s}, b^{(k)}_{2s}]$, the discrepancy can be defined by a linear scaling $\iota_k: \mathrm{Q}_k \to [0, 1]^{2s}$,
\begin{equation}\label{scaling}
\begin{split}
D_{P_k}^\ast(\mathcal{S}_k^+) &= D_{P_k}^\ast(\iota_{k}(\bx_{1}^{k, +}, \bk_{1}^{k, +}), \dots, \iota_{k}(\bx_{P_k}^{k, +}, \bk_{P_k}^{k, +})), \\
D_{M_k}^\ast(\mathcal{S}_k^-) &= D_{M_k}^\ast(\iota_{k}(\bx_{1}^{k, -}, \bk_{1}^{k, -}), \dots, \iota_{k}(\bx_{M_k}^{k, -}, \bk_{M_k}^{k, -})),
\end{split}
\end{equation}
with $S_k^+ = \{(\bx_{1}^{k, +}, \bk_{1}^{k, +}), \dots, (\bx_{P_k}^{k, +}, \bk_{P_k}^{k, +})\}$ and $S_k^- = \{(\bx_{1}^{k, -}, \bk_{1}^{k, -}), \dots, (\bx_{M_k}^{k, -}, \bk_{M_k}^{k, -})\}$ being the positive and negative particles located in $\mathrm{Q}_k$, respectively, and
\begin{equation}
\iota_{k}(\bx, \bk) = \left(\frac{x_1 - a^{(k)}_{1}}{b^{(k)}_{1} - a^{(k)}_{1}}, \dots, \frac{x_{s} - a^{(k)}_{s}}{b^{(k)}_{s} - a^{(k)}_{s}}, \frac{k_1 - a^{(k)}_{s+1}}{b^{(k)}_{s+1} - a^{(k)}_{s+1}}, \dots, \frac{k_{s} - a^{(k)}_{2s}}{b^{(k)}_{2s} - a^{(k)}_{2s}}\right).
\end{equation}
The $k$-th bin $\mathrm{Q}_k$ continues to be split until both discrepancy bounds are satisfied, 
\begin{equation}\label{def.discrepancy_bound}
D_{P_k}^\ast(\mathcal{S}_k^+) \le \frac{\vartheta \sqrt{N_0}}{\max(P_k, M_k)}, \quad D_{M_k}^\ast(\mathcal{S}_k^-)\le \frac{\vartheta \sqrt{N_0}}{\max(P_k, M_k)},
\end{equation}
where the sole parameter $\vartheta$ adjusts the depth of partition.

{\bf Where to split}: For $\mathrm{Q}_k = [a^{(k)}_{1}, b^{(k)}_{1}] \times \dots \times [a^{(k)}_{2s}, b^{(k)}_{2s}]$, we shall select a node $c^{(k)}_j$ in the $j$-th dimension and split $\mathrm{Q}_k$ into $\mathrm{Q}_k^{(1)}$  and $\mathrm{Q}_k^{(2)}$:
\begin{equation}\label{split_subrectangle}
 \mathrm{Q}_k^{(1)} = \prod_{i=1}^{j-1} [a^{(k)}_{i}, b^{(k)}_{i}] \times [a^{(k)}_{j}, c^{(k)}_{j}] \times \prod_{i=j+1}^{2s}  [a^{(k)}_{i}, b^{(k)}_{i}], \quad  \mathrm{Q}_k^{(2)} = \mathrm{Q}_k \setminus \mathrm{Q}_k^{(1)}.
\end{equation}

Denote by $P_k^{(1)}$ and $M_k^{(1)}$ the counts of positive and negative particles in $\mathrm{Q}_k^{(1)}$, respectively. It suggests to choose $c_j^{(k)}$ to optimize
the {\bf difference gap} 
\begin{equation}\label{difference_gap}
\frac{1}{2}\max_{\mathrm{Q}_k^{(1)}}\left(\Big | \frac{P_k^{(1)}}{P_k}  - \frac{M_k^{(1)}}{M_k}  \Big |\right) = \frac{1}{2} \max_{\mathrm{Q}_k^{(2)}}\left(\Big | \frac{P_k^{(2)}}{P_k}  - \frac{M_k^{(2)}}{M_k}  \Big |\right).
\end{equation}
The physical intuition behind is to dig out the nodal surfaces that divide positive and negative particles. When $P_k^{(1)}/P_k$ is much larger than $M_k^{(1)}/M_k$,  positive particles are concentrated in $\mathrm{Q}_k^{(1)}$. At the same time, $P_k^{(2)}/P_k$ shall be smaller than $M_k^{(2)}/M_k$ so that negative particles are concentrated in $\mathrm{Q}_k^{(2)}$. 

A practical way to obtain a (sub)-optimal $c_j^{(k)}$ is to pick up the $j$-th dimension and $m$ equidistant points $c^{(k)}_{j, l} = a^{(k)}_{j} + \frac{l}{m}(b^{(k)}_{j} - a^{(k)}_{j})$ in $[a^{(k)}_{j}, b^{(k)}_{j}]$, $l = 1, \dots, m-1$, to maximize the gap functions. For sufficiently large $m$, it can approximate well to the true gap. According to our tests, too small $m$ might lead to a large deviation of total energy, while $m= 512$ seems to achieve a good compromise in accuracy and cost.





{\bf Star discrepancy:} Calculation of the star discrepancy is a NP-hard problem and essentially difficult to solve exactly. In a sense, SPADE tries to convert the NP-hard sign problem into another NP-hard combinatorial problem. Fortunately, the star discrepancy can be approximated by some heuristic algorithms, such as the improved version of threshold accepting algorithm (TA-improved), which is the state-of-the-art algorithm for moderately large dimension (D $\le$ 60) \cite{GnewuchWahlstromWinzen2012}. According to our tests, running the TA-improved algorithm once can produce a reliable approximation to the star discrepancy of a  6-D sequence under the iteration times $I = 64$, while $I=100$ is suggested for a 12-D sequence (see our arXiv note for more details).

\subsection{Demonstration of SPADE}

An illustrative example is given to demonstrate the intuition behind SPADE. Suppose one needs to draw samples according to a determinental function with $\psi_\pm(x) = \frac{1}{\sqrt{2\pi}}\me^{-\frac{(x\pm1)^2}{2}}$,
\begin{equation}
\psi(x_1, x_2) = 
\text{det}
\begin{pmatrix}
\psi_-(x_1) & \psi_+(x_1) \\
\psi_-(x_2) & \psi_+(x_2)
\end{pmatrix}=\psi_-(x_1)\psi_+(x_2)-\psi_+(x_1)\psi_-(x_2),
\end{equation}
 One can draw samples from two Gaussian functions and take the minus sign as the particle weight (here we set $N_0 = 2000$).  

\begin{figure}[!h]
  \centering
 \subfigure[Continuous function. \label{SPADE_continuous}]{\includegraphics[width=0.32\textwidth,height=0.22\textwidth]{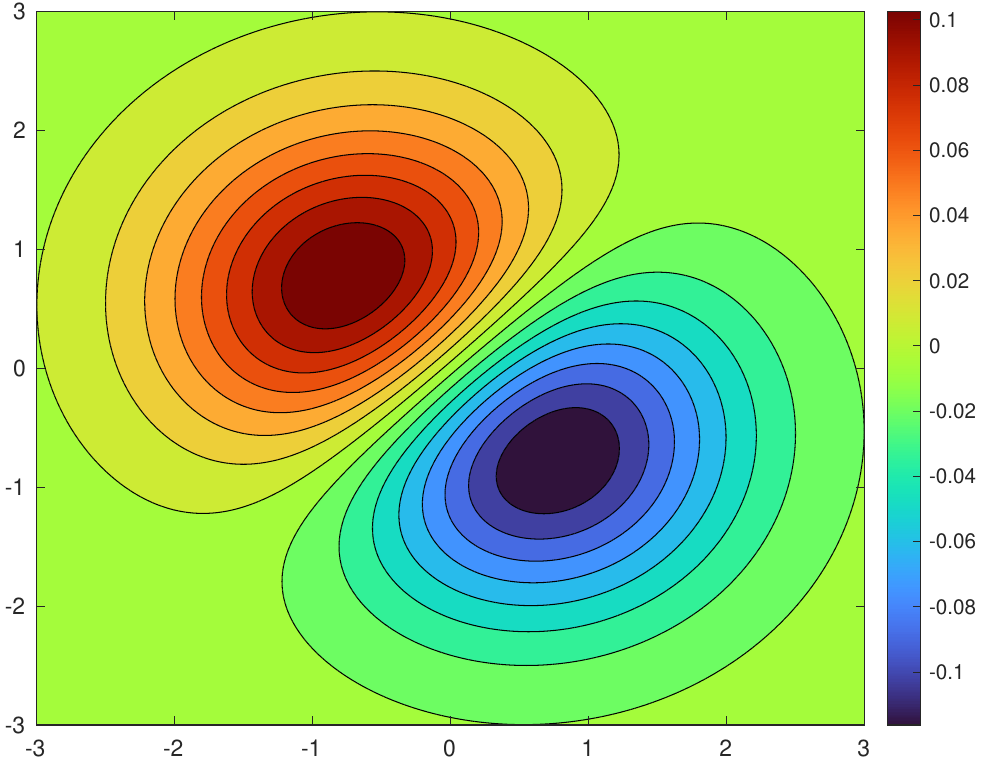}}
 \subfigure[Partition, $\vartheta=0.4, K=86$. \label{adaptive_04}]{\includegraphics[width=0.32\textwidth,height=0.22\textwidth]{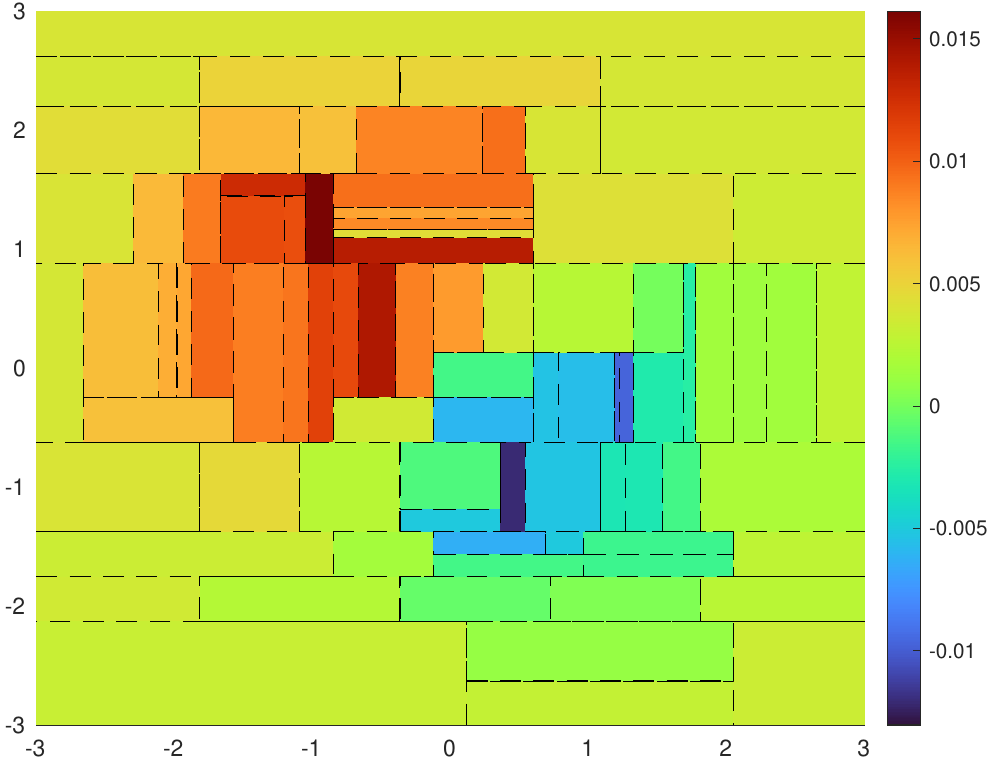}}
 \subfigure[Partition, $\vartheta=0.1, K=399$. \label{adaptive_01}]{\includegraphics[width=0.32\textwidth,height=0.22\textwidth]{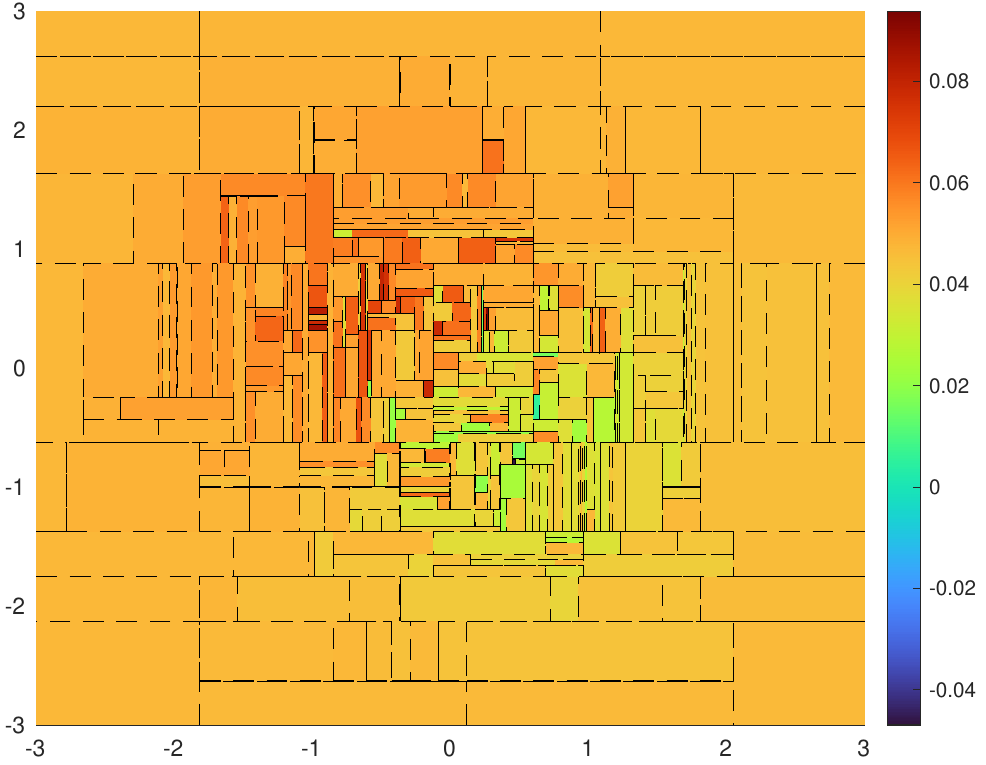}} 
    \\
  \centering
 \subfigure[Before PA, $P = M = 2000$. \label{discrete_particle}]{\includegraphics[width=0.32\textwidth,height=0.22\textwidth]{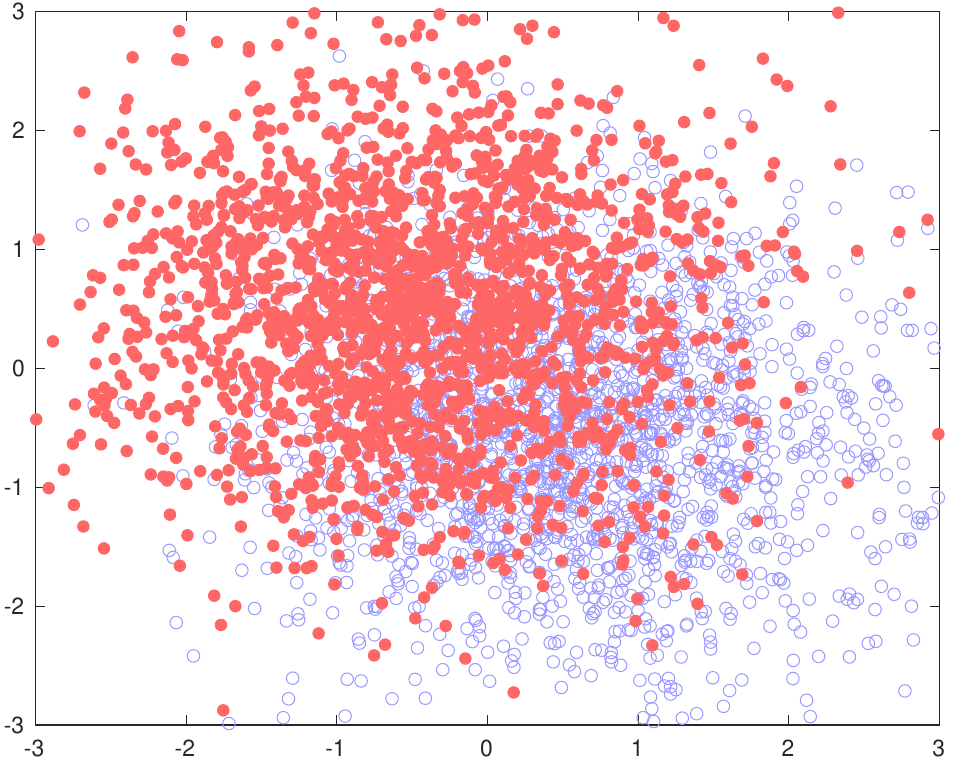}}
 \subfigure[After PA, $P = M = 999$ ($\vartheta=0.4$) or  $1143$ ($\vartheta=0.1$). \label{after_PA_t04}]{\includegraphics[width=0.32\textwidth,height=0.22\textwidth]{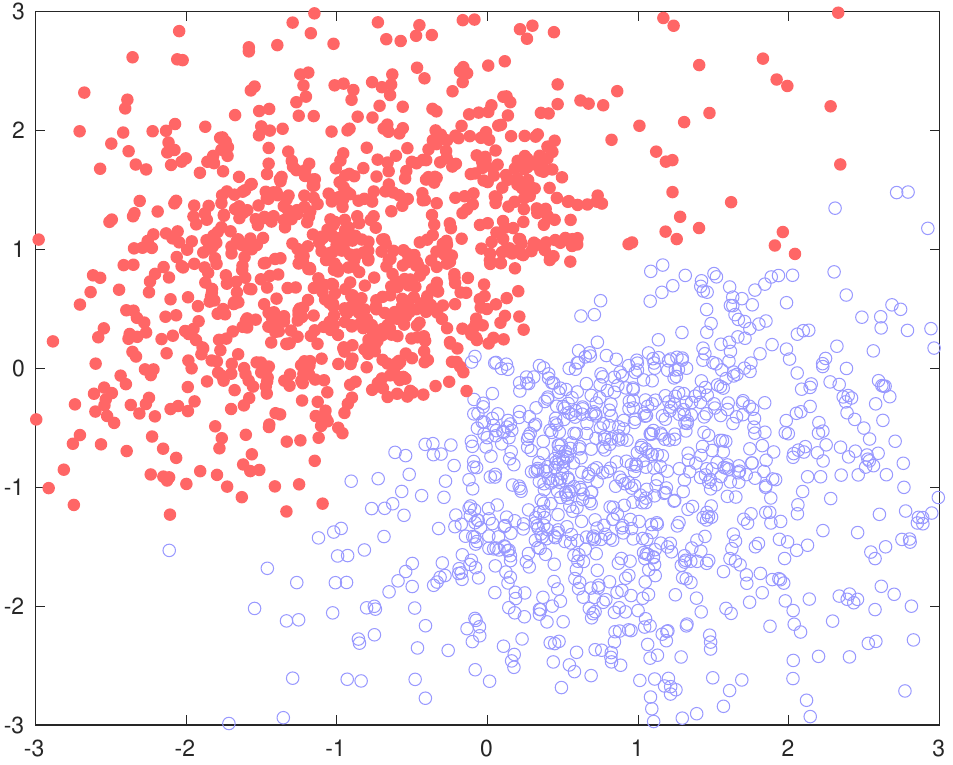}
{\includegraphics[width=0.32\textwidth,height=0.22\textwidth]{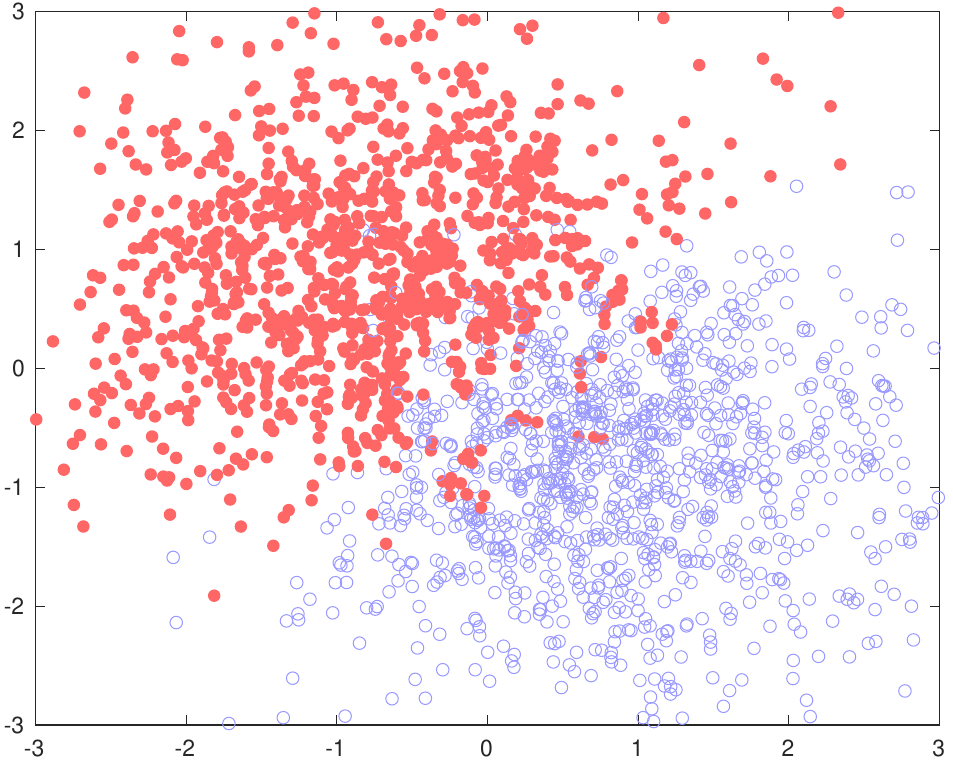}}}
    \caption{\small Adaptive particle annihilation via discrepancy estimation. Initially, there are 2000 positive particles and 2000 negative particles. When $\vartheta$ is smaller, the partition level $K$ increases and consequently fewer particles are removed. \label{comparison_deltaR}}
\end{figure} 

As seen in Fig.~\ref{SPADE_continuous}, the overlap of two Gaussians almost cancels out. Thus for 2000 positive particles (red dot) and 2000 negative particles (blue circle) in Fig.~\ref{discrete_particle}, we also want to cancel out the particles in the central region carrying opposite weights. Figs.~\ref{adaptive_04} and \ref{adaptive_01} plot the adaptive partitions under $\vartheta = 0.4$ and $\vartheta = 0.1$, respectively. Choosing a smaller $\vartheta$  leads to a refinement of partition, and consequently more particles are left uncanceled (see Fig.~\ref{after_PA_t04}). The partition is refined in the region where samples are concentrated and ceases to be split further when points are sparsely distributed, without a priori knowledge of underlying sparse structure. 



\subsection{Bounds for the partition level}


A major drawback of PAUM is that the partition level $K$ scales as $L^\tD$ with $L$ the mesh size in each direction. By contrast, as stated in the following theorem, the partition level $K$ in SPADE must be less than $P+M$ so that CoD can be partially alleviated.
\begin{theorem}\label{thm_1}
For finite $P$ positive particles $\mathcal{S}^+$ and $M$ negative particles $\mathcal{S}^-$, suppose all overlapped particles carrying opposite sign have been removed, and each bin $\mathrm{Q}_k$ in $\mathcal{P}$ ceases to be split when either the discrepancy bounds \eqref{def.discrepancy_bound} or $\min(P_k, M_k) = 0$ is satisfied. Then $\mathcal{P}$ must end with finite level $K$. Moreover, when $\max(P_k, M_k) \le \vartheta \sqrt{N_0}$ holds in each bin, it has 
\begin{equation}\label{K_bound}
 \frac{D(S^+, S^-)}{\vartheta\sqrt{N_0}}\frac{PM}{P+M} \le K \le P+M - \vartheta \sqrt{N_0},
\end{equation}
where $D(S^+, S^-) = \sup_{\bv \in [\bm{a}, \bm{b}]}  | \frac{1}{P} \sum_{i=1}^P \mone_{\{S_i^+ \subseteq [\bm{0}, \bv)\}} - \frac{1}{M} \sum_{i=1}^M \mone_{\{S_i^- \subseteq [\bm{0}, \bv)\}}  |$. 


\end{theorem}
\begin{proof}
Since $D_{P_k}^\ast(\mathcal{S}_k^+) \le 1$ and $D_{M_k}^\ast(\mathcal{S}_k^-) \le 1$, the bin $\mathrm{Q}_k$ ceases to split under either 
$\frac{\vartheta \sqrt{N_0}}{\max(P_k, M_k)} > 1$ or $\min(P_k, M_k) = 0$.
Now we pick up a bin $\mathrm{Q}_k = [\bm{a}_k, \bm{b}_k]$ containing $P_k$ positive particles and $M_k$ negative particles. When it needs to be split, it shall choose a split node to attain the maximum of difference gap. 
We claims that each sub-bin must have at least one particle, namely, $\min(P_k^{(1)}, M_k^{(1)}) \ge 1$ and  $\min(P_k^{(2)}, M_k^{(2)}) \ge 1$. If not, it suffices to take $P_k^{(1)} = P_k, M_k^{(1)} = M_k, P_k^{(2)}  = M_k^{(2)} =0$, then $\frac{P_k^{(1)}}{P_k}  - \frac{M_k^{(1)}}{M_k} = 0$,
and then Eq.~\eqref{difference_gap} implies $\frac{A(\mathcal{S}_k^+, P_k, \Omega)}{P_k} = \frac{A(\mathcal{S}_k^-, M_k, \Omega)}{M_k}$ for all hyper-rectangles $\Omega$ anchored at $\bm{a}_k$,
where $A(\mathcal{S}_k^{+}, P_k, \Omega)$ and $A(\mathcal{S}_k^{-}, M_k, \Omega)$ are numbers  of particles in $\mathcal{S}_k^{+}$ and $\mathcal{S}_k^{-}$ in $\Omega$, respectively. Now pick a positive particle $\bx^+_{r}$ with minimal $|\bx^+ - \bm{a}_k|^2$ and a negative particle $\bx^-_{s}$ with minimal $|\bx^- - \bm{a}_k|^2$. If $|\bx_r^+ - \bm{a}_k|^2 \le |\bx_s^- - \bm{a}_k|^2$ and $\bx^+_{r}$ and $\bx^-_{s}$ are not overlapped, there are at least one coordinate $j$ such that $x^+_{r, j} < x^-_{s, j}$. Choosing $\Omega =  \prod_{k=1}^{j-1}[a_k, b_k] \times [a_j, x_{r,j}^+] \times \prod_{k=j+1}^{2s}[a_k, b_k]$,  it yields $\frac{A(\mathcal{S}_k^+, P_k, \Omega)}{P_k} > 0$, $ \frac{A(\mathcal{S}_k^-, M_k, \Omega)}{M_k} = 0$ and arrives at a contradiction. It is similar for $|\bx_r^+ - \bm{a}_k|^2 > |\bx_s^- - \bm{a}_k|^2$. Hence, every time at least one particle is dropped, then either $\max\{P_k, M_k\} < \vartheta \sqrt{N_0}$ or $\min\{P_k, M_k\} = 0$ shall hold  after finite steps, so that the bin ceases to be split. This arrives at the upper bound $K \le P+M - \vartheta \sqrt{N_0}$.  

For the lower bound, it starts from 
\begin{equation*}
\begin{split}
\tilde{D}_{P_k}^\ast(\mathcal{S}_k^+) &:=  D_{P_k}^\ast(\iota_0(\bx_{1}^{k, +}, \bk_{1}^{k, +}), \dots, \iota_0(\bx_{P_k}^{k, +}, \bk_{P_k}^{k, +})) \le 1 \le \frac{\vartheta \sqrt{N_0}}{\max(P_k, M_k)} \\
\tilde{D}_{M_k}^\ast(\mathcal{S}_k^-) &:= D_{M_k}^\ast(\iota_0(\bx_{1}^{k, -}, \bk_{1}^{k, -}), \dots, \iota_0(\bx_{M_k}^{k, -}, \bk_{M_k}^{k, -}))  \le 1 \le \frac{\vartheta \sqrt{N_0}}{\max(P_k, M_k)},
\end{split}
\end{equation*}
for the scaling $\iota_0(\bx) = (\frac{x_1-a_1}{b_1 - a_1}, \dots, \frac{x_{2s}-a_{2s}}{b_{2s} - a_{2s}})$. By the triangular inequality, 
\begin{equation*}
\begin{split}
& |\frac{A(S^+, P, [\bm{a}, \bv))}{P}  -  \frac{A(S^-, M, [\bm{a}, \bv))}{M}  |= 
 |\frac{A(\iota_0(S^+), P, [\bm{0}, \bu))}{P}  -  \frac{A(\iota_0(S^-), M, [\bm{0}, \bu))}{M}  | \\
&\le \sum_{k=1}^K \frac{P_k}{P}  | \frac{A(\iota_0(S_k^+), P_k, [\bm{0}, \bu))}{P_k}  -  \textup{vol}[\bm{0}, \bu)  | + \frac{M_k}{M}  | \frac{A(\iota_0(S_k^-), M_k, [\bm{0}, \bu))}{M_k}  -  \textup{vol} [\bm{0}, \bu)   |,
\end{split}
\end{equation*}
where $\bu = \iota_0(\bv)$. Now taking supremum of $\bv \in [\bm{a}, \bm{b}]$ on both side, it yields that
\begin{equation}\label{K_bound0}
D(S^+, S^-) \le  \sum_{k=1}^K \frac{P_k}{P} \tilde{D}_{P_k}^\ast(\mathcal{S}_k^+) + \sum_{k=1}^K \frac{M_k}{M} \tilde{D}_{M_k}^\ast(\mathcal{S}_k^-) \le K \vartheta \sqrt{N_0} \left(\frac{1}{P} + \frac{1}{M}\right),
\end{equation}
which gives the lower bound of $K$.
\end{proof}

\begin{remark}
The storage complexity in SPADE scales as $(P+M) \times (3\tD+2)$, including storing all particles by a $(P+M)\times \tD$ matrix and an adaptive partition by a $K\times (2\tD+2)$ matrix (upper and lower bounds of bins and numbers of particles.
\end{remark}

\subsection{Particle annihilation outside domain}

For dynamical problems, particles that move outside the computational domain may result in loss of total mass, which is inconsistent with a conservative quantum system. To fix it, we use an outer pointset $\mathcal{S}_{\textup{out}}$ to store the particles outside the computational domain $\Omega$. Once the particles in $\mathcal{S}_{\textup{out}}$ reach its maximal size, we make random matching among positive and negative particles and directly remove the redundant particles carrying opposite weights in pair. As the positive and negative particles are also generated in pair, the total mass can be rigorously conserved in the simulations (see Fig.~\ref{comparison_SPADE_PAUM} below). However, it is still difficult to conserve the total energy rigorously as the cancelation of positive and negative particles may bring in some small shifts in both kinetic and potential parts, which deserves a further investigation.

\section{Particle simulations of 6-D Wigner-Coulomb dynamics}
\label{sec.num}

From this section, we are about to perform a series of benchmarks on simulating 6-D Wigner-Coulomb dynamics \cite{BenamBallicchiaWeinbubSelberherrNedjalkov2021,GrazianiBauerMurillo2014}, with the atomic units  $\hbar = m_e = \gamma =1$ and  $m_p \approx 1836 m_e$ adopted.  Additional 4-D and 6-D benchmarks are provided in our arXiv note. 
\begin{example}\label{example1}
\textup{
Suppose the initial electron Wigner function $f_e(\bx, \bk, 0)$ is
\begin{equation}
f_e(\bx, \bk, 0) = \frac{1}{\pi^{3}} \me^{-\frac{1}{2}|\bx - \bR|^2} \me^{-2|\bk|^2}, \quad \bR = (1, 0, 0),
\end{equation}
interacting with a proton fixed at $\bx_A = (0, 0, 0)$ under the attractive Coulomb potential.  The motivation comes from the quantum optics as the coherent state is usually described by a Gaussian wavepacket.
}
\end{example}

Our performance evaluation is two-pronged: First, we make a thorough comparison between PAUM and SPADE, i.e., WBRW-SPA-PAUM v.s. WBRW-SPA-SPADE. Second, we investigate how the sample size $N_0$, the parameter $\lambda_0$ in SPA and the parameter $\vartheta$ in SPADE influence the accuracy, energy conservation, growth of particles and the partition level $K$. The latter is towards a comprehensive understanding of SPADE and a guiding principle for systematically improving its accuracy, which is pivotal to the rigorous numerical analysis.

\begin{table}[!h]
  \centering
  \caption{\small Several important notations and their impacts on the performance of particle simulations. \label{rule}}
\label{notation}
 \begin{lrbox}{\tablebox}
  \begin{tabular}{ccc}
\hline\hline
Notation & What the notation stands for & Relation with other quantities\\
\hline
$N_0$ &  Initial effective sample size & $N_0 \uparrow \Rightarrow$ Errors $\downarrow$ and $K \uparrow$ \\
$P(t)$ & Number of positive particles at time $t$ &  $P(t)-M(t) = N_0$\\ 
$M(t)$ & Number of  negative particles at time $t$ & $P(t)-M(t) = N_0$\\ 
$\mathcal{N}(t)$ &  Particle number after PA at time $t$ & $P(t) + M(t) > \mathcal{N}(t)$\\ 
$\mathcal{N}^b(t)$ & Particle number before PA at time $t$ & $P(t) + M(t) = \mathcal{N}^b(t)$\\ 
$K(t)$ & Total partition level at time $t$ & $\frac{PM \vartheta^{-1}}{(P+M)\sqrt{N_0}}\lesssim K \le \mathcal{N}^b - \vartheta \sqrt{N_0}$ \\
$\vartheta$ & Parameter in discrepancy bounds & $\vartheta \downarrow \Rightarrow$  $K \uparrow$ and partition is refined \\
\hline\hline
 \end{tabular}
\end{lrbox}
\scalebox{0.94}{\usebox{\tablebox}}
\end{table} 
Our main findings are summarized as follows (see Table \ref{rule}). 
\begin{itemize}

\item[(1)] The overall accuracy is limited by both the sampling error and the asymptotic error in SPA.

\item[(2)] Too small $\vartheta$ may lead to the over-partitioning problem and make many particles uncanceled \cite{YanCaflisch2015}, because few particles are located in the same bin when the partition level largely exceeds the sample size.  A direct consequence is the oversampling, say, a rapid growth of particle number.

\item[(3)] By increasing the sample size $N_0$, it can diminish the bias induced by SPADE and alleviate the oversampling problem simultaneously.

\item[(4)] According to Theorem \ref{thm_1}, the over-partitioning problem can be avoided if the partition level $K(t)$ approaches to its lower bound, that is, $K(t)$ is expected to be proportional to $\frac{P(t)M(t)}{(P(t)+M(t))\sqrt{N_0}}$ and inversely proportional to $\vartheta$.

\end{itemize}

The reference solutions are produced by a characteristic-spectral-mixed scheme, where the Wigner function defined in the domain $[-10.8, 10.8]^3 \times [-4, 4]^3$ is expanded as the tensor product of $75^3$ cubic spline basis (with spacing $\Delta x = 0.3$)  and $80^3$ Fourier spectral basis (with spacing $\Delta k = 0.1$) and integrated by the Lawson predictor-corrector scheme (with time step $\Delta t = 0.025$a.u.) to ensure its accuracy \cite{XiongZhangShao2022}.

Several groups of stochastic simulations are performed under $N_0 = 4\times 10^6, 1\times 10^7, 4\times 10^7, 1\times10^8$ and $\vartheta$ ranging from $0.004$ to $0.04$. Here we adopt $\gamma_0 = 50$, a finite $\bk$-domain $[-3, 3]^3$ in Algorithm \ref{WBRW_SPA}, and annihilate particles every 1 a.u. The reduced Wigner function $W_1(x_1, k_1, t)$ \eqref{reduced_Wigner_function} and spatial marginal distribution $P_{xy}(x_1, x_2, t)$ \eqref{spatial_distribution} are obtained by the histogram approximation \eqref{histogram} under a uniform grid mesh $[-9,9] \times [-3, 3]$ with $N_x = 61$, $N_k = 60$, with the same spacing as the deterministic solver adopts. It allows both visualization of quantum Coulomb interaction and a quantitative comparison with deterministic counterparts. The performance metrics include the normalized $l^2$-errors $\mathcal{E}_{2}[W_1](t)$ and $\mathcal{E}_{2}[P_{xy}](t)$ to monitor the stochastic variances,
\begin{equation}\label{def.L2error}
\begin{split}
\mathcal{E}_{2}[W_1](t) &= \{\frac{1}{N_x N_k}\sum_{i=1}^{N_x} \sum_{j=1}^{N_k} (W_1^{\textup{ref}}(x_{1}^{(i)}, k_{1}^{(j)},t)- W_1^{\textup{num}}(x_{1}^{(i)}, k_{1}^{(j)},t))^2\}^{1/2}, \\
\mathcal{E}_{2}[P_{xy}](t) &= \{\frac{1}{N_x^2}\sum_{i=1}^{N_x} \sum_{j=1}^{N_x} (P_{xy}^{\textup{ref}}(x_{1}^{(i)}, x_{2}^{(j)},t)- P_{xy}^{\textup{num}}(x_{1}^{(i)}, x_{2}^{(j)}, t))^2\}^{1/2},
\end{split}
\end{equation}
where $W_1^{\textup{ref}}$ and $W_1^{\textup{num}}$ denote the reference and stochastic solution for $W_1$, respectively (similar for $P_{xy}$), as well as the deviation of total energy $\mathcal{E}_{\textup{H}}(t)$
\begin{equation}\label{def.Herr}
\mathcal{E}_{H}(t) = |H(t) - H(0)|, ~~ H(t) = \iint_{\mathbb{R}^{3N} \times \mathbb{R}^{3N}} \left(\frac{\hbar^2 |\bk|^2}{2\bm{m}} + V(\bx) \right) f(\bx, \bk, t)\D \bx \D \bk.
\end{equation}
In addition, the growth ratio of total particle $\mathcal{N}(t)/N_0$ is closely related to the computational complexity, while $P(t) - M(t)$ is always conserved.  We also run WBRW-SPA-PAUM with a $61^3 \times 60^3$ uniform grid mesh ($K\approx 4.9\times 10^{10}$) and $N_0 = 10^8$ to ensure a side-by-side comparison with SPADE. With this, we show that SPADE is able to control the growth of both errors and particle number more efficiently than PAUM, especially when $N_0$ is not very large.

\begin{remark}
For the purpose of benchmark tests, we try to annihilate particles by establishing a new adaptive partition every 1a.u., without utilizing the partition in the previous step. However, it is possible to further reduce the computational cost by using some tricks. For example, one can refine the old partition for adapting to new particles, which borrows the idea from the adaptive mesh refinement technique.
\end{remark}

\subsection{Sampling error}

First, we need to emphasize that the overall accuracy of WBRW-SPA-SPADE is still limited by both the sampling error and the asymptotic error in SPA.  In order to illustrate how it depends on the filter $\lambda_0$, we investigate the $l^2$-errors $\mathcal{E}[W_1](t)$, the deviation of energy and the growth of particle number under $N_0 = 4\times10^7$, $\vartheta = 0.01$ or $0.02$, and $\lambda_0 = 3, 4, 4.65, 5$.
\begin{figure}[!h]
    \centering
    \subfigure[$N_0=4\times10^6$. \label{optimal_lambda_N400}]{\includegraphics[width=0.49\textwidth,height=0.24\textwidth]{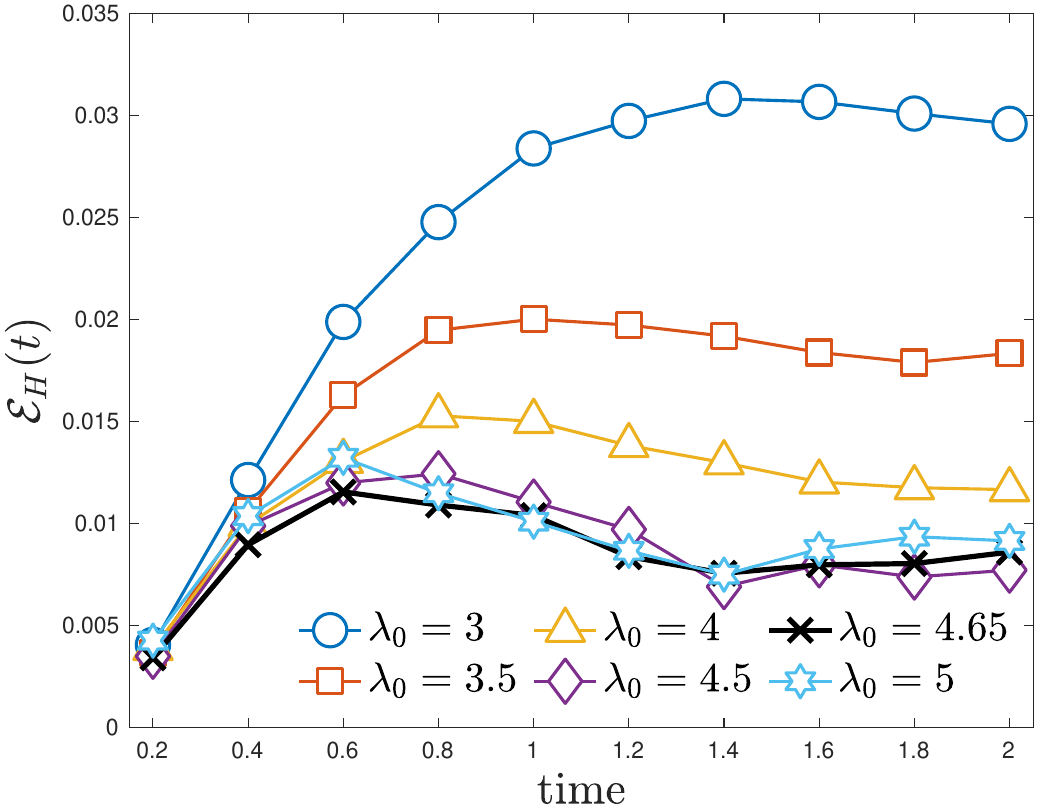}}
        \subfigure[$N_0=1\times10^7$. \label{optimal_lambda_N1000}]{\includegraphics[width=0.49\textwidth,height=0.24\textwidth]{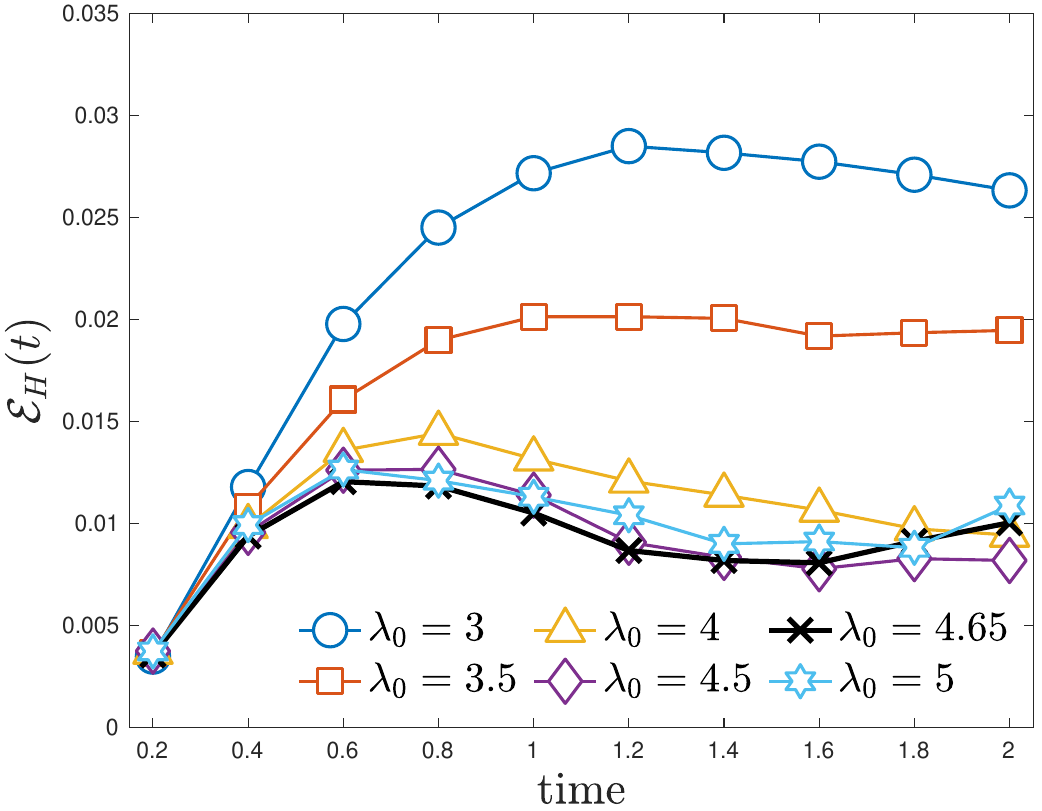}}
        \\
    \centering
    \subfigure[$N_0=4\times10^7$. \label{optimal_lambda_N4000}]{\includegraphics[width=0.49\textwidth,height=0.24\textwidth]{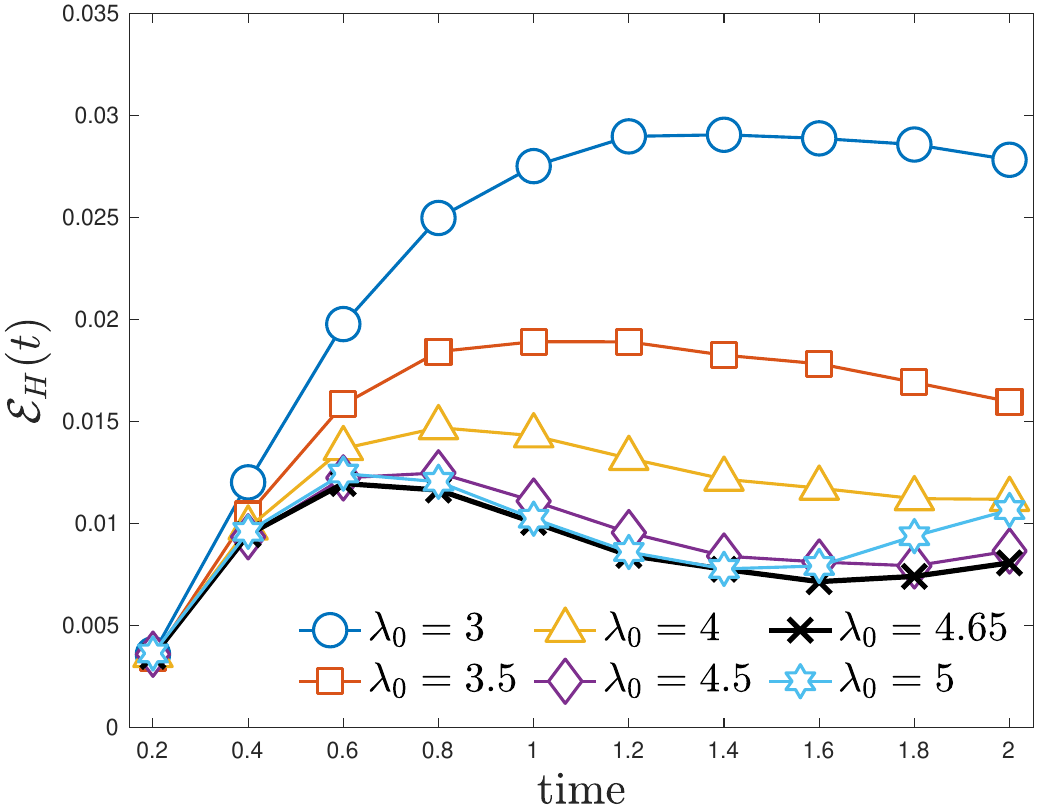}}
        \subfigure[$N_0=1\times10^8$. \label{optimal_lambda_N10000}]{\includegraphics[width=0.49\textwidth,height=0.24\textwidth]{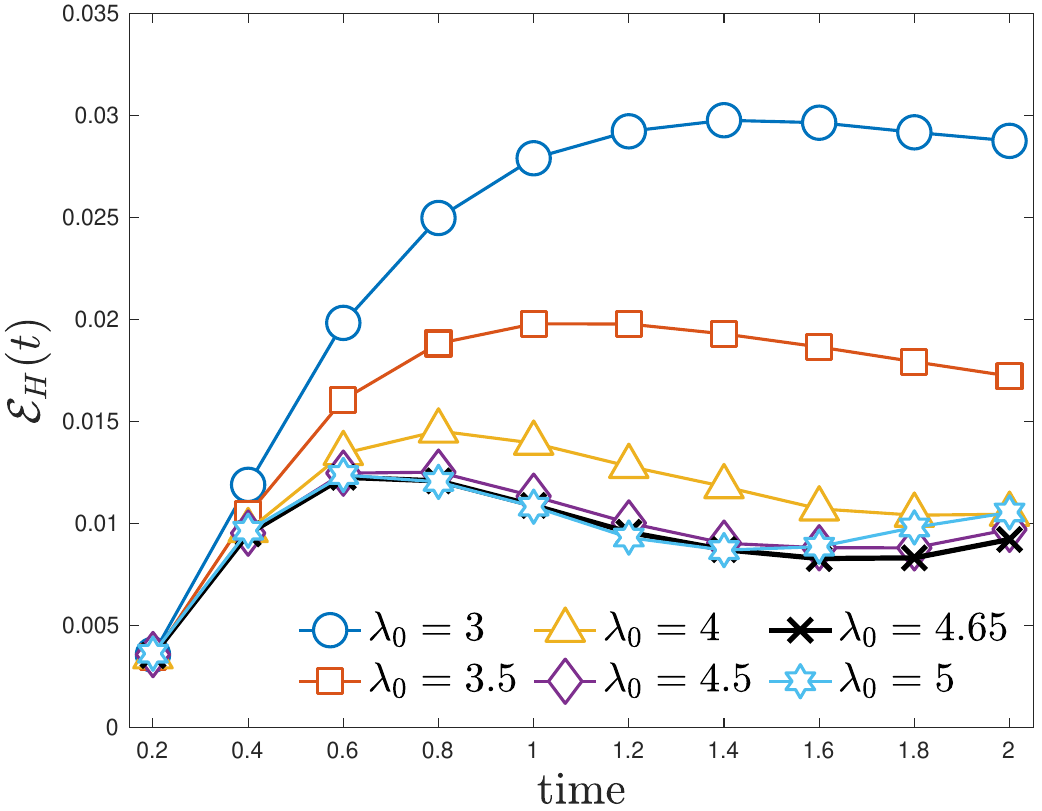}}
\caption{\small The filter $\lambda_0$ in WBRW-SPA: For a conservative Wigner-Coulomb system, $\lambda_0 = 4.65$ is chosen as it achieves smallest deviation of total energy.}
\label{optimal_lambda0}
\end{figure}

\begin{figure}[!h]
    \centering
    \subfigure[$\vartheta = 0.01$ (left: $l^2$-error for $W_1$, middle: deviation of energy, right: growth of particle).]{
    {\includegraphics[width=0.32\textwidth,height=0.22\textwidth]{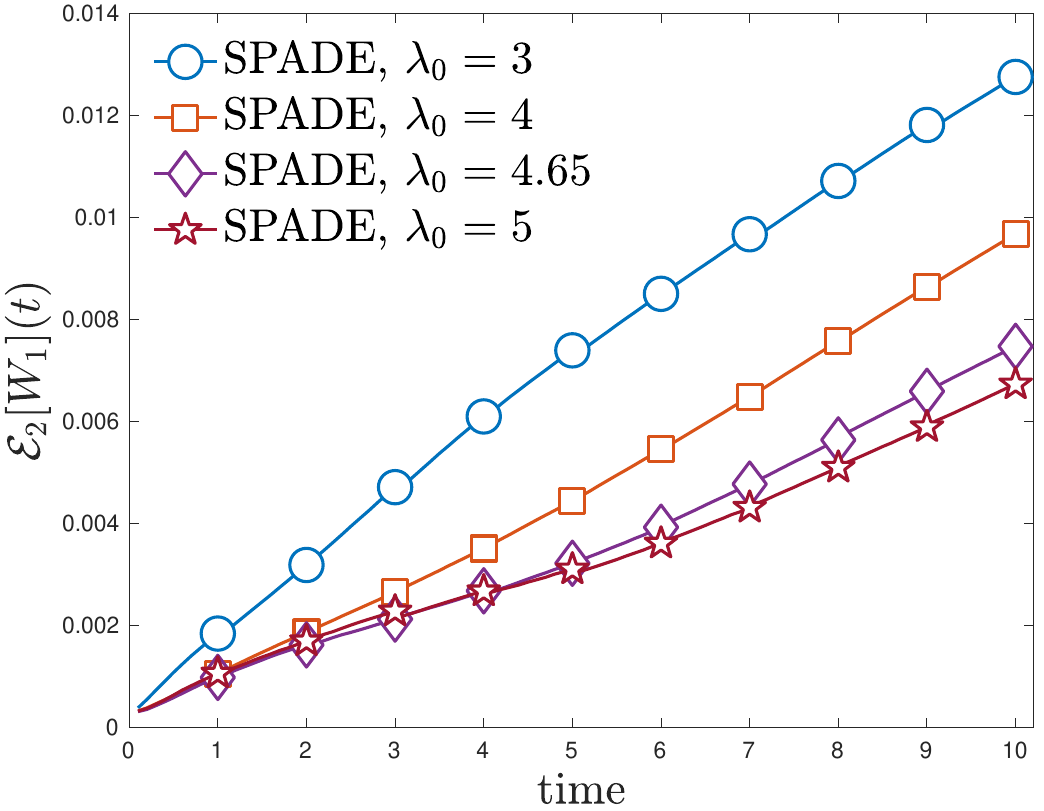}}
    {\includegraphics[width=0.32\textwidth,height=0.22\textwidth]{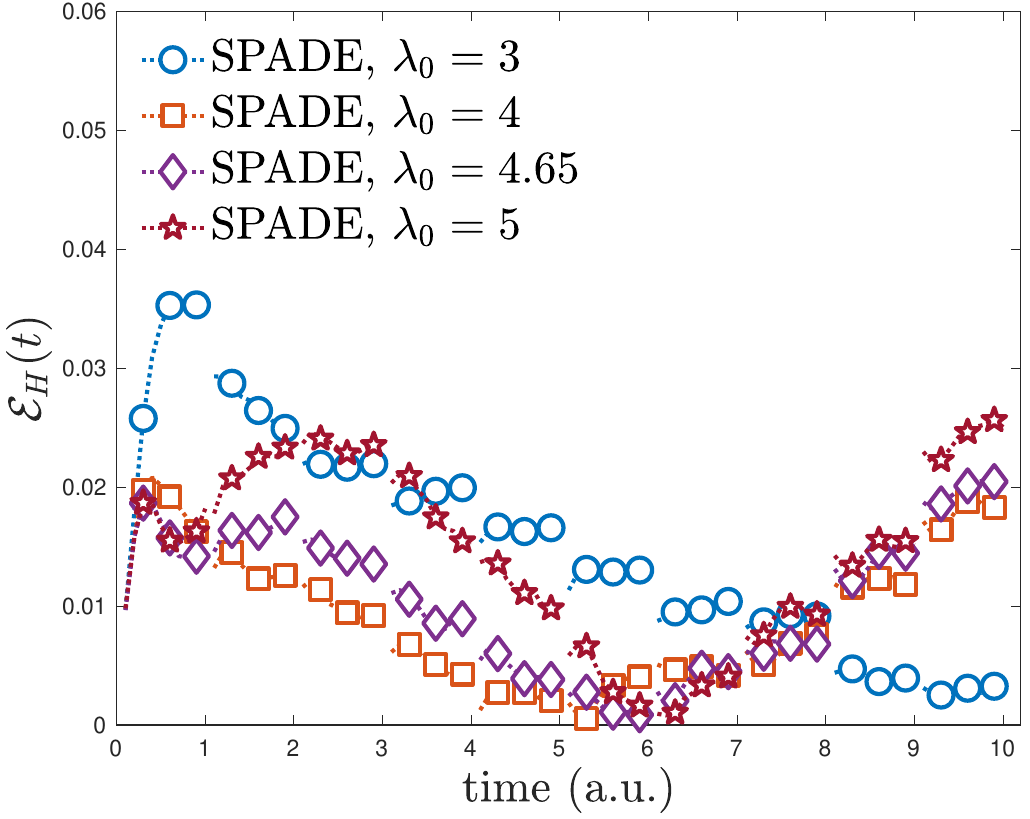}}
    {\includegraphics[width=0.32\textwidth,height=0.22\textwidth]{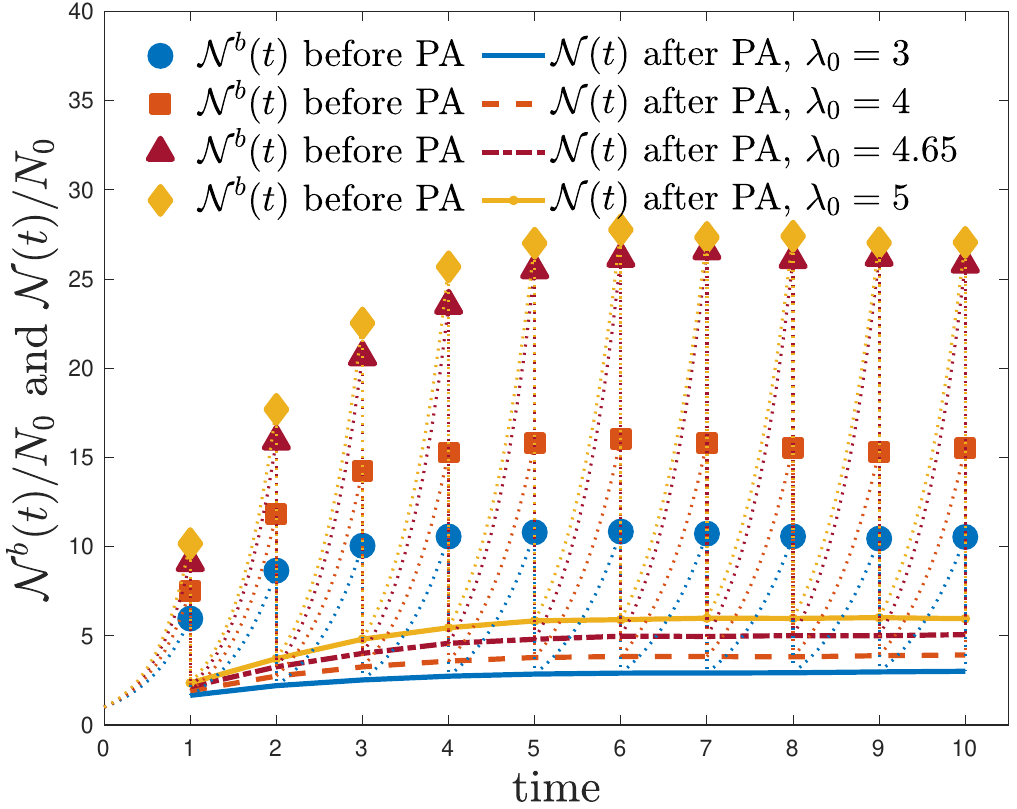}}}
     \\
     \centering
     \subfigure[$\vartheta = 0.02$ (left: $l^2$-error for $W_1$, middle: deviation of energy, right: growth of particle).]{
     {\includegraphics[width=0.32\textwidth,height=0.22\textwidth]{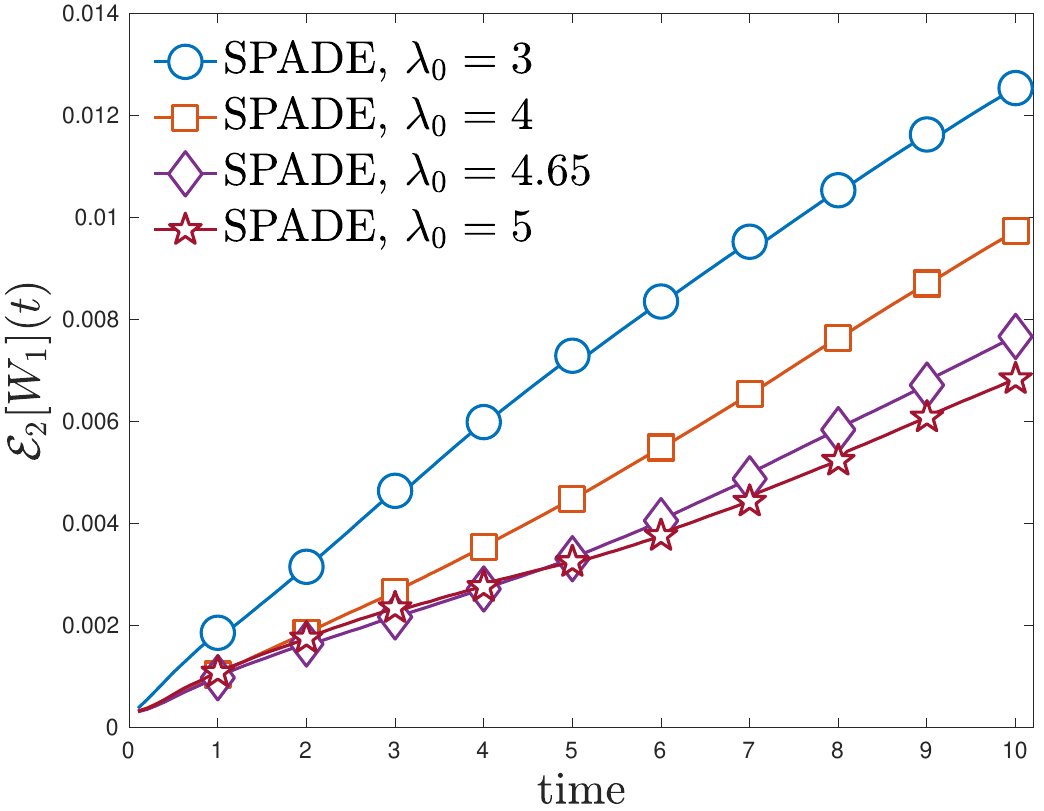}}
     {\includegraphics[width=0.32\textwidth,height=0.22\textwidth]{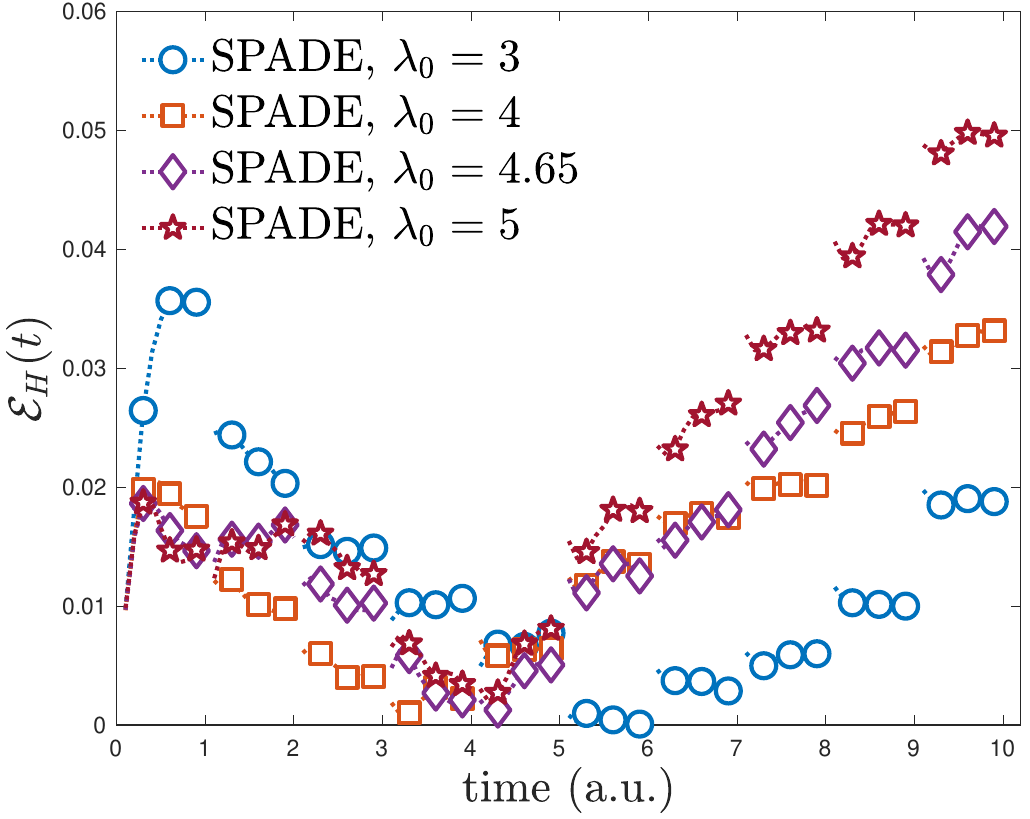}} 
     {\includegraphics[width=0.32\textwidth,height=0.22\textwidth]{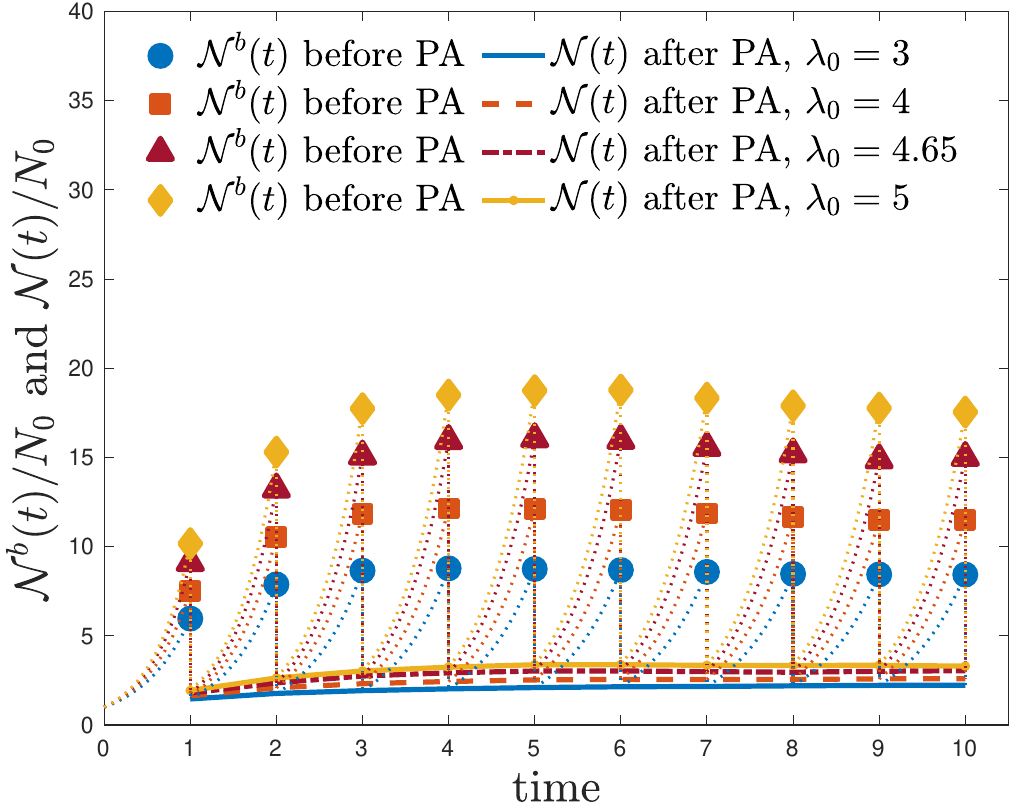}}}
    \caption{\small  The overall accuracy of WBRW-SPA-SPADE is limited by both the sampling error and the asymptotic error in SPA. Here $N_0$ is fixed to be $4\times 10^7$.\label{comparison_lambda}}
\end{figure} 

{\bf Accuracy}:  Using Algorithm \ref{optimal_lambda}, it is found that $\lambda_0 = 4.65$ can achieve the smallest deviation of total energy up to $2$a.u. under $N_0 = 4\times10^6,1\times10^7, 4\times10^7$ and $1\times10^8$ (see Fig.~\ref{optimal_lambda0}). But such choice only achieves a compromise between efficiency and accuracy. According to Fig.~\ref{comparison_lambda}, the results under $\lambda_0 = 4.65$ seem to outperform other groups before $t= 4$a.u. But the $l^2$-error $\mathcal{E}_{2}[W_1]$ after $t= 4$a.u. can be further improved under  $\lambda_0 = 5$ due to the reduction in the asymptotic errors. The price to pay is that more particles are generated and left uncanceled.

 {\bf Sign problem}: As observed in Fig.~\ref{comparison_lambda}, both $\mathcal{E}[W_{1}](t)$ and $\mathcal{E}_H(t)$ are augmented during the time intervals in which two successive PAs are performed. This provides an evidence that PA can only alleviate the sign problem, instead of eliminating it.

\subsection{Comparison between PAUM and SPADE}

\begin{figure}[!h]
    \subfigure[$l^2$-errors for $W_1(x_1, k_1, t)$ (left) and $P_{xy}(x_1, x_2, t)$ (right), $N_0= 1\times 10^7$. \label{comparison_N1000}]{
    {\includegraphics[width=0.49\textwidth,height=0.26\textwidth]{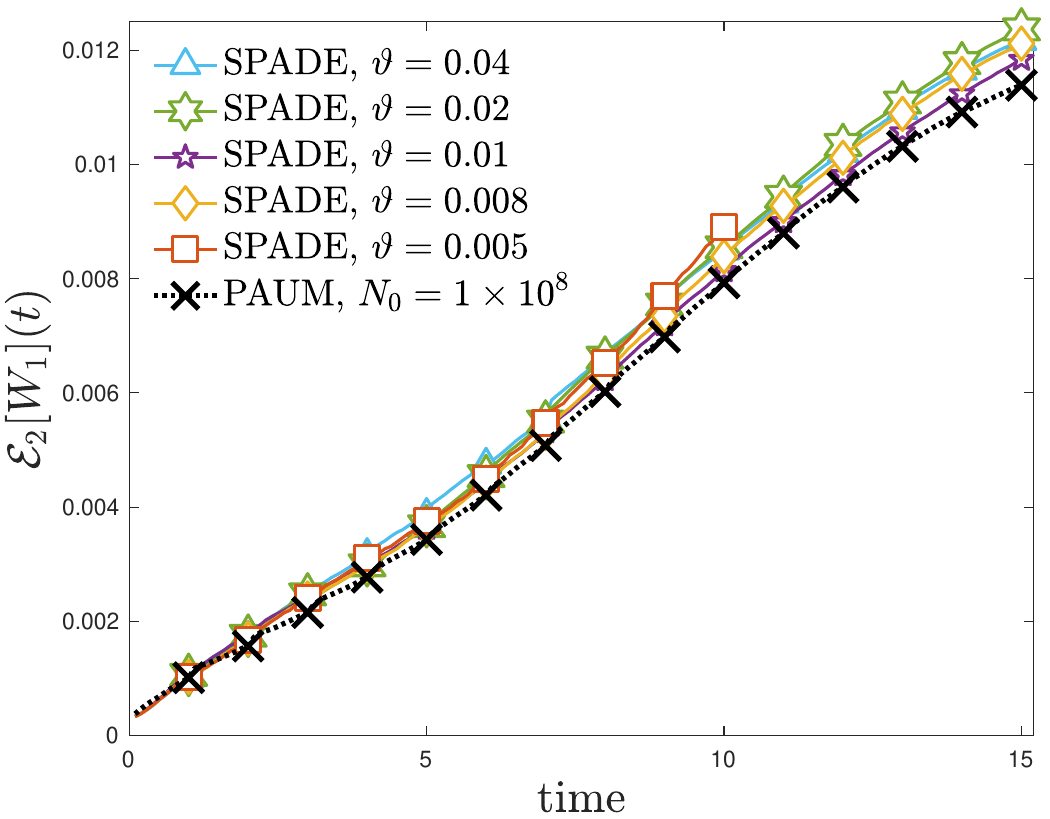}}
    {\includegraphics[width=0.49\textwidth,height=0.26\textwidth]{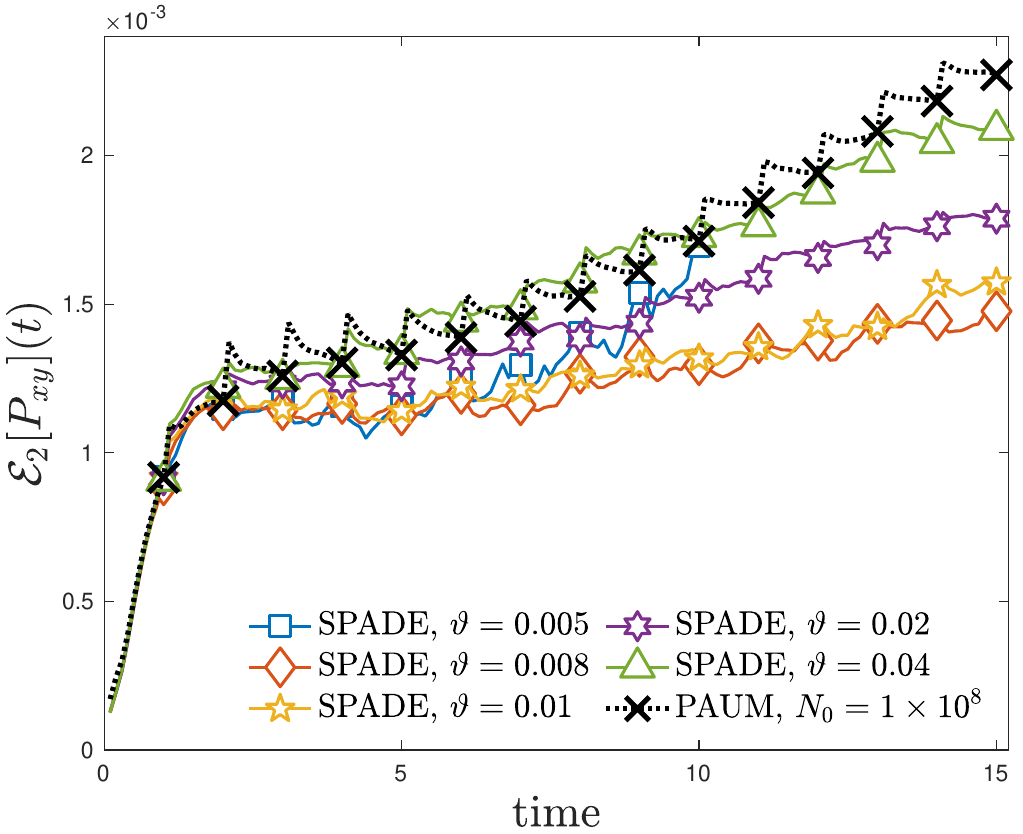}}}
  \\
    \subfigure[$l^2$-errors for $W_1(x_1, k_1, t)$ (left) and $P_{xy}(x_1, x_2, t)$ (right), $N_0= 4\times 10^7$.\label{comparison_N4000}]{
   {\includegraphics[width=0.49\textwidth,height=0.26\textwidth]{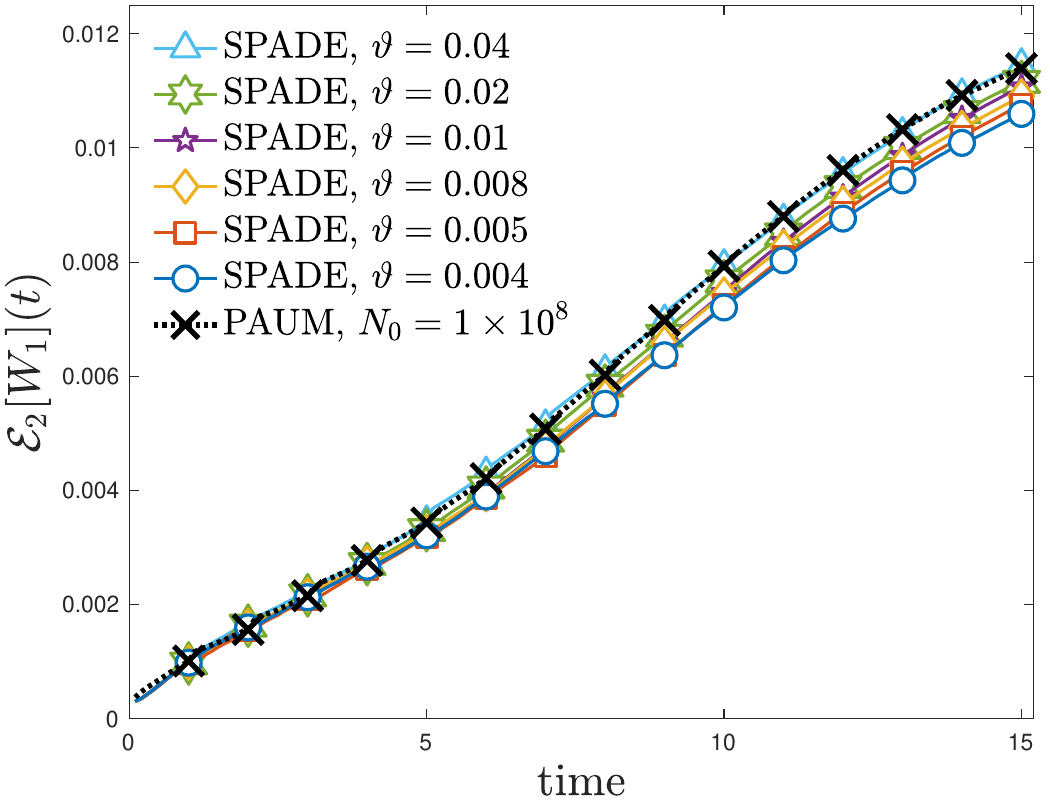}}
   {\includegraphics[width=0.49\textwidth,height=0.26\textwidth]{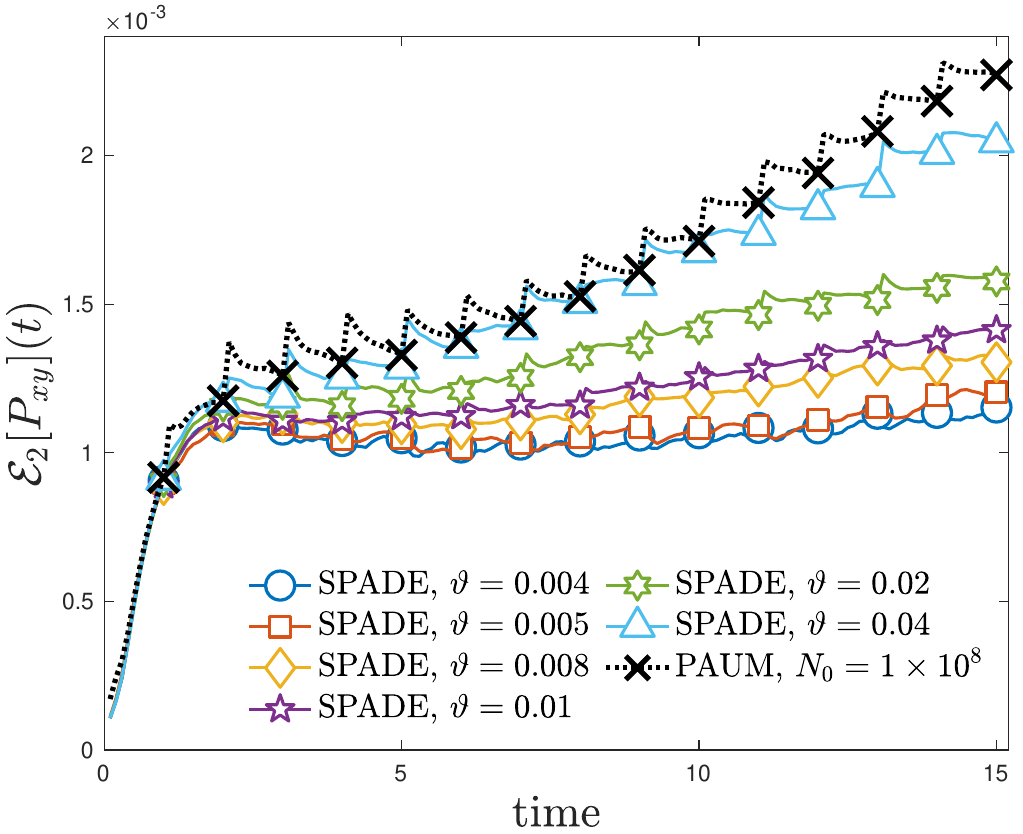}}}
     \\
    \subfigure[$l^2$-errors for $W_1(x_1, k_1, t)$ (left) and $P_{xy}(x_1, x_2, t)$ (right), $N_0= 1\times 10^8$.\label{comparison_N10000}]{
   {\includegraphics[width=0.49\textwidth,height=0.26\textwidth]{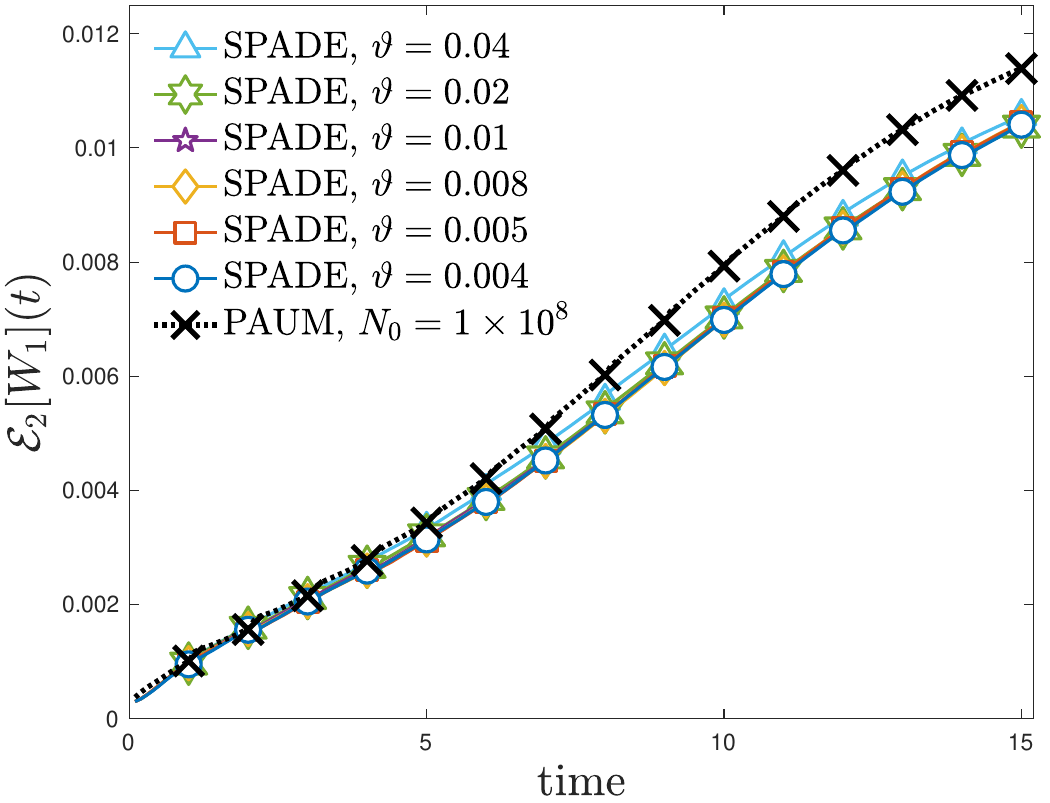}}
   {\includegraphics[width=0.49\textwidth,height=0.26\textwidth]{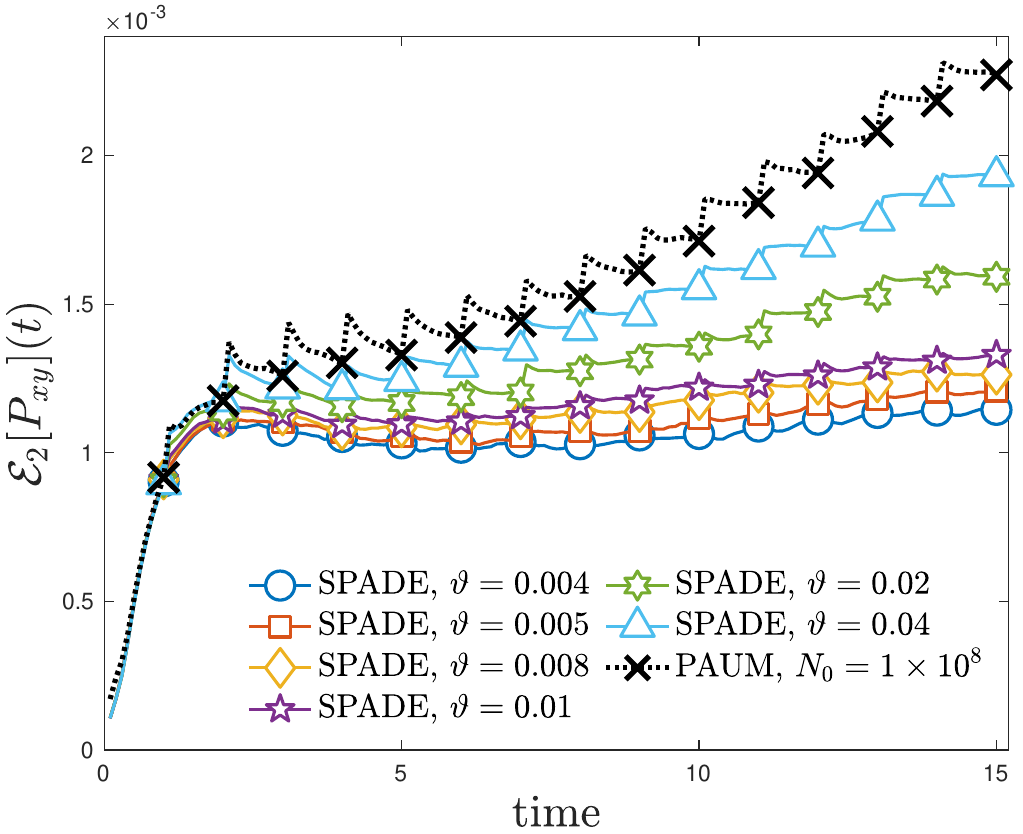}}}
\caption{\small The time evolution of the $l^2$-error $\mathcal{E}_2[W_1](t)$ and $\mathcal{E}_2[P_{xy}](t)$ under different parameter $\vartheta$. SPADE significantly alleviates the rapid growth of stochastic variances, and outperforms PAUM when $N_0 \ge 4\times 10^7$. The oversampling problem is observed in the group $N_0 = 1\times 10^7$, $\vartheta = 0.005$, where errors are accumulated more rapidly.\label{comparison_SPADE_PAUM_N}}
\end{figure} 

\begin{figure}[!h]
    \subfigure[$\vartheta = 0.008$ (left: $l^2$-error for $W_1$, middle: deviation of energy, right: growth of particle).\label{comparison_t0008}]{
    {\includegraphics[width=0.32\textwidth,height=0.22\textwidth]{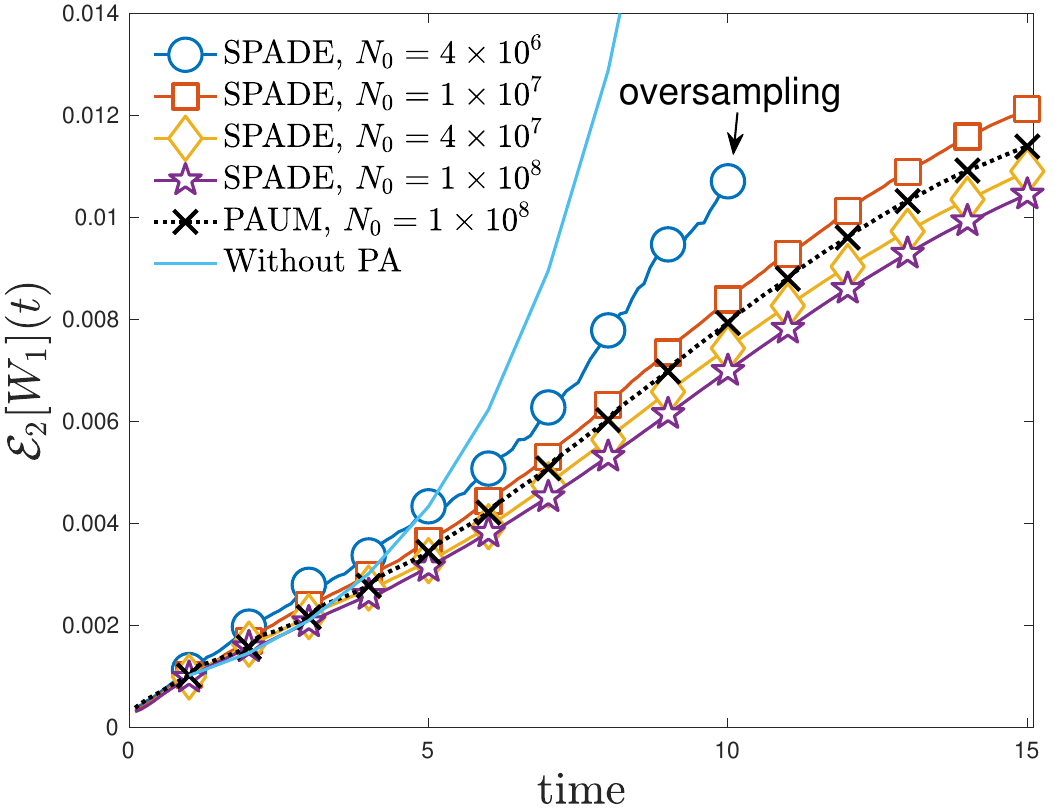}}
    {\includegraphics[width=0.32\textwidth,height=0.22\textwidth]{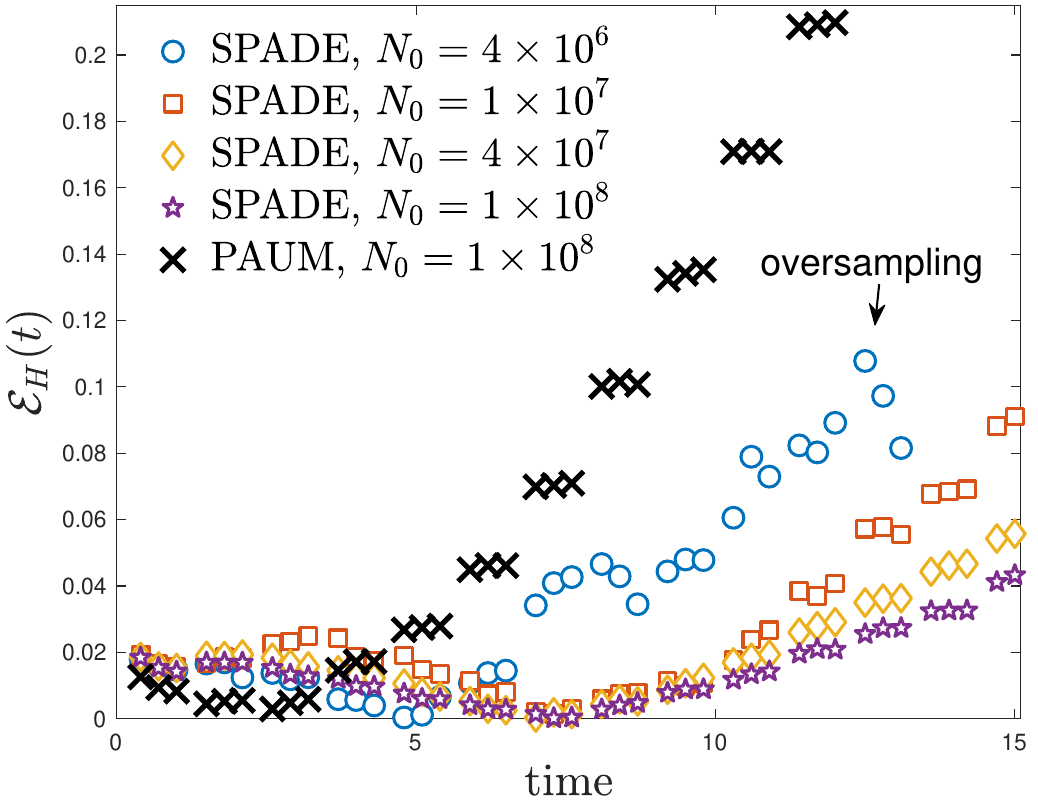}}
    {\includegraphics[width=0.32\textwidth,height=0.22\textwidth]{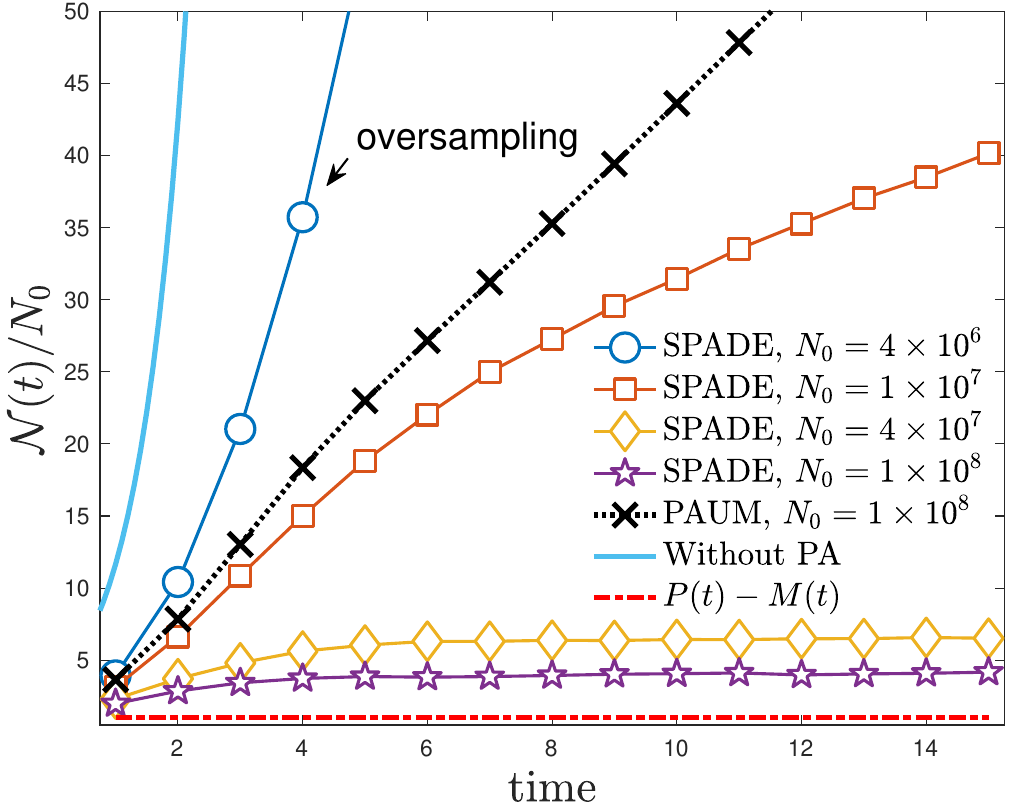}}}
     \\
    \subfigure[$\vartheta = 0.01$ (left: $l^2$-error for $W_1$, middle: deviation of energy, right: growth of particle).\label{comparison_t001}]{
    {\includegraphics[width=0.32\textwidth,height=0.22\textwidth]{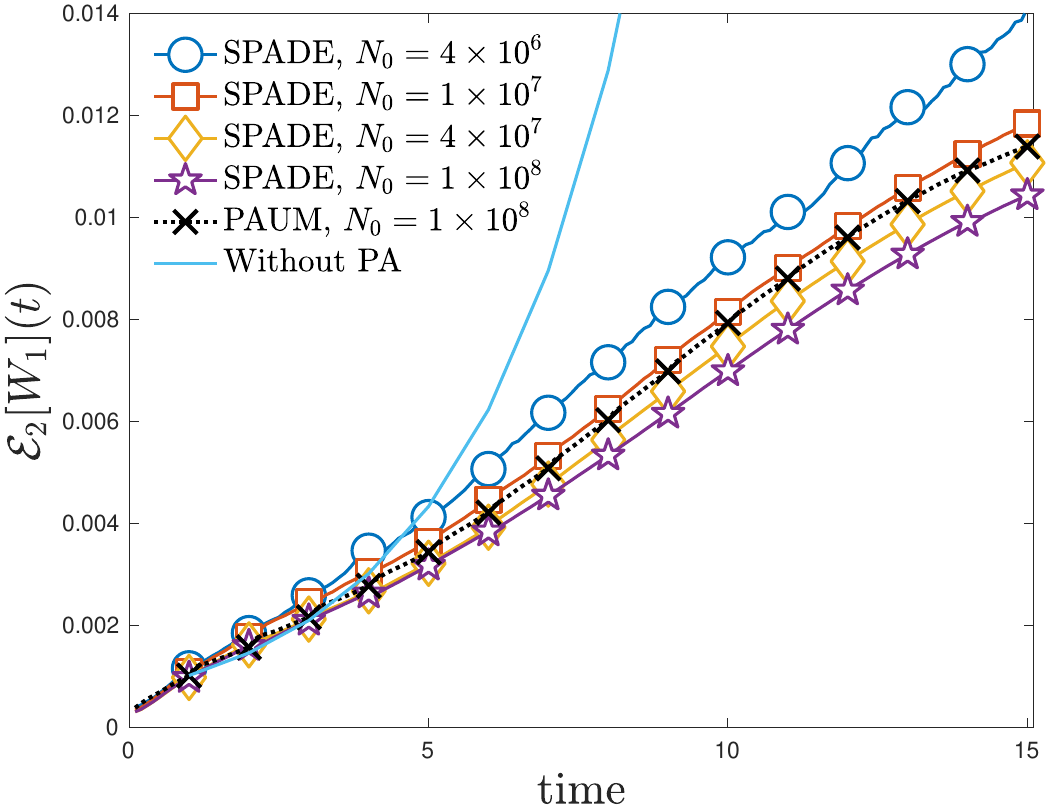}}
    {\includegraphics[width=0.32\textwidth,height=0.22\textwidth]{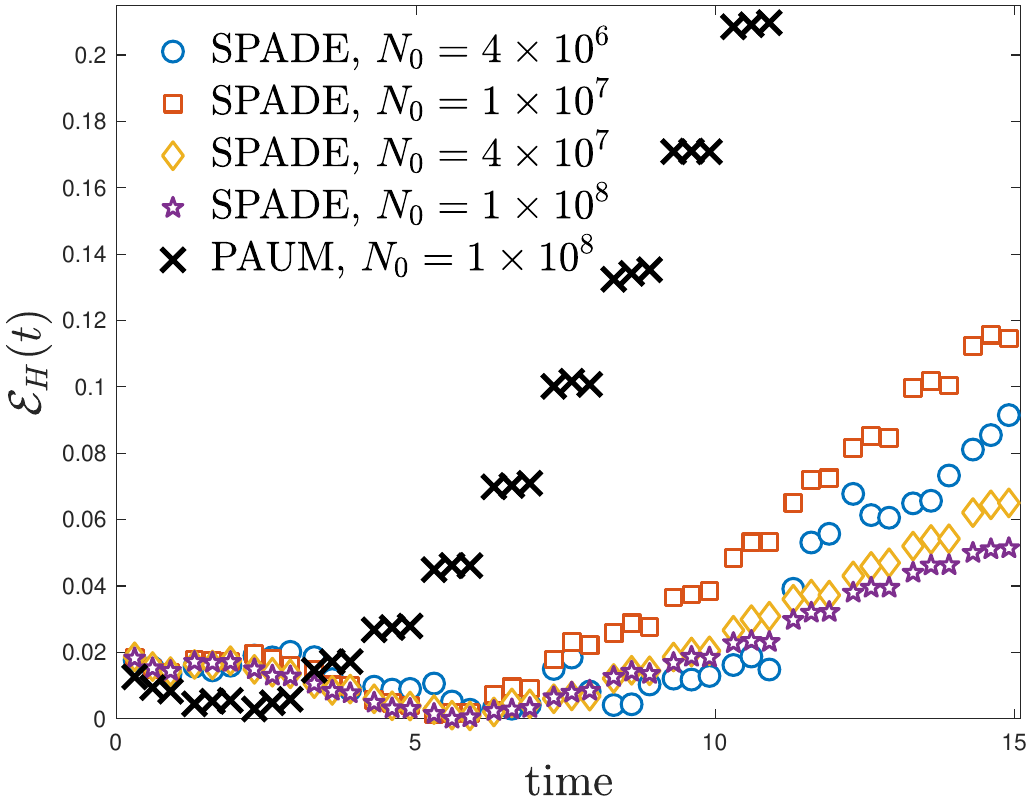}}
    {\includegraphics[width=0.32\textwidth,height=0.22\textwidth]{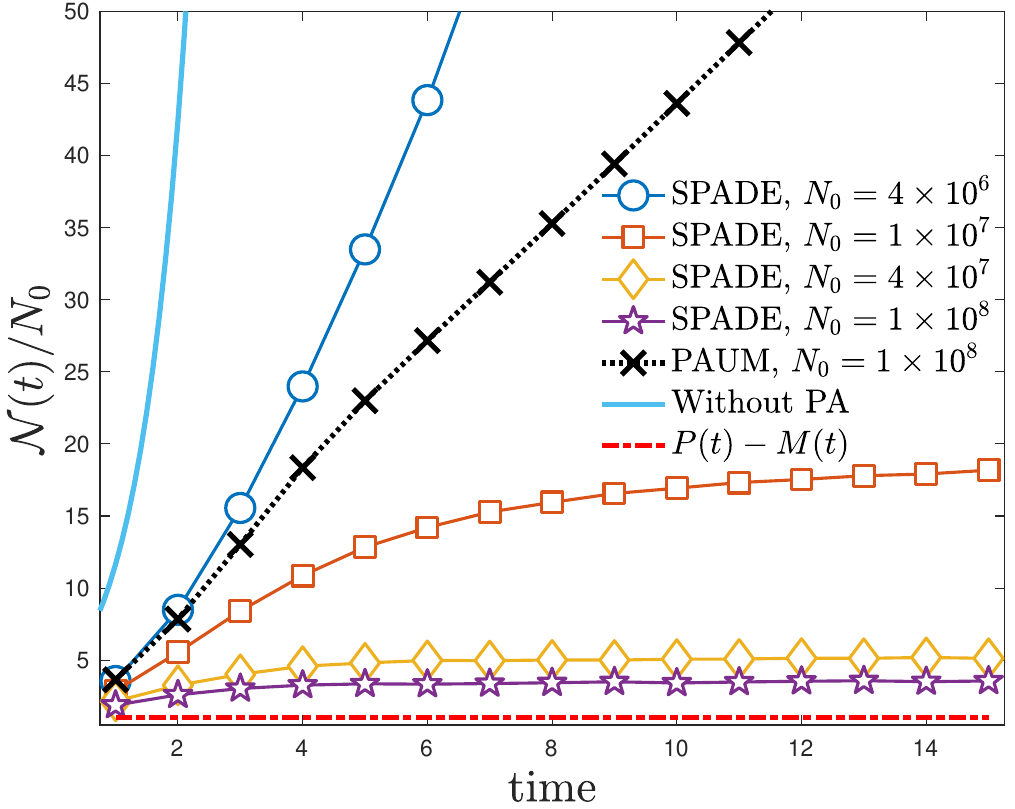}}}
     \\
     \subfigure[$\vartheta = 0.02$ (left: $l^2$-error for $W_1$, middle: deviation of energy, right: growth of particle).\label{comparison_t002}]{
     {\includegraphics[width=0.32\textwidth,height=0.22\textwidth]{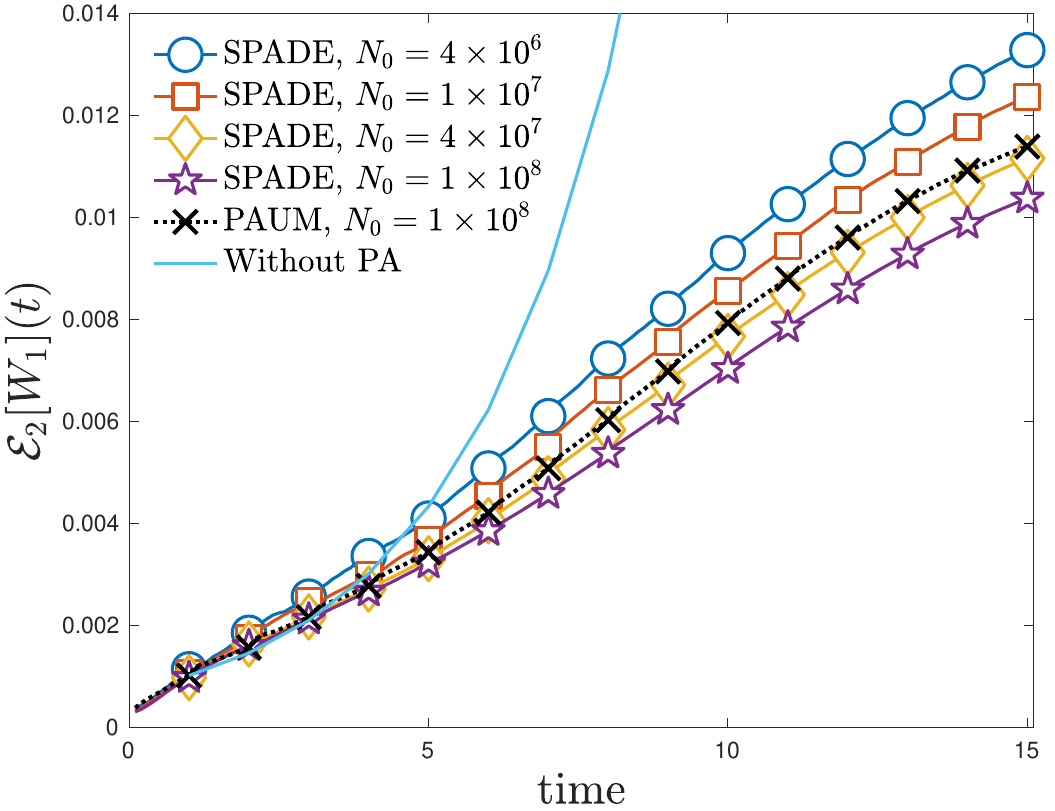}}
     {\includegraphics[width=0.32\textwidth,height=0.22\textwidth]{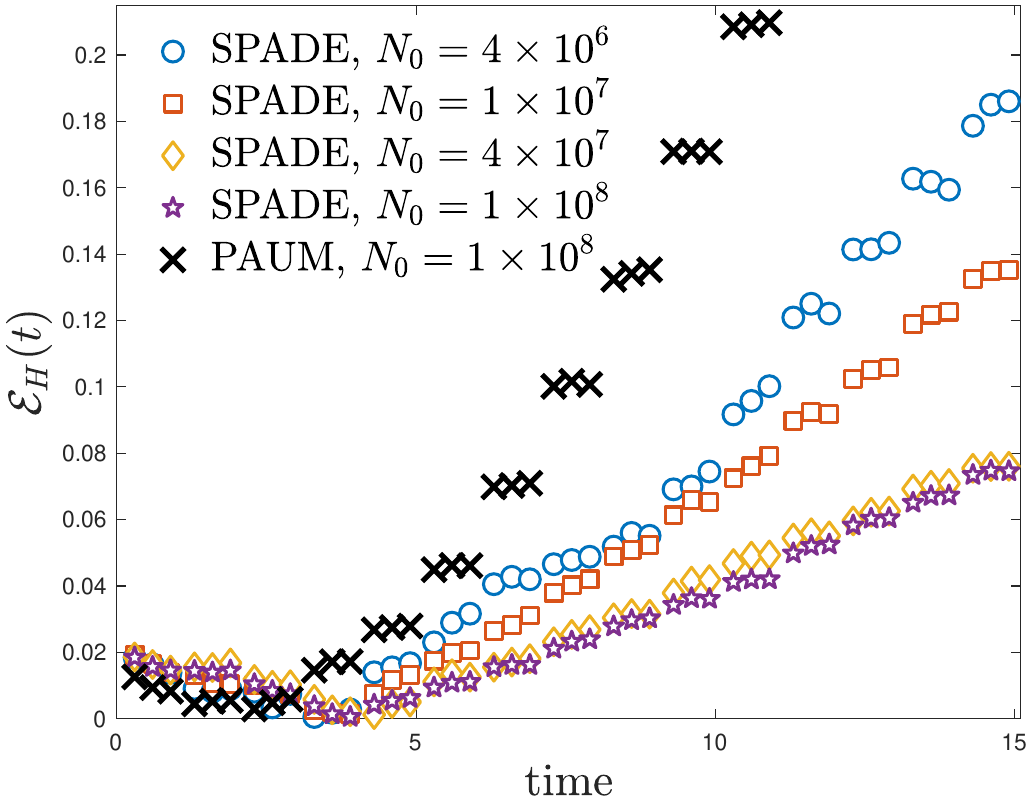}} 
     {\includegraphics[width=0.32\textwidth,height=0.22\textwidth]{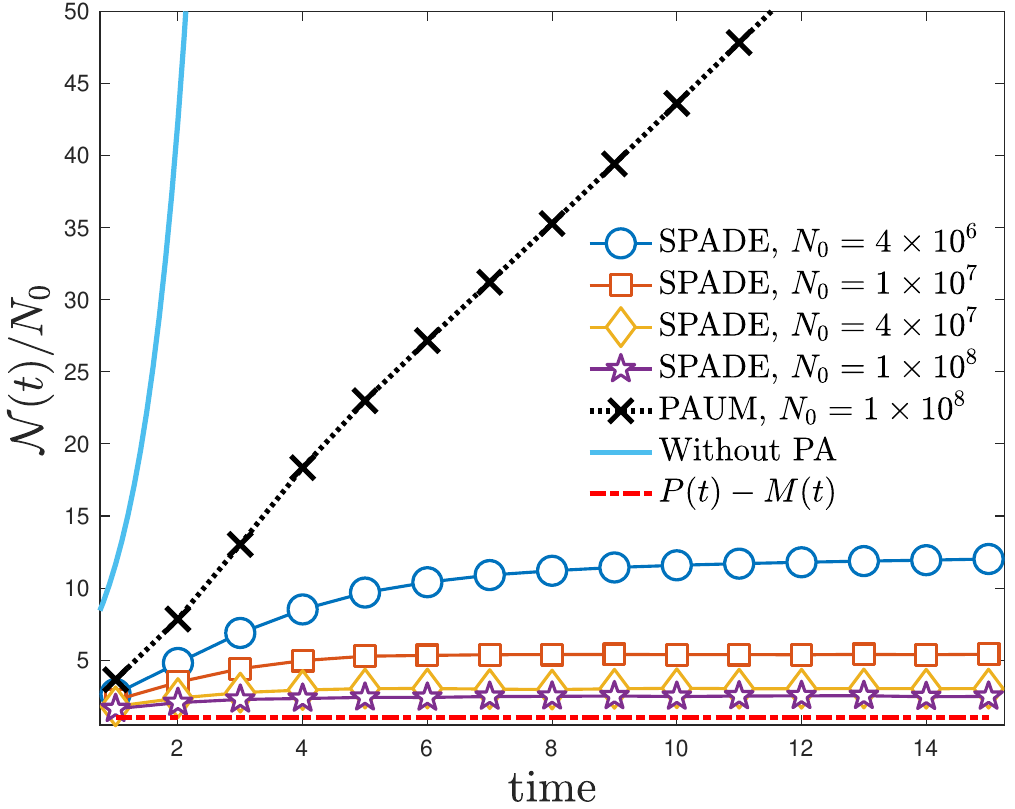}}}
          \\
     \subfigure[$\vartheta = 0.04$ (left: $l^2$-error for $W_1$, middle: deviation of energy, right: growth of particle).\label{comparison_t004}]{
     {\includegraphics[width=0.32\textwidth,height=0.22\textwidth]{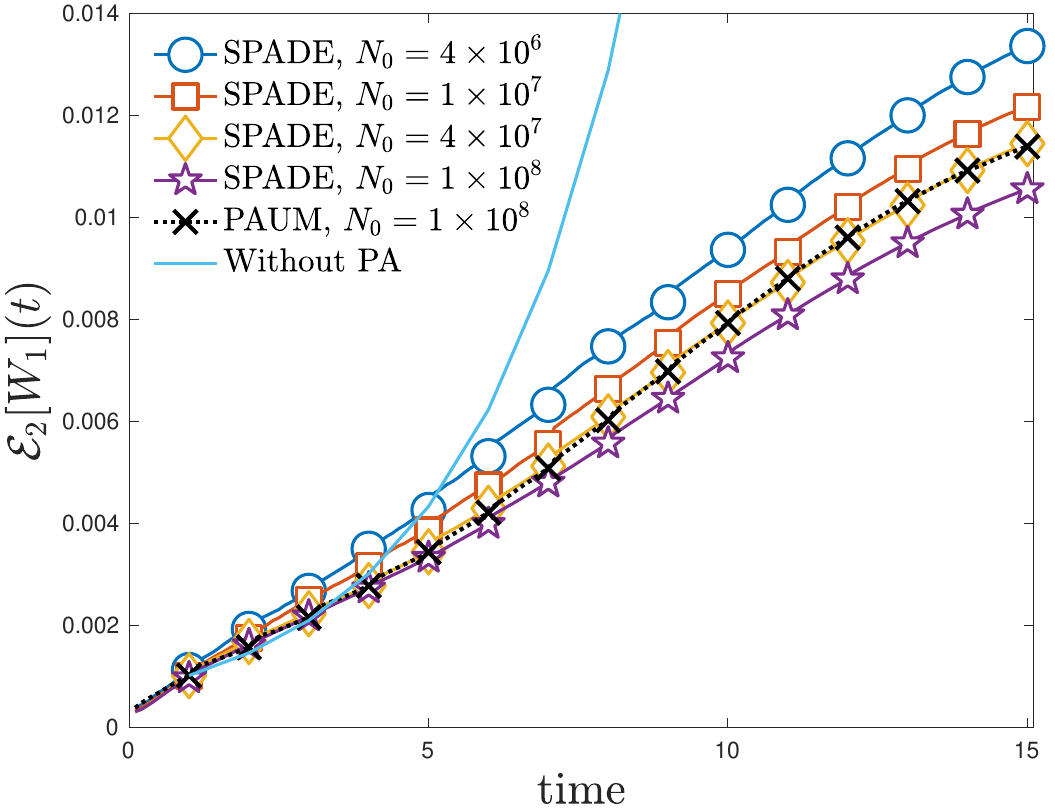}}
     {\includegraphics[width=0.32\textwidth,height=0.22\textwidth]{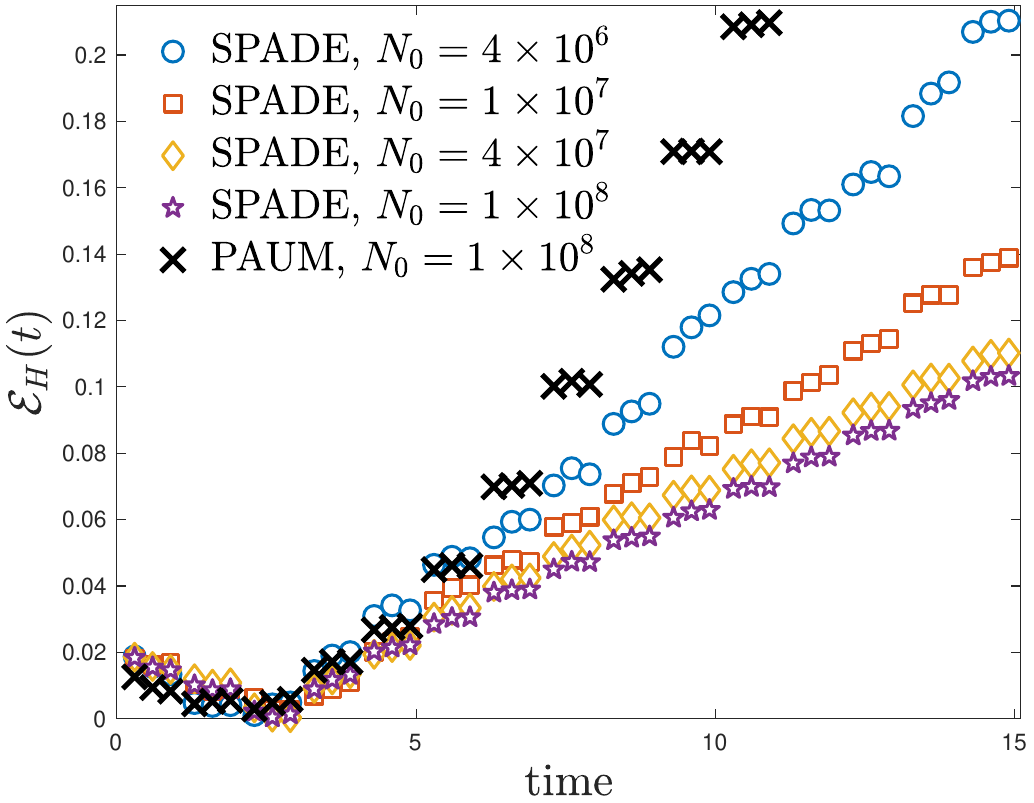}} 
     {\includegraphics[width=0.32\textwidth,height=0.22\textwidth]{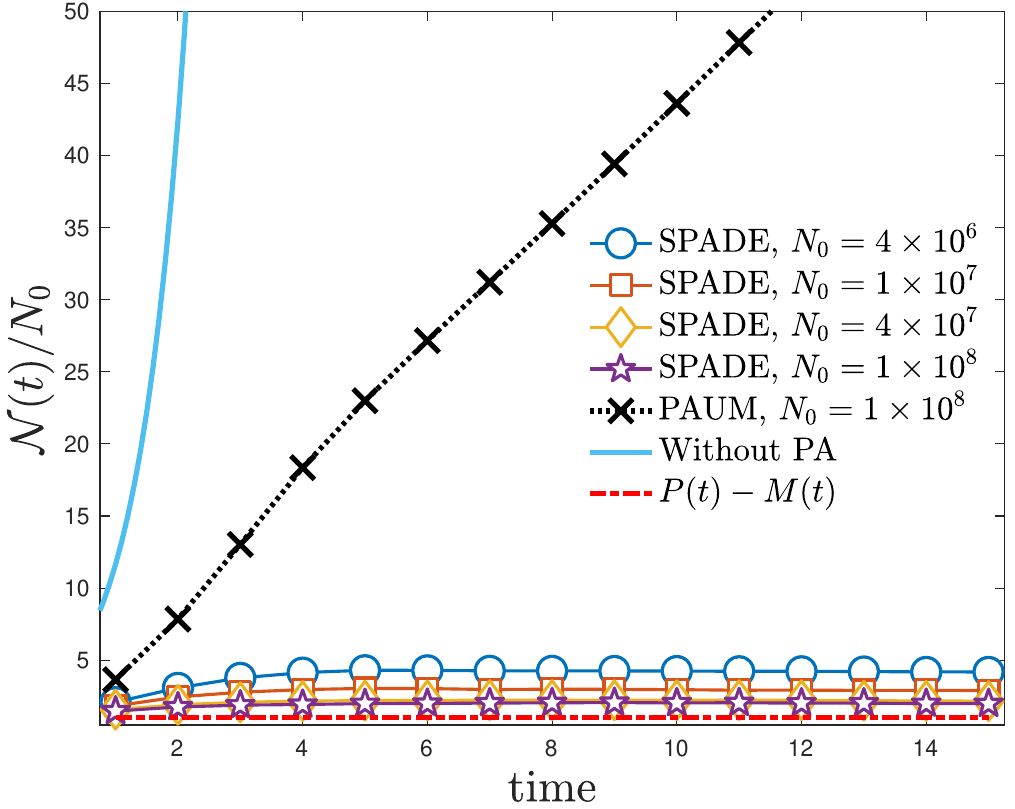}}}
    \caption{\small A comparison of the $l^2$-error of $\mathcal{E}_2[W_1](t)$ (left), the deviation of total energy $\mathcal{E}_H(t)$ (middle) and the growth of particle number after PA (right). SPADE is able to control particle number more efficiently for moderately large $N_0\ge 4\times 10^7$, but still suffers from oversampling problem when both $N_0$ and $\vartheta$ are too small (see $N_0 = 4\times10^6, \vartheta =0.008$).  \label{comparison_SPADE_PAUM}}
\end{figure} 

\begin{figure}[!h]
\centering
\subfigure[$t=4$a.u. (left: deterministic, middle: PAUM, right: SPADE, $\vartheta = 0.008$).]{
{\includegraphics[width=0.32\textwidth,height=0.22\textwidth]{./redist_CHASM_T4.pdf}}
{\includegraphics[width=0.32\textwidth,height=0.22\textwidth]{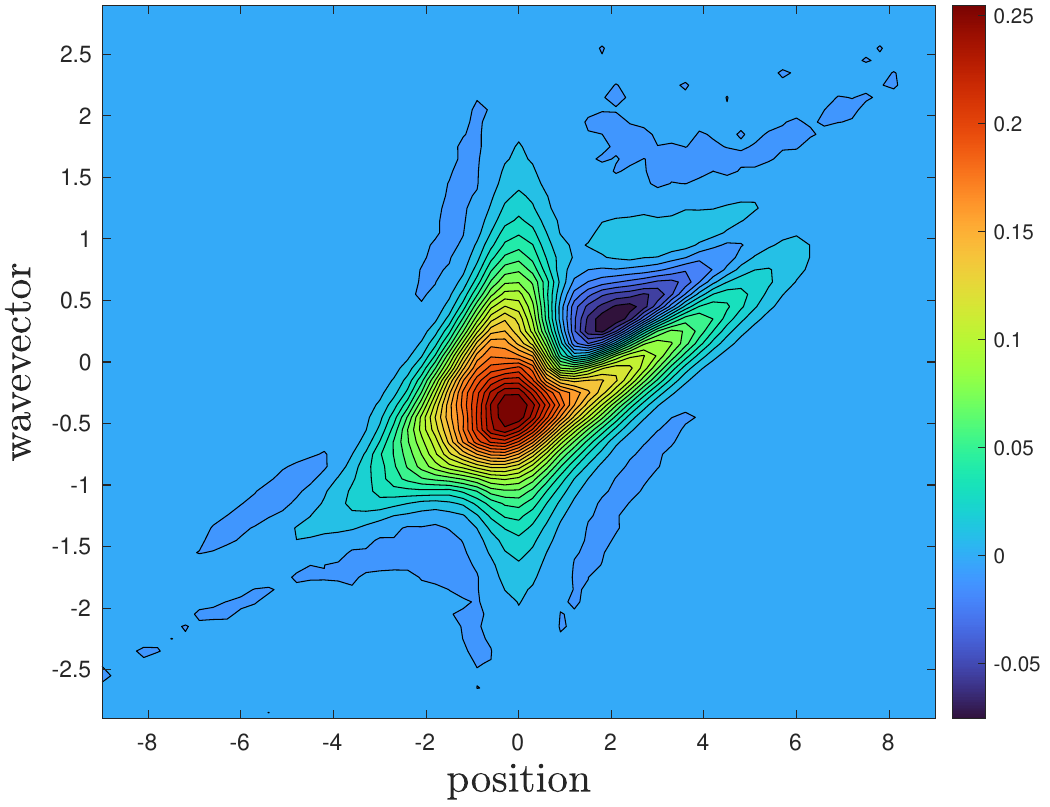}}
{\includegraphics[width=0.32\textwidth,height=0.22\textwidth]{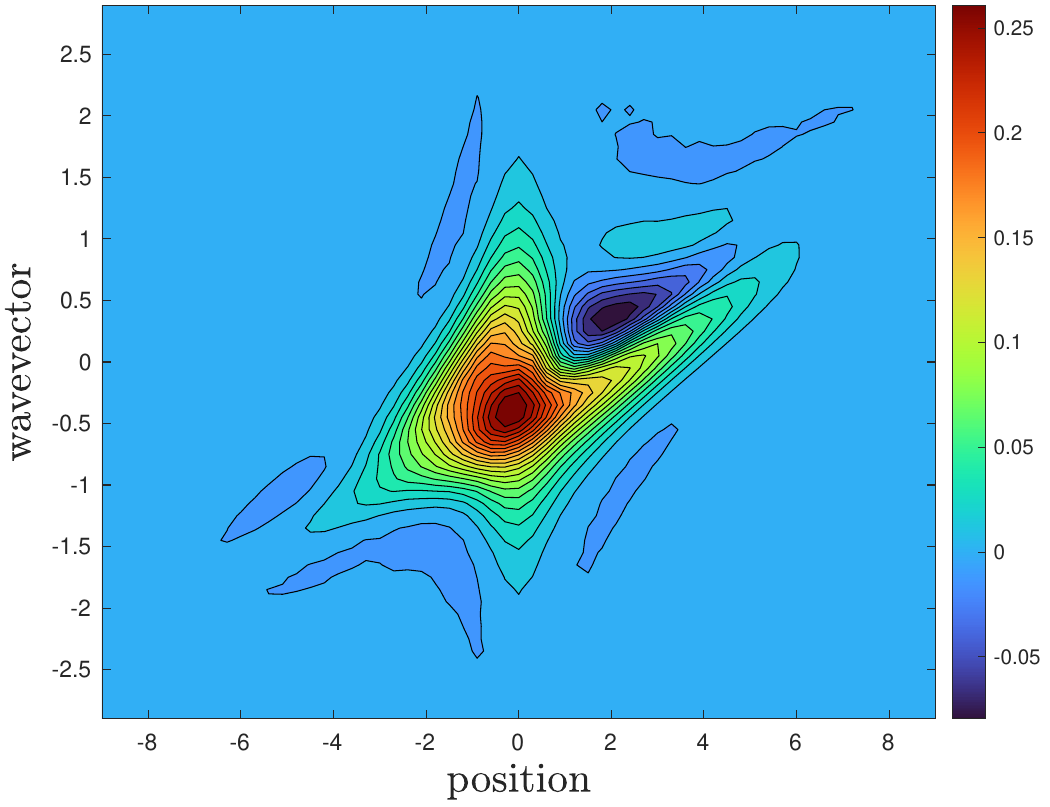}}}
\\
\centering
\subfigure[$t=8$a.u. (left: deterministic, middle: PAUM, right: SPADE, $\vartheta = 0.008$).]{
{\includegraphics[width=0.32\textwidth,height=0.22\textwidth]{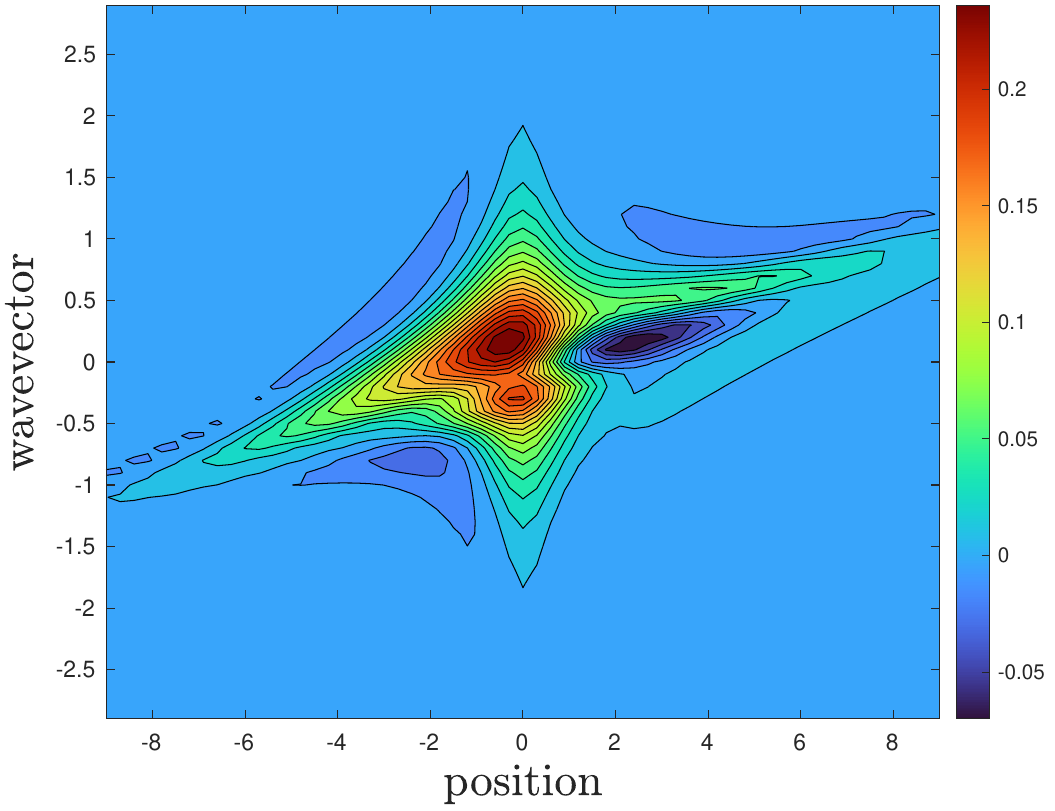}}
{\includegraphics[width=0.32\textwidth,height=0.22\textwidth]{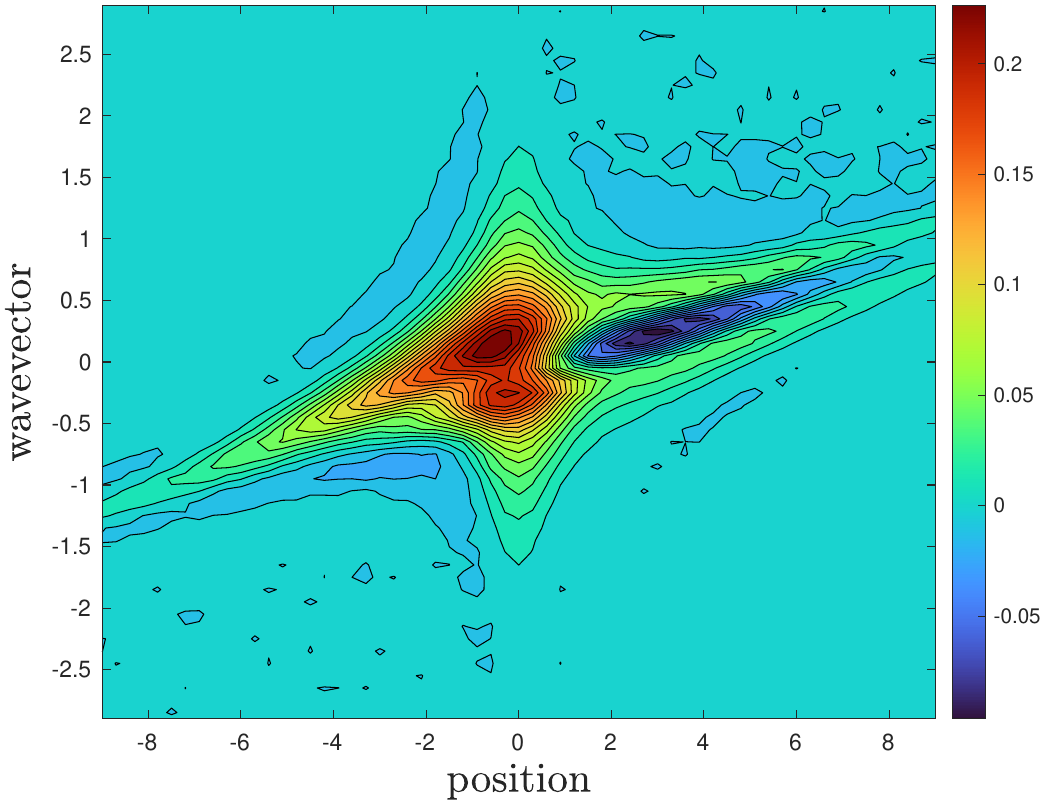}}
{\includegraphics[width=0.32\textwidth,height=0.22\textwidth]{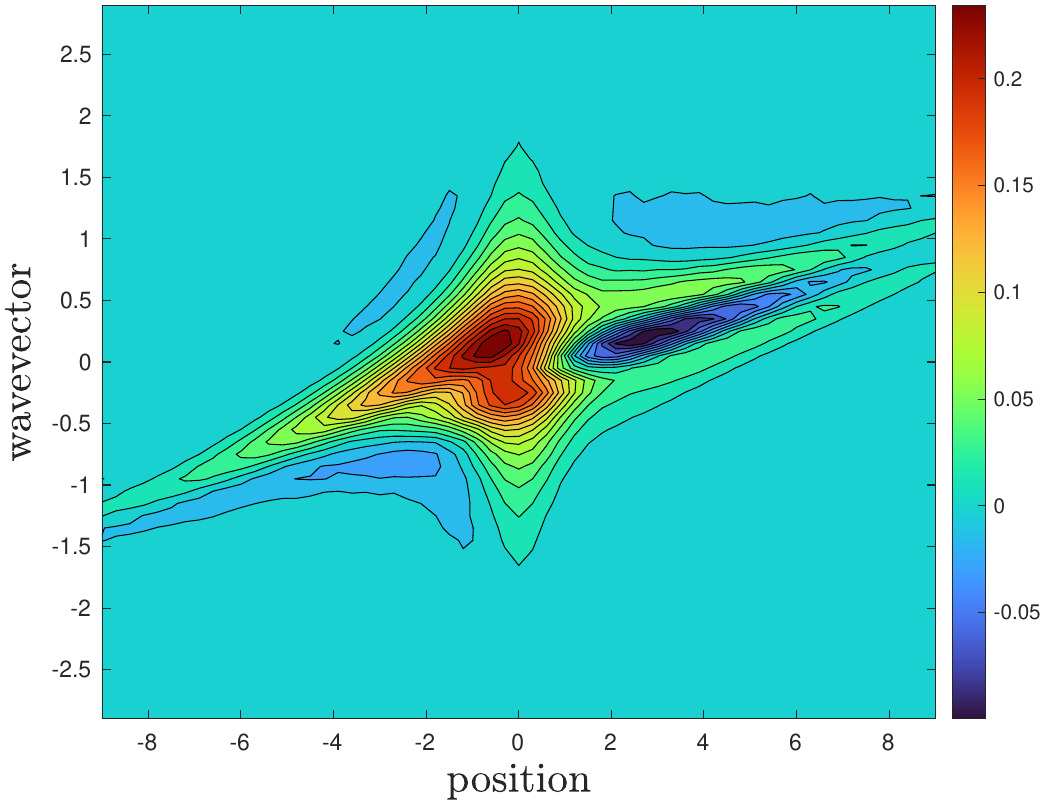}}}
\\
\centering
\subfigure[$t=15$a.u. (left: deterministic, middle: PAUM, right: SPADE, $\vartheta = 0.008$).]{
{\includegraphics[width=0.32\textwidth,height=0.22\textwidth]{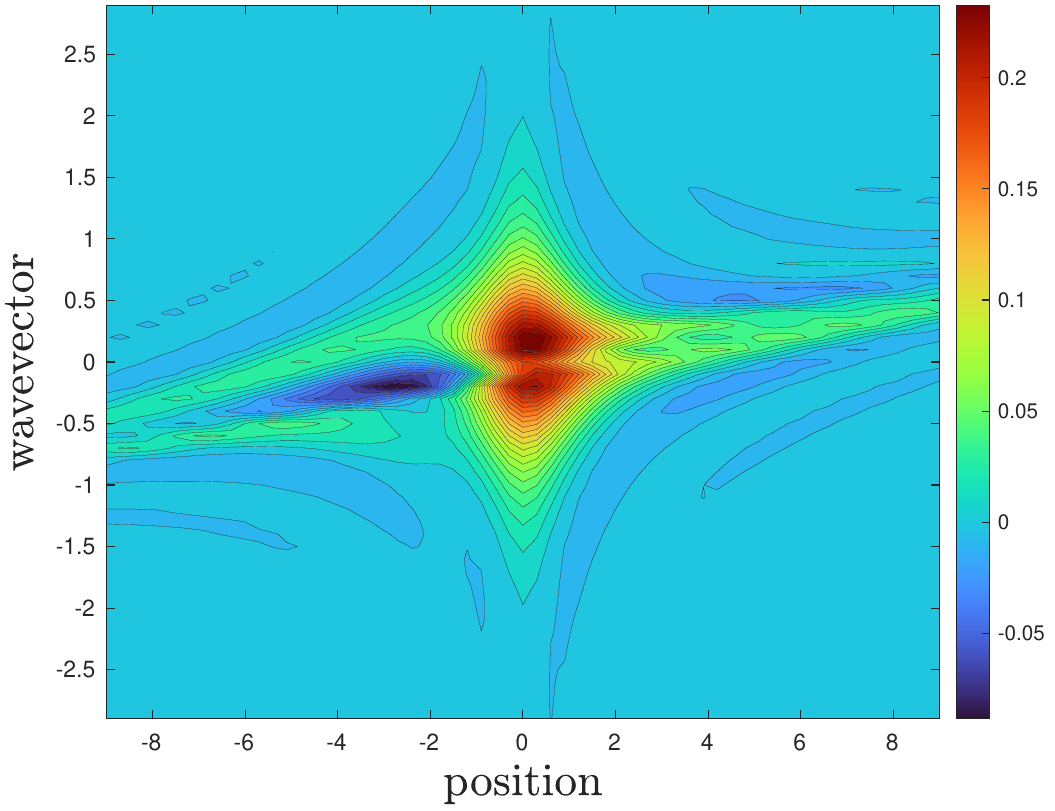}}
{\includegraphics[width=0.32\textwidth,height=0.22\textwidth]{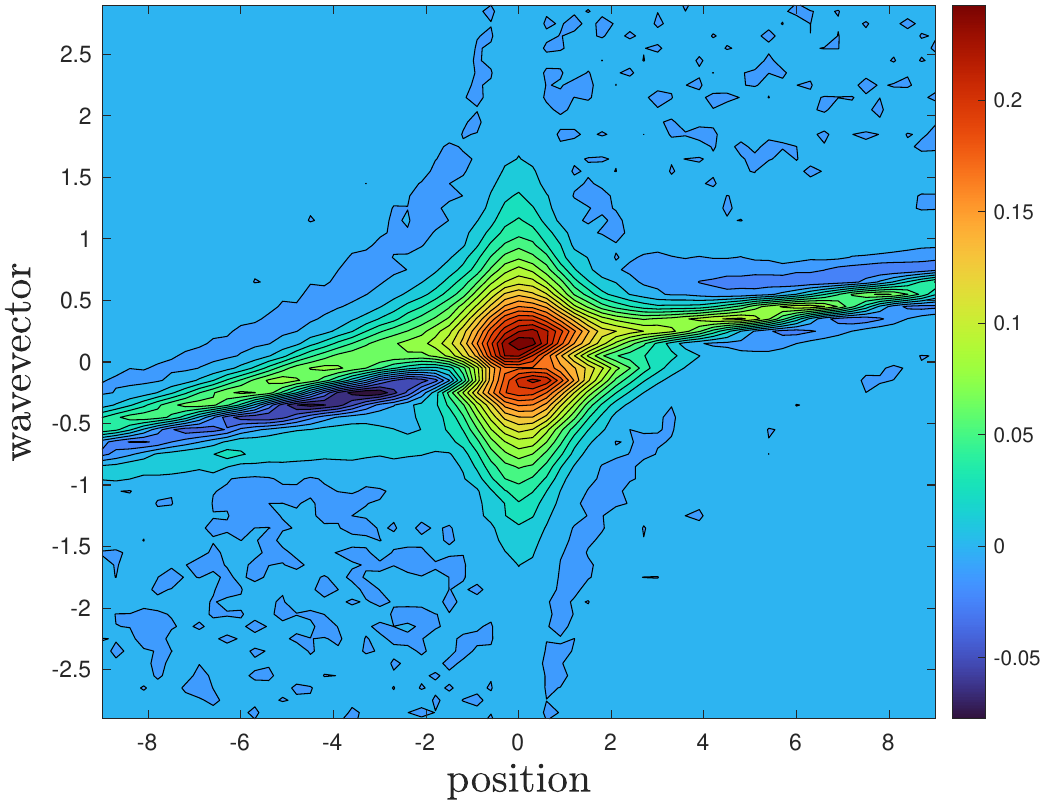}}
{\includegraphics[width=0.32\textwidth,height=0.22\textwidth]{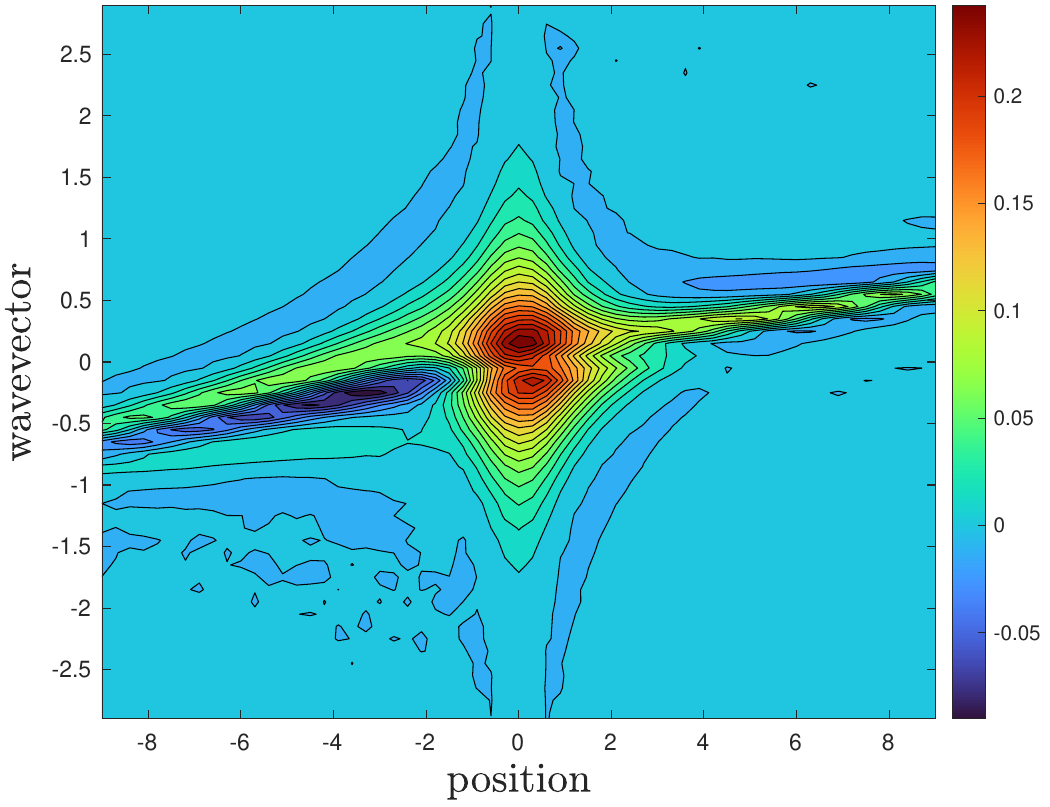}}}
\caption{\small  Snapshots of the reduced Wigner function $W_1(x_1, k_1, t)$ under $N_0 = 10^8$. The particle-based stochastic algorithms can properly capture the double-peak structure (Coulomb collision) and negative valley (uncertainty principle). The oscillating tails can also be reconstructed, albeit with some random noises.}
\label{qcs_time_evolution}
\end{figure}

\begin{figure}[!h]
    \subfigure[$t = 8$a.u. (left: $\vartheta = 0.008$, right: $\vartheta=0.02$).\label{xdist_t8}]{
    {\includegraphics[width=0.49\textwidth,height=0.26\textwidth]{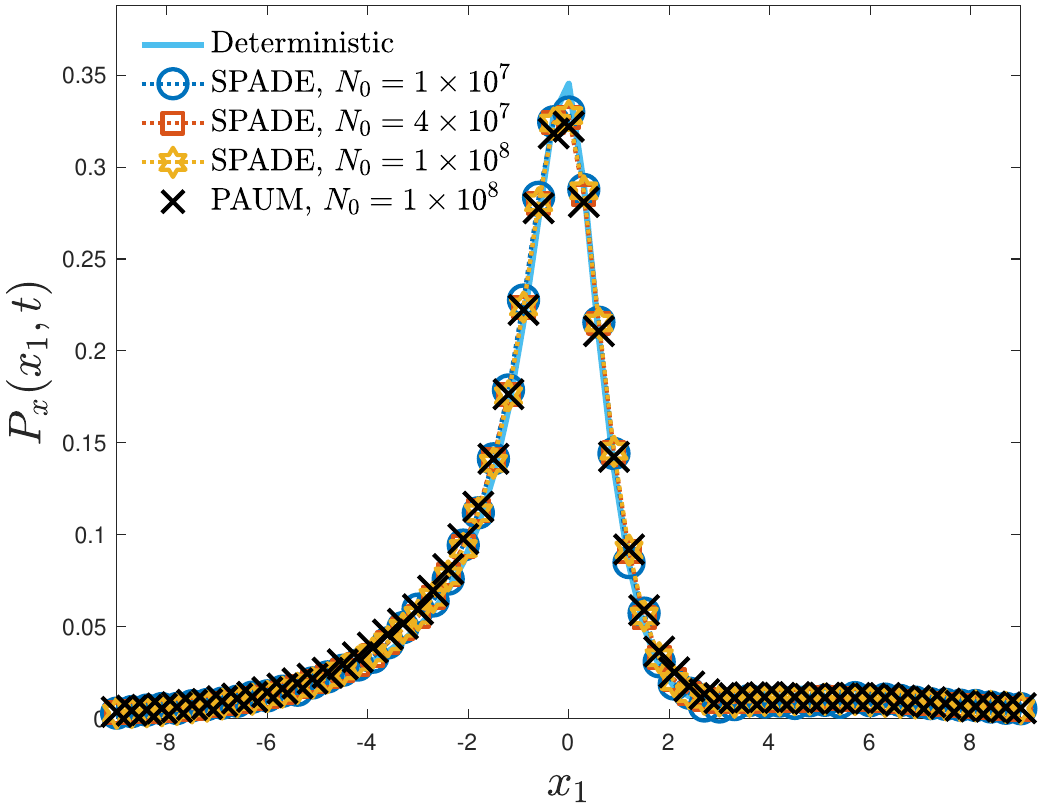}}
   {\includegraphics[width=0.49\textwidth,height=0.26\textwidth]{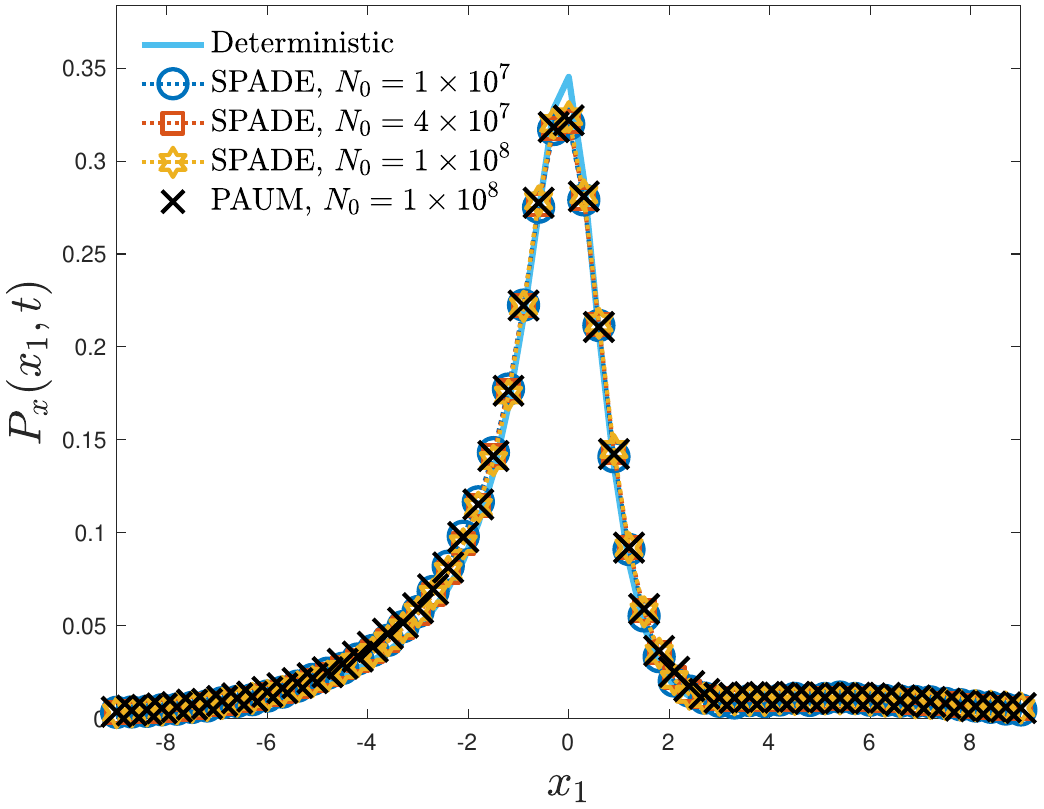}}}
  \\
    \subfigure[$t = 15$a.u. (left: $\vartheta = 0.008$, right: $\vartheta=0.02$).\label{xdist_t15}]{
   {\includegraphics[width=0.49\textwidth,height=0.26\textwidth]{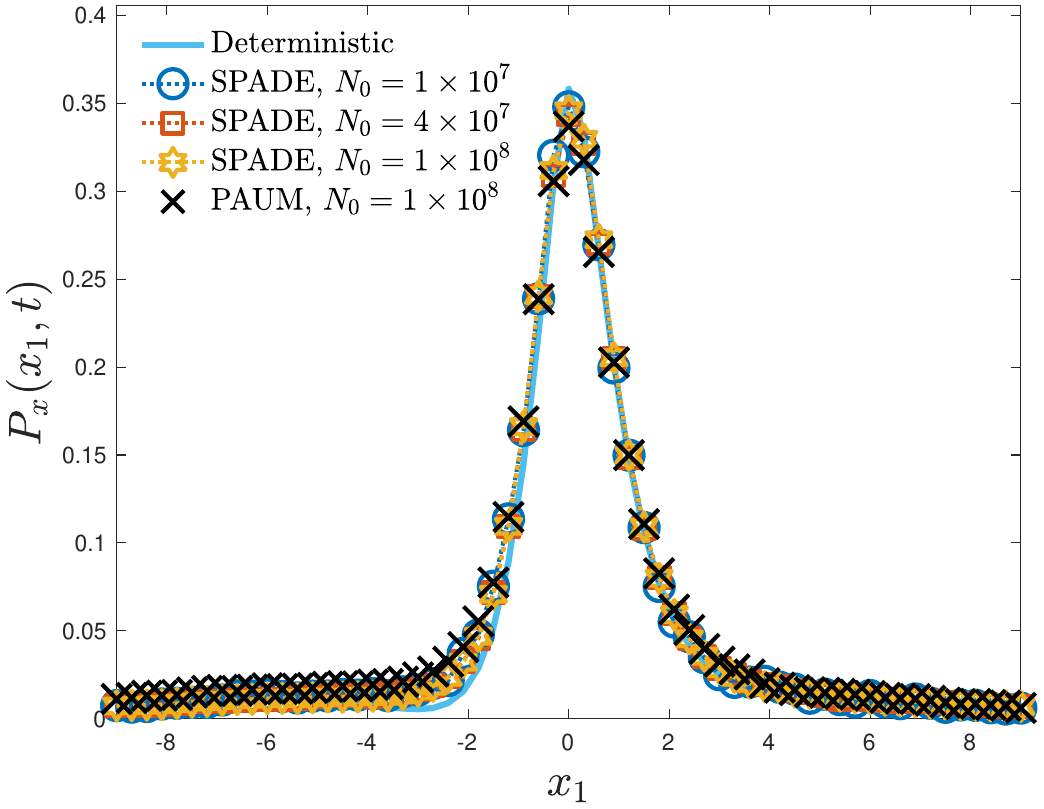}}
   {\includegraphics[width=0.49\textwidth,height=0.26\textwidth]{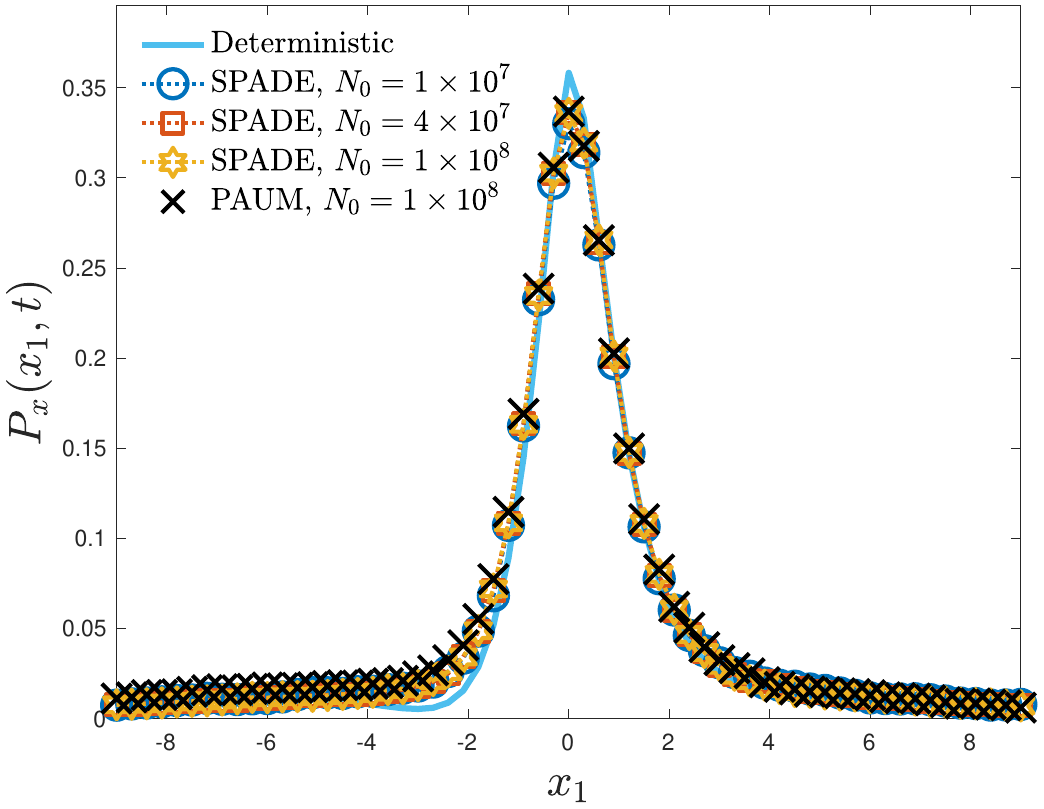}}}
    \caption{\small Comparison of the spatial marginal distribution $P_x(x_1, t)$ produced by the deterministic solver, WBRW-SPA-PAUM and WBRW-SPA-SPADE.  The discrepancies in the crests can be compensated by increasing the sample size or refining the adaptive partition. \label{xdist_time_evolution}}
\end{figure} 

For the sake of comparison, a fixed parameter $\lambda_0 = 4.65$ will be adopted in SPA hereafter. The stochastic errors are presented in 
Fig.~\ref{comparison_SPADE_PAUM_N} by monitoring $\mathcal{E}_2[W_1](t)$ and $\mathcal{E}_2[P_{xy}](t)$, as well as in Fig.~\ref{comparison_SPADE_PAUM} by monitoring $\mathcal{E}_2[W_1](t)$ and $\mathcal{E}_H(t)$. The growth of particle number is plotted in Fig.~\ref{comparison_SPADE_PAUM}. Visualizations of the reduced Wigner function $W_1(x_1, k_1, t)$ and the spatial marginal distribution $P_x(x_1, t)$ are given in Figs.~\ref{qcs_time_evolution} and \ref{xdist_time_evolution}, respectively.  Based on them, we make the following observations.


{\bf Accuracy}: In Figs.~\ref{comparison_SPADE_PAUM_N} and \ref{comparison_SPADE_PAUM}, the results produced by PAUM are set as the baseline (black cross). The sign problem can be largely alleviated when PA is adopted, and can be further suppressed under larger sample size $N_0$. An inspiring finding is that $\mathcal{E}_2[W_1](t)$ and $\mathcal{E}_2[P_{xy}](t)$ under SPADE with $N_0 = 4\times10^7$ or $N_0 = 10^8$ always outperform those under PAUM with $N_0=10^8$, regardless of the choice of $\vartheta$. This actually manifests the advantage of adaptive partitioning over a uniform one, especially when the sample size $N_0$ is relatively small.

{\bf Fluctuation in total energy}: The deviations of total energy are observed in all particle simulations. Although it seems difficult to eliminate them completely due to the mixture of stochastic noises and bias by PA, the fluctuations can be considerably ameliorated when SPADE is used, and can be further improved as either sample size $N_0$ or the partition level $K$ increases. This provides another evidence on the convergence of SPADE.

{\bf Snapshots}: The particle methods with either PAUM or SPADE (under $\vartheta = 0.008$) can properly capture some fine structures of wavepackets as seen in Fig.~\ref{qcs_time_evolution}, including the location of negative valley that manifest the uncertainty principle, and the double-peak structure of wavepacket induced by the Coulomb collisions. Even small oscillating tails can be recovered by SPADE, albeit with small random noises, while the particle solutions by PAUM are evidently more noisy. In Fig.~\ref{xdist_time_evolution}, we also compare the projection $P_{x}(x_1, t)$ and find  their coincidence with the deterministic solutions. Although some discrepancies are observed in their crests and near the left shoulder, they can be alleviated by refining the adaptive partition under smaller $\vartheta$.

It is also observed that the particle reconstructions of $W_1(x_1, k_1, t)$ might not perfectly match the deterministic reference solutions at low contour values, where large errors are concentrated. This is possibly induced by the mixture of errors from (1) the sampling process, (2) the asymptotic error in SPA, (3) the cancelation of particles and (4) the histogram reconstruction. Actually, the particle estimator only approximates the Wigner function by finite weighted points in the weak sense, so that both $W_1(x_1, k_1, t)$ and $P_{xy}(x_1, x_2, t)$ have to be reconstructed by a piecewise constant function \eqref{histogram} with spacings $\Delta x = 0.3, \Delta k = 0.1$. This brings in a smoothing effect as the values of wavepackets are averaged in each bin \cite{Raviart1985}. Nonetheless, there is still a quantitative coincidence between deterministic and stochastic solutions and the difference can be gradually compensated under larger sample size, e.g., see the convergence trend in Figs.~\ref{comparison_SPADE_PAUM_N} and \ref{comparison_SPADE_PAUM}.

{\bf Growth of particle number}:   In Fig.~\ref{comparison_SPADE_PAUM}, the particle number after PAUM grows from $10^8$ to  $6.4\times 10^9$ at $t = 15$a.u. ($K \approx 4.9 \times 10^{10}$). By contrast, SPADE can annihilate particles more efficiently for $N_0\ge 4\times 10^7$. However, for $N_0 \le 10^7$ and  too small $\vartheta$, the particle number still grows rapidly. Such oversampling problem may even hamper the accuracy as the redundant particles carrying stochastic noises are not removed (see the group $N_0= 10^7, \vartheta =0.005$ in Fig.~\ref{comparison_N1000}).


\subsection{How to avoid the oversampling problem} 
\begin{figure}[!h]
    \centering
    \subfigure[Particle number before and after SPADE (left: $\vartheta = 0.01$, right: $\vartheta = 0.02$). \label{bottom_theta}]{{\includegraphics[width=0.49\textwidth,height=0.26\textwidth]{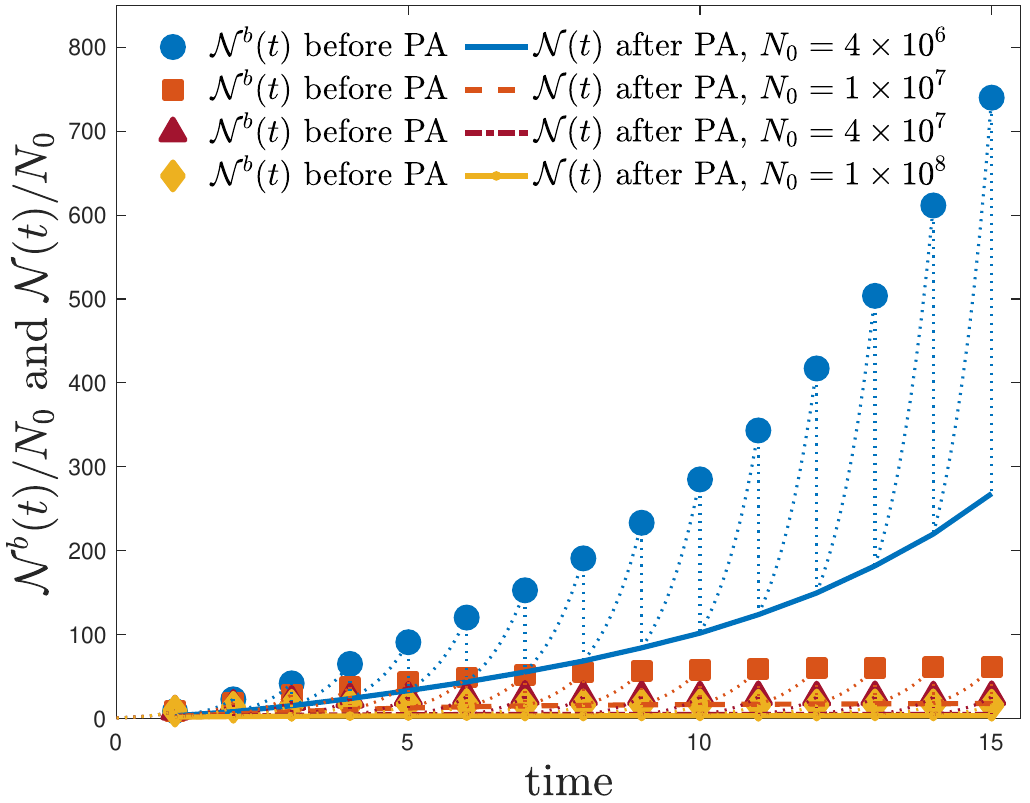}}
     {\includegraphics[width=0.49\textwidth,height=0.26\textwidth]{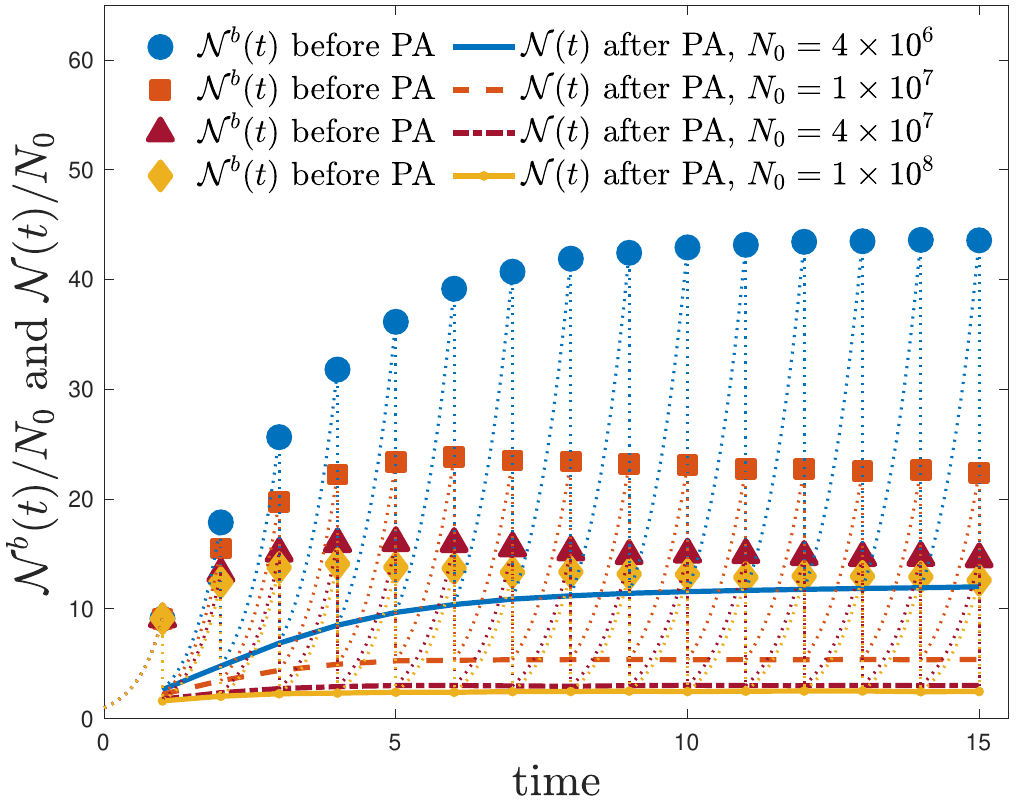}}}
     \\
    \centering
    \subfigure[Particle number before and after SPADE (left: $N_0 = 4\times 10^7$, right: $N_0 = 1\times 10^8$). \label{bottom_N}]{{\includegraphics[width=0.49\textwidth,height=0.26\textwidth]{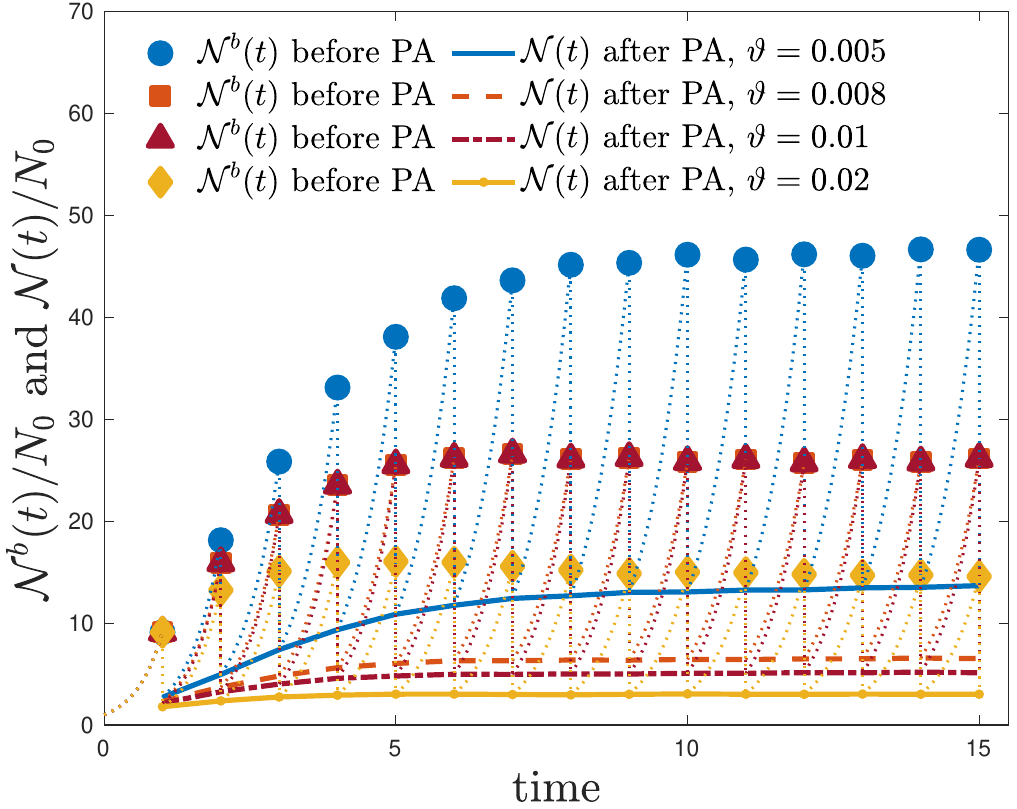}}
     {\includegraphics[width=0.49\textwidth,height=0.26\textwidth]{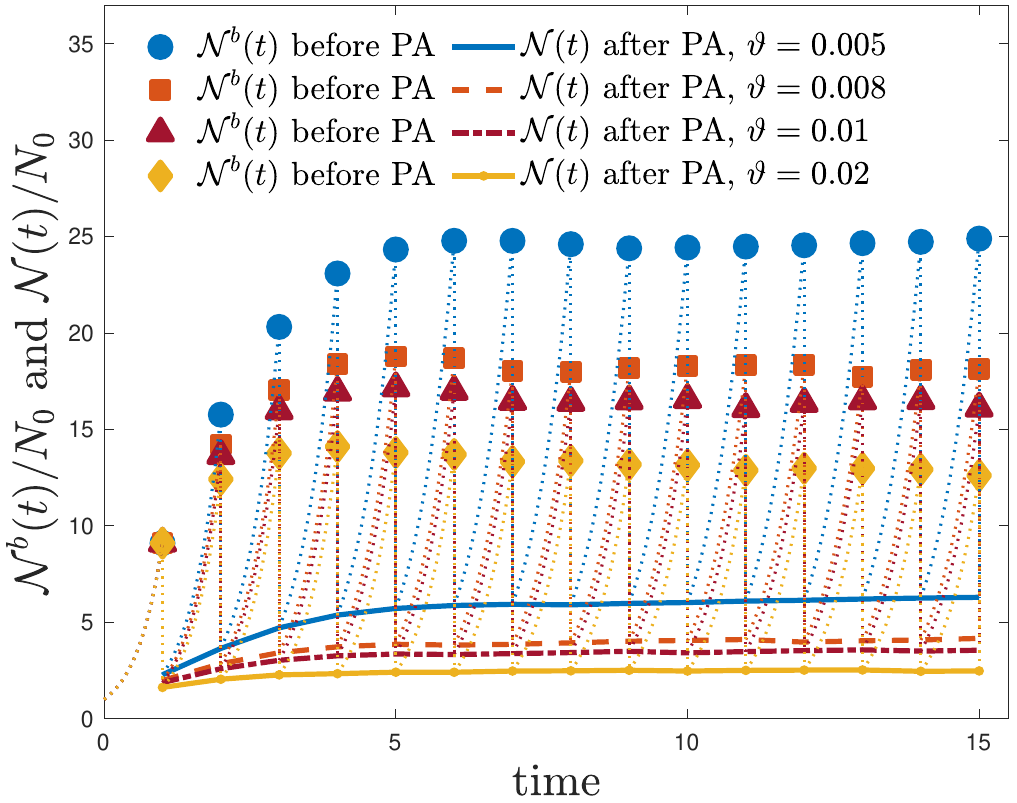}}}
         \centering
    \subfigure[$K$ is inversly proportional to $\vartheta$. \label{relation_K_theta}]{\includegraphics[width=0.49\textwidth,height=0.26\textwidth]{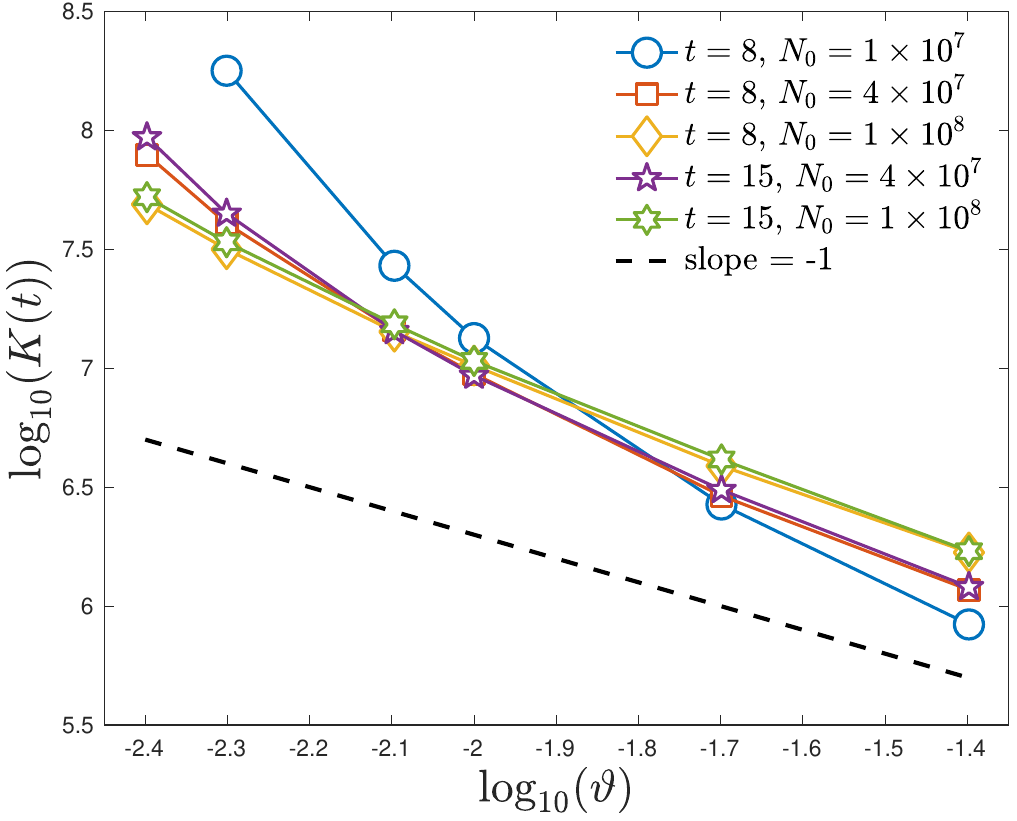}}
    \subfigure[$K$ is proportional to $\frac{P(t)M(t)}{(P(t)+M(t))\sqrt{N_0}}$. \label{convergence_K}]{\includegraphics[width=0.49\textwidth,height=0.26\textwidth]{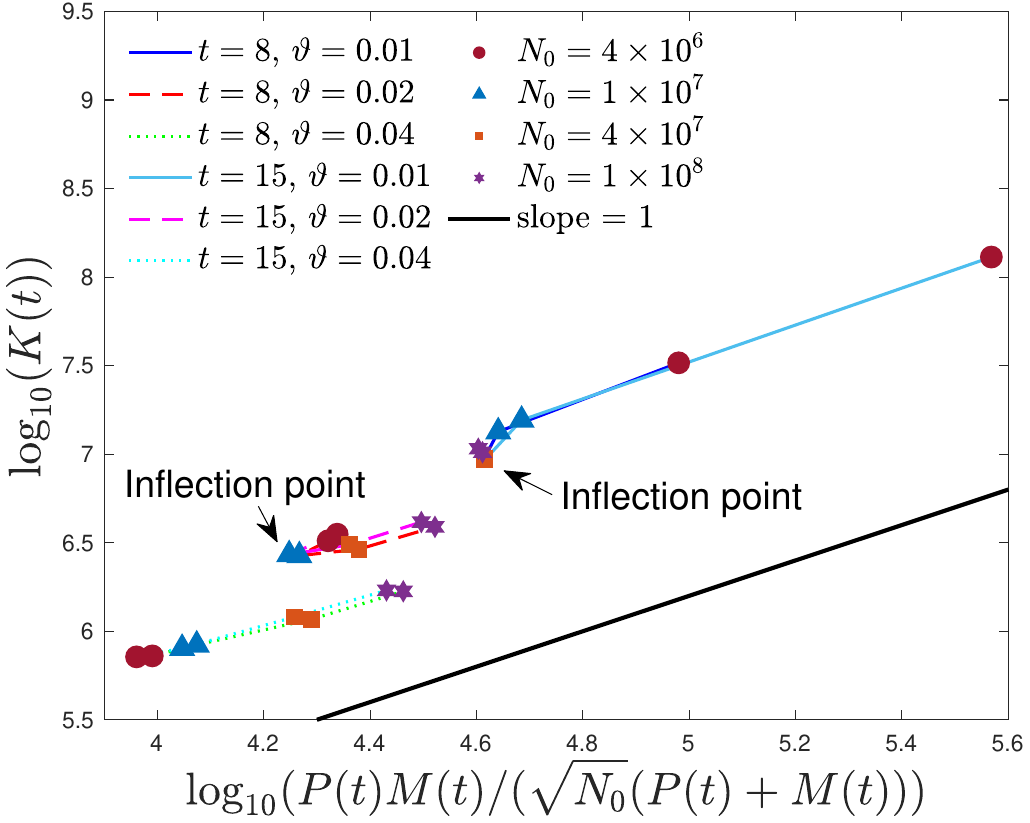}}
    \caption{\small Growth ratio of particle number before and after PA: The bottom line structure indicates that SPADE learns the minimal amount of particles that accurately captures the non-classicality of the Wigner function under the prescribed sample size $N_0$. The partition level $K(t)$ is expected to be inversely proportional to $\vartheta$  and proportional to $\frac{P(t)M(t)}{(P(t)+M(t))\sqrt{N_0}}$, while the inflection point is an indicator of the oversampling problem.
    \label{bottomline_structure}}
\end{figure} 

As already seen in Fig. \ref{comparison_SPADE_PAUM}, SPADE is capable to control the numerical sign problem if an appropriate parameter $\vartheta$ is adopted, and the total particle number after annihilation can stay at a stable level. But SPADE still  suffers from the oversampling problem under too small $N_0$ and  $\vartheta$. Thus it is mandatory to propose a strategy to avoid the oversampling wisely, which in turn requires to understand how the partition level $K(t)$ relies on the particle number $P(t)$, $M(t)$, the parameters $\vartheta$ and $N_0$.

{\bf Bottom line for particle number}: The efficiency of SPADE can be characterized by its capability to maintain the accuracy, as well as the non-classicality, with minimal amount of particles. The physical motivation is that the Wigner function is bounded below and above and only allows finite negative values \cite{bk:CurtrightFairlieZachos2013}. Thus the size of signed particles reflecting non-classicality should be limited within a stable level. 

According to Figs.~\ref{bottom_theta} and \ref{bottom_N}, the particle number, although growing exponentially at each step, always returns back to a stable level after annihilation (except the group $N_0 = 4\times 10^6$, $\vartheta = 0.008$). This is called the bottom line structure as also observed in PAUM \cite{ShaoSellier2015}. Definitely, the bottom line must be larger than $N_0$ to properly account for the negative part of the Wigner function. When the particle number after PA attains the bottom line, the exponential growth of numerical errors can also be successfully suppressed (see Fig.~\ref{comparison_SPADE_PAUM_N}).

{\bf Oversampling}: The exceptions are the groups $N_0= 4\times10^6,1\times 10^7, \vartheta < 0.01$, where particle numbers grow even faster than PAUM. For instance, when $N_0 = 4\times 10^6$, $\vartheta = 0.01$, the growth ratio almost reaches $800$ at $15$a.u. and the bottom line also dramatically ascends. Such oversampling problem is induced by the over-partitioning of the adaptive clustering, so that more and more particles are generated without being canceled out. However, this do not necessarily bring in improvements in  accuracy. As presented in Fig.~\ref{comparison_SPADE_PAUM_N},  $\mathcal{E}_2[P_{xy}]$ can be gradually improved by refining the partitioning, but $\mathcal{E}_2[W_1]$ seem to reach its limit and even become slightly worse as $\vartheta$ increases. Definitely, the oversampling of signed particles may result in a rapid increase in computational time (see Table \ref{cpu_time}) and should be avoided.

 {\bf Large sample size alleviates oversampling}:  Fortunately, SPADE can get rid of the oversampling problem by simply increasing the sample size $N_0$.  According to Figs.~\ref{comparison_SPADE_PAUM} and \ref{bottomline_structure}, the particle number always remains at a stable level under $N_0 = 10^8$ regardless of $\vartheta$ (even for $\vartheta = 0.004$), indicating that the redundant sampling can be avoided. From Table \ref{cpu_time}, the computational time of the group  $\vartheta = 0.01, N_0 = 1\times 10^8$ is even less than that under $\vartheta = 0.01, N_0 = 4\times 10^6$, while the accuracy of the former significantly outperforms the latter. 

\begin{table}[!h]
  \centering
  \caption{\small Total wall time (in hours) of SPADE and the average partition levels (avg $K$) for 6-D simulations up to $15$a.u. The groups $N_0=4\times 10^6, \vartheta = 0.008$ and $N_0=1\times 10^7, \vartheta=0.005$ are simulated up to $10$a.u. because of the oversampling problem. \label{cpu_time}}
 \begin{lrbox}{\tablebox}
  \begin{tabular}{c|c|c|c|c|c|c|c|c}
\hline\hline
$N_0$		&      \multicolumn{2}{c|}{$4\times10^6$} & \multicolumn{2}{c|}{$1\times10^7$} & \multicolumn{2}{c|}{$4\times10^7$} & \multicolumn{2}{c}{$1\times10^8$}\\
\hline
$\vartheta$ &	Time		&	avg $K$ 		&	Time		&	avg $K$		&	Time		&	 avg $K$ 		&	Time		&	 avg $K$\\
\hline
0.004	&	-		&	-				&	-		&		-			&	46.17 	&	6.79$\times10^7$	&	48.61 	&	4.48$\times10^7$	\\
0.005	&	-		&	-				&	(22.76) 	&	(9.0$\times10^7$)	&	27.17 	&	3.56$\times10^7$	&	36.93 	&	2.96$\times10^7$	\\
0.008	&	(9.85) 	&	(4.2$\times10^7)$	&	14.39 	&	2.49$\times10^7$	&	13.92 	&	1.29$\times10^7$	&	24.69 	&	1.34$\times10^7$	\\
0.01		&	21.02 	&	4.43$\times10^7$	&	7.73	 	&	1.17$\times10^7$	&	10.95 	&	8.61$\times10^6$	&	21.42 	&	9.66$\times10^6$	\\
0.02		&	2.04 		&	2.85$\times10^6$	&	2.67 		&	2.46$\times10^6$	&	7.22 		&	2.97$\times10^6$	&	15.55 	&	3.81$\times10^6$	\\
0.04		&	0.86 		&	6.79$\times10^5$	&	1.71 		&	7.88$\times10^5$	&	5.84 		&	1.18$\times10^6$	&	13.88 	&	1.65$\times10^6$	\\
\hline\hline
 \end{tabular}
\end{lrbox}
\scalebox{0.92}{\usebox{\tablebox}}
\end{table}

{\bf Efficiency of adaptive partition}: In order to dig out the relation between $K(t)$ and the parameter $\vartheta$, we plot $\vartheta$-$K(t)$ curve in Fig.~\ref{relation_K_theta}. When $N_0$ becomes larger, $K(t)$ tends to be inversely proportional to $\vartheta$. In addition, we also plot the relation between $K(t)$ and $\frac{P(t)M(t)}{(P(t)+M(t))\sqrt{N_0}}$ under different $\vartheta$ in Fig.~\ref{convergence_K} and find that they are almost linearly dependent for $\vartheta = 0.04$ and $\vartheta = 0.02$ (except $N_0=4\times10^6$). These observations explain the meaning of the lower and upper bounds of $K(t)$. The oversampling  is avoided when $K(t)$ is close to the lower bound (see the line $K \propto \frac{P(t)M(t)}{(P(t)+M(t)\sqrt{N_0}}$), but occurs when $K(t)$ approaches the upper bound. The inflection points of the V-shape curve in Fig.~\ref{convergence_K}, say, $N_0=1\times10^7$ for $\vartheta =0.02$ and $N_0=4\times10^7$ for $\vartheta =0.01$,  are indicators for the presence of oversampling, which coincides with the trend in Fig.~\ref{bottom_theta}.

\subsection{Parallel implementation}

Domain decomposition $\Omega = \bigcup_{p=1}^{N_p} \Omega_p$ is a pretreatment for distributed-memory implementation. By dividing a tree into a forest composed of $N_p$ independent trees, the adaptive partitions can be established independently in $N_p$ processors. This also splits particles into $N_p$ batches, and $\mathcal{S}_{\textup{out}} = \bigcup_{p = 1}^{N_p} \mathcal{S}_{\textup{out}}^{(p)}$ with $ \mathcal{S}_{\textup{out}}^{(p)}$ particles in $\Omega_p$ manipulated by the $p$-th processor. A relevant point is to strike a balance in overload. To this end, one shall keep the particle number in each $\Omega_p$ more or less the same.

All simulations via our Fortran implementations run on the High-Performance Computing Platform of Peking University: 2*Intel Xeon E5-2697A-v4 (2.60GHz, 40MB Cache, 9.6GT/s QPI Speed, 16 Cores, 32 Threads) with 256GB Memory $\times$ 16. To the best of our knowledge, this is also the first attempt to simulate the 6-D Wigner dynamics via the massively parallel deterministic solver or the stochastic particle method with PAUM.

\begin{itemize}

\item[(1)] The deterministic Wigner simulation was realized via a mixture of MPI and OpenMP library. The domain was decomposed to $4\times 4\times 4$ patches and each task used $7$ threads (448 cores).  It spent about 15 days to reach $T= 15$a.u.

\item[(2)] WBRW-SPA-PAUM under $N_0 = 10^8$ was also realized via a mixture of MPI and OpenMP library. The domain was decomposed to $4\times 4\times 4$ patches and each task used $7$ threads (448 cores). It spent about $50$ hours to reach $T= 15$a.u. 

\item[(3)] Each task of WBRW-SPA-SPADE used $128$ cores and was realized via MPI library. The wall time for SPADE up to $15$a.u., which occupies more than $95\%$ of total wall time, is recorded in Table \ref{cpu_time}. For 6-D problems, the advantage of particle-based stochastic methods over grid-based deterministic counterparts becomes prominent. For $\vartheta \ge 0.01$,  the wall time scales almost linearly on the sample size $N_0$. But it grows rapidly when the oversampling problem occurs.

\end{itemize}

\section{Particle simulations of 12-D Wigner quantum dynamics}
\label{sec.num_12d}

The readers may be curious about whether SPADE is applicable in higher dimensional problem. Here we would like to demonstrate the potential of SPADE for  D $=$ 12 by solving the proton-electron Wigner equation \eqref{eq.Wigner} with $\pdo$ \eqref{pdo_scatter_manybody}.  

\subsection{The localized proton-electron Wigner dynamics} 
For the convenience of benchmarks, we first consider a specific model that has a quasi-analytical solution for a short time.
 \begin{example}\label{example3}
   \textup{ Consider a system composed of one proton and one electron interacting under the Coulomb potential. The initial Wigner function $f(\bx_e, \bx_p, \bk_e, \bk_p, 0)$ is an uncorrelated Gaussian function, with centers $ \bR = (1, 0, 0)$ and $\bx_A = (0, 0, 0)$,
\begin{equation}
f(\bx_e, \bx_p, \bk_e, \bk_p, 0) = \frac{1}{\pi^{6}} \me^{-\frac{1}{2}|\bx_e - \bR|^2 - 2|\bk_e|^2} \me^{-\frac{1}{2\varepsilon^2} |\bx_p - \bx_A|^2 - 2\varepsilon^2 |\bk_p|^2}.
\end{equation}
}
\end{example}
Suppose the proton is strongly localized in $\bx_p$-space and omit small terms for $m_p \approx 1836 m_e$, then $f_e(\bx_e, \bk_e, t)$ in Eq.~\eqref{single_body_Wigner} becomes a quasi-analytical solution to the reduced electron Wigner function for a short time. Detailed derivations are put in our arXiv note.

\begin{figure}[!h]
    \centering
    \subfigure[$l^2$-error for $W_1$. \label{12d_error_W}]{\includegraphics[width=0.32\textwidth,height=0.22\textwidth]{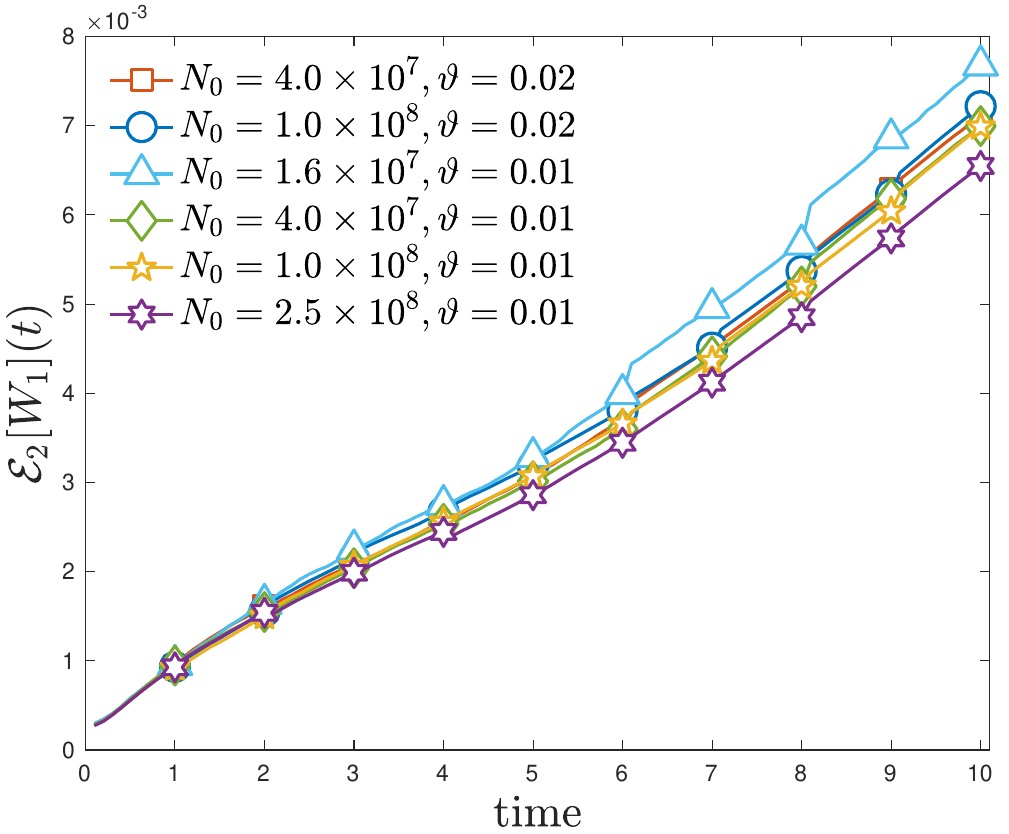}}
    \subfigure[$l^2$-error for $P_{xy}$.\label{12d_error_P}]{\includegraphics[width=0.32\textwidth,height=0.22\textwidth]{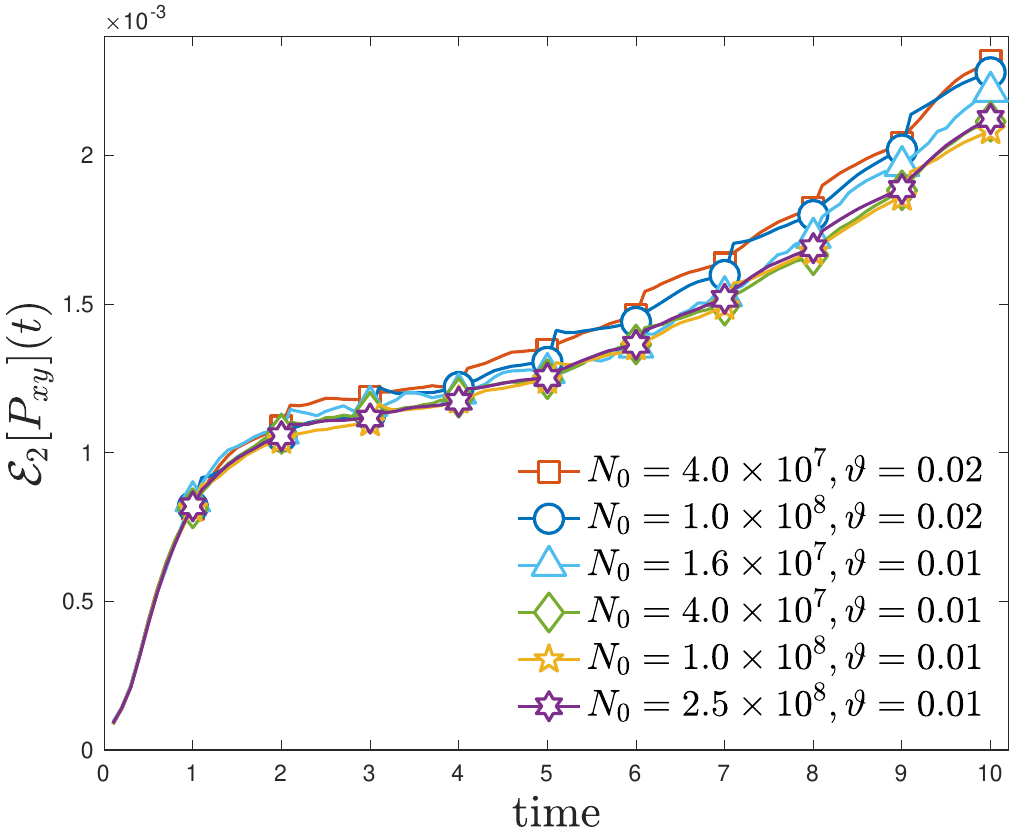}}
         \subfigure[Deviation in energy. \label{12d_energy}]{\includegraphics[width=0.32\textwidth,height=0.22\textwidth]{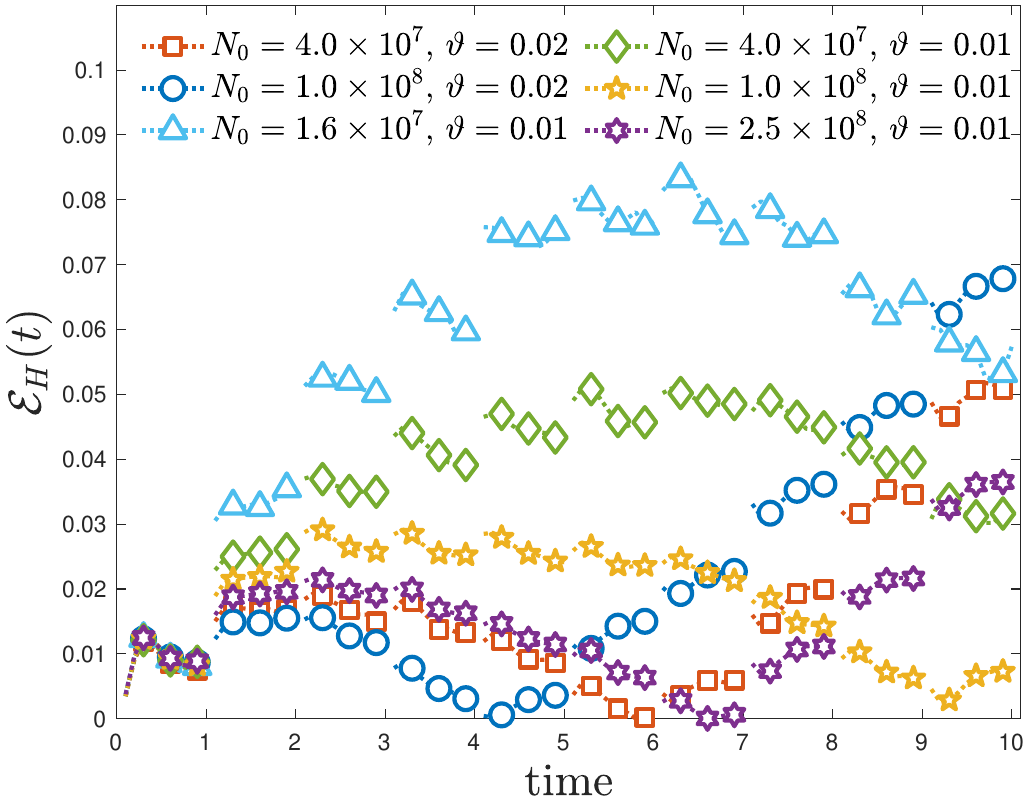}}
    \\
    \subfigure[$N^b(t)$ and $\mathcal{N}(t)$.\label{Gbeam12d_particle}]{\includegraphics[width=0.49\textwidth,height=0.26\textwidth]{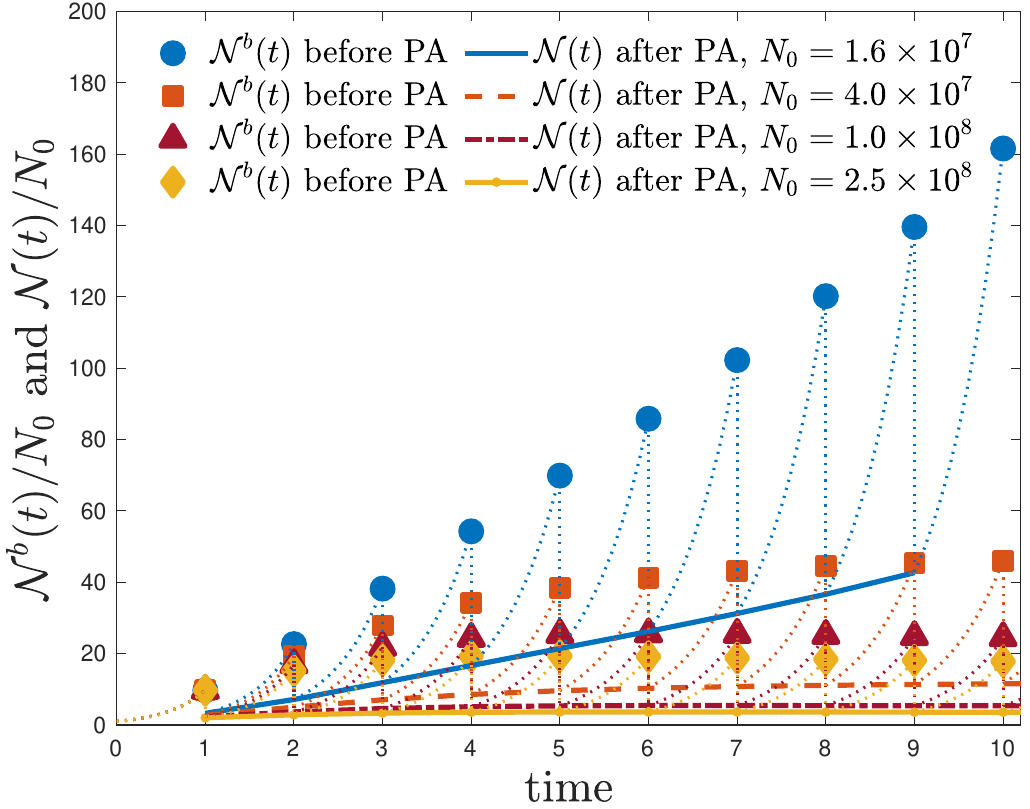}}
    \subfigure[$K(t)$ and $\frac{P(t)M(t)}{(P(t)+M(t))\sqrt{N_0}}$. \label{Gbeam12_K}]{\includegraphics[width=0.49\textwidth,height=0.26\textwidth]{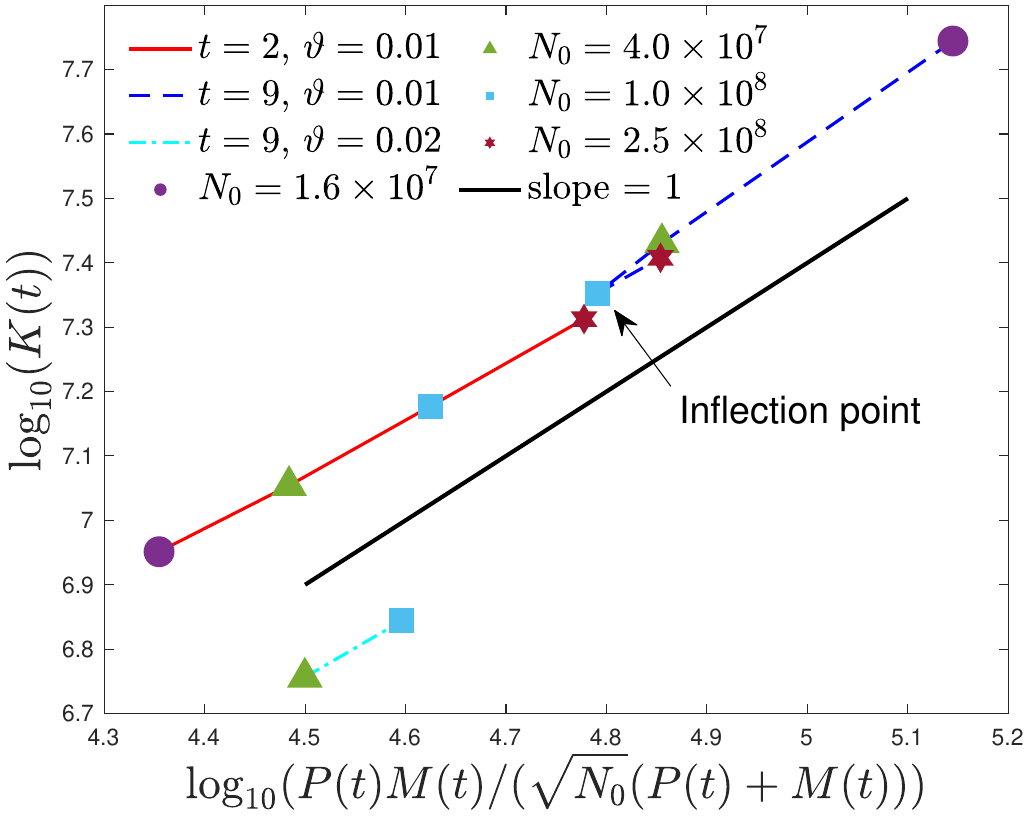}}
    \caption{\small The time evolution of the $l^2$-errors $\mathcal{E}_2[W_1](t)$, $\mathcal{E}_2[P_{xy}](t)$ and the deviation of total energy (the initial value is $1.735$a.u.) in 12-D simulations, as well as the growth of particle number and partition level. The oversampling problem can be avoided as $N_0$ increases. \label{12d_result}}
\end{figure}
\begin{figure}[!h]
\centering
\subfigure[$t=4$a.u.  Fixed proton (left) and $W_1(x_{e,1}, k_{e,1}, t)$ under $N_0=4\times10^7$, $10^8$ and $2.5\times10^8$.]{
{\includegraphics[width=0.23\textwidth,height=0.18\textwidth]{./redist_CHASM_T4.pdf}}
{\includegraphics[width=0.23\textwidth,height=0.18\textwidth]{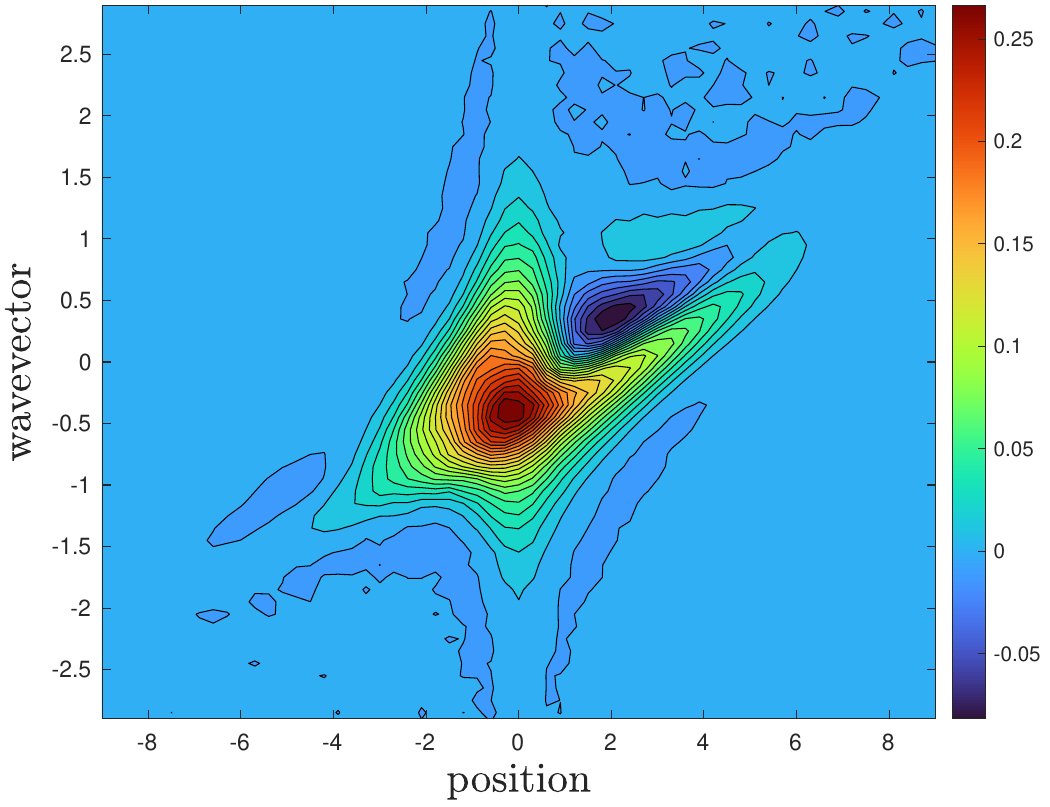}}
{\includegraphics[width=0.23\textwidth,height=0.18\textwidth]{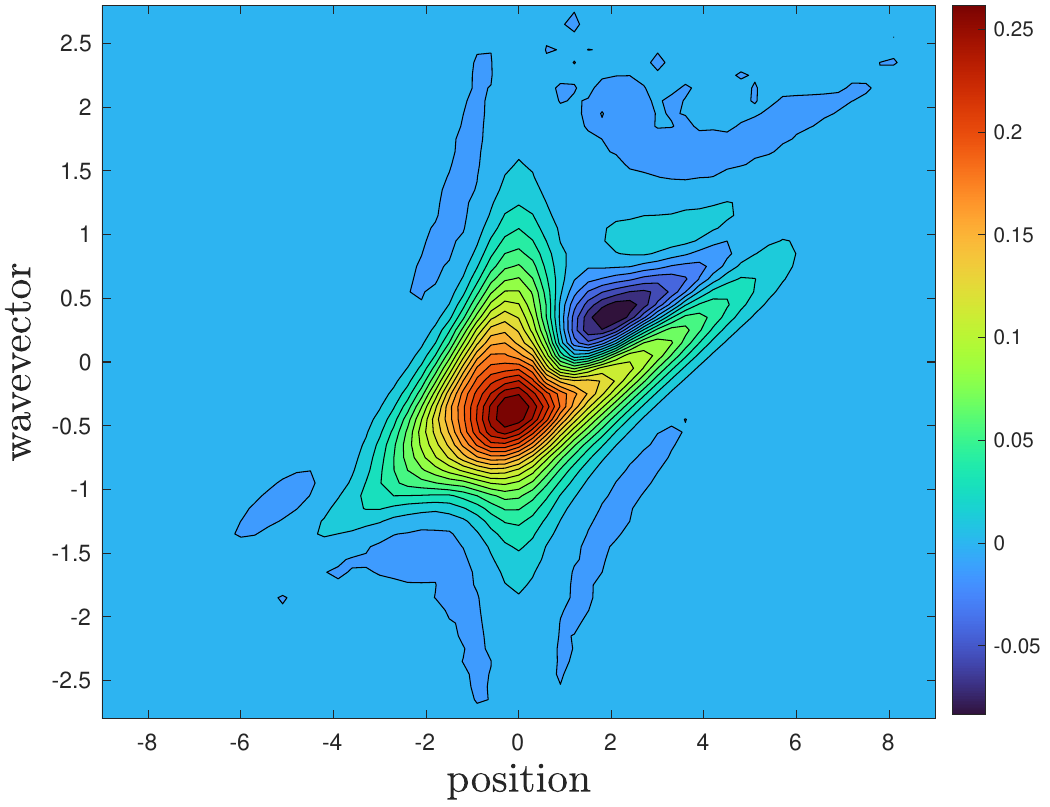}}
{\includegraphics[width=0.23\textwidth,height=0.18\textwidth]{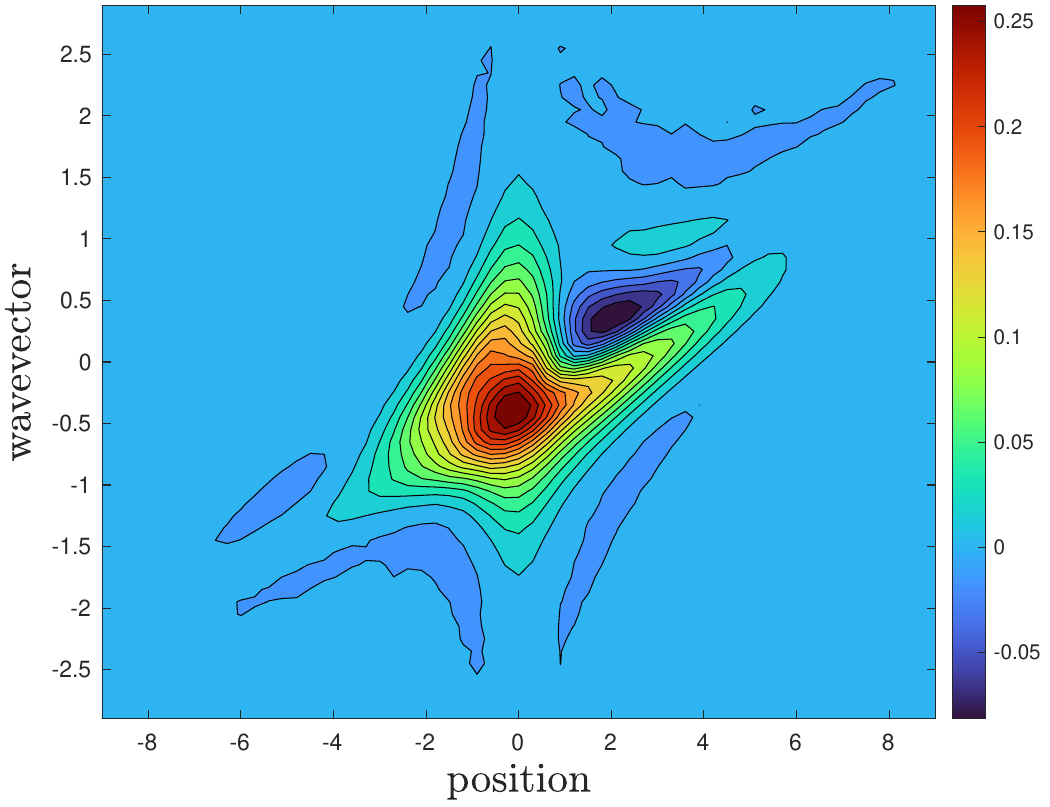}}}
\\
\centering
\subfigure[$t=8$a.u.  Fixed proton (left) and $W_1(x_{e,1}, k_{e,1}, t)$ under $N_0=4\times10^7$, $10^8$ and $2.5\times10^8$.]{
{\includegraphics[width=0.23\textwidth,height=0.18\textwidth]{./redist_CHASM_T8.pdf}}
{\includegraphics[width=0.23\textwidth,height=0.18\textwidth]{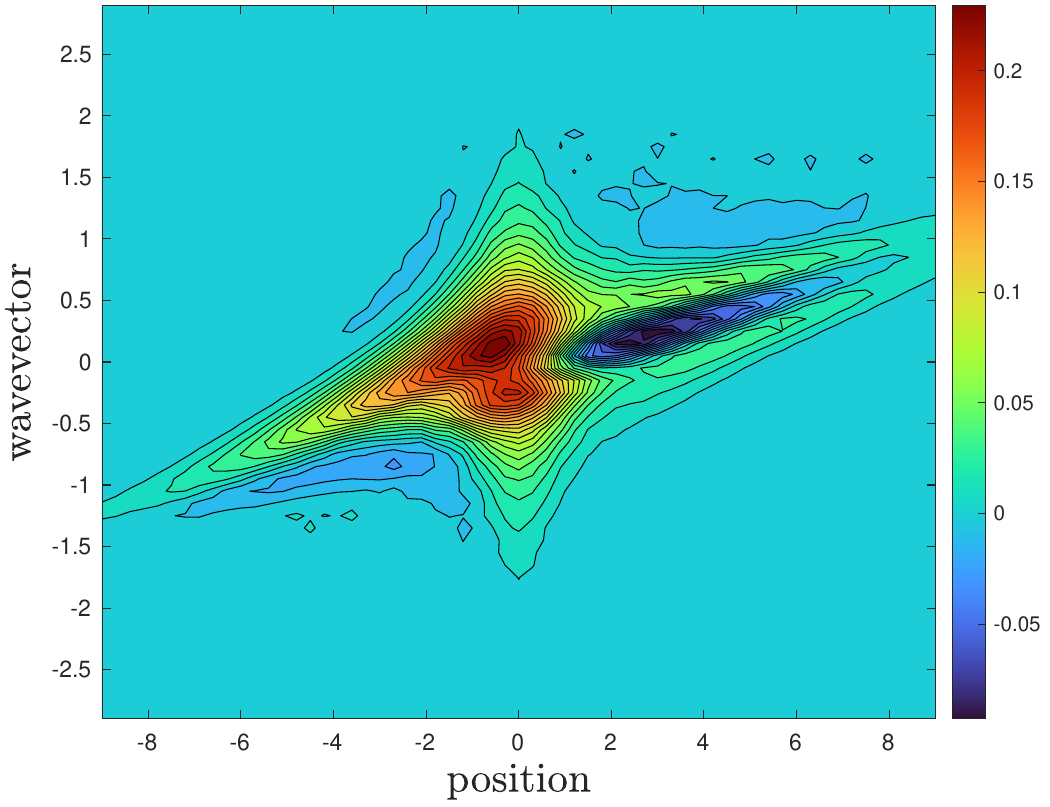}}
{\includegraphics[width=0.23\textwidth,height=0.18\textwidth]{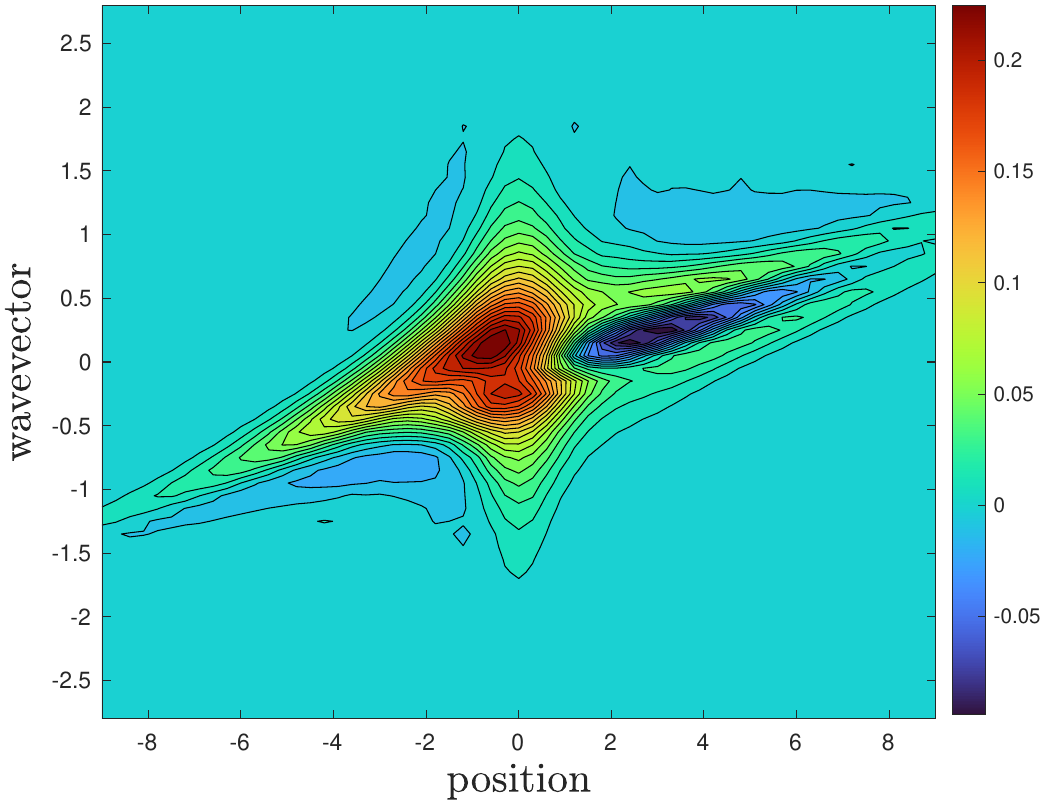}}
{\includegraphics[width=0.23\textwidth,height=0.18\textwidth]{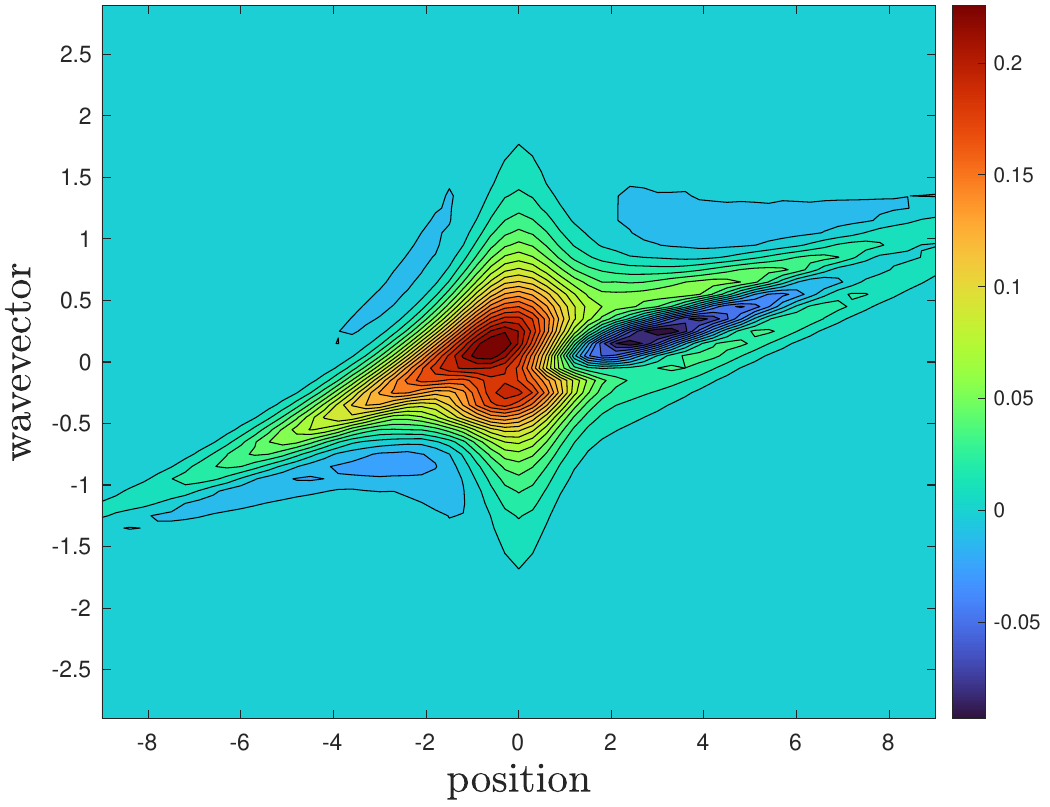}}}
\\
\centering
\subfigure[$t=10$a.u.  Fixed proton (left) and $W_1(x_{e,1}, k_{e,1}, t)$ under $N_0=4\times10^7$, $10^8$ and $2.5\times10^8$.]{
{\includegraphics[width=0.23\textwidth,height=0.18\textwidth]{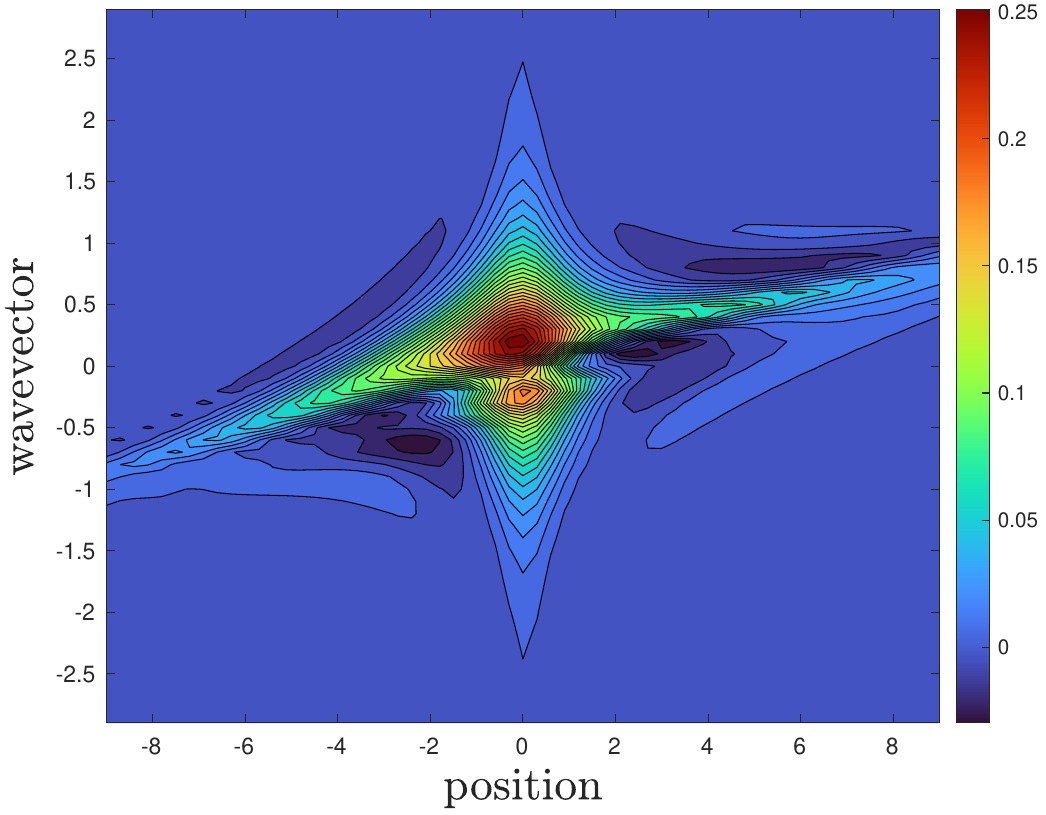}}
{\includegraphics[width=0.23\textwidth,height=0.18\textwidth]{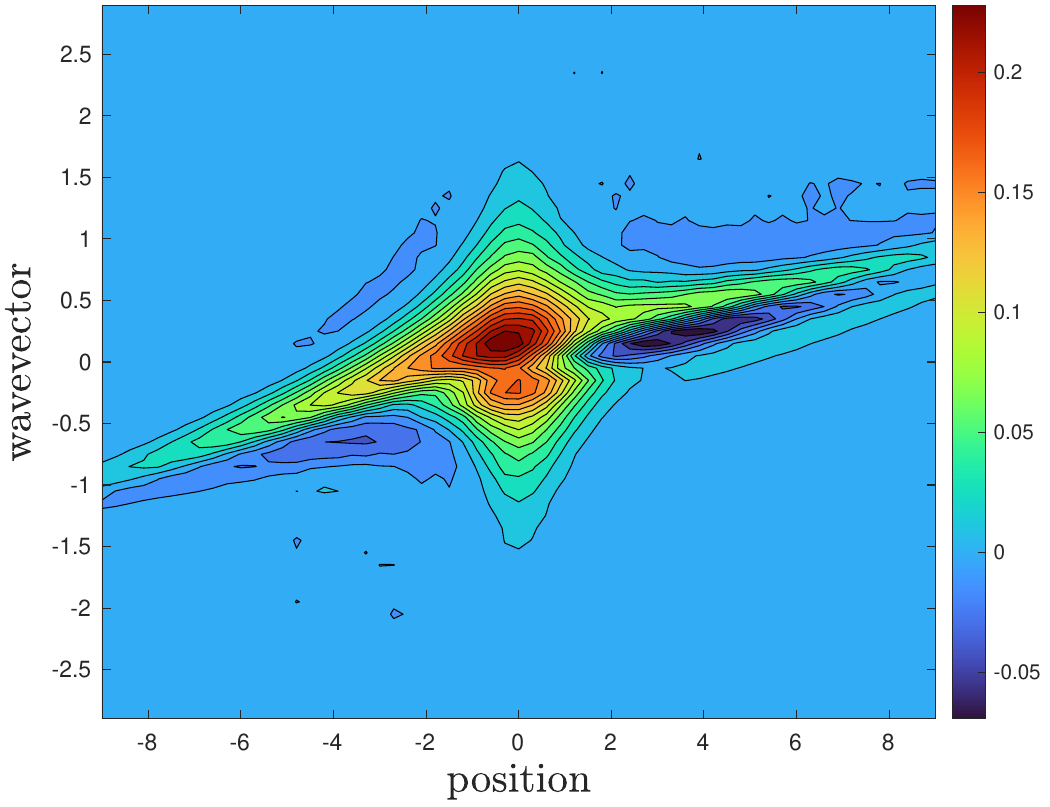}}
{\includegraphics[width=0.23\textwidth,height=0.18\textwidth]{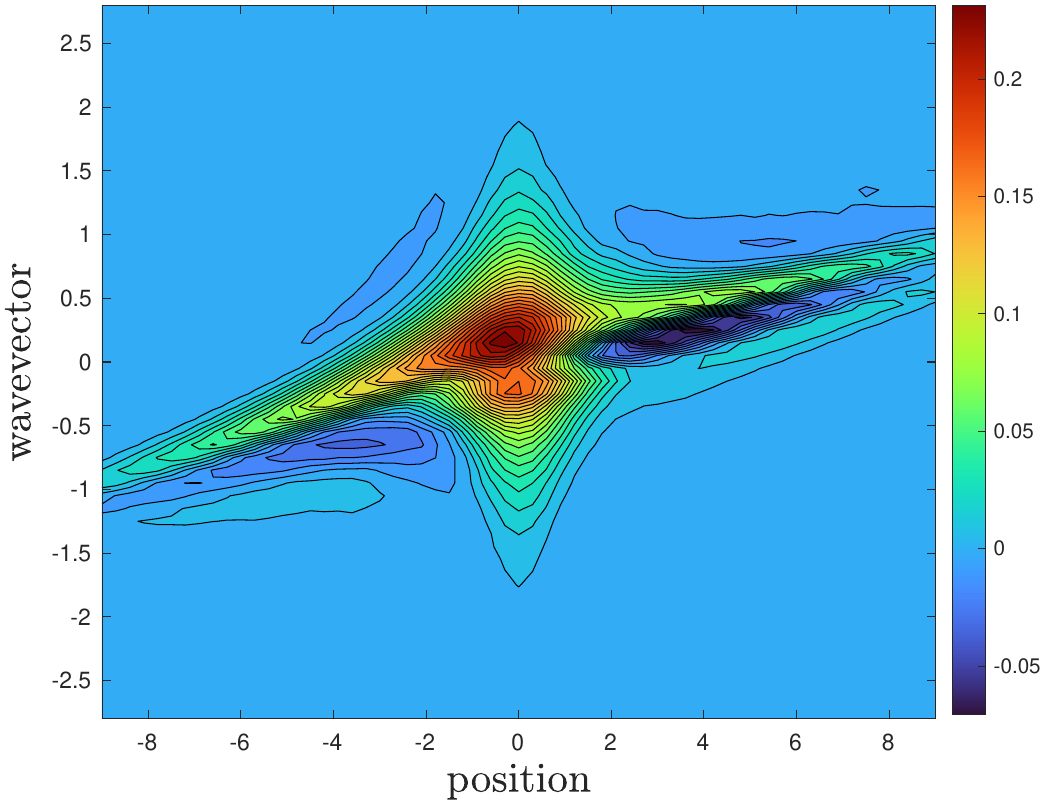}}
{\includegraphics[width=0.23\textwidth,height=0.18\textwidth]{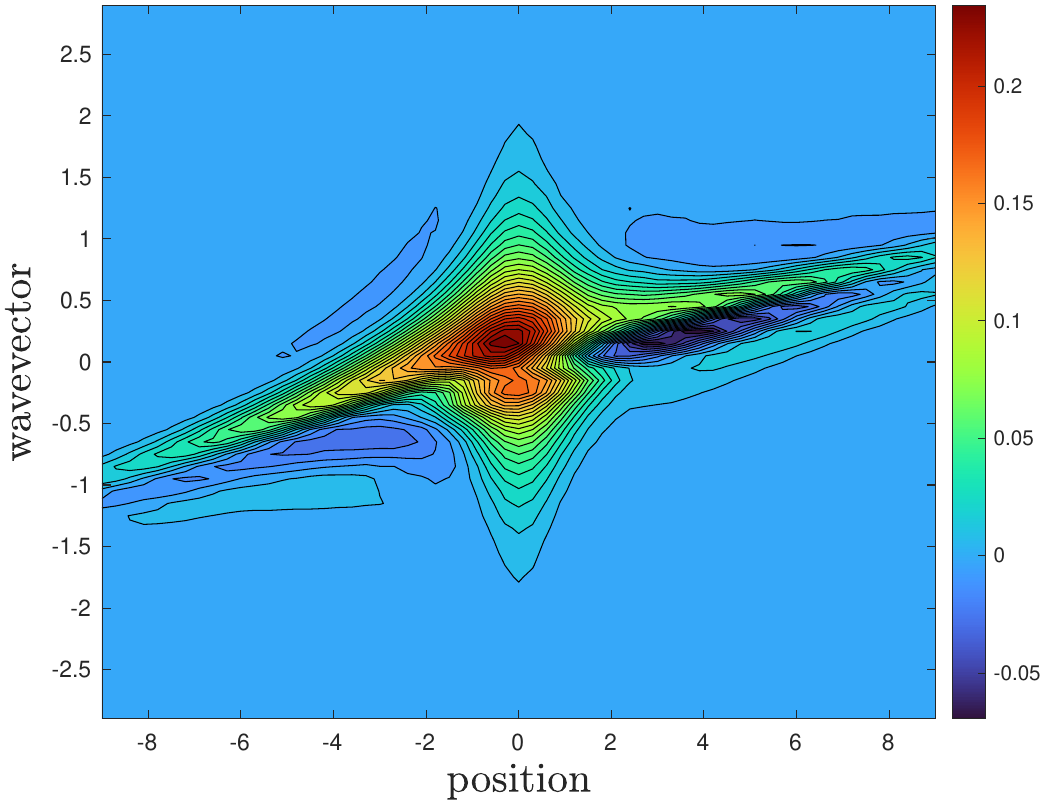}}}
\\
\centering
\subfigure[$P_{xy}(x_{e,1}, x_{e,2},t)$ and $P_x(x_{e,1},t )$ at $1$a.u.]{
{\includegraphics[width=0.23\textwidth,height=0.18\textwidth]{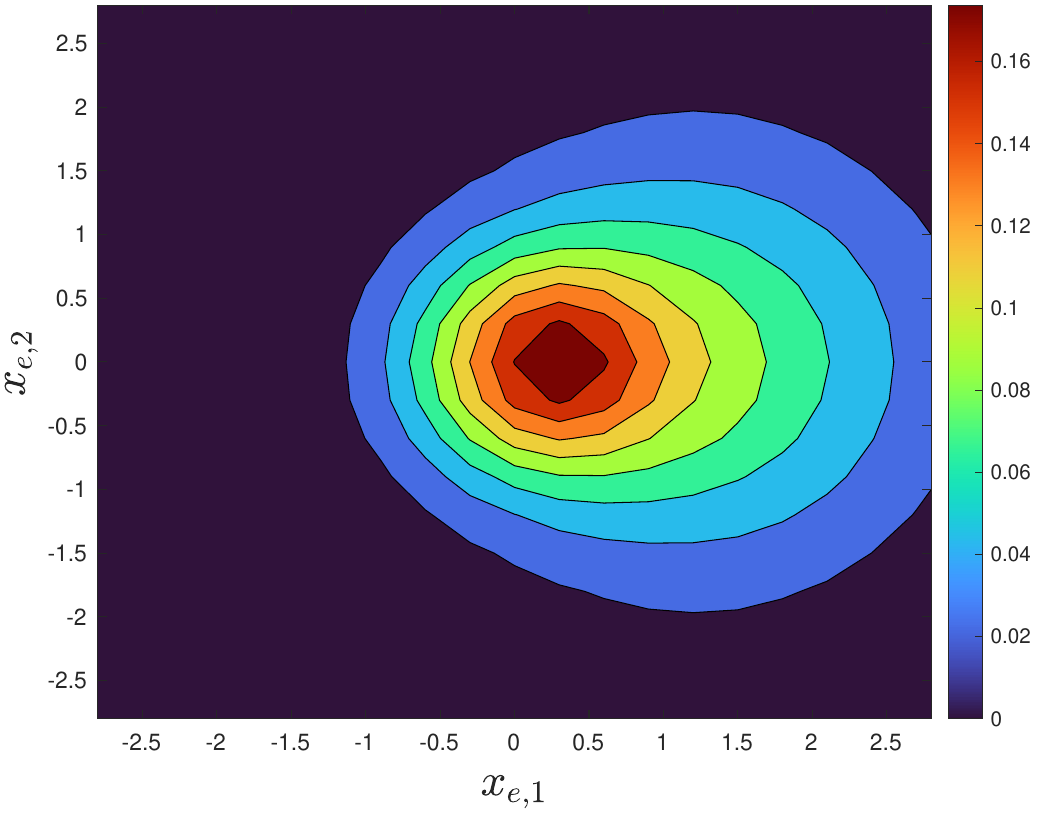}}
{\includegraphics[width=0.23\textwidth,height=0.18\textwidth]{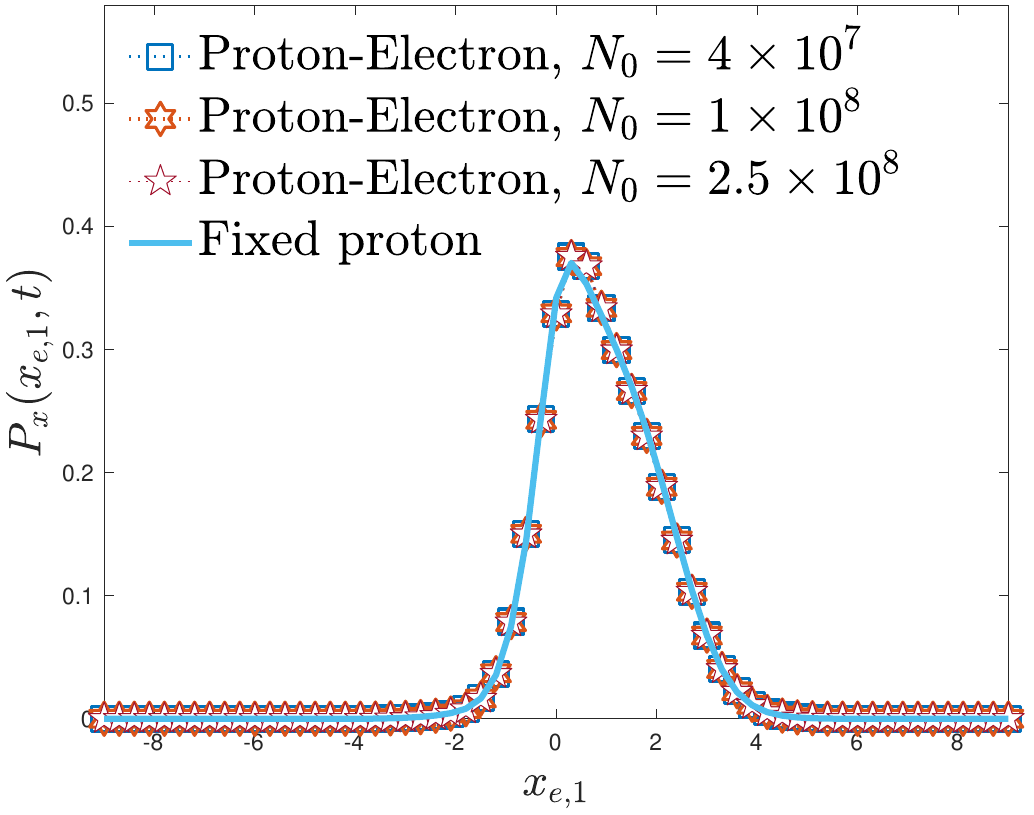}}}
\subfigure[$P_{xy}(x_{e,1}, x_{e,2},t)$ and $P_x(x_{e,1},t )$ at $2.5$a.u.]{
{\includegraphics[width=0.23\textwidth,height=0.18\textwidth]{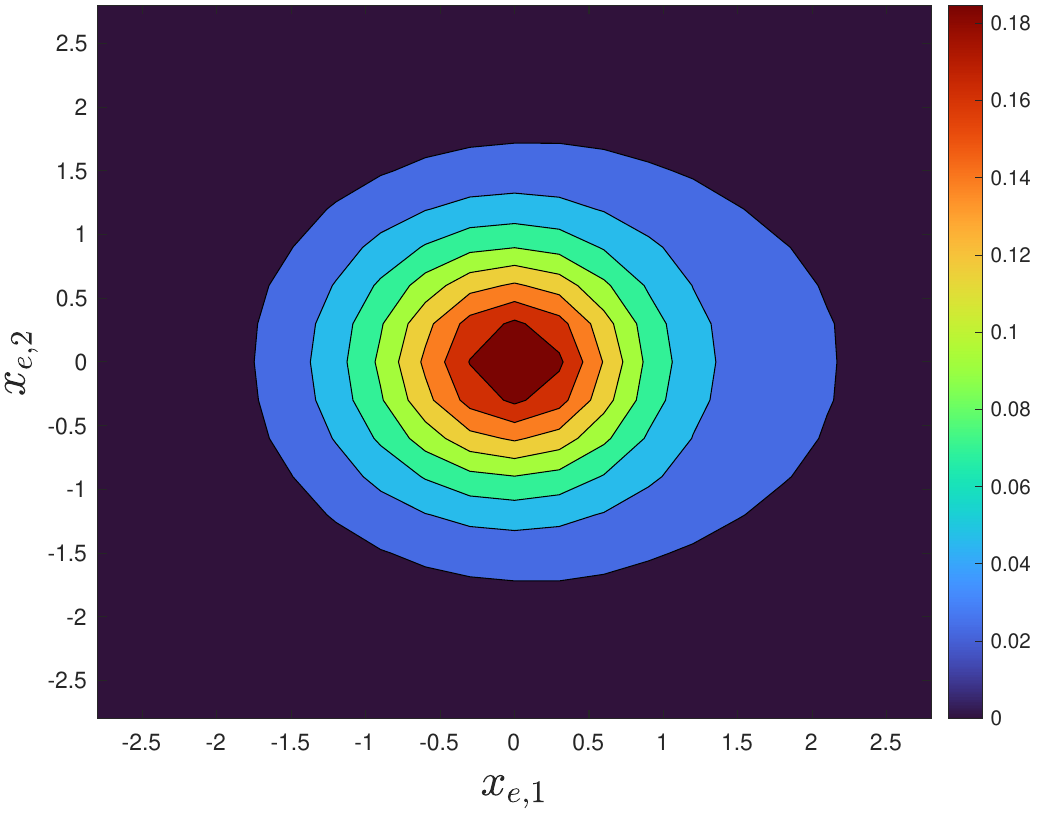}}
{\includegraphics[width=0.23\textwidth,height=0.18\textwidth]{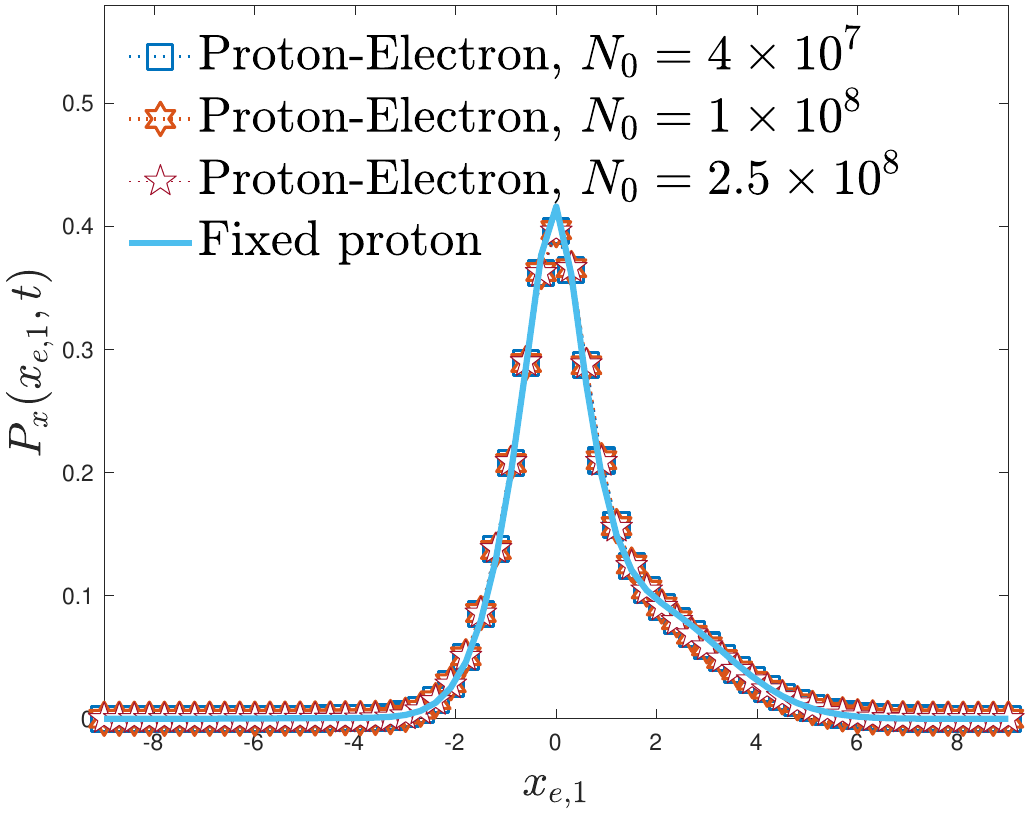}}}
\\
\centering
\subfigure[$P_{xy}(x_{e,1}, x_{e,2},t)$ and $P_x(x_{e,1},t )$ at $4$a.u.]{
{\includegraphics[width=0.23\textwidth,height=0.18\textwidth]{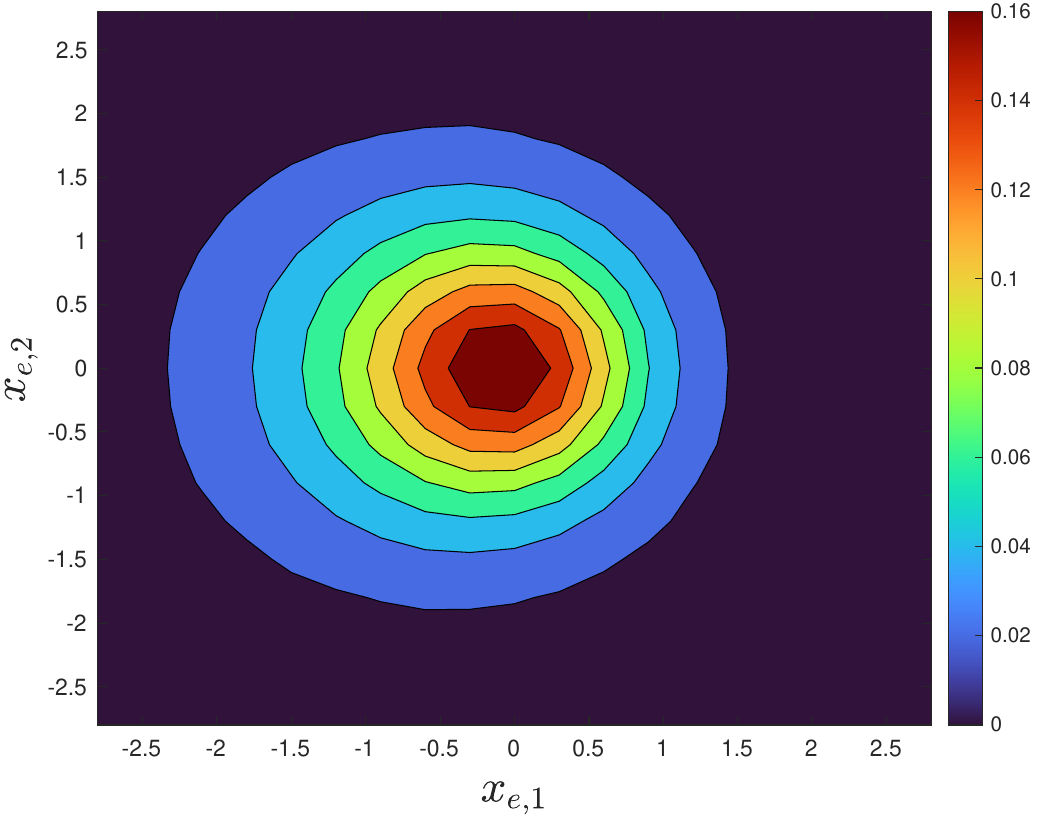}}
{\includegraphics[width=0.23\textwidth,height=0.18\textwidth]{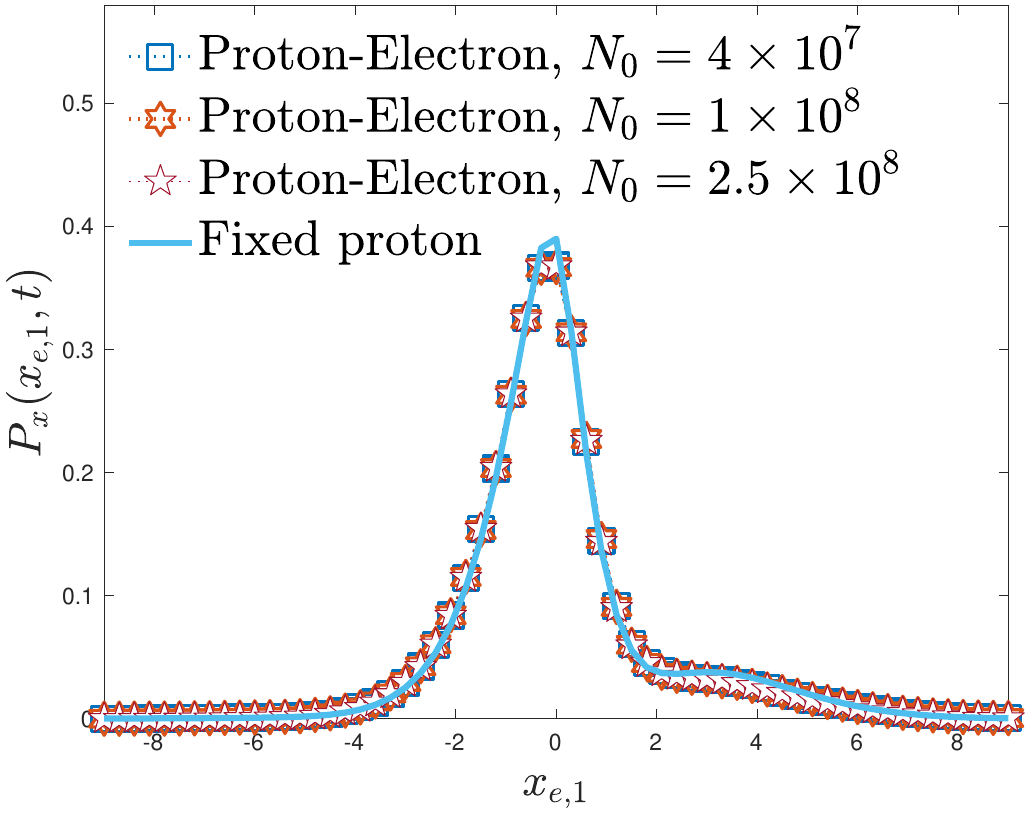}}}
\subfigure[$P_{xy}(x_{e,1}, x_{e,2},t)$ and $P_x(x_{e,1},t )$ at $8$a.u.]{
{\includegraphics[width=0.23\textwidth,height=0.18\textwidth]{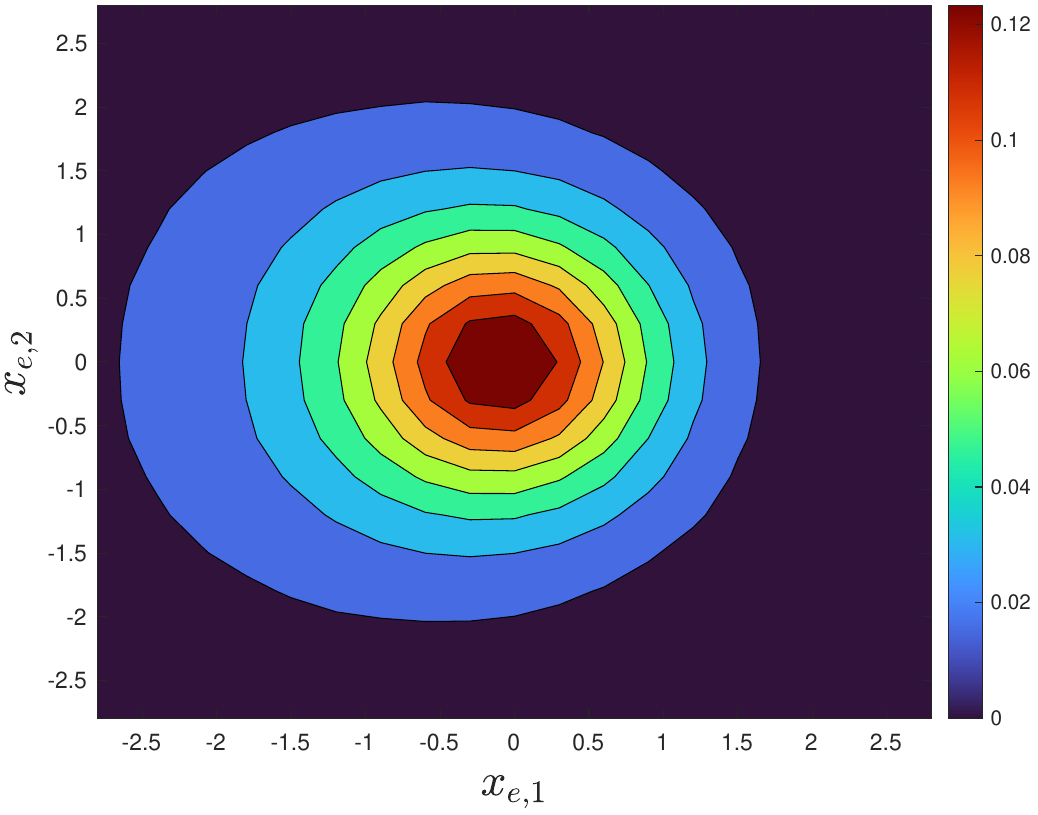}}
{\includegraphics[width=0.23\textwidth,height=0.18\textwidth]{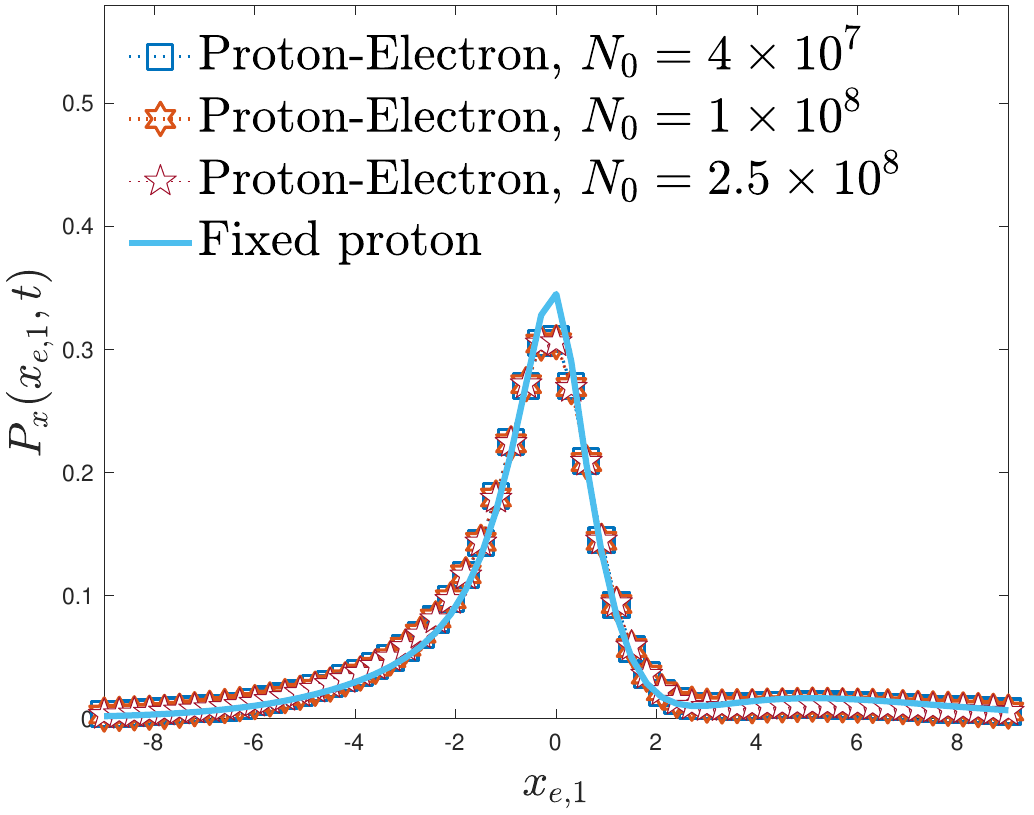}}}
\caption{\small  The localized proton-electron Wigner dynamics: Snapshots of the reduced electron Wigner function $W_1(x_{e,1}, k_{e,1}, t)$ and the spatial marginal distributions $P_{xy}(x_{e,1}, x_{e,2},t)$, $P_x(x_{e,1},t )$ under the parameter $\vartheta = 0.01$. The particle method can recover the double-peak structure and oscillating tails, and noises can be suppressed as the sample size increases.
 \label{12d_wigner_snapshots}}
\end{figure}

Now we take $\varepsilon = 1/100$. By Algorithm \ref{optimal_lambda}, it is suggested to choose $\lambda_0 = 4.85$. Other parameters are: $\gamma_0 = 50$, a finite $\bk$-domain $[-3, 3]^3 \times [-240, 240]^3$ and particles are annihilated every 1 a.u. The reduced electron Wigner function $W_1(x_{e,1}, k_{e,1}, t)$ and the spatial distribution $P_{xy}(x_{e,1}, x_{e,2}, t)$ are reconstructed by the histogram  \eqref{histogram} with $\mathcal{X} = [-15, 15]$, $\mathcal{K} = [-3, 3]$, $\Delta x = 0.3$, $\Delta k = 0.1$. The $l^2$-errors $\mathcal{E}_2[W_1](t)$, $\mathcal{E}_2[P_{xy}](t)$ and the deviation of total energy $\mathcal{E}_H(t)$ are provided in Fig.~\ref{12d_result}. The snapshots of $W_1(x_{e,1}, k_{e,1}, t)$, $P_{xy}(x_{e, 1}, x_{e, 2}, t)$ and $P_x(x_{e, 1},t)$ under different $N_0$ are plotted in Fig.~\ref{12d_wigner_snapshots}. The computational time, the average partition level $K$ and the growth ratio of particle number $\mathcal{N}(t)/N_0$ at $t=10$a.u. are recorded in Table \ref{12d_cpu_time}. 


\begin{table}[!h]
  \centering
  \caption{\small Total wall time of SPADE (in hours), average partition level $K$ and growth ratio of total particles for 12-D simulations up to $10$a.u. \label{12d_cpu_time}}
 \begin{lrbox}{\tablebox}
  \begin{tabular}{c|c|c|c|c|c|c|c}
\hline\hline
\multicolumn{2}{c|}{Parameters}	&      \multicolumn{3}{c|}{$\vartheta = 0.01$} & \multicolumn{3}{c}{$\vartheta = 0.02$} \\
\hline
cores	&	$N_0$ 	&	Time	&	 avg $K$ 	&	$\mathcal{N}(10)/N_0$		&	Time		&	avg $K$ 	&	$\mathcal{N}(10)/N_0$	\\
\hline
128	&	$1.6\times10^7$	&	34.81 	&	2.82$\times10^7$	&	42.69 	&	$$-$$	&	$$-$$			&	$$-$$	\\
128	&	$4.0\times10^7$	&	42.18 		&	2.07$\times10^7$	&	11.55 	&	22.29 	&	5.33$\times10^6$	&	4.06 	\\
128	&	$1.0\times10^8$	&	67.86 		&	2.01$\times10^7$	&	5.37 	&	48.94 		&	6.78$\times10^6$	&	2.97 	\\
256	&	$2.5\times10^8$	&	68.44 	&	2.41$\times10^7$	&	3.58 	&	$$-$$		&	$$-$$			&	$$-$$	\\
\hline\hline
 \end{tabular}
\end{lrbox}
\scalebox{0.95}{\usebox{\tablebox}}
\end{table} 

{\bf Accuracy}: The growth of $l^2$-errors $\mathcal{E}_2[W_1]$ and $\mathcal{E}_2[P_{xy}]$ in 12-D simulations presented in Figs.~\ref{12d_error_W} and \ref{12d_error_P} is similar to the trend in Fig.~\ref{comparison_SPADE_PAUM_N} for $t \le 5$a.u. 
Since the difference between the two-body solutions and the single-body quasi-analytical ones 
gradually increases in time evolution, both $\mathcal{E}_2[W_1]$ and $\mathcal{E}_2[P_{xy}]$ grow faster compared with those in 6-D cases. The fluctuation of total energy can be suppressed by either increasing $N_0$ or refining the partition (see Fig.~\ref{12d_energy}).

The snapshots of $W_1(x_{e,1}, k_{e,1}, t)$ are visualized in Fig.~\ref{12d_wigner_snapshots}. The random noises are suppressed as $N_0$ increases, and the double-peak structure and oscillating tails can be recovered. From the snapshots, errors are still concentrated at the negative valley. 

{\bf Oversampling:} The oversampling problem is still observed in the group $N_0 = 1.6\times 10^7$, $\vartheta = 0.01$. As shown in Fig.~\ref{Gbeam12d_particle} and Table \ref{12d_cpu_time}, the growth ratio of particle number reaches $161.60$ before PA and $42.69$ after PA at $t = 10$a.u., leading to a severe fluctuation of total energy in Fig.~\ref{12d_energy}. As expected,  as the sample size $N_0$ increases, the oversampling problem is alleviated and the bottom line structure emerges in Fig.~\ref{Gbeam12d_particle} as the particle number remains stable after annihilation. These observations also support our findings in Table \ref{rule}.

{\bf Efficiency of SPADE}: The relation between $K(t)$ and $\frac{P(t)M(t)}{(P(t)+M(t))\sqrt{N_0}}$ is given in Fig.~\ref{Gbeam12_K}. The inflection point is $N_0 = 1\times 10^8$ for $\vartheta= 0.01$. Oversampling occurs when smaller $N_0$ is adopted, and can be completely avoided when $N_0 \ge 10^8$. This coincides with the records in Table \ref{12d_cpu_time}. In the meantime, the computational cost almost scales linearly on $N_0$. The bottom line structure in Fig.~\ref{Gbeam12d_particle} and lower bound of $K(t)$ observed in Fig.~\ref{Gbeam12_K} partially explain the reason why SPADE seems to be less affected by CoD, as the lower bound of $K$ does not depend on the dimensionality $\tD$ due to its combinatorial nature.


\subsection{The delocalized proton-electron Wigner dynamics}

\begin{figure}[!h]
    \centering
    \subfigure[Deviation of energy.\label{H2plus_energy}]{\includegraphics[width=0.49\textwidth,height=0.26\textwidth]{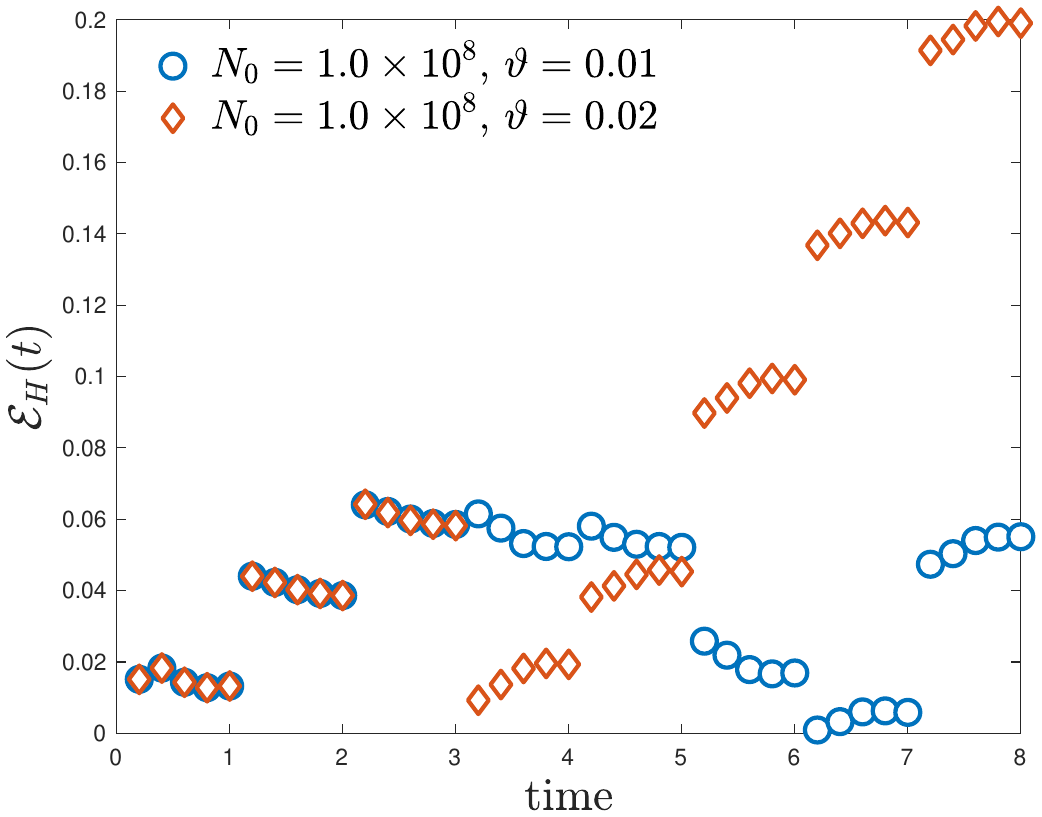}}
     \subfigure[Particle number before and after PA.]{\includegraphics[width=0.49\textwidth,height=0.26\textwidth]{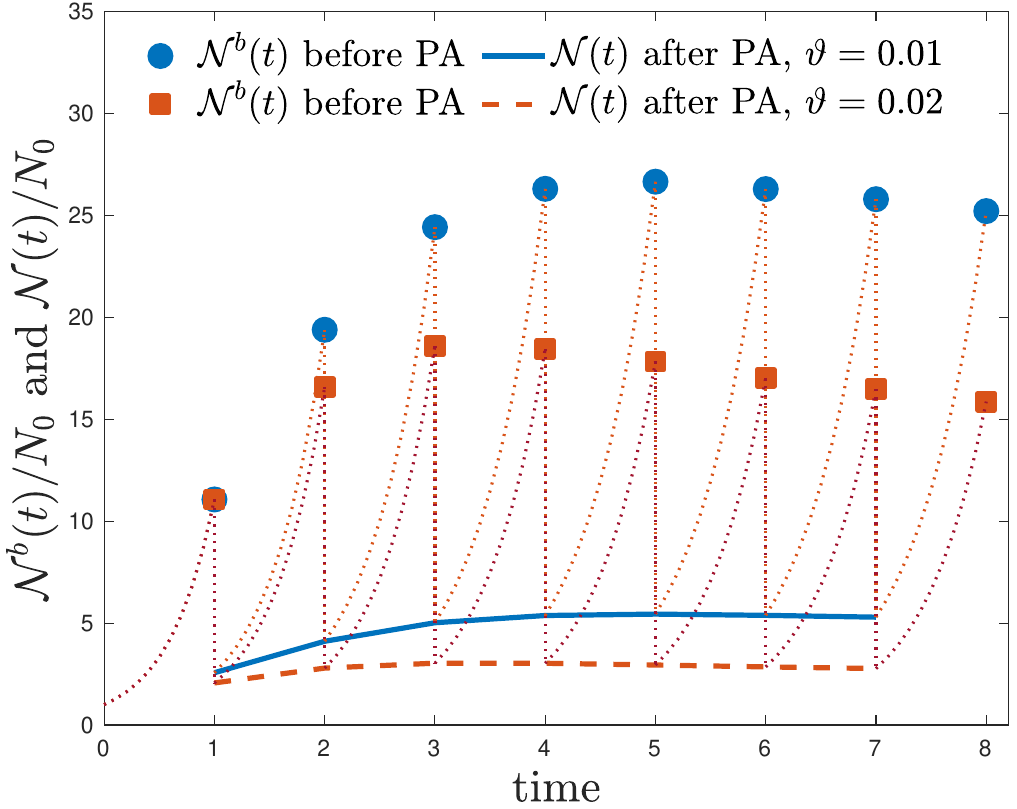}}
    \caption{\small  The delocalized proton-electron Wigner dynamics: The deviation of total energy (the initial value is $7.747$a.u.) and the growth of particle number  in 12-D simulations.\label{12d_H2plus_result}}
\end{figure}

Finally, we try to simulate the dynamics of a delocalized proton-electron Wigner function \cite{PakHammesSchiffer2004} to further demonstrate the capabilities of SPADE. In general, there is no analytical solution to such non-equilibrium dynamics of the proton-electron correlation \cite{bk:CurtrightFairlieZachos2013}.  
\begin{example}
\textup{
Consider a system composed of one proton and one electron interacting under the Coulomb potential, where both the proton and the electron are delocalized in $\bx$-space, with $\bm{R}=(2, 0, 0)$ and $\varepsilon= 1/200$,
\begin{equation}
\begin{split}
f(\bx_e, \bk_e, \bx_p, \bk_p, 0) = &\frac{1}{2\pi^{6}}  \me^{-\frac{1}{2}|\bx_e - \bR|^2 - 2|\bk_e|^2} \me^{-\frac{1}{2\varepsilon^2}|\bx_p - \bm{R}|^2 -2\varepsilon^2|\bk_p|^2} \\
&+ \frac{1}{2\pi^{6}}  \me^{-\frac{1}{2}|\bx_e + \bR|^2 - 2|\bk_e|^2} \me^{-\frac{1}{2\varepsilon^2}|\bx_p + \bm{R}|^2 - 2\varepsilon^2|\bk_p|^2}.
\end{split}
\end{equation}
}
\end{example}
Here the parameters are: $\lambda_0 = 4.65$, $\gamma_0 = 50$,  a finite $\bk$-domain $[-3, 3]^3 \times [-480, 480]^3$, the final time $T=8$. Particles are annihilated every 1 a.u. As shown in Fig.~\ref{12d_H2plus_result}, the growth of particle number can be controlled efficiently by SPADE. Only a slight deviation of total energy is observed under $\vartheta = 0.01$.

\begin{figure}[!h]
\centering
\subfigure[The reduced electron Wigner function $W_1(x_{e,1}, k_{e,1}, t)$ at $2 \to 4 \to 6 \to 8$a.u.\label{H2plus_electron}]{
{\includegraphics[width=0.23\textwidth,height=0.18\textwidth]{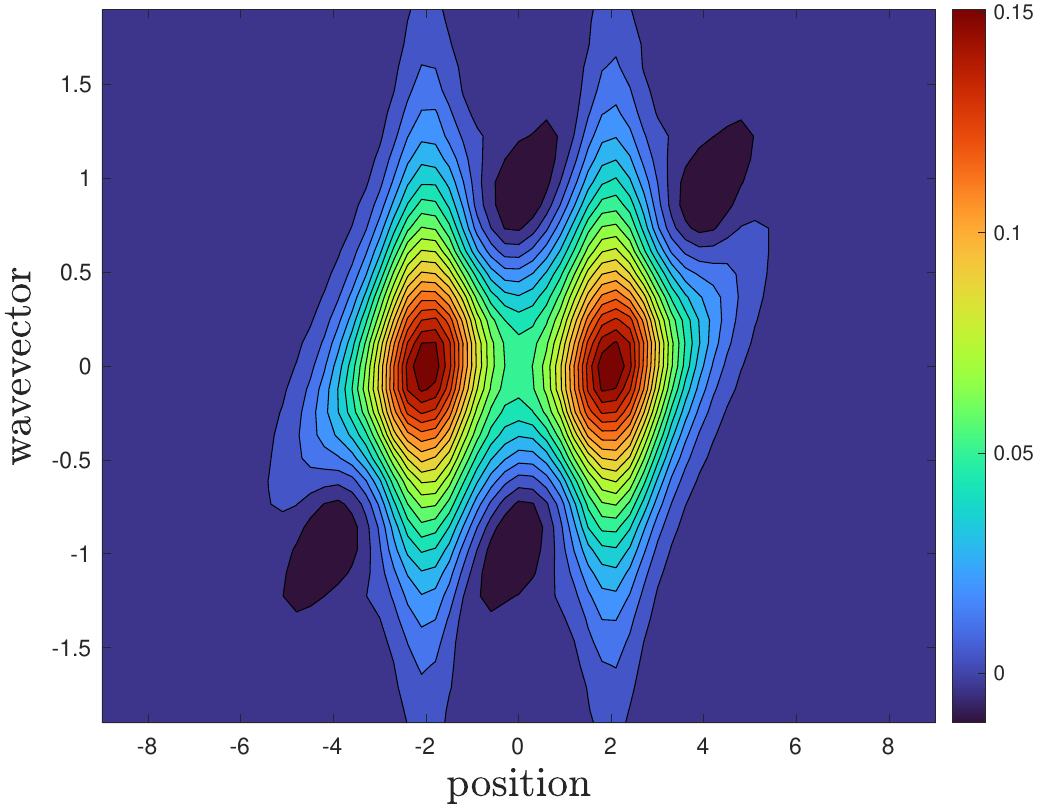}}
{\includegraphics[width=0.23\textwidth,height=0.18\textwidth]{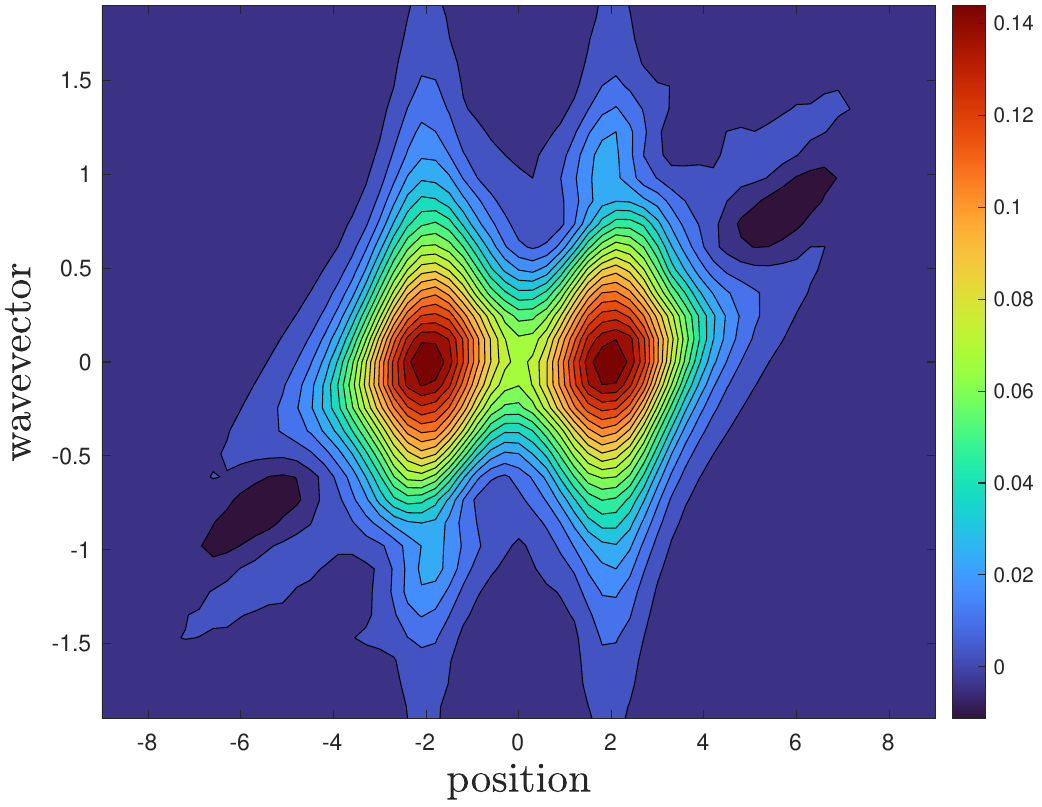}}
{\includegraphics[width=0.23\textwidth,height=0.18\textwidth]{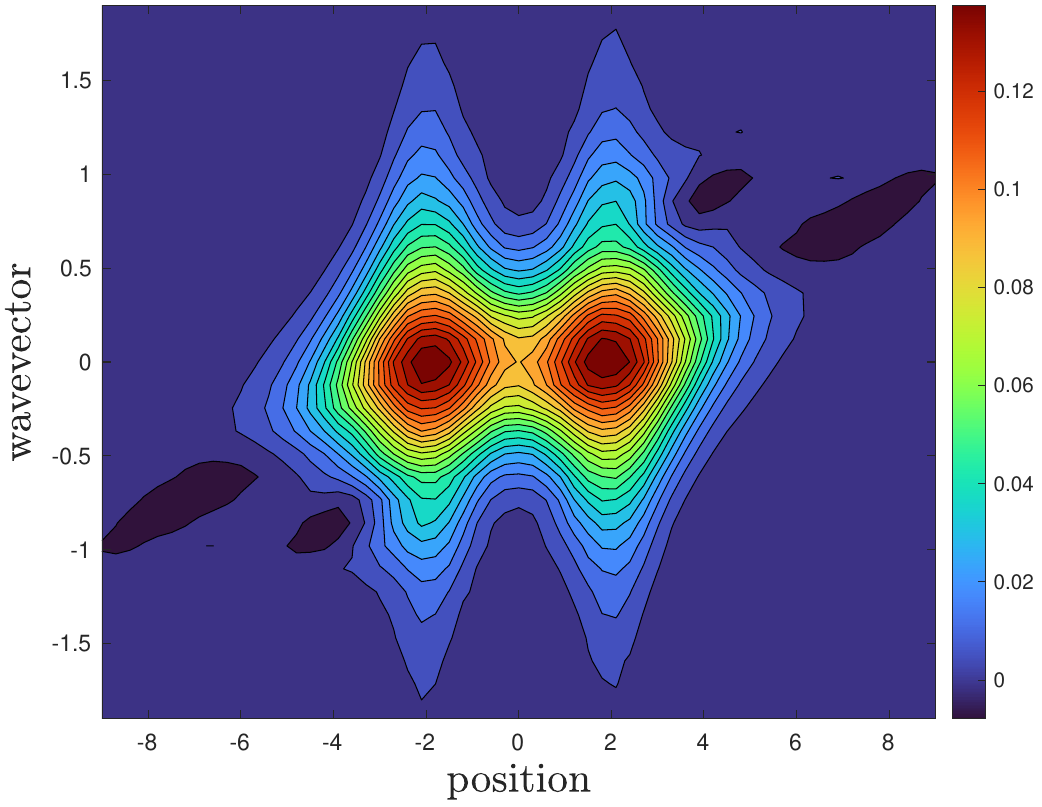}}
{\includegraphics[width=0.23\textwidth,height=0.18\textwidth]{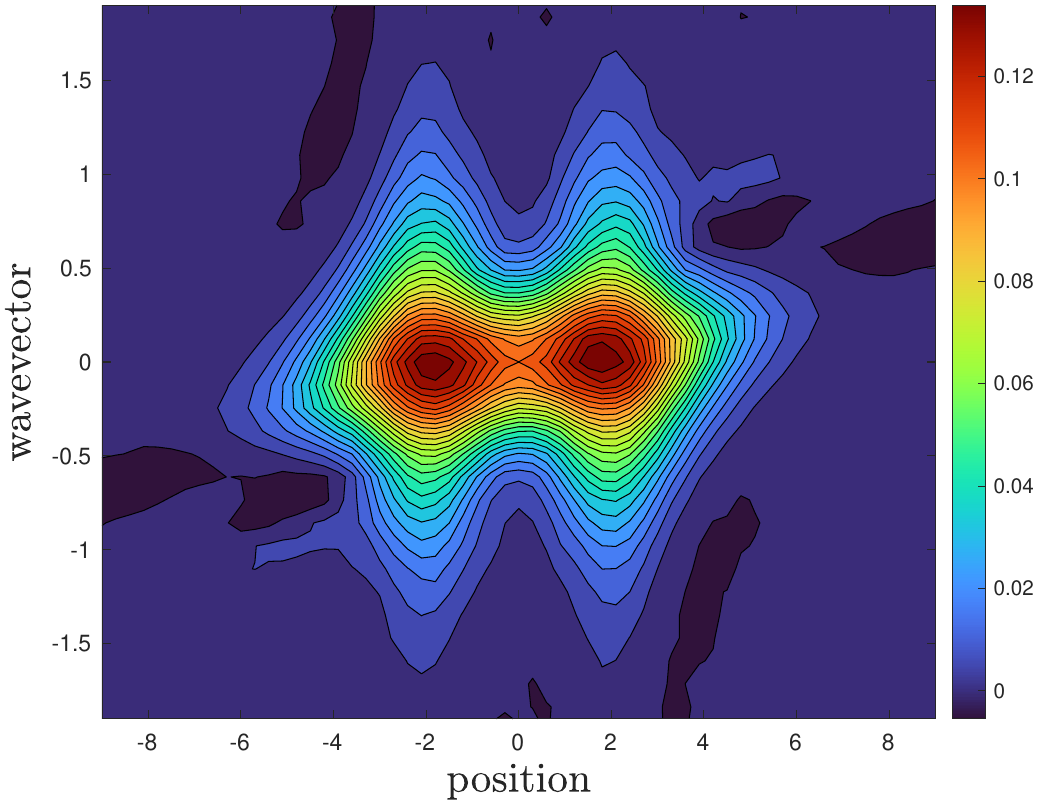}}}
\\
\centering
\subfigure[The reduced proton Wigner function $W_4(x_{p,1}, k_{p,1}, t)$ at $2 \to 4 \to 6 \to 8$a.u.\label{H2plus_proton}]{
{\includegraphics[width=0.23\textwidth,height=0.18\textwidth]{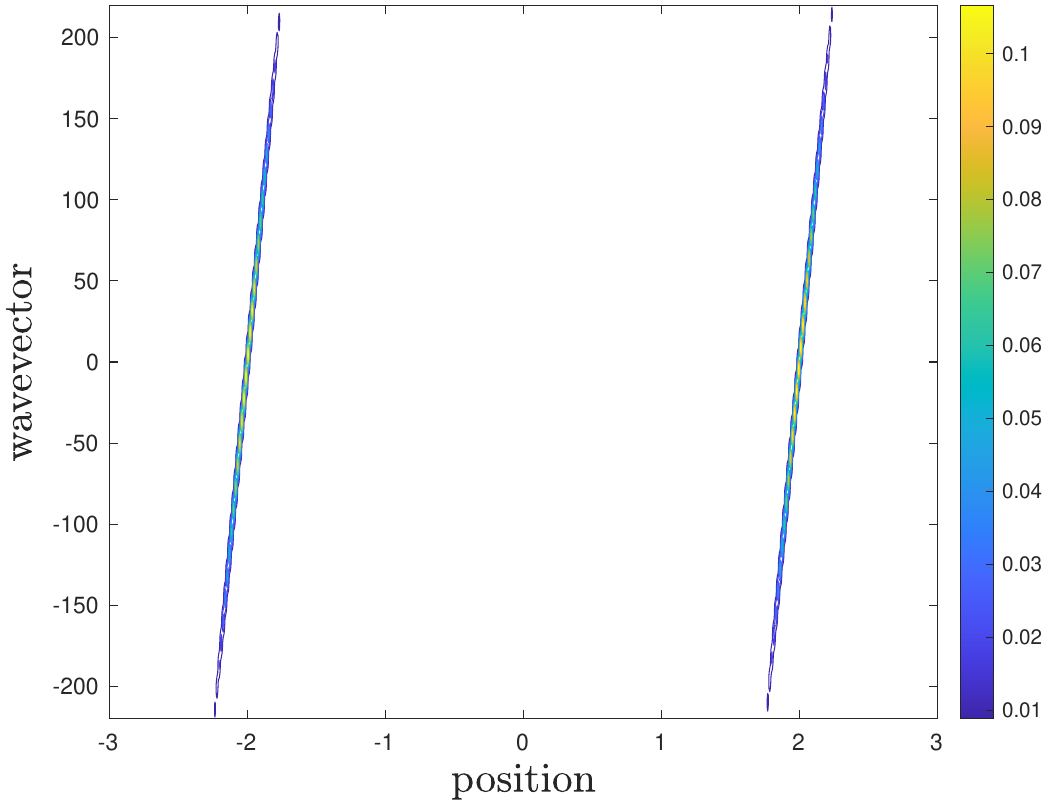}}
{\includegraphics[width=0.23\textwidth,height=0.18\textwidth]{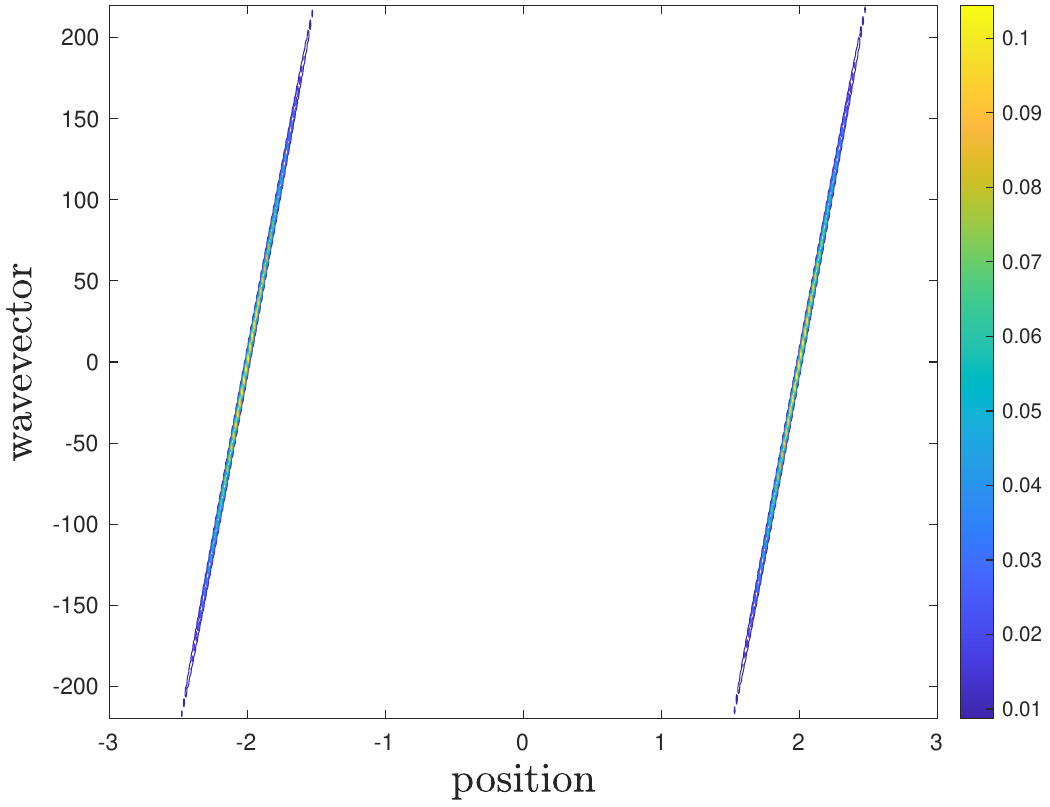}}
{\includegraphics[width=0.23\textwidth,height=0.18\textwidth]{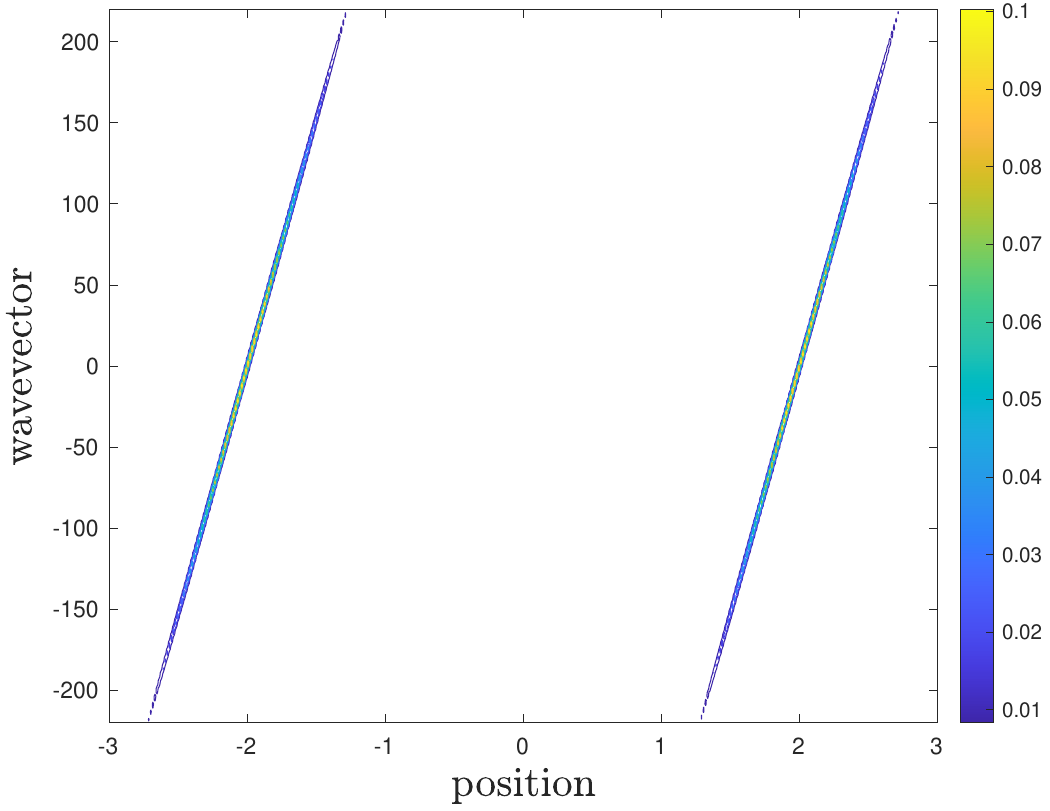}}
{\includegraphics[width=0.23\textwidth,height=0.18\textwidth]{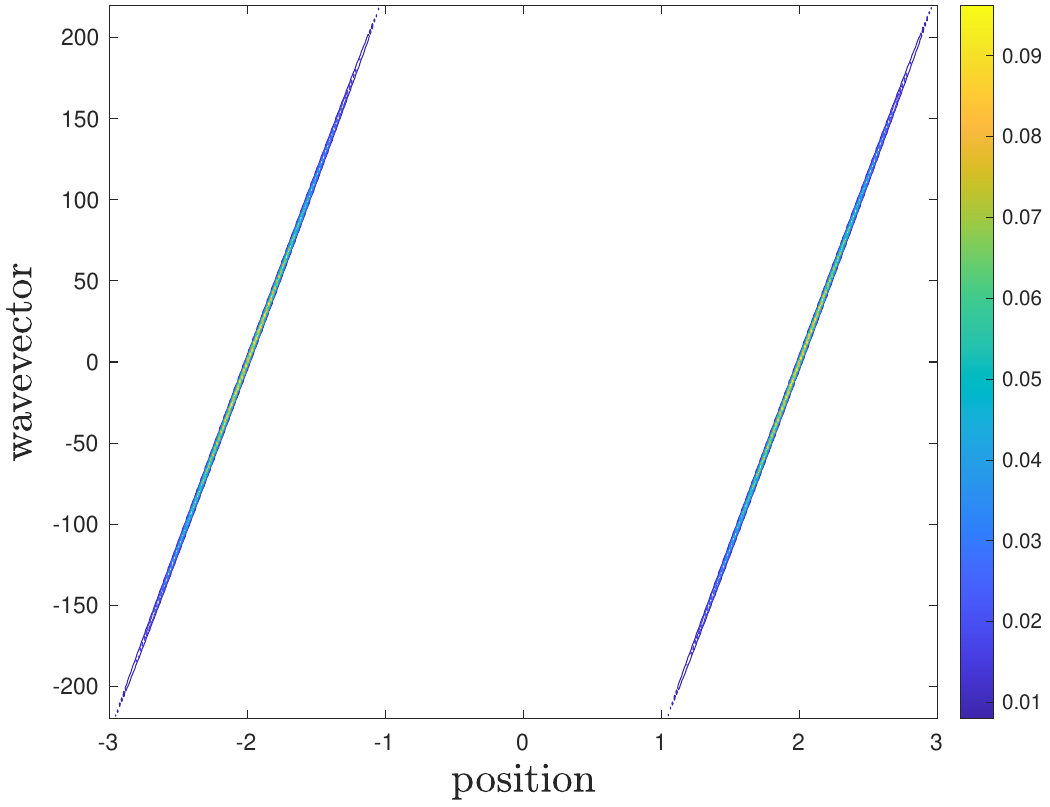}}}
\\
\centering
\subfigure[$P_{xy}(x_{e, 1}, x_{e,2}, t)$, $P_{x}(x_{e, 1}, t)$ at $t=0.5$a.u.]{
{\includegraphics[width=0.23\textwidth,height=0.18\textwidth]{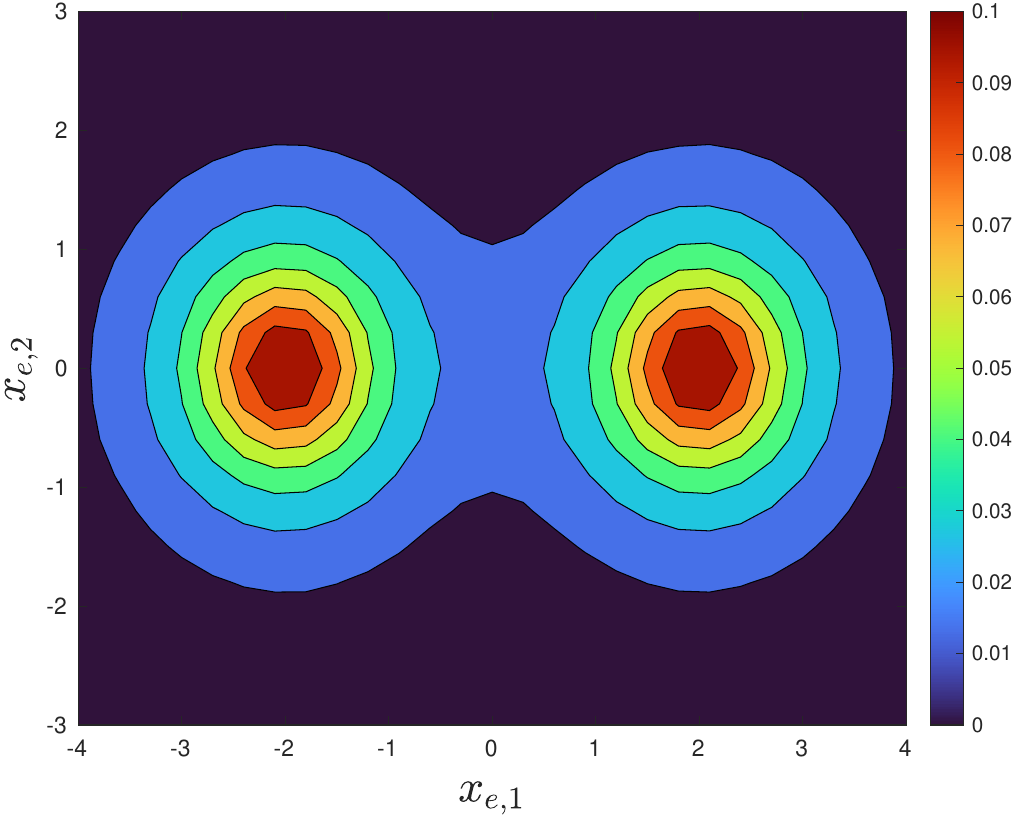}}
{\includegraphics[width=0.23\textwidth,height=0.18\textwidth]{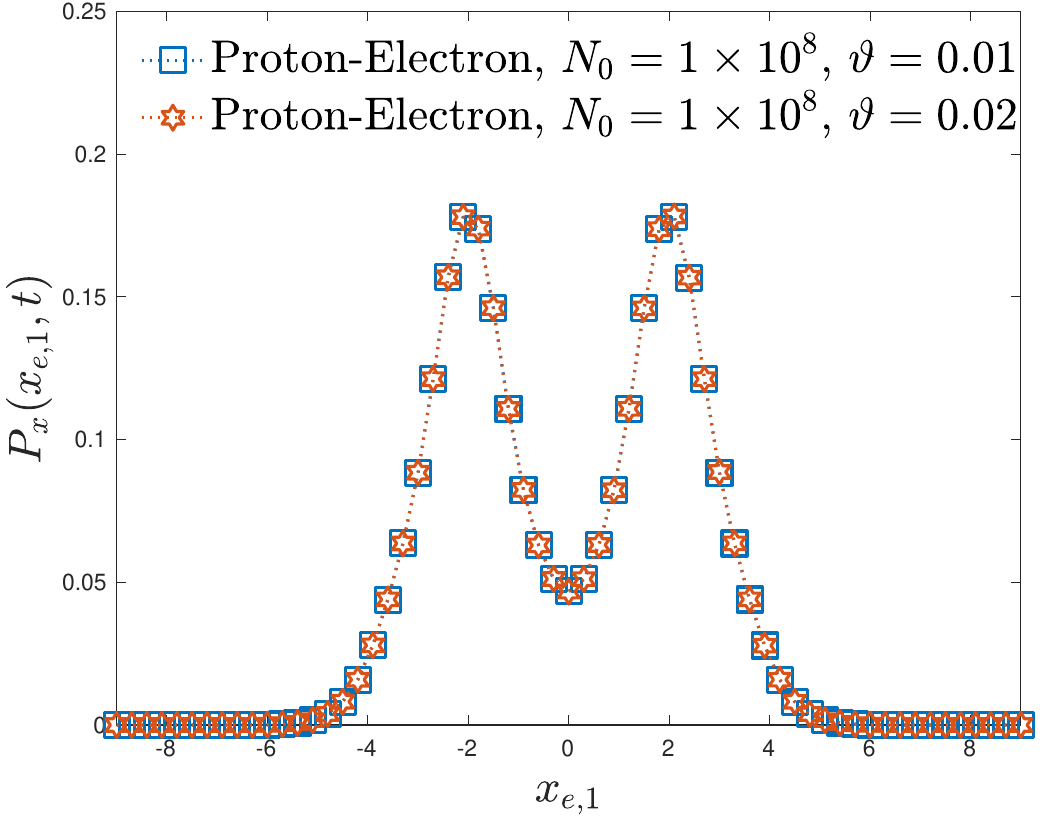}}}
\subfigure[$P_{xy}(x_{e, 1}, x_{e,2}, t)$, $P_{x}(x_{e, 1}, t)$ at $t=1.5$a.u.]{
{\includegraphics[width=0.23\textwidth,height=0.18\textwidth]{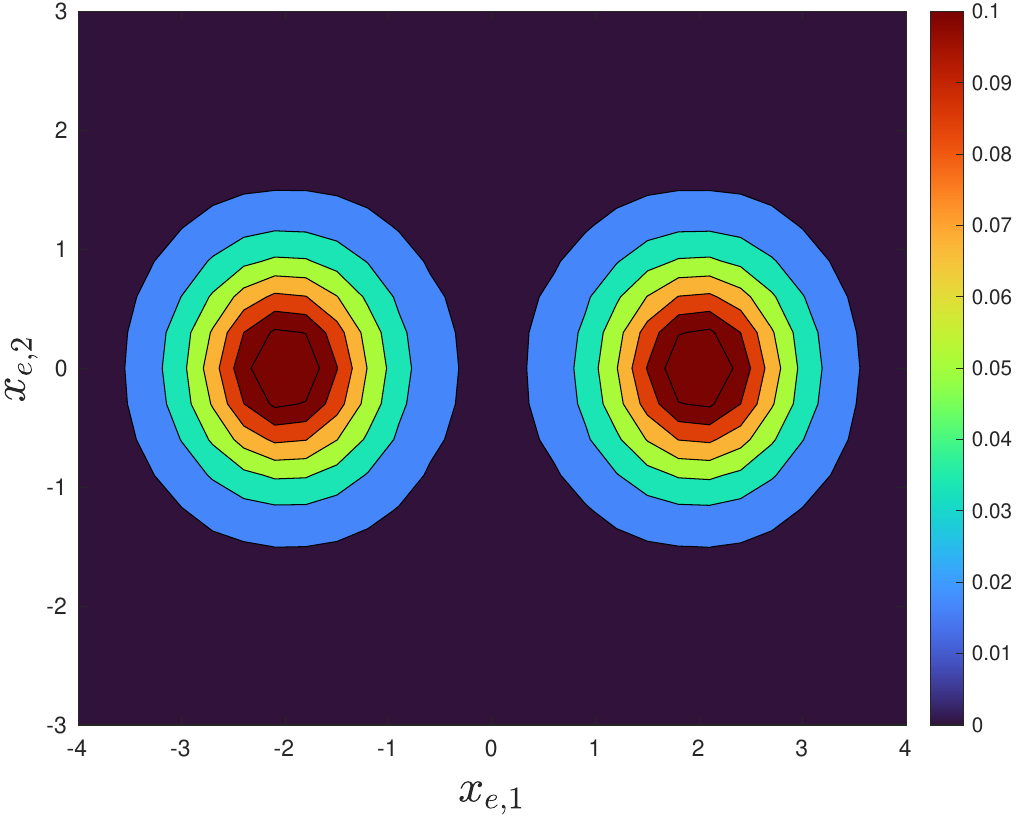}}
{\includegraphics[width=0.23\textwidth,height=0.18\textwidth]{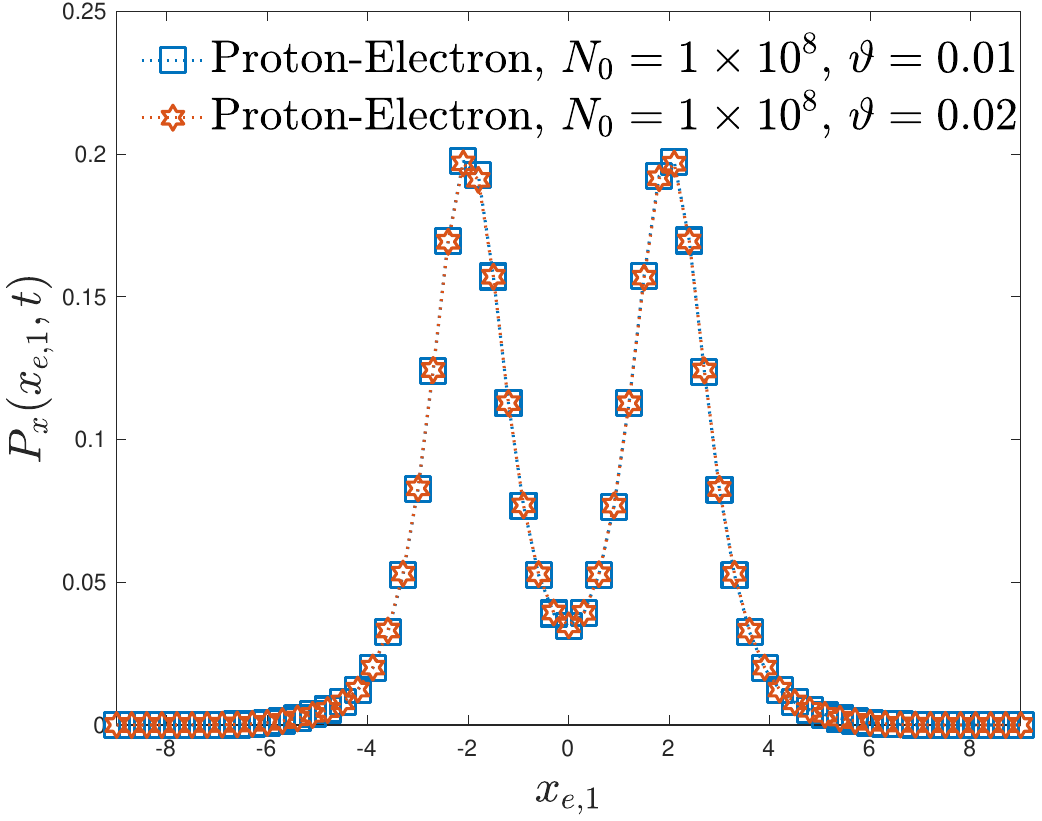}}}
\\
\centering
\subfigure[$P_{xy}(x_{e, 1}, x_{e,2}, t)$, $P_{x}(x_{e, 1}, t)$ at $t=3$a.u.]{
{\includegraphics[width=0.23\textwidth,height=0.18\textwidth]{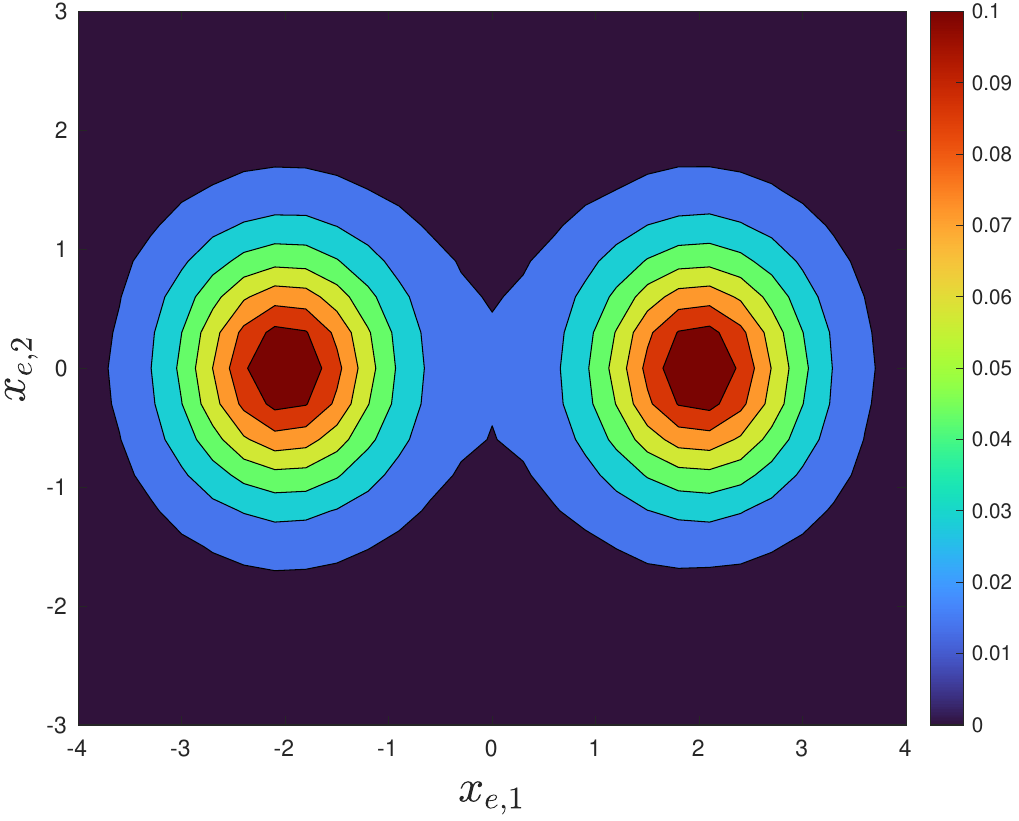}}
{\includegraphics[width=0.23\textwidth,height=0.18\textwidth]{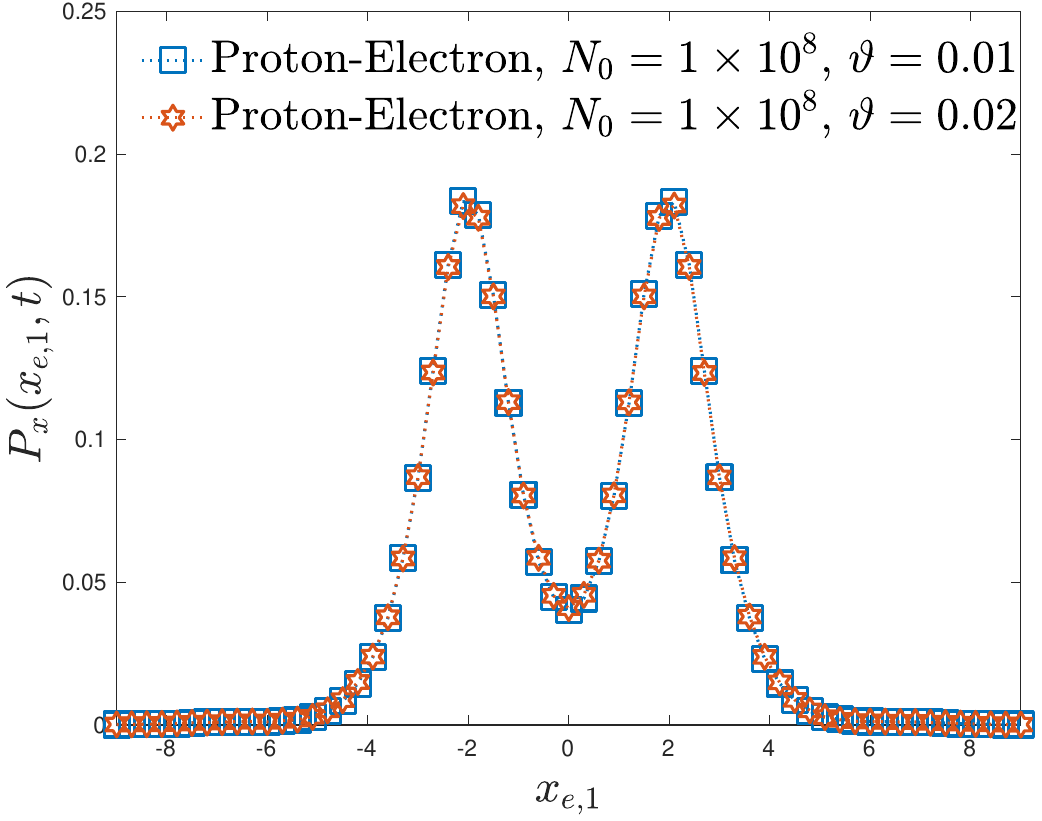}}}
\subfigure[$P_{xy}(x_{e, 1}, x_{e,2}, t)$, $P_{x}(x_{e, 1}, t)$ at $t=5$a.u.]{
{\includegraphics[width=0.23\textwidth,height=0.18\textwidth]{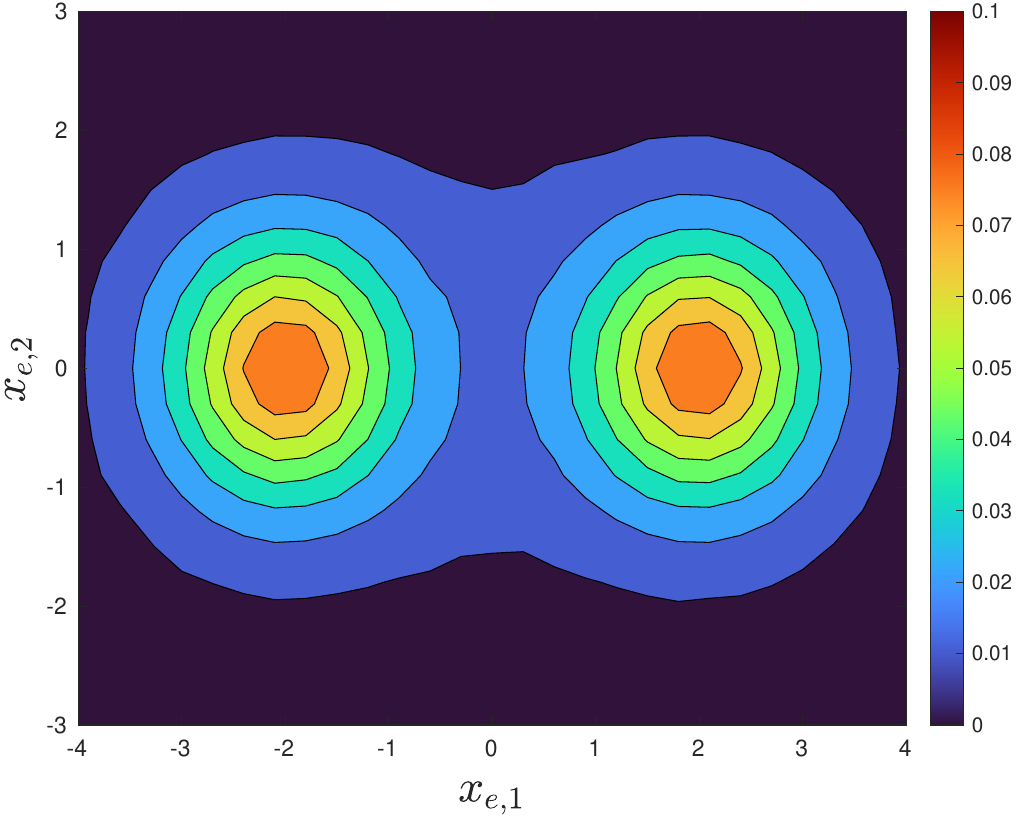}}
{\includegraphics[width=0.23\textwidth,height=0.18\textwidth]{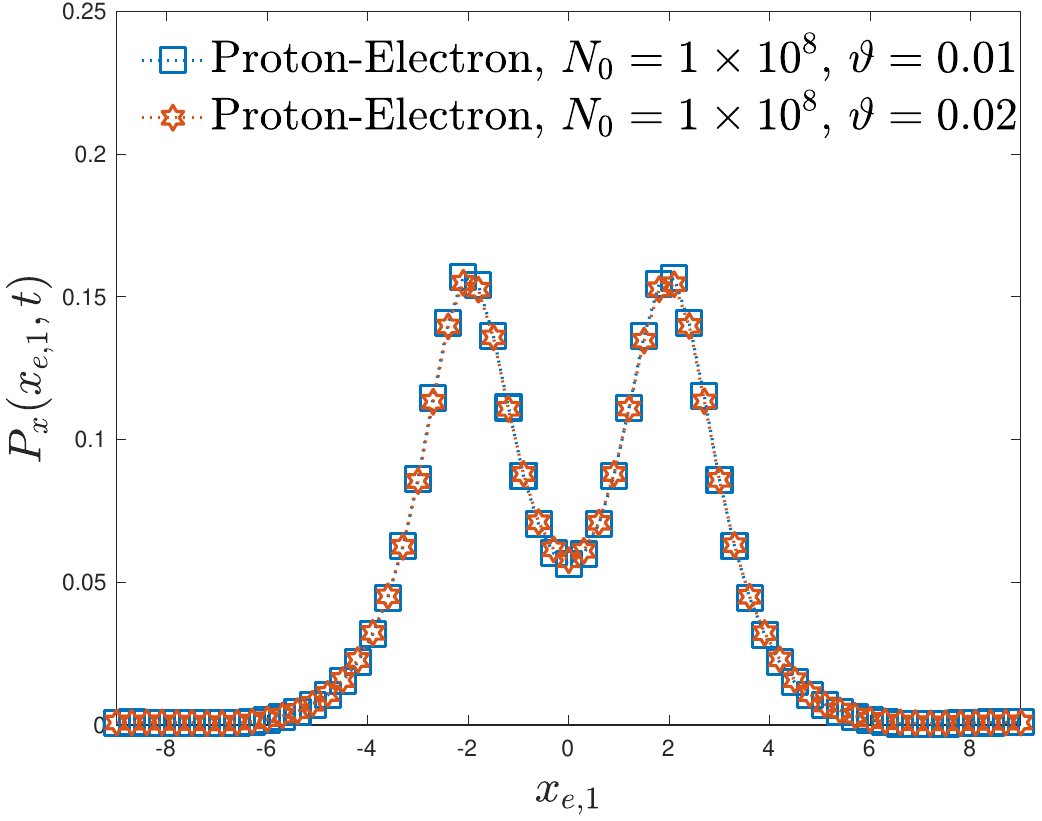}}}
\caption{\small  The delocalized proton-electron Wigner dynamics:  Snapshots of the reduced electron and proton Wigner functions on $(x_1$-$k_1)$ plane and the spatial marginal distribution. \label{H2plus_snapshots}}
\end{figure}

The snapshots of the reduced electron and proton Wigner functions onto $(x_1$-$k_1$) plane, as well as the spatial marginal distributions $P_{xy}(x_{e,1}, x_{e,2}, t)$ and $P_x(x_{e,1}, t)$, are plotted in Fig.~\ref{H2plus_snapshots}.  The quantum Coulomb interactions produce some negative regions in the Wigner function. For $t \le 2$a.u., the negative parts of the Wigner function are observed near the origin, which forbid the electron with certain momentum to occupy the central region. As a consequence, the electron wavepackets are ``squeezed'' and become polarized in the spatial space. After $2$a.u., the negative Wigner function disappears in the origin and emerges near the tail, so that the electron wavepackets gradually merge in the spatial space. This may provide some insights on the non-classicality of a quantum system under the Coulomb interaction.


\section{Conclusion and discussion}
\label{sec.conclusion}

This paper discusses the adaptive particle annihilation algorithms to overcome the numerical sign problem in stochastic Wigner simulations. The Sequential-clustering Particle Annihilation via Discrepancy Estimation (SPADE) is proposed for breaking the curse of dimensionality (CoD) in existing particle annihilation via uniform mesh. By performing a series of benchmark tests on 6-D electron-proton coupling and a thorough comparison with our massively parallel deterministic solver, we can conclude that (1) SPADE may potentially alleviate the sign problem in 6-D cases and can learn the minimal amount of particles that capture the non-classicality of the Wigner function under arbitrary sample size $N_0$; (2) the oversampling problem under small $\vartheta$ can be surmounted by increasing the sample size $N_0$;  (3) the oversampling problem might be avoided when the partition level $K(t)$ approaches its lower bound. It follows by an attempt to simulate 12-D proton-electron Wigner dynamics.  Experimental results demonstrate the potential of SPADE to overcome CoD in higher dimensional Wigner simulations. Our ongoing work is to explore the extension of WBRW-SPA-SPADE to the quantum BBGKY hierarchy \cite{CarruthersZachariasen1983,GrazianiBauerMurillo2014,bk:CurtrightFairlieZachos2013}, which paves a pivotal step for the interlacement of kinetic theory and  molecular dynamics in high energy density physics \cite{GrazianiBauerMurillo2014}, as well as lays the foundation
for studying the Hydrogen tunneling via the Wigner approach \cite{PakHammesSchiffer2004}.

\section*{Acknowledgement}
This research was supported by the National Natural Science Foundation of China (Nos.11822102, 1210010642, 12288101), the Projects funded by China Postdoctoral Science Foundation (Nos.~2020TQ0011, 2021M690227) and the High-performance Computing Platform of Peking University. 
SS is partially supported by Beijing Academy of Artificial Intelligence (BAAI). The authors are sincerely grateful to the handling editor and referees for their patience and valuable suggestions. They would like also to thank Haoyang Liu and Shuyi Zhang at Peking University for their technical supports on computing environment, which greatly facilitate both stochastic and deterministic Wigner simulations.

\appendix

\numberwithin{figure}{section}
\numberwithin{table}{section}
\numberwithin{algorithm}{section}

\renewcommand{\theequation}{\Alph{section}.\arabic{equation}}

\section{Flowchart of the Wigner Monte Carlo}

\begin{figure}[!h]
\centering
\includegraphics[width = 1\textwidth]{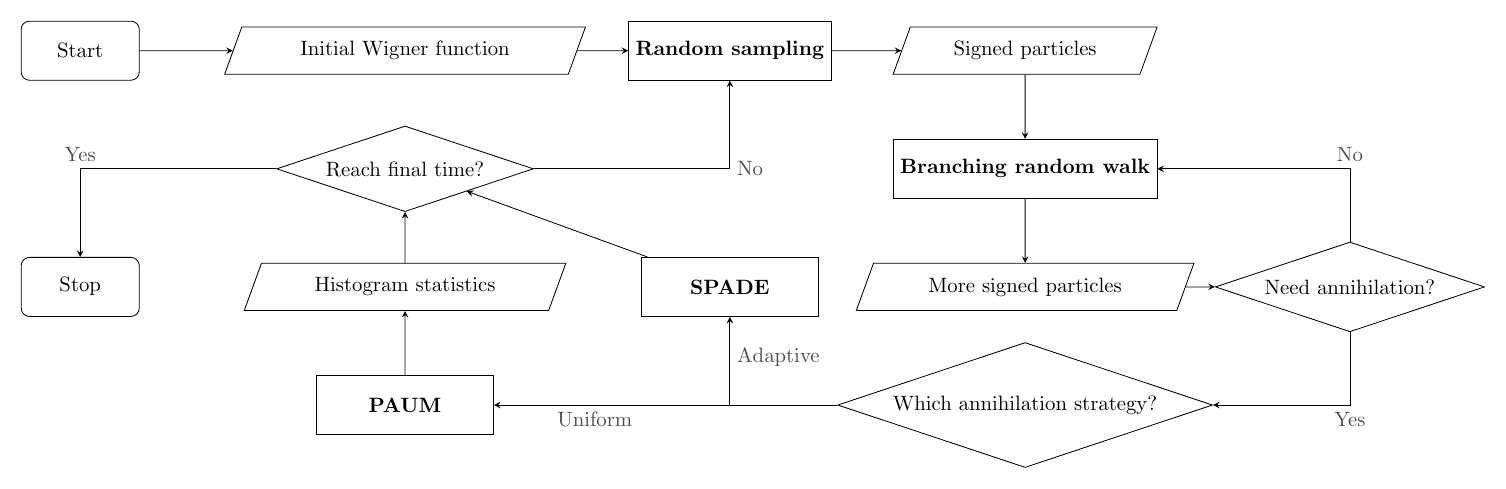}
\caption{\small A complete flow chart of the stochastic Wigner simulations.}
\label{supp_flow_chart}
\end{figure}

A complete flowchart of the stochastic Wigner simulations, as depicted in Figure \ref{supp_flow_chart}, consists of three cornerstones: Probabilistic interpretation to the Wigner equation, sequential importance sampling and particle resampling (either particle annihilation via uniform mesh (PAUM) \cite{KosinaSverdlovGrasser2006,MuscatoWagner2016,XiongShao2019} or adaptive particle annihilation algorithm SPADE \cite{ShaoXiong2020_arXiv}). 

\begin{itemize}

\item[1.] {\bf Probabilistic interpretation} represents the solution of the deterministic Wigner equation by the expectation of a branching random walk model. It defines the probabilistic rules of particle motions and generations.

\item[2.] {\bf Sequential importance sampling} represents the Wigner function by a finite collection of weighted particles. First, the initial data is approximated by a weighted empirical measure. Second, particles move under the probabilistic rules, including  deterministic motions, random jumps and random generations.

\item[3.] {\bf Particle resampling} intends to adjust the particle weights by combinatorial techniques, such as clustering, matching and sampling with or without replacement. Specially, for a finite particle system involving both positive and negative particles, the particle resampling, also termed particle annihilation, cancels out those carrying opposite weights, thereby controlling the rapid growth of particle number and stochastic variances.

\end{itemize}

The fundamental obstacle to the stochastic Wigner algorithms is the numerical sign problem, which  is believed to be NP-hard in nature \cite{TroyerWiese2005,IazziSoluyanovTroyer2016}. In fact, we can prove that both particle number and stochastic variances in such branching particle system may grow exponentially in time, thereby dramatically hampering the efficiency of the sampling process \cite{ShaoXiong2020}. Fortunately, the sign problem can be alleviated to a large extent if one carefully cancels out  the stochastic trajectories that contribute little to the expectation but amplify the variance.

\section{Typical examples}

Here we provide some examples about the Wigner equation under different interacting potentials $V(\bx)$.

\subsection{The single-body Wigner equation under 2-D Morse potential}

The single-body Wigner equation plays a fundamental role in describing quantum mechanics in phase space. It describes the dynamics of the Wigner function $f(\bx, \bk, t)$,
\begin{equation}\label{supp_eq.Wigner}
\begin{split}
\frac{\partial }{\partial t}f(\bm{x}, \bm{k}, t)+\frac{\hbar \bm{k}}{{m}} \cdot \nabla_{\bm{x}} f(\bm{x},\bm{k}, t)  = \Theta_V[f](\bx, \bk, t),
\end{split}
\end{equation}
where $m$ is the mass, $\hbar$ is the reduced Planck constant and the pseudodifferential operator (hereafter $\pdo$ for brevity) describes the nonlocal quantum interaction under a potential function $V(\bx)$,
\begin{equation*}
\Theta_V[f](\bx, \bk, t) = \frac{1}{\mi \hbar (2\pi)^d} \iint_{\mathbb{R}^{d}\times \mathbb{R}^d} \me^{-\mi (\bk - \bk^{\prime}) \cdot \by }(V(\bx + \frac{\by}{2}) - V(\bx - \frac{\by}{2}))f(\bx, \bk^{\prime}, t) \D \by \D \bk^{\prime}.
\end{equation*}
where $d$ is the dimensionality.

The first example is the electron Wigner dynamics in 4-D phase space under the 2-D Morse potential ($d = 2$), 
\begin{equation}\label{supp_morse}
V(\bx) = - 2\me^{-\kappa(|\bx-\bx_A|- r_0)} + \me^{-2\kappa(|\bx-\bx_A|- r_0)}.
\end{equation}
Under this potential, an equivalent form of $\pdo$ reads that 
\begin{equation}\label{supp_pdo_2d_morse}
\begin{split}
\Theta_V[\varphi](\bx, \bk, t) =  &  - \frac{\kappa \me^{\kappa r_0} c_2}{\hbar} \int_{0}^{2\pi} \D \vartheta \int_{0}^{+\infty} \D r~ \frac{r \sin(2(\bx - \bx_A) \cdot \bk^{\prime})}{\sqrt{r^2+(\kappa/2)^2}}  \frac{\Delta_{r\sigma}[\varphi](\bx, \bk, t)}{r^2 + (\kappa/2)^2}  \\
&+ \frac{\kappa \me^{2\kappa r_0} c_2}{\hbar} \int_{0}^{2\pi} \D \vartheta \int_{0}^{+\infty} \D r~ \frac{r \sin(2(\bx-\bx_A) \cdot \bk^{\prime})}{\sqrt{r^2+\kappa^2}}  \frac{\Delta_{r\sigma}[\varphi](\bx, \bk, t)}{r^2 + \kappa^2} ,
\end{split}
\end{equation}
where  $\bk^{\prime} = r\sigma = (r \cos \vartheta, r\sin \vartheta)$ and $c_2 = \Gamma(3/2)\pi^{-3/2} \approx 0.1592$, and the difference operator $\Delta_{r \sigma}[\varphi](\bx, \bk, t) = \varphi(\bx, \bk - {r\sigma}/{2}, t) - \varphi(\bx, \bk + {r\sigma}/{2}, t)$.
Although the eigenfunctions of the quantum Hamiltonian operator can be solved, its phase space solution from the first principle is less than straightforward or complete, especially for its non-equilibrium dynamics \cite{bk:CurtrightFairlieZachos2013}.

\subsection{The single-body Wigner equation under 3-D Coulomb potential}

The second example is the single-body electron Wigner dynamics in 6-D phase space,
\begin{equation}\label{supp_single_body_Wigner}
\begin{split}
\frac{\partial }{\partial t}f_e(\bm{x}_e, \bm{k}_e, t) &+\frac{\hbar \bm{k}_e}{m_e} \cdot \nabla_{\bm{x}_e} f_e(\bm{x}_e,\bm{k}_e, t)  = \Theta_V[f_e](\bx_e, \bk_e, t)
\end{split}
\end{equation}
 under the the interacting potential with fixed $\bx_A$, 
\begin{equation}
V(\bx) = -\frac{\gamma}{|\bx_e - \bx_A|},
\end{equation}
where $\gamma = e^2/(4\pi \epsilon_0)$ with the point charge $e$ and the dielectric constant $\epsilon_0$. $\pdo$ reads
\begin{equation}\label{supp_pdo_scatter}
\Theta_V[f_e](\bx_e, \bk_e, t) =\frac{\gamma}{\mi \hbar c_{3,1}}  \int_{\mathbb{R}^3}\frac{ \me^{\mi \bk^{\prime} \cdot (\bx_e - \bx_A)}}{|\bk^{\prime}|^2}(f_e(\bx_e, \bk_e - \frac{\bk^{\prime}}{2}, t) - f_e(\bx_e, \bk_e + \frac{\bk^{\prime}}{2}, t)) \D \bk^{\prime}.
\end{equation}

\subsection{Proton-electron Wigner dynamics in 12-D phase space}

Now consider one proton and one electron interacting under the Coulomb potential, where proton has finite mass $m_p \approx 1836 m_e$,
\begin{equation}
V(\bx_e, \bx_p) = -\frac{\gamma}{|\bx_e - \bx_p|},
\end{equation}
then the proton-electron (two-body) Wigner equation in 12-D phase space reads that
\begin{equation*}
\begin{split}
&\frac{\partial }{\partial t}f(\bm{x}_e, \bm{x}_p, \bm{k}_e, \bm{k}_p, t)+\frac{\hbar \bm{k}_e}{m_e} \cdot \nabla_{\bm{x}_e} f(\bm{x}_e, \bm{x}_p, \bm{k}_e, \bm{k}_p, t) + \frac{\hbar \bm{k}_p}{m_p} \cdot \nabla_{\bm{x}_p} f(\bm{x}_e, \bm{x}_p, \bm{k}_e, \bm{k}_p, t)  \\
&= \frac{\gamma}{\mi \hbar (2\pi)^6}\iiiint_{\mathbb{R}^{6} \times \mathbb{R}^{6}} \frac{\me^{-\mi (\bk_e - \bk_e^{\prime}) \cdot \by_e - \mi (\bk_p - \bk_p^{\prime}) \cdot \by_p} }{|\bx_e -\frac{\by_e}{2} - \bx_p + \frac{\by_p}{2}|}  f(\bm{x}_e, \bm{x}_p, \bm{k}^{\prime}_e, \bm{k}^{\prime}_p, t) \D \by_e  \D \by_p \D \bk_e^{\prime}  \D \bk_p^{\prime} \\
& \quad - \frac{\gamma}{\mi \hbar (2\pi)^6} \iiiint_{\mathbb{R}^{6} \times \mathbb{R}^{6}} \frac{ \me^{-\mi (\bk_e - \bk_e^{\prime}) \cdot \by_e - \mi (\bk_p - \bk_p^{\prime}) \cdot \by_p}}{|\bx_e +\frac{\by_e}{2} - \bx_p - \frac{\by_p}{2}|}  f(\bm{x}_e, \bm{x}_p, \bm{k}^{\prime}_e, \bm{k}^{\prime}_p, t)\D \by_e  \D \by_p \D \bk_e^{\prime}  \D \bk_p^{\prime}.
\end{split}
\end{equation*}

By the conversion $\by_e - \by_p = \bm{\xi}_1$, $\frac{\by_e + \by_p}{2} = \bm{\xi}_2$, it yields that
\begin{equation*}
\begin{split}
&\textup{RHS}  =  \iiiint_{\mathbb{R}^{12}} \frac{\me^{-\mi (\bk_e - \bk_e^{\prime}) \cdot (\bxi_2 + \frac{\bxi_1}{2}) - \mi (\bk_p - \bk_p^{\prime}) \cdot (\bxi_2 - \frac{\bxi_1}{2}}) }{\mi \hbar \gamma^{-1} (2\pi)^6 (|\bx_e - \bx_p - \frac{\bxi_1}{2} |)}  f(\bm{x}_e, \bm{x}_p, \bm{k}^{\prime}_e, \bm{k}^{\prime}_p, t) \D \bxi_1  \D \bxi_2 \D \bk_e^{\prime}  \D \bk_p^{\prime} \\
 & \quad - \iiiint_{\mathbb{R}^{12}} \frac{\me^{-\mi (\bk_e - \bk_e^{\prime}) \cdot (\bxi_2 + \frac{\bxi_1}{2}) - \mi (\bk_p - \bk_p^{\prime}) \cdot (\bxi_2 - \frac{\bxi_1}{2}}) }{\mi  \hbar \gamma^{-1} (2\pi)^6 (|\bx_e - \bx_p + \frac{\bxi_1}{2} |)}  f(\bm{x}_e, \bm{x}_p, \bm{k}^{\prime}_e, \bm{k}^{\prime}_p, t) \D \bxi_1  \D \bxi_2 \D \bk_e^{\prime}  \D \bk_p^{\prime}. 
\end{split}
\end{equation*}
Using the Fourier completeness relation,
\begin{equation*}
\int_{\mathbb{R}^3} \me^{-\mi (\bk_e - \bk_e^{\prime})\cdot \bxi_2  -\mi (\bk_p - \bk_p^{\prime})\cdot \bxi_2} \D \bxi_2 = (2\pi)^3 \delta(\bk_e - \bk^{\prime}_e - \bk_p + \bk^{\prime}_p),
\end{equation*}
it further yields that 
\begin{equation*}
\begin{split}
\textup{RHS} = &~\frac{\gamma}{\mi \hbar (2\pi)^3 }\iint_{\mathbb{R}^6}  \frac{\me^{-\mi (\bk_e - \bk_e^{\prime}) \cdot \bxi_1} }{ |\bx_e - \bx_p - \frac{\bxi_1}{2} |}  f(\bm{x}_e, \bm{x}_p, \bk^{\prime}_e,  \bk_p - \bk_e +  \bm{k}^{\prime}_e, t) \D \bxi_1  \D \bk_e^{\prime} \\
& - \frac{\gamma}{\mi \hbar (2\pi)^3 }\iint_{\mathbb{R}^6}  \frac{\me^{-\mi (\bk_e - \bk_e^{\prime}) \cdot \bxi_1} }{ |\bx_e - \bx_p + \frac{\bxi_1}{2} |} f(\bm{x}_e, \bm{x}_p, \bk^{\prime}_e,  \bk_p - \bk_e + \bm{k}^{\prime}_e, t) \D \bxi_1  \D \bk_e^{\prime} \\
= &~\frac{\gamma}{\mi \hbar (2\pi)^3 }\iint_{\mathbb{R}^6}  \frac{\me^{-\mi \bk_e^{\prime} \cdot \bxi_1} }{ |\bx_e - \bx_p - \frac{\bxi_1}{2} |}  f(\bm{x}_e, \bm{x}_p, \bk_p + \bk^{\prime}_e,  \bk_p -  \bm{k}^{\prime}_e, t) \D \bxi_1  \D \bk_e^{\prime} \\
& - \frac{\gamma}{\mi \hbar (2\pi)^3 }\iint_{\mathbb{R}^6}  \frac{\me^{-\mi \bk_e^{\prime} \cdot \bxi_1} }{ |\bx_e - \bx_p + \frac{\bxi_1}{2} |} f(\bm{x}_e, \bm{x}_p,\bk_p + \bk^{\prime}_e,  \bk_e - \bm{k}^{\prime}_e, t) \D \bxi_1  \D \bk_e^{\prime}.
\end{split}
\end{equation*}

Finally, by changing the variables $\bxi_1 \to \bm{\eta} + 2\bx_p - 2\bx_e$ for the first line and $\bxi_1 \to \bm{\eta}  - 2\bx_p + 2\bx_e$ for the second, it arrives at
\begin{equation*}
\begin{split}
\Theta_V[f] = ~&\frac{2\gamma}{\mi \hbar (2\pi)^3} \iint_{\mathbb{R}^6} \frac{\me^{2\mi \bk^{\prime}_e \cdot (\bx_e - \bx_p) - \mi \bk^{\prime}_e \cdot \bm{\eta}}}{|\bm{\eta}|} f(\bm{x}_e, \bm{x}_p, \bk_p + \bk^{\prime}_e,  \bk_p -  \bm{k}^{\prime}_e, t) \D \bm{\eta}  \D \bk_e^{\prime} \\
 & - \frac{2\gamma}{\mi \hbar (2\pi)^3} \iint_{\mathbb{R}^6} \frac{\me^{-2\mi \bk^{\prime}_e \cdot (\bx_e - \bx_p) - \mi \bk^{\prime}_e \cdot \bm{\eta}}}{|\bm{\eta}|} f(\bm{x}_e, \bm{x}_p, \bk_p + \bk^{\prime}_e,  \bk_p -  \bm{k}^{\prime}_e, t) \D \bm{\eta}  \D \bk_e^{\prime}.
\end{split}
\end{equation*}
Since the Fourier conjugate of $\frac{1}{|\bm{\eta}|}$ is $\frac{(2\pi)^3}{c_{3, 1}} \frac{1}{|\bk|^2}$, $c_{n, \alpha} = \pi^{n/2} 2^\alpha \frac{\Gamma(\frac{\alpha}{2})}{\Gamma(\frac{n-\alpha}{2})}$, and $\bk_e^{\prime} \to \pm \bk^{\prime}/2$ for the first and second line, respectively, we obtain that 
\begin{equation}\label{supp_6d_wigner_c}
\begin{split}
\Theta_V[f](\bx, \bk,t) = &\frac{\gamma}{\mi \hbar c_{3,1}}\int_{\mathbb{R}^3} \me^{\mi \bk^{\prime} \cdot (\bx_e - \bx_p)} \frac{1}{|\bk^{\prime}|^2} f(\bx_e, \bx_p, \bk_e - \frac{\bk^{\prime}}{2}, \bk_p + \frac{\bk^{\prime}}{2}, t) \D \bk^{\prime} \\
& - \frac{\gamma}{\mi \hbar c_{3,1}}\int_{\mathbb{R}^3}  \me^{\mi \bk^{\prime} \cdot (\bx_e - \bx_p)}  \frac{1}{|\bk^{\prime}|^2}  f(\bx_e, \bx_p, \bk_e + \frac{\bk^{\prime}}{2}, \bk_p - \frac{\bk^{\prime}}{2}, t) \D \bk^{\prime}.
\end{split}
\end{equation}
This arrives at the proton-electron Wigner equation:
\begin{equation}\label{supp_12d_wigner_c}
\begin{split}
&\frac{\partial }{\partial t}  f(\bm{x}_e, \bm{x}_p, \bk_e, \bk_p, t)+\frac{\hbar \bk_e}{m_e} \cdot \nabla_{\bm{x}_e}  f(\bm{x}_e, \bm{x}_p,\bk_e, \bk_p, t) + \frac{\hbar \bk_p}{m_p} \cdot  \nabla_{\bm{x}_p} f(\bm{x}_e, \bm{x}_p, \bk_e, \bk_p, t) \\
~&= \frac{\gamma}{\mi \hbar c_{3,1}}\int_{\mathbb{R}^3} \me^{\mi \bk^{\prime} \cdot (\bx_e - \bx_p)} \frac{1}{|\bk^{\prime}|^2}  f(\bx_e, \bx_p, \bk_e - \frac{\bk^{\prime}}{2}, \bk_p +\frac{\bk^{\prime}}{2}, t) \D \bk^{\prime} \\
&\quad - \frac{\gamma}{\mi \hbar c_{3,1}}\int_{\mathbb{R}^3}  \me^{\mi \bk^{\prime} \cdot (\bx_e - \bx_p)}  \frac{1}{|\bk^{\prime}|^2}  f(\bx_e, \bx_p, \bk_e +\frac{\bk^{\prime}}{2}, \bk_p - \frac{\bk^{\prime}}{2}, t) \D \bk^{\prime}.
\end{split}
\end{equation}

\subsection{Exact solution of the proton-electron Wigner equation}

Now we introduce the electron and proton velocities $\bv = (\bv_e, \bv_p)$, $\bv_e = \hbar \bk_e/m_e$, $\bv_p = \hbar \bk_p/m_p$, the scaled Wigner function is
\begin{equation}
\widetilde{f}(\bx_e, \bx_p, \bv_e, \bv_p, t) = \widetilde{f}(\bx_e, \bk_p, \frac{\hbar \bk_e}{m_e}, \frac{\hbar \bk_p}{m_p}, t) = f(\bx_e, \bx_p, \bk_e, \bk_p, t).
\end{equation}
We will show that the proton-electron Wigner equation \eqref{supp_12d_wigner_c} can be solved exactly provided that 
\begin{equation}\label{supp_exact_init}
\widetilde f(\bx_e, \bx_p, \bv_e, \bv_p, 0) = \widetilde f_c(\bx_c, \bv_c, t) \widetilde f_r(\bx_r, \bv_r, 0).
\end{equation}

First, we use  the center-of-mass coordinate
\begin{equation}\label{supp_centre_of_mass}
\left\{
\begin{split}
& \bx_c = \frac{m_p \bx_p + m_e \bx_e}{m_e + m_p}, \\
& \bx_r = \bx_e - \bx_p, \\
& \bv_c = \frac{m_p \bv_p + m_e \bv_e}{m_e + m_p}, \\
& \bv_r = \bv_e - \bv_p,
\end{split}
\right.
\end{equation}
A simple calculation yields that
\begin{equation*}
\begin{split}
&\bv_e \cdot \nabla_{\bx_e}  = \frac{m_e}{m_e + m_p} \bv_e \cdot \nabla_{\bx_c} + \bv_e \cdot \nabla_{\bx_r} , \\
&\bv_p \cdot \nabla_{\bx_p}  = \frac{m_p}{m_e + m_p} \bv_p \cdot \nabla_{\bx_c} - \bv_p \cdot \nabla_{\bx_r},
\end{split}
\end{equation*}
so that the kinetic part is obtained,
\begin{equation}
\bv_e \cdot \nabla_{\bx_e}  + \bv_p \cdot \nabla_{\bx_p} = \bv_c \cdot \nabla_{\bx_c} +  \bv_r \cdot \nabla_{\bx_r}. 
\end{equation}

For $\pdo$, it reads that
\begin{equation*}
\begin{split}
\Theta_V[\widetilde f](\bx_e, \bx_p, \bv_e, \bx_p, &t) = \frac{\gamma}{ \mi \hbar c_{3,1}} \int_{\mathbb{R}^3}  \frac{ \me^{\mi \bk^{\prime} \cdot (\bx_e - \bx_p)}}{|\bk^{\prime}|^2} \widetilde f(\bx_e, \bx_p, \bv_e - \frac{\hbar\bk^{\prime}}{2m_e}, \bv_p + \frac{\hbar\bk^{\prime}}{2m_p}, t) \D \bk^{\prime} \\
& - \frac{\gamma}{\mi \hbar c_{3,1}} \int_{\mathbb{R}^3} \frac{\me^{\mi \bk^{\prime} \cdot (\bx_e - \bx_p)}}{|\bk^{\prime}|^2}  \widetilde f(\bx_e, \bx_p, \bv_e + \frac{\hbar \bk^{\prime}}{2m_e}, \bv_p - \frac{\hbar \bk^{\prime}}{2m_p}, t) \D \bk^{\prime}.
\end{split}
\end{equation*}

Now we will show that Eq.~\eqref{supp_12d_wigner_c} can be solved by separation of variables. By taking the ansatz
\begin{equation}\label{supp_assumption}
\widetilde f(\bx_e, \bx_p, \bv_e, \bv_p, t) = \widetilde f_c(\bx_c, \bv_c, t) \widetilde f_r(\bx_r, \bv_r, t),
\end{equation}
it has that
\begin{equation}
\begin{split}
&\widetilde f(\bx_e, \bx_p, \bv_e + \frac{\hbar \bk^{\prime}}{2m_e}, \bv_p - \frac{\hbar \bk^{\prime}}{2m_p}, t) =  \widetilde f_c(\bx_c, \bv_c, t) \widetilde f_r(\bx_r, \bv_r + \frac{\hbar \bk^{\prime}}{2m_e} +  \frac{\hbar \bk^{\prime}}{2m_p}, t), \\
&\widetilde f(\bx_e, \bx_p, \bv_e - \frac{\hbar \bk^{\prime}}{2m_e}, \bv_p + \frac{\hbar \bk^{\prime}}{2m_p}, t) =  \widetilde f_c(\bx_c, \bv_c, t) \widetilde f_r(\bx_r, \bv_r - \frac{\hbar \bk^{\prime}}{2m_e} -  \frac{\hbar \bk^{\prime}}{2m_p}, t). 
\end{split}
\end{equation}
As a consequence, $\pdo$ becomes 
\begin{equation*}
\begin{split}
\Theta_V[\widetilde f](\bx_c,\bx_r, \bv_c, \bv_r, t) = &\frac{\gamma\widetilde f_c(\bx_c, \bv_c, t)}{ \mi \hbar c_{3,1}} \int_{\mathbb{R}^3}  \frac{ \me^{\mi \bk^{\prime} \cdot \bx_r}}{|\bk^{\prime}|^2}  \widetilde f_r(\bx_r, \bv_r + \frac{\hbar \bk^{\prime}}{2m_e} +  \frac{\hbar \bk^{\prime}}{2m_p}, t) \D \bk^{\prime} \\
& - \frac{\gamma\widetilde f_c(\bx_c, \bv_c, t)}{\mi \hbar c_{3,1}} \int_{\mathbb{R}^3} \frac{\me^{\mi \bk^{\prime} \cdot \bx_r }}{|\bk^{\prime}|^2} \widetilde f_r(\bx_r, \bv_r - \frac{\hbar \bk^{\prime}}{2m_e} -  \frac{\hbar \bk^{\prime}}{2m_p}, t) \D \bk^{\prime}.
\end{split}
\end{equation*}

Finally, we can integrate $\bx_c$ and $\bv_c$ variables and obtain the Wigner equation in $(\bx_r, \bv_r)$-space
\begin{equation*}
\begin{split}
\left(\frac{\partial }{\partial t}   + \bv_r \cdot  \nabla_{\bm{x}_r}\right) \widetilde f_r(\bm{x}_r, \bv_r, &t) = \frac{\gamma}{ \mi \hbar c_{3,1}} \int_{\mathbb{R}^3}  \frac{ \me^{\mi \bk^{\prime} \cdot \bx_r}}{|\bk^{\prime}|^2} \widetilde f_r(\bx_r, \bv_r- \frac{\hbar \bk^{\prime}}{2m_e} - \frac{\hbar \bk^{\prime}}{2m_p}, t) \D \bk^{\prime}  \\
&- \frac{\gamma}{\mi \hbar c_{3,1}} \int_{\mathbb{R}^3} \frac{\me^{\mi \bk^{\prime} \cdot \bx_r}}{|\bk^{\prime}|^2}  \widetilde f_r(\bx_r, \bv_r + \frac{\hbar \bk^{\prime}}{2m_e} + \frac{\hbar \bk^{\prime}}{2m_p}, t) \D \bk^{\prime},
\end{split}
\end{equation*}
which obeys the single-body Wigner equation \eqref{supp_6d_wigner_c}. Similarly, by integrating $\bx_r$ and $\bv_r$ variables, it yields that
\begin{equation}
\left(\frac{\partial }{\partial t}   + \bv_c \cdot  \nabla_{\bm{x}_c}\right) \widetilde f_c(\bm{x}_c, \bv_c, t)  = 0.
\end{equation}

\subsection{Asymptotic approximation to the reduced Wigner function}

When the proton-electron Wigner function is strongly localized in $\bx_p$-space, it is possible to derive an asymptotic approximation to the reduced Wigner function via the single-body Wigner-Coulomb dynamics, which uses the fact that $m_p \approx 1836m_e$. This may facilitate our subsequent benchmarks.

\begin{theorem} Suppose there exists a fixed $\bx_A \in \mathbb{R}^3$ and a small $\varepsilon >0 $ such that  for $t \le T$, $T = \mathcal{O}(1)$,
\begin{equation}
f(\bx_e, \bx_p, \bk_e, \bk_p, t) = F(\bx_e, \bx_p, \bk_e, \bk_p, t) \me^{-|\bx_p - \bx_A - \frac{\hbar \bk_p t}{m_p}|^2/\varepsilon}
\end{equation}
with $F(\bx_e, \bx_p, \bk_e, \bk_p, t) = \mathcal{O}(1)$. Then  it has that
\begin{equation}\label{supp_eq.asymptotic_approximation}
 P_e(\bx_e, \bk_e, t)  =   f_e(\bx_e, \bk_e, t) + \mathcal{O}( m_p^{-1} \varepsilon^{3/2}),
\end{equation}
 where the reduced electron Wigner function $P_e(\bx_e, \bk_e, t)$ is defined by 
\begin{equation}
 P_e(\bx_e, \bk_e, t) = \iint_{\mathbb{R}^3 \times \mathbb{R}^3}  f(\bx_e, \bx_p, \bk_e, \bk_p, t) \D \bx_p \D \bk_p,
\end{equation}
and $f_e(\bx_e, \bk_e, t)$ is the solution of a single-body Wigner equation \eqref{supp_single_body_Wigner}-\eqref{supp_pdo_scatter}.
\end{theorem}


\begin{proof}
To illustrate its derivation, we first start from one branch
\begin{equation*}
\begin{split}
&\int_{\mathbb{R}^3} \D \bx_p \int_{\mathbb{R}^3} \me^{\mi \bk^{\prime} \cdot (\bx_e - \bx_p)} \frac{1}{|\bk^{\prime}|^2}  f(\bx_e, \bx_p, \bk_e -\frac{\bk^{\prime}}{2}, \bk_p + \frac{\bk^{\prime}}{2}, t) \D \bk^{\prime}\\
&= \underbrace{\int_{\mathbb{R}^3} \D \bx_p \int_{\mathbb{R}^3} \me^{\mi \bk^{\prime} \cdot (\bx_e - \bx_A)} \frac{1}{|\bk^{\prime}|^2}  f(\bx_e, \bx_p, \bk_e - \frac{\bk^{\prime}}{2}, \bk_p + \frac{\bk^{\prime}}{2}, t) \D \bk^{\prime}}_{\textup{I}} \\
& ~+ \underbrace{\int_{\mathbb{R}^3} \D \bx_p \int_{\mathbb{R}^3} \left(\me^{-\mi \bk^{\prime} \cdot (\bx_p - \bx_A)} -1 \right)\frac{ \me^{\mi \bk^{\prime} \cdot (\bx_e - \bx_A)}}{|\bk^{\prime}|^2} f(\bx_e, \bx_p, \bk_e -\frac{\bk^{\prime}}{2}, \bk_p + \frac{\bk^{\prime}}{2}, t) \D \bk^{\prime}}_{\textup{II}}.
\end{split}
\end{equation*}

For the first term, using the variable conversion $\bk_p \to \bk_p - \bk^{\prime}/2$, it has that
\begin{equation*}
\begin{split}
&\int_{\mathbb{R}^3} \D \bk_p \int_{\mathbb{R}^3} \D \bx_p \int_{\mathbb{R}^3} \me^{\mi \bk^{\prime} \cdot (\bx_e - \bx_A)} \frac{1}{|\bk^{\prime}|^2}  f(\bx_e, \bx_p, \bk_e -\frac{\bk^{\prime}}{2}, \bk_p + \frac{\bk^{\prime}}{2}, t) \D \bk^{\prime} \\
& = \int_{\mathbb{R}^3} \D \bk_p \int_{\mathbb{R}^3} \D \bx_p \int_{\mathbb{R}^3} \me^{\mi \bk^{\prime} \cdot (\bx_e - \bx_A)} \frac{1}{|\bk^{\prime}|^2} f(\bx_e, \bx_p, \bk_e - \frac{\bk^{\prime}}{2}, \bk_p, t) \D \bk^{\prime} \\
& = \int_{\mathbb{R}^3} \me^{\mi \bk^{\prime} \cdot (\bx_e - \bx_A)} \frac{1}{|\bk^{\prime}|^2} P_e(\bx_e, \bk_e -\frac{\bk^{\prime}}{2}, t) \D \bk^{\prime}. 
\end{split}
\end{equation*}

For the second term, using the Laplace asymptotic expansion, it yields that
\begin{equation*}
\begin{split}
\textup{II} = \int_{\mathbb{R}^3} &\D \bx_p \int_{\mathbb{R}^3} (\me^{-\mi \bk^{\prime} \cdot (\bx_p - \bx_A)} -1 ) \frac{\me^{\mi \bk^{\prime} \cdot (\bx_e - \bx_A)}}{|\bk^{\prime}|^2}  f(\bx_e, \bx_p, \bk_e - \frac{\bk^{\prime}}{2}, \bk_p + \frac{\bk^{\prime}}{2}, t) \D \bk^{\prime}\\
= \int_{\mathbb{R}^3} &\me^{-\frac{|\bx_p - \bx_A  - \frac{\hbar \bk_p t}{m_p}|^2}{\varepsilon}}\D \bx_p \int_{\mathbb{R}^3}  (\me^{-\mi \bk^{\prime} \cdot (\bx_p - \bx_A)} -1 ) \\
&\times\frac{\me^{\mi \bk^{\prime} \cdot (\bx_e - \bx_A)}}{|\bk^{\prime}|^2}  F(\bx_e, \bx_p, \bk_e - \frac{\bk^{\prime}}{2}, \bk_p + \frac{\bk^{\prime}}{2}, t) \D \bk^{\prime}\\
\sim\int_{\mathbb{R}^3} &  \left(2\pi\varepsilon\right)^{\frac{3}{2}}(\me^{-\mi \bk^{\prime} \cdot  \frac{\hbar \bk_p t}{m_p}} -1 )\frac{ \me^{\mi \bk^{\prime} \cdot (\bx_e - \bx_A)}}{|\bk^{\prime}|^2} \\
&\times F(\bx_e, \bx_A + \frac{\hbar \bk_p t}{m_p}, \bk_e - \frac{\bk^{\prime}}{2}, \bk_p + \frac{\bk^{\prime}}{2}, t) \D \bk^{\prime}+ \mathcal{O}(\varepsilon^{{5}/{2}}) \sim \mathcal{O}(\hbar m_p^{-1} \varepsilon^{3/2}).
\end{split}
\end{equation*}
Now further integrating in $\bk_p$-space, we have that 
\begin{equation}
\begin{split}
\frac{\partial }{\partial t} &P_e(\bm{x}_e, \bm{x}_p, t) +  \frac{\hbar \bk_e}{m_e} \cdot \nabla_{\bm{x}_e} P_e(\bm{x}_e, \bm{x}_p, t)  \\
&=  \frac{\gamma}{\mi \hbar c_{3,1}}\int_{\mathbb{R}^3} \me^{\mi \bk^{\prime} \cdot (\bx_e - \bx_A)} \frac{1}{|\bk^{\prime}|^2} P_e(\bx_e, \bk_e - \frac{\bk^{\prime}}{2}, t) \D \bk^{\prime} + \mathcal{O}( m_p^{-1}\varepsilon^{3/2}).
\end{split}
\end{equation}
Another branch can be tackled in a similar way. Finally, by omitting the asymptotic error terms, we arrive at Eq.~\eqref{supp_eq.asymptotic_approximation}.
\end{proof}

\section{Particle generation and numerical sign problem}
\label{supp_sec.wbrw}
The Wigner Monte Carlo can be constructed by the formal Neumann series expansion of the Wigner equation. Detailed derivations can be found in many literatures, e.g.,  \cite{KosinaNedjalkovSelberherr2003,bk:NedjalkovQuerliozDollfusKosina2011,Wagner2016}, and rigorous mathematical proofs have also been established via the framework of the theory of the continuous Markov branching process \cite{Wagner2016,ShaoXiong2020}. 

Despite its success, the accuracy of the Wigner Monte Carlo is still hampered by the numerical sign problem, which can be greatly alleviated by the stationary phase approximation (SPA). In the following, we will use the Wigner equation under 2-D Morse potential as an example to illustrate this point.

\subsection{Neumann series expansion}

Consider the inner product problem 
\begin{equation}
\langle f, g\rangle = \iint_{\mathbb{R}^d\times \mathbb{R}^d} f(\bx, \bk) g(\bx, \bk) \D \bx \D \bk,
\end{equation}
then for $f(\bx, \bk, 0) = f_0(\bx, \bk)$, it has 
\begin{equation*}
\begin{split}
&\langle \varphi(\bx, \bk), f(\bx, \bk, t) \rangle = \underbrace{\me^{-\gamma_0 t} \langle \varphi(\bx(t), \bk), f_0(\bx, \bk) \rangle}_{\textup{frozen state}} \\
&- \int_0^t \underbrace{\gamma_0 \me^{-\gamma_0 (t-t^{\prime})}}_{\textup{particle life}} \langle \underbrace{ \frac{\Theta_V[\varphi](\bx(t-t^{\prime}), \bk^{\prime}, t^{\prime})}{\gamma_0} - \varphi(\bx(t-t^{\prime}), \bk, t^{\prime})}_{\textup{transition of states}}, f(\bx, \bk, t^{\prime}) \rangle \D t^{\prime},
\end{split}
\end{equation*}
where $(\bx(\tau), \bk) = (\bx + \frac{\hbar \bk \tau}{m}, \bk)$. One can further expand $\langle \varphi(\bx, \bk), f(\bx, \bk, t^{\prime})\rangle$ and rewrite it by an iterative integral,
\begin{equation}\label{supp_Neumann_series}
\begin{split}
&\langle \varphi(\bx, \bk), f(\bx, \bk, t) \rangle=  \underbrace{\me^{-\gamma_0 t} \langle \varphi(\bx(t), \bk), f_0(\bx, \bk) \rangle}_{\textup{zeroth expansion}}  \\
&+ \underbrace{\int_0^t \D t_1 \langle \me^{-\gamma_0(t - t_1)}   (\Theta_V[\varphi] - \gamma_0\varphi)(\bx(t-t_1), \bk, t_1),  \me^{-\gamma_0 t_1} f_0(\bx(t), \bk)\rangle }_{\textup{first-order expansion}} \\
&+ \underbrace{\int_{0}^{t} \D t_1 \int_{0}^{t_1} \D t_2\langle \me^{-\gamma_0(t - t_1)}  (\Theta_V[\varphi]- \gamma_0\varphi)(\bx(t-t_1), \bk, t_1)}_{\textup{second-order expansion}}\\
&\quad \underbrace{\times \me^{-\gamma_0(t_1-t_2)} (\Theta_V[\varphi]- \gamma_0\varphi)(\bx(t-t_2), \bk, t_2), \me^{-\gamma_0 t_2}f_0(\bx(t), \bk)\rangle }_{\textup{second-order expansion}}   + \cdots 
\end{split}
\end{equation}
The $n$-th expansion ($n \ge 1$) corresponds to $n$-th jump in a Markov process.

The key is to endow $\pdo$ with a probabilistic interpretation, which can be understood by the basic idea of particle splitting \cite{KosinaNedjalkovSelberherr2003},
\begin{equation}\label{supp_pdo_branch}
\begin{split}
\Theta_V[\varphi]&(\bx, \bk, t) =    - \frac{\kappa \me^{\kappa r_0} c_2}{\hbar} \int_{0}^{2\pi} \D \phi \int_{0}^{+\infty} \D r~ \frac{r \sin(2(\bx-\bx_A) \cdot \bk^{\prime})}{\sqrt{r^2+(\kappa/2)^2}}  \frac{\Delta_{r\sigma}[\varphi](\bx, \bk, t)}{r^2 + (\kappa/2)^2}  \\
& + \frac{\kappa \me^{2\kappa r_0} c_2}{\hbar} \int_{0}^{2\pi} \D \phi \int_{0}^{+\infty} \D r~ \frac{r \sin(2(\bx-\bx_A) \cdot \bk^{\prime})}{\sqrt{r^2+\kappa^2}}  \frac{\Delta_{r\sigma}[\varphi](\bx, \bk, t)}{r^2 + \kappa^2} \\
& =   \int_0^{2\pi} \D \phi \int_{0}^{+\infty} \D r \underbrace{\psi(\bx, r, \sigma)}_{\textup{particle weight}} \underbrace{\frac{\kappa}{\pi} \frac{r}{r^2 + \kappa^2}}_{\textup{jump}} ( \underbrace{\varphi(\bx, \bk - \frac{r\sigma}{2}, t)}_{\textup{left branch}} - \underbrace{\varphi(\bx, \bk + \frac{r\sigma}{2}, t)}_{\textup{right branch}}),
\end{split}
\end{equation}
where $\sigma = (\cos \phi, \sin \phi)$ and
\begin{equation*}
\psi(\bx, r, \sigma) =  \frac{ \pi \me^{\kappa r_0} c_2}{\hbar} \left[- \frac{r^2 + \kappa^2}{(r^2 + \kappa^2/4)^{3/2}} +  \frac{\me^{\kappa r_0}}{(r^2 +\kappa^2)^{1/2}}\right] \sin(2r \sigma \cdot (\bx - \bx_A)).
\end{equation*}

Combining Eqs.~\eqref{supp_Neumann_series} and \eqref{supp_pdo_branch}, we can perform the following simulation. Each time we can pick up one particle and sample a random life-length $\tau\propto \gamma_0 \me^{-\gamma_0 t}$. When $t + \tau > T$, it moves to the state $(\bx(T - t), \bk)$ and becomes frozen. Otherwise, it moves to the state $(\bx(\tau), \bk)$ and is killed, and generate three new particles at states $(\bx(\tau), \bk)$, $(\bx(\tau), \bk - \frac{r\sigma}{2}, t)$ and $(\bx(\tau), \bk + \frac{r\sigma}{2}, t)$ with  random state ${r\sigma}/{2}$ generated from the Cauchy distribution with particle weight multiplied by $\psi(\bx(\tau), r, \sigma)$ for one offspring and $-\psi(\bx(\tau), r, \sigma)$ for another offspring. The simulation continues until all particles become frozen, with more details put in Algorithm \ref{supp_WBRW_SPA}.

 \begin{figure}[!h]
 \centering
\subfigure[Variances in $W_1(x, k, t)$.]{\includegraphics[width=0.32\textwidth,height=0.22\textwidth]{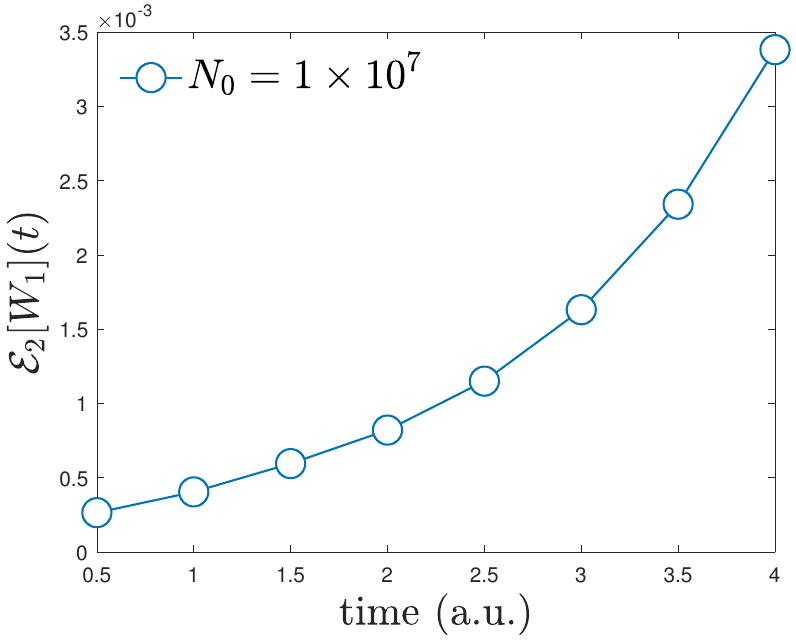}}
\subfigure[Variances in $P(x_1, x_2, t)$.]{\includegraphics[width=0.32\textwidth,height=0.22\textwidth]{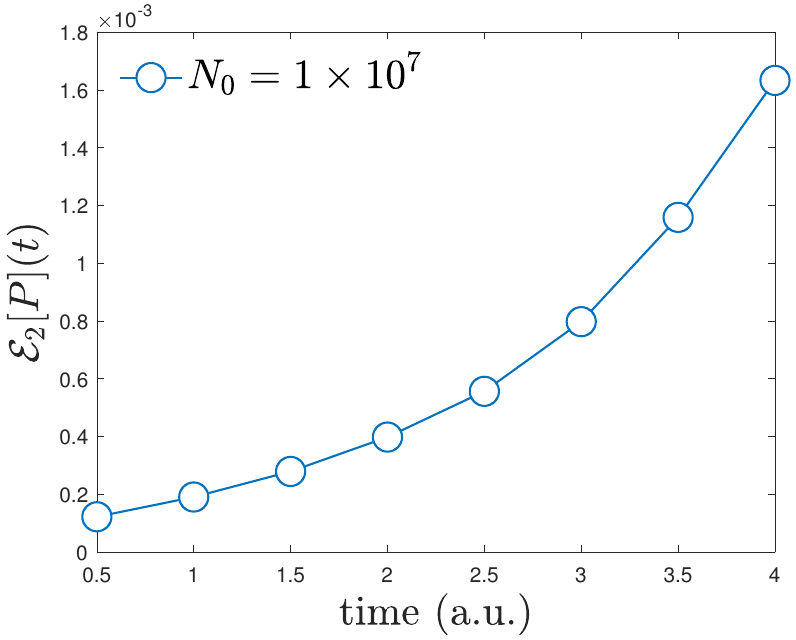}}
\subfigure[Growth of particles.]{\includegraphics[width=0.32\textwidth,height=0.22\textwidth]{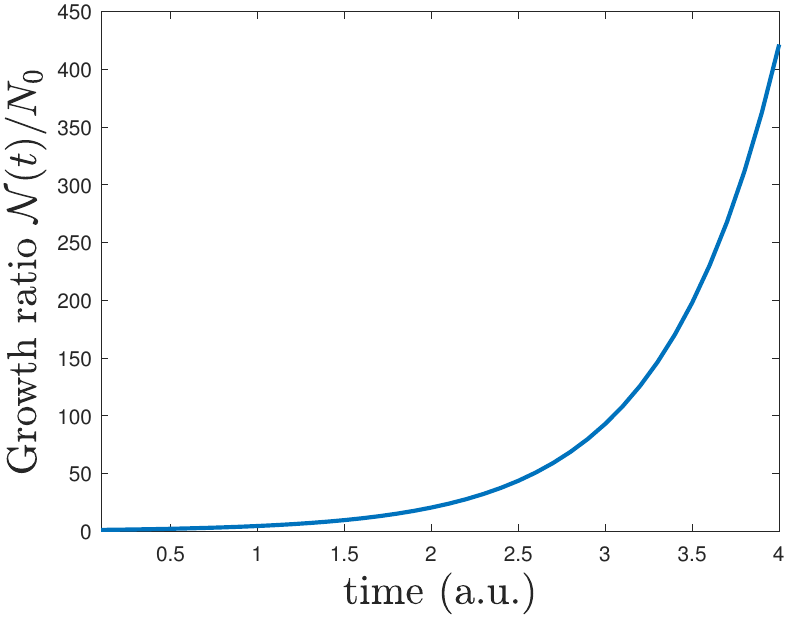}}
\\
\centering
\subfigure[$W_1(x, k, t)$ at $t=4$a.u., produced by deterministic scheme (left) and MC (right).\label{supp_plot_noise}]{
{\includegraphics[width=0.48\textwidth,height=0.27\textwidth]{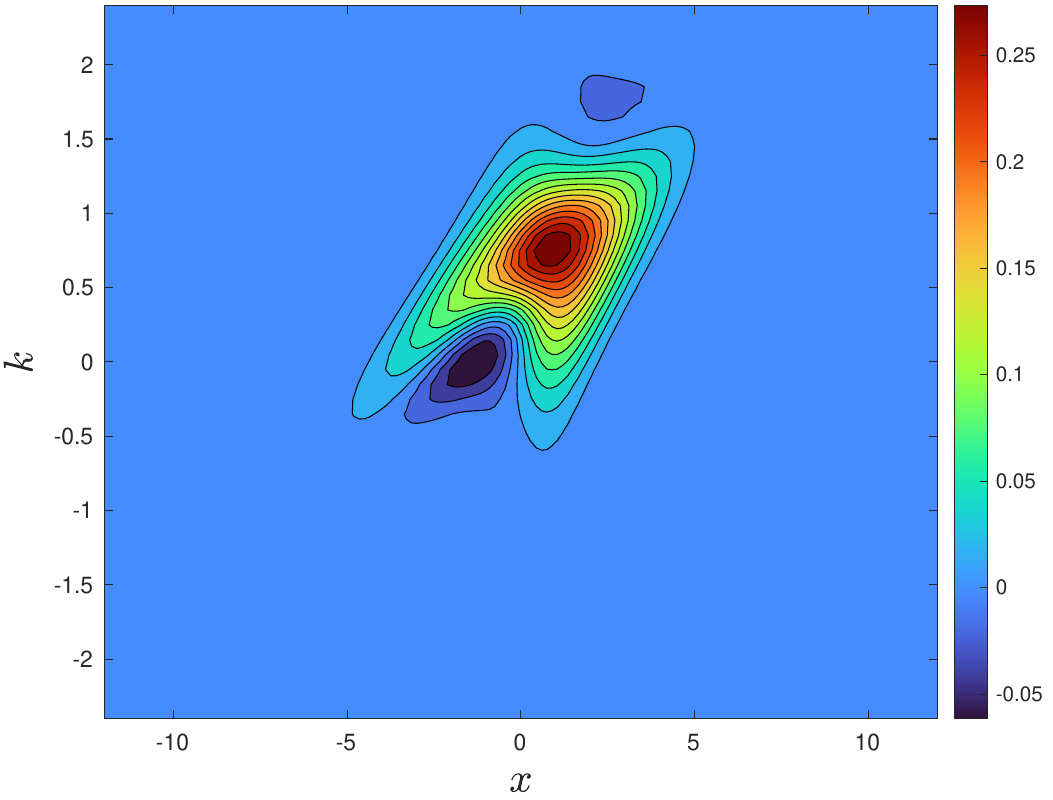}}
{\includegraphics[width=0.48\textwidth,height=0.27\textwidth]{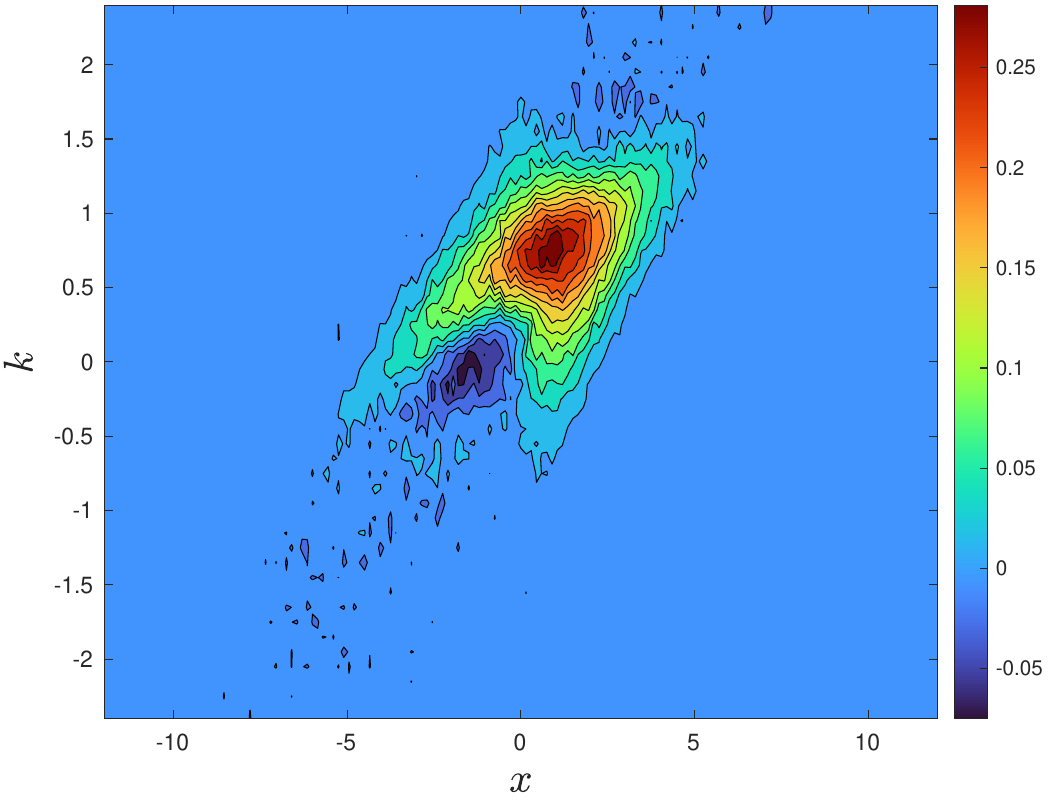}}}
\caption{\small  The 4-D Morse system: A visualization of sign problem and the exponential growth of errors (stochastic variances).  }
\label{supp_2dm_sign_problem}
\end{figure}
Unfortunately, direct particle splitting simulations may suffer from a sharp deterioration in accuracy due to the exponential growth of stochastic variances, as it ignores the decay property of $\pdo$ \cite{ShaoXiong2020}. The sign problem can be visualized by the stochastic noises in the reduced Wigner functions at $t = 4$a.u. (see Figure \ref{supp_plot_noise}), where the effective sample size is $N_0 = 1\times 10^7$. Here the model parameters are: $\bx_A = (0, 0), r_0 = 0.5, \kappa = 0.5$, $\hbar = m = 1$.

Actually, we have monitored the stochastic variances by $l^2$-errors $\mathcal{E}_{2}[W_1](t)$ and $\mathcal{E}_{2}[P_1](t)$, defined by \eqref{supp_def.L2error} and \eqref{supp_def.L2error_xdist}, respectively. In Figure \ref{supp_2dm_sign_problem}, it is seen that both $\mathcal{E}_{2}[W_1](t)$ and $\mathcal{E}_{2}[P_1](t)$ grow exponentially in time.  Besides, the particle number increases from $1 \times 10^7$ at $t= 0$ to $4.21 \times 10^9$ at $t = 4$a.u. The growth ratio is about $\exp(1.51t)$, which reaches $3.6\times 10^{6}$ at $t = 10$a.u. In other words, it is prohibitive to perform long-time stochastic simulations due to the numerical sign problem.

\subsection{The stationary phase approximation}

In our recent work \cite{ShaoXiong2020}, we have analyzed the stochastic variances and found that the numerical sign problem is actually induced by the particle splitting technique. Although it gives a practical stochastic interpretation to $\pdo$, the splitting of the oscillatory integral ignores the near-cancelation of its high-frequency components, and consequently leads to an exponential increases of variances.
The remedy is the stationary phase approximation (SPA) to $\pdo$. The leading terms of the asymptotic expansion capture the major contribution of the oscillatory integrals.

First, we introduce a filter $\lambda_0$ and a ball $B(r)$ with radius $r$, and try to replace the components outside the ball by an integral over a line,
\begin{equation*}
\begin{split}
&\Theta^{\lambda_0}_V[\varphi](\bx, \bk, t)  =  \int_{B(\frac{\lambda_0}{|\bx-\bx_A|})} \me^{\mi (\bx - \bx_A) \cdot \bk^{\prime}} \psi(\bk^{\prime}) ( \underbrace{\varphi(\bx, \bk - \frac{\bk^{\prime}}{2}, t)}_{\textup{left branch}} - \underbrace{\varphi(\bx, \bk+\frac{\bk^{\prime}}{2}, t)}_{\textup{right branch}} ) \D \bk^{\prime} \\
&+2\int_{\frac{\lambda_0}{|\bx-\bx_A|}}^{+\infty} \textup{Im} \left(\frac{\sqrt{2\pi}\me^{\mi r |(\bx - \bx_A)|}}{\sqrt{\mi r |\bx - \bx_A|}}\right) r{\psi}(r\sigma_\ast)( \underbrace{\varphi(\bx, \bk - \frac{r\sigma_\ast}{2}, t)}_{\textup{left branch}} - \underbrace{\varphi(\bx, \bk + \frac{r\sigma_\ast}{2}, t)}_{\textup{right branch}}) \D r,
\end{split}
\end{equation*}
where the amplitude function reads 
\begin{equation}
\psi(\bk) = \frac{1}{\mi \hbar} \left[  - \frac{2\kappa \me^{\kappa r_0}c_2 }{(|\bk|^2 + \kappa^2)^{3/2}} + \frac{2\kappa \me^{2\kappa r_0}c_2}{(|\bk|^2 + 4\kappa^2)^{3/2}} \right],
\end{equation}
with $\sigma_\ast = (\cos \phi_\ast, \sin \phi_\ast)$ and $\phi_\ast =\textup{atan2}(\frac{x_2 - x_{A,2}}{x_1-x_{A,1}})$. One can prove  
\begin{equation}
\Theta_V[\varphi](\bx, \bk, t) = \Theta_V^{\lambda_0}[\varphi](\bx, \bk, t) + \mathcal{O}(\lambda_0^{-1}),
\end{equation}
in the sense that
\begin{equation}
\Vert \Theta_V[\varphi](\bx, \bk, t) - \Theta_V^{\lambda_0}[\varphi](\bx, \bk, t) \Vert_{L^2_{\bx}L^2_{\bk}} \lesssim \lambda_0^{-1} \Vert \varphi(t) \Vert_{L^2_{\bx} H^1_{\bk}},
\end{equation}
and $\Vert \varphi(t) \Vert_{L^2_{\bx} H^1_{\bk}} = \Vert \varphi(t) \Vert_{L^2_{\bx}L^2_{\bk}} + \Vert  \nabla_{\bk} \varphi(t)\Vert_{L^2_{\bx}L^2_{\bk}}$. When $\lambda_0$ is larger than $1$, the asymptotic error term $\mathcal{O}(\lambda_0^{-1
})$ decays as $\lambda_0$ increases.

 Again, we take the 4-D Wigner equation under the Morse potential as an example. The implementation of the Wigner Branching Random Walk associated with SPA (termed WBRW-SPA for short) is illustrated in Algorithm \ref{supp_WBRW_SPA}, starting from the initial instant $t_l$ and stopping at the final instant $t_{l+1}$. In this way, the particle method resolves the Wigner dynamics by simulating the deterministic motions, random jumps, random generation of superparticles in the phase space.

\begin{algorithm}[!h]
\caption{WBRW-SPA for the 2-D Morse system.}
\label{supp_WBRW_SPA}
\vspace{1mm} 

{\bf Input parameters}: The initial time $t_l$ and final time $t_{l+1}$, the constant rate $\gamma_0$, the filter $\lambda_0$, $\bk$-domain $\mathcal{K}$ and the upper band $r_{\max} > 4|\mathcal{K}|$.

{\bf Sampling processes}: Suppose each particle in the branching particle system, carrying an initial weight $w$ either $1$ or $-1$, starts at state $(\bx, \bk)$ at time $t_l$ and moves until $t_{l+1} = t_l +\Delta t$ according to the following rules.  

\begin{description}

\item[1.] ({\bf Frozen}) Generate a random $\tau \propto \gamma_0 \me^{-\gamma_0 t}$. For a particle at $(\bx, \bk)$ at instant $t \in [t_{l}, t_{l+1}]$, if $t+ \tau \ge t_{l+1}$, it becomes frozen at $(\bx +  \frac{\hbar \bk(t_{l+1} - t)}{m}, \bk, t_{l+1})$.

\item[2.] ({\bf Death}) If $\tau < \Delta t$, the particle moves to $(\bx +  \frac{\hbar \bk \tau}{m}, \bk, t+\tau)$ and is killed. 

\item[3.] ({\bf Branching}) When the particle is killed, it produces at most three offsprings at states $(\bx^{(1)}, \bk^{(1)}, t+\tau)$, $(\bx^{(2)}, \bk^{(2)}, t+\tau)$ and $(\bx^{(3)}, \bk^{(3)}, t+\tau)$. The third offspring is produced at state $(\bx^{(3)}, \bk^{(3)}) = (\bx +  \frac{\hbar \bk \tau}{m}, \bk)$ with probability $1$, carrying the weight $w$.

\item[4.] ({\bf Scattering}) Generate a random number $r$ from the Cauchy distribution $\frac{1}{\pi}\frac{r}{r^2 + \kappa^2}$.

\begin{description}

\item[(1)] If $r < \lambda_0/|\bx^{(3)} - \bx _A|$, generate random numbers $\phi$ uniformly in $[0, 2\pi]$, yielding vectors $\sigma =  (\cos \phi, \sin \phi)$, $\bk^{\prime} = r\sigma$. It produces two offsprings with probability $\Pr(1)$, $\Pr(2)$ at states $(\bx^{(1)}, \bk^{(1)})$, $(\bx^{(2)}, \bk^{(2)})$ endowed with updated weights $w_1$ and $w_2$, respectively.  
\begin{align*}
&\Pr(1) = \Pr(2) =  \frac{|\psi(\bx^{(3)}, r, \sigma)| }{\gamma_0}, \\
 &\bx^{(1)}  = \bx^{(2)} =  \bx^{(3)} = \bx +  \frac{\hbar \bk \tau}{m}, ~~ \bk^{(1)} = \bk - \frac{\bk^{\prime}}{2}, ~~ \bk^{(2)} = \bk + \frac{\bk^{\prime}}{2}, \\
&w_i = w \cdot \frac{(-1)^{i-1} \psi(\bx^{(3)}, r, \sigma)}{|\psi(\bx^{(3)}, r, \sigma)|} \cdot \mone_{\{\bk^{(i)} \in \mathcal{K}\}}, ~~ i = 1, 2.
\end{align*}

\item[(2)] If $r \ge \lambda_0/|\bx^{(3)} - \bx _A|$, it produces two offsprings with the probability $\Pr(1)$, $\Pr(2)$ at states $(\bx^{(1)}, \bk^{(1)})$, $(\bx^{(2)}, \bk^{(2)})$ endowed with updated weights $w_1$ and $w_2$, respectively.
\begin{align*}
&\Pr(1) = \Pr(2) = \frac{2}{\gamma_0} \Big |\textup{Im}[\me^{\mi r |(\bx^{(3)} - \bx_A)|} \left(\frac{2\pi}{\mi r |\bx^{(3)} - \bx_A|}\right)^{\frac{1}{2}}] r{\psi}(r\sigma_\ast) \Big |, \\
&\bx^{(1)}  = \bx^{(2)} =  \bx^{(3)}, ~~ \bk^{(1)} = \bk - \frac{r\sigma_\ast}{2}, ~~ \bk^{(2)} = \bk +  \frac{r\sigma_\ast}{2},\\
&w_i = w \cdot \frac{(-1)^{i-1}\textup{Im}[\me^{\mi r |(\bx^{(3)} - \bx_A)|} (\frac{2\pi}{\mi r |\bx^{(3)} - \bx_A|})^{\frac{1}{2}}] r{\psi}(r\sigma_\ast)}{\Big |\textup{Im}[\me^{\mi r |(\bx^{(3)} - \bx_A)|} (\frac{2\pi}{\mi r |\bx^{(3)} - \bx_A|})^{\frac{1}{2}}] r{\psi}(r\sigma_\ast) \Big |} \cdot \mone_{\{\bk^{(i)} \in \mathcal{K}\}},
\end{align*}  
$\sigma_\ast = \sigma_\ast(\bx^{(3)}) = (\cos \phi_\ast, \sin \phi_\ast)$ and $\phi_\ast =\textup{atan2}\left(\frac{x_2^{(3)} - x_{A,2}}{x_1^{(3)}-x_{A,1}}\right)$, $i = 1, 2$.

\end{description}

\item[5.] ({\bf Independence}) The offsprings continue to move independently.

\end{description}

{\bf Termination condition}: All particles in the branching particle system are frozen.

\end{algorithm}

\begin{figure}[!h]
 \subfigure[Deterministic.]{
{\includegraphics[width=0.32\textwidth,height=0.22\textwidth]{./redist_asm_0040.pdf}}}
\subfigure[MC without SPA.]{
{\includegraphics[width=0.32\textwidth,height=0.22\textwidth]{./redist_mc_HJD_0040.pdf}}}
\subfigure[MC with SPA, $\lambda_0 = 1$.]{
{\includegraphics[width=0.32\textwidth,height=0.22\textwidth]{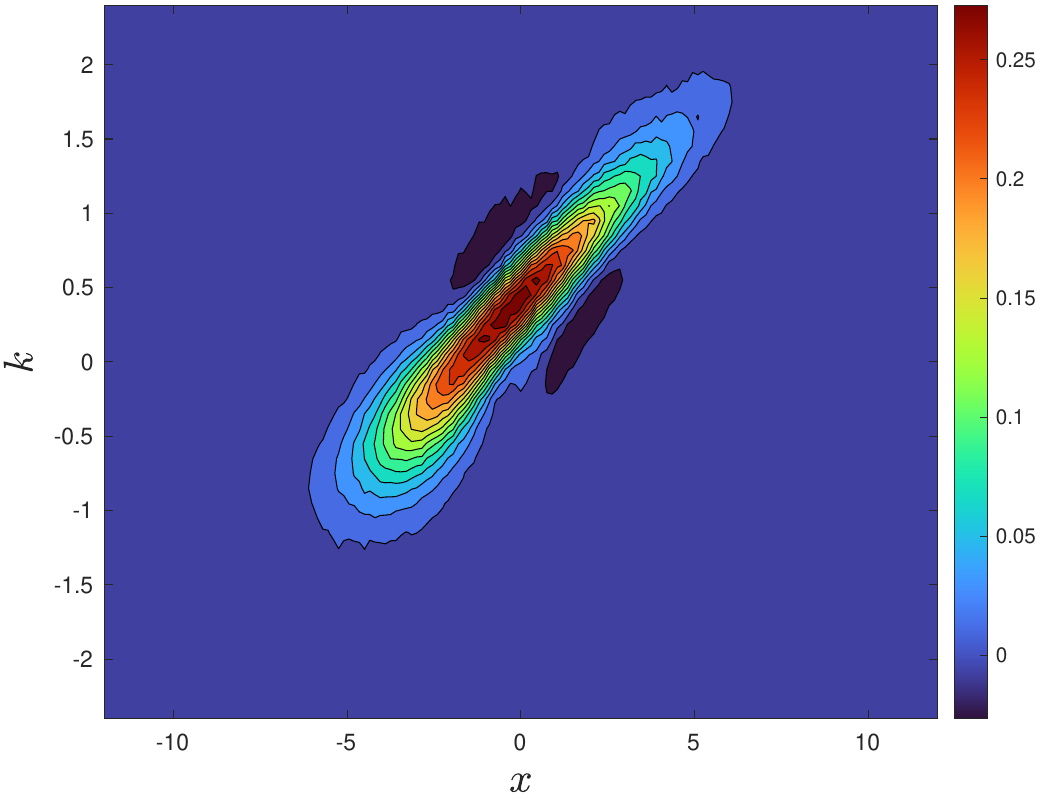}}}
\\
\subfigure[MC with SPA, $\lambda_0 = 2$.]{
{\includegraphics[width=0.32\textwidth,height=0.22\textwidth]{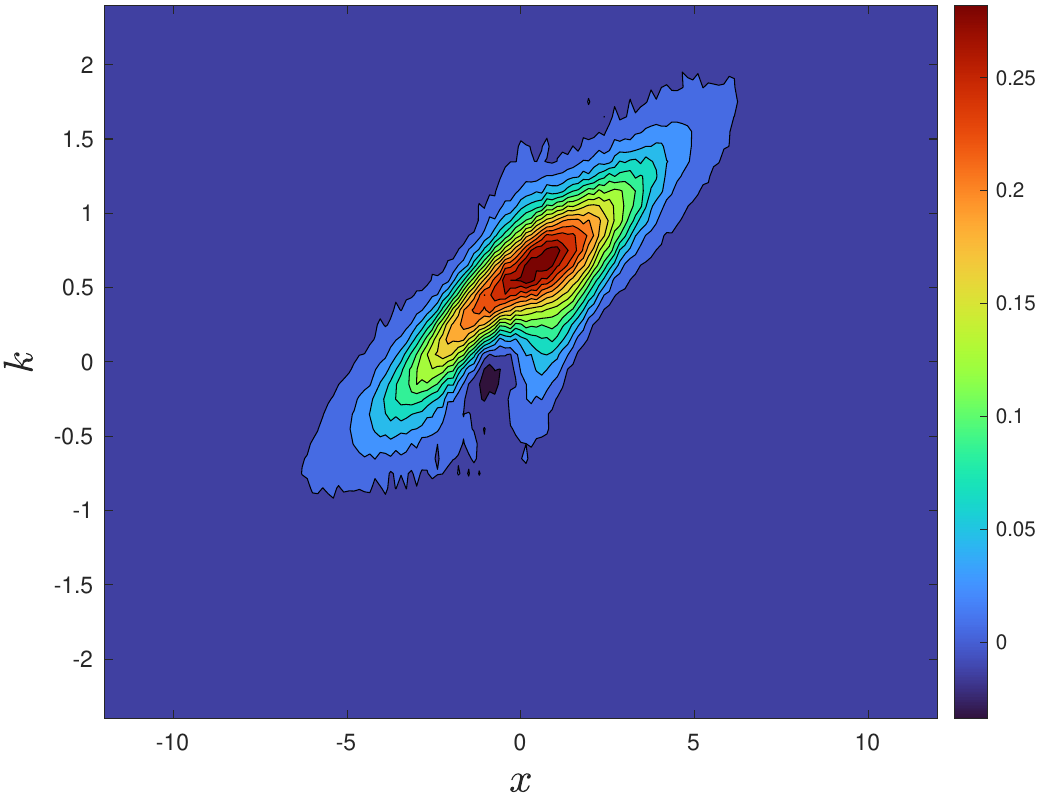}}}
\subfigure[MC with SPA, $\lambda_0 = 6$.]{
{\includegraphics[width=0.32\textwidth,height=0.22\textwidth]{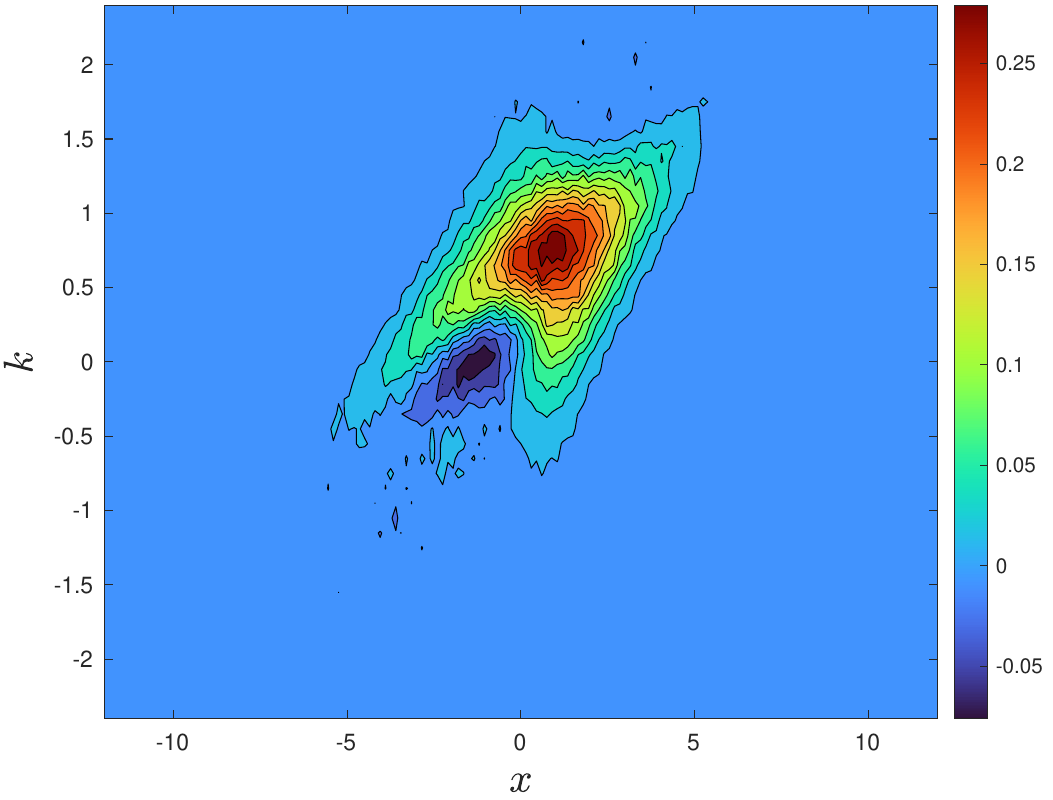}}}
\subfigure[MC with SPA, $\lambda_0 = 16$.]{
{\includegraphics[width=0.32\textwidth,height=0.22\textwidth]{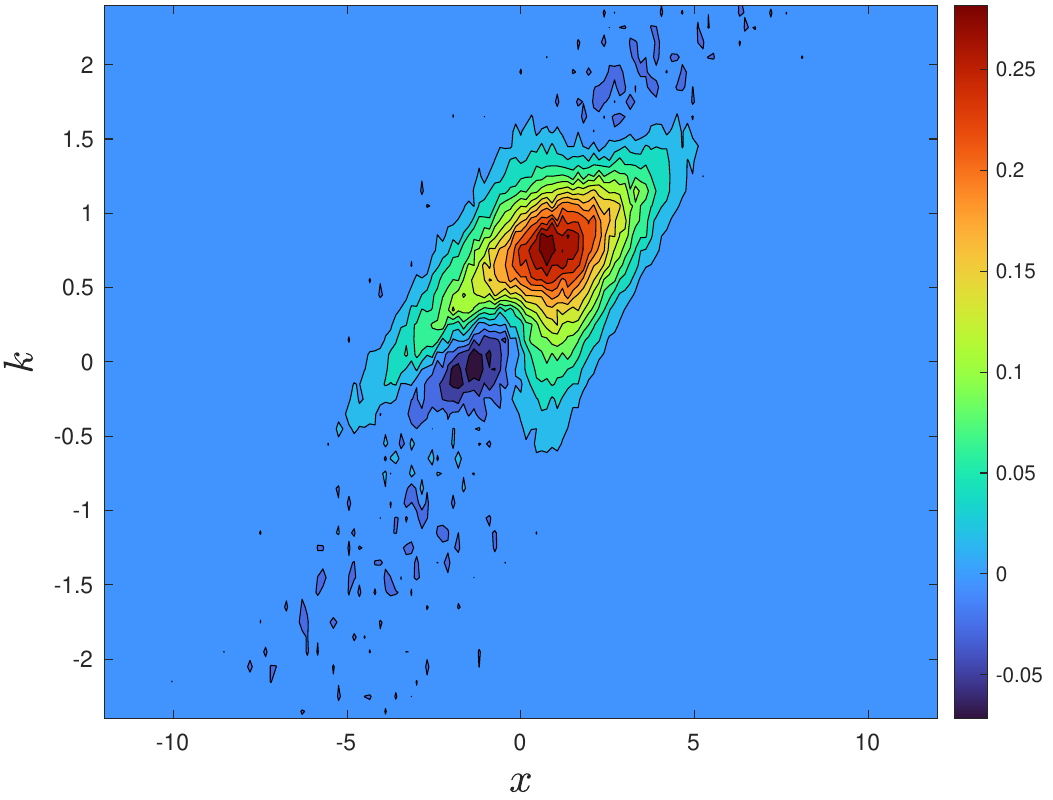}}}
\caption{\small  The 4-D Morse system: A visualization of the reduced Wigner function $W_1(x, k, t)$ produced by WBRW-SPA, under different filter $\lambda_0$.}
\label{supp_2dm_SPA}
\end{figure}

The remaining problem is how to choose the filter $\lambda_0$. From the theoretical results, $\lambda_0$ must not be too small, otherwise the asymptotic errors will dominate. A visualization of the reduced Wigner function at $t=4$a.u., produced by WBRW-SPA under the initial effective sample size $1 \times 10^7$, is presented in Figure \ref{supp_2dm_SPA}. It seems that $\lambda_0 =6$ achieves the best performance in controlling the random noises. Apparently, SPA under $\lambda_0 = 1$ or $2$ fails to produce correct results, and the noises seems to be amplified when $\lambda_0$ is too large ($\lambda_0 = 16$).

\begin{figure}[!h]
\subfigure[Stochastic variances in $W_1(x, k)$.  \label{supp_SPA_redist}]{
{\includegraphics[width=0.48\textwidth,height=0.27\textwidth]{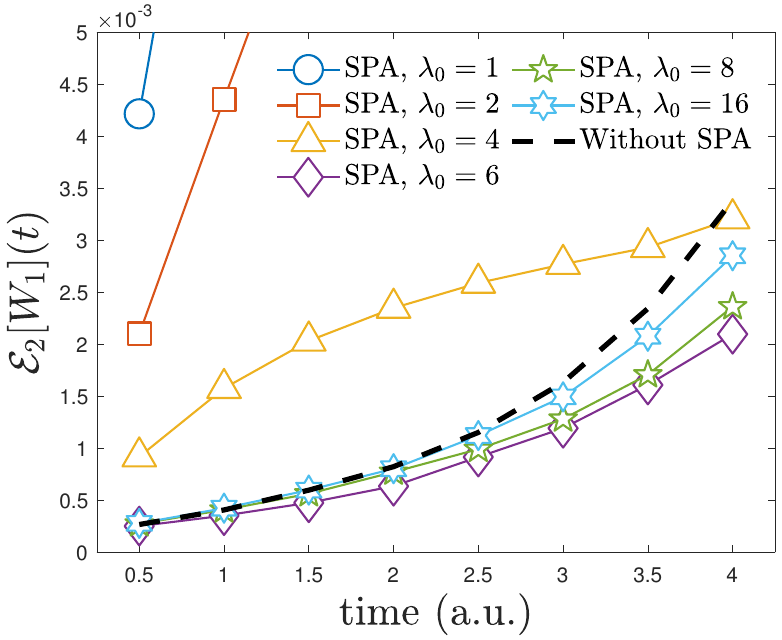}}}
\subfigure[Stochastic variances in $P(x_1, x_2)$.  \label{supp_SPA_xdist}]{
{\includegraphics[width=0.48\textwidth,height=0.27\textwidth]{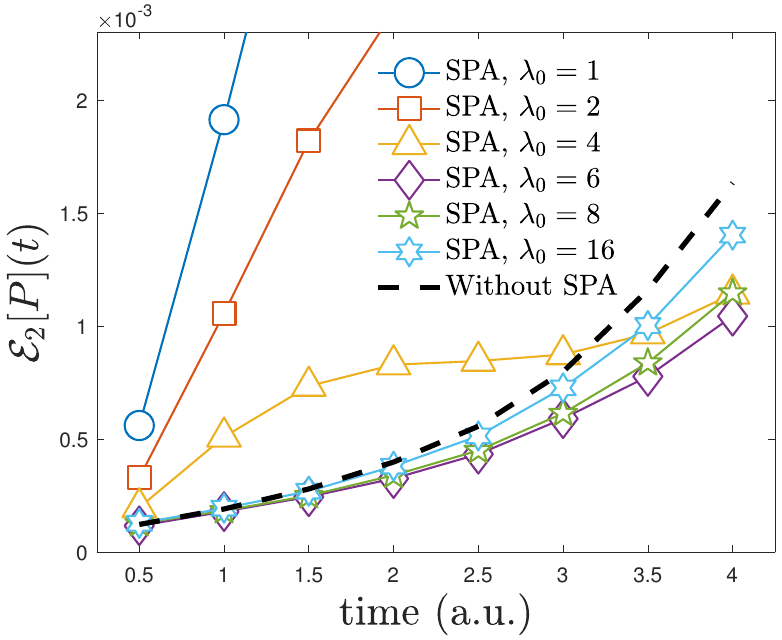}}}
\\
\subfigure[Deviation in Hamiltonian. \label{supp_SPA_energy}]{
{\includegraphics[width=0.48\textwidth,height=0.27\textwidth]{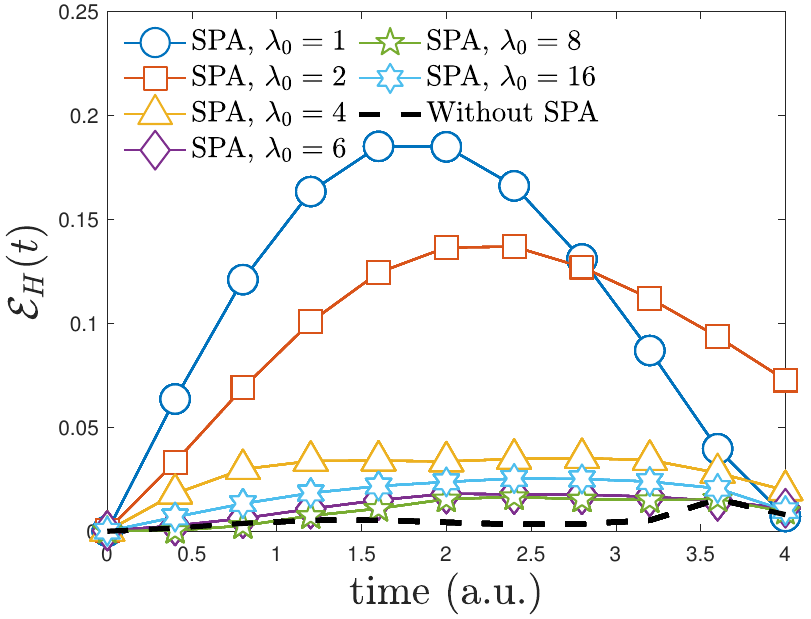}}}
\subfigure[Growth of particle number. \label{supp_SPA_particle}]{
{\includegraphics[width=0.48\textwidth,height=0.27\textwidth]{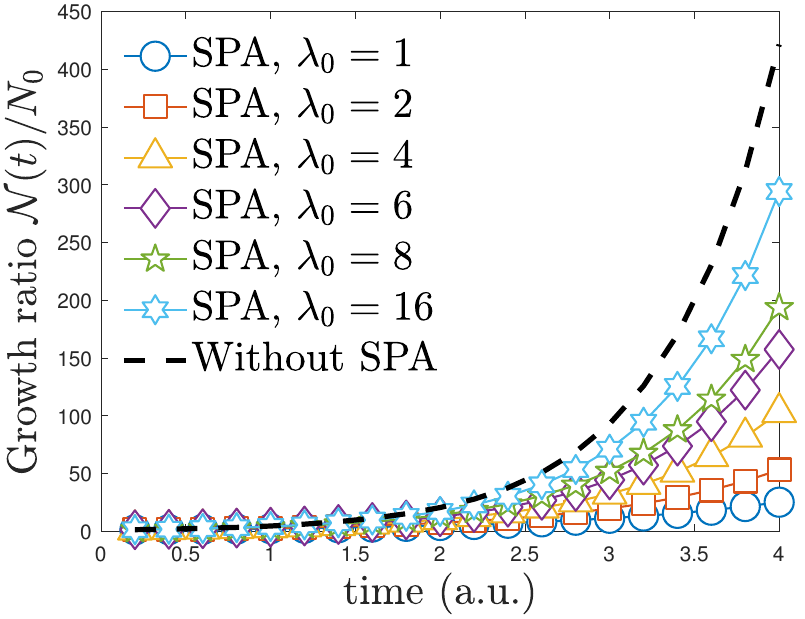}}}
\caption{\small  The 4-D Morse system: The growth of stochastic variances, the deviations of total energy and the particle number in stochastic Wigner simulations. Under appropriate $\lambda_0$, SPA can alleviate the exponential growth of particle number and variances simultaneously. }
\label{supp_2dm_spa_accuracy}
\end{figure}

To qualify the variances, we measure the $l^2$-error of $W_1(x, k, t)$ and $P(x_1, x_2, t)$, the deviation of total energy $\varepsilon_H(t)$ as defined in Eq.~\eqref{supp_def.Herr}. According to Figure \ref{supp_2dm_spa_accuracy}, SPA under $\lambda_0 =6$ indeeds alleviates the exponential growth of particle number and variances simultaneously. The growth ratio of total particle with SPA is $157$ at $t = 4$a.u., compared to 193 without SPA. Too small $\lambda_0$ kills the accuracy due to the large asymptotic errors, while too large  $\lambda_0$ may fail to kill redundant particles. 

In practice, the choice of $\lambda_0$ can be determined by monitoring the deviation in Hamiltonian. In Figure \ref{supp_SPA_energy}, a large fluctuation of the total Hamiltonian is observed under $\lambda_0 = 1, 2, 16$, while the deviation becomes very small under $\lambda_0 = 6$. This is consistent with the trends in Figures \ref{supp_SPA_redist} and \ref{supp_SPA_xdist}. An adaptive choice of optimal $\lambda_0$ is discussed in the main body of our paper.

\section{Performance evaluation of SPADE in 4-D phase space}

From this section, we are about to make a thorough benchmark on SPADE by simulating the 4-D Wigner equation under the Morse potential, the purpose of which is  is two-pronged. First, we make a thorough comparison between the particle annihilation via uniform mesh (PAUM) and SPADE. Second, we would like to investigate how the parameter  $\vartheta$ in SPADE and sample size $N_0$ influence the numerical accuracy, energy conservation, particle number and the partition level $K$. The latter is towards a comprehensive understanding of SPADE and a guiding principle for improving accuracy systematically, and is pivotal to rigorous numerical analysis.

To visualize the quantum dynamics in phase space, we adopt the reduced Wigner function 
\begin{equation}\label{supp_reduced_density}
W_1(x, k, t) =  \iint_{\mathbb{R}^2} f(x, x_2, k, k_2, t)  \D x_2  \D k_2,
\end{equation}
and the spatial marginal distribution 
\begin{equation}\label{supp_spatial_density}
P(x_1, x_2, t) =   \iint_{\mathbb{R}^2} f(x_1, x_2, k_1, k_2, t) \D k_1 \D k_2.
\end{equation} 
The initial Wigner function is a Gaussian wavepacket
\begin{equation}\label{supp_def.Gaussian_wave_packet}
f(x_1, x_2, k_1, k_2, 0) = \frac{1}{\pi^2}\me^{-0.5(x_1-8)^2 - 0.5(x_2-12)^2 - 2(k_1 - 0.5)^2 - 2(k_2 + 0.5)^2}.
\end{equation}

The model parameters are: $\bx_A = (0, 0), r_0 = 0.5, \kappa = 0.5$, $\hbar = m = 1$.  The reference solutions are produced by a highly accurate deterministic advective-spectral-mixed scheme \cite{XiongChenShao2016}, where the Wigner function defined in a 4-D computational domain $[-12, 12]^2 \times \left[-\frac{5\pi}{3}, \frac{5\pi}{3}\right]^2 $ is expanded as the tensor product of $163^2$ cubic spline basis and $128^2$ Chebyshev spectral basis (with $8\times8$ cells), with $\Delta y_{\nu} = \Delta y_{\mu} = 0.3$ and $y_{\nu}$, $y_{\mu}$ truncated at $\mathcal{Y} = \left[-15, 15\right]^2$. The three-step Lawson scheme is used for temporal integration, with time step $\Delta t = 0.02$a.u.

\begin{figure}[!h]
    \subfigure[$t = 2.5$a.u.]{
    {\includegraphics[width=0.49\textwidth,height=0.23\textwidth]{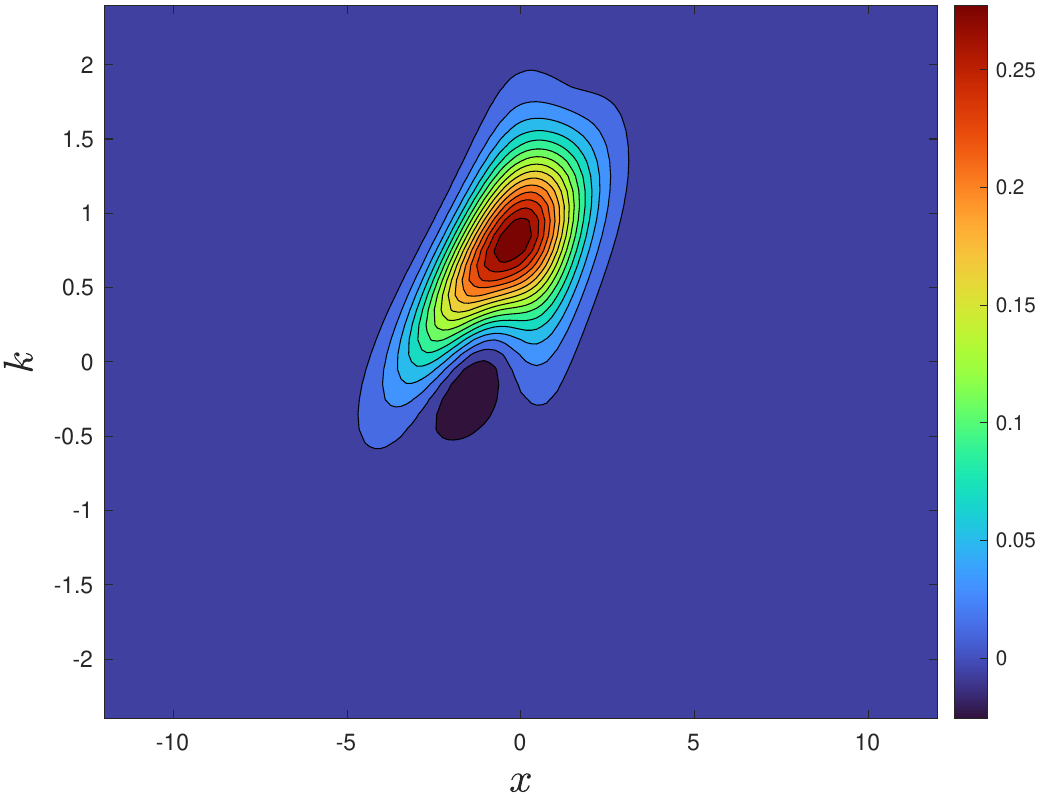}}
    {\includegraphics[width=0.49\textwidth,height=0.23\textwidth]{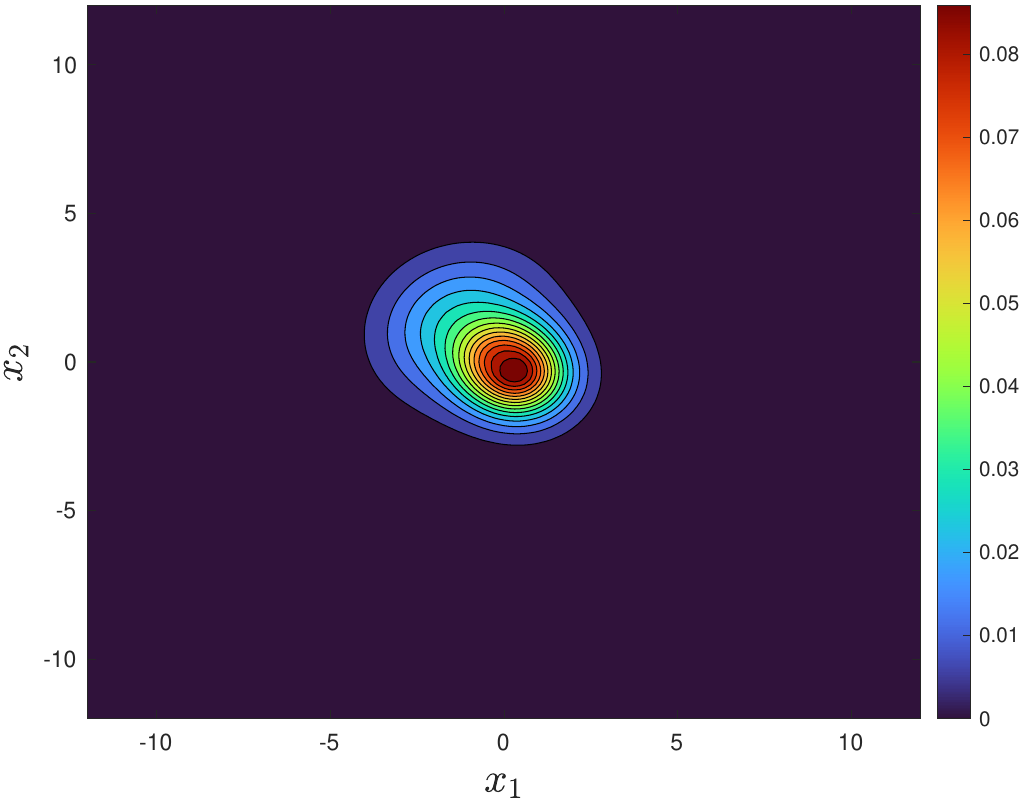}}}
     \\
    \subfigure[$t = 5$a.u.]{
    {\includegraphics[width=0.49\textwidth,height=0.23\textwidth]{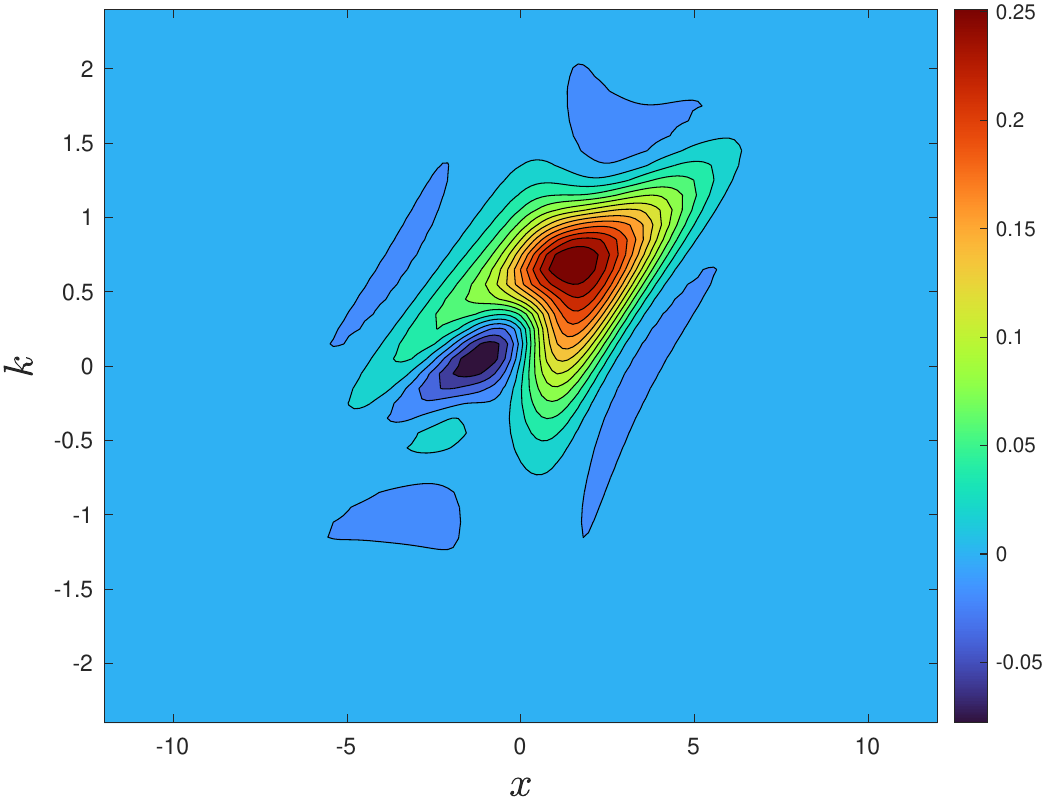}}
    {\includegraphics[width=0.49\textwidth,height=0.23\textwidth]{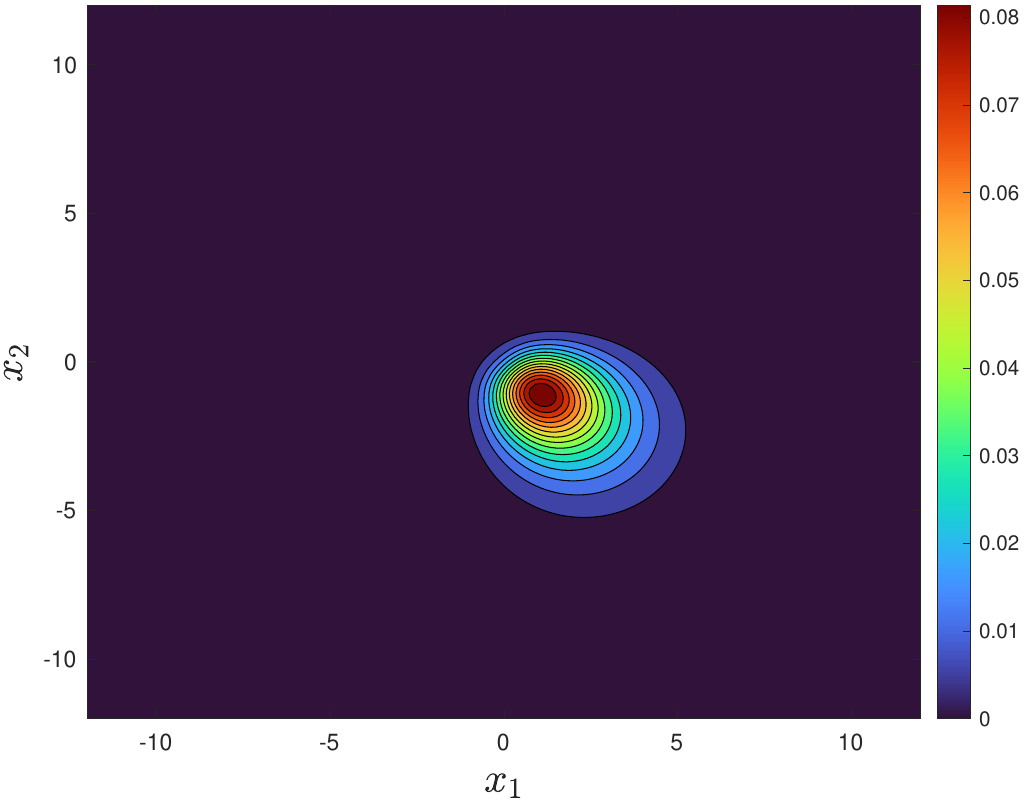}}}
     \\
    \subfigure[$t = 7.5$a.u.]{
    {\includegraphics[width=0.49\textwidth,height=0.23\textwidth]{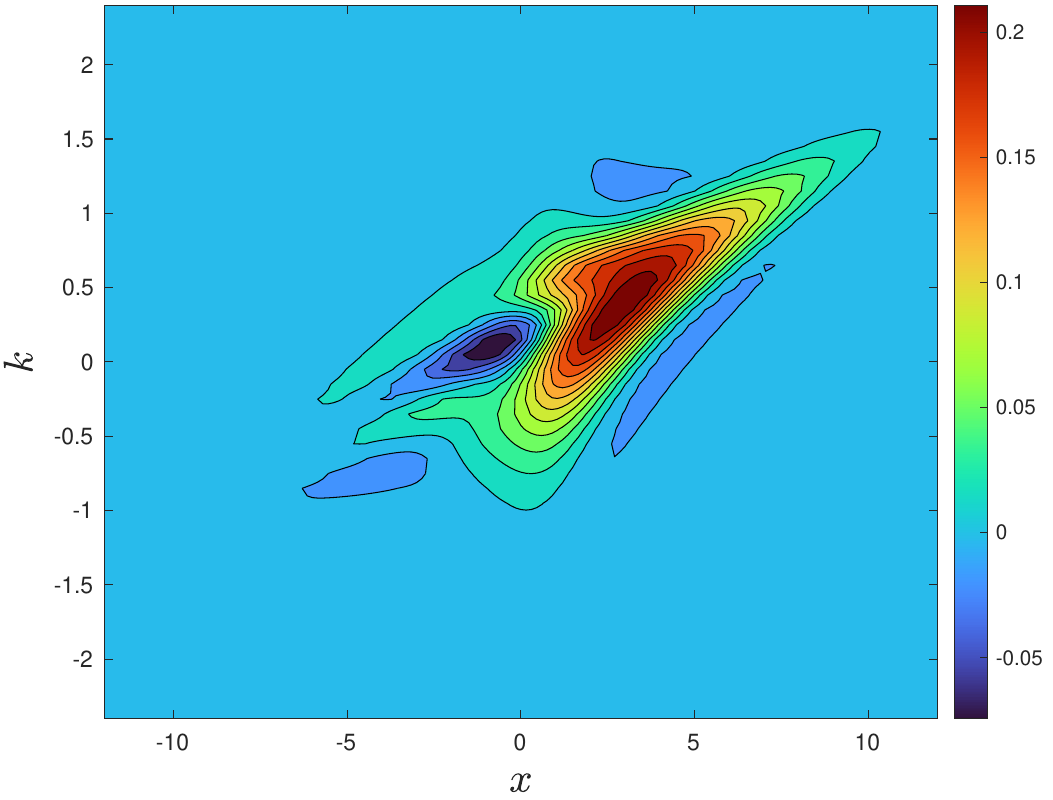}}
    {\includegraphics[width=0.49\textwidth,height=0.23\textwidth]{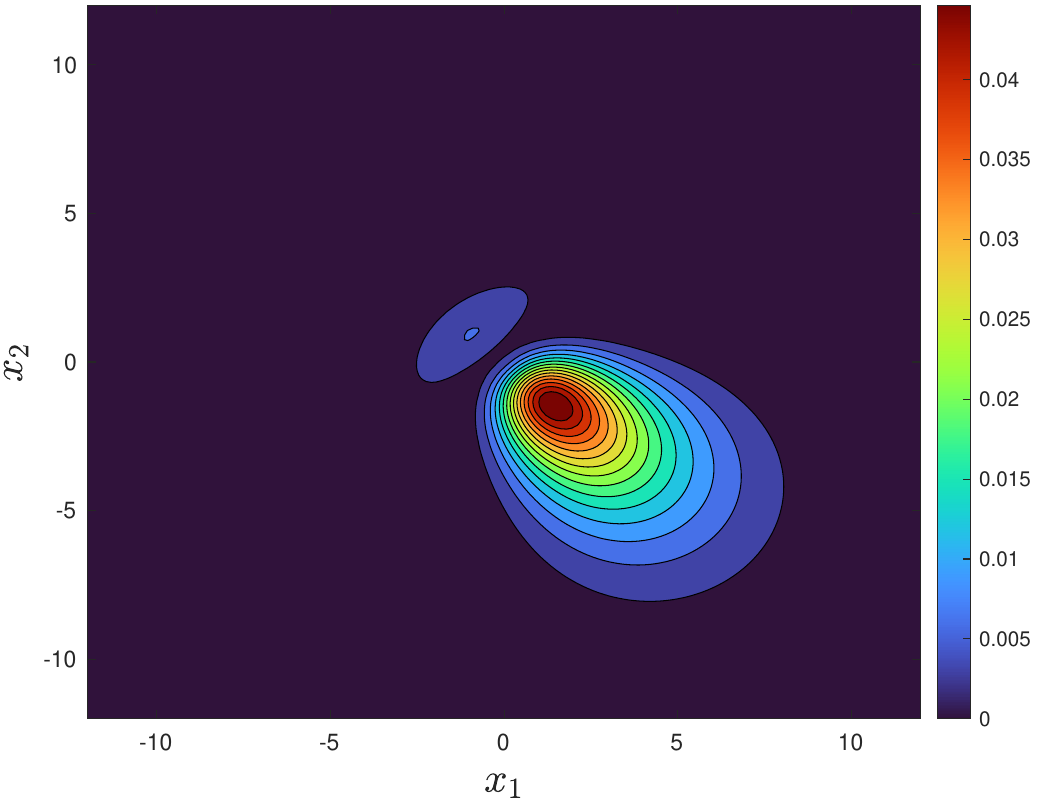}}}
    \\
     \subfigure[$t = 10$a.u.]{
     {\includegraphics[width=0.49\textwidth,height=0.23\textwidth]{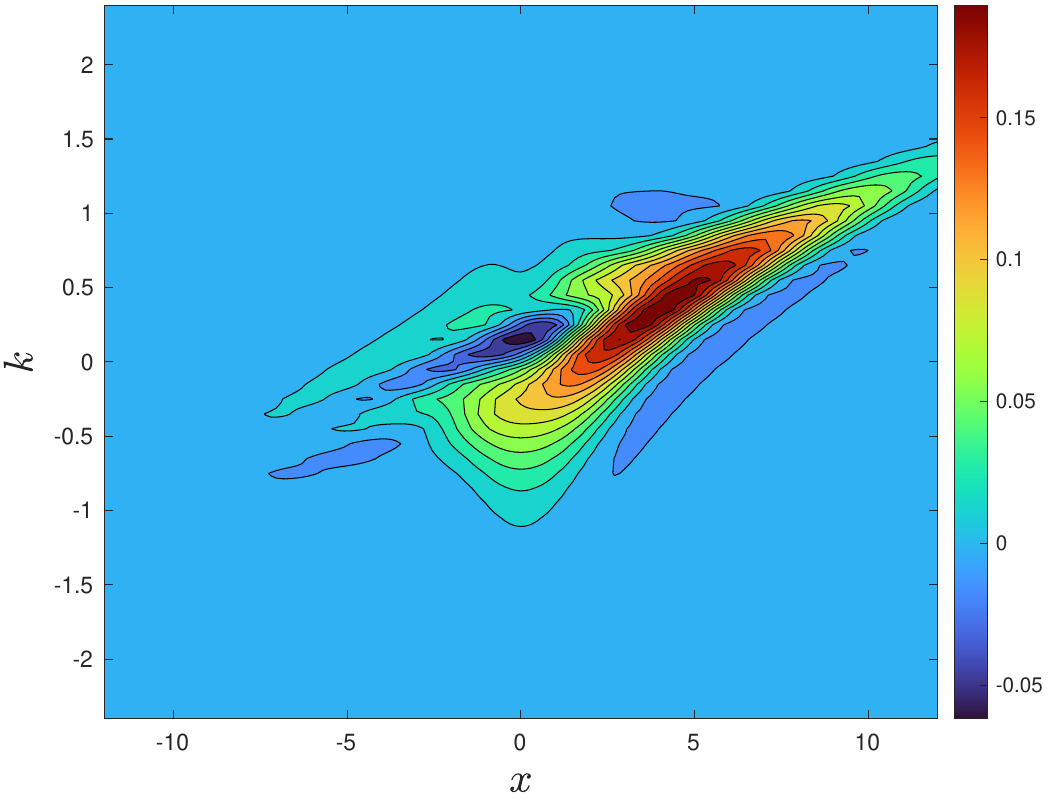}}
     {\includegraphics[width=0.49\textwidth,height=0.23\textwidth]{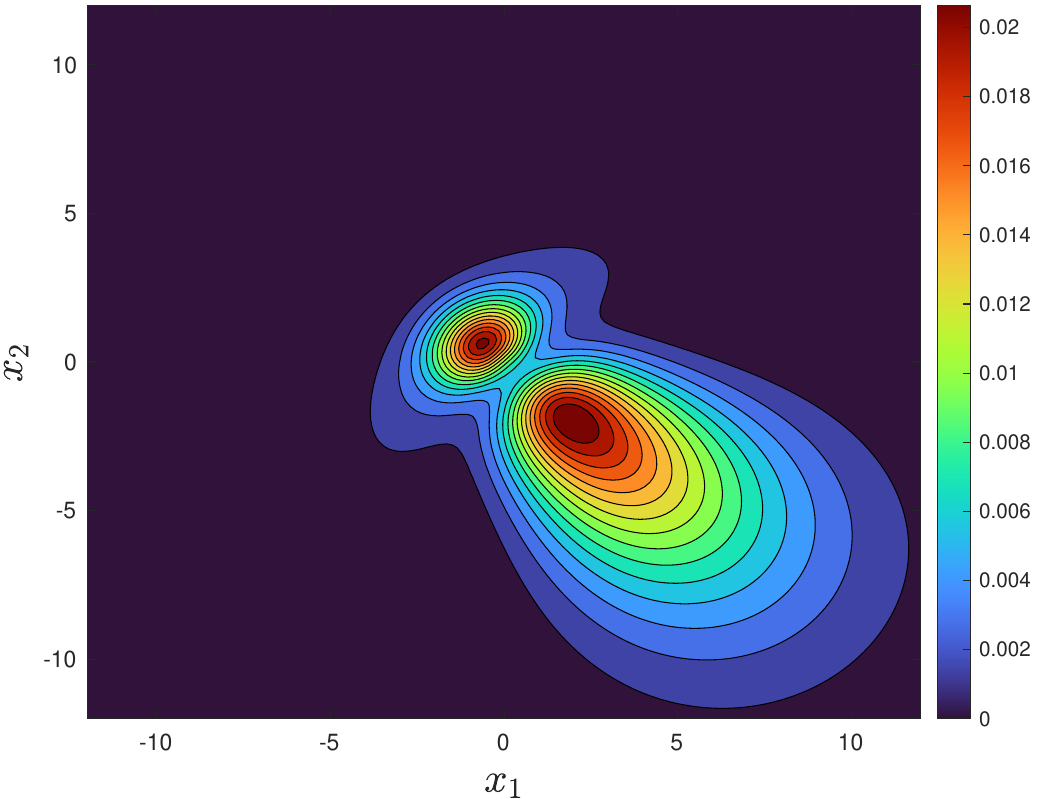}}}    
    \caption{\small The 4-D Morse system: Snapshots of the reduced Wigner function $W_1(x, k, t)$ (left) and the spatial marginal distribution $P(x_1, x_2, t)$ (right) produced  by the highly accurate deterministic Wigner solver. \label{supp_snap_asm}}
\end{figure}

The snapshots of $W_1(x, k, t)$ and $P(x_1, x_2, t)$ up to $t = 10$a.u.  are plotted in Figure \ref{supp_snap_asm}.  The negative components and oscillatory structure of the Wigner function are clearly seen in phase space. In the spatial direction, the Gaussian wavepacket is first attracted by the interacting body at the origin, and then oscillates near the origin $\bx_A$.

For stochastic particle simulations, we adopt $\gamma_0 = 2.59$, $\lambda_0 = 6$ and a finite $\bk$-domain $[-5, 5]^2$ in Algorithm \ref{supp_WBRW_SPA}, and annihilate particles every $1$a.u. The reduced Wigner function $W_1(x, k, t)$ and spatial marginal distribution $P(x_1, x_2, t)$ can be readily obtained by histogram reconstruction under a uniform grid mesh $[-12,12] \times [-5, 5]$ with $N_x = 161$, $N_k = 100$, $\Delta x = 0.15$, $\Delta k = 0.1$. For instance, for $\mathcal{X}_\mu = [-12+(\mu-1)\Delta x, -12+\mu\Delta x]$, $\mathcal{K}_\nu = [-5+(\nu-1)\Delta k, -5+\nu\Delta k]$, the reduced Wigner function $W_1(x, k, t)$ can be reconstructed by
\begin{equation}\label{supp_histogram}
W_1(x, k, t) \approx \sum_{\mu=1}^{N_x} \sum_{\nu=1}^{N_k} (\sum_{i=1}^{P(t)}\mone_{\mathcal{X}_\mu \times \mathcal{K}_\nu}(\bx_i^+, \bk_i^+) - {\sum_{i=1}^{M(t)}}\mone_{\mathcal{X}_\mu \times \mathcal{K}_\nu}(\bx_i^-, \bk_i^-)) \frac{ \mone_{\mathcal{X}_\mu \times \mathcal{K}_\nu}(x, k)}{N_0 |\mathcal{X}_\mu | |\mathcal{K}_\nu|}. 
\end{equation}
As a comparison, we also perform the stochastic Wigner simulations and  annihilate particles via PAUM with a $161^3 \times 100^3$ uniform grid mesh. The partition level is $K = 2.592\times 10^8$, which is even larger than the sample size.

In order to measure the stochastic variances, we use the $l^2$-errors $\mathcal{E}_{2}[W_1](t)$ and $\mathcal{E}_{2}[P_1](t)$
\begin{align}
&\mathcal{E}_{2}[W_1](t) = \{\frac{1}{N_x N_k}\sum_{i=1}^{N_x} \sum_{j=1}^{N_k} (W_1^{\textup{ref}}(x_i, k_j,t)- W_1^{\textup{num}}(x_i, k_j,t))^2\}^{1/2},\label{supp_def.L2error}\\
&\mathcal{E}_{2}[P](t) = \{\frac{1}{N_x^2}\sum_{i=1}^{N_x} \sum_{j=1}^{N_x} (P^{\textup{ref}}(x_i, x_j,t)- P^{\textup{num}}(x_i, x_j,t))^2\}^{1/2}. \label{supp_def.L2error_xdist}
\end{align}
with $W_1^{\text{ref}}(x, k,t)$ and $W_1^{\text{num}}(x, k,t)$ reference and numerical solutions, respectively (similar for $P$). 
In addition, since the system is mass-conservative and energy-conservative, the particle method should keep the effective sample size $P(t) - M(t)$ invariant in time, while the deviation of the total Hamiltonian can be used to measure the numerical accuracy,
\begin{equation}\label{supp_def.Herr}
\mathcal{E}_{\textup{H}}(t) = |H(t) - H(0)|, ~~ H(t) = \iint_{\mathbb{R}^{2} \times \mathbb{R}^{2}} \left(\frac{\hbar^2 |\bk|^2}{2\bm{m}} + V(\bx) \right) f(\bx, \bk, t)\D \bx \D \bk.
\end{equation}


\subsection{Comparison between PAUM and SPADE}

\begin{figure}[!h]
\centering
\subfigure[$t = 2.5$a.u.]{{\includegraphics[width=0.32\textwidth,height=0.22\textwidth]{./redist_asm_0025.pdf}}
{\includegraphics[width=0.32\textwidth,height=0.22\textwidth]{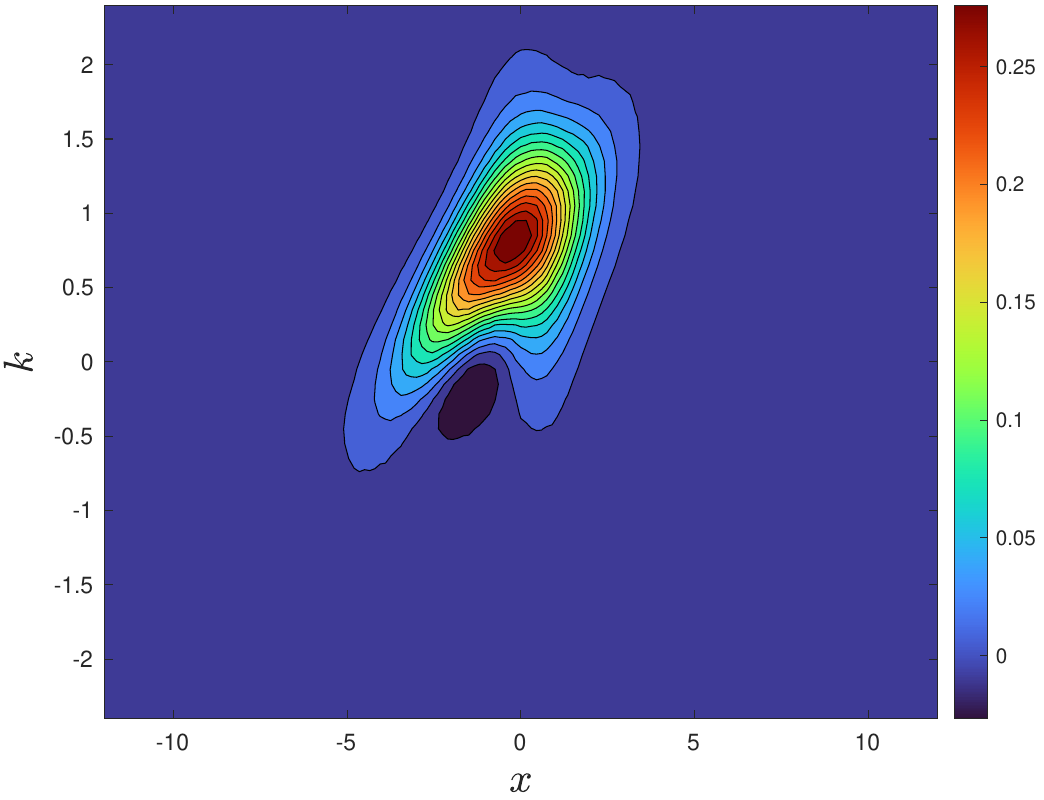}}
{\includegraphics[width=0.32\textwidth,height=0.22\textwidth]{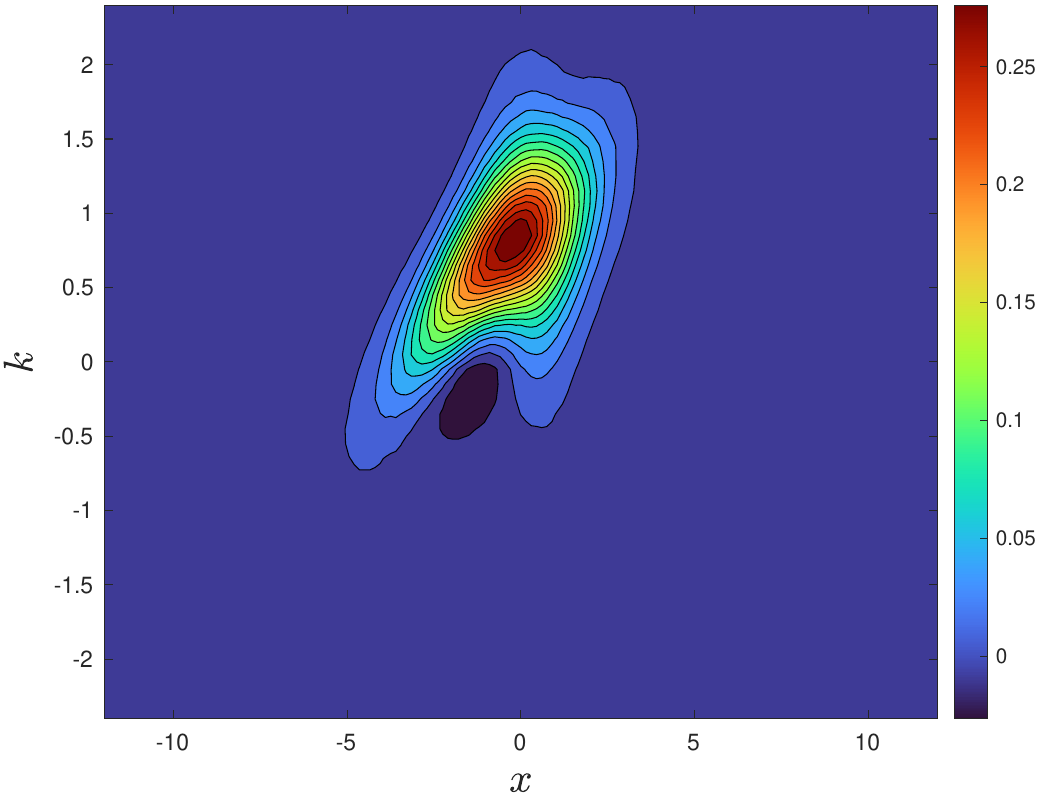}}}
\\
\centering
\subfigure[$t = 5$a.u.]{{\includegraphics[width=0.32\textwidth,height=0.22\textwidth]{./redist_asm_0050.pdf}}
{\includegraphics[width=0.32\textwidth,height=0.22\textwidth]{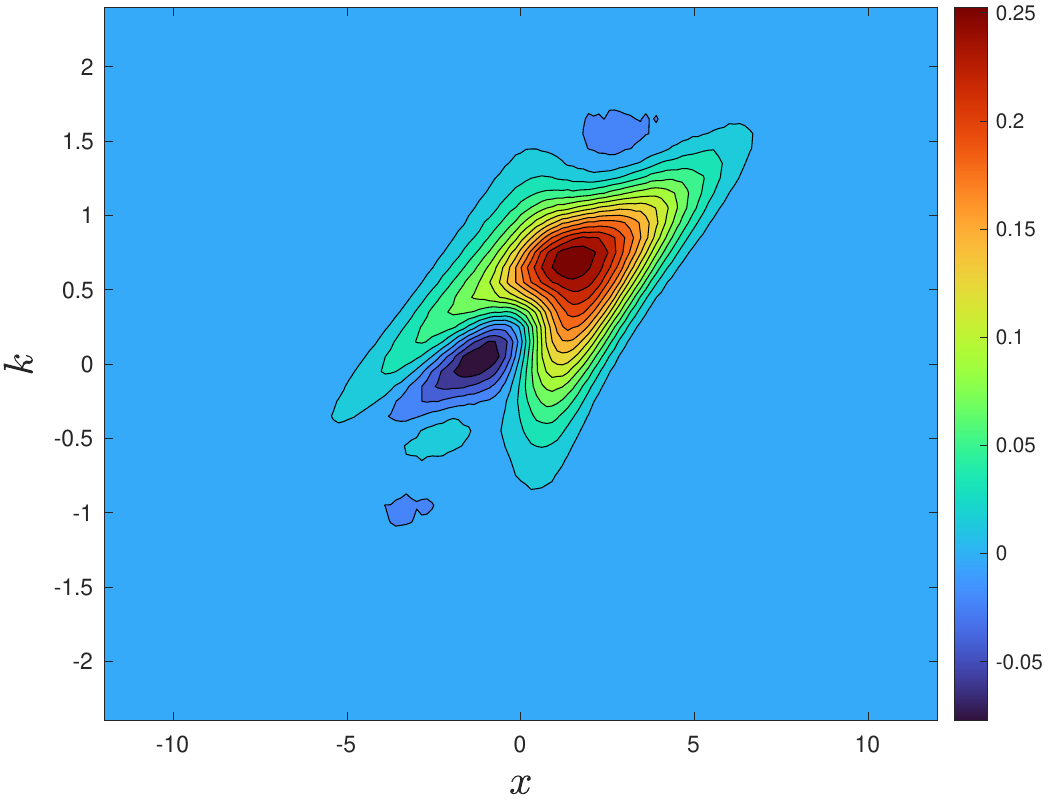}}
{\includegraphics[width=0.32\textwidth,height=0.22\textwidth]{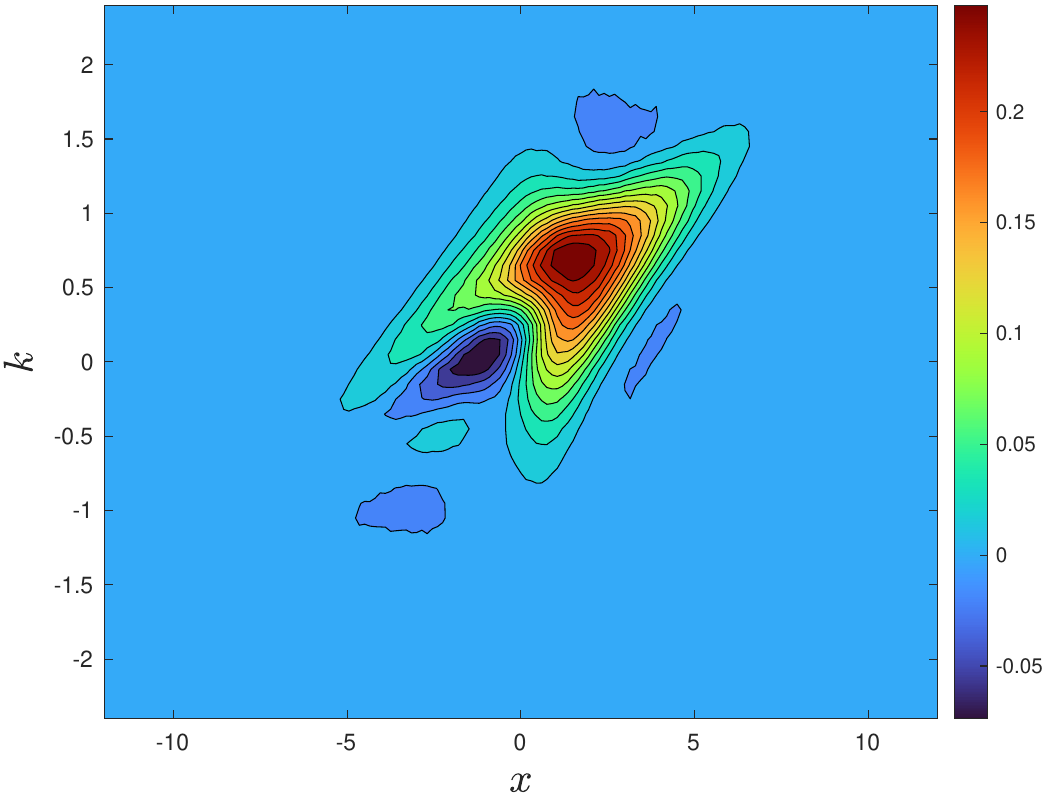}}}
\\
\centering
\subfigure[$t = 7.5$a.u.]{{\includegraphics[width=0.32\textwidth,height=0.22\textwidth]{./redist_asm_0075.pdf}}
{\includegraphics[width=0.32\textwidth,height=0.22\textwidth]{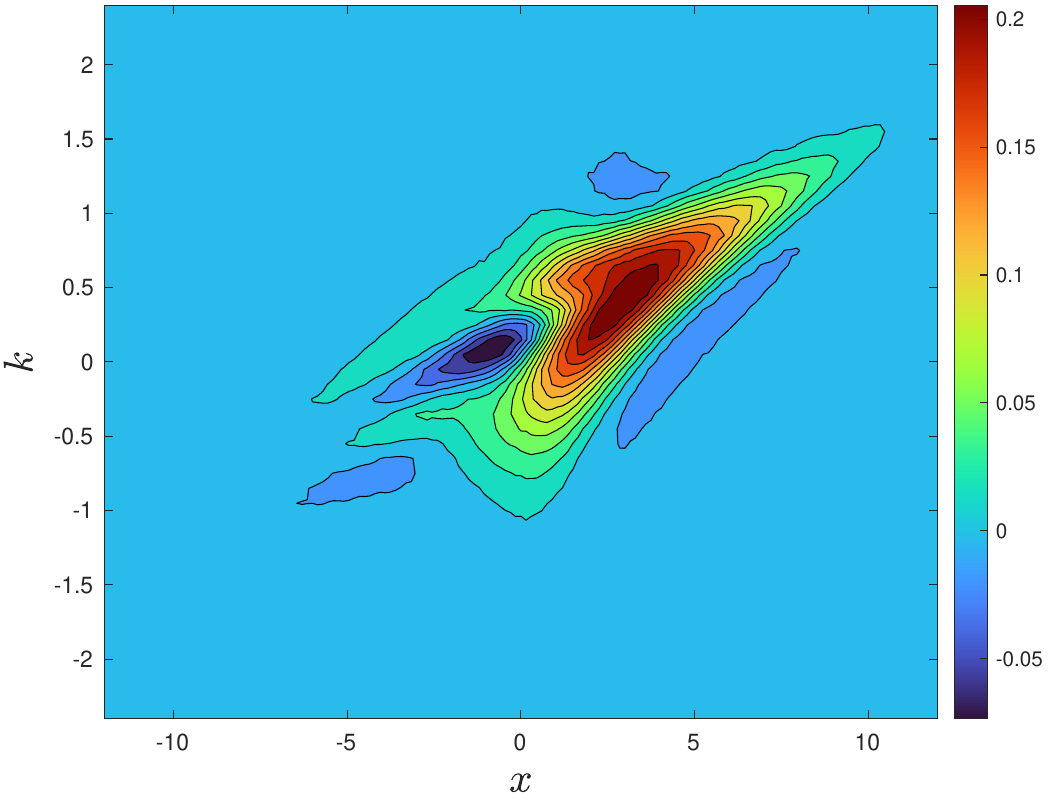}}
{\includegraphics[width=0.32\textwidth,height=0.22\textwidth]{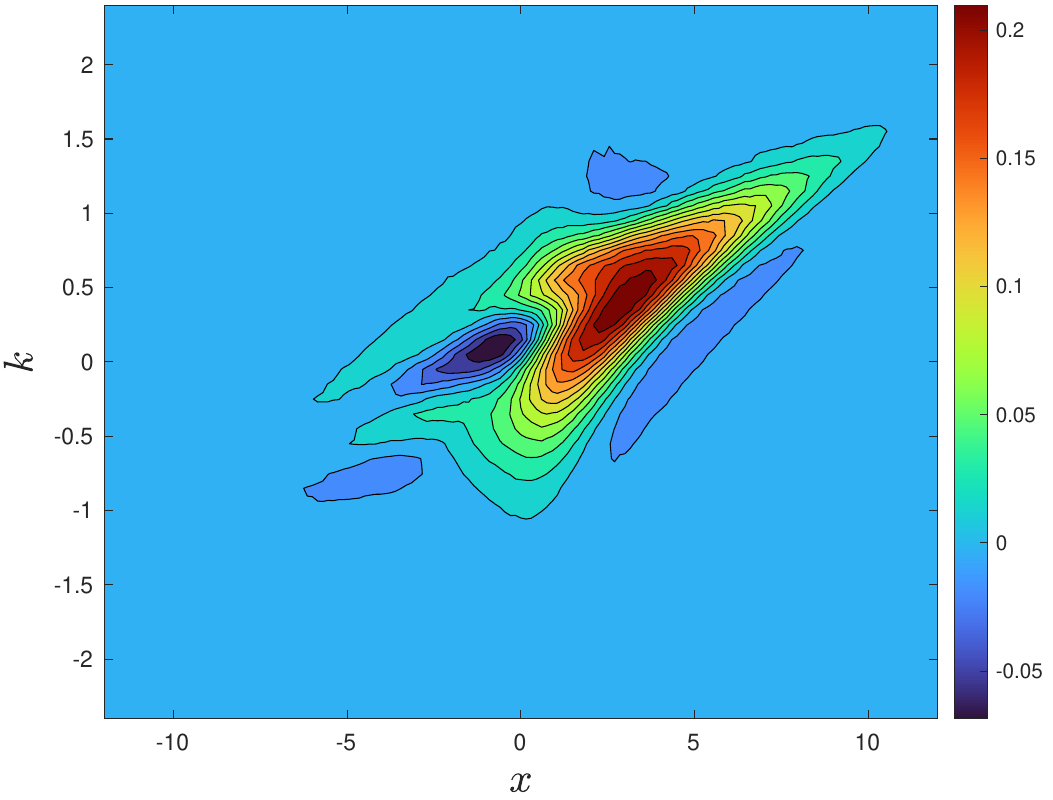}}}
\\
\centering
\subfigure[$t = 10$a.u.]{{\includegraphics[width=0.32\textwidth,height=0.22\textwidth]{./redist_asm_0100.pdf}}
{\includegraphics[width=0.32\textwidth,height=0.22\textwidth]{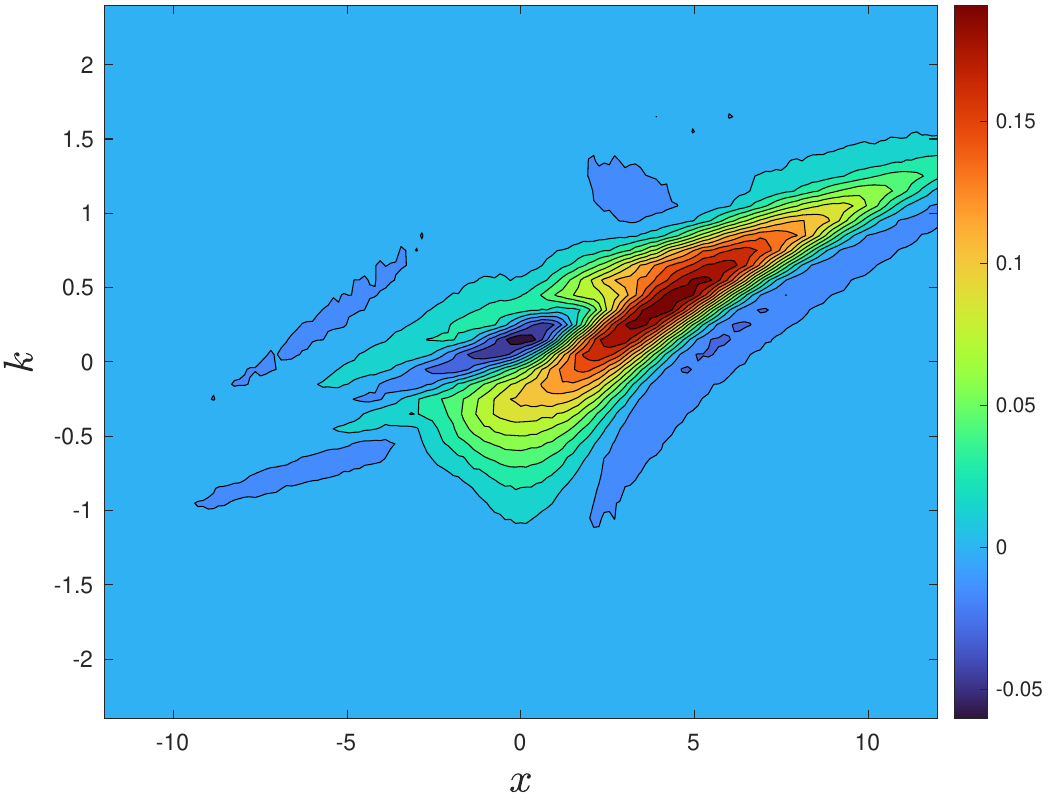}}
{\includegraphics[width=0.32\textwidth,height=0.22\textwidth]{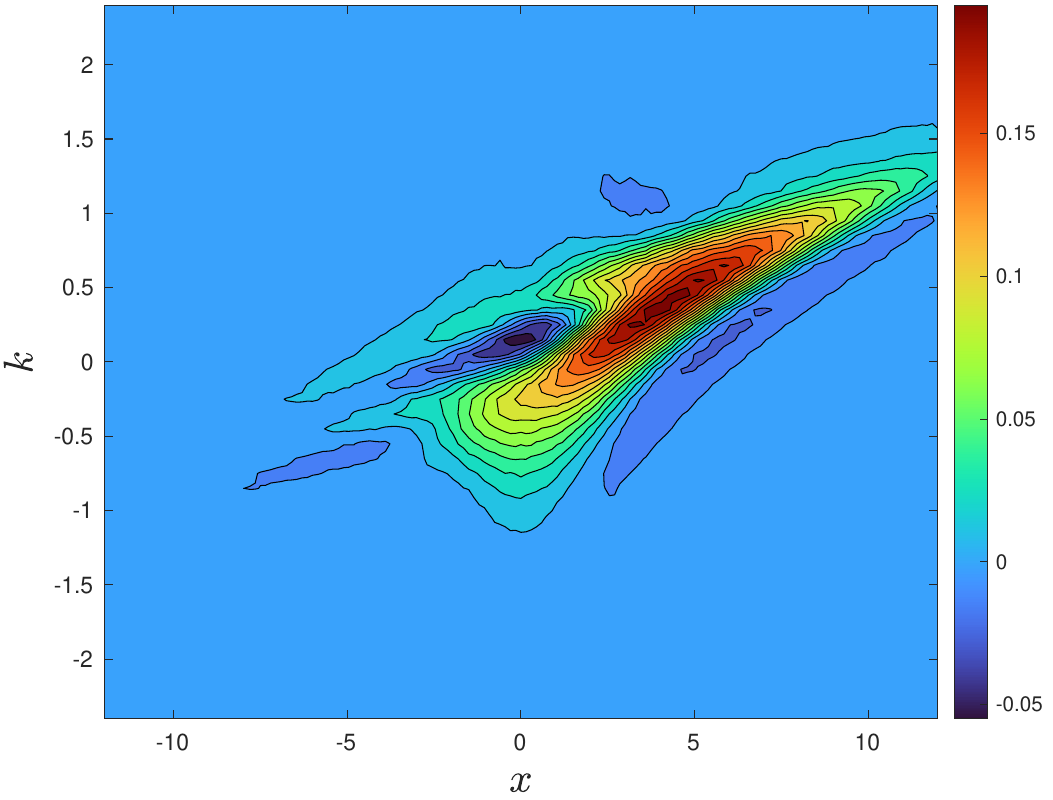}}}
\caption{The 4-D Morse system: Visualization of the reduced Wigner function $W_1(x, k, t)$ produced by the deterministic scheme (left), WBRW-SPA-PAUM under $N_0 = 4\times 10^7$ (middle) and WBRW-SPA-SPADE under $N_0 = 4\times 10^7$, $\vartheta= 0.003$ (right).}
\label{supp_fig_asm_redist}
\end{figure}

\begin{figure}[!h]
\centering
\subfigure[$t = 2.5$a.u.]{{\includegraphics[width=0.32\textwidth,height=0.22\textwidth]{./xdist_asm_0025.pdf}}
{\includegraphics[width=0.32\textwidth,height=0.22\textwidth]{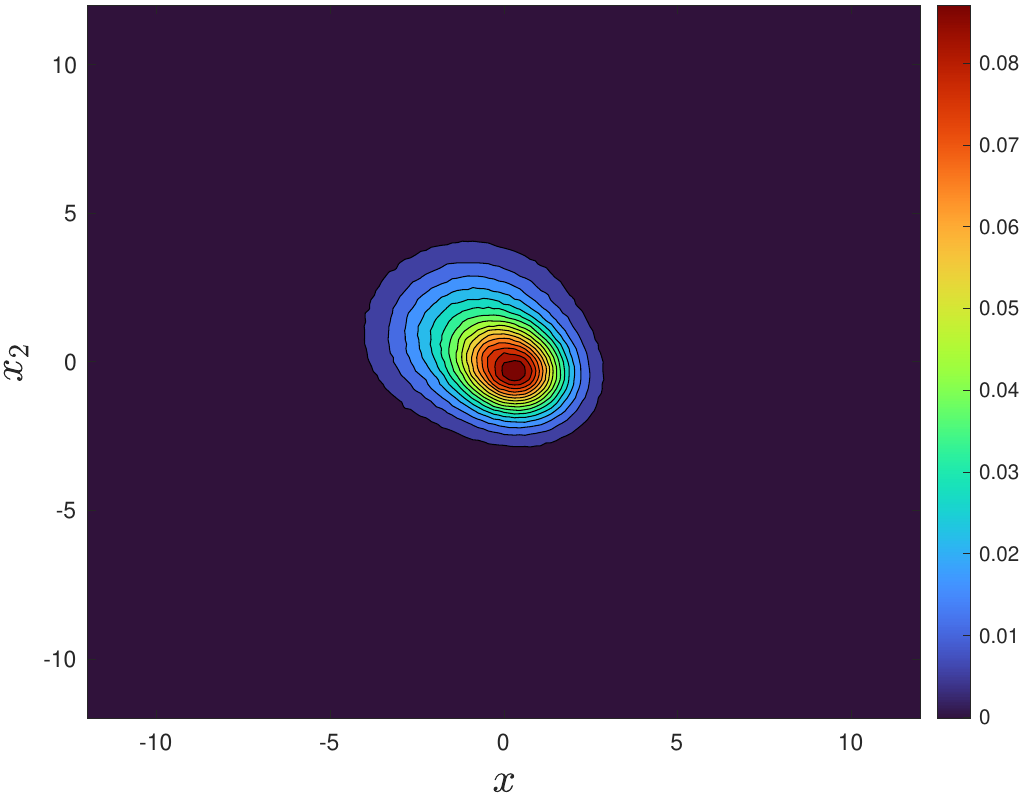}}
{\includegraphics[width=0.32\textwidth,height=0.22\textwidth]{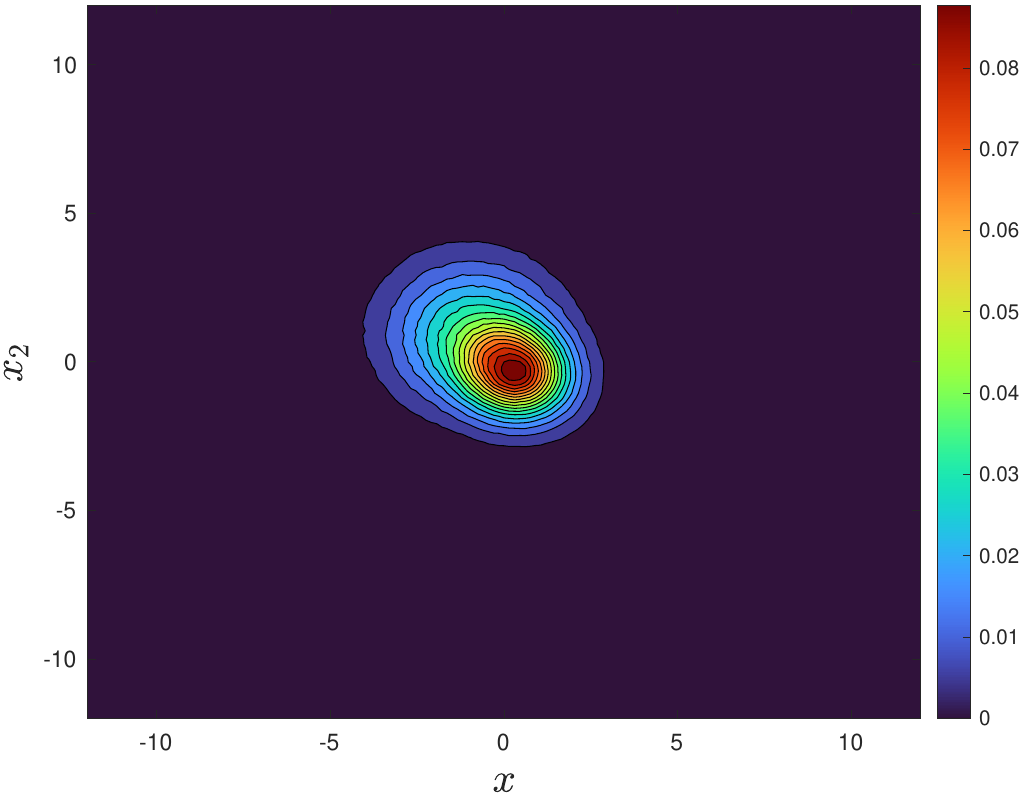}}}
\\
\centering
\subfigure[$t = 5$a.u.]{{\includegraphics[width=0.32\textwidth,height=0.22\textwidth]{./xdist_asm_0050.pdf}}
{\includegraphics[width=0.32\textwidth,height=0.22\textwidth]{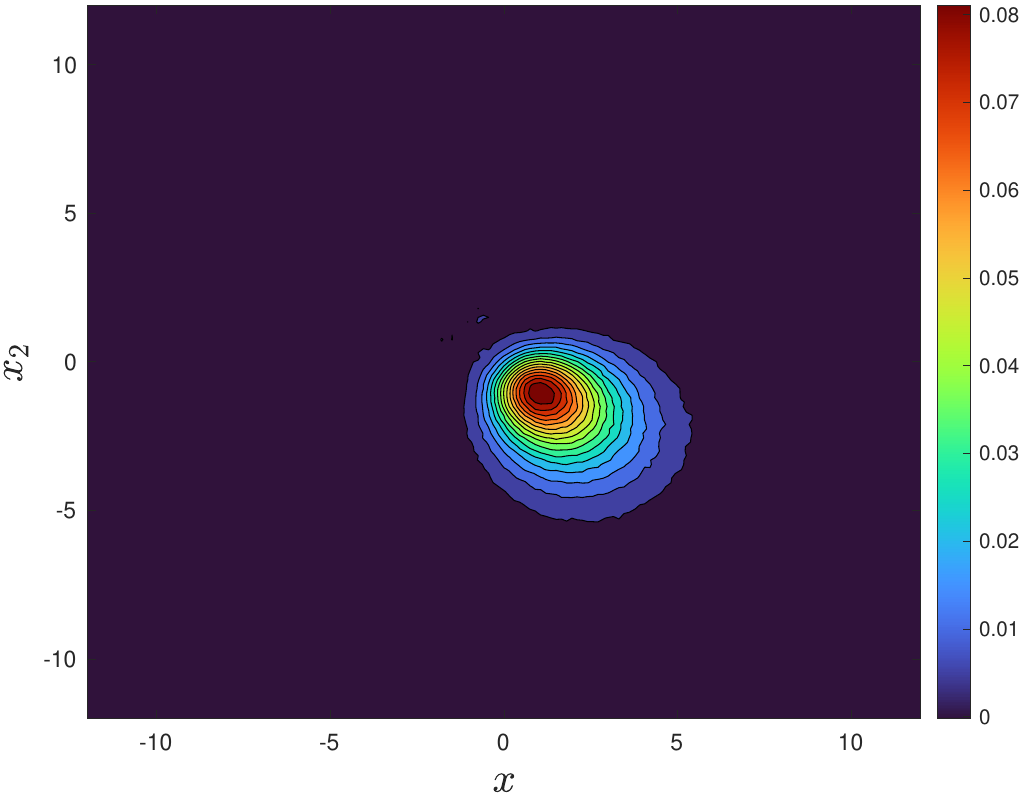}}
{\includegraphics[width=0.32\textwidth,height=0.22\textwidth]{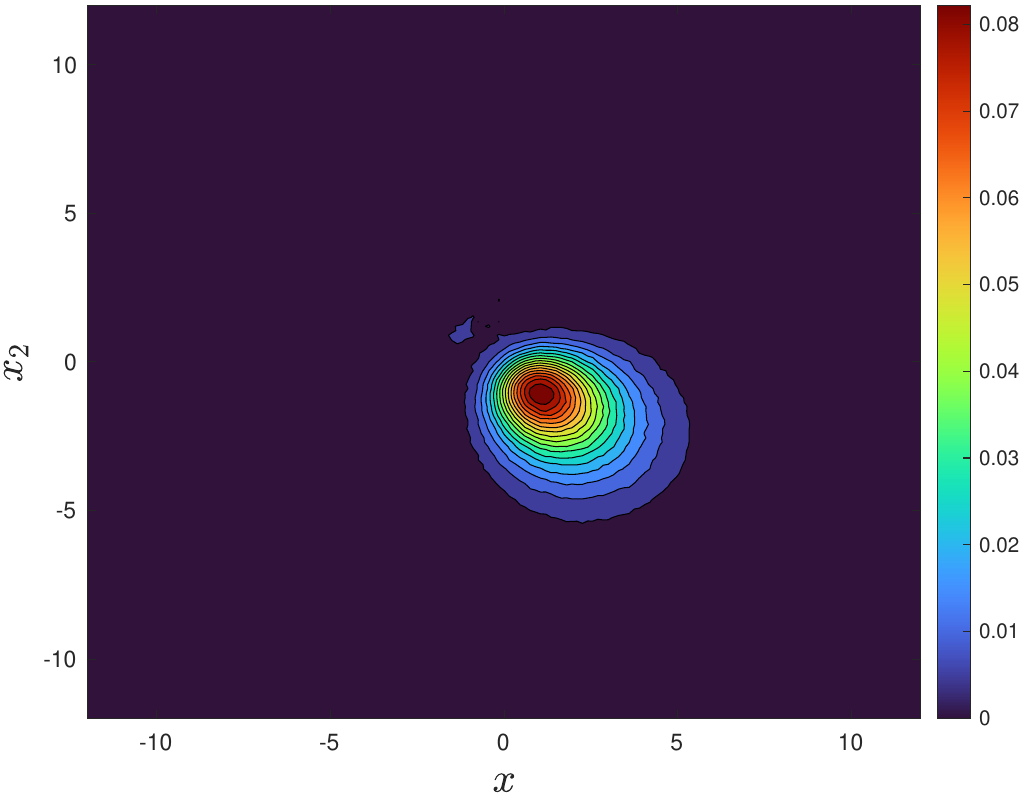}}}
\\
\centering
\subfigure[$t = 7.5$a.u.]{{\includegraphics[width=0.32\textwidth,height=0.22\textwidth]{./xdist_asm_0075.pdf}}
{\includegraphics[width=0.32\textwidth,height=0.22\textwidth]{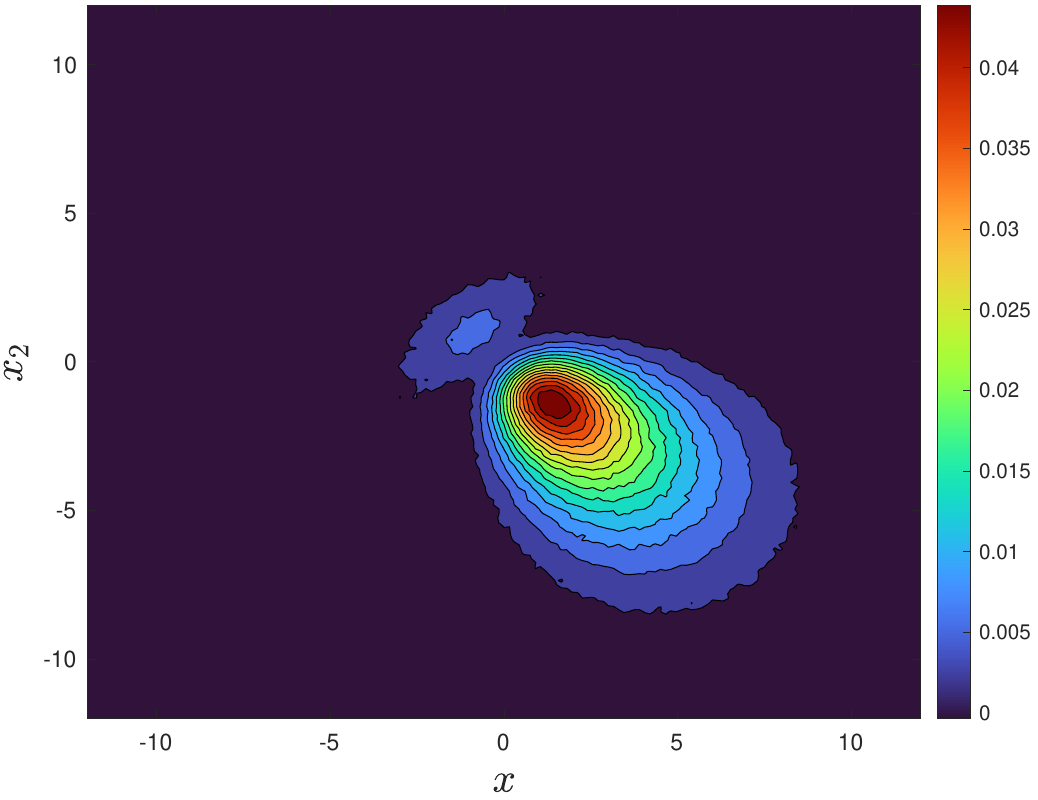}}
{\includegraphics[width=0.32\textwidth,height=0.22\textwidth]{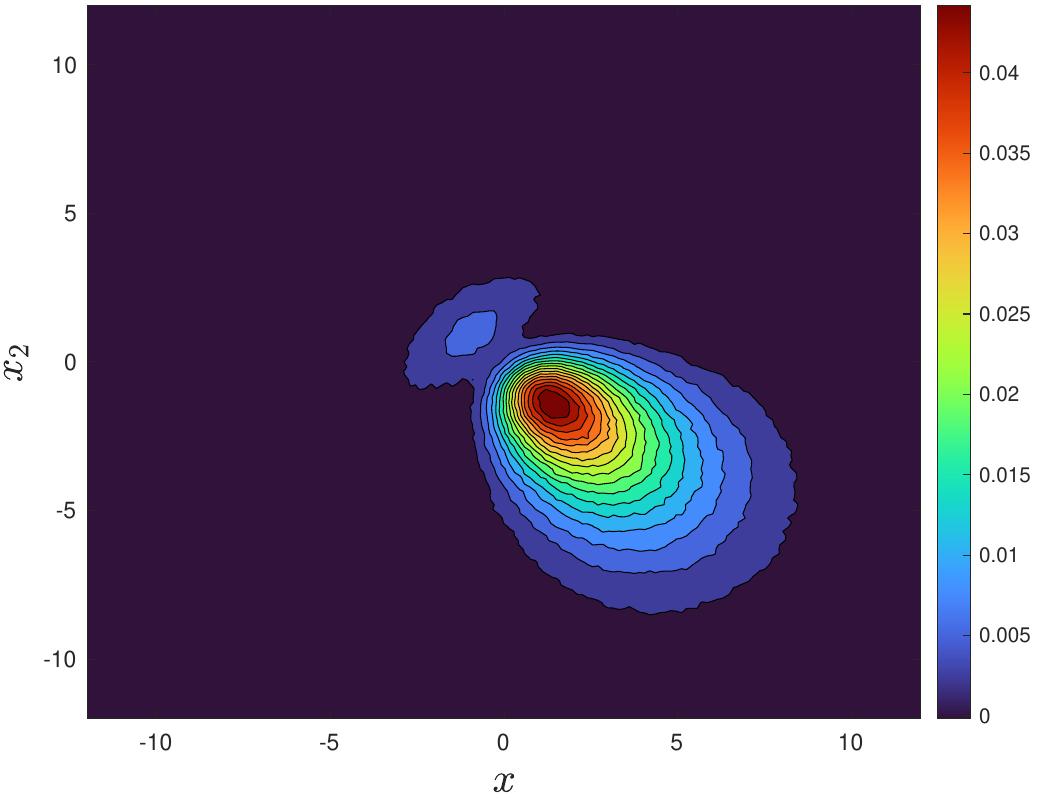}}}
\\
\centering
\subfigure[$t = 10$a.u.]{{\includegraphics[width=0.32\textwidth,height=0.22\textwidth]{./xdist_asm_0100.pdf}}
{\includegraphics[width=0.32\textwidth,height=0.22\textwidth]{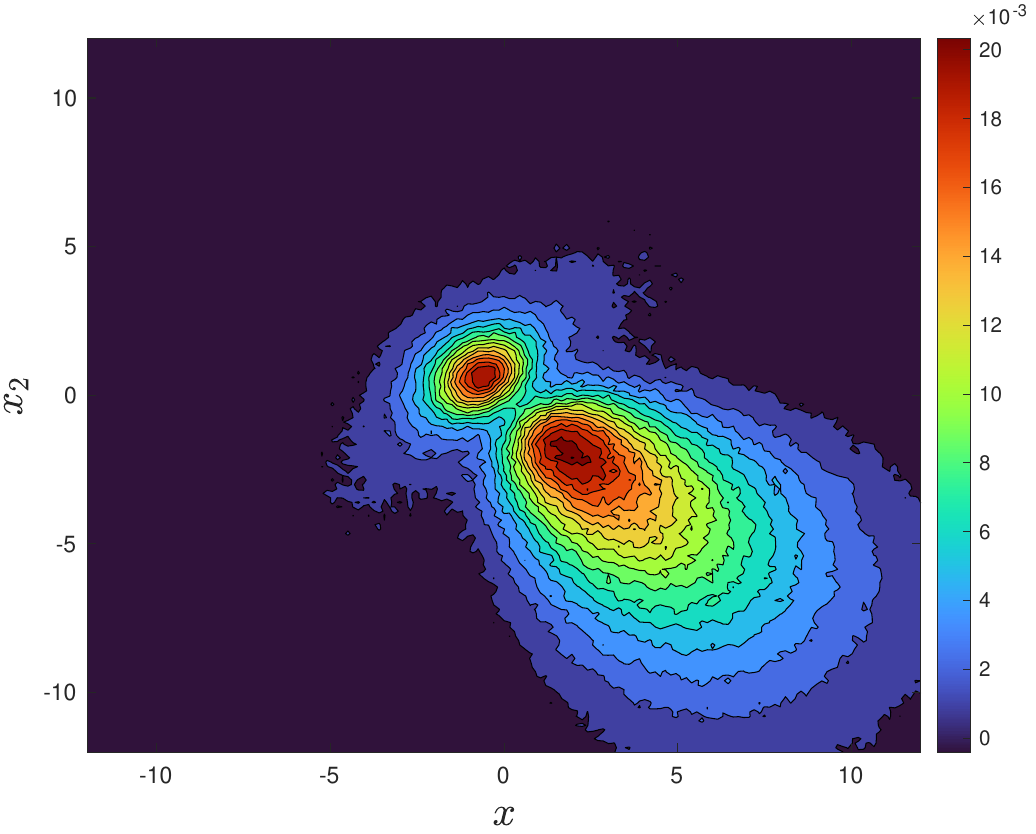}}
{\includegraphics[width=0.32\textwidth,height=0.22\textwidth]{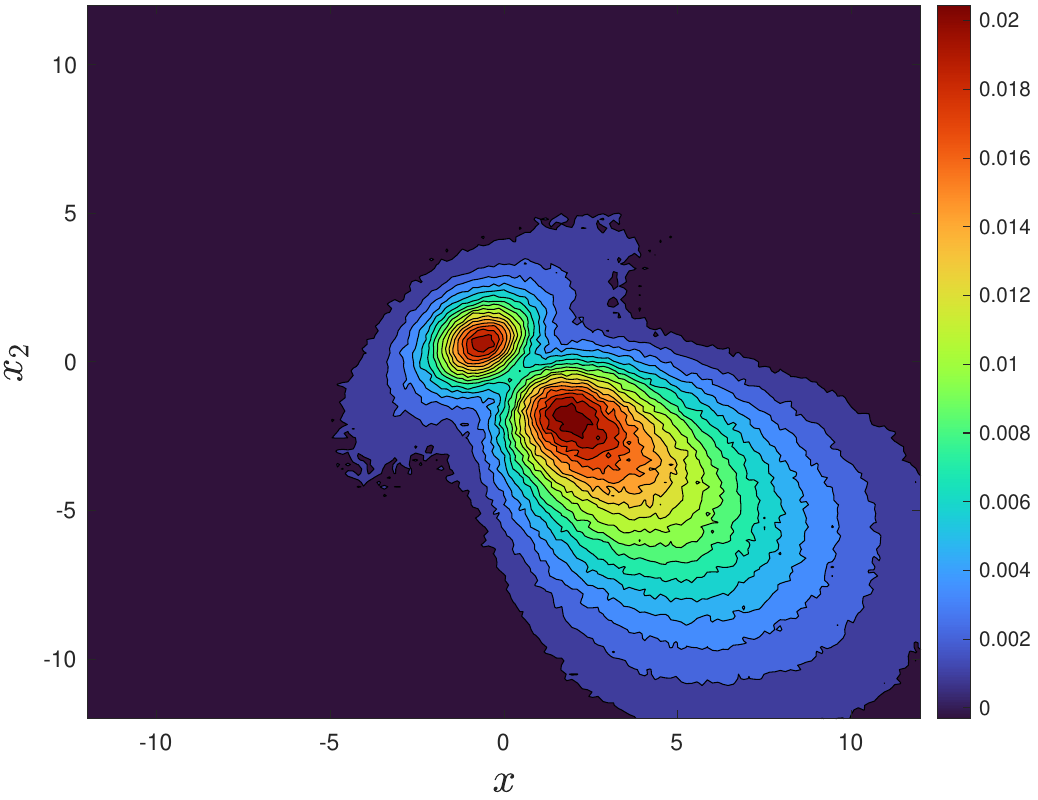}}}
\caption{The 4-D Morse system: Visualization of the spatial marginal distribution $P(x_1, x_2, t)$ produced by the deterministic scheme (left), WBRW-SPA-PAUM under $N_0 = 4\times 10^7$ (middle) and WBRW-SPA-SPADE under $N_0 = 4\times 10^7$, $\vartheta= 0.003$ (right).}
\label{supp_fig_asm_xdist}
\end{figure}

 \begin{figure}[!h]
 \subfigure[PAUM, $N_0 = 1\times10^6$.]{
{\includegraphics[width=0.48\textwidth,height=0.27\textwidth]{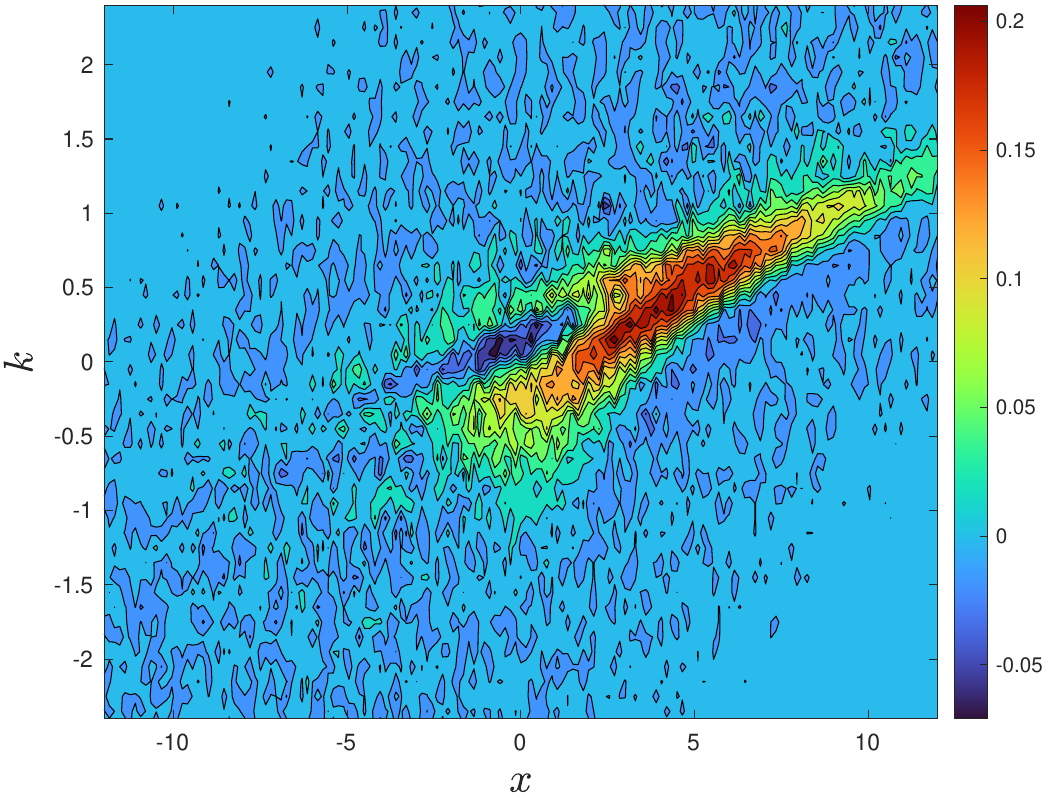}}}
\subfigure[SPADE, $N_0 = 1\times10^7$, $\vartheta = 0.01$.]{
{\includegraphics[width=0.48\textwidth,height=0.27\textwidth]{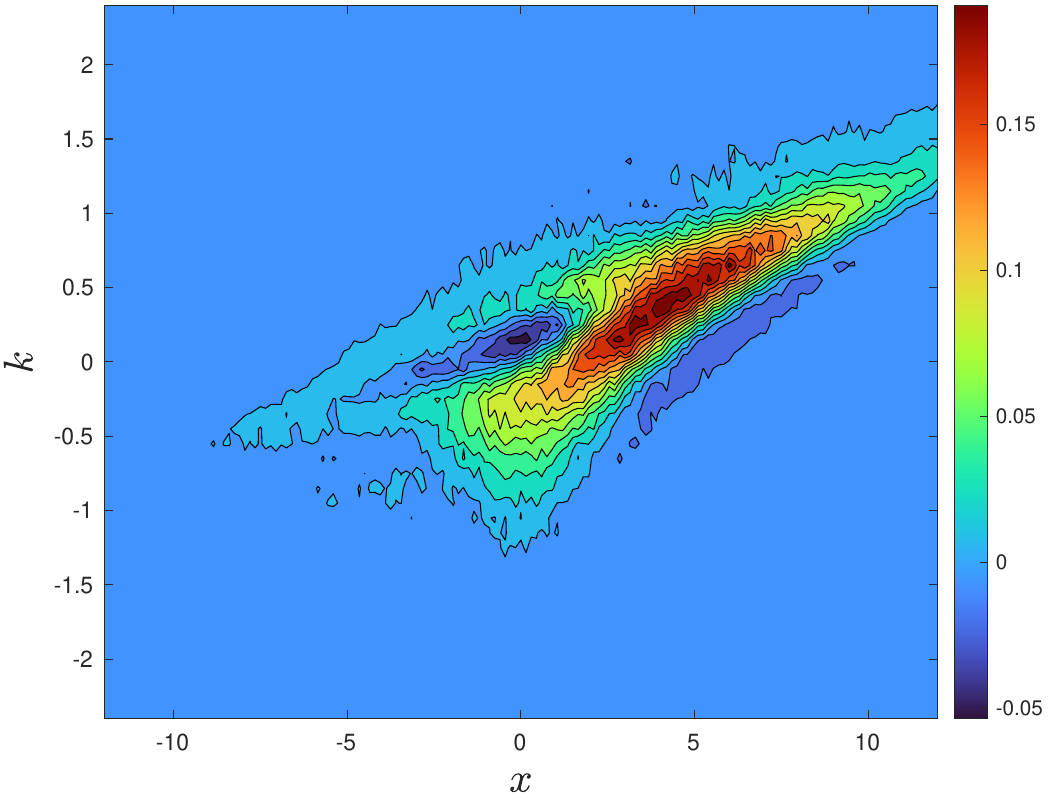}}}
\\
\subfigure[PAUM, $N_0 = 1\times10^7$.]{
{\includegraphics[width=0.48\textwidth,height=0.27\textwidth]{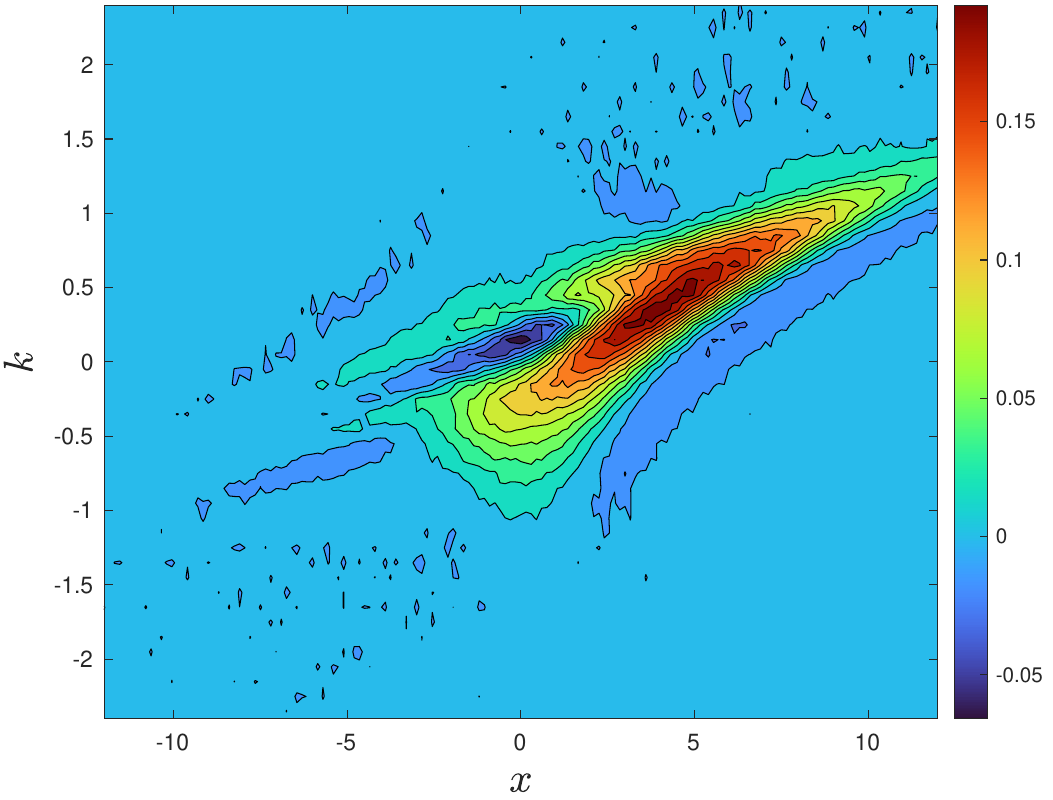}}}
\subfigure[SPADE, $N_0 = 1\times10^7$, $\vartheta = 0.003$.]{
{\includegraphics[width=0.48\textwidth,height=0.27\textwidth]{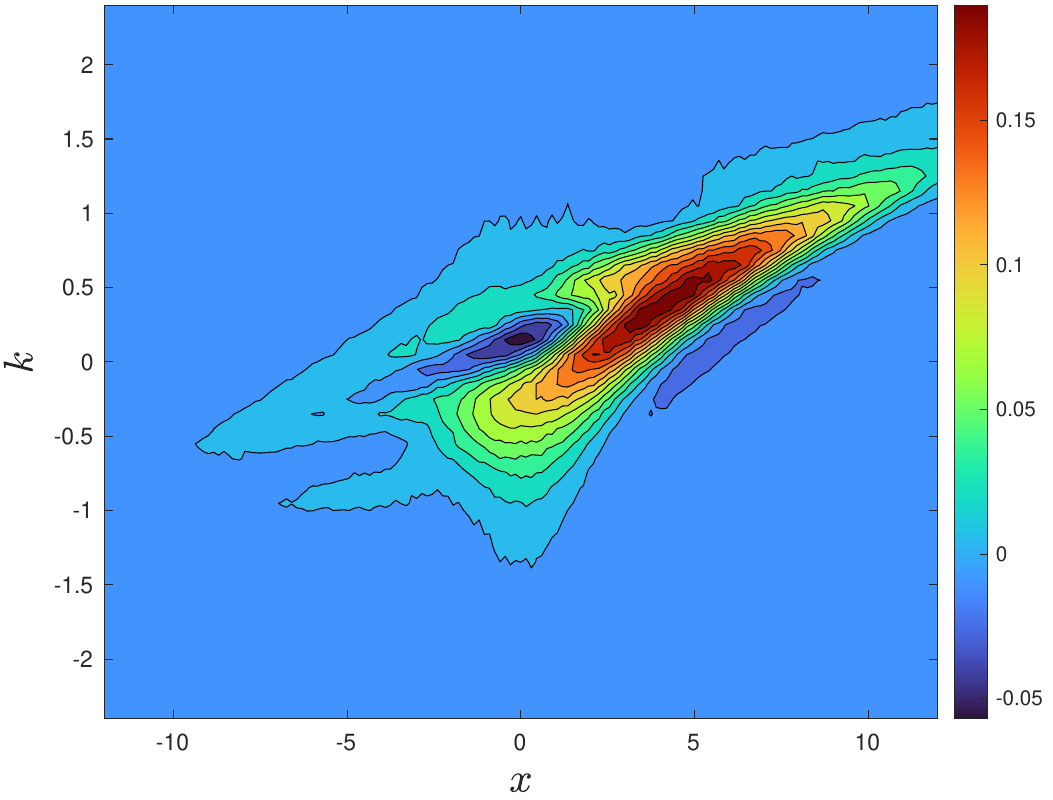}}}
\\
\subfigure[PAUM, $N_0 = 4\times10^7$.]{
{\includegraphics[width=0.48\textwidth,height=0.27\textwidth]{./redist_uhist_n4000_0100.pdf}}}
\subfigure[SPADE, $N_0 = 4\times10^7$, $\vartheta = 0.003$.]{
{\includegraphics[width=0.48\textwidth,height=0.27\textwidth]{./redist_spade_0100.pdf}}}
\caption{\small  The 4-D Morse system: A comparison of reduced Wigner function $W_1(x, k, t)$ at $t=10$a.u. produced by either PAUM (left) or SPADE (right) under relatively small sample size $N_0 = 1\times 10^6, 1\times 10^7$. This gives the first impression on the overfitting problem.}
\label{supp_2dm_particle_comp_1}
\end{figure}

\begin{figure}[!h]
 \centering
\subfigure[$\vartheta = 0.003$.]{
{\includegraphics[width=0.32\textwidth,height=0.17\textwidth]{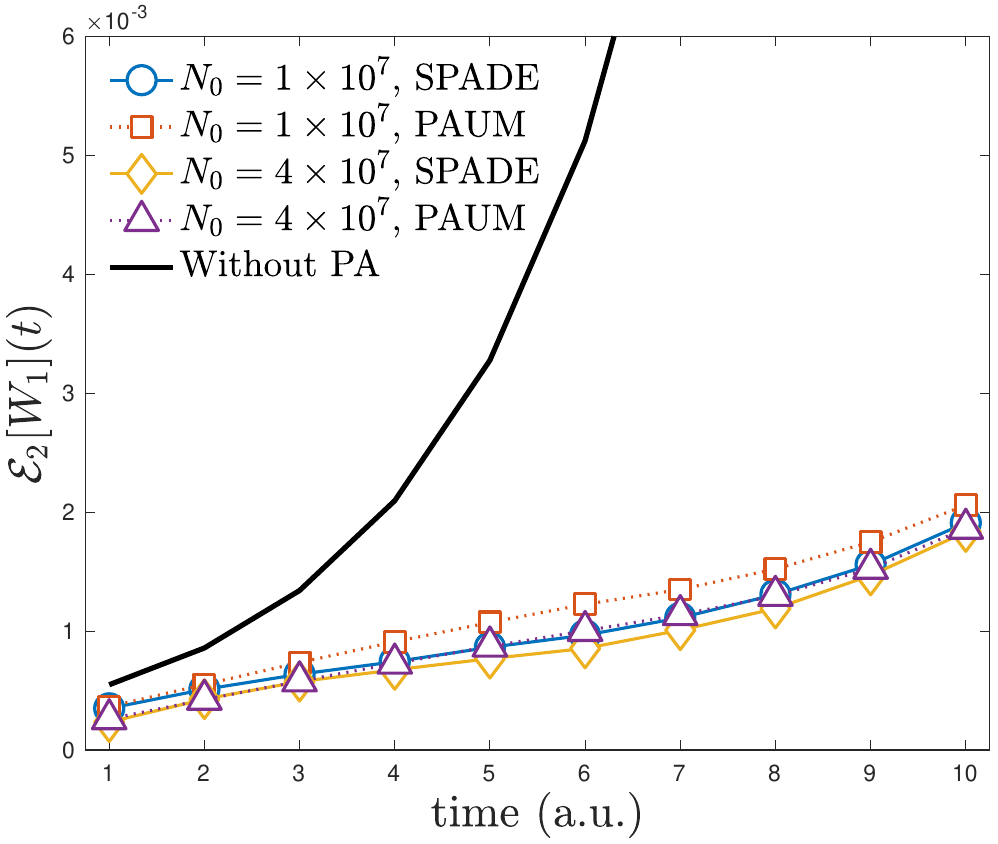}}
{\includegraphics[width=0.32\textwidth,height=0.17\textwidth]{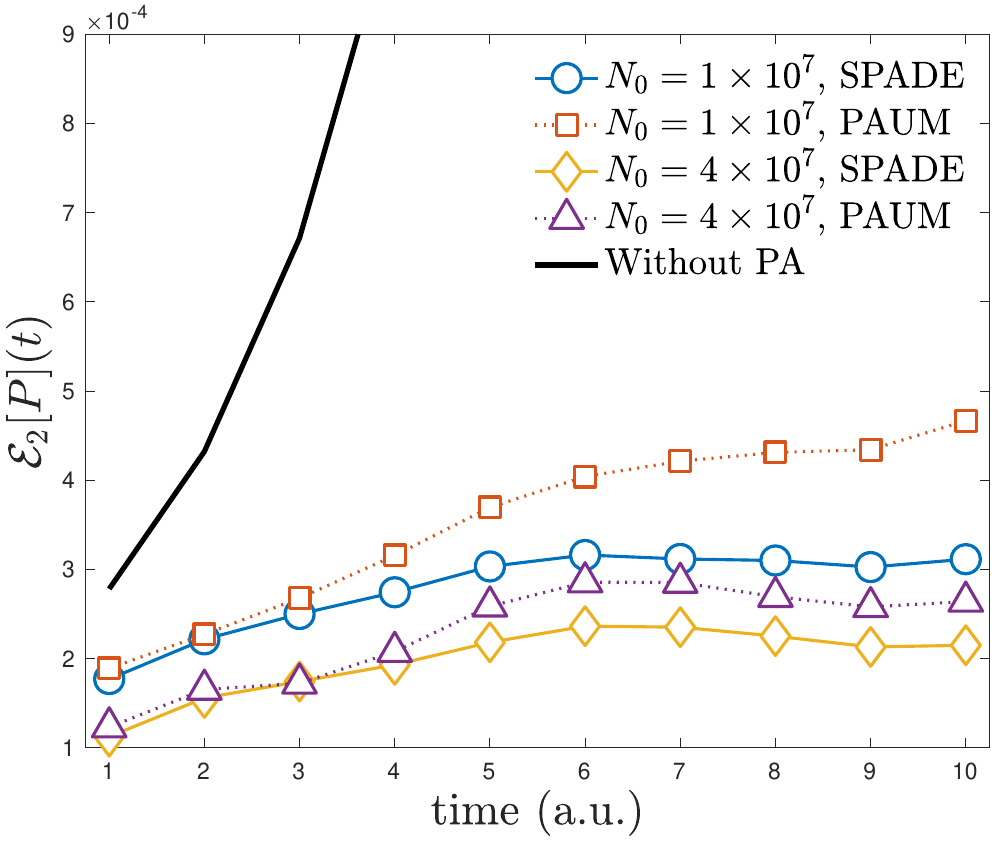}}
{\includegraphics[width=0.32\textwidth,height=0.17\textwidth]{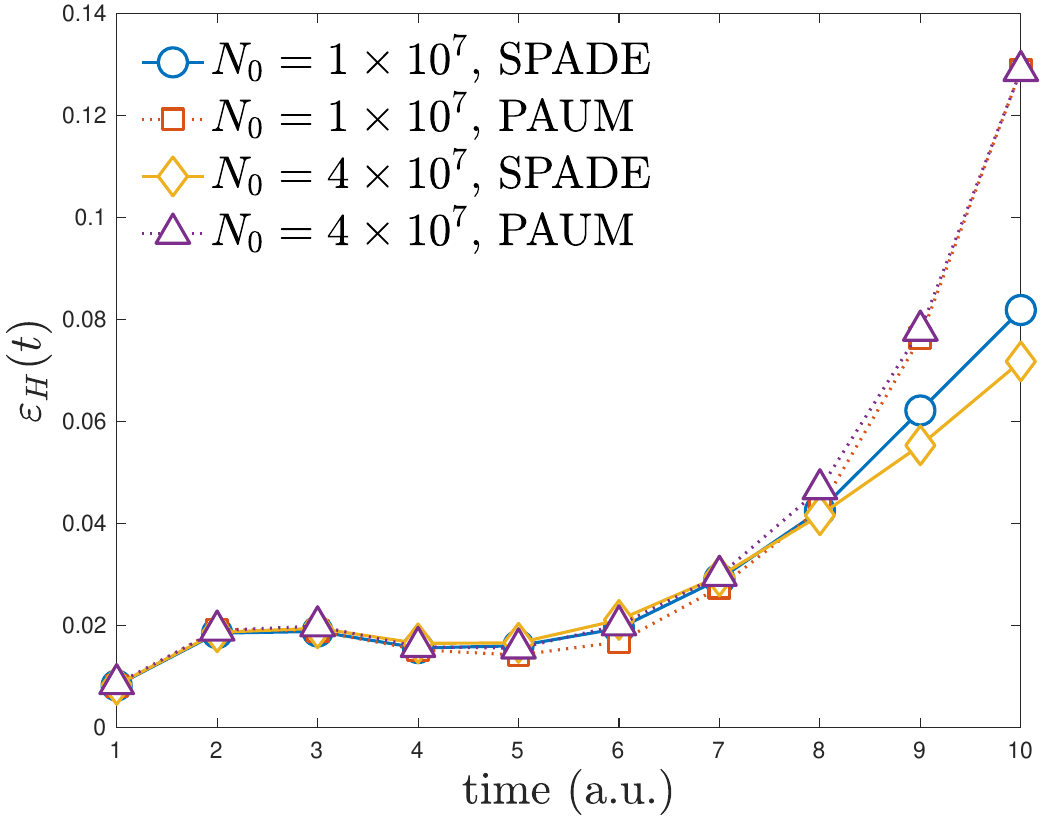}}}
\\
\subfigure[$\vartheta = 0.005$.]{
{\includegraphics[width=0.32\textwidth,height=0.17\textwidth]{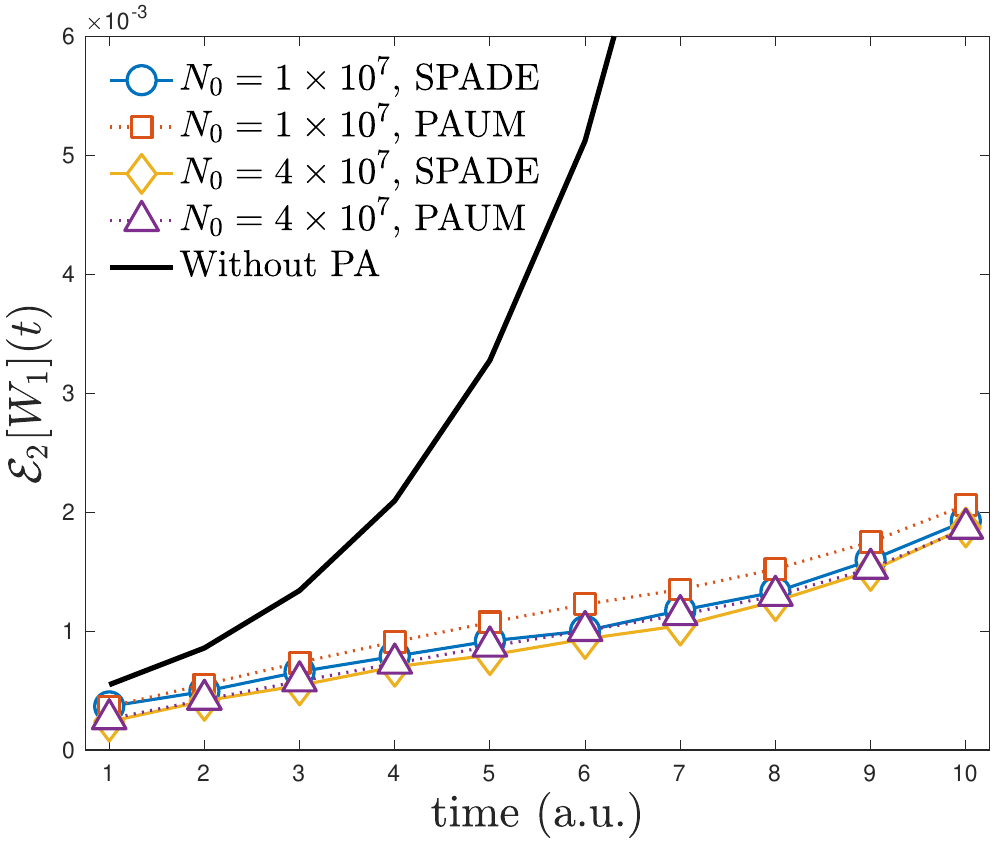}}
{\includegraphics[width=0.32\textwidth,height=0.17\textwidth]{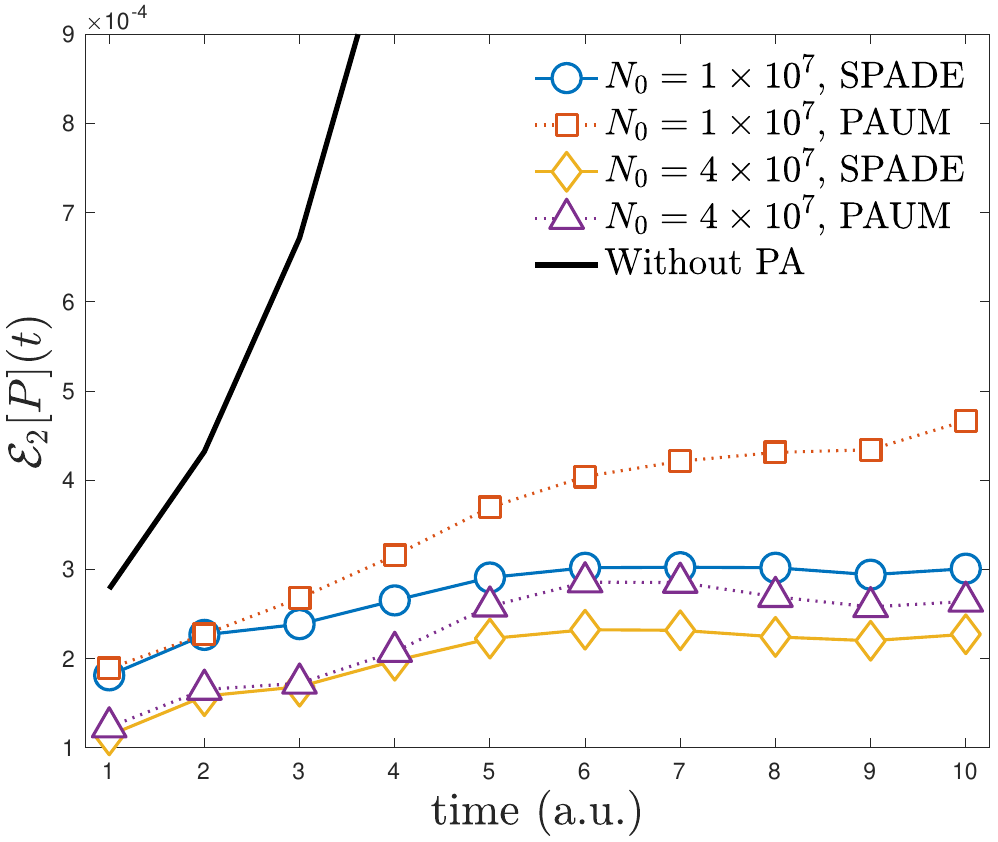}}
{\includegraphics[width=0.32\textwidth,height=0.17\textwidth]{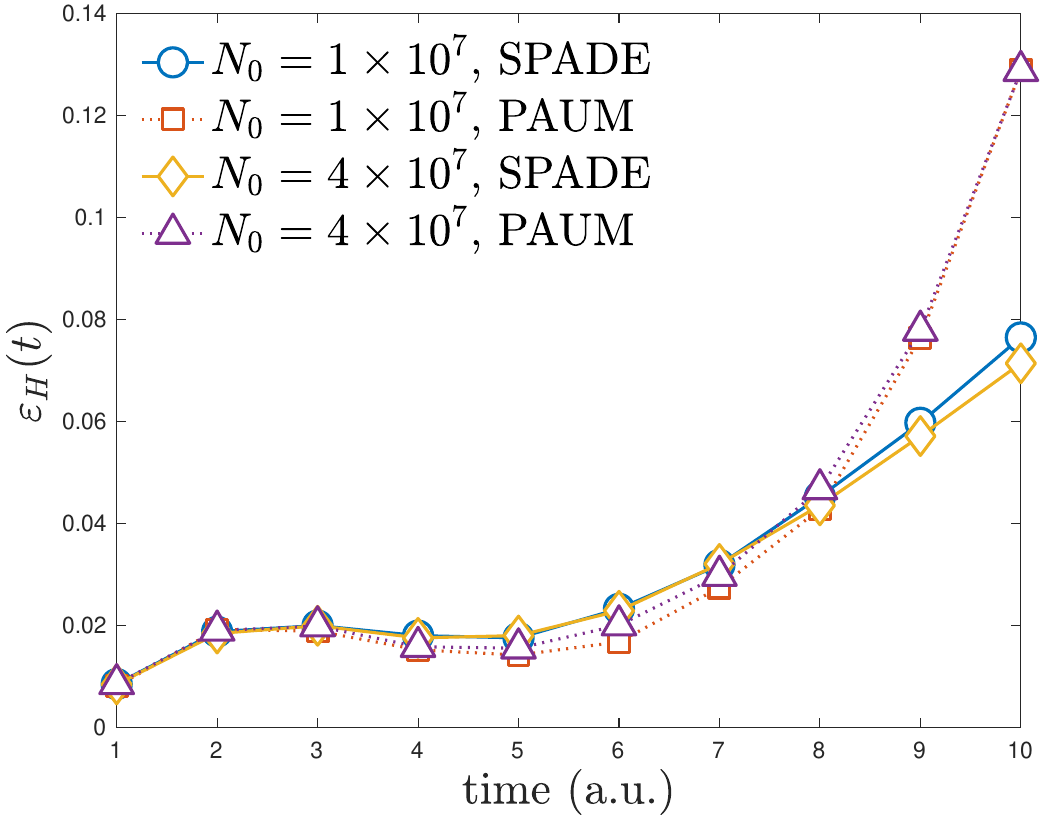}}}
\\
\subfigure[$\vartheta = 0.01$.]{
{\includegraphics[width=0.32\textwidth,height=0.17\textwidth]{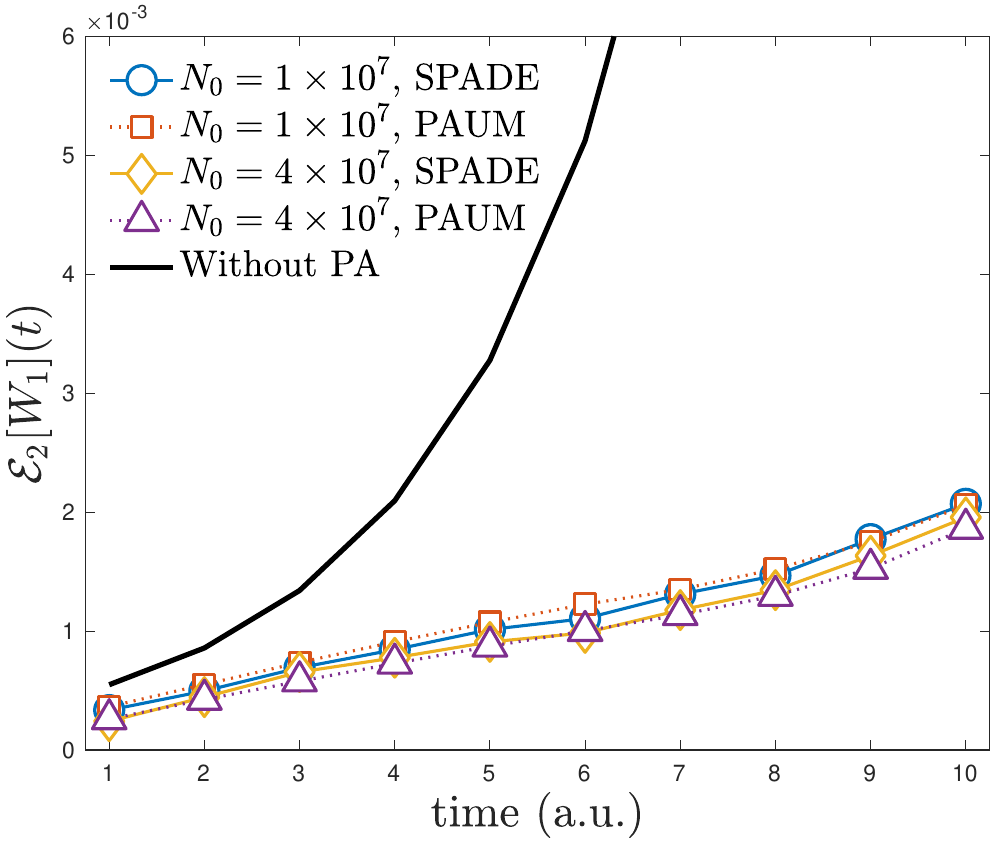}}
{\includegraphics[width=0.32\textwidth,height=0.17\textwidth]{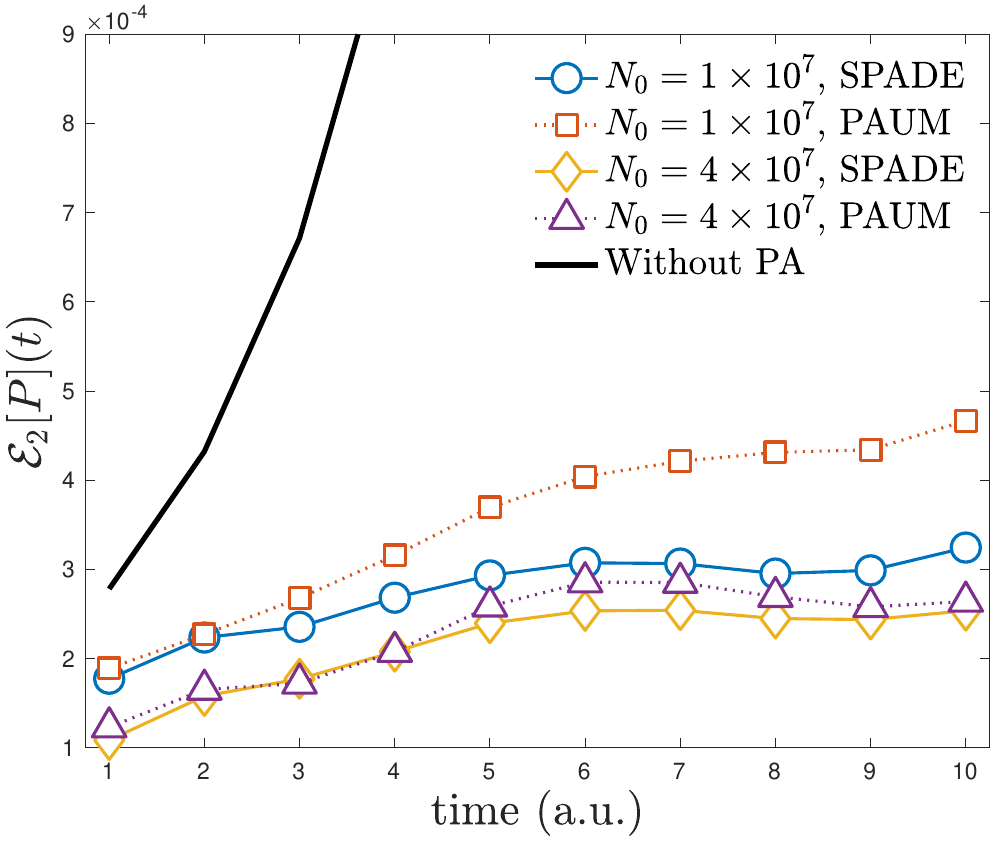}}
{\includegraphics[width=0.32\textwidth,height=0.17\textwidth]{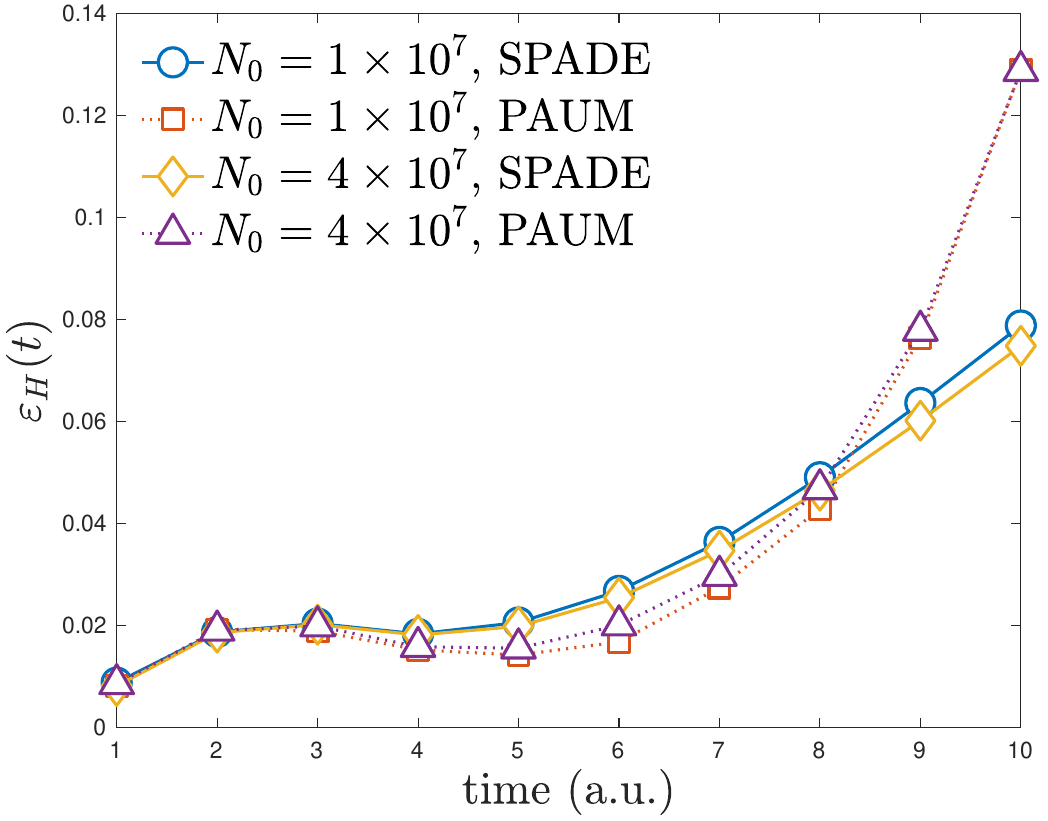}}}
\\
\subfigure[$\vartheta = 0.02$.]{
{\includegraphics[width=0.32\textwidth,height=0.17\textwidth]{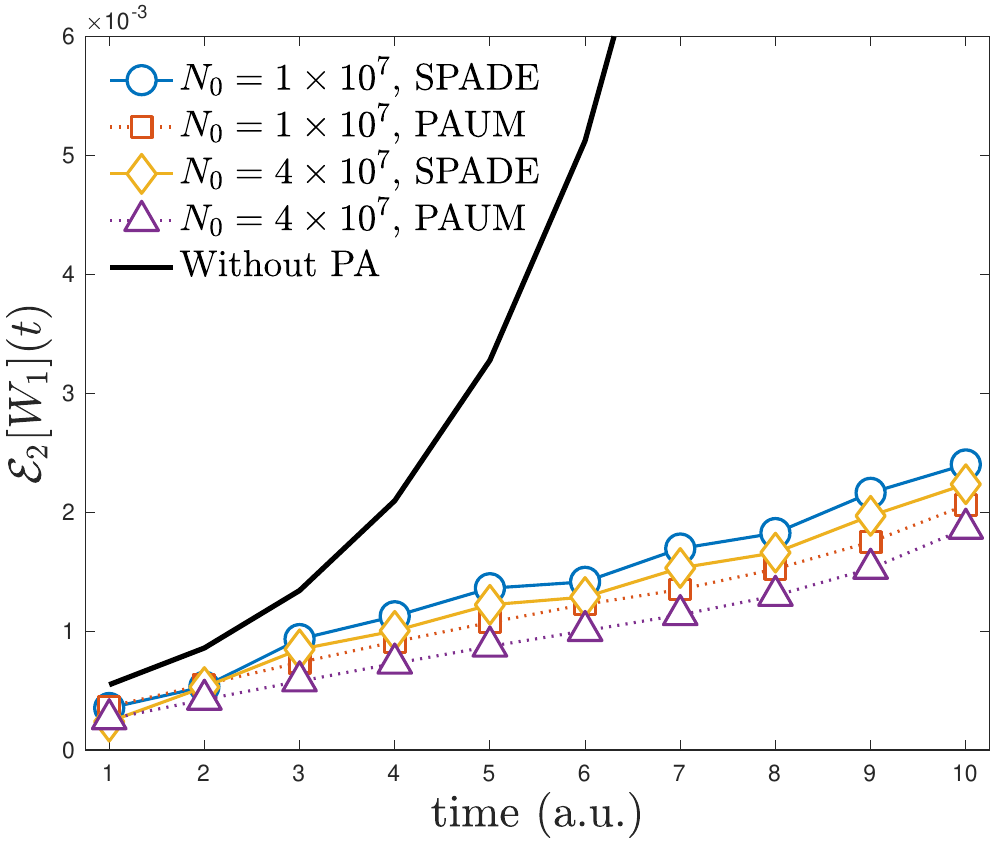}}
{\includegraphics[width=0.32\textwidth,height=0.17\textwidth]{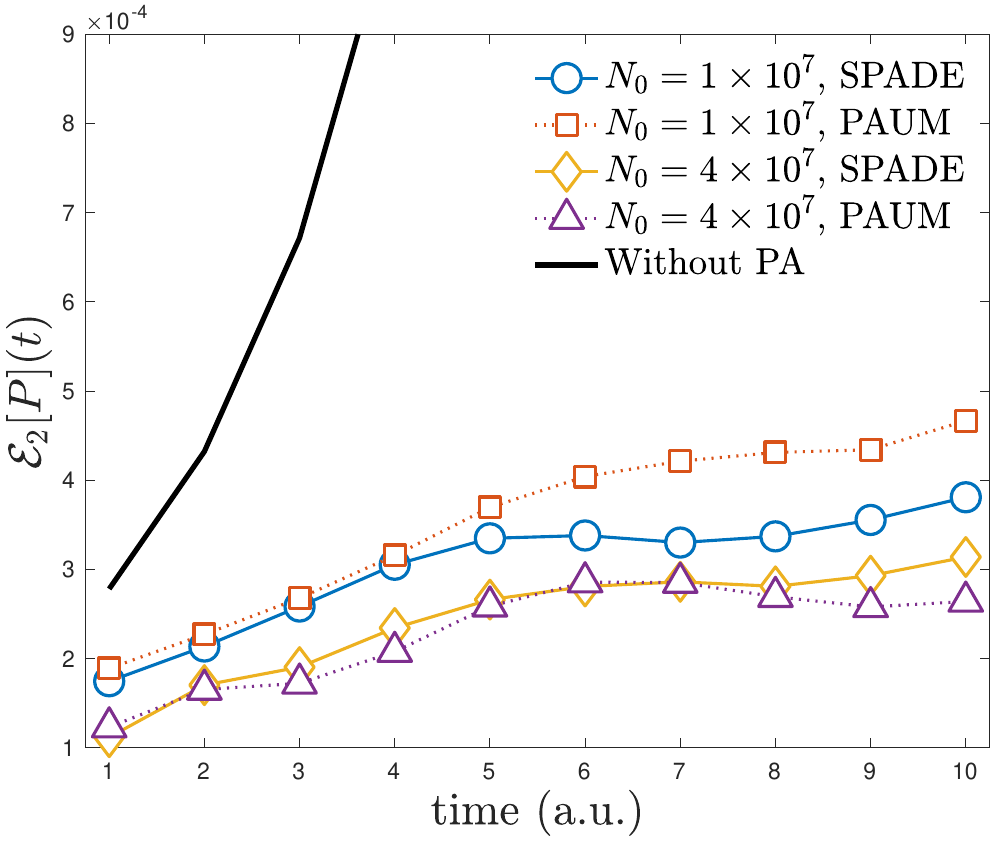}}
{\includegraphics[width=0.32\textwidth,height=0.17\textwidth]{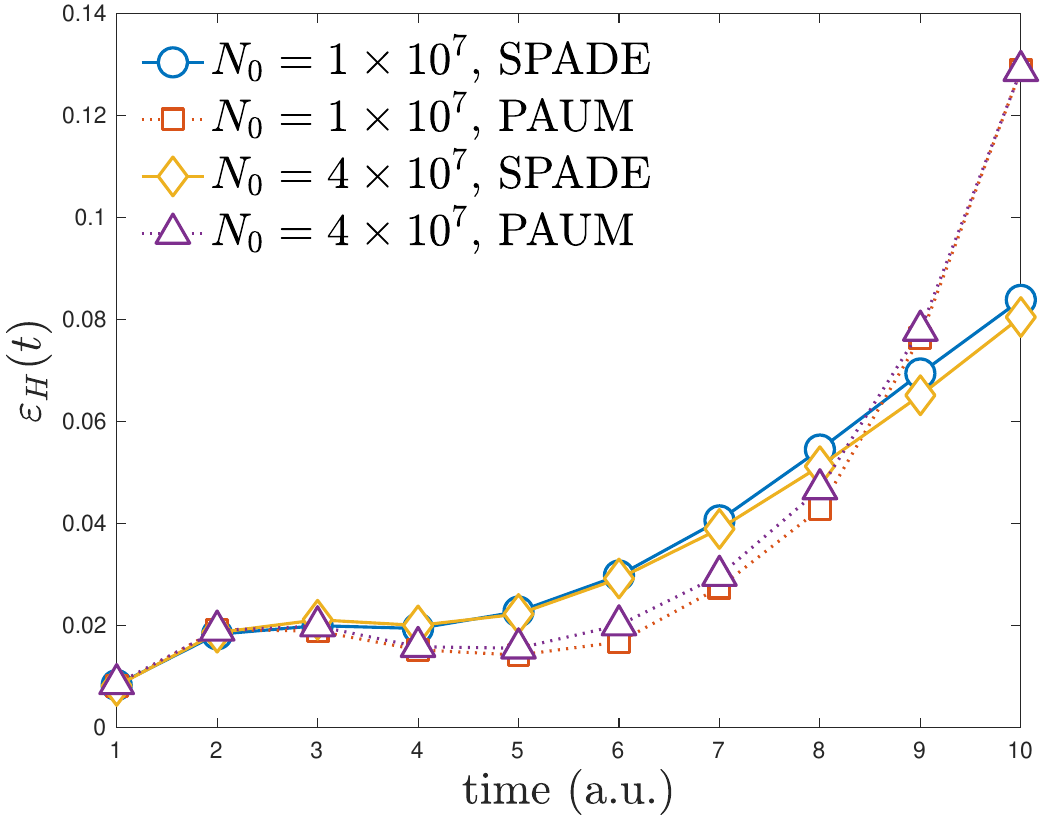}}}
\\
\subfigure[$\vartheta = 0.04$.]{
{\includegraphics[width=0.32\textwidth,height=0.17\textwidth]{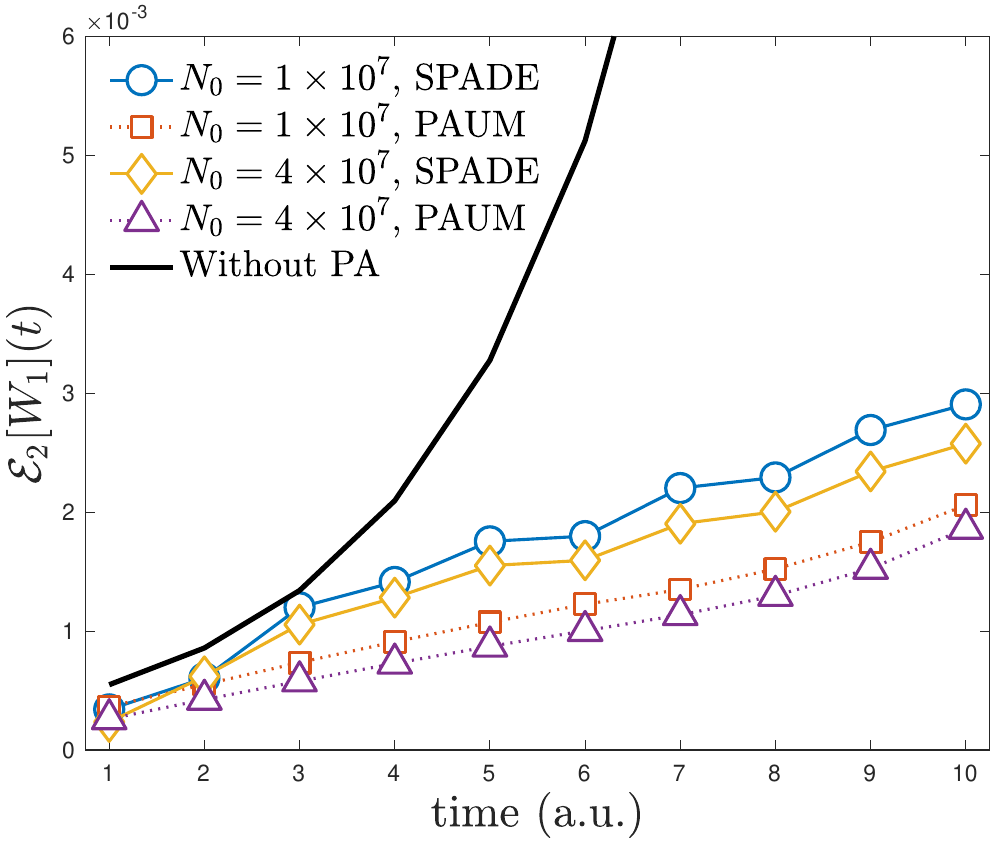}}
{\includegraphics[width=0.32\textwidth,height=0.17\textwidth]{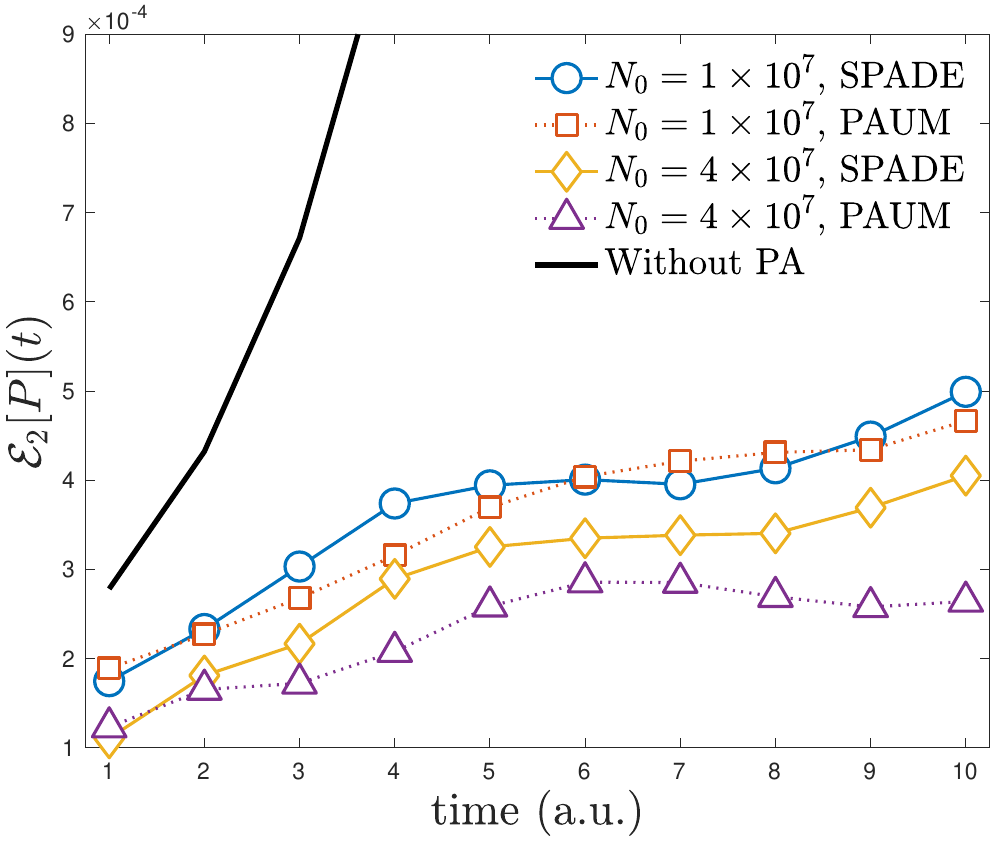}}
{\includegraphics[width=0.32\textwidth,height=0.17\textwidth]{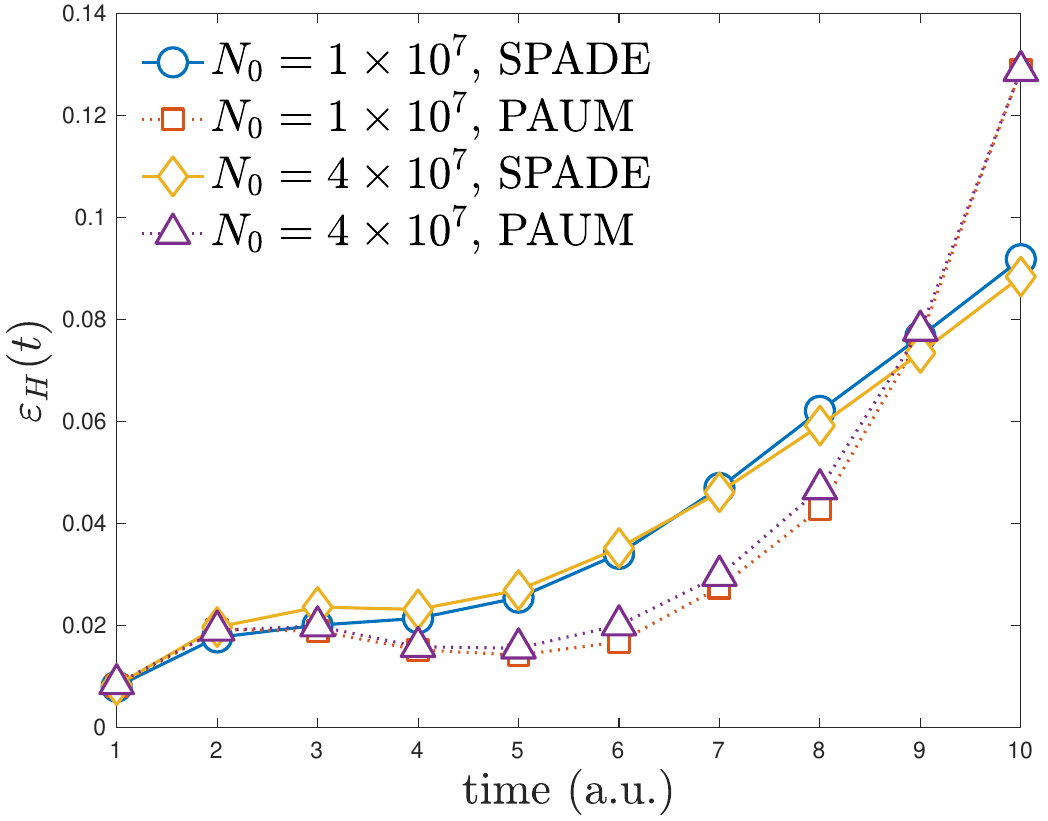}}}
\\
\subfigure[$\vartheta = 0.08$.]{
{\includegraphics[width=0.32\textwidth,height=0.17\textwidth]{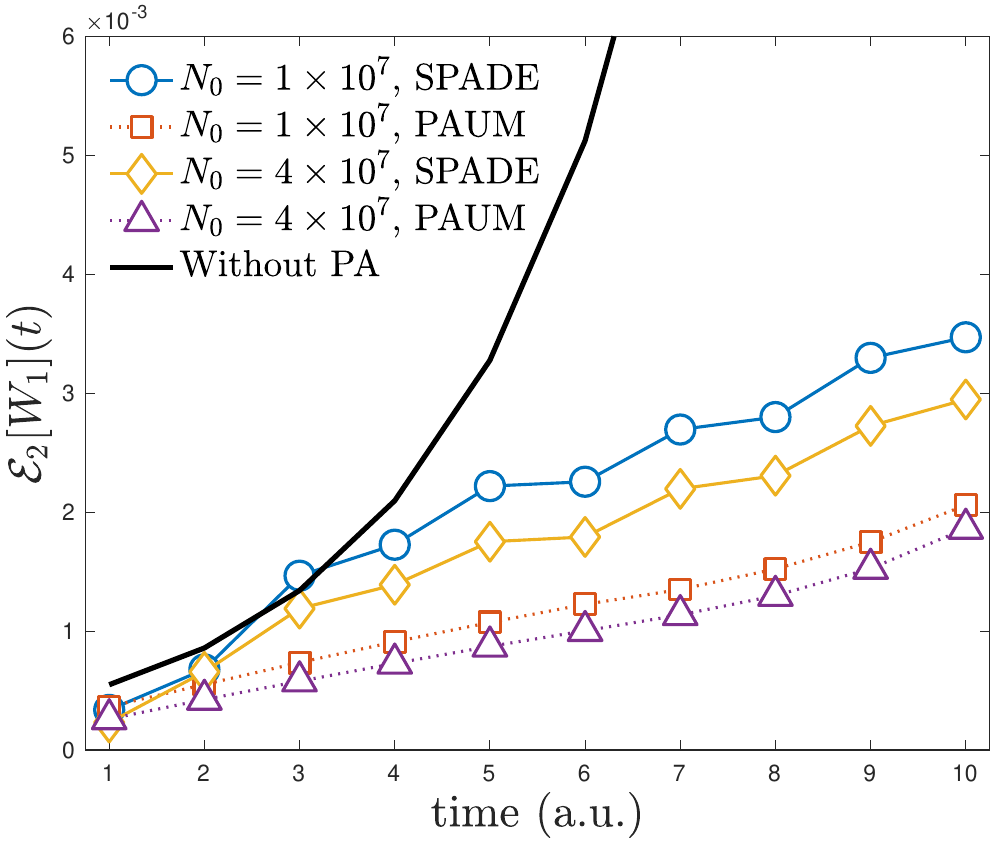}}
{\includegraphics[width=0.32\textwidth,height=0.17\textwidth]{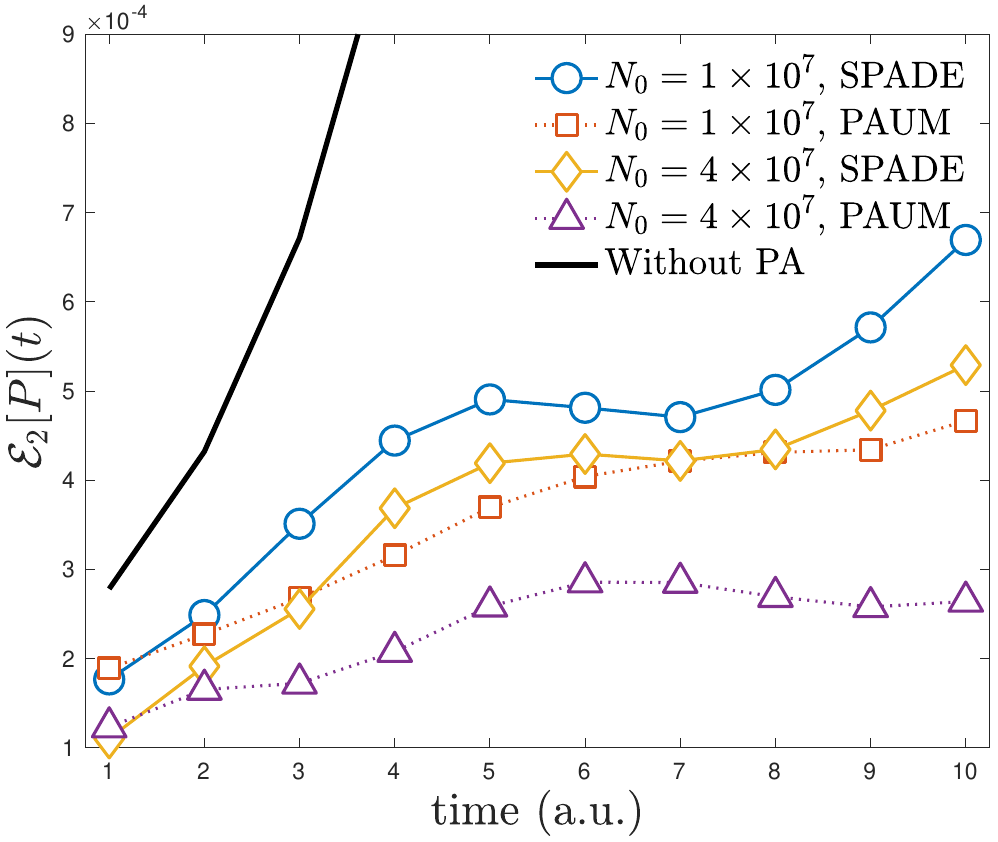}}
{\includegraphics[width=0.32\textwidth,height=0.17\textwidth]{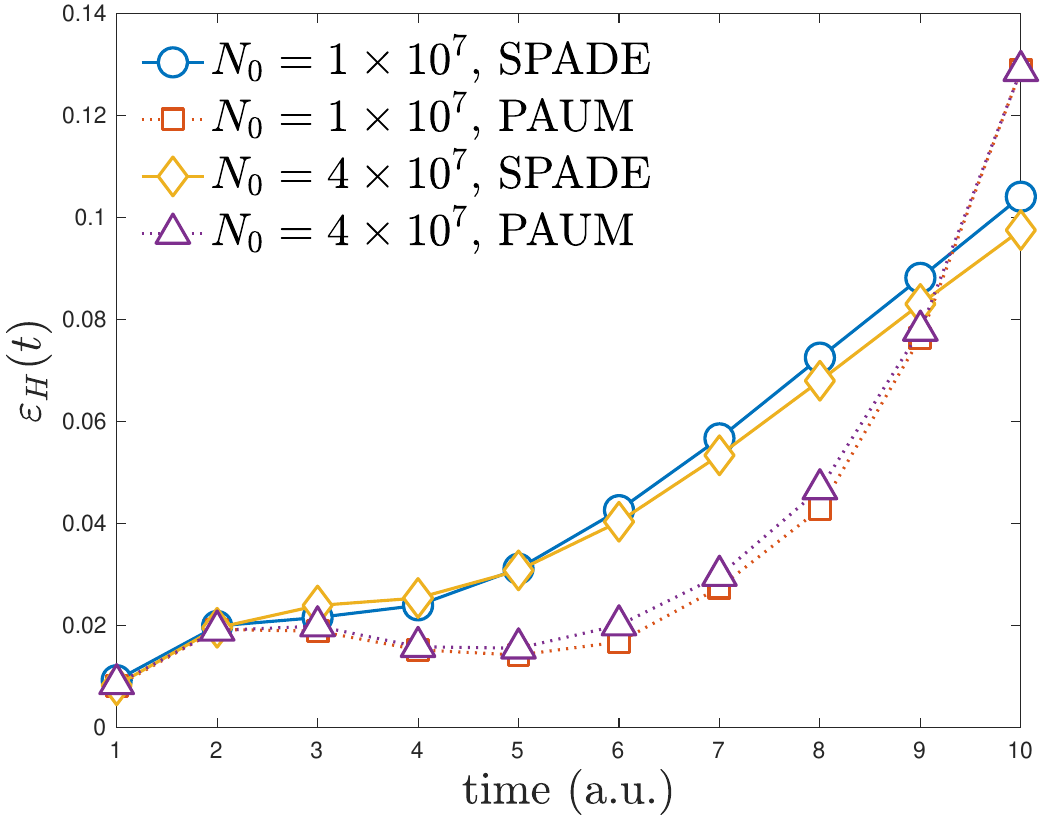}}}
\caption{\small   The 4-D Morse system: Evolution of errors (left: reduced Wigner function, middle: spatial distribution, right: deviations in energy). SPADE is inferior to PAUM when $\vartheta= 0.04, 0.08, 0.16$ but outperforms PAUM when $\vartheta = 0.003, 0.005$. The efficiency of PAUM is maintained only when $N_0$ is sufficiently large.} 
\label{supp_2dm_error}
\end{figure}

\begin{figure}[!h]
 \subfigure[$N_0 = 4\times10^6$.\label{supp_2dm_np_400m}]{
{\includegraphics[width=0.32\textwidth,height=0.22\textwidth]{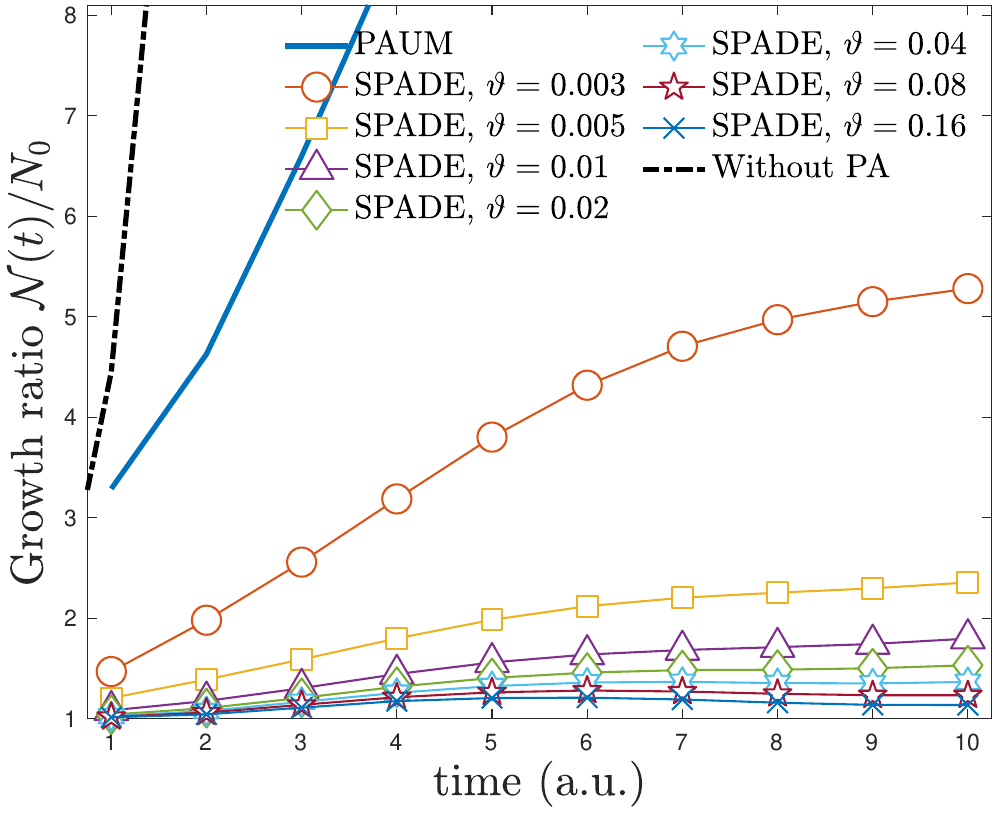}}}
\subfigure[$N_0 = 1\times10^7$.]{
{\includegraphics[width=0.32\textwidth,height=0.22\textwidth]{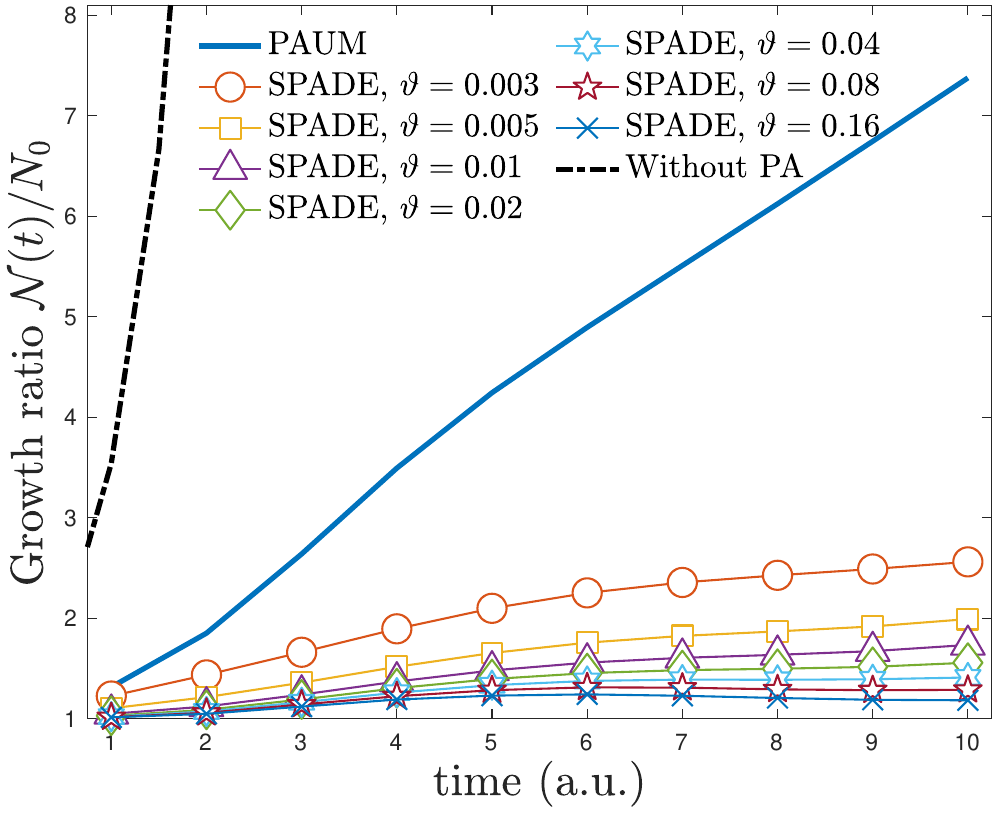}}}
\subfigure[ $N_0 = 4\times10^7$.]{
{\includegraphics[width=0.32\textwidth,height=0.22\textwidth]{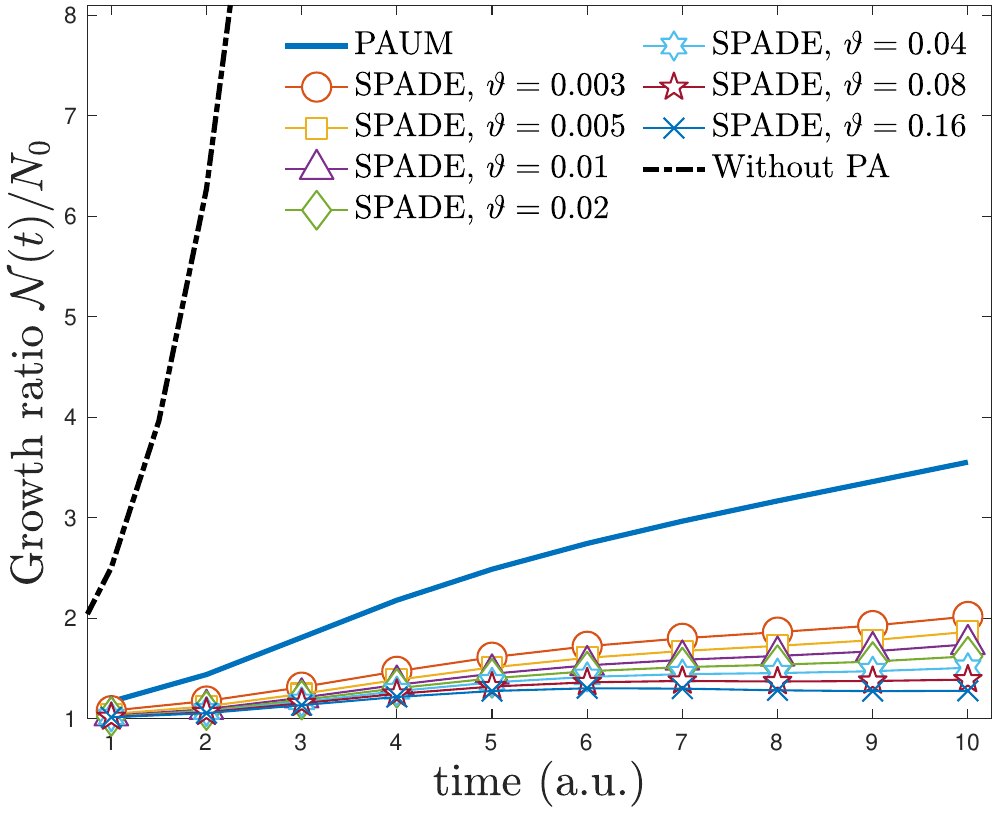}}}
\caption{\small  The 4-D Morse system: Growth ratio of particle number under PAUM and SPADE. The efficiency of particle annihilation may be hampered when the partition level is much larger than sample size as many particle are uncanceled, which is known as the overfitting problem.}
\label{supp_2dm_particle_Np}
\end{figure}

Now we begin to make a thorough comparison between SPADE and PAUM. A visualization of the reduced Wigner function $W_1(x, k, t)$ and spatial distribution $P(x_1, x_2, t)$ at $t = 10$a.u. is demonstrated  in Figures \ref{supp_fig_asm_redist}-\ref{supp_2dm_particle_comp_1}, respectively. The time evolution of $l^2$-errors, as well as the deviations of total energy, are plotted in Figure \ref{supp_2dm_error}. The growth ratio of total particle number $P(t)+M(t)$ is plotted in Figure \ref{supp_2dm_particle_Np}. Based on these results, we have the following observations.

{\bf Snapshots}: The snapshots demonstrate the capability of stochastic Wigner algorithm, with either PAUM or SPADE,  to recover the fine oscillating structure of the Wigner function for sufficiently large sample size ($N_0=4\times 10^7$). However, when the sample size decreases to $N_0 = 1\times 10^7$, as visualized in Figure \ref{supp_2dm_particle_comp_1}, the solutions produced by PAUM are evidently more noisy than those by SPADE. What is worse, when $N_0 = 1\times 10^6$, PAUM might fail to produce reliable results, while SPADE still works in this situation. This is because of the overfitting problem. If the partition level $K$ is much larger than the sample size, many particles might be left uncanceled, so that the stochastic noises cannot be suppressed efficiently.

 {\bf Comparison between PAUM and SPADE}: According to Figure \ref{supp_2dm_error}, the rapid growth of stochastic variances can be dramatically suppressed when the particle annihilation is used. PAUM outperforms SPADE when $\vartheta \ge 0.04$, while their accuracy seems to be comparable when $\vartheta = 0.02$. By further decreasing $\vartheta$ to $0.003$ or $0.005$,  the accuracy of SPADE even outperforms that of PAUM, which coincides with the observation in Figure \ref{supp_2dm_particle_comp_1}. In fact,  the performance of PAUM is sensitive to the sample-to-partition ratio $N_0/K$. When  $N_0/K$ is too small, PAUM fails to kill redundant particles, so that might not suppress the random noises efficiently. This phenomenon has also been observed in our previous work \cite{XiongShao2019}. By contrast,  the average partition level in SPADE is $6.83\times 10^5$ under $N_0 = 1\times 10^7$ and $\vartheta = 0.005$, which ensures the efficiency of particle annihilation.

{\bf Particle growth}: The particle growth is presented in Figure \ref{supp_2dm_particle_Np}. When $N_0 = 1\times10^6$, the particle number after PAUM reaches $4.5\times 10^7$ until $t = 10$a.u. (growth ratio is $11.3$). Meanwhile, when $N_0 = 1\times10^7$, the particle number after PAUM reaches $7.4\times 10^7$ (growth ratio is $7.4$). This accounts for the reason why PAUM only works when $N_0$ is comparable to $K$ but soon becomes inefficient when $N_0$ is much smaller than $K$. By contrast, particle number after SPADE almost remains at a stable level. The exception is the group $N_0 = 4\times 10^6$, $\vartheta = 0.003$ in Figure \ref{supp_2dm_np_400m}, where too small $\vartheta$ may lead to over-partitioning when sample size is not enough and  hamper the efficiency of SPADE.

\subsection{Deep partition improves SPADE}

\begin{figure}[!h]
 \subfigure[$N_0 = 4\times10^6$.\label{supp_error_n400}]{
{\includegraphics[width=0.48\textwidth,height=0.27\textwidth]{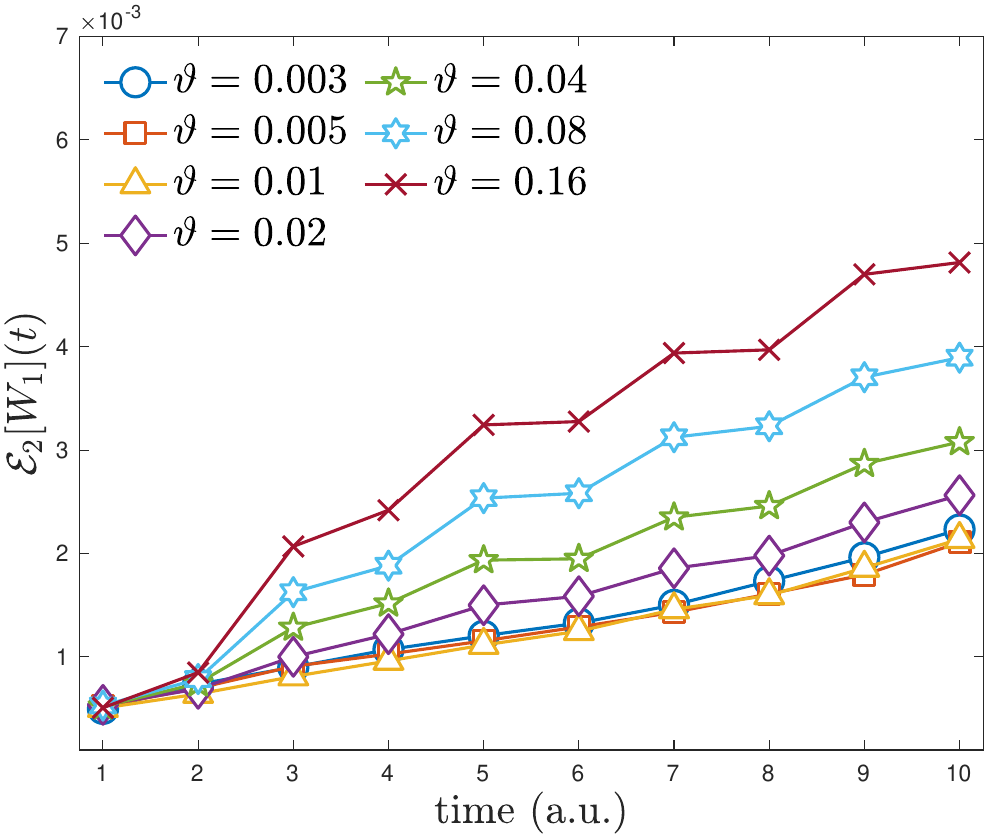}}
{\includegraphics[width=0.48\textwidth,height=0.27\textwidth]{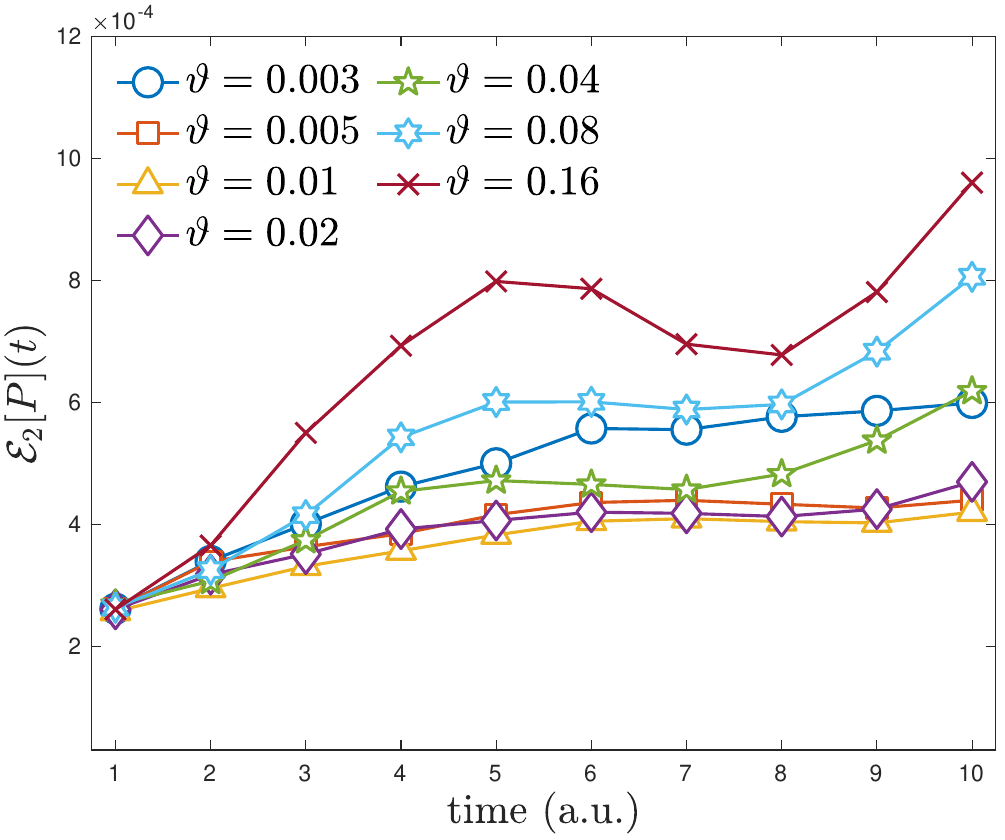}}}
\\
\subfigure[$N_0 = 1\times10^7$.]{
{\includegraphics[width=0.48\textwidth,height=0.27\textwidth]{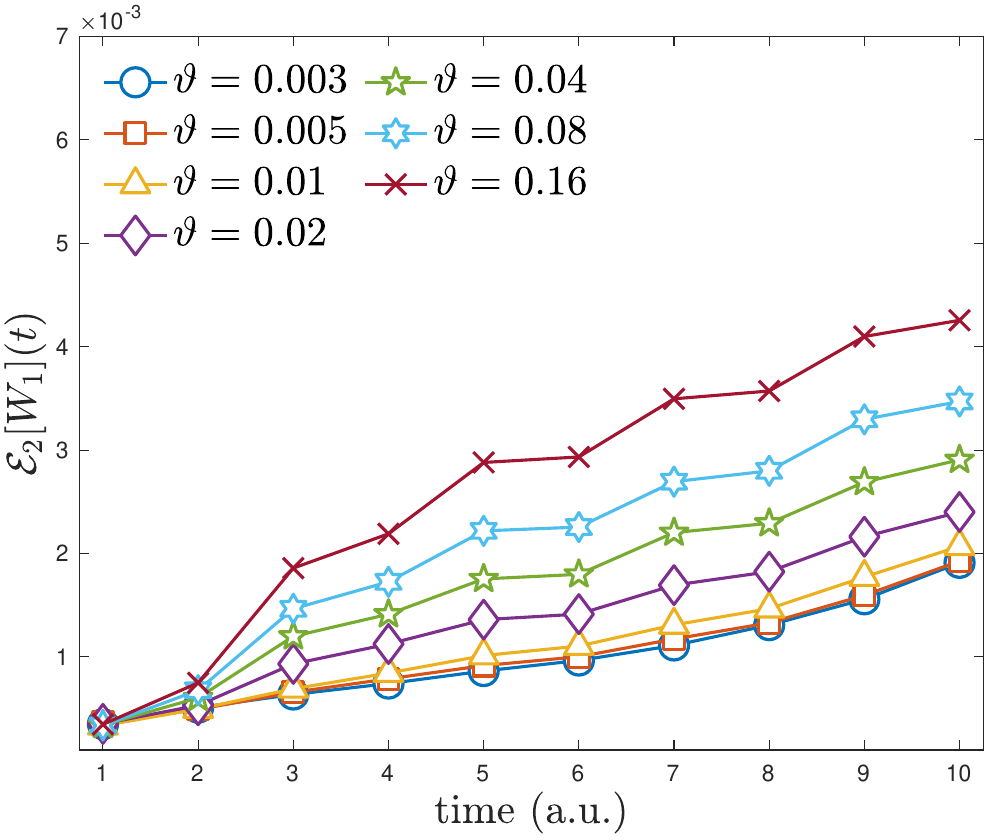}}
{\includegraphics[width=0.48\textwidth,height=0.27\textwidth]{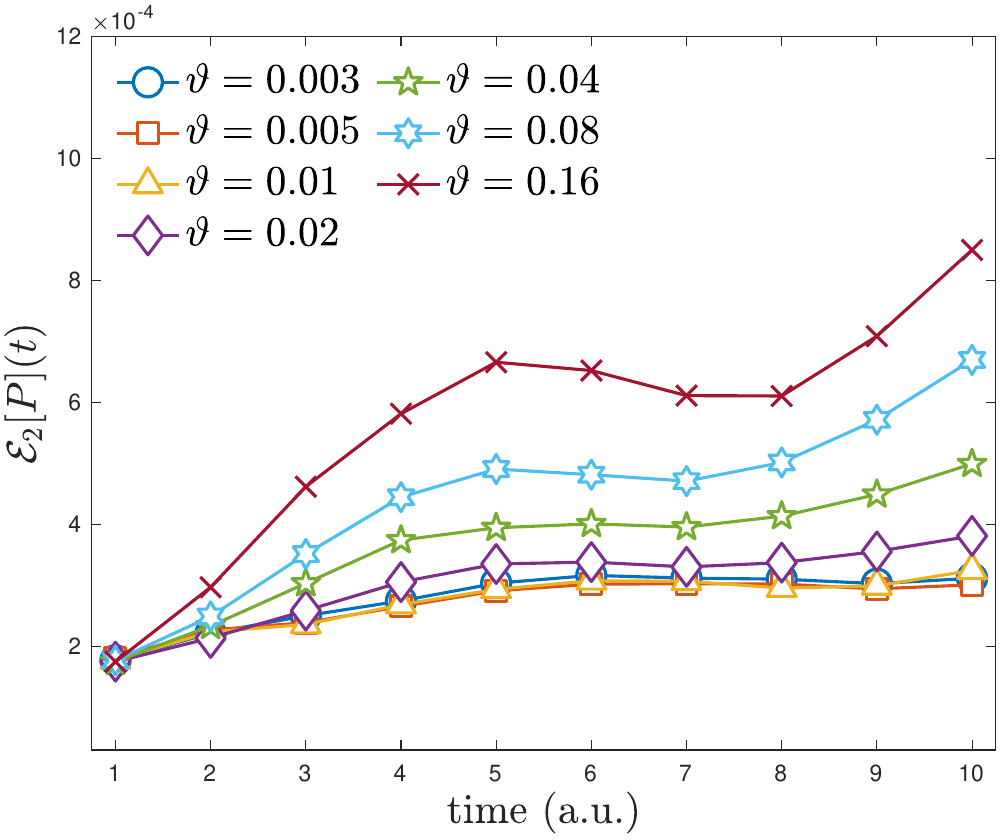}}}
\\
\subfigure[$N_0 = 4\times10^7$.]{
{\includegraphics[width=0.48\textwidth,height=0.27\textwidth]{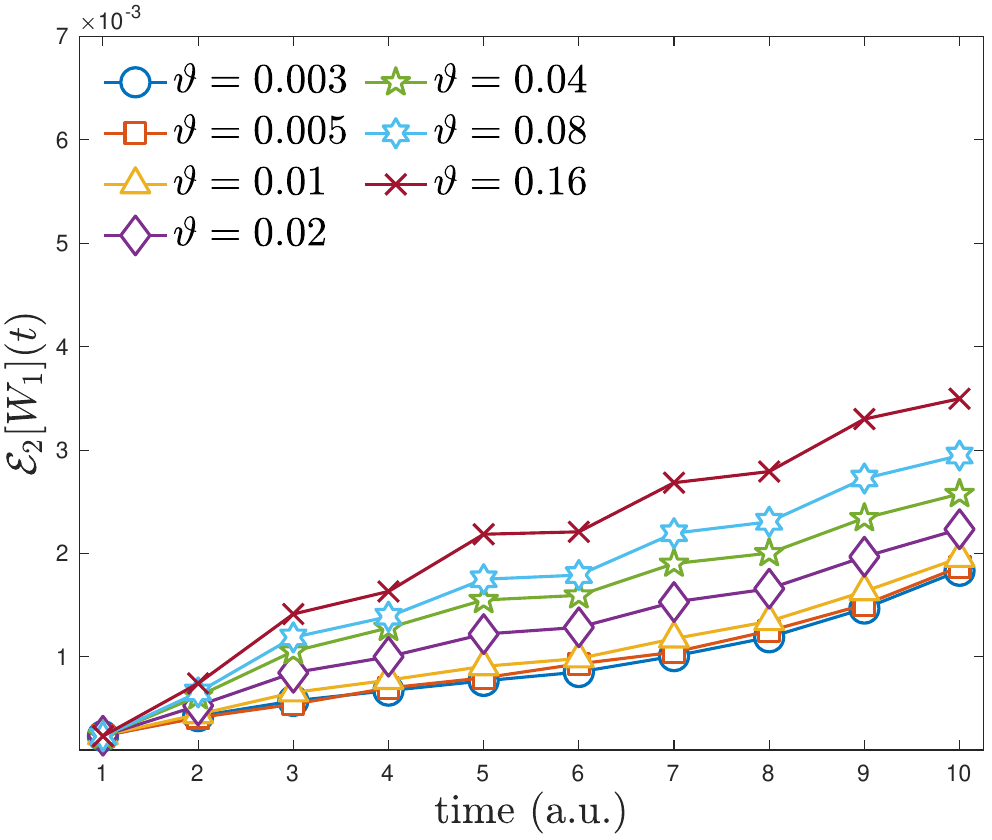}}
{\includegraphics[width=0.48\textwidth,height=0.27\textwidth]{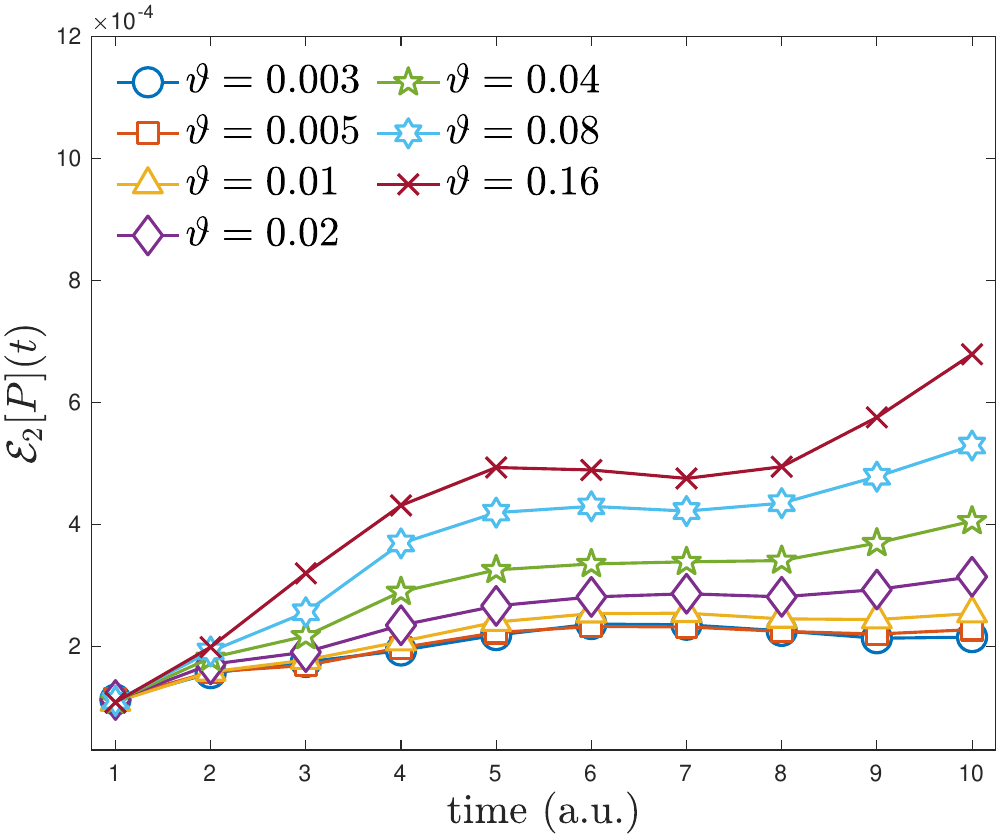}}}
\caption{\small The 4-D Morse system: $l^2$-errors under different $N_0$ (left: reduced Wigner function, right: spatial distribution). The accuracy of SPADE can be systematically improved by choosing smaller parameter $\vartheta$ and deepening the partitioning. }
\label{supp_2dm_spade_t}
\end{figure}

\begin{figure}[!h]
 \centering
\subfigure[$N_0 = 1\times10^7$.]{
{\includegraphics[width=0.48\textwidth,height=0.27\textwidth]{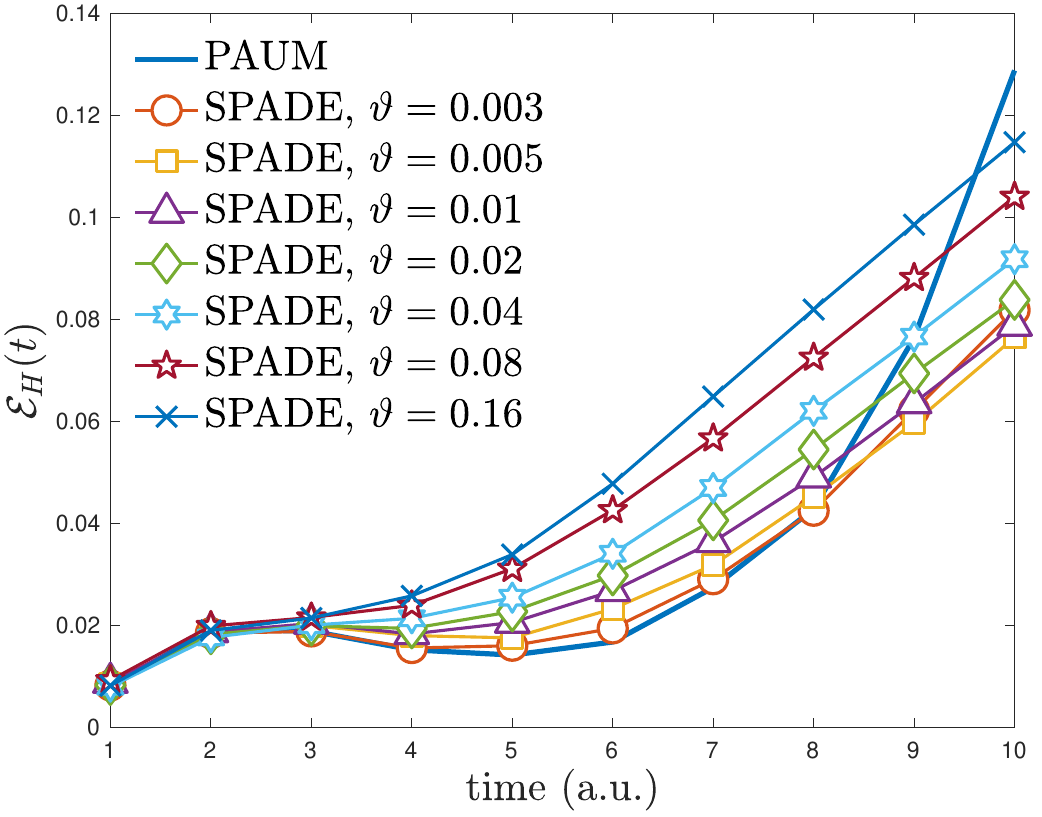}}}
\subfigure[$N_0 = 4\times10^6$.]{
{\includegraphics[width=0.48\textwidth,height=0.27\textwidth]{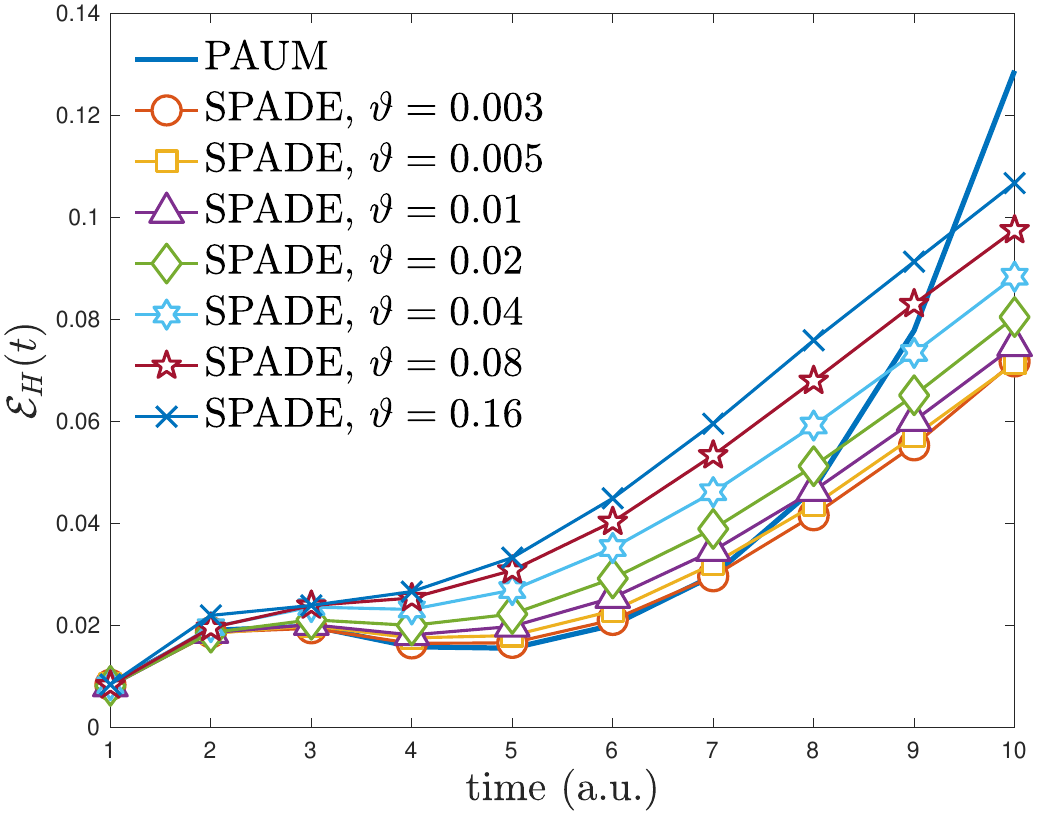}}}
\caption{\small  The 4-D Morse system: Deviation of total energy can be ameliorated by choosing small $\vartheta$ and deepening the partitioning.  }
\label{supp_2dm_E_t}
\end{figure}

\begin{figure}[!h]
 \centering
\subfigure[Smaller $\vartheta$ leads to larger $K$. \label{supp_relation_K_theta}]{
{\includegraphics[width=0.48\textwidth,height=0.27\textwidth]{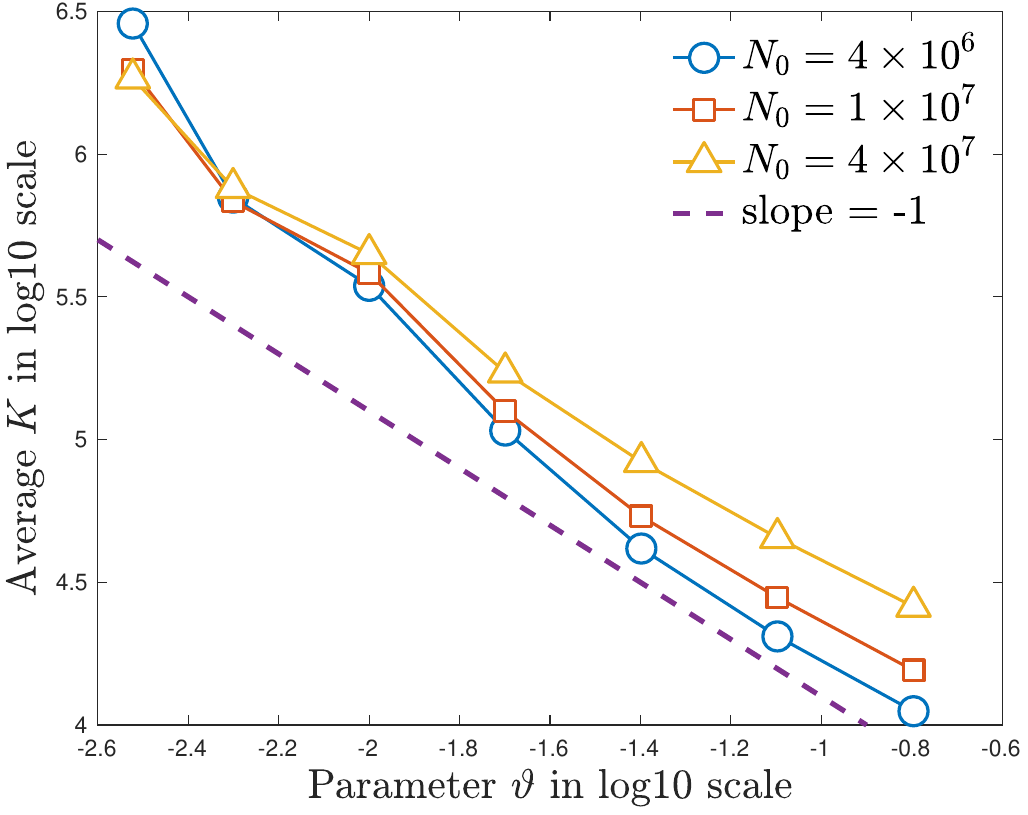}}}
\subfigure[Larger $K$ improves accuracy of SPADE.  \label{supp_convergence_K}]{
{\includegraphics[width=0.48\textwidth,height=0.27\textwidth]{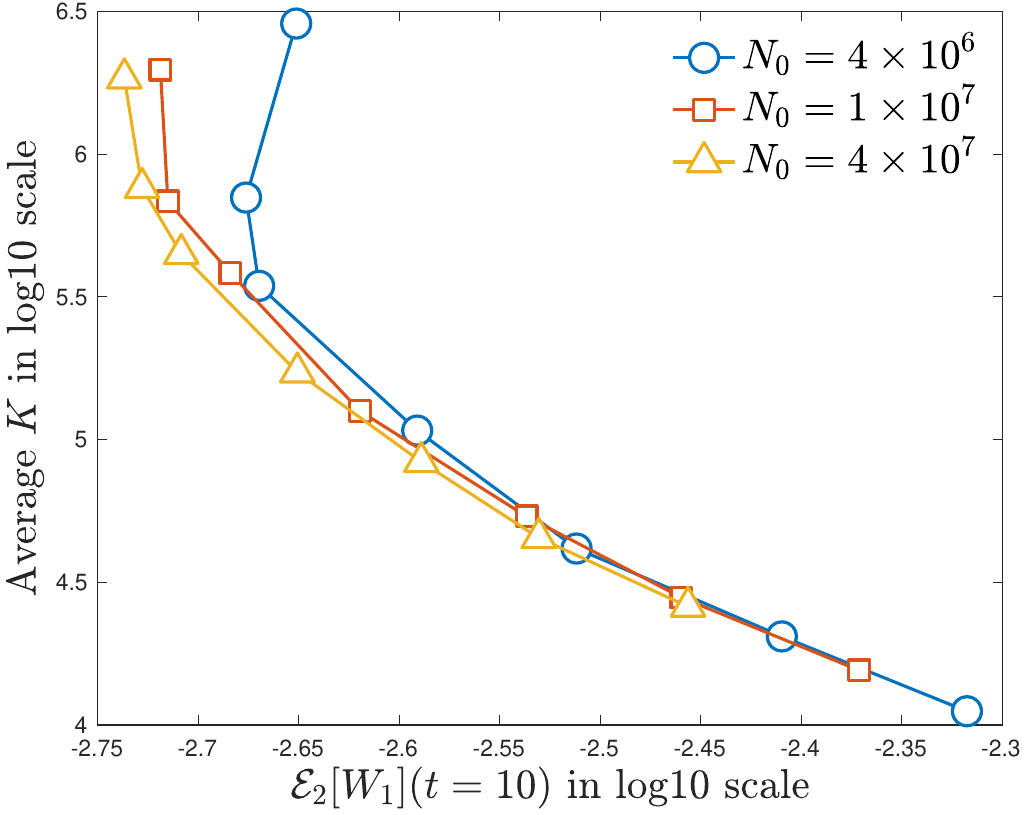}}}
\caption{\small  The 4-D Morse system: Choosing small $\vartheta$ may lead to an exponential growth of partition level $K$, and consequently improve the accuracy of SPADE systematically. An exception is the group $\vartheta = 0.003$, $N_0 = 4\times10^6$ as  too many redundant particles are uncanceled, which may hamper the accuracy of PA. }
\label{supp_2dm_partition_t}
\end{figure}

Now we would like to demonstrate that the accuracy of SPADE can be systematically improved by deepening the partition, which is realized by choosing smaller $\vartheta$. To this end, we fix the sample size $N_0$ and evaluate the performance of SPADE under $\vartheta = 0.003, 0.005, 0.01, 0.02, 0.04, 0.08, 0.16$. The time evolutions of $l^2$-errors and deviation in energy are plotted in  Figures \ref{supp_2dm_spade_t} and \ref{supp_2dm_E_t}, respectively. The partition level $K$ is recorded in Figure \ref{supp_2dm_partition_t}. Based on the numerical results, we have the following observations.

{\bf Convergence with respect to $\vartheta$}: According to Figure \ref{supp_2dm_spade_t}, the numerical errors can be gradually improved by  decreasing $\vartheta$ from $0.16$ to $0.003$, indicating that refinement in the adaptive partition can systematically improve the accuracy. The numerical energy may slightly increase due to the bias induced by SPADE. Fortunately, the deviations can be alleviated when the partition is deepened. 

{\bf Partition level $K$ with respect to $\vartheta$}: In Figure \ref{supp_2dm_partition_t}, the average partition level $K$ increases along with the decrease of $\vartheta$, and consequently leads to a reduction in stochastic variances. It is observed that $K$ is inversely proportional to $\vartheta$, which verifies the lower bound \eqref{K_bound} of $K$. This actually gives us a hint to postulate the partition level $K$ by first performing some tests under relatively larger $\vartheta$. An exception is still the group $N_0 = 4\times 10^6$, $\vartheta = 0.003$ due to the overfitting problem. From Figure \ref{supp_error_n400}, the over-refinement in partition may lead to large errors. 

\subsection{Large sample size improves SPADE}

\begin{figure}[!h]
 \subfigure[$\vartheta = 0.003$.]{
{\includegraphics[width=0.48\textwidth,height=0.27\textwidth]{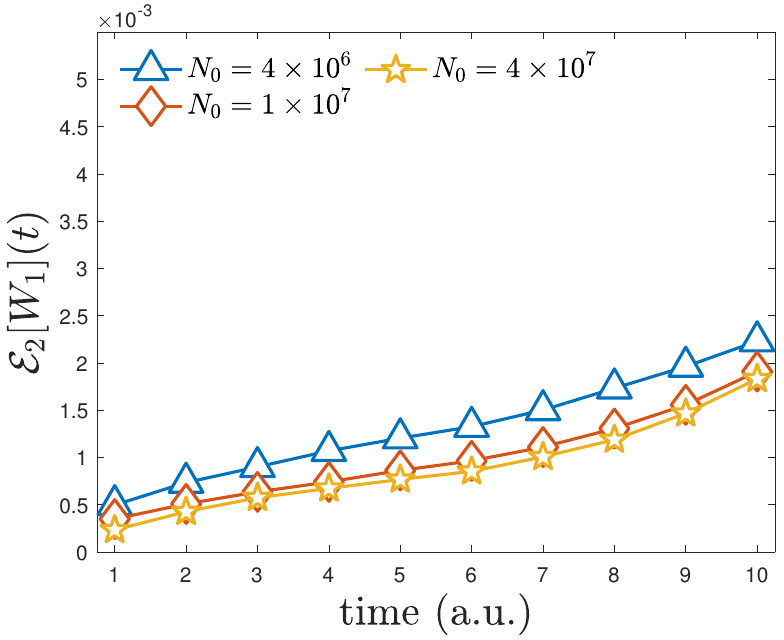}}
{\includegraphics[width=0.48\textwidth,height=0.27\textwidth]{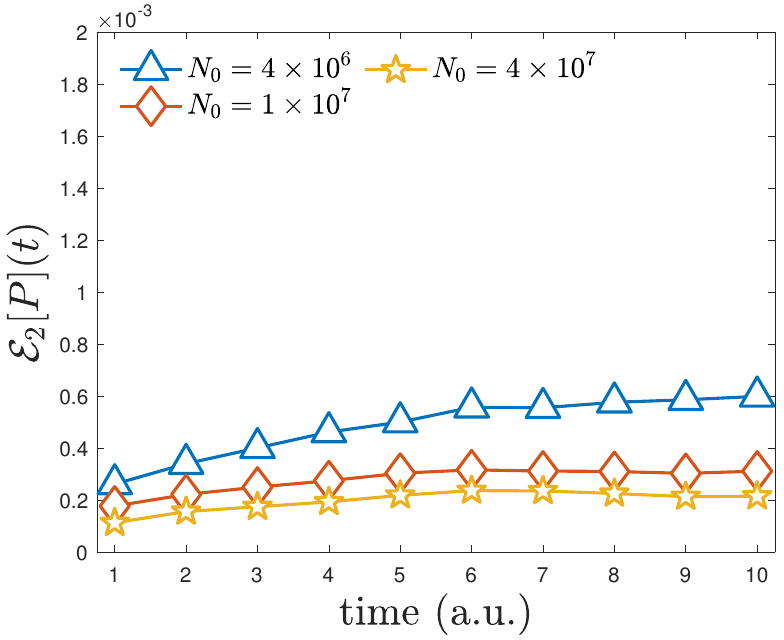}}}
\\
 \subfigure[$\vartheta = 0.005$.]{
{\includegraphics[width=0.48\textwidth,height=0.27\textwidth]{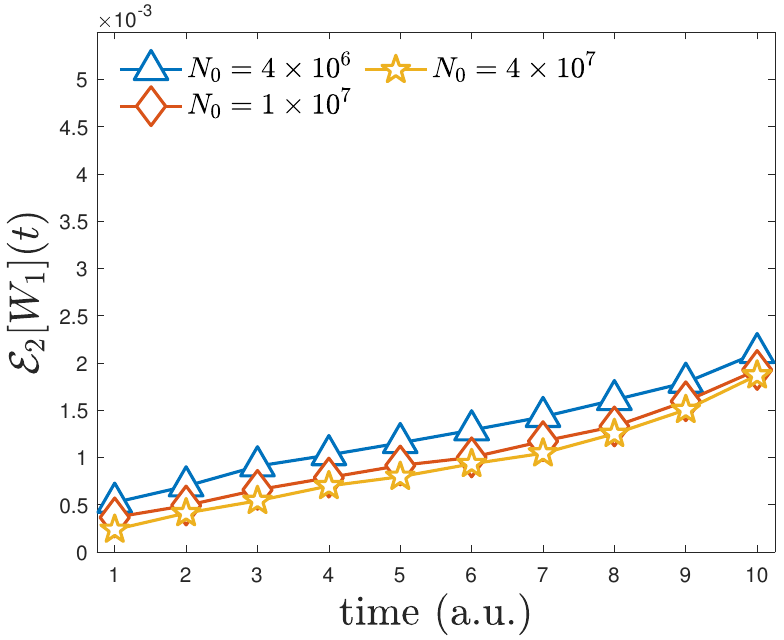}}
{\includegraphics[width=0.48\textwidth,height=0.27\textwidth]{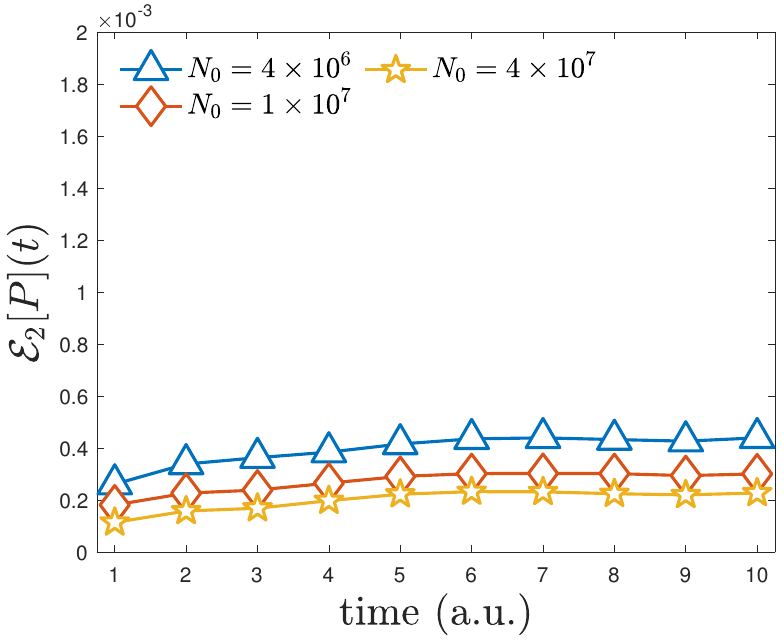}}}
\\
 \subfigure[$\vartheta = 0.01$.]{
{\includegraphics[width=0.48\textwidth,height=0.27\textwidth]{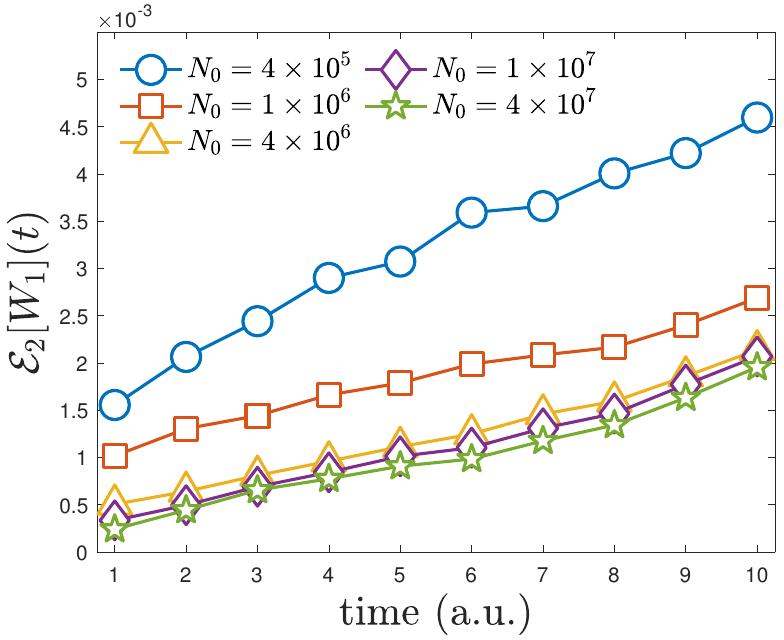}}
{\includegraphics[width=0.48\textwidth,height=0.27\textwidth]{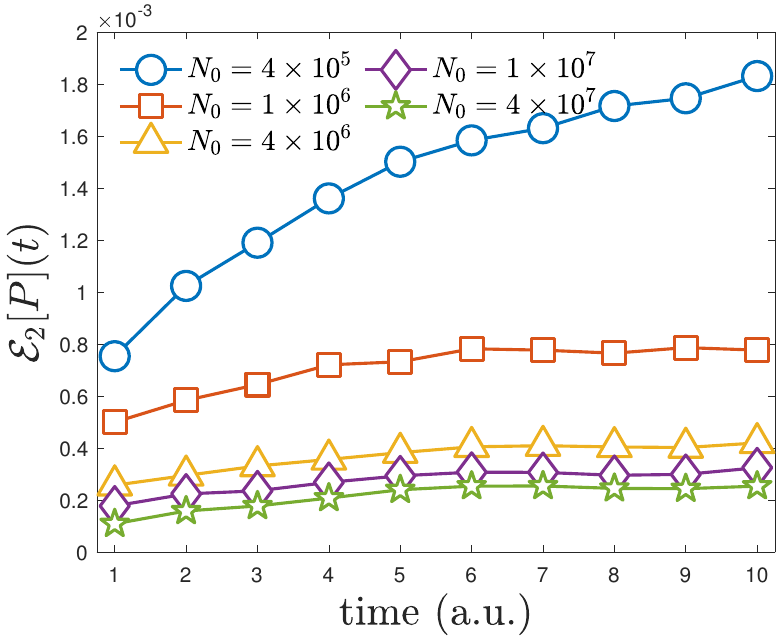}}}
\caption{\small The 4-D Morse system: $l^2$-errors under different $N_0$ (left: reduced Wigner function, right: spatial distribution). The accuracy of SPADE is improved as the sample size increases. }
\label{supp_2dm_spade_N}
\end{figure}

\begin{figure}[!h]
 \centering
 \subfigure[Convergence with respect to $N_0$.\label{supp_convergence_N}]{
{\includegraphics[width=0.48\textwidth,height=0.27\textwidth]{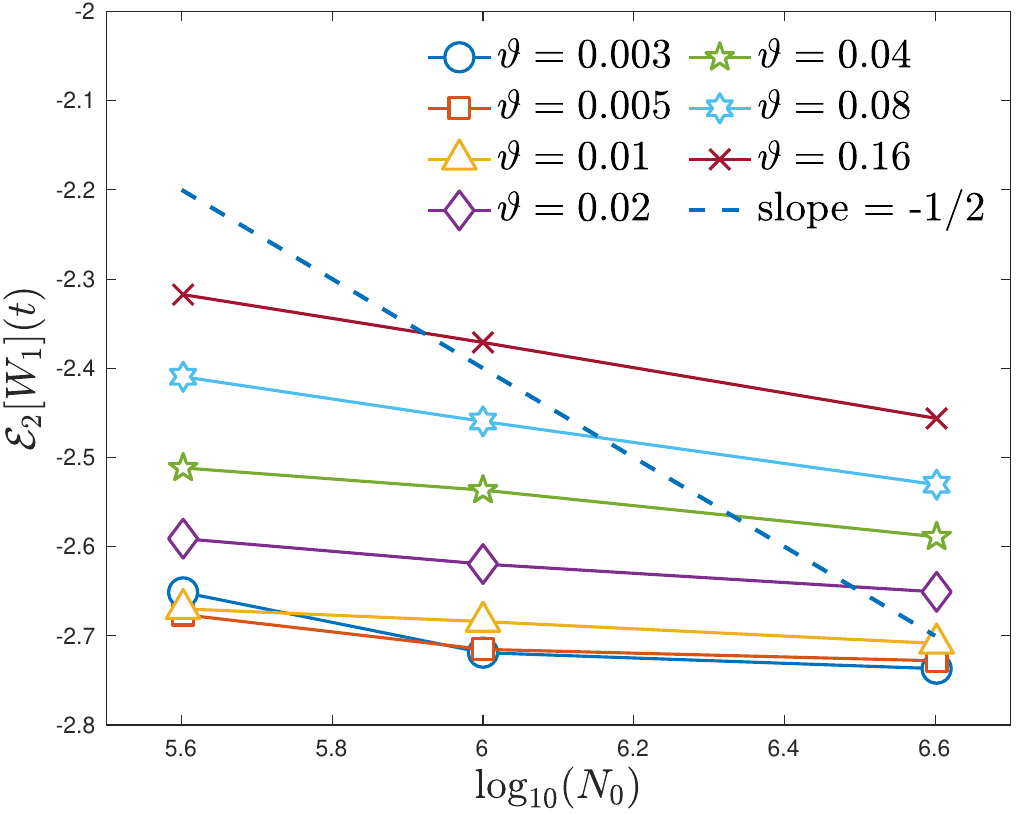}}}
\subfigure[The relation between $N_0$ and $K$. \label{supp_K_and_N}]{
{\includegraphics[width=0.48\textwidth,height=0.27\textwidth]{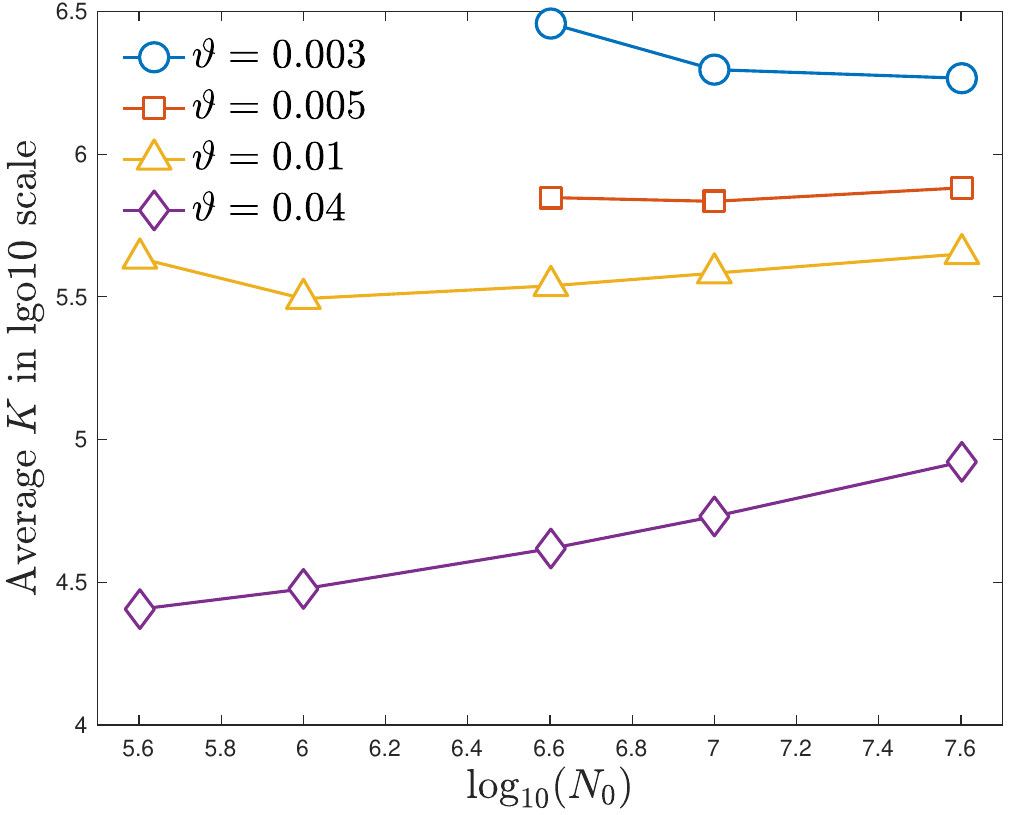}}}
\caption{\small  The 4-D Morse system: The convergence  with respect to $N_0$ and the partition level.  }
\label{supp_2dm_partition_N}
\end{figure}

The accuracy of SPADE can also be improved by increasing the effective sample size $N_0$. Five groups of simulations are performed under the sample size $N_0= 4\times 10^5$, $1\times 10^6$, $4\times 10^6$,  $1\times 10^7$ and $4\times 10^7$. The time evolution of $l^2$-errors is plotted in Figure \ref{supp_2dm_spade_N} and the partition level $K$ is recorded in Figure \ref{supp_2dm_partition_N}.

 {\bf Convergence with respect to $N_0$}: According to Figure \ref{supp_2dm_spade_N}, the numerical accuracy can be systematically improved by increasing $N_0$ from $2\times 10^5$ to $4\times 10^7$, which validates the convergence of stochastic Wigner algorithm.  However, the convergence rate largely deviates from $-1/2$ as seen in Figure \ref{supp_convergence_N} due to the mixture of MC errors and bias induced by SPADE. Again,  the deviation in total energy can be suppressed when $N_0$ becomes larger.

 {\bf Partition level $K$ with respect to $N_0$}:  As shown in Figure \ref{supp_K_and_N}, SPADE can work under a wide spectrum of sample sizes. It deserves to mention that too small $\vartheta$ is NOT recommended to be used when sample size is not large, as the partition level may increase rapidly and even exceed the sample size, and consequently leads to the overfitting problem and hamper the efficiency of SPADE (see the group $N_0=4\times 10^6$, $\vartheta = 0.003$ in Figure \ref{supp_convergence_K}).

\section{Performance evaluation of SPADE in 6-D phase space}
\label{supp_6d_wigner}

\begin{figure}[!h]
\centering
\subfigure[$W_1(x_1, k_1, t)$ (left) and $W_3(x_3, k_3, t)$ (right) at $t=1$a.u.]{
{\includegraphics[width=0.32\textwidth,height=0.17\textwidth]{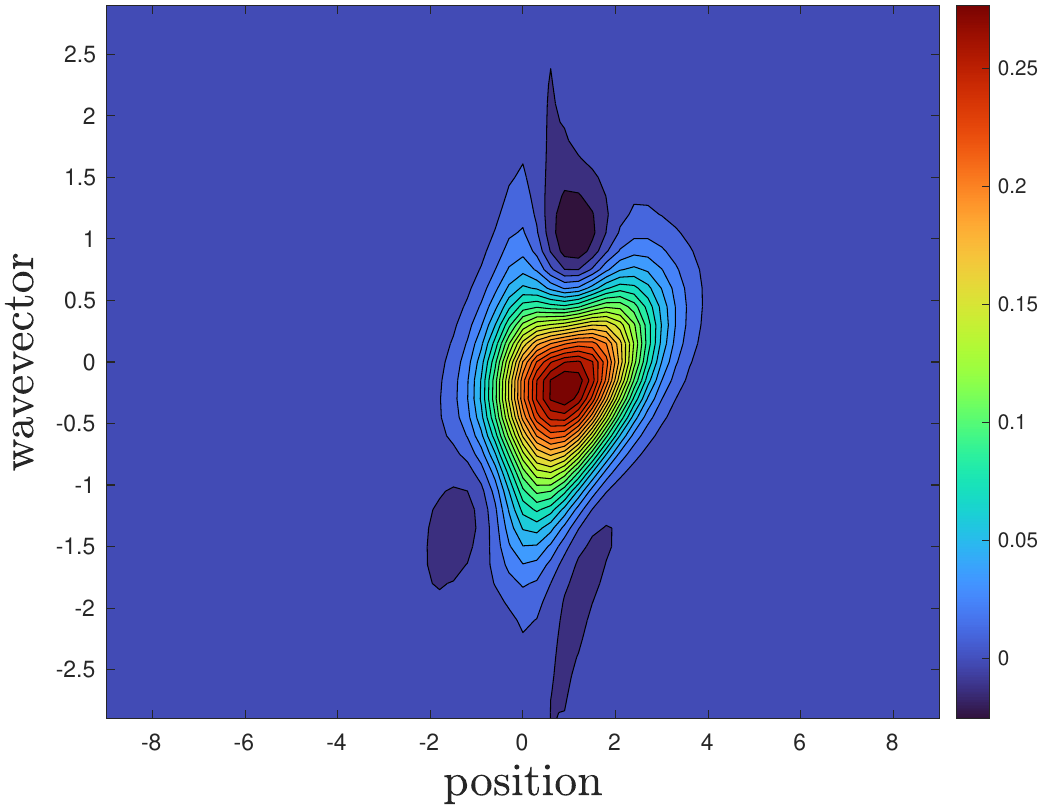}}
{\includegraphics[width=0.32\textwidth,height=0.17\textwidth]{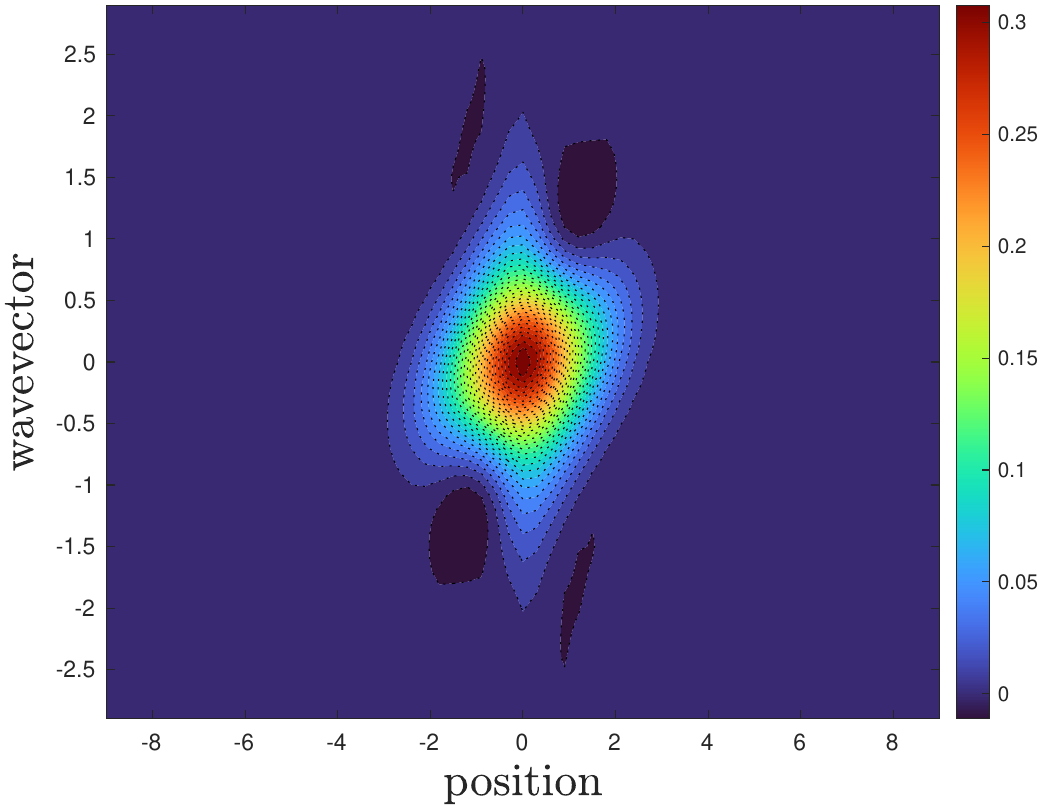}}}
\subfigure[$P_{xy}(x_1, x_2, t)$ at $t=0.5$a.u.\label{supp_xdist_t005}]{
{\includegraphics[width=0.32\textwidth,height=0.17\textwidth]{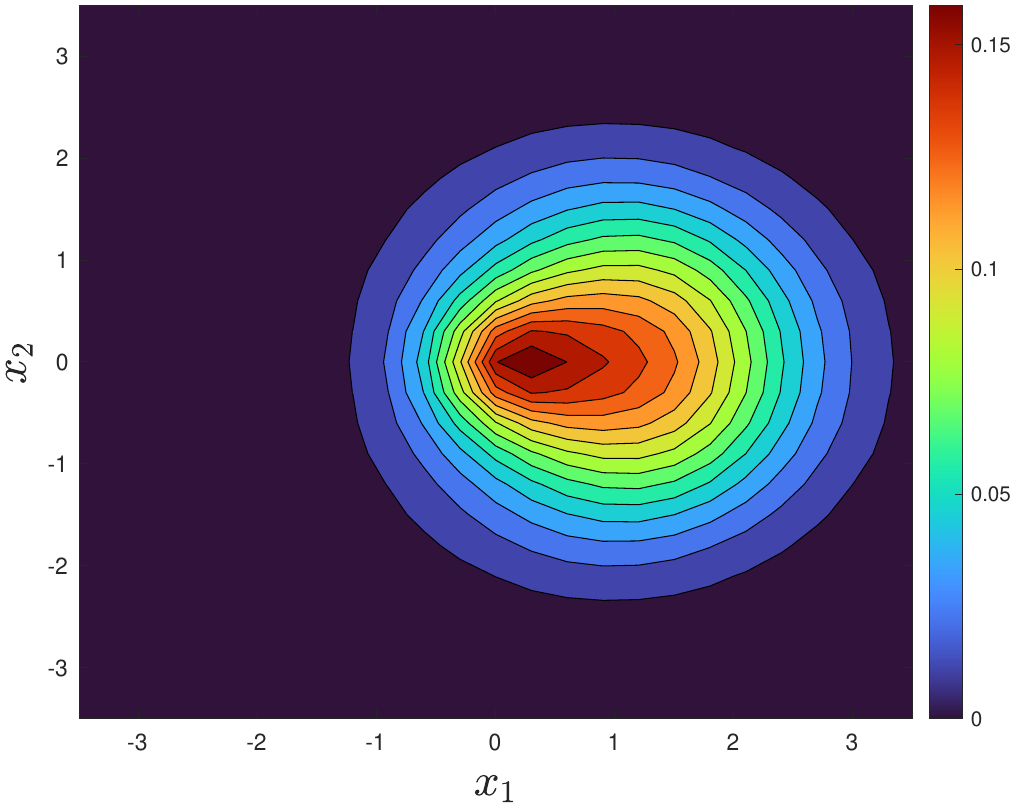}}}
\centering
\subfigure[$W_1(x_1, k_1, t)$ (left) and $W_3(x_3, k_3, t)$ (right) at $t=2$a.u.]{
{\includegraphics[width=0.32\textwidth,height=0.17\textwidth]{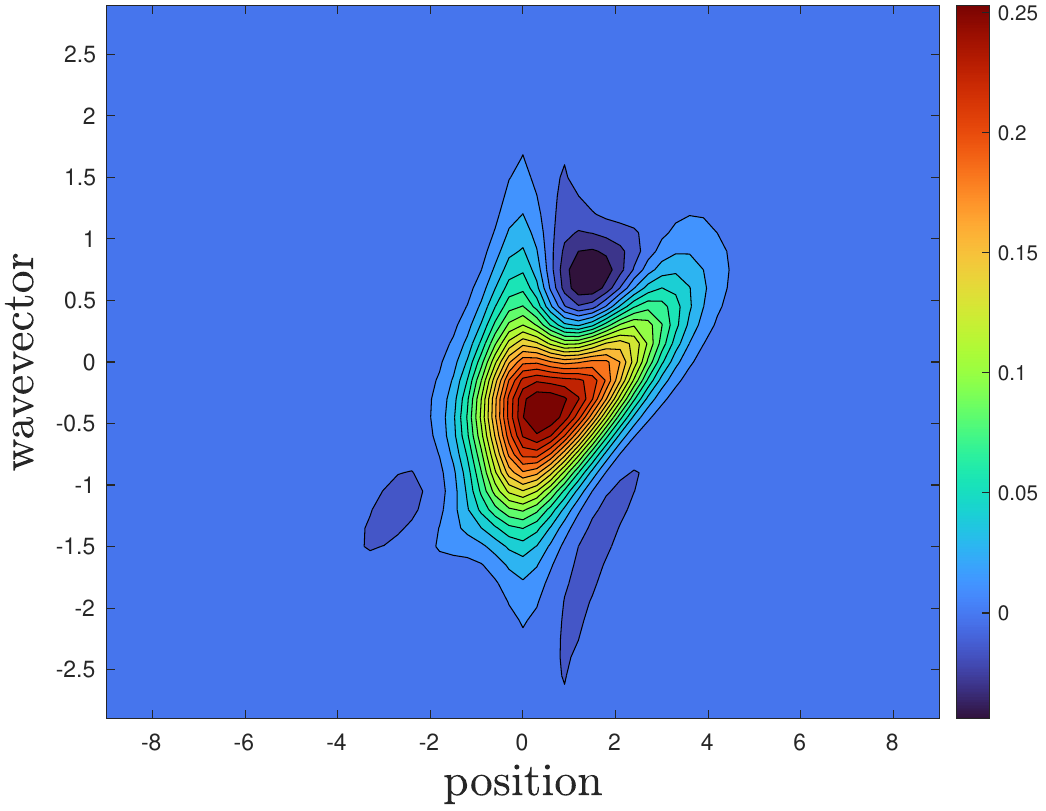}}
{\includegraphics[width=0.32\textwidth,height=0.17\textwidth]{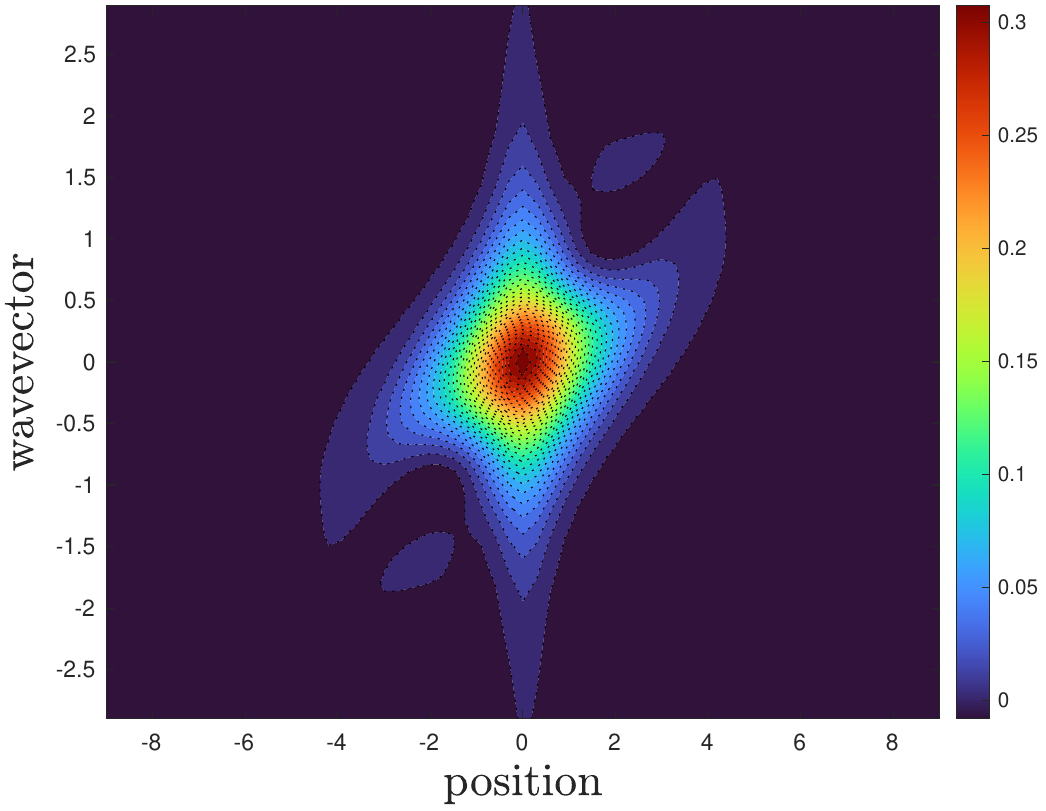}}}
\subfigure[$P_{xy}(x_1, x_2, t)$ at $t=1$a.u.]{
{\includegraphics[width=0.32\textwidth,height=0.17\textwidth]{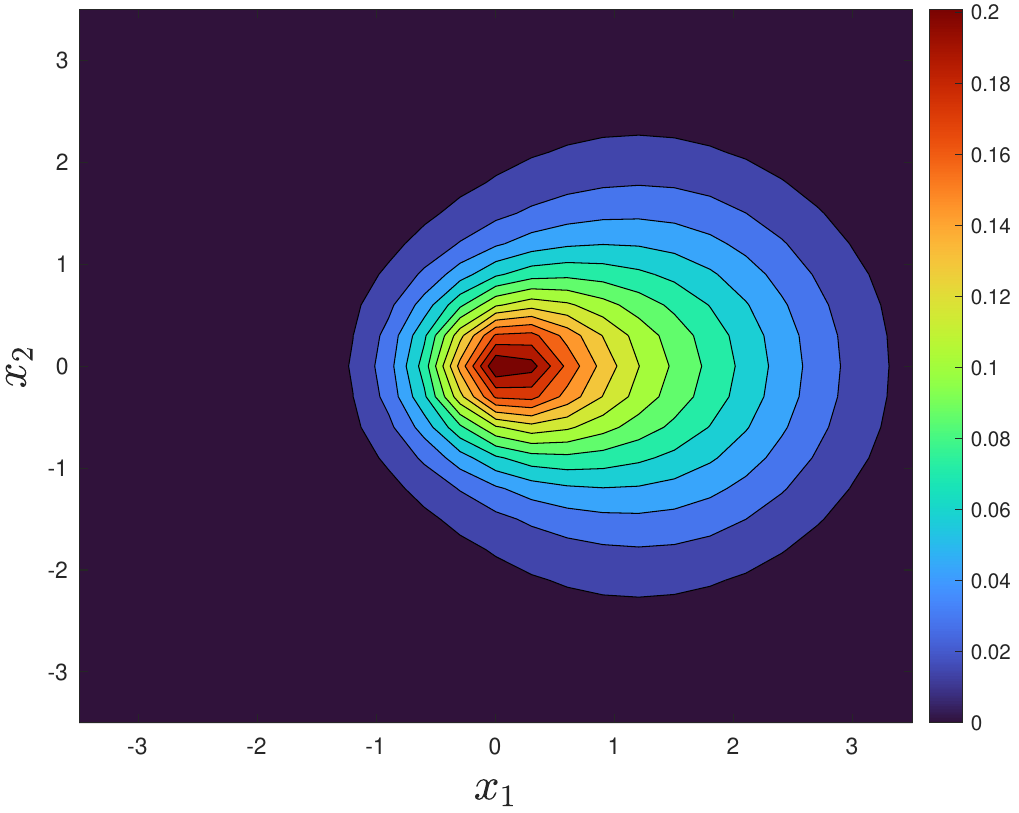}}}
\\
\centering
\subfigure[$W_1(x_1, k_1, t)$ (left) and $W_3(x_3, k_3, t)$ (right) at $t=4$a.u.]{
{\includegraphics[width=0.32\textwidth,height=0.17\textwidth]{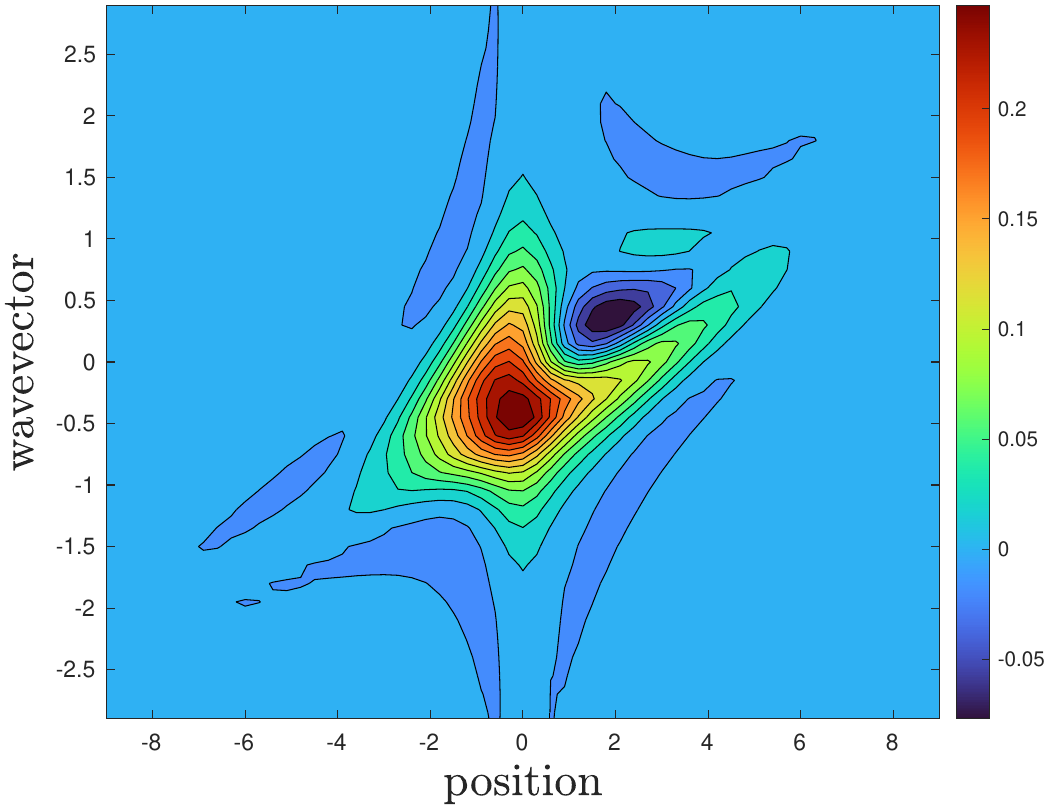}}
{\includegraphics[width=0.32\textwidth,height=0.17\textwidth]{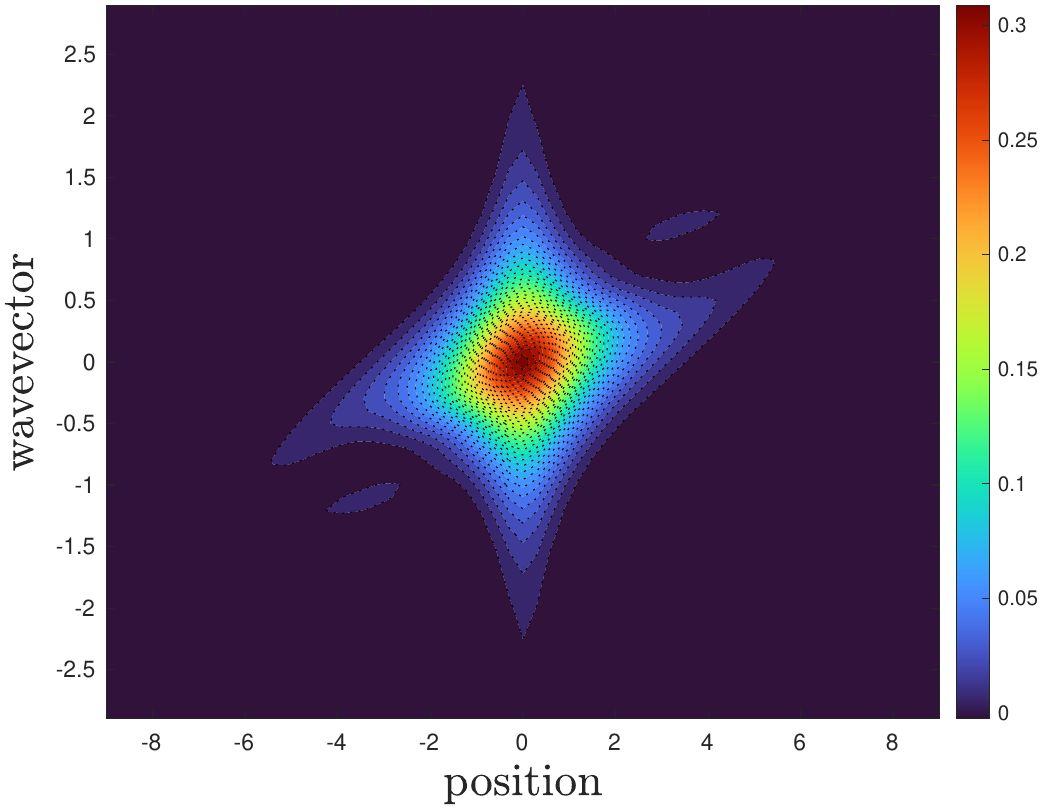}}}
\subfigure[$P_{xy}(x_1, x_2, t)$ at $t=2$a.u.]{
{\includegraphics[width=0.32\textwidth,height=0.17\textwidth]{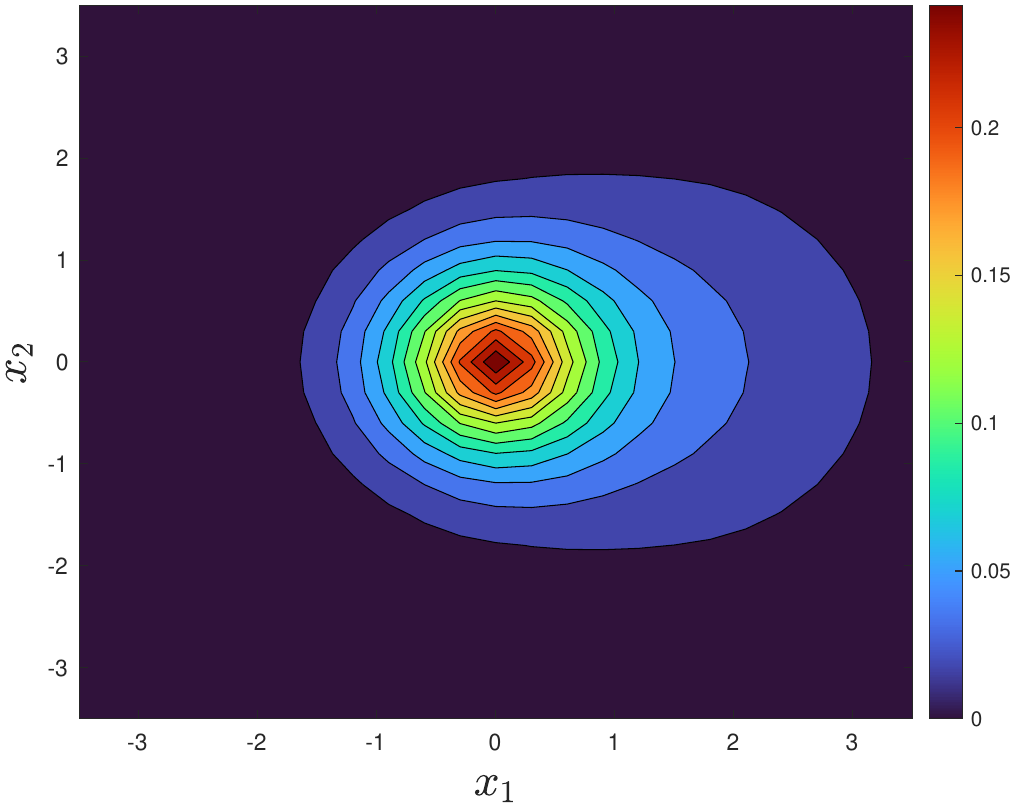}}}
\\
\centering
\subfigure[$W_1(x_1, k_1, t)$ (left) and $W_3(x_3, k_3, t)$ (right) at $t=8$a.u.]{
{\includegraphics[width=0.32\textwidth,height=0.17\textwidth]{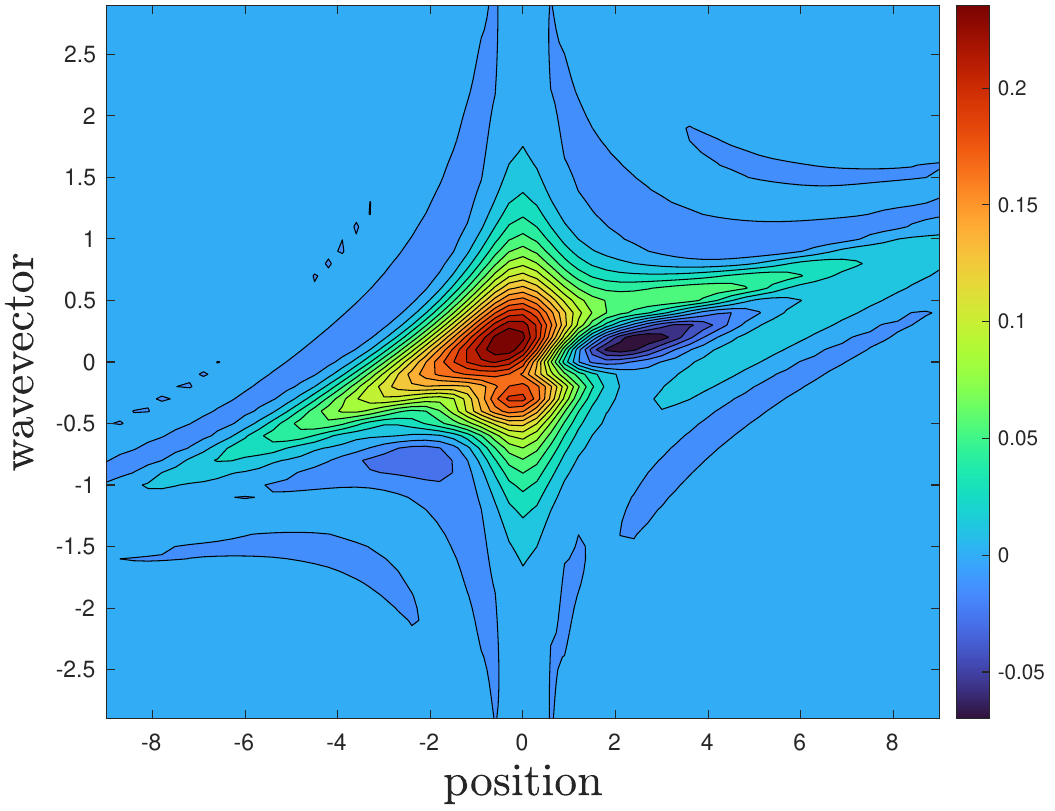}}
{\includegraphics[width=0.32\textwidth,height=0.17\textwidth]{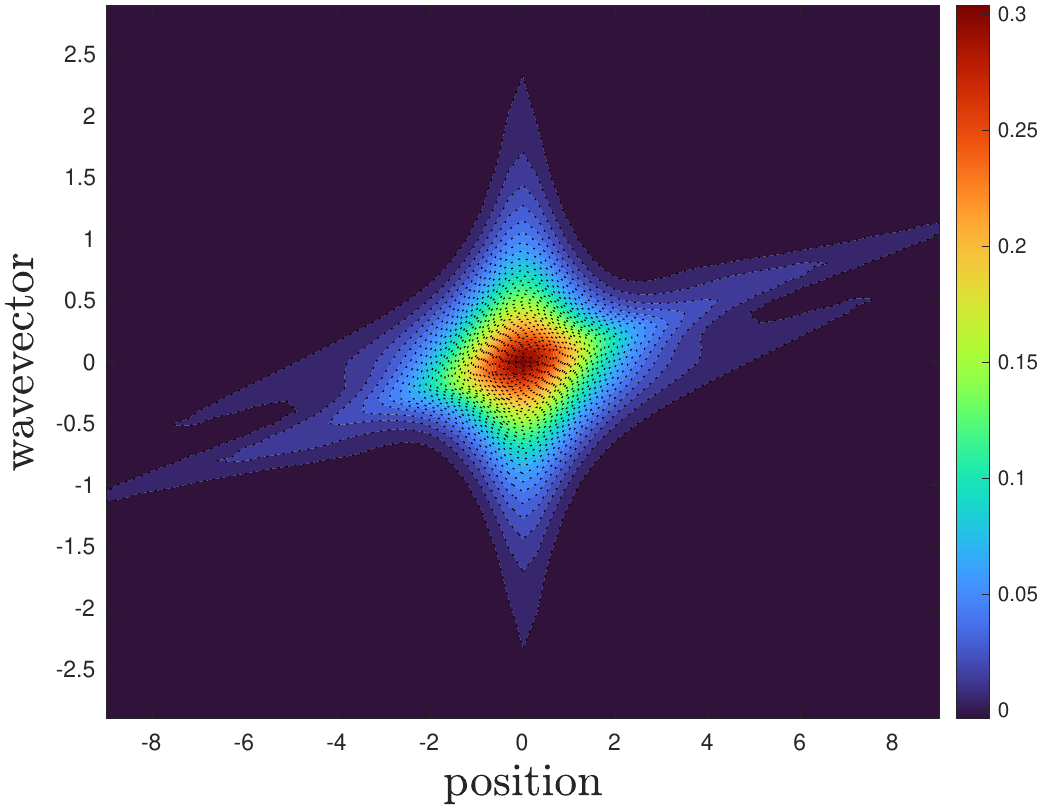}}}
\subfigure[$P_{xy}(x_1, x_2, t)$ at $t=5$a.u.]{
{\includegraphics[width=0.32\textwidth,height=0.17\textwidth]{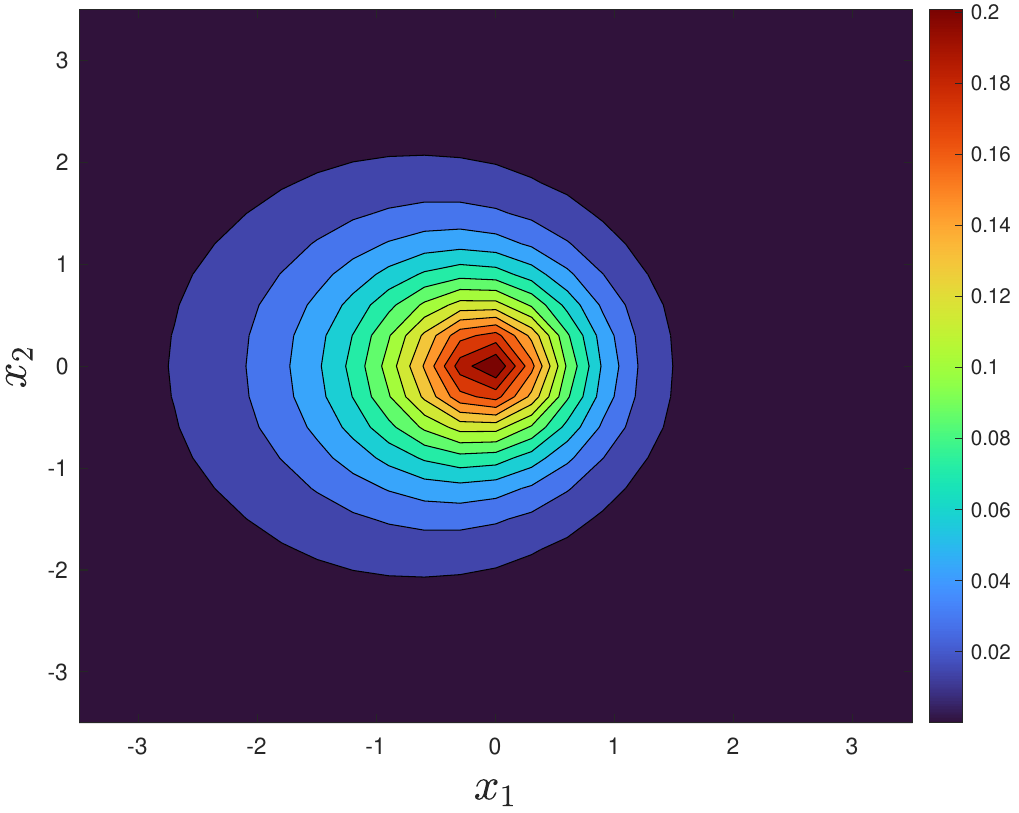}}}
\\
\centering
\subfigure[$W_1(x_1, k_1, t)$ (left) and $W_3(x_3, k_3, t)$ (right) at $t=12$a.u.]{
{\includegraphics[width=0.32\textwidth,height=0.17\textwidth]{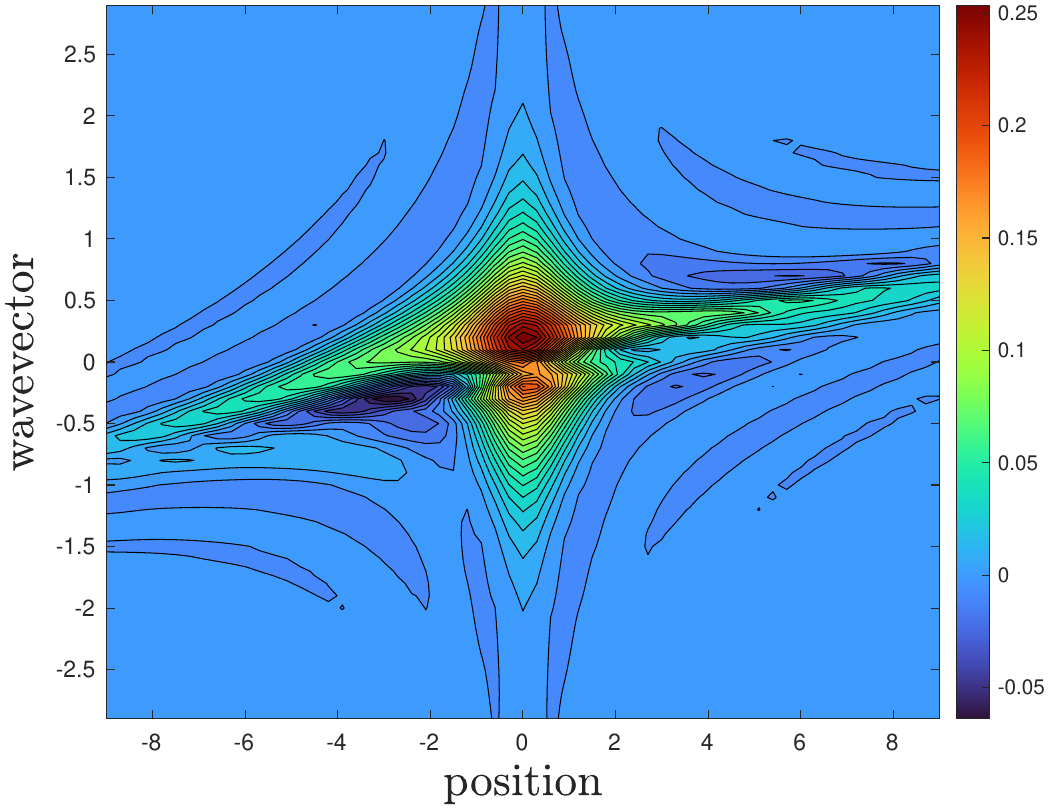}}
{\includegraphics[width=0.32\textwidth,height=0.17\textwidth]{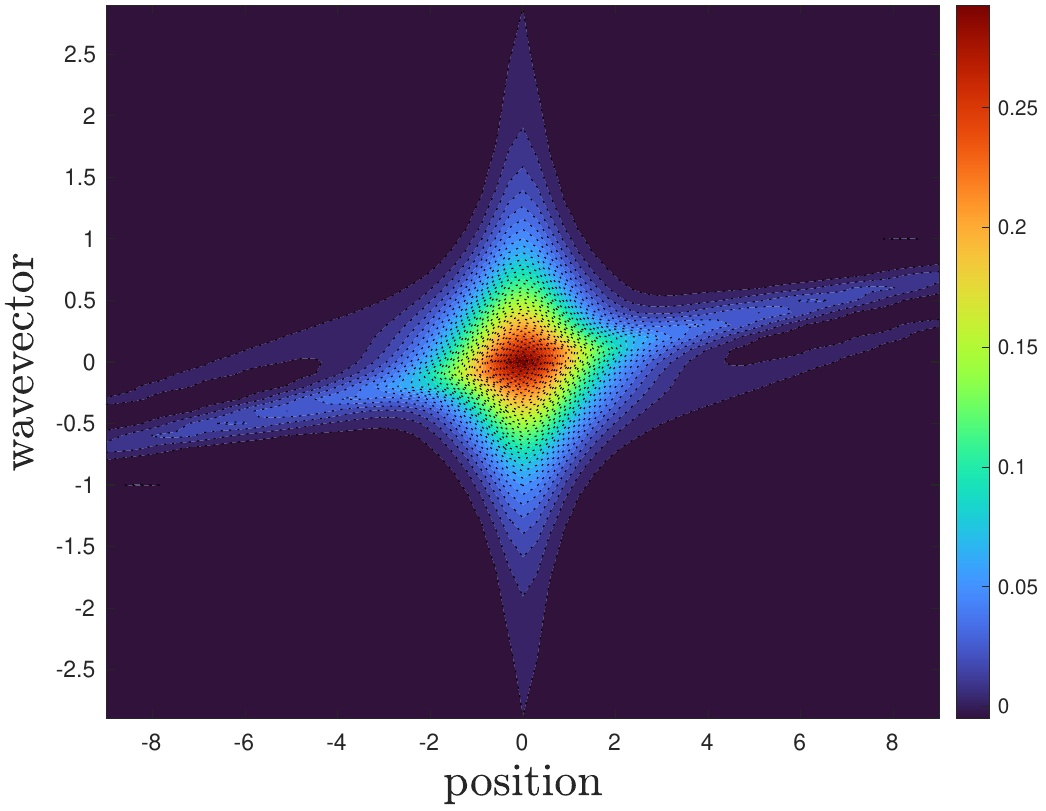}}}
\subfigure[$P_{xy}(x_1, x_2, t)$ at $t=10$a.u.]{
{\includegraphics[width=0.32\textwidth,height=0.17\textwidth]{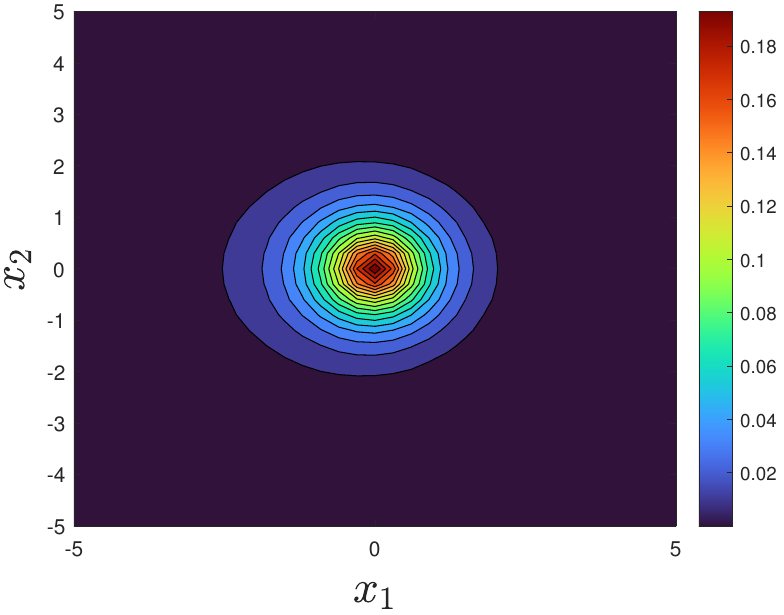}}}
\\
\centering
\subfigure[$W_1(x_1, k_1, t)$ (left) and $W_3(x_3, k_3, t)$ (right) at $t=15$a.u.]{
{\includegraphics[width=0.32\textwidth,height=0.17\textwidth]{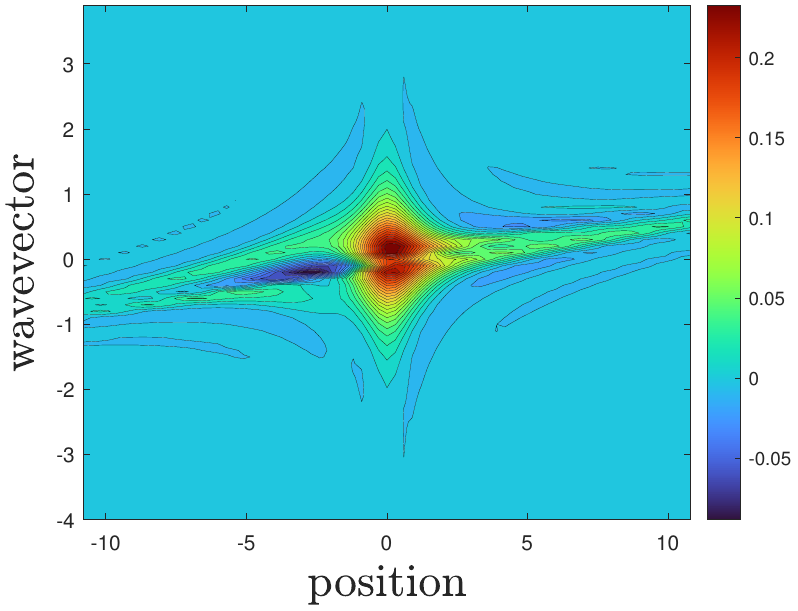}}
{\includegraphics[width=0.32\textwidth,height=0.17\textwidth]{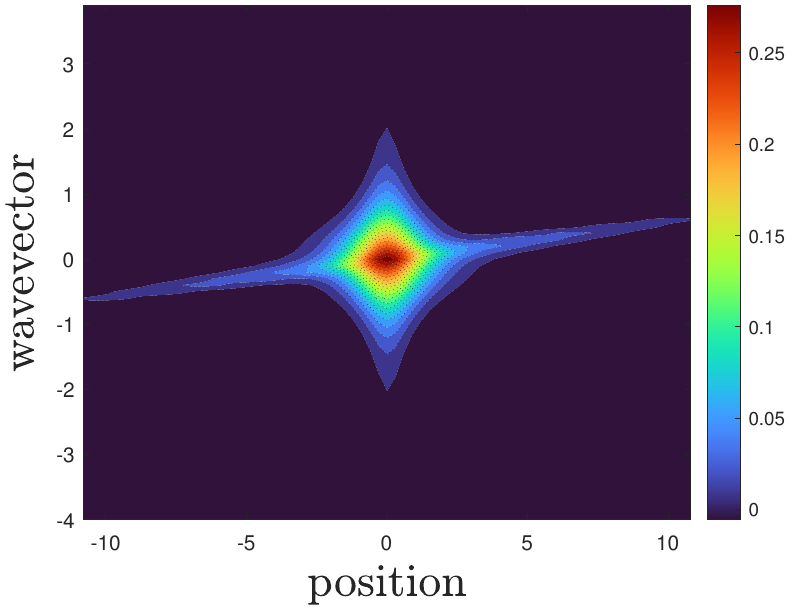}}}
\subfigure[$P_{xy}(x_1, x_2, t)$ at $t=15$a.u.]{
{\includegraphics[width=0.32\textwidth,height=0.17\textwidth]{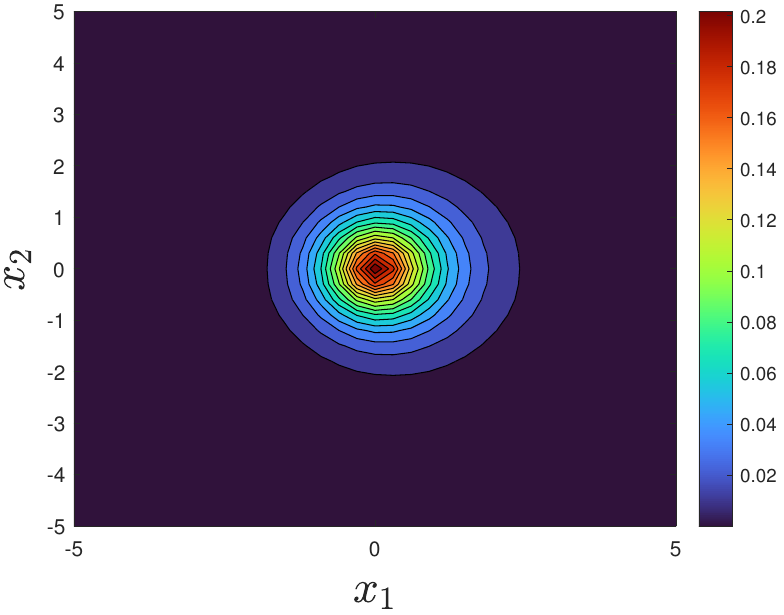}}}
\caption{\small  Electron-proton interaction:  Snapshots of the reduced Wigner functions on $(x_1$-$k_1)$ plane (left) and on $(x_2$-$k_2)$ plane (middle), the spatial marginal distribution (right) produced by the deterministic characteristic-spectral-mixed scheme. \label{supp_qcs_time_evolution}}
\end{figure}

Now we turn to the simulation of the single-body Wigner-Coulomb equation \eqref{supp_single_body_Wigner} with $\pdo$ \eqref{supp_pdo_scatter}, with the initial condition
\begin{equation}
f_e(\bx, \bk, 0) = \frac{1}{\pi^{3}} \me^{-\frac{1}{2}|\bx - R|^2} \me^{-2|\bk|^2}, \quad R = (1, 0, 0).
\end{equation} 
The snapshots of the reduced Wigner functions $W_1(x_1, k_1, t)$, $W_3(x_3, k_3, t)$ and the spatial marginal density $P_{xy}(x_1, x_2, t)$, produced by a deterministic characteristic-spectral-mixed scheme \cite{XiongZhangShao2023}, are plotted in Figure \ref{supp_qcs_time_evolution}, where
\begin{equation}
\begin{split}
&W_1(x_1, k_1, t) = \iiiint_{\mathbb{R}^{2} \times \mathbb{R}^{2}} f(\bx, \bk, t) \D x_2 \D x_3  \D k_2 \D k_3, \\
&W_3(x_3, k_3, t) = \iiiint_{\mathbb{R}^{2} \times \mathbb{R}^{2}} f(\bx, \bk, t) \D x_1 \D x_2  \D k_2 \D k_3, 
\end{split}
\end{equation}
and the spatial marginal distribution $P_{xy}$ projected onto $(x_1$-$x_2$) plane is 
\begin{equation}
P_{xy}(x_1, x_2, t) = \iint_{\mathbb{R} \times \mathbb{R}^{3}} f(\bx, \bk, t) \D x_{3} \D \bk.
\end{equation}

The performance metrics include the  $l^2$-errors $\mathcal{E}_{2}[W_1](t)$ and $\mathcal{E}_{2}[P_{xy}](t)$ to monitor the stochastic variances,
\begin{equation}
\begin{split}
\mathcal{E}_{2}[W_1](t) &= \{\frac{1}{N_x N_k}\sum_{i=1}^{N_x} \sum_{j=1}^{N_k} (W_1^{\textup{ref}}(x_{1}^{(i)}, k_{1}^{(j)},t)- W_1^{\textup{num}}(x_{1}^{(i)}, k_{1}^{(j)},t))^2\}^{1/2}, \\
\mathcal{E}_{2}[P_{xy}](t) &= \{\frac{1}{N_x^2}\sum_{i=1}^{N_x} \sum_{j=1}^{N_x} (P_{xy}^{\textup{ref}}(x_{1}^{(i)}, x_{2}^{(j)},t)- P_{xy}^{\textup{num}}(x_{1}^{(i)}, x_{2}^{(j)}, t))^2\}^{1/2},
\end{split}
\end{equation}
as well as the deviation of total Hamiltonian $\mathcal{E}_{\textup{H}}(t)$. Here $W_1^{\textup{ref}}$ and $W_1^{\textup{num}}$ denote the reference and stochastic solution for $W_1$, respectively (similar for $P_{xy}$).

Our subsequent simulations are organized as follows. Several parameters that may influence the accuracy of the stochastic Wigner simulations are investigated. A good filter $\lambda_0$ is crucial in suppressing the stochastic variances. The choice of gap functions and the parameter $m$ in approximating gaps determines the accuracy and efficiency of SPADE. 
\begin{description}

\item[(1)] Section \ref{supp_sec.filter} investigates the impact  of the parameter $\lambda_0$ in SPA on the overall accuracy. 

\item[(2)] Section \ref{supp_sec.split} investigates the impact of the parameter $m$ in SPADE on the overall accuracy. 

\item[(3)] Section \ref{supp_sec.PAUM} performs a benchmark on PAUM in 6-D simulations.

\item[(4)] Section \ref{supp_sec.gap} compares two gap functions in SPADE.

\end{description}

\subsection{How to choose the filter $\lambda_0$ in SPA}
\label{supp_sec.filter}

The filter $\lambda_0$ in SPA deserves a careful investigation because the accuracy of particle method is  limited by the Monte Carlo sampling errors. Thus optimization of sampling process is always the first step. 

\begin{figure}[!h]
\subfigure[$l^2$-errors of the reduced Wigner function $W_1$ (left: up to $T=6$a.u., right: up to $T=20$a.u.) \label{supp_err_redist_msplit}]{
{\includegraphics[width=0.49\textwidth,height=0.26\textwidth]{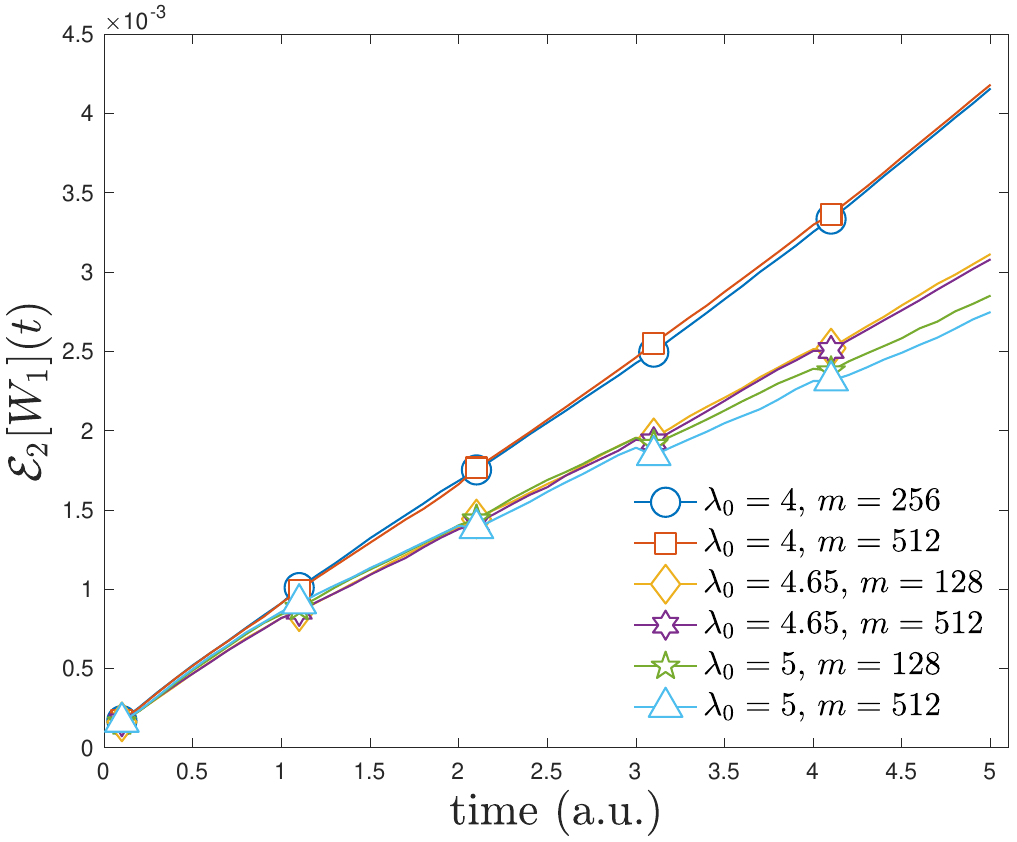}}
{\includegraphics[width=0.49\textwidth,height=0.26\textwidth]{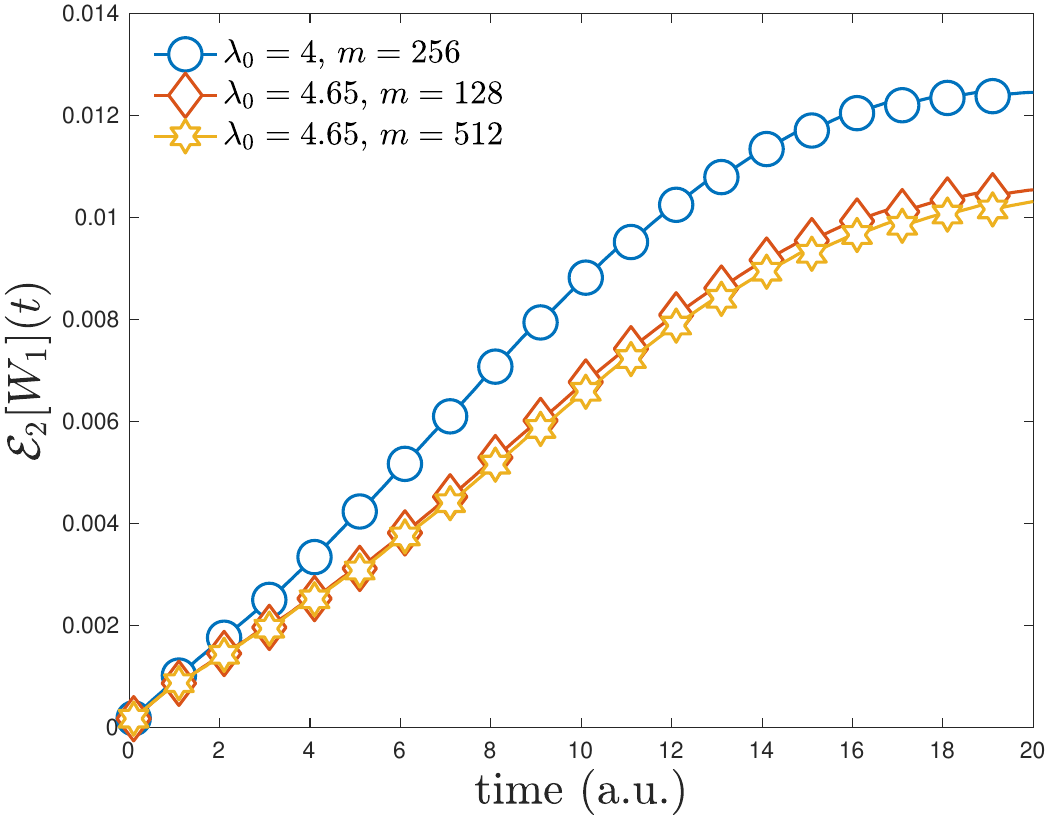}}}
\\
\subfigure[$l^2$-errors of the spatial distribution $P_{xy}$ (left: up to $T=6$a.u., right: up to $T=20$a.u.) \label{supp_err_xdist_msplit}]{
{\includegraphics[width=0.49\textwidth,height=0.26\textwidth]{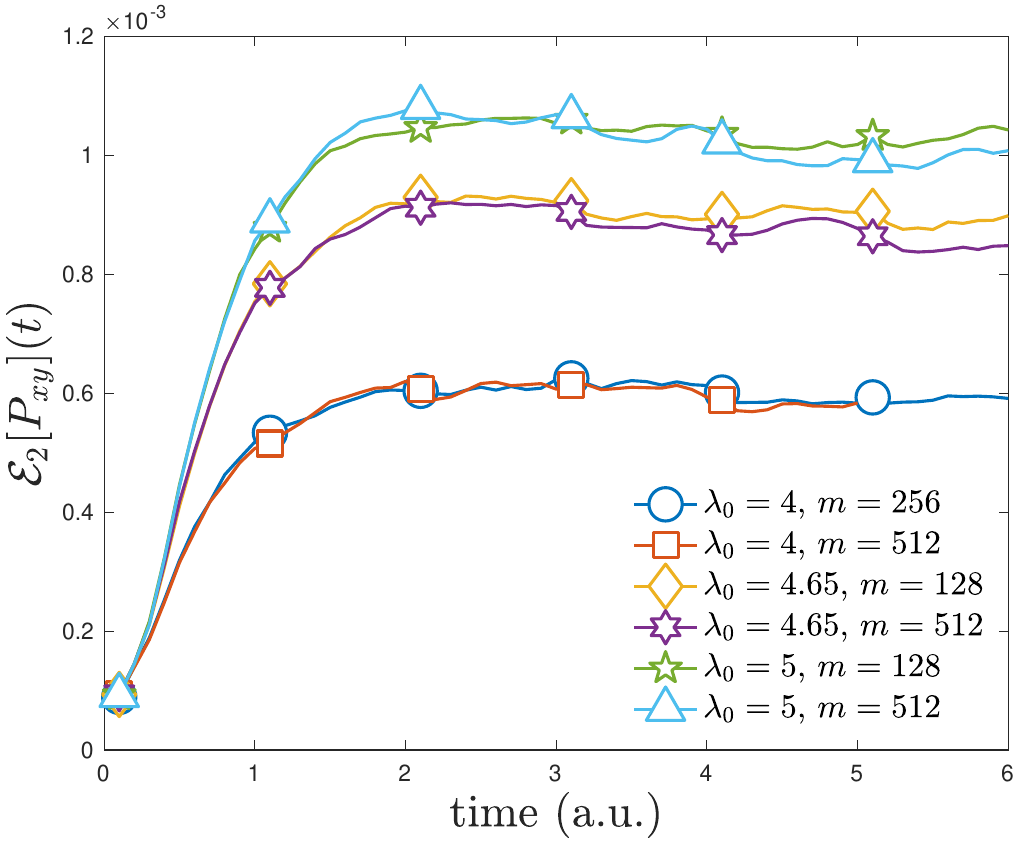}}
{\includegraphics[width=0.49\textwidth,height=0.26\textwidth]{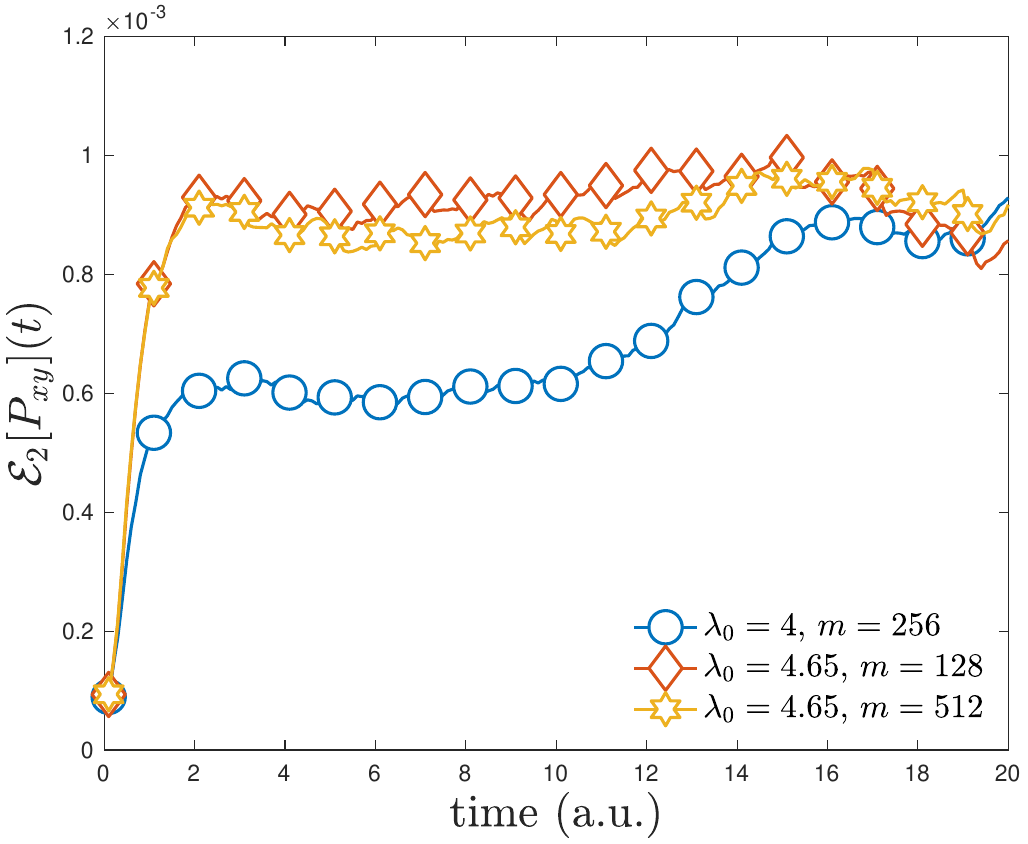}}}
\\
\subfigure[Energy fluctuation under $\lambda_0 = 4$.\label{supp_err_energy_l4}]{
{\includegraphics[width=0.49\textwidth,height=0.26\textwidth]{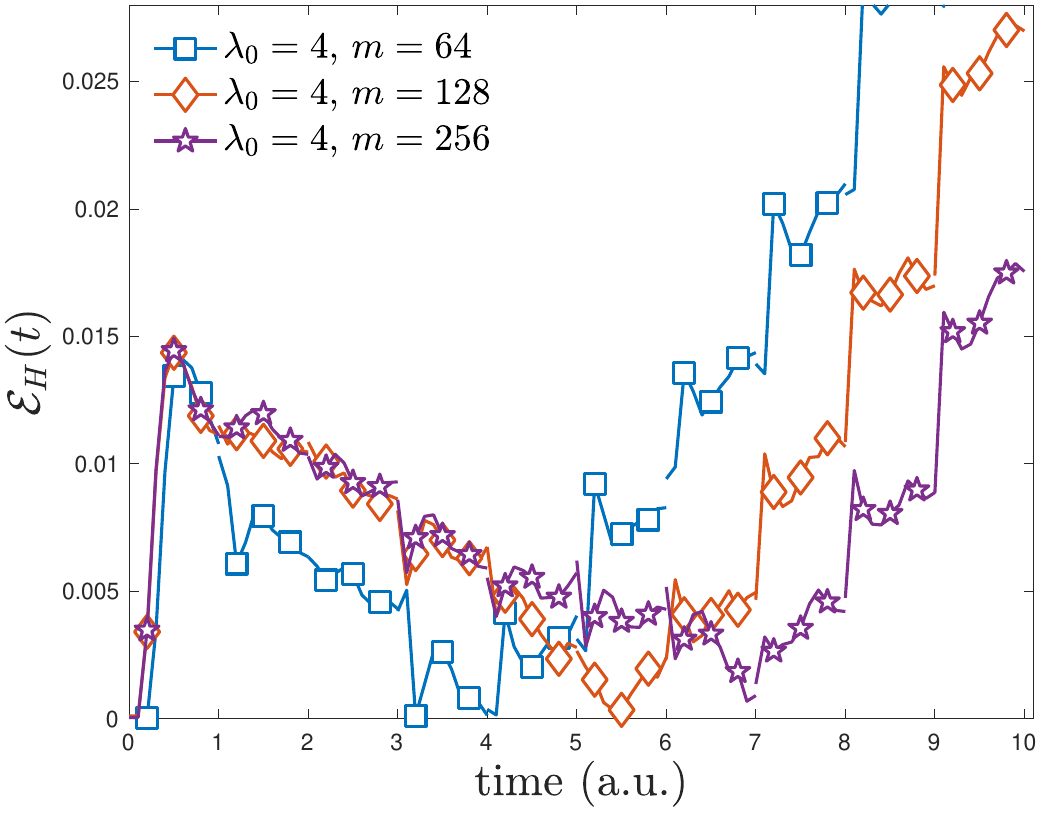}}}
\subfigure[Energy fluctuation under $\lambda_0 = 4.65$.\label{supp_err_energy_lopt}]{
{\includegraphics[width=0.48\textwidth,height=0.26\textwidth]{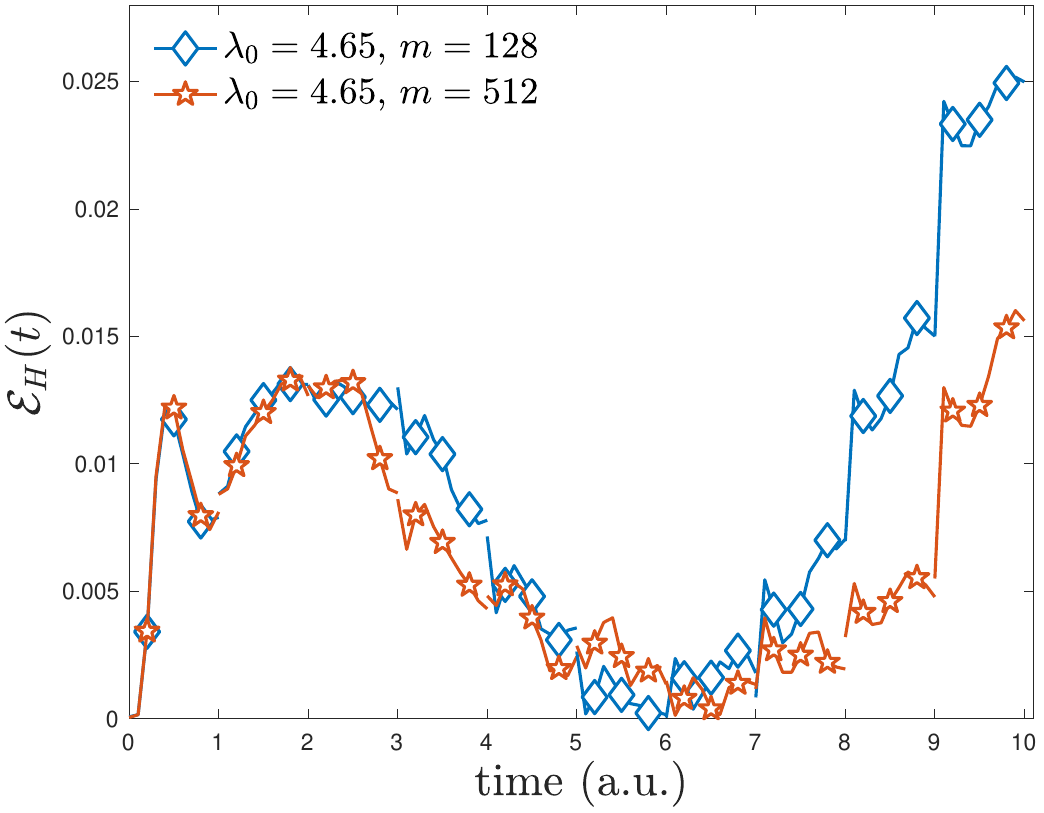}}}
\caption{\small A comparison of $l^2$-errors $\mathcal{E}_2[W_1](t)$, $\mathcal{E}_2[P_{xy}](t)$ and deviation of total energy under different parameters $\lambda_0$ in SPA and $m$ in SPADE. Here $N_0 = 4\times 10^7$, $\vartheta = 0.005$ and the difference gap is adopted.}
\end{figure}

In order to choose a good filter $\lambda_0$, we suggest to monitor the deviation of total energy and find that $\lambda_0 = 4.65$ is the optimal.  But we need to emphasize that {\bf this might only achieves a balanced accuracy}.  In Figures \ref{supp_err_redist_msplit} and \ref{supp_err_xdist_msplit}, it is shown that the $l^2$-error of the reduced Wigner function can be diminished by choosing a larger $\lambda_0$. On the contrary, the $l^2$-error of the spatial marginal distribution might be augmented as $\lambda_0$ increases. This is caused by the accumulation of sampling errors (see the curve in the time interval $0\le t \le1$ as particle annihilation is absent). Fortunately, for a long-time evolution up to $T= 20$a.u., both $W_1$ and $P_{xy}$ under $\lambda_0 = 4.65$ seem to be better than those under $\lambda_0 = 4$. Figures \ref{supp_err_energy_l4} and \ref{supp_err_energy_lopt} present the fluctuation of total energy up to $10$a.u. (before some particles move outside the domain). It is also verified $\lambda_0 = 4.65$ can achieve relatively less fluctuation of energy.

\begin{figure}[!h]
\subfigure[Partition level  $K(t)$ under different $\lambda_0$ and its relation with $\mathcal{N}^b(t)/\sqrt{N_0}$.\label{supp_Gbeam_K_lambda0}]{
{\includegraphics[width=0.49\textwidth,height=0.26\textwidth]{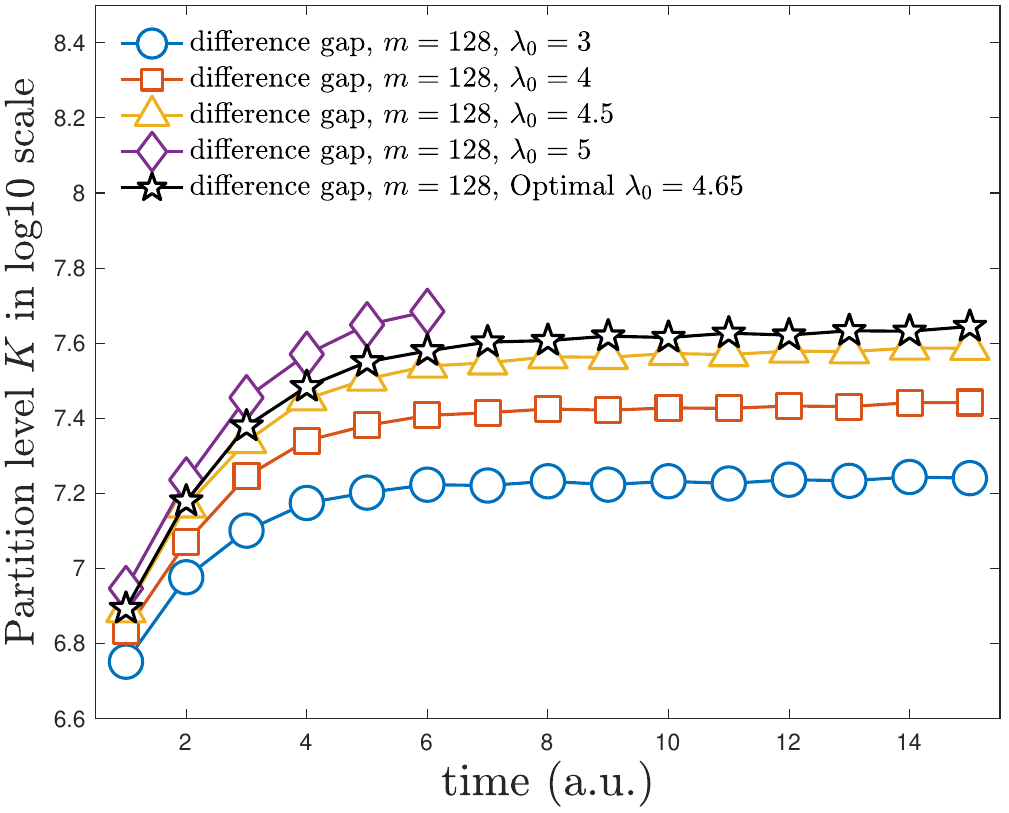}}
{\includegraphics[width=0.49\textwidth,height=0.26\textwidth]{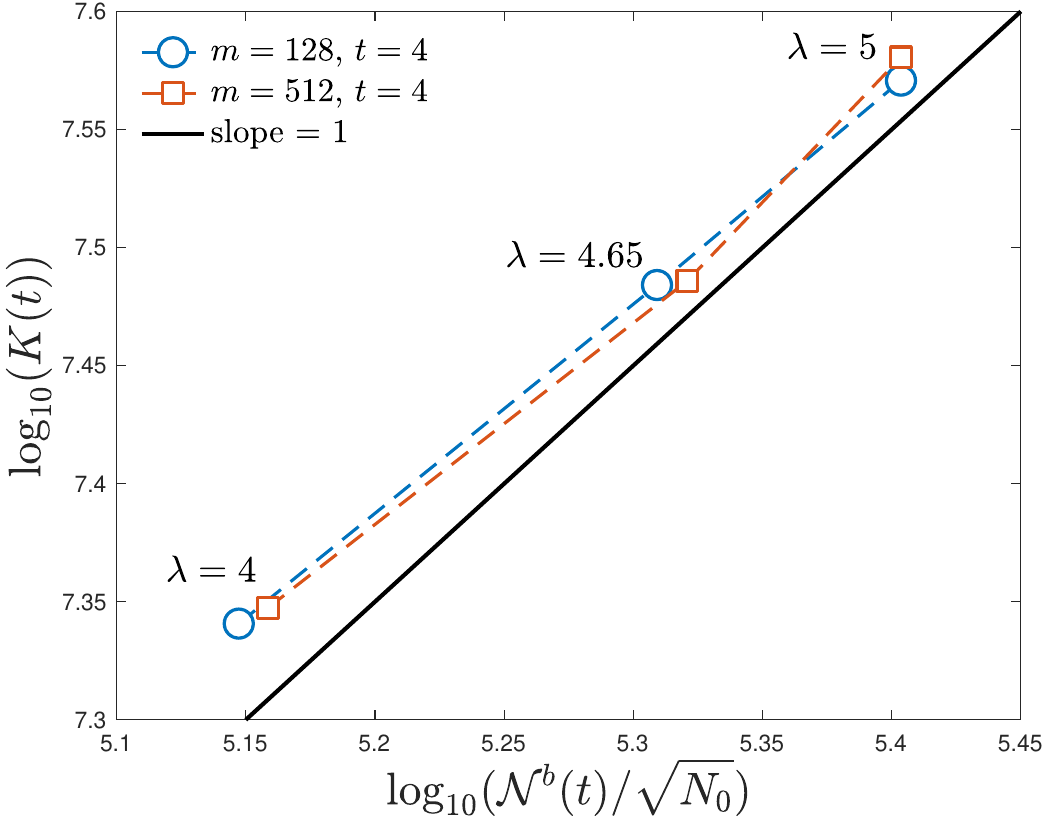}}}
\\
\centering
\subfigure[Partition level  $K(t)$ under different $m$ and its relation with $\mathcal{N}^b(t)/\sqrt{N_0}$.\label{supp_Gbeam_K_m}]{
{\includegraphics[width=0.49\textwidth,height=0.26\textwidth]{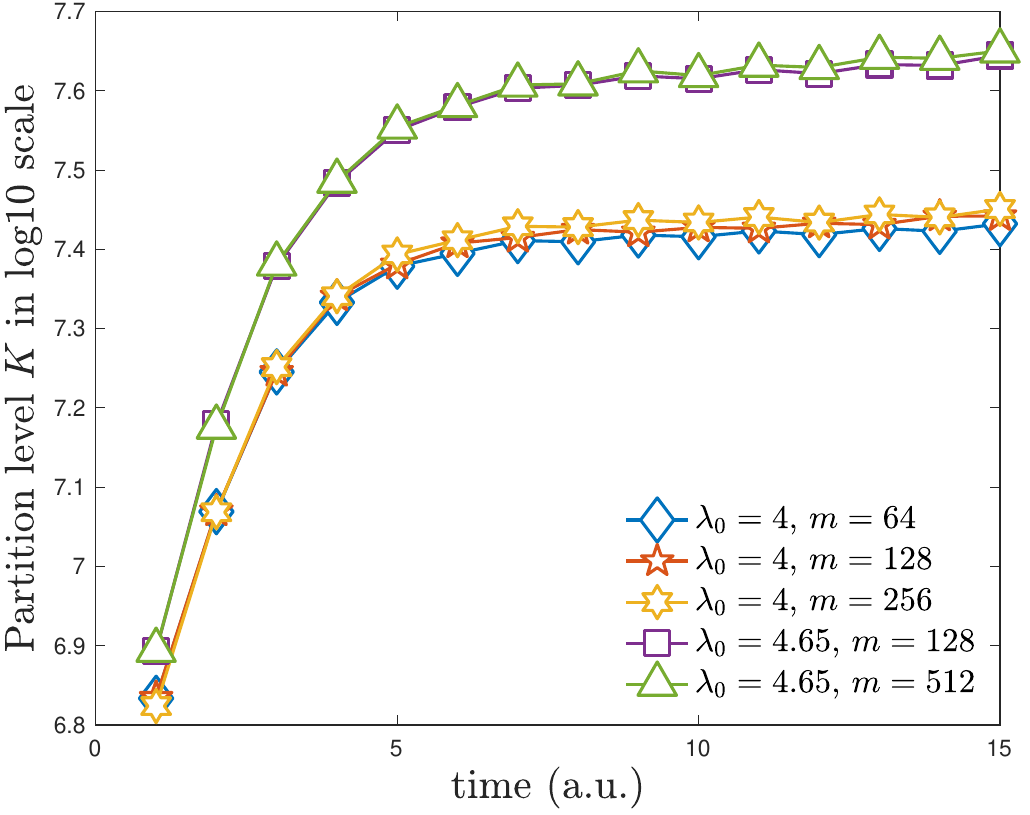}}
{\includegraphics[width=0.49\textwidth,height=0.26\textwidth]{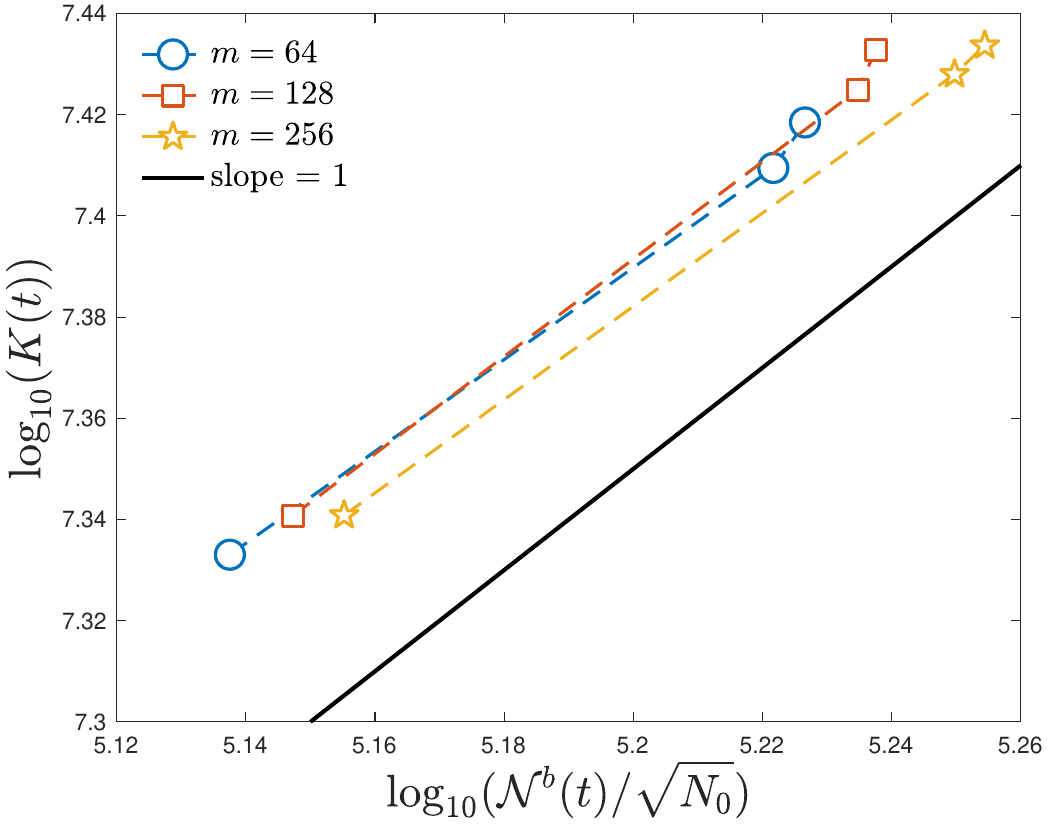}}}
\caption{\small The partition level $K(t)$ under different parameters $\lambda$ and $m$. Here $K$ is almost linearly proportional to $\mathcal{N}^b/\sqrt{N_0}$.\label{supp_comparison_SPADE_PAUM_N}}
\end{figure}

Figure \ref{supp_Gbeam_K_lambda0} presents the relation between partition level $K(t)$ and $\lambda_0$. When $\lambda_0$ increases, the partition level $K(t)$ also increases, which coincides with the growth of particle number. We also plot the relation between $K(t)$ and $\mathcal{N}^b(t)/\sqrt{N_0}$, where $\mathcal{N}^b(t) = P(t) + M(t)$ denotes the particle number before PA at time $t$.  We find that under different $\lambda_0$, $K(t)$ is linearly proportional to $\mathcal{N}^b(t)/\sqrt{N_0}$. This provides some evidence to support our lower bound \eqref{K_bound} of partition level $K$.

\subsection{How to choose the parameter $m$ in SPADE}
\label{supp_sec.split}

The next part is devoted to studying the parameter $m$ in the decision of cuts in SPADE, which determines how well the true gap function is approximated. The following observations are made.

\begin{description}

\item[(1)] The parameter $m$ has a great influence on the deviation of energy. From Figures \ref{supp_err_energy_l4} and \ref{supp_err_energy_lopt}, it is seen that {\bf the fluctuation of energy can be evidently suppressed by choosing a larger $m$}.

\item[(2)] From Figures \ref{supp_err_redist_msplit} and \ref{supp_err_xdist_msplit}, it is verified that the $l^2$-errors of $W_1$ and $P_{xy}$ can be slightly improved when larger $m$ is adopted.

\item[(3)] According to Figures \ref{supp_Gbeam_K_lambda0} and \ref{supp_Gbeam_K_m}, the parameter $m$ has only a slight influence on the partition level $K(t)$. Again  $K(t)$ seems to be linearly dependent on $\mathcal{N}^b(t)/\sqrt{N_0}$, regardless of the choice $m$.  This verifies the lower bound \eqref{K_bound} of $K$.

\end{description}

To conclude, choosing a sufficiently large $m$, e.g., $m\ge 512$, is highly desirable. 

\subsection{Accuracy of PAUM}
\label{supp_sec.PAUM}

We have tested PAUM under $N_0=1\times 10^8, 4\times 10^8, 1\times 10^9$ with fixed grid size $K  = 61^3 \times 60^3 \approx 4.9\times 10^{10}$. The results are collected in Figure \ref{supp_fig_6d_PAUM} below. PAUM still works in 6-D simulations provided that sample size is sufficiently large. However, for the group with $N_0 = 10^8$, PAUM might fail to annihilate particle very efficiently. The particle number after PAUM grows from $1\times 10^8$ initially to $6.4\times 10^{9}$ at $15$a.u. Meanwhile, it might not be able to capture the tail distribution of the Wigner function accurately due to inadequate sampling.

Fortunately, the oversampling problem in PAUM can be alleviated by simply increasing the sample size $N_0$, and the noises near the tail distribution can be suppressed. But the overall accuracy is still limited by the bias induced by the finite bin size.

\begin{figure}[!h]
    \subfigure[$l^2$-error for $W_1$.]{\includegraphics[width=0.32\textwidth,height=0.22\textwidth]{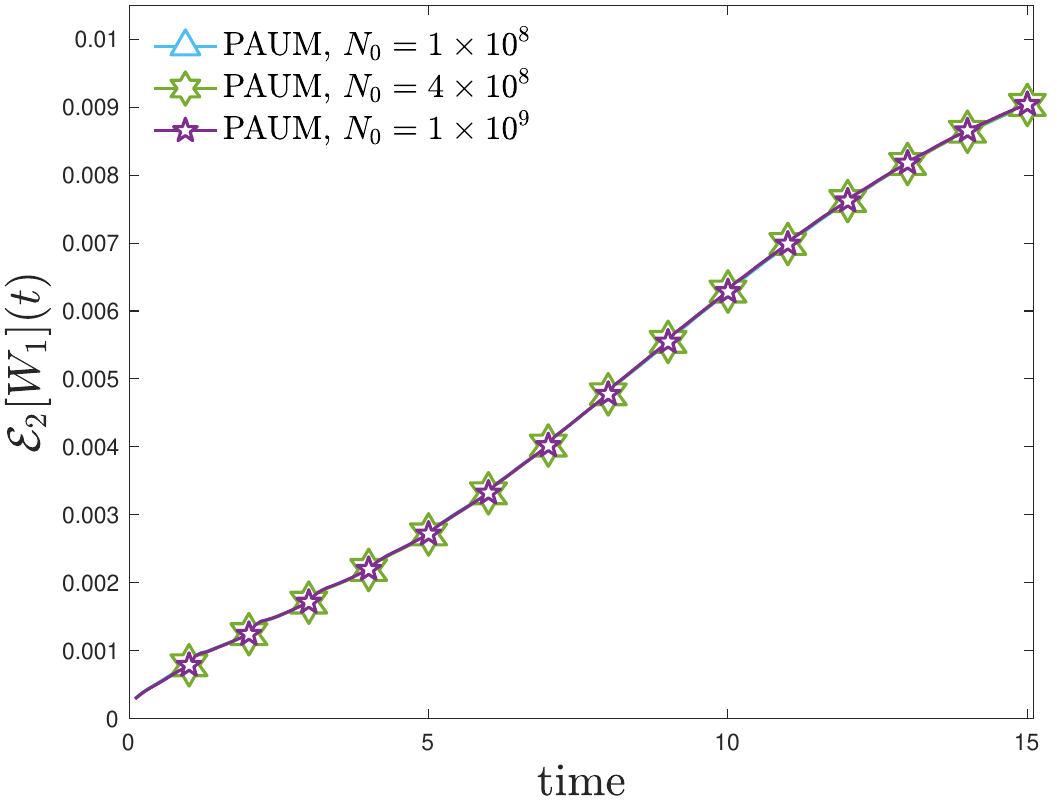}}
    \subfigure[Deviation in energy.]{\includegraphics[width=0.32\textwidth,height=0.22\textwidth]{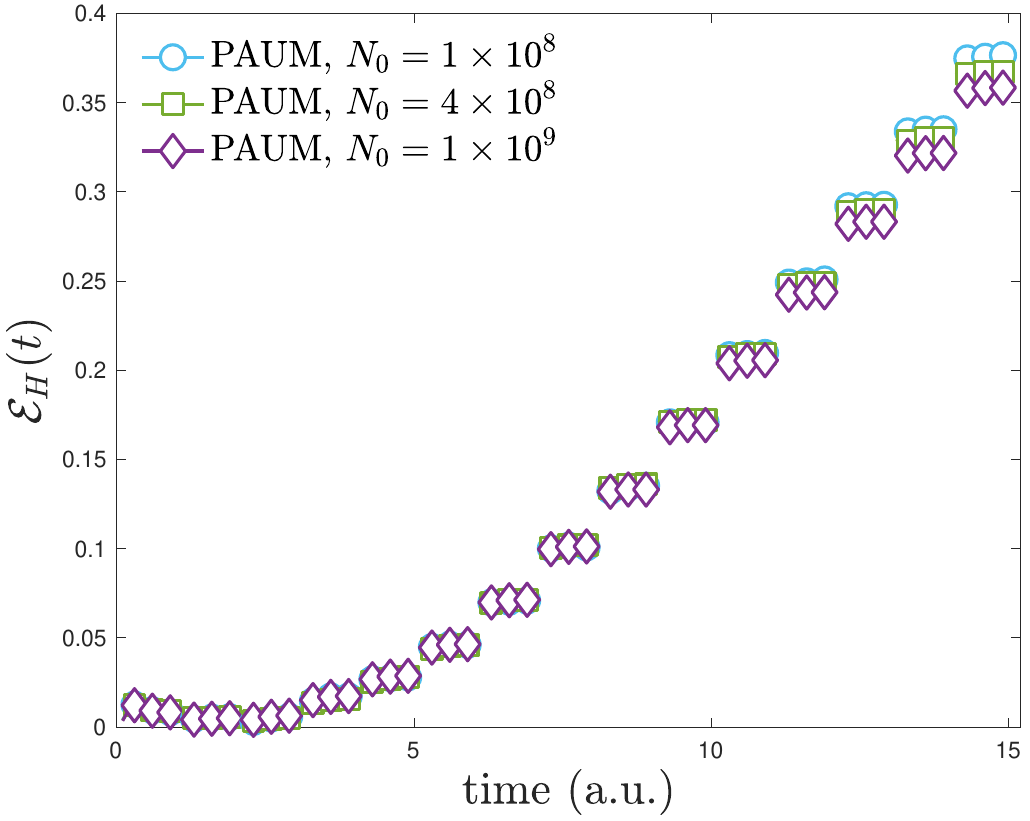}}
     \subfigure[Growth of particle number.]{\includegraphics[width=0.32\textwidth,height=0.22\textwidth]{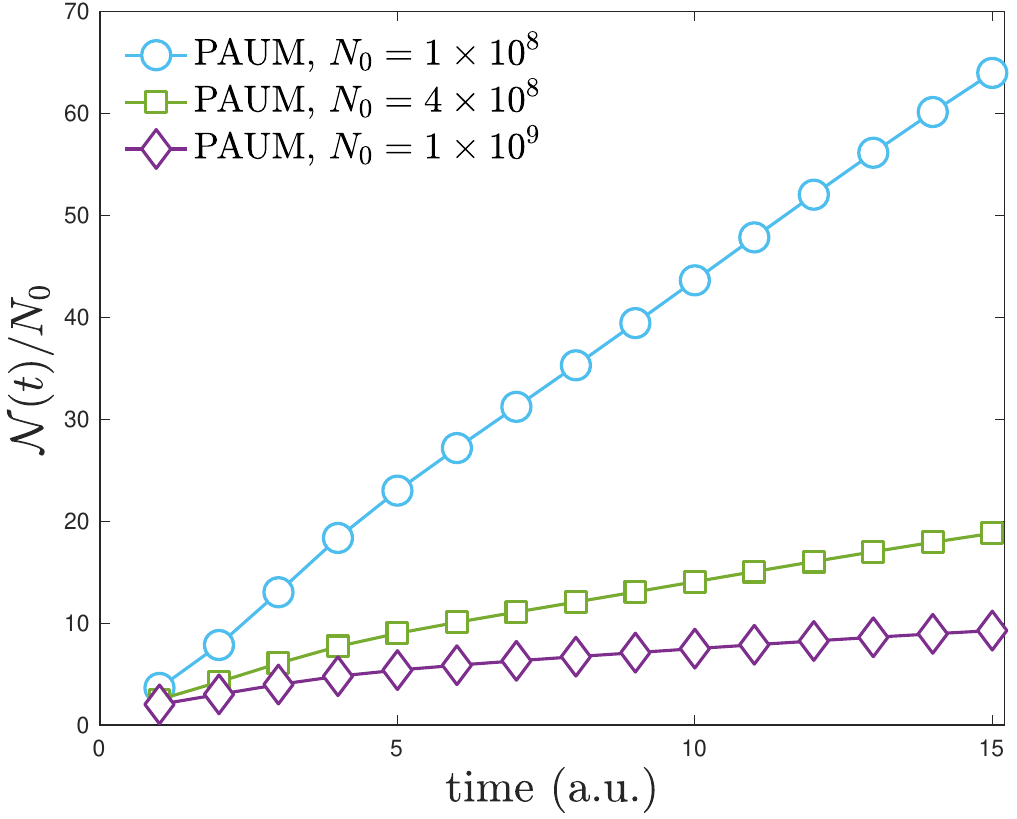}}
    \\
    \subfigure[Snapshots of $W_1(x_{1}, k_{1}, t)$ at $4$a.u. (left: $N_0 = 10^8$, middle: $N_0 = 4\times 10^8$, right: $N_0 =  10^9$). ]{{\includegraphics[width=0.32\textwidth,height=0.22\textwidth]{./redist_uhist_T4_n100m}}
    {\includegraphics[width=0.32\textwidth,height=0.22\textwidth]{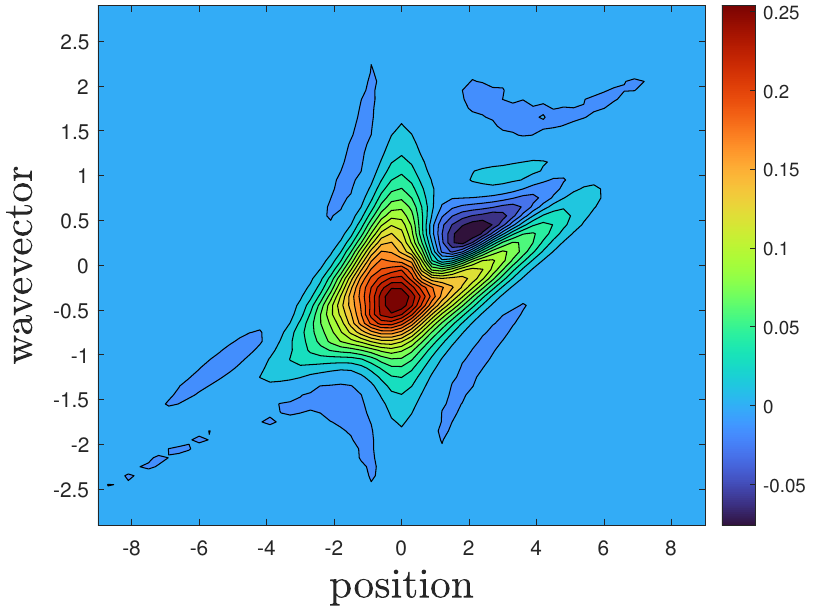}}
     {\includegraphics[width=0.32\textwidth,height=0.22\textwidth]{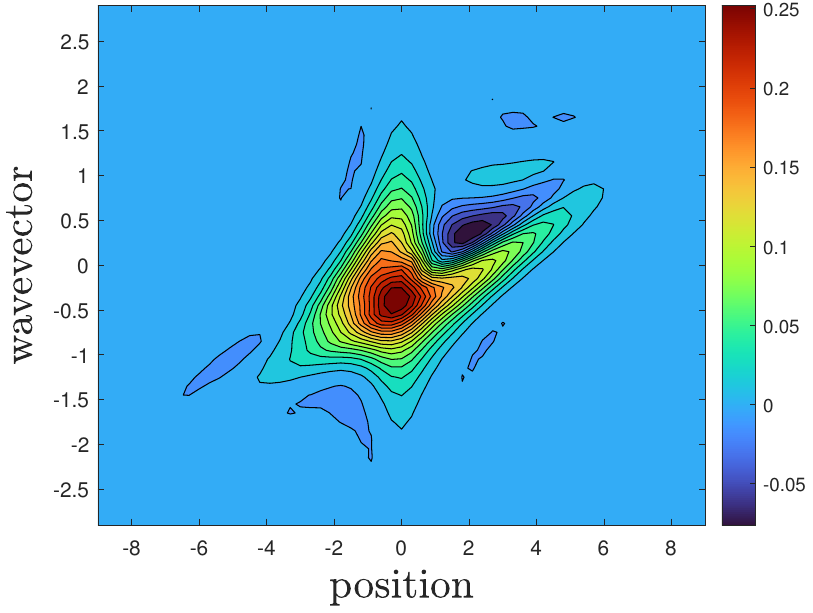}}}
     \\
    \subfigure[Snapshots of $W_1(x_{1}, k_{1}, t)$ at $8$a.u. (left: $N_0 = 10^8$, middle: $N_0 = 4\times 10^8$, right: $N_0 =  10^9$). ]{
    {\includegraphics[width=0.32\textwidth,height=0.22\textwidth]{./redist_uhist_T8_n100m}}
    {\includegraphics[width=0.32\textwidth,height=0.22\textwidth]{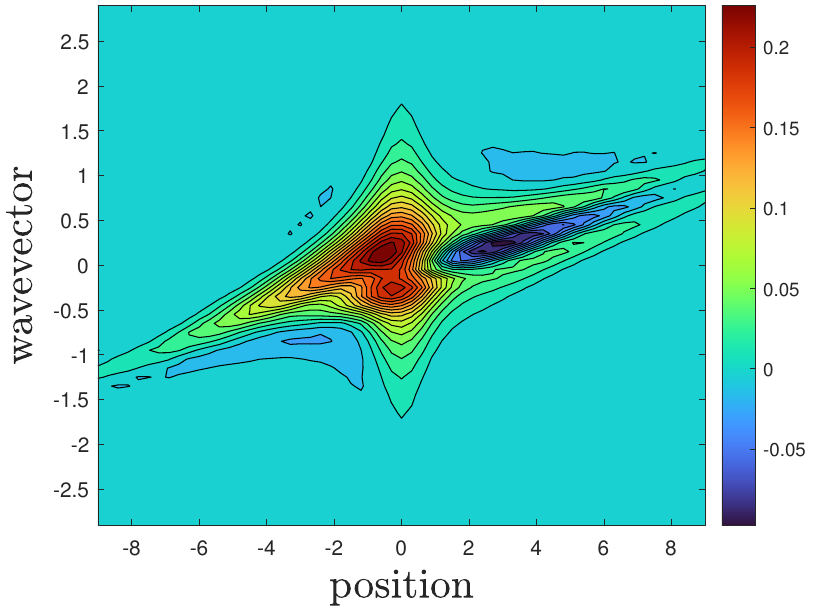}}
     {\includegraphics[width=0.32\textwidth,height=0.22\textwidth]{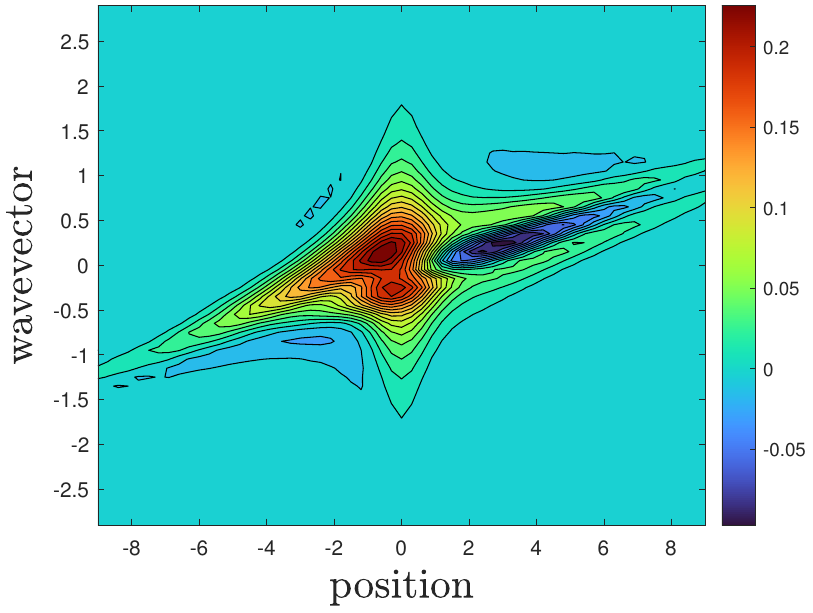}}}
     \\
    \subfigure[Snapshots of $W_1(x_{1}, k_{1}, t)$ at $15$a.u. (left: $N_0 = 10^8$, middle: $N_0 = 4\times 10^8$, right: $N_0 =  10^9$). ]{{\includegraphics[width=0.32\textwidth,height=0.22\textwidth]{./redist_uhist_T15_n100m}}
    {\includegraphics[width=0.32\textwidth,height=0.22\textwidth]{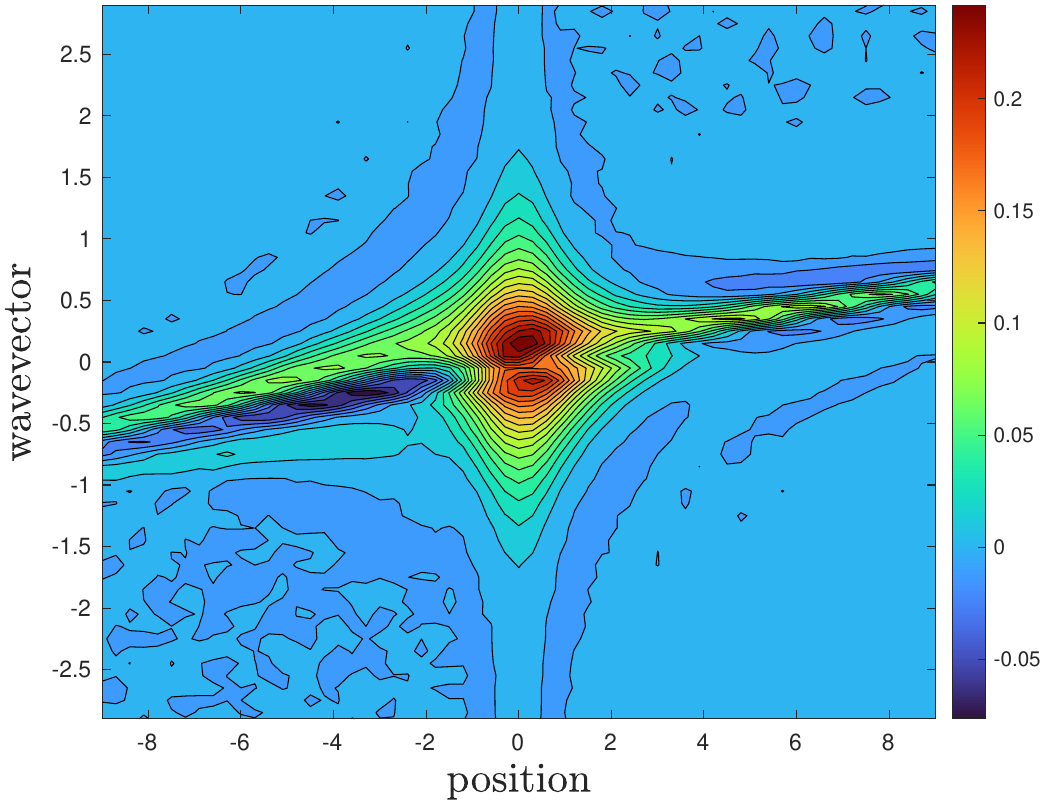}}
     {\includegraphics[width=0.32\textwidth,height=0.22\textwidth]{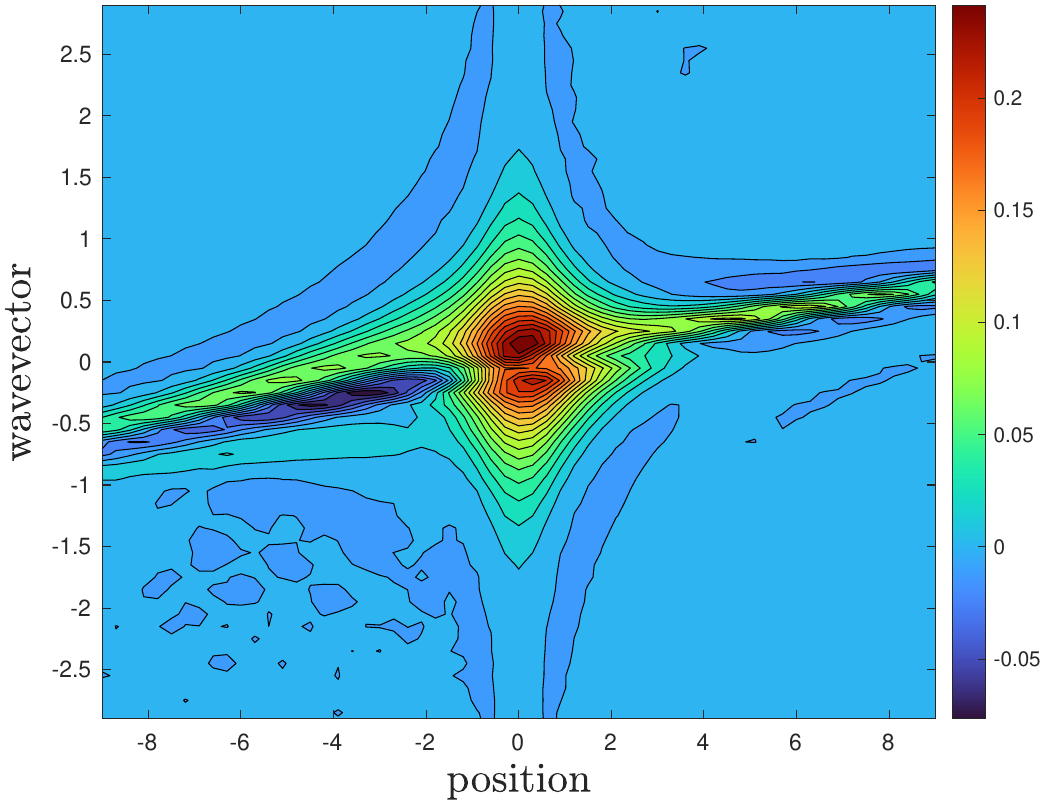}}}
    \caption{\small The performance of PAUM under different $N_0$. PAUM still works in 6-D simulations when the sample size is sufficiently large. \label{supp_fig_6d_PAUM}}
\end{figure} 

\subsection{Comparison of difference gap and maximal gap in SPADE}
\label{supp_sec.gap}

A key ingredient in SPADE (Algorithm \ref{algorithm_SPADE}) is to decide where to split the bin. We would like to show that both gap functions are applicable for 6-D problem, but {\bf the difference gap is more preferable} as it suffers less from the smoothing effect and has less fluctuations in the total energy.

For each $\mathrm{Q}_k = [a^{(k)}_{1}, b^{(k)}_{1}] \times \dots \times [a^{(k)}_{6}, b^{(k)}_{6}]$,  it can be split by selecting a node $c^{(k)}_j$ in the $j$-th dimension and split $\mathrm{Q}_k$ into $\mathrm{Q}_k^{(1)}$  and $\mathrm{Q}_k^{(2)}$:
\begin{equation}
 \mathrm{Q}_k^{(1)} = \prod_{i=1}^{j-1} [a^{(k)}_{i}, b^{(k)}_{i}] \times [a^{(k)}_{j}, c^{(k)}_{j}] \times \prod_{i=j+1}^{6}  [a^{(k)}_{i}, b^{(k)}_{i}], \quad  \mathrm{Q}_k^{(2)} = \mathrm{Q}_k \setminus \mathrm{Q}_k^{(1)}.
\end{equation}
Denote by $P_k^{(1)}$ and $M_k^{(1)}$ the counts of positive and negative particles in $\mathrm{Q}_k^{(1)}$, respectively. It suggests to choose $c_j^{(k)}$ to optimize either the {\bf maximal gap} \cite{LiYangWong2016}
\begin{equation}
\max_{\mathrm{Q}_k^{(1)}}\left(\Big | \frac{P_k^{(1)}}{P_k} -  \frac{\textup{vol}(\mathrm{Q}_k^{(1)})}{\textup{vol}(\mathrm{Q}_k)} \Big |, \Big | \frac{M_k^{(1)}}{M_k} -  \frac{\textup{vol}(\mathrm{Q}_k^{(1)})}{\textup{vol}(\mathrm{Q}_k)}  \Big |\right),
\end{equation}
or the {\bf difference gap} \cite{ShaoXiong2020_arXiv}.
\begin{equation}
\frac{1}{2}\max_{\mathrm{Q}_k^{(1)}}\left(\Big | \frac{P_k^{(1)}}{P_k}  - \frac{M_k^{(1)}}{M_k}  \Big |\right) = \frac{1}{2} \max_{\mathrm{Q}_k^{(2)}}\left(\Big | \frac{P_k^{(2)}}{P_k}  - \frac{M_k^{(2)}}{M_k}  \Big |\right).
\end{equation}

\begin{figure}[!h]
    \subfigure[$\vartheta = 0.008$.  (left: $l^2$-error for $W_1$, middle: deviation of energy, right: growth of particle)\label{supp_comparison_t0008}]{
    {\includegraphics[width=0.32\textwidth,height=0.22\textwidth]{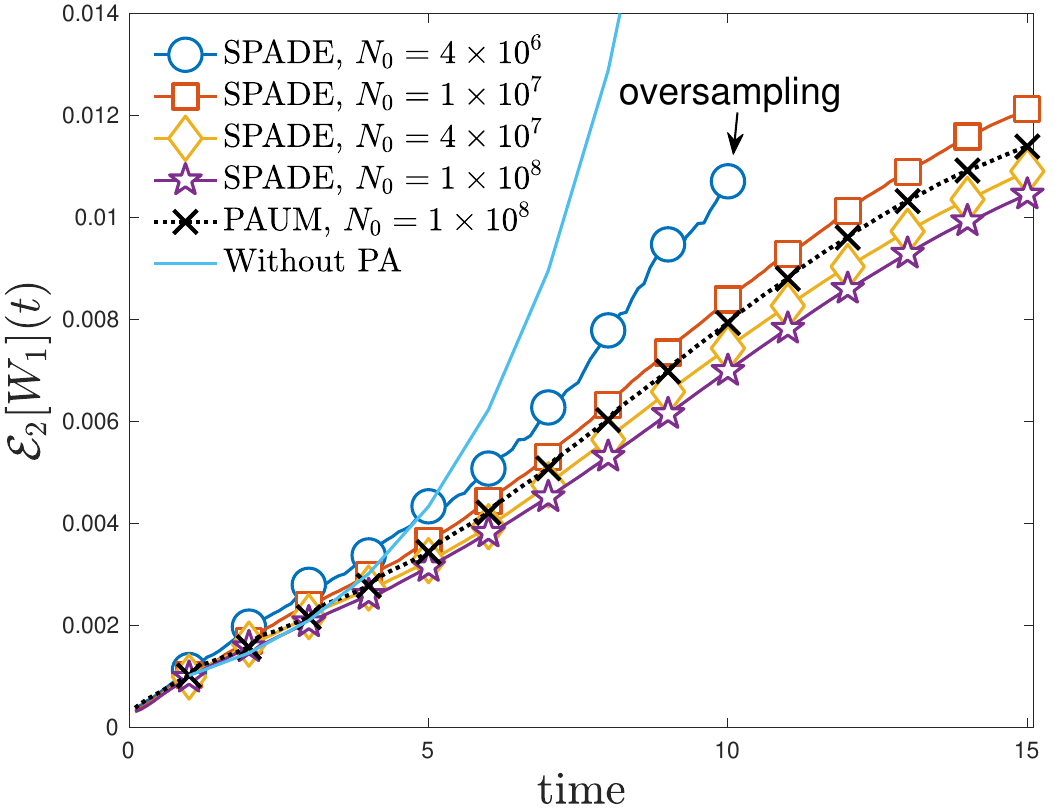}}
    {\includegraphics[width=0.32\textwidth,height=0.22\textwidth]{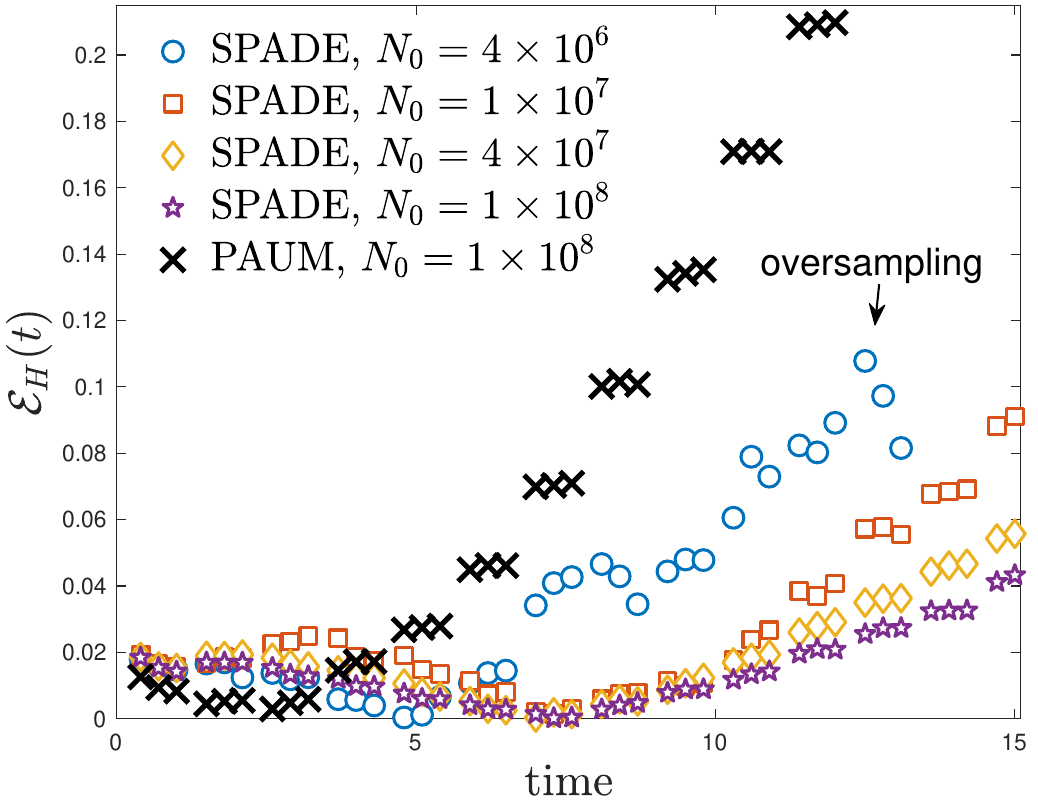}}
    {\includegraphics[width=0.32\textwidth,height=0.22\textwidth]{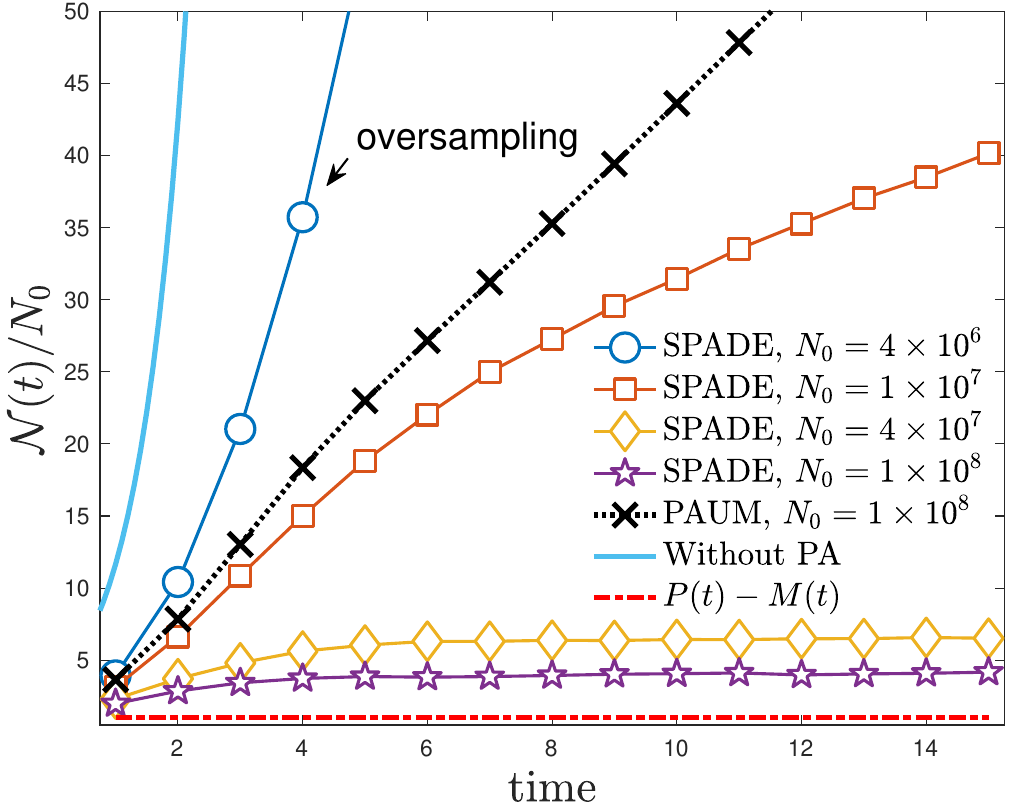}}}
     \\
    \subfigure[$\vartheta = 0.01$.  (left: $l^2$-error for $W_1$, middle: deviation of energy, right: growth of particle)\label{supp_comparison_t001}]{
    {\includegraphics[width=0.32\textwidth,height=0.22\textwidth]{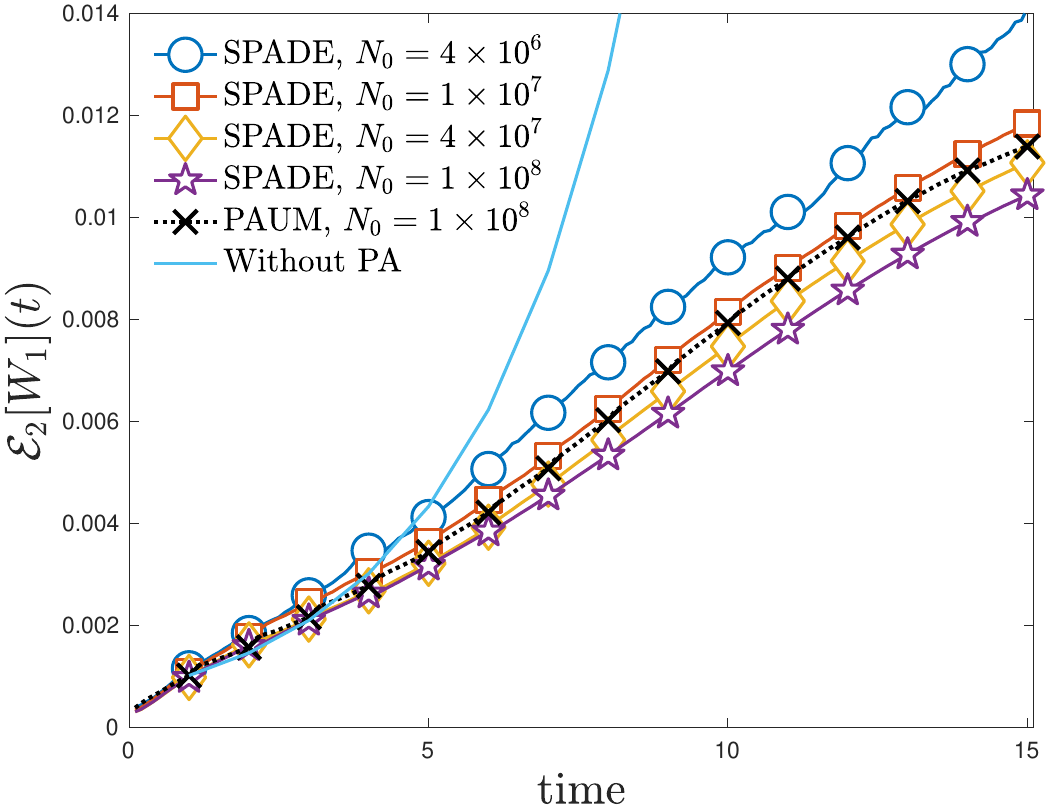}}
    {\includegraphics[width=0.32\textwidth,height=0.22\textwidth]{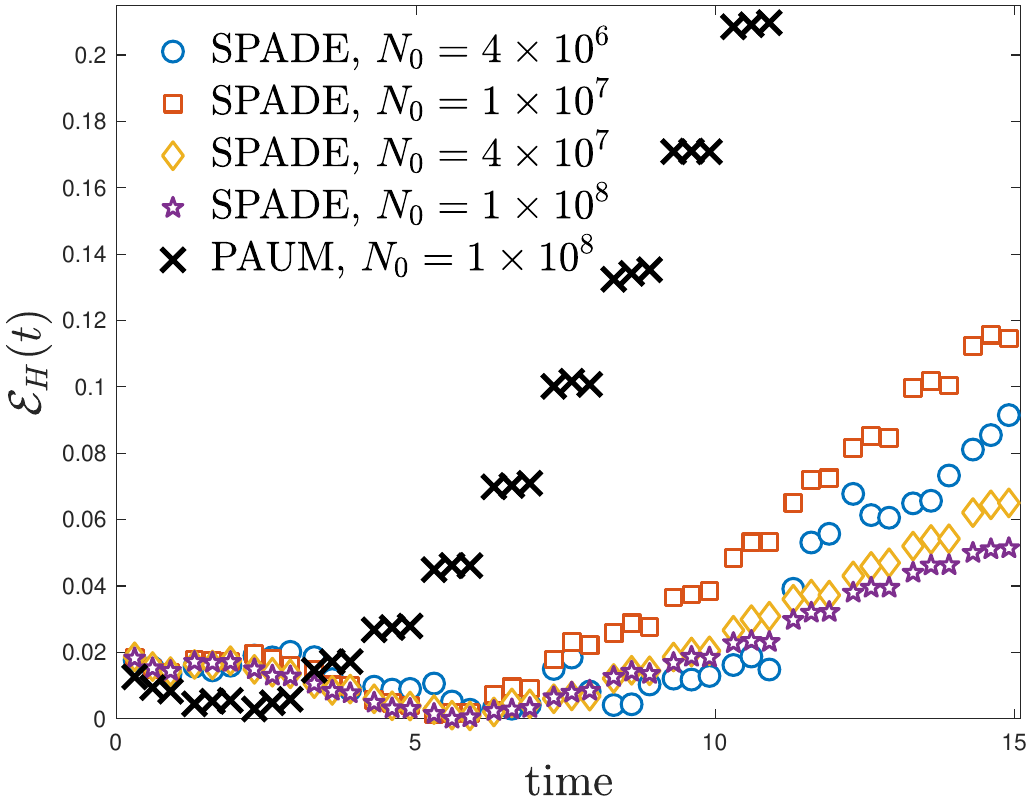}}
    {\includegraphics[width=0.32\textwidth,height=0.22\textwidth]{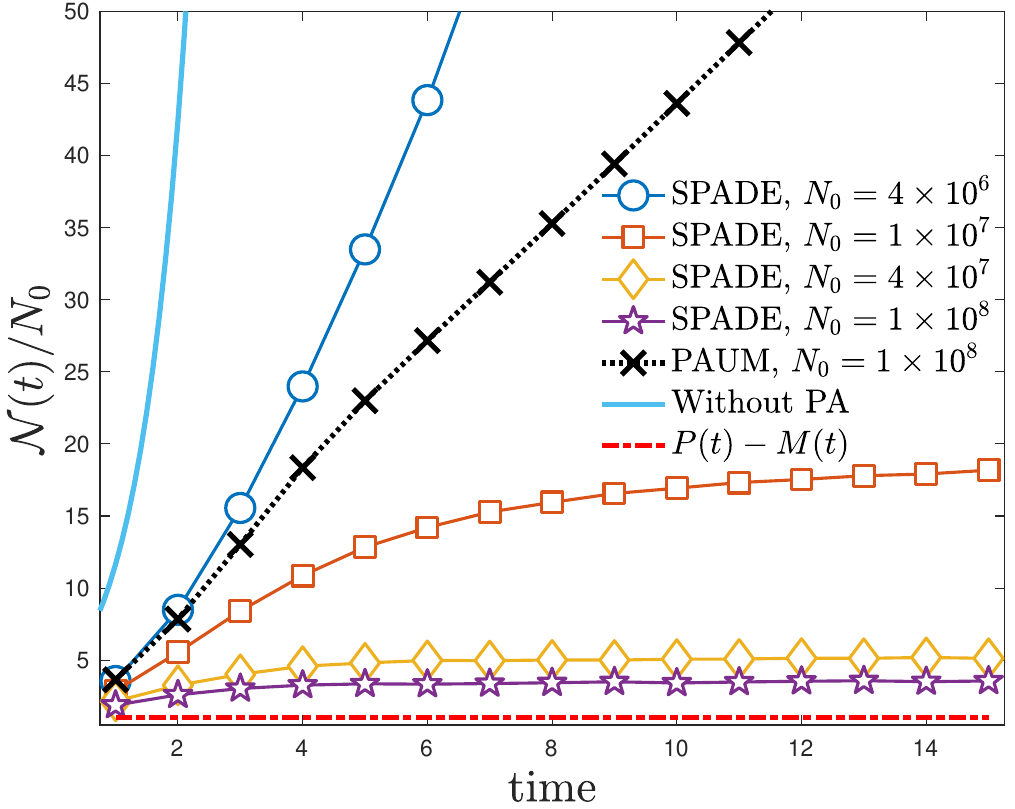}}}
     \\
     \subfigure[$\vartheta = 0.02$.  (left: $l^2$-error for $W_1$, middle: deviation of energy, right: growth of particle)\label{supp_comparison_t002}]{
     {\includegraphics[width=0.32\textwidth,height=0.22\textwidth]{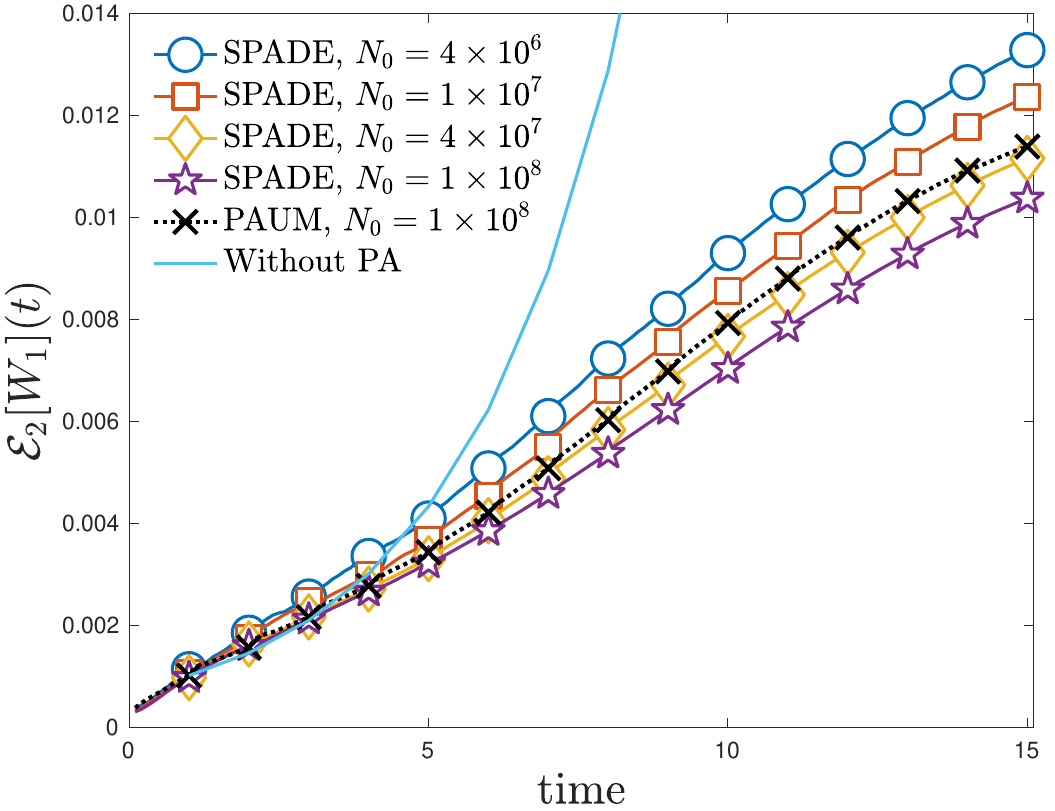}}
     {\includegraphics[width=0.32\textwidth,height=0.22\textwidth]{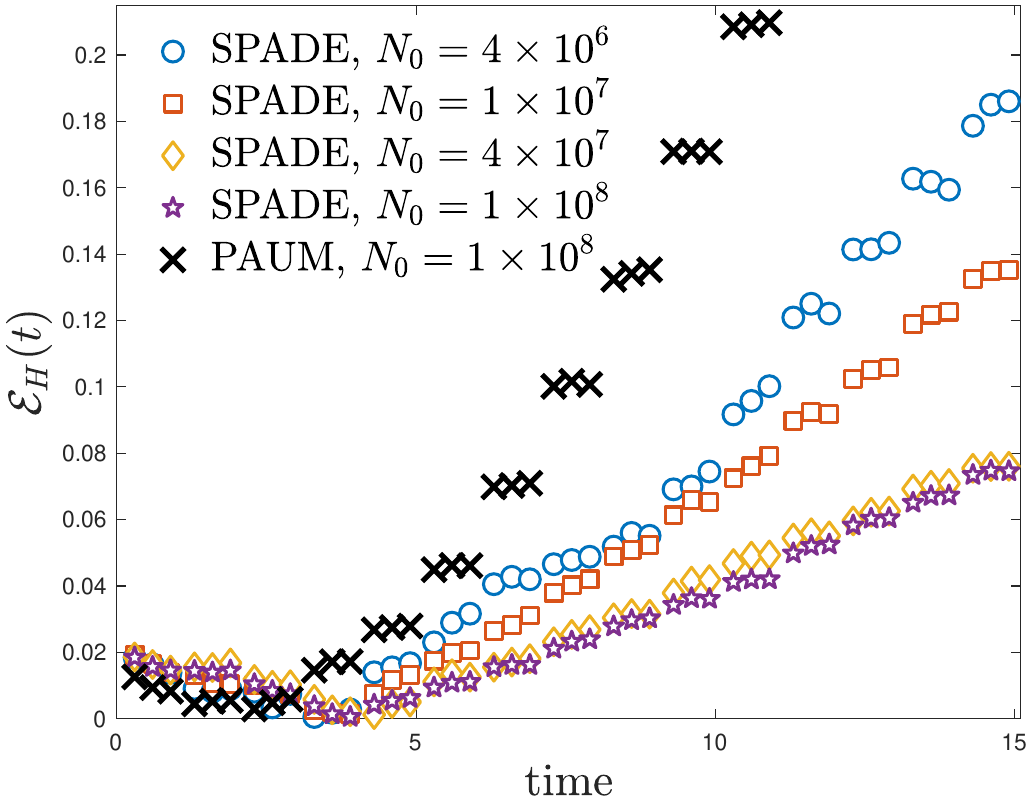}} 
     {\includegraphics[width=0.32\textwidth,height=0.22\textwidth]{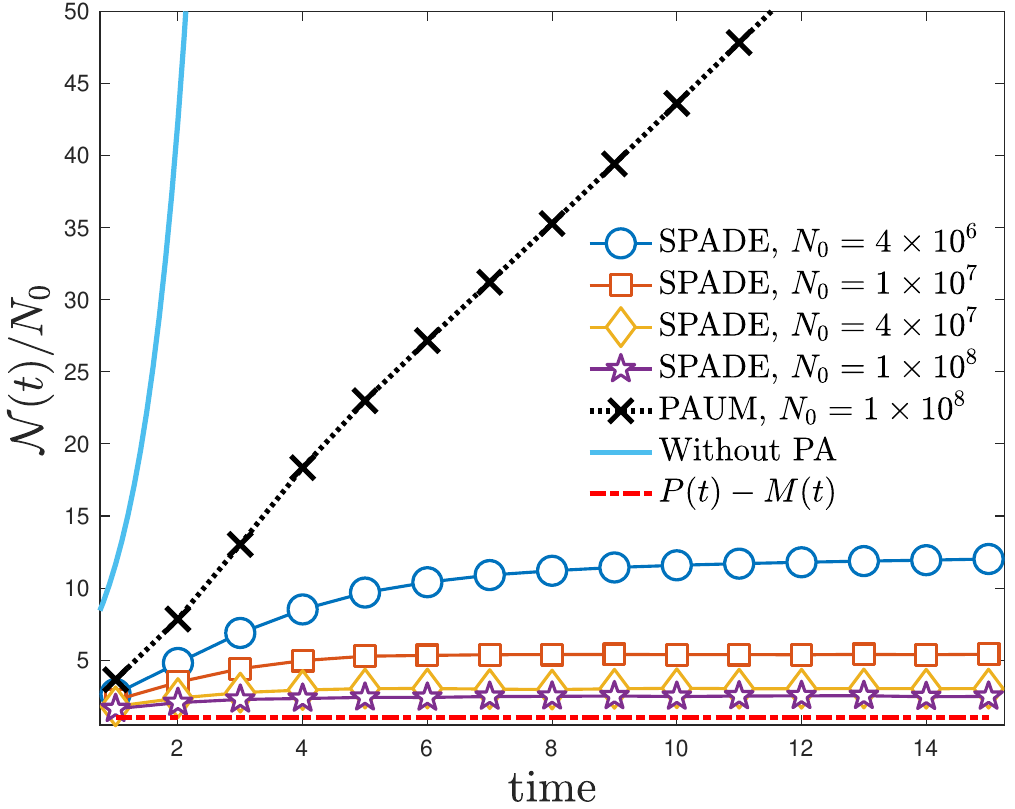}}}
          \\
     \subfigure[$\vartheta = 0.04$. (left: $l^2$-error for $W_1$, middle: deviation of energy, right: growth of particle)\label{supp_comparison_t004}]{
     {\includegraphics[width=0.32\textwidth,height=0.22\textwidth]{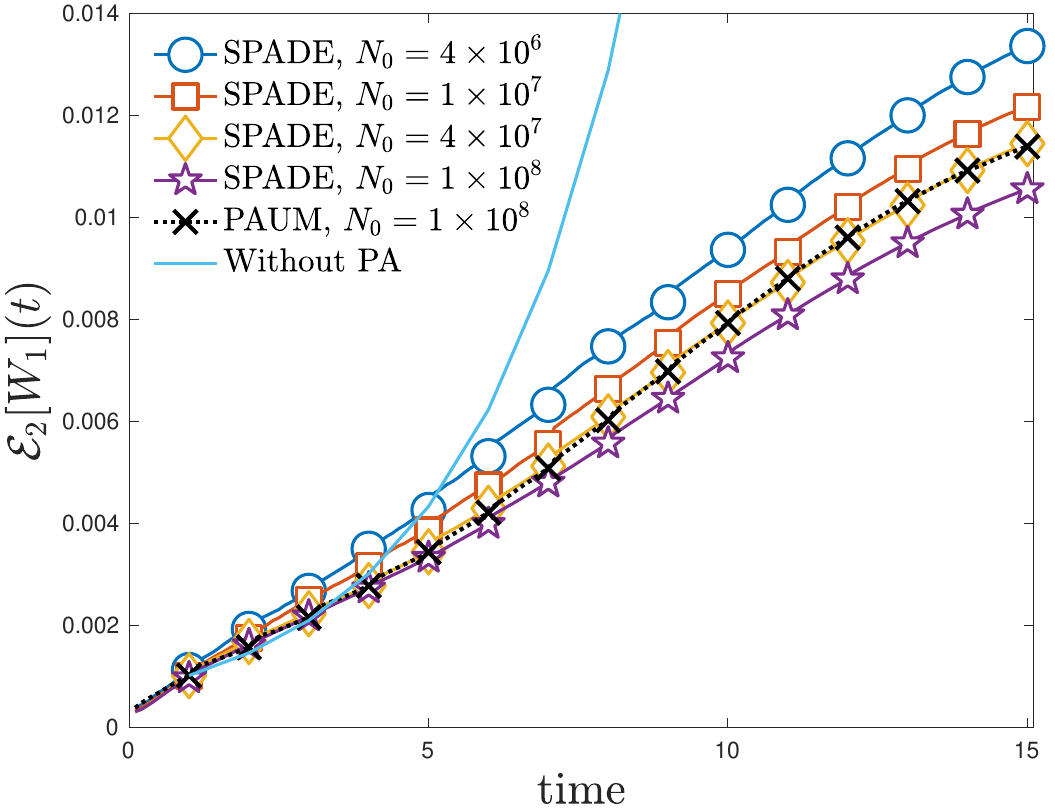}}
     {\includegraphics[width=0.32\textwidth,height=0.22\textwidth]{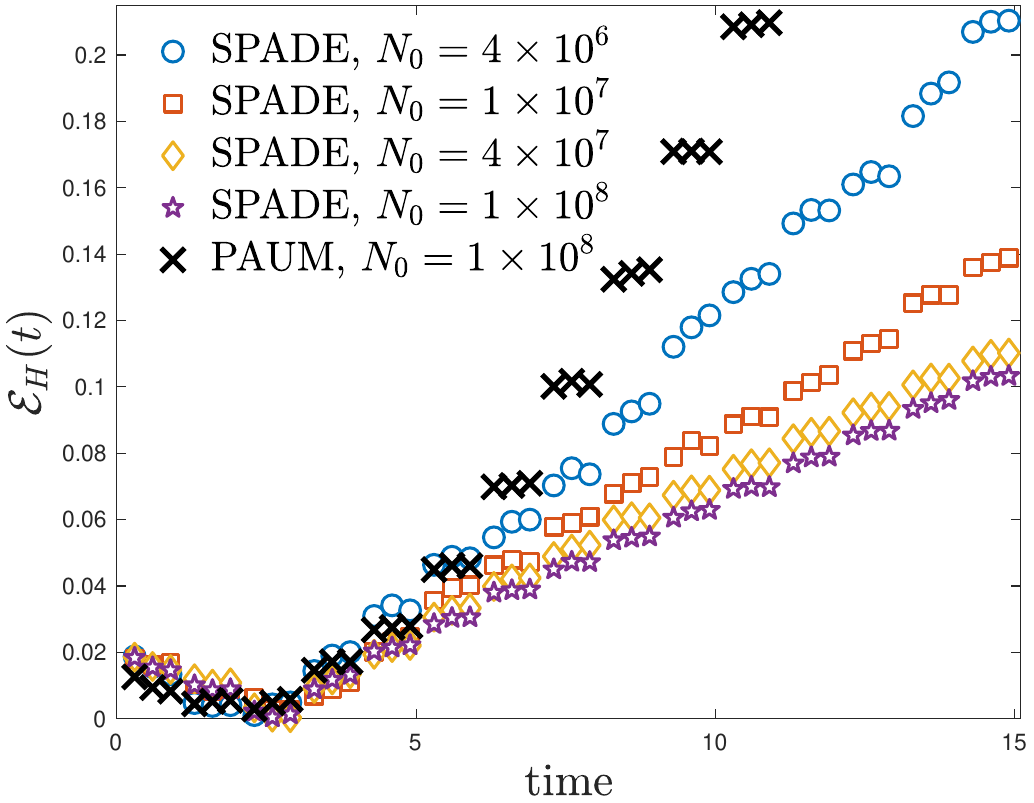}} 
     {\includegraphics[width=0.32\textwidth,height=0.22\textwidth]{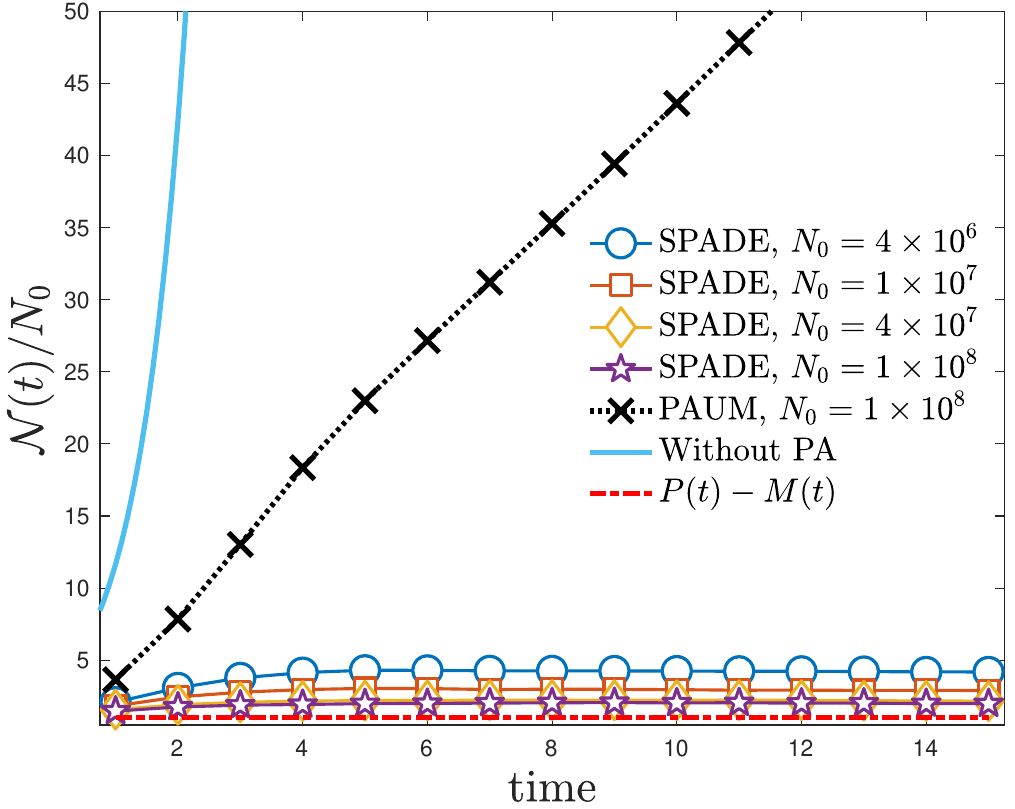}}}
    \caption{\small  A comparison of the $l^2$-errors of $\mathcal{E}_2[W_1](t)$ (left) and the deviation of total energy $\mathcal{E}_H(t)$ (middle) to monitor the numerical errors, and growth of particle number after PA (right). The accuracy of SPADE outperforms PAUM when $N_0 = 4\times 10^7$ or $N_0 = 1\times 10^8$. In addition, SPADE is able to control particle number more efficiently for moderately large $N_0$, but still suffers from oversampling problem when $N_0$ is too small.  \label{supp_comparison_SPADE_PAUM_max}}
\end{figure}

\begin{figure}[!h]
   \subfigure[$l^2$-error for $P_{xy}$, $\vartheta=0.008$.]{
   \includegraphics[width=0.48\textwidth,height=0.27\textwidth]{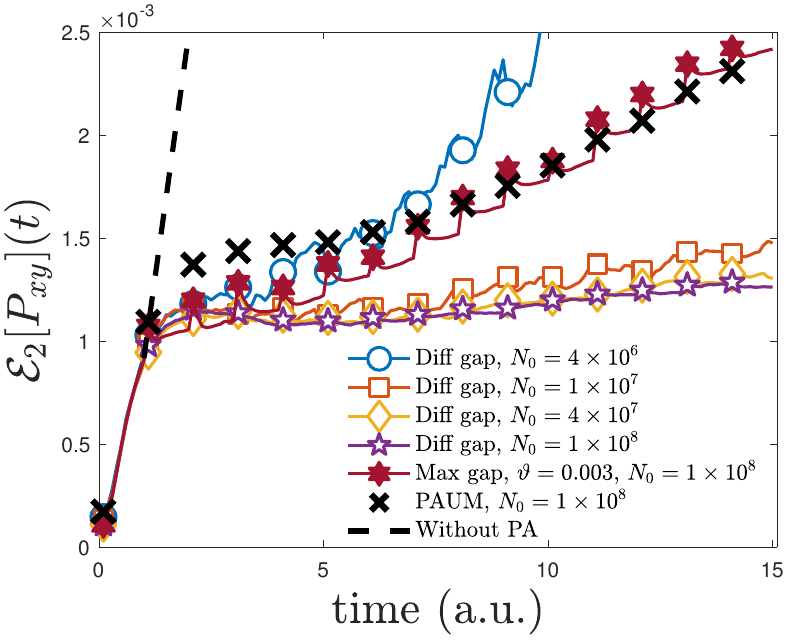}}
   \subfigure[$l^2$-error for $P_{xy}$, $\vartheta=0.01$.]{
   {\includegraphics[width=0.48\textwidth,height=0.27\textwidth]{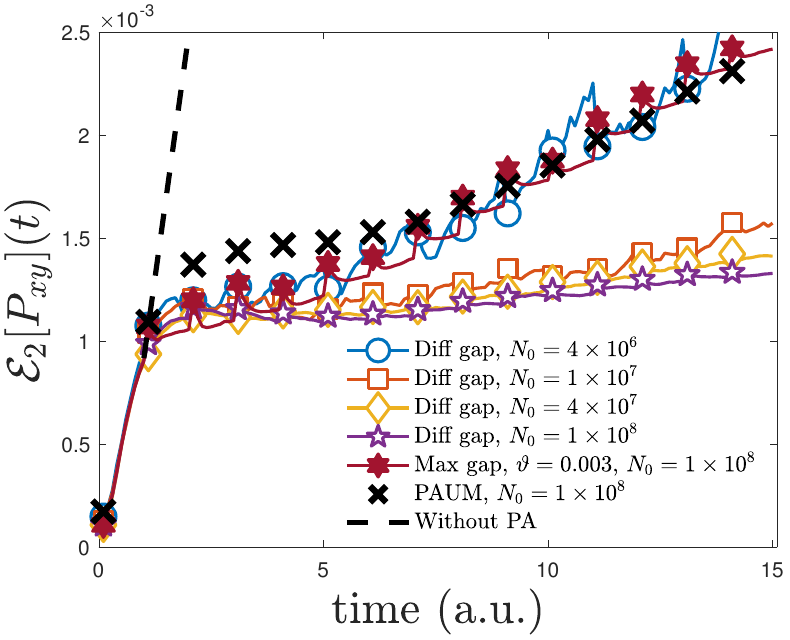}}}
  \\
   \subfigure[$l^2$-error for $P_{xy}$, $\vartheta=0.02$.]{
    {\includegraphics[width=0.48\textwidth,height=0.27\textwidth]{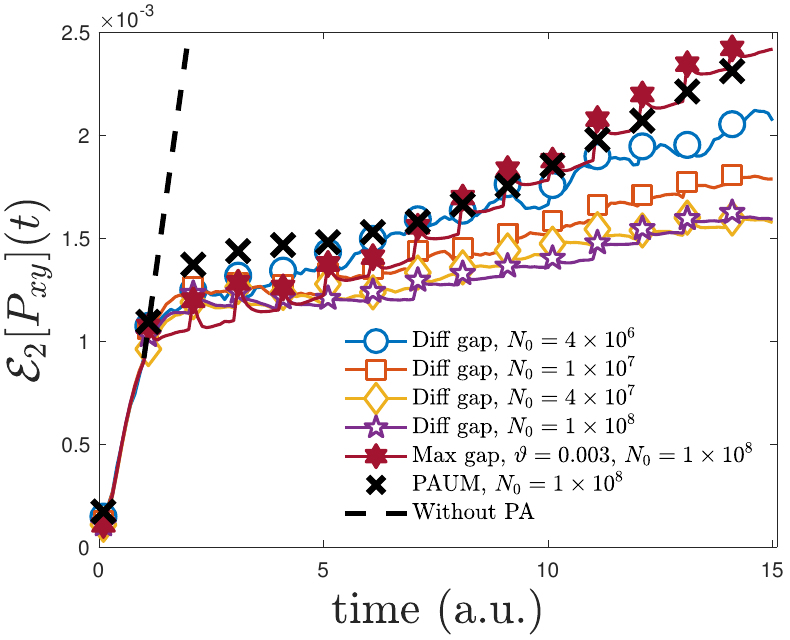}}}
   \subfigure[$l^2$-error for $P_{xy}$, $\vartheta=0.04$.]{
   {\includegraphics[width=0.48\textwidth,height=0.27\textwidth]{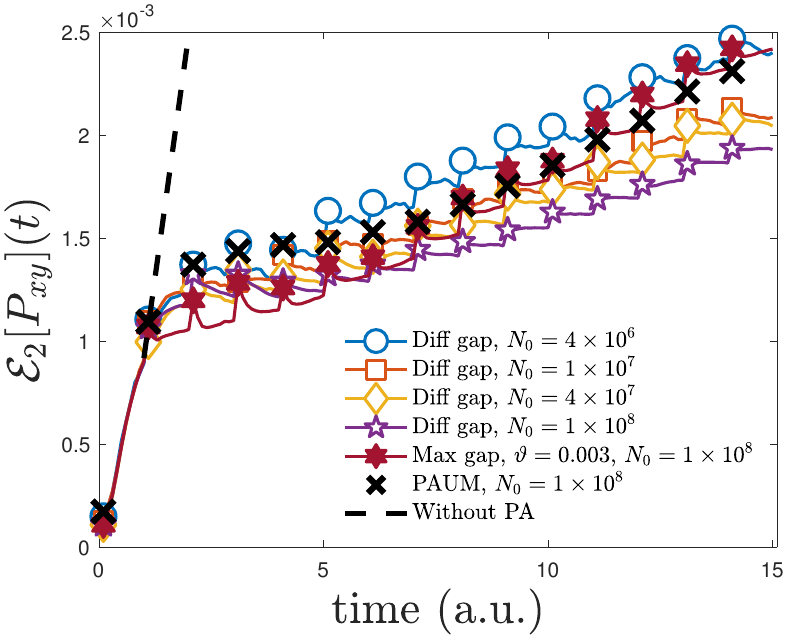}}}
   \\
   \centering
\subfigure[The spatial marginal distribution $P_x(x, t)$ at $t=4 \to 8 \to 15$a.u., $\vartheta= 0.008$.]{
{\includegraphics[width=0.32\textwidth,height=0.21\textwidth]{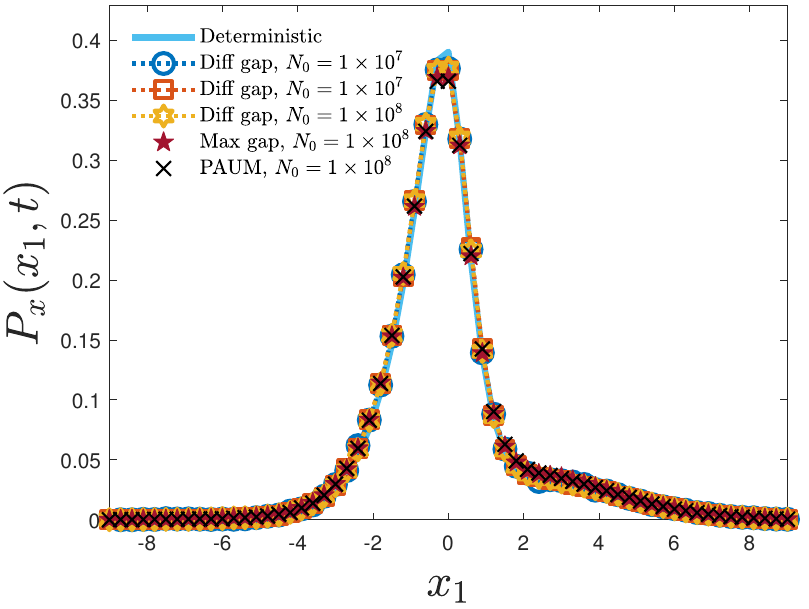}}
{\includegraphics[width=0.32\textwidth,height=0.21\textwidth]{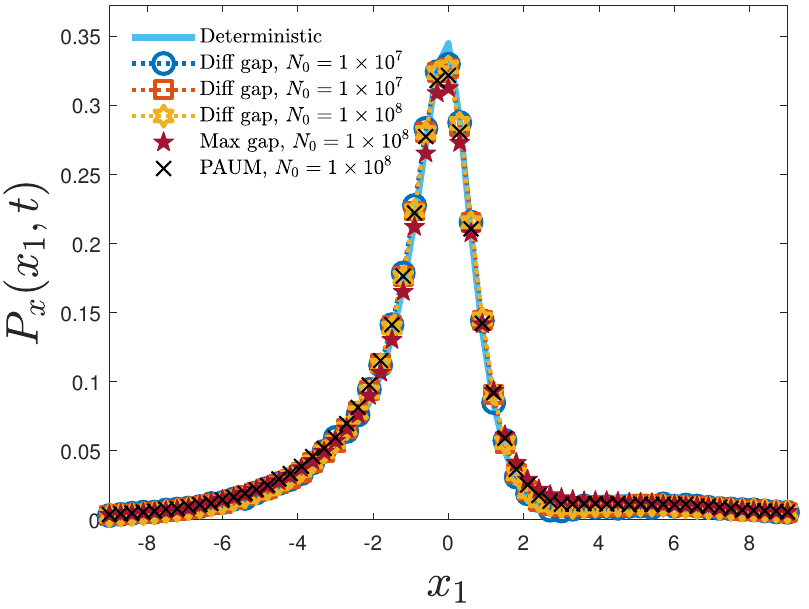}}
{\includegraphics[width=0.32\textwidth,height=0.21\textwidth]{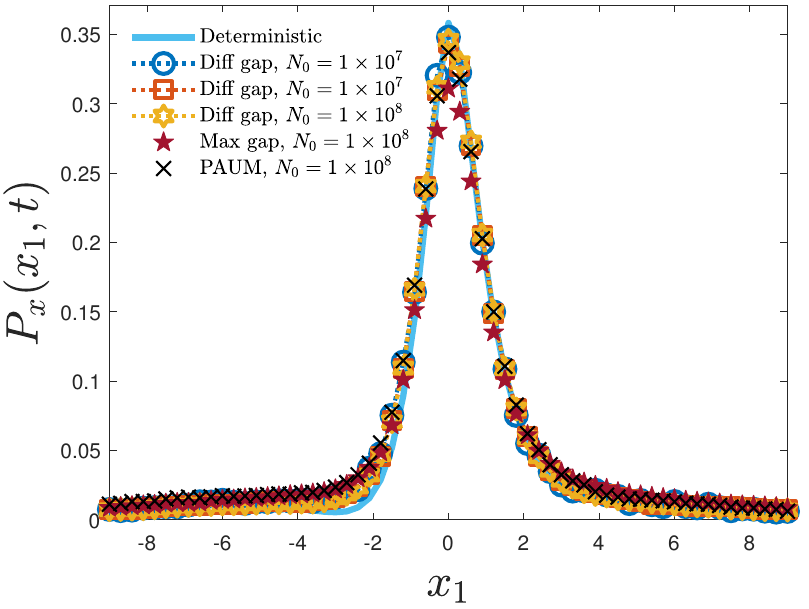}}}
\\
\centering
\subfigure[The spatial marginal distribution $P_x(x, t)$ at $t=4 \to 8 \to 15$a.u., $\vartheta= 0.02$.]{
{\includegraphics[width=0.32\textwidth,height=0.21\textwidth]{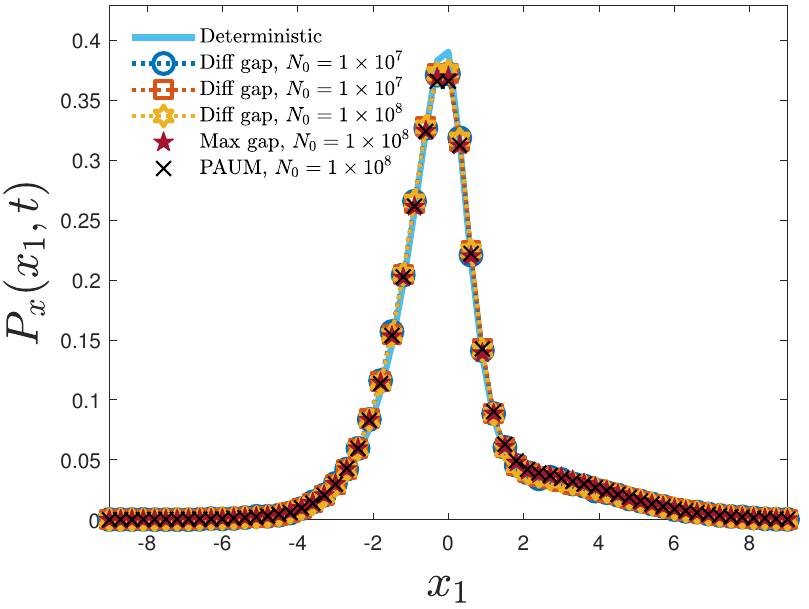}}
{\includegraphics[width=0.32\textwidth,height=0.21\textwidth]{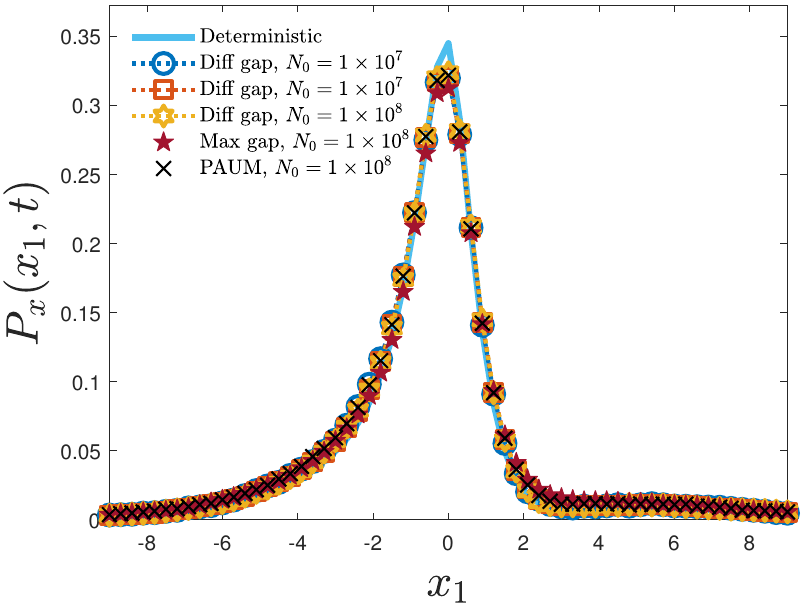}}
{\includegraphics[width=0.32\textwidth,height=0.21\textwidth]{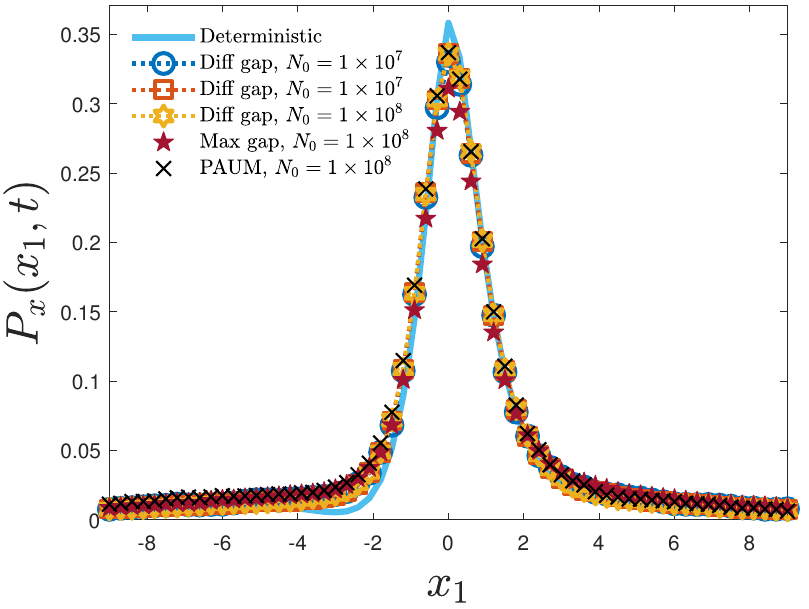}}}
    \caption{\small The spatial marginal distribution $P_x(x, t)$ produced by the deterministic scheme, WBRW-SPA with either difference gap or the maximal gap adopted.  There are some evident discrepancies observed in the crest and left shoulder of the wavepacket when the maximal gap is adopted, which is an indicator of  the smoothing effect. Fortunately, this can be alleviated when the difference gap is adopted. \label{supp_xdist_time_evolution_max}}
\end{figure}

Figures \ref{supp_comparison_SPADE_PAUM_max} and \ref{supp_xdist_time_evolution_max} provide a comparison between the maximal gap and the difference gap. The sample size is $N_0 = 1\times 10^8$ and the filter is $\lambda_0 = 4.65$ in SPA. When the maximal gap is adopted, the parameters $\vartheta = 0.003, m = 512$ are fixed. The curve marked by red hexagram denotes the results with the maximal gap, while the curve marked by black cross denotes the results using PAUM. Several observations are made.
\begin{description}

\item[(1)] When the partition level is comparable (see Table \ref{supp_6d_cpu_time}), the maximal gap ($\vartheta = 0.003$) seems to outperform the difference gap ($\vartheta = 0.008$) regarding the $l^2$-errors of the reduced Wigner function $W_1$ (see Figure \ref{supp_comparison_SPADE_PAUM_max}). However, the difference gap becomes superior to the maximal gap in consideration of the spatial marginal distribution $P_{xy}$ (see Figure \ref{supp_xdist_time_evolution_max}). 

\item[(2)]  More particles are left uncanceled when the difference gap is adopted (see Table \ref{supp_6d_cpu_time}), although the partition levels are comparable. In fact, the intuition behind the difference gap is to dig out the nodal surfaces that divide positive and negative particles. As a result, fewer particles are matched and canceled out. Fortunately, we find that  {\bf the computational cost of SPADE associated with difference gap is less than that with  the maximal gap} because it spends less time in calculation of the star discrepancy. The group with difference gap produces more cuts, but still saves a lot of computational time. 

\item[(3)] The deviation of total energy can be evidently suppressed when the difference gap is adopted. By contrast, a severe fluctuation is observed when the maximal gap is used.

\begin{table}[!h]
  \centering
  \caption{\small Computational time (in hours) of SPADE, average partition level $K$ and growth ratio of total particles for 6-D simulations up to $15$a.u.  Here $m=512$ is fixed and $128$ cores are used for each task. \label{supp_6d_cpu_time}}
 \begin{lrbox}{\tablebox}
  \begin{tabular}{c|c|c|c|c|c|c}
\hline\hline
	&      \multicolumn{3}{c|}{Maximal gap, $\vartheta = 0.003$} & \multicolumn{3}{c}{Difference gap, $\vartheta = 0.008$} \\
\hline
$N_0$ 	&	Time(h)	&	Average $K$ 	&	$\mathcal{N}(15)/N_0$		&	Time(h)		&	Average $K $	&	$\mathcal{N}(15)/N_0$	\\
\hline
$4\times10^7$		&	23.95	&	1.07$\times 10^7$	&	 3.05		&	13.92	&	1.29$\times 10^7$	&	6.52	\\
$1\times10^8$		&	48.26	&	1.15$\times 10^7$	&	 2.46		&	24.69	&	1.34$\times 10^7$	&	4.17	\\
\hline\hline
 \end{tabular}
\end{lrbox}
\scalebox{0.95}{\usebox{\tablebox}}
\end{table} 

\end{description}

\begin{figure}[!h]
\centering
\subfigure[Deterministic scheme, $t=4 \to 8 \to 15$a.u.]{
{\includegraphics[width=0.32\textwidth,height=0.21\textwidth]{./redist_CHASM_T4.pdf}}
{\includegraphics[width=0.32\textwidth,height=0.21\textwidth]{./redist_CHASM_T8.pdf}}
{\includegraphics[width=0.32\textwidth,height=0.21\textwidth]{./redist_CHASM_T15.pdf}}}
\\
\centering
\subfigure[PAUM, $N_0 = 1\times 10^8$, $t=4 \to 8 \to 15$a.u.]{
{\includegraphics[width=0.32\textwidth,height=0.21\textwidth]{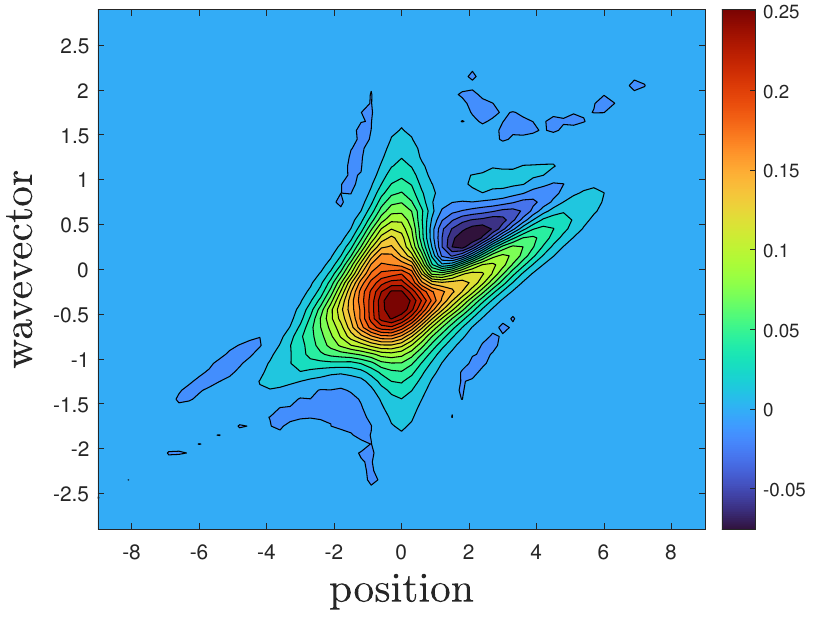}}
{\includegraphics[width=0.32\textwidth,height=0.21\textwidth]{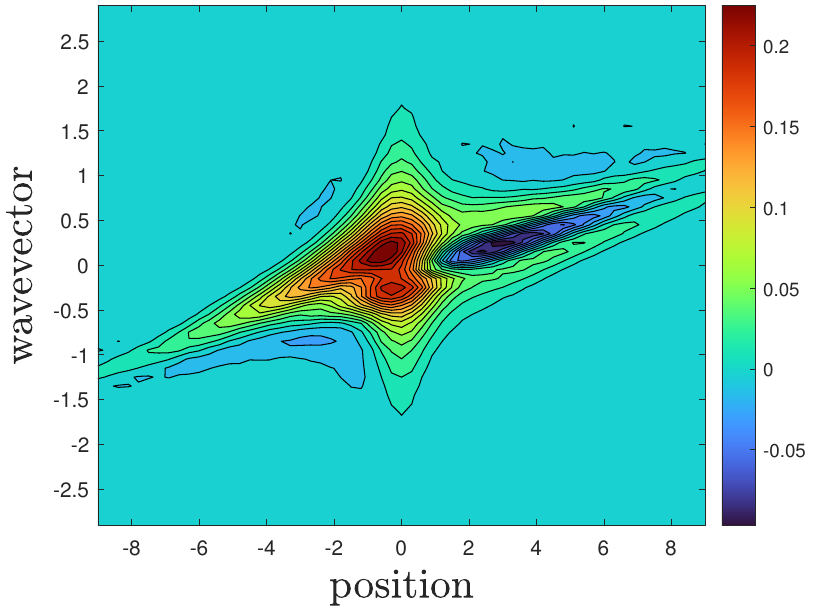}}
{\includegraphics[width=0.32\textwidth,height=0.21\textwidth]{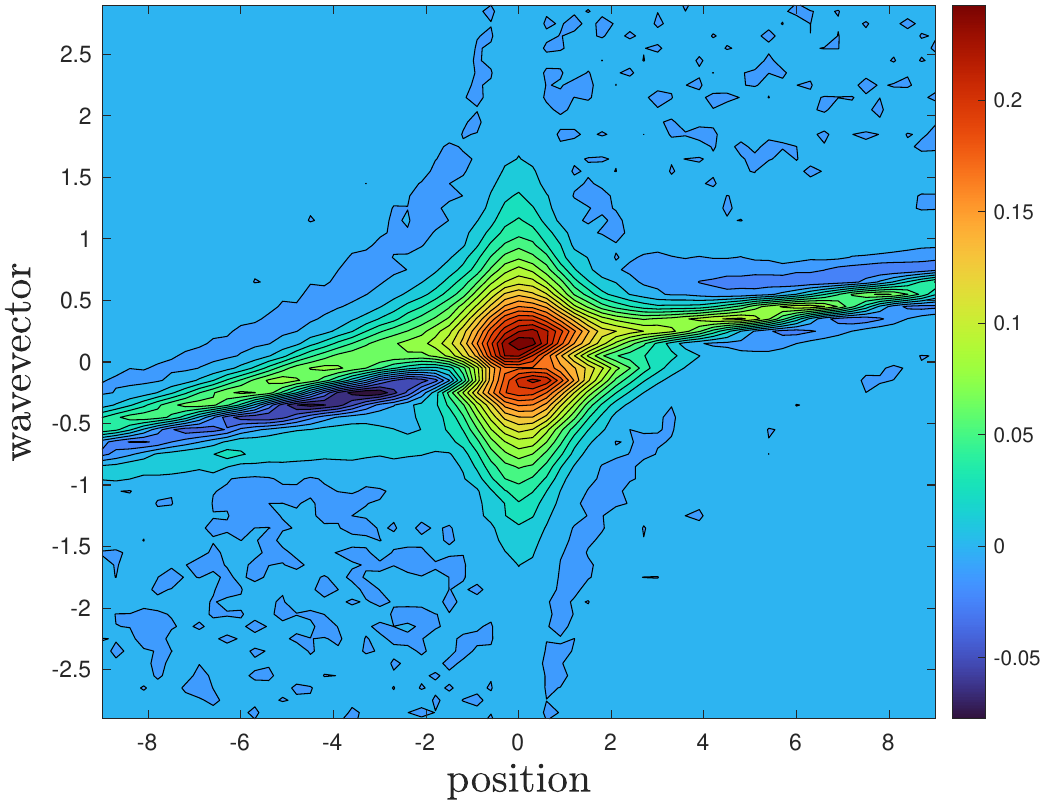}}}
\\
\centering
\subfigure[SPADE with difference gap adopted, $N_0 = 1\times 10^8$, $\vartheta = 0.008$, $t=4 \to 8 \to 15$a.u.]{
{\includegraphics[width=0.32\textwidth,height=0.21\textwidth]{./redist_SPADE_T4.pdf}}
{\includegraphics[width=0.32\textwidth,height=0.21\textwidth]{./redist_SPADE_T8.pdf}}
{\includegraphics[width=0.32\textwidth,height=0.21\textwidth]{./redist_SPADE_T15.pdf}}}
\\
\centering
\subfigure[SPADE with maximal gap adopted, $N_0 = 1\times 10^8$, $\vartheta = 0.003$, $t=4 \to 8 \to 15$a.u.]{
{\includegraphics[width=0.32\textwidth,height=0.21\textwidth]{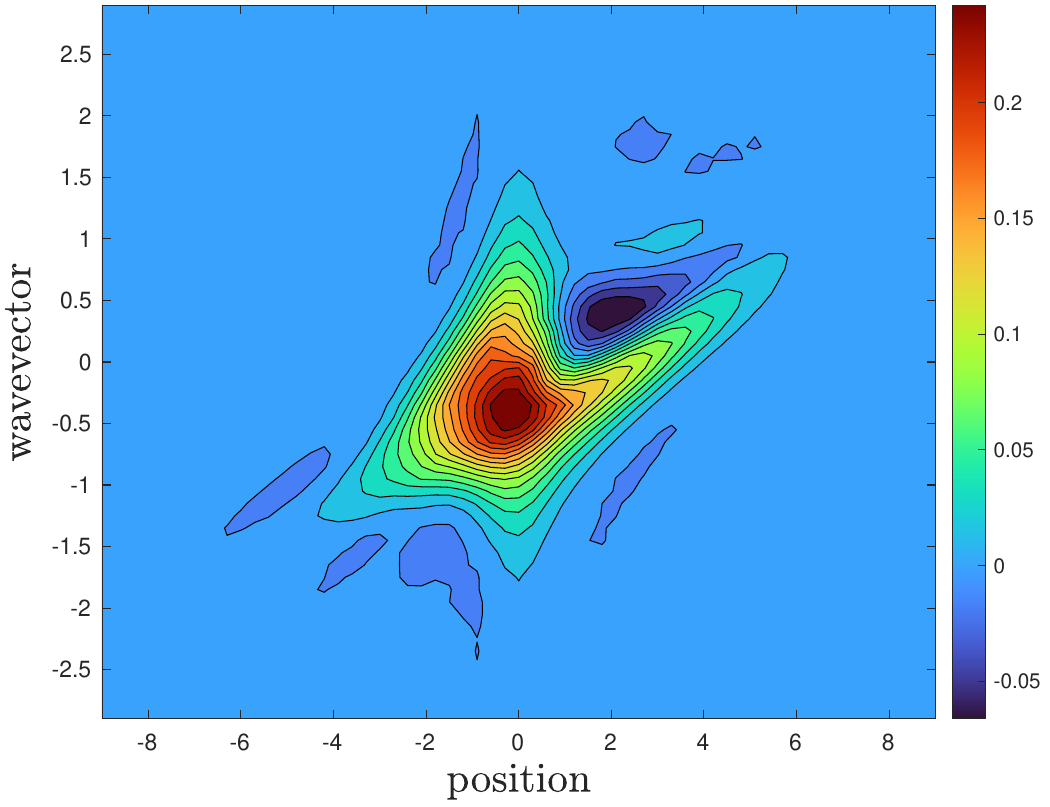}}
{\includegraphics[width=0.32\textwidth,height=0.21\textwidth]{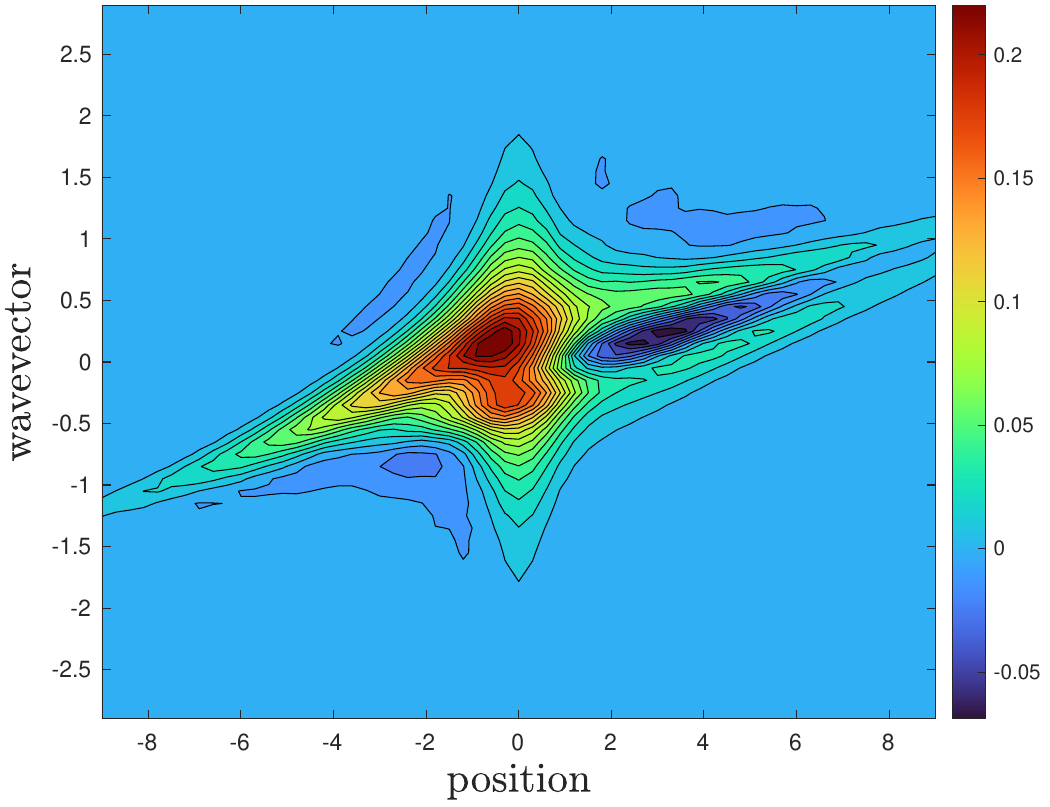}}
{\includegraphics[width=0.32\textwidth,height=0.21\textwidth]{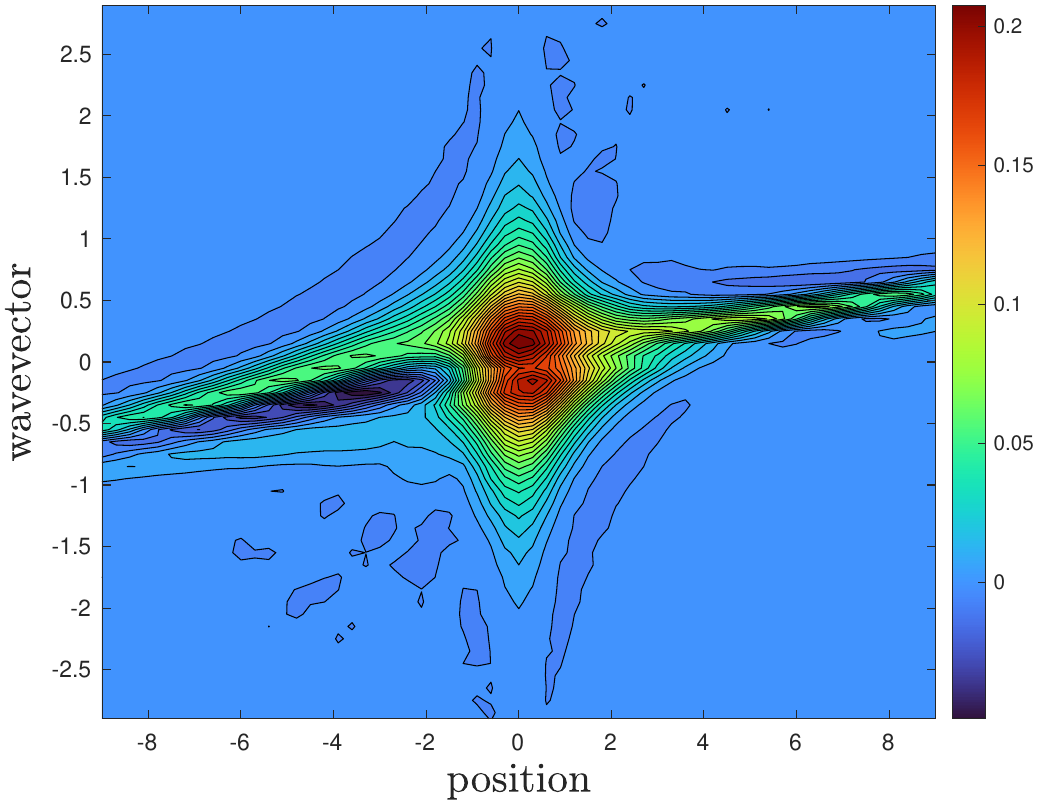}}}
\\
\centering
\subfigure[Stochastic noises are observed near the tail of the Wigner function, $t=4 \to 8 \to 15$a.u.]{
{\includegraphics[width=0.32\textwidth,height=0.21\textwidth]{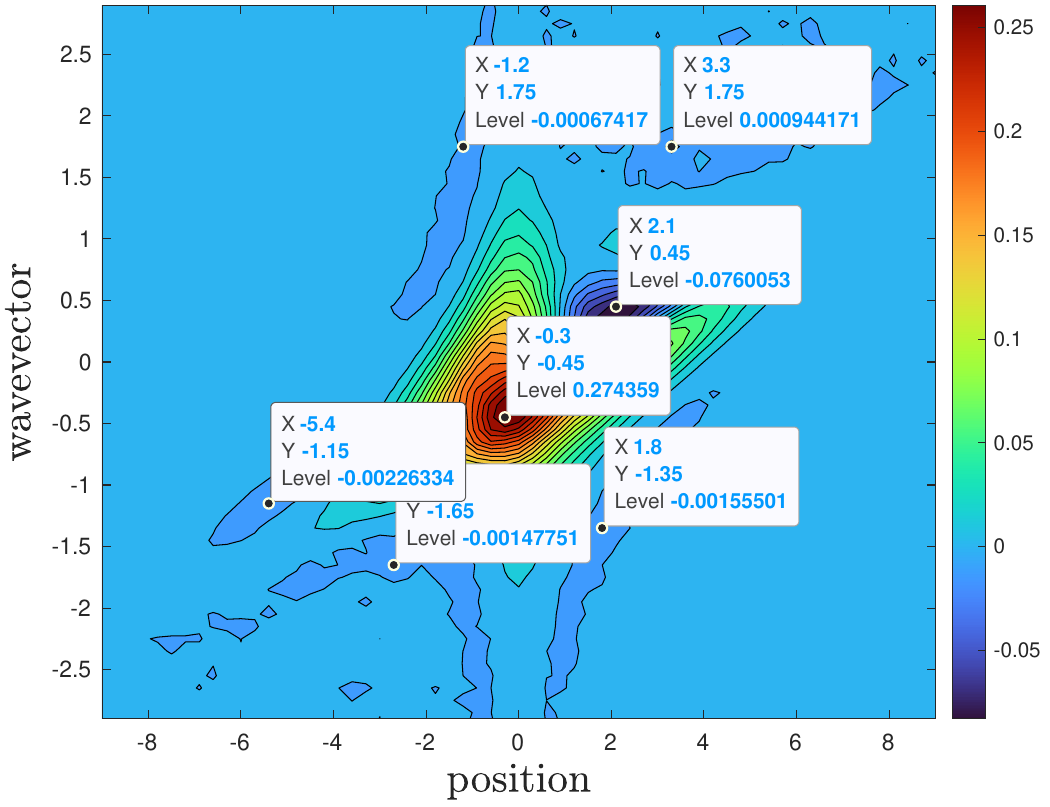}}
{\includegraphics[width=0.32\textwidth,height=0.21\textwidth]{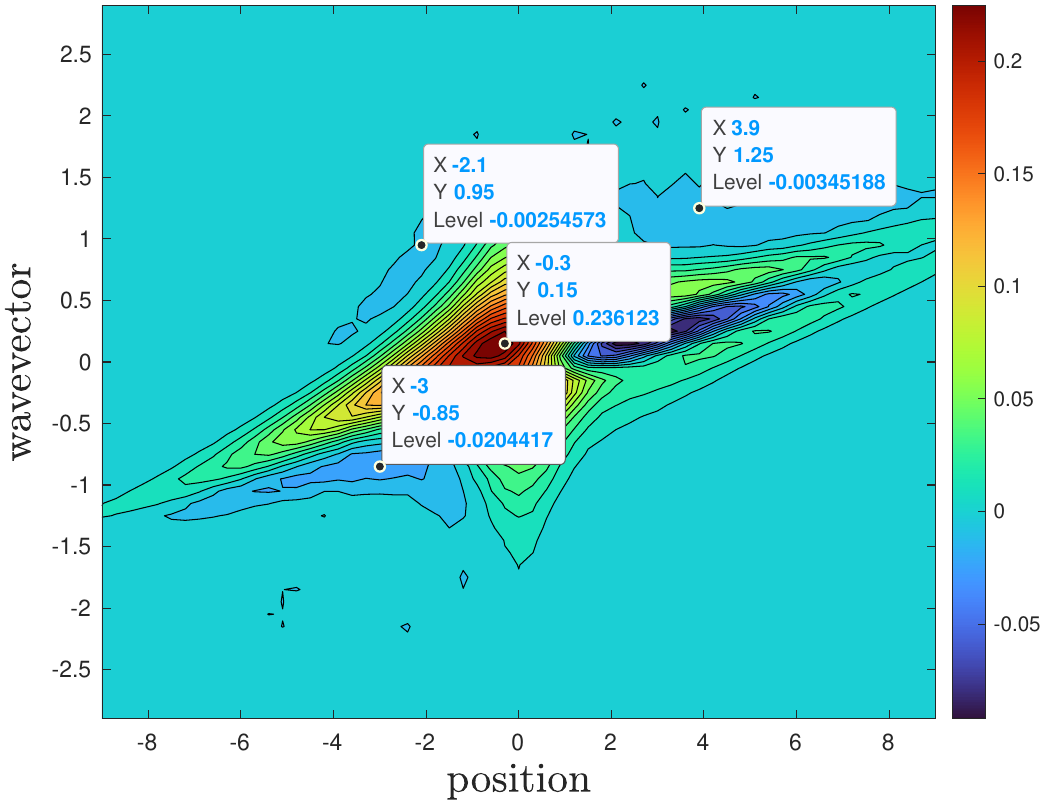}}
{\includegraphics[width=0.32\textwidth,height=0.21\textwidth]{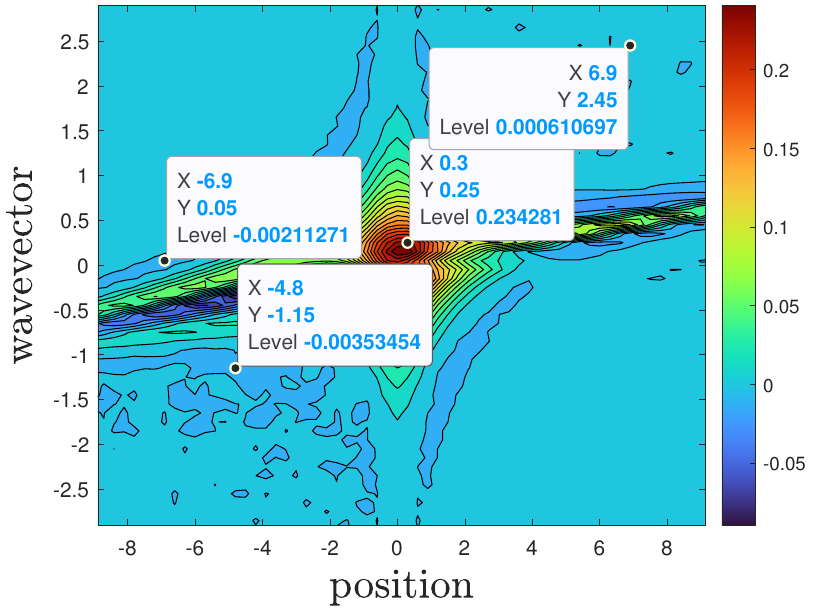}}}
\caption{\small  Snapshots of the reduced Wigner function $W_1(x_1, k_1, t)$  by the deterministic scheme (left) and WBRW-SPA-SPADE with $N_0 = 10^8$ under either the maximal gap ($\vartheta = 0.003$) (middle) or the difference gap ($\vartheta= 0.02$) (right). The particle-based stochastic algorithms can properly capture the double-peak structure (Coulomb collision) and negative valley (uncertainty principle).}
\label{supp_redist_time_evolution_max}
\end{figure}

For visualization of numerical results, we plot the snapshots of $W_1(x_1, k_1, t)$ at the instants $t= 4, 8, 15$a.u. in Figure \ref{supp_redist_time_evolution_max}, as well as the projection of $P_{xy}$ in the first direction in Figure \ref{supp_xdist_time_evolution_max}, where
\begin{equation}
P_x(x,t) = \int_{\mathbb{R}} P_{xy}(x, x_2, t) \D x_2.
\end{equation}
The parameters for the group with the maximal gap are $N_0 = 1 \times 10^8, \vartheta = 0.003, m = 512$ to ensure the accuracy and avoid the oversampling problem.  From Figure \ref{supp_comparison_SPADE_PAUM_max}, it is verified that SPADE under both gaps can capture the double-peak structure (Coulomb collision) and negative valley (uncertainty principle). 
 Even the tail parts can be reconstructed, albeit with some random noises. Numerical errors are mainly concentrated near the negative valley.   

The smoothing effect is observed in the spatial marginal distribution (see Figure \ref{supp_xdist_time_evolution_max}). Small errors near the peak and left shoulder are observed at $t = 15$a.u. Fortunately, the difference can be compensated when the partition is refined (as $\vartheta$ goes down). It seems that SPADE under the maximal gap may suffer more from the  smoothing effects since there is an evident collapse at the peak. 

\section{Performance evaluation of SPADE in 12-D phase space}
\label{supp_12d_wigner}
Now we would like to demonstrate the potential of SPADE in resolving the first-principle solution to non-equilibrium proton-electron coupling, where both proton and electron are treated quantum mechanically. This requires to evolve the Wigner function $f(\bx_e, \bx_p, \bk_e, \bk_p, t)$ in 12-D phase space.

In principle, the proton-electron Wigner equation can be solved by separation of variables. However, it is somehow difficult to make a direct comparison between the stochastic Wigner algorithm and deterministic reference solution in the centre-of-mass coordinate. To facilitate the benchmark, we simply choose an initial uncorrelated Wigner function, say, $f_{e}(\bx_e, \bk_e, 0) f_p(\bx_p, \bk_p, 0)$ with
\begin{equation}
\begin{split}
f_e(\bx_e, \bk_e, 0) &= \pi^{-3} \me^{-\frac{1}{2}((x_{e,1} - 1)^2 + x_{e,2}^2 + x_{e,3}^2)} \me^{-2(k_{e,1}^2 + k_{e,2}^2 + k_{e,3}^2)}, \\
f_p(\bx_p, \bk_p, 0) &= \pi^{-3} \me^{-\frac{100^2}{2}(x_{p,1}^2 + x_{p,2}^2 + x_{p,3}^2)} \me^{-\frac{2}{100^2}(k_{p,1}^2 + k_{p,2}^2 + k_{p,3}^2)}.
\end{split}
\end{equation}
The asymptotic approximation \eqref{supp_eq.asymptotic_approximation} allows us to make a quantitative comparison between the particle-based stochastic algorithm and the deterministic characteristic-spectral-mixed scheme \cite{XiongZhangShao2022}, where the Wigner function is represented as a tensor product of $75^3$ cubic spline basis in $\bx$-space and $80^3$ Fourier basis in $\bk$-space (with mesh size $73^3\times 80^3 \approx 2\times10^{11}$) to attain high accuracy.

\begin{figure}[!h]
\centering
\subfigure[Deterministic (left) and particle solutions at $1$a.u. under $N_0 = 4\times 10^7$ (middle) and $10^8$ (right).]{
{\includegraphics[width=0.32\textwidth,height=0.22\textwidth]{./redist_QCS_x_0010.pdf}}
{\includegraphics[width=0.32\textwidth,height=0.22\textwidth]{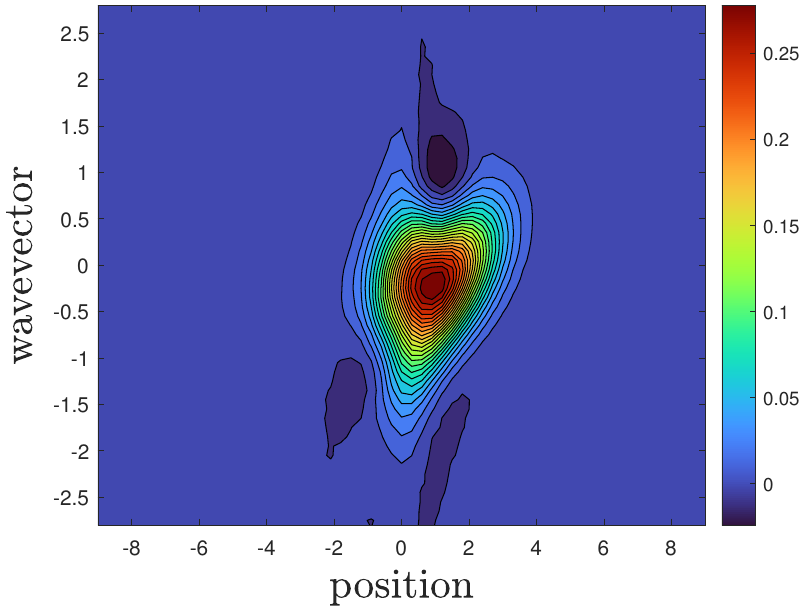}}
{\includegraphics[width=0.32\textwidth,height=0.22\textwidth]{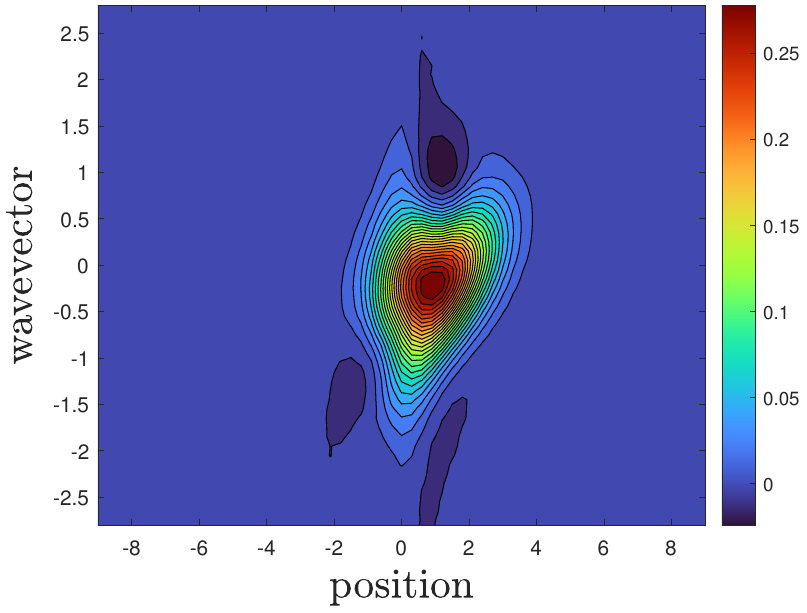}}}
\\
\centering
\subfigure[Deterministic (left) and particle solutions at $2$a.u. under $N_0 = 4\times 10^7$ (middle) and $10^8$ (right).]{
{\includegraphics[width=0.32\textwidth,height=0.22\textwidth]{./redist_QCS_x_0020.pdf}}
{\includegraphics[width=0.32\textwidth,height=0.22\textwidth]{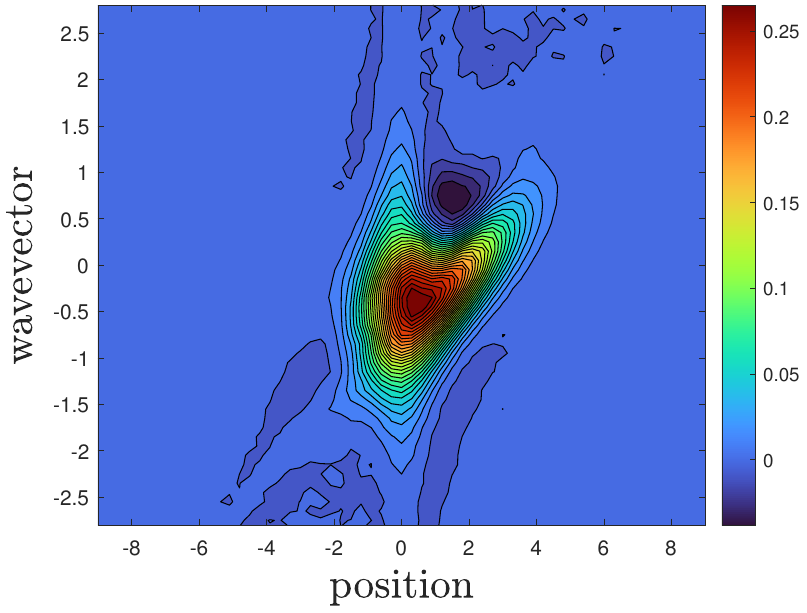}}
{\includegraphics[width=0.32\textwidth,height=0.22\textwidth]{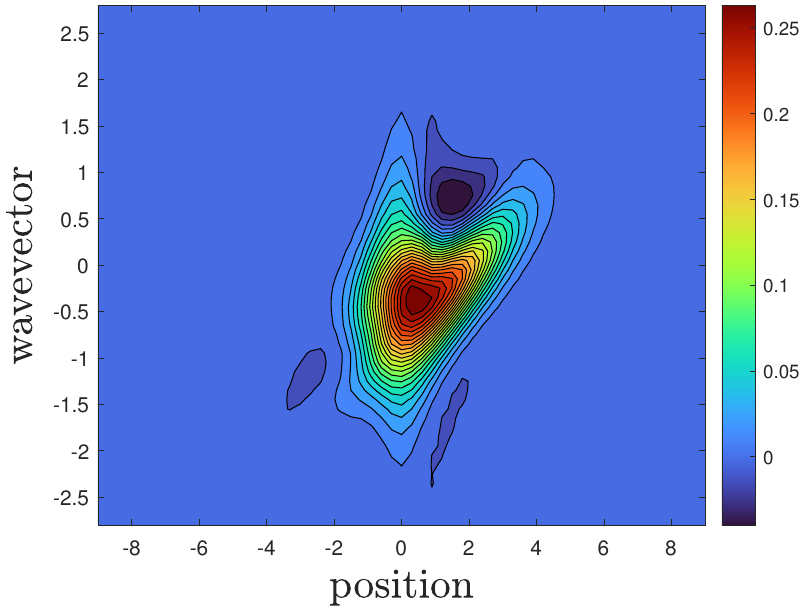}}}
\\
\centering
\subfigure[Deterministic (left) and particle solutions at $4$a.u. under $N_0 = 4\times 10^7$ (middle) and $10^8$ (right).]{
{\includegraphics[width=0.32\textwidth,height=0.22\textwidth]{./redist_QCS_x_0040.pdf}}
{\includegraphics[width=0.32\textwidth,height=0.22\textwidth]{./redist6d_SPADE_n4000_T4.pdf}}
{\includegraphics[width=0.32\textwidth,height=0.22\textwidth]{./redist6d_SPADE_n10000_T4.pdf}}}
\\
\centering
\subfigure[Deterministic (left) and particle solutions at $8$a.u. under $N_0 = 4\times 10^7$ (middle) and $10^8$ (right).]{
{\includegraphics[width=0.32\textwidth,height=0.22\textwidth]{./redist_QCS_x_0080.pdf}}
{\includegraphics[width=0.32\textwidth,height=0.22\textwidth]{./redist6d_SPADE_n4000_T8.pdf}}
{\includegraphics[width=0.32\textwidth,height=0.22\textwidth]{./redist6d_SPADE_n10000_T8.pdf}}}
\\
\centering
\subfigure[Deterministic (left) and particle solutions at $12$a.u. under $N_0 = 4\times 10^7$ (middle) and $10^8$ (right).]{
{\includegraphics[width=0.32\textwidth,height=0.22\textwidth]{./redist_QCS_x_0120.pdf}}
{\includegraphics[width=0.32\textwidth,height=0.22\textwidth]{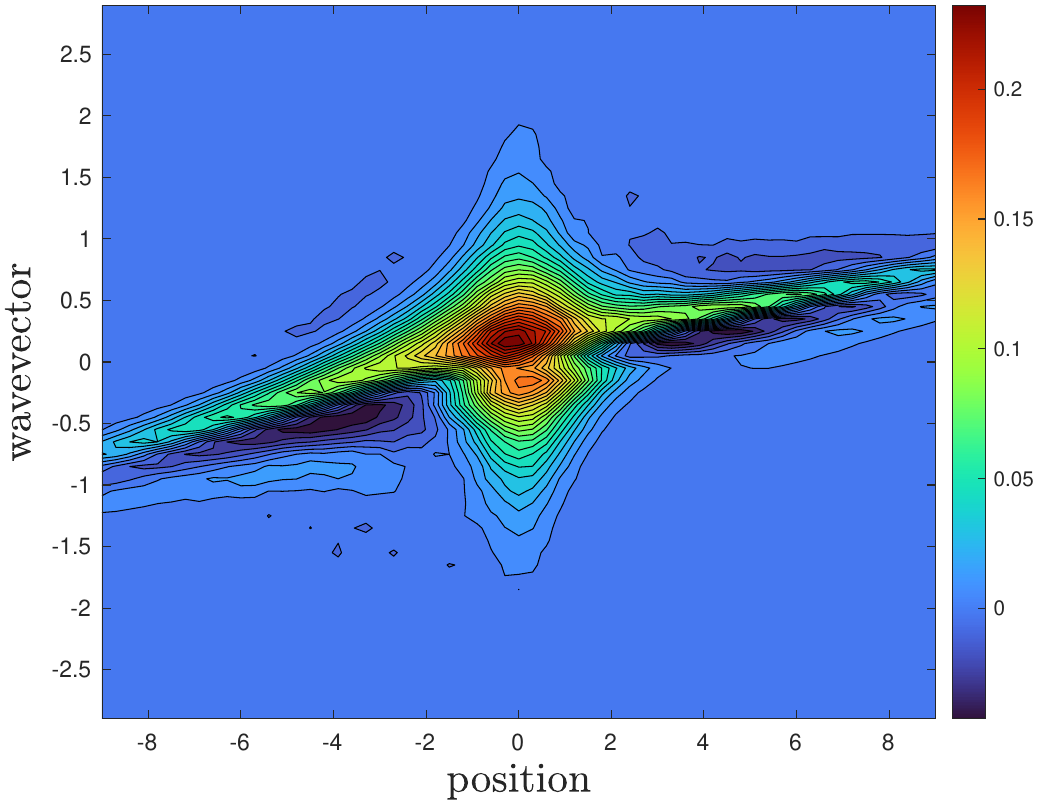}}
{\includegraphics[width=0.32\textwidth,height=0.22\textwidth]{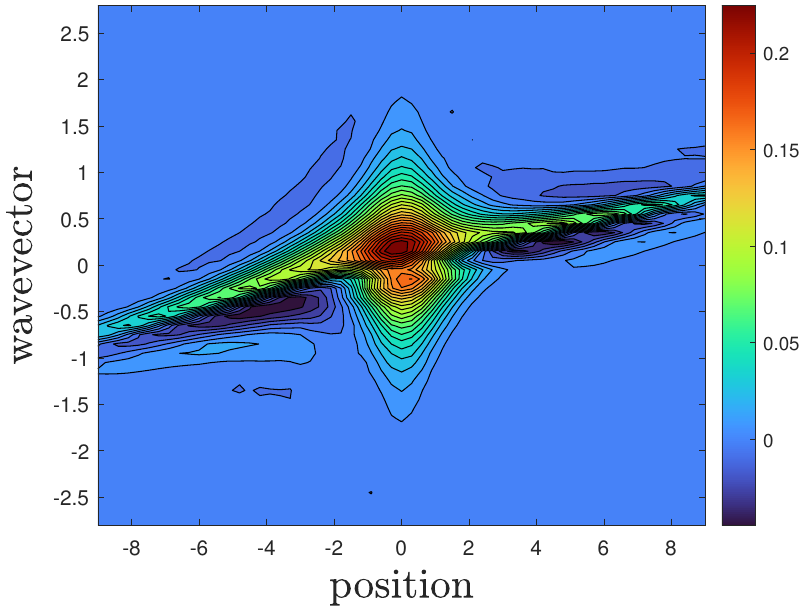}}}
\caption{\small  The 12-D proton-electron coupling: Visualization of the reduced electron Wigner function $W_1(x, k, t)$ from $t=1$a.u. to $5$a.u., produced by the deterministic scheme (left), the 12-D stochastic simulations under $N_0 = 4\times 10^7$ (middle) and $N_0 = 1\times 10^8$(right), where $\vartheta = 0.01$ is fixed. The projection of many-body Wigner function seems to coincide with the single-body counterpart under the asymptotic approximation.
 \label{supp_12d_wigner_snapshots_reduced_1}}
\end{figure}
\begin{figure}[!h]
\centering
\subfigure[$P_{xy}$ at $t=1$a.u., deterministic (left) and particle (middle)]{
{\includegraphics[width=0.32\textwidth,height=0.22\textwidth]{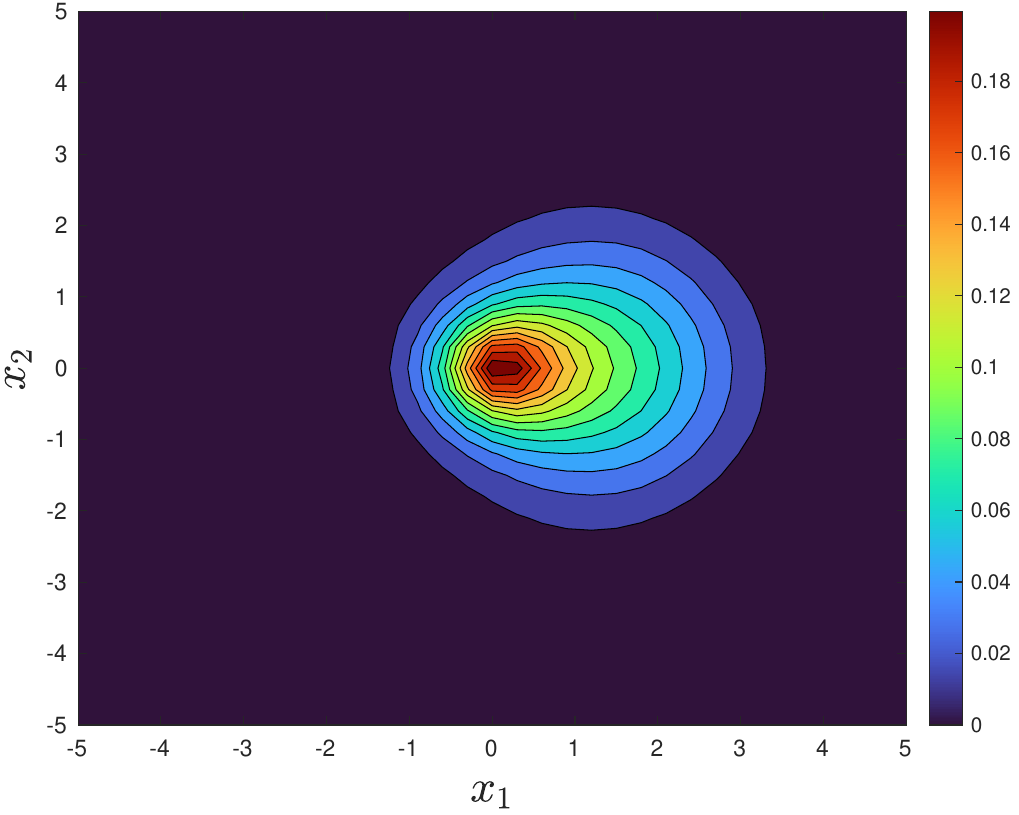}}
{\includegraphics[width=0.32\textwidth,height=0.22\textwidth]{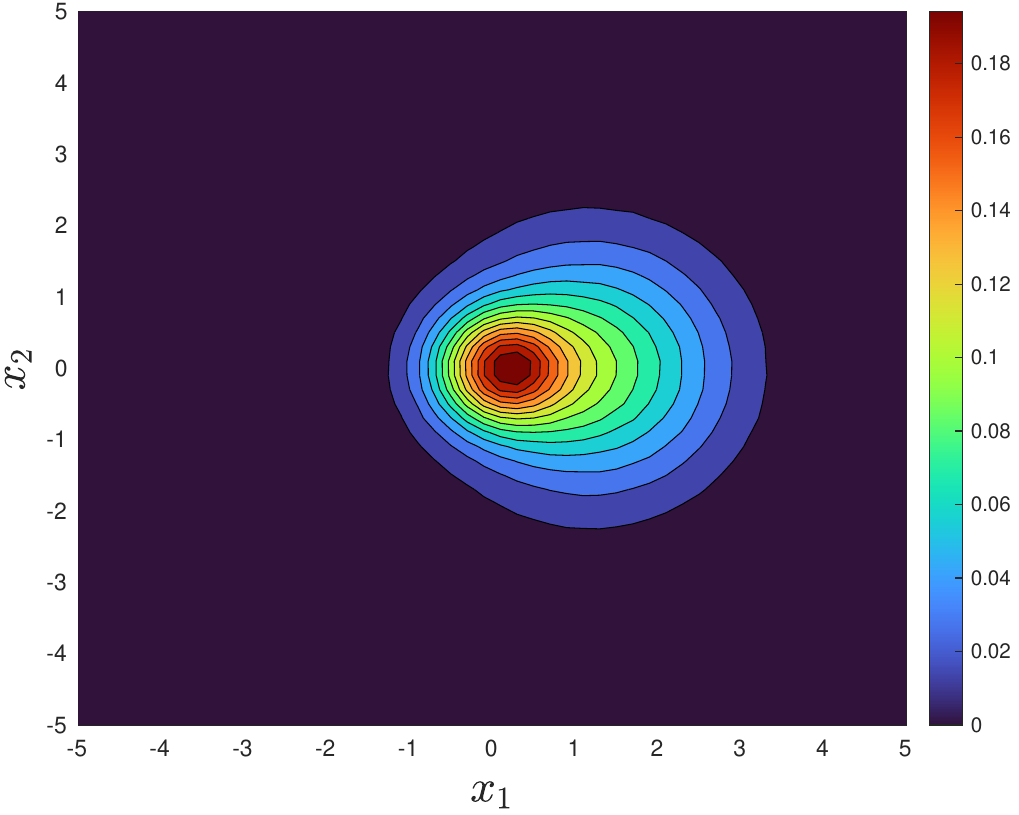}}}
\subfigure[$P(x, t)$ at $t=1$a.u.]
{\includegraphics[width=0.32\textwidth,height=0.22\textwidth]{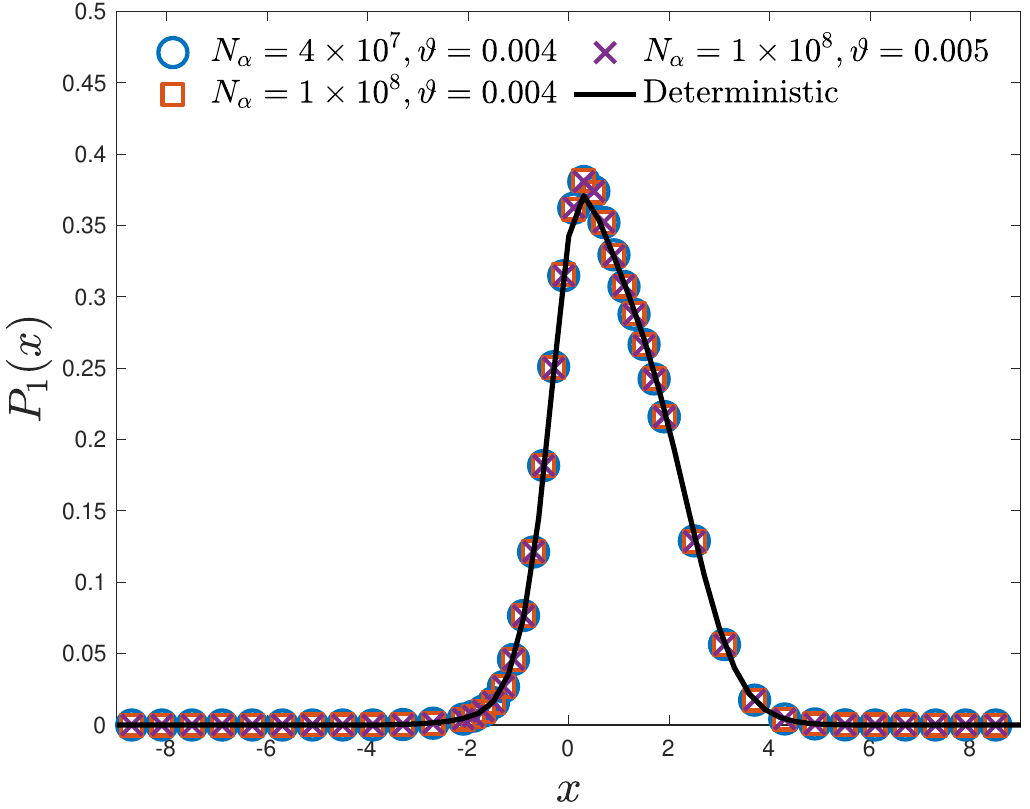}}
\\
\centering
\subfigure[$P_{xy}$ at $t=2$a.u., deterministic (left) and particle (middle)]{
{\includegraphics[width=0.32\textwidth,height=0.22\textwidth]{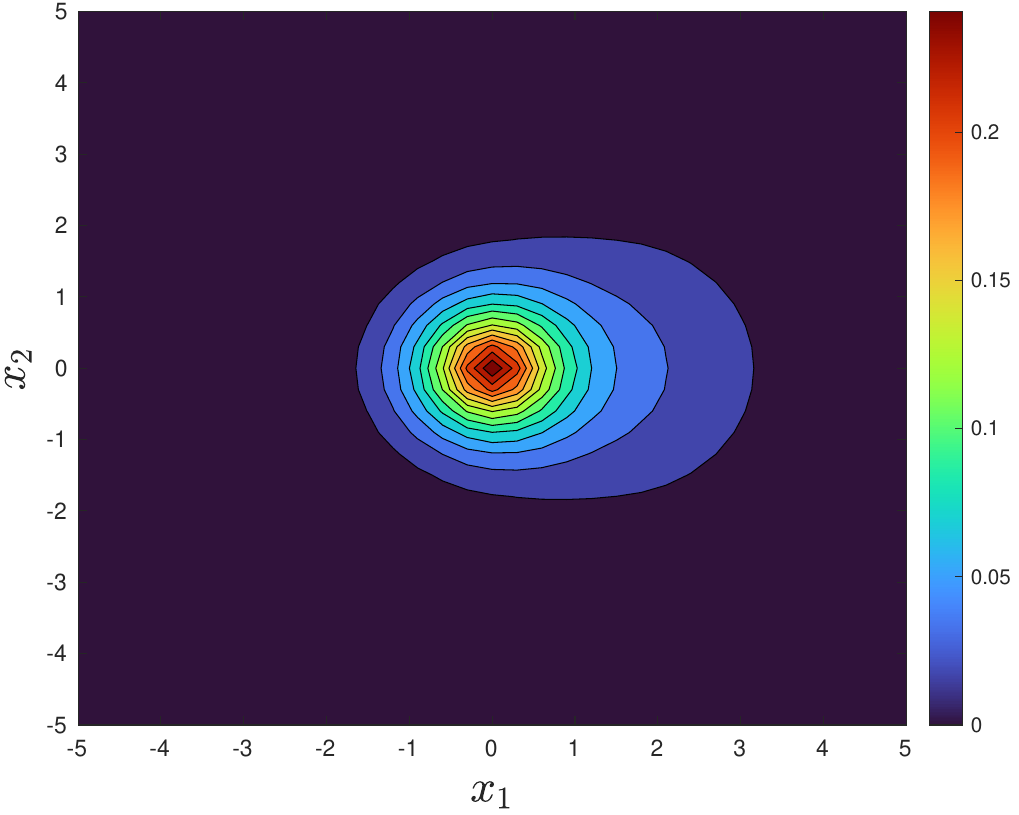}}
{\includegraphics[width=0.32\textwidth,height=0.22\textwidth]{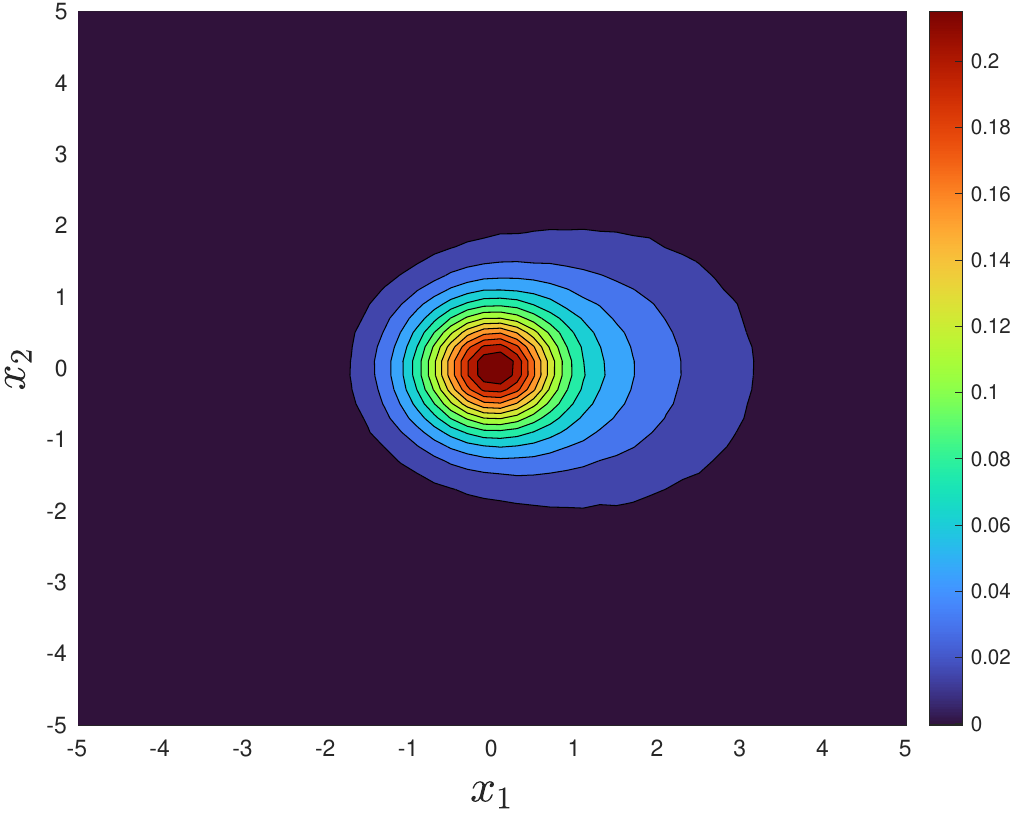}}}
\subfigure[$P_1(x, t)$ at $t=2$a.u.]
{\includegraphics[width=0.32\textwidth,height=0.22\textwidth]{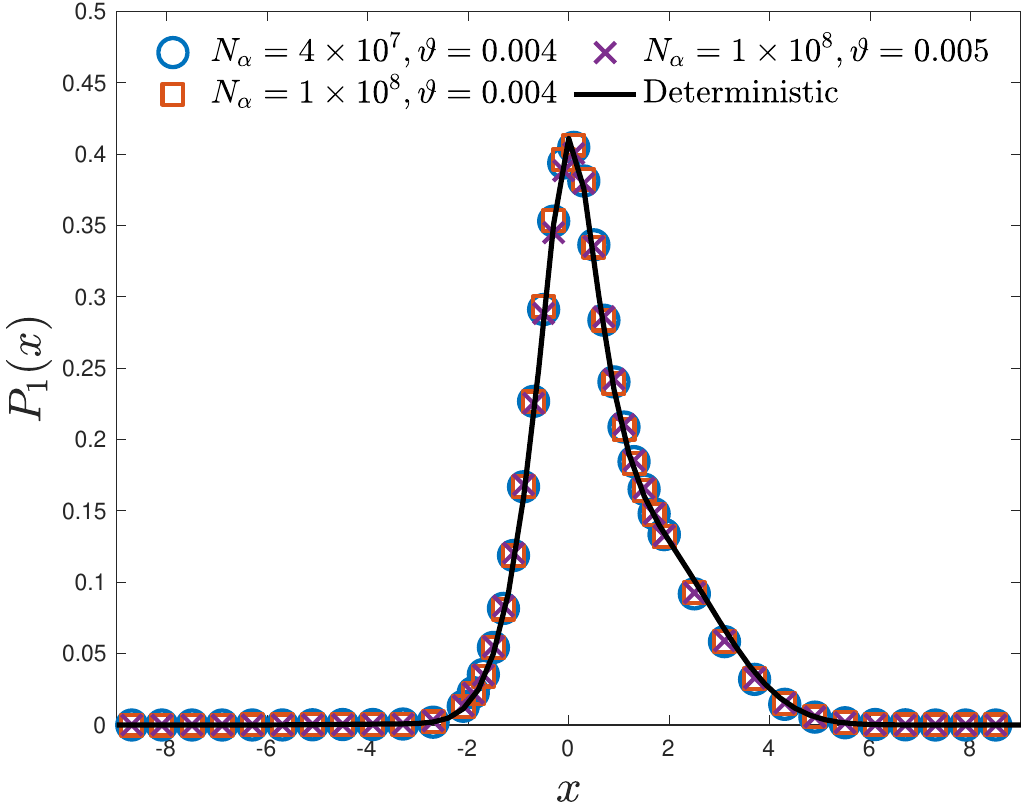}}
\\
\centering
\subfigure[$P_{xy}$ at $t=4$a.u., deterministic (left) and particle (middle)]{
{\includegraphics[width=0.32\textwidth,height=0.22\textwidth]{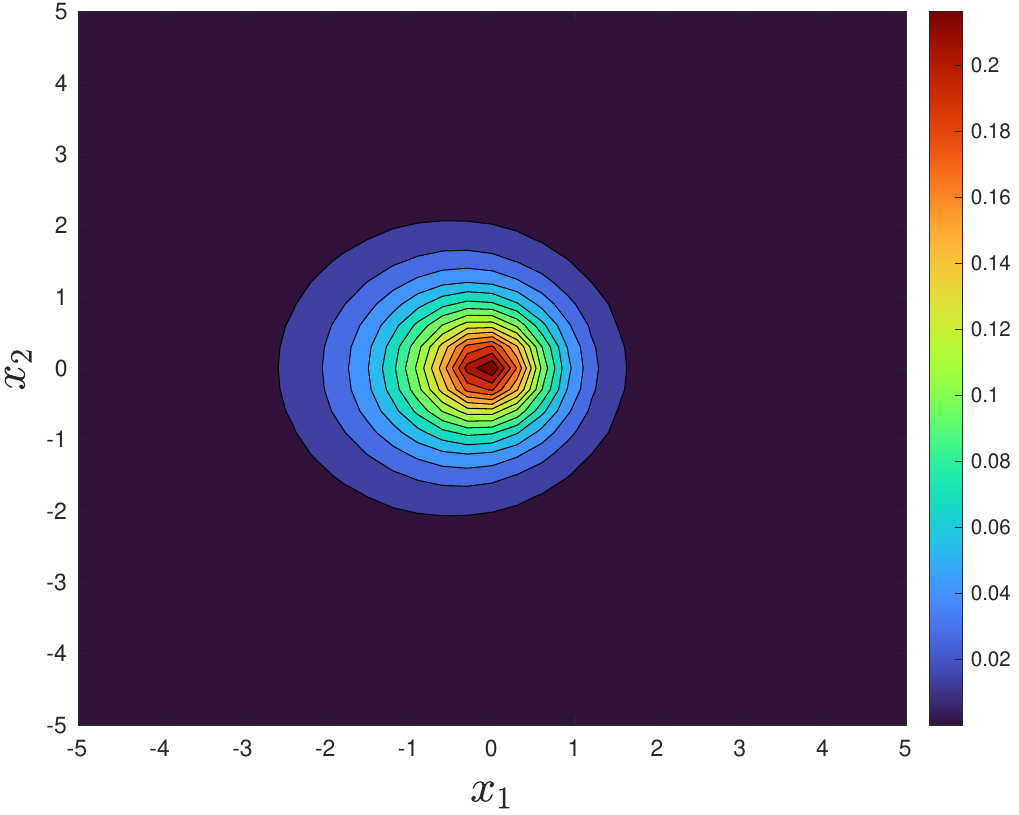}}
{\includegraphics[width=0.32\textwidth,height=0.22\textwidth]{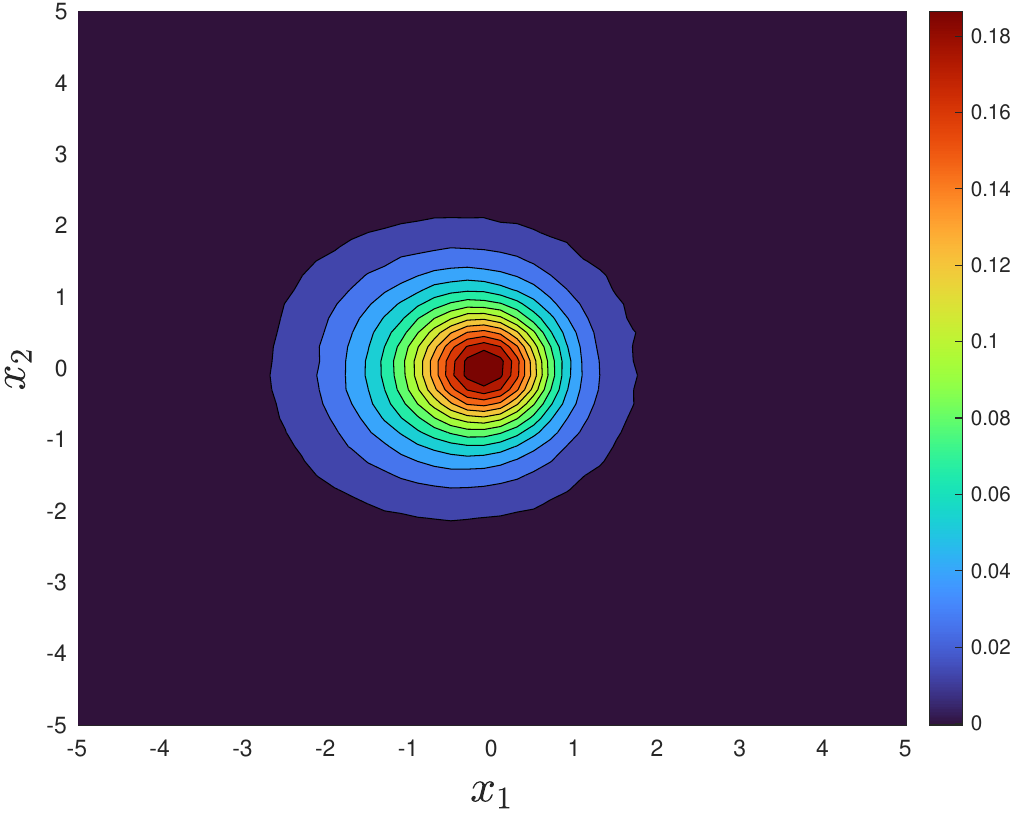}}}
\subfigure[$P_1(x, t)$ at $t=4$a.u.]
{\includegraphics[width=0.32\textwidth,height=0.22\textwidth]{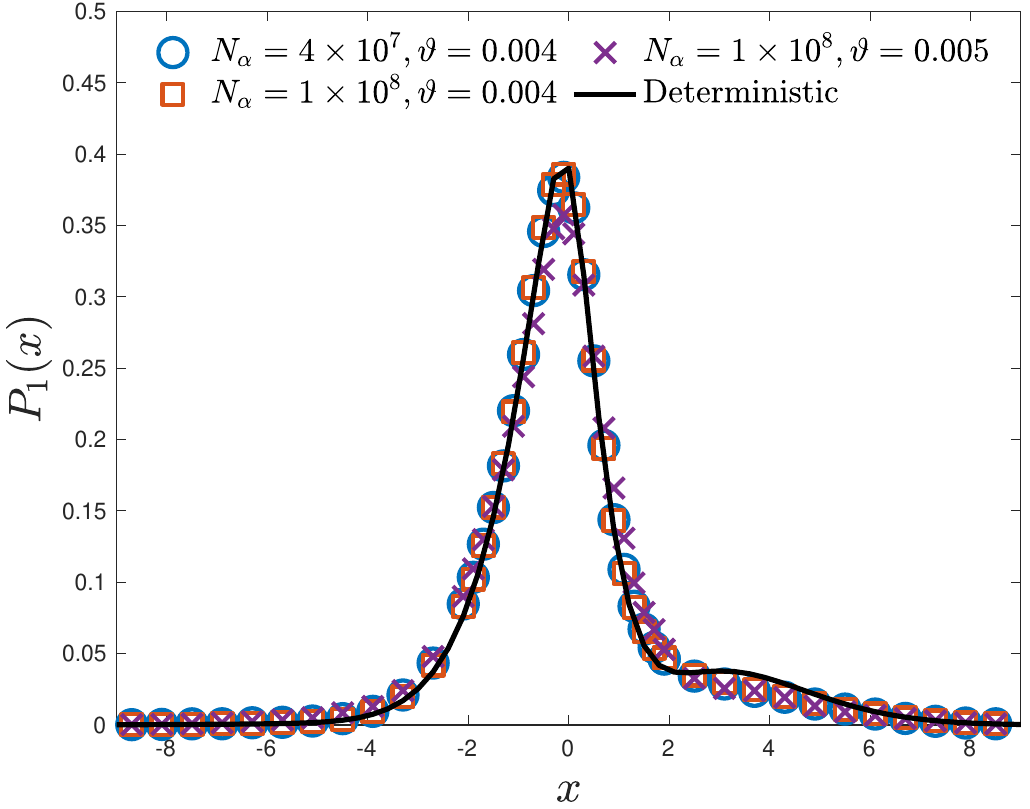}}
\\
\centering
\subfigure[$P_{xy}$ at $t=6$a.u., deterministic (left) and particle (middle)]{
{\includegraphics[width=0.32\textwidth,height=0.22\textwidth]{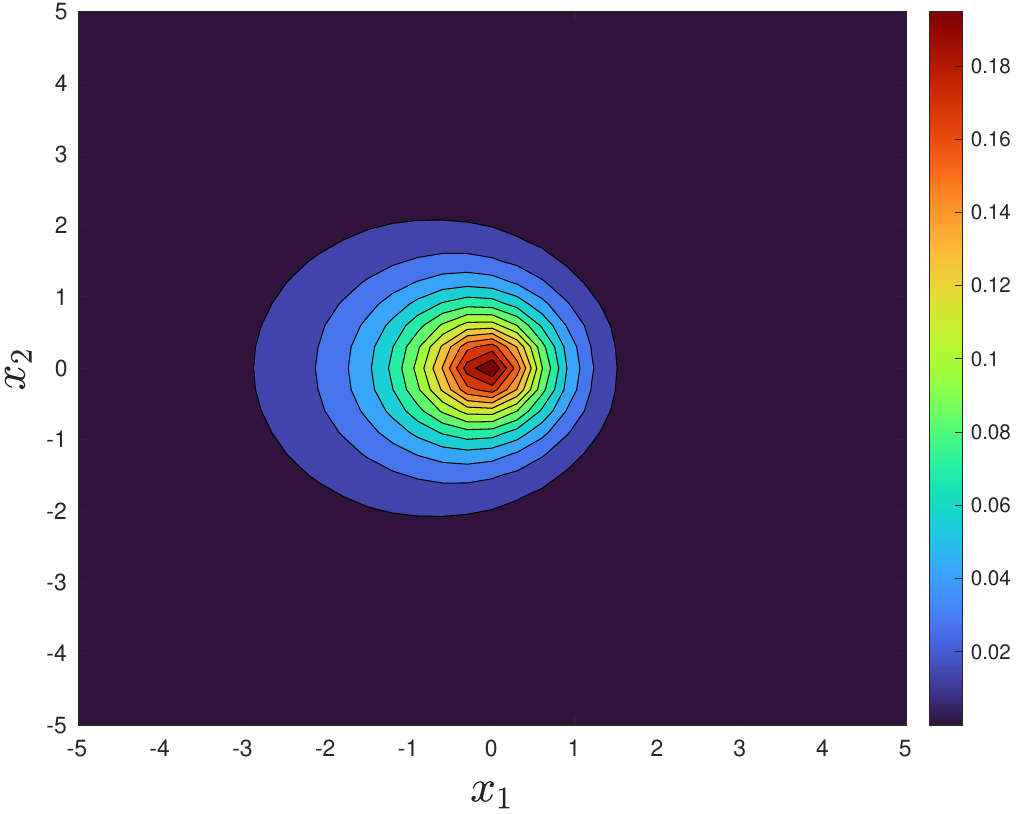}}
{\includegraphics[width=0.32\textwidth,height=0.22\textwidth]{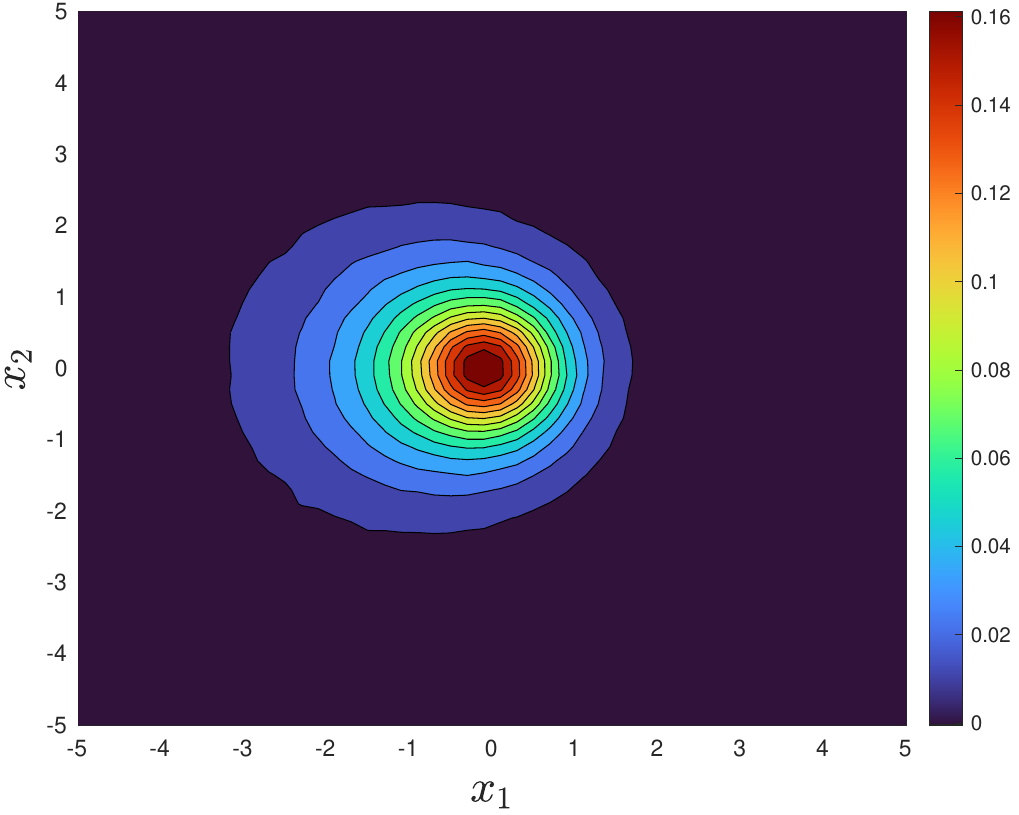}}}
\subfigure[$P_1(x, t)$ at $t=6$a.u.]
{\includegraphics[width=0.32\textwidth,height=0.22\textwidth]{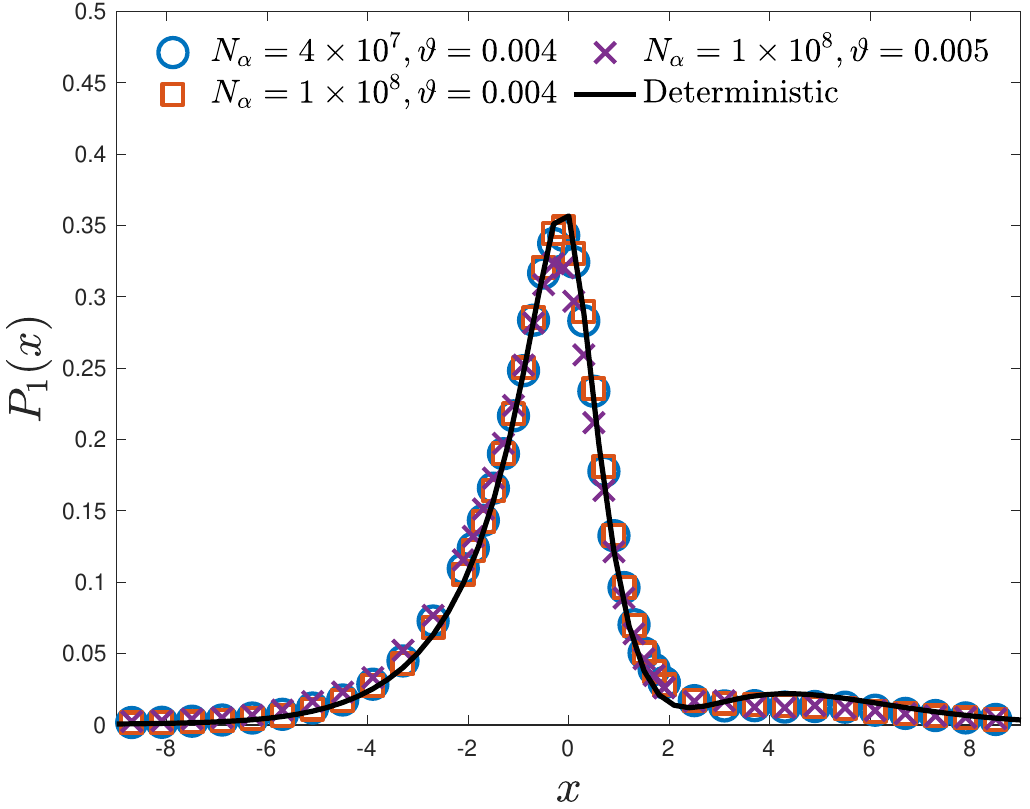}}
\\
\centering
\subfigure[$P_{xy}$ at $t=9$a.u., deterministic (left) and particle (middle)]{
{\includegraphics[width=0.32\textwidth,height=0.22\textwidth]{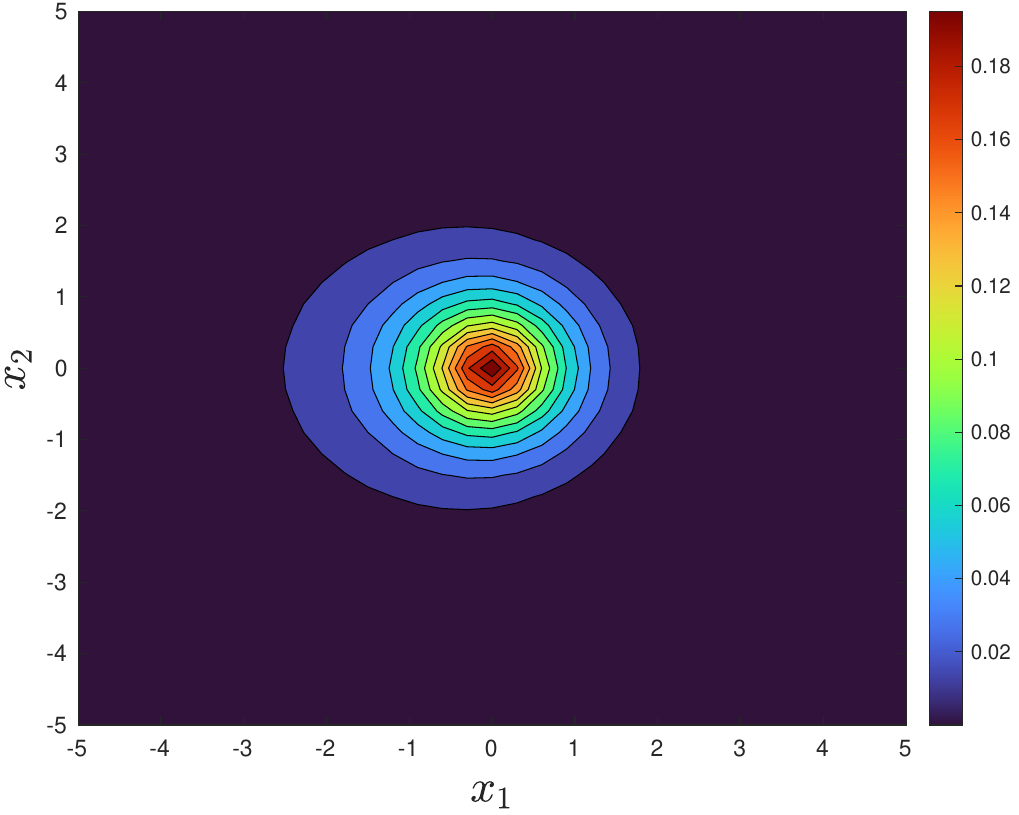}}
{\includegraphics[width=0.32\textwidth,height=0.22\textwidth]{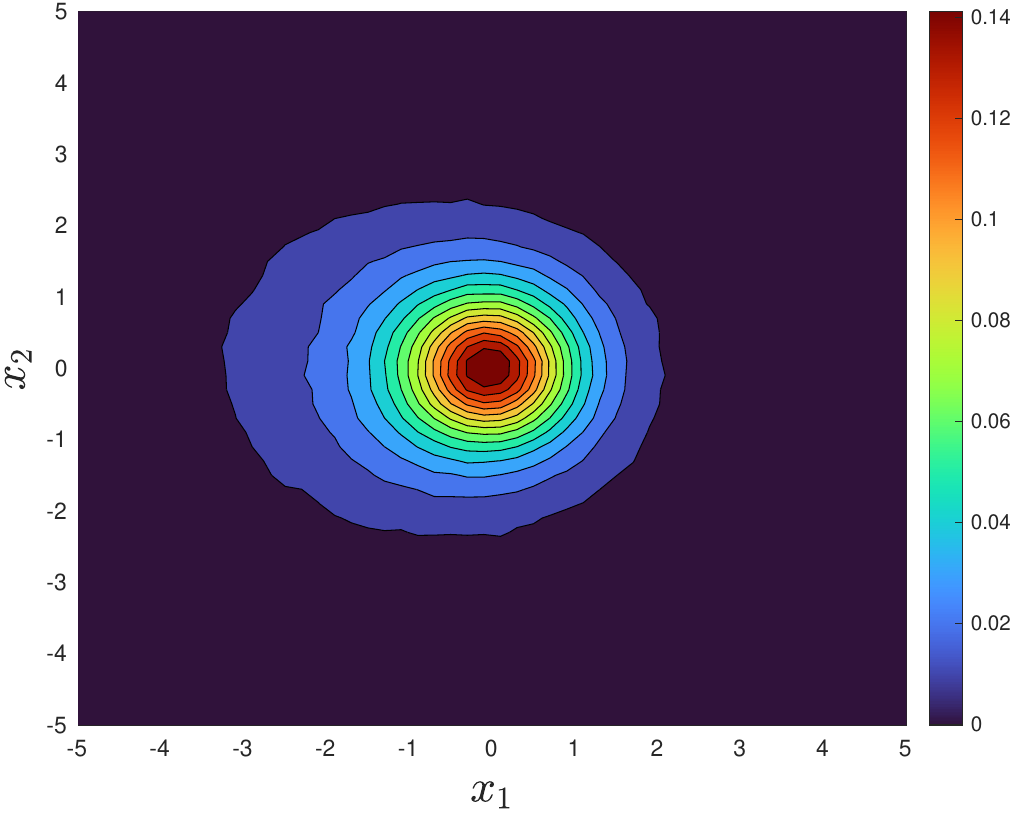}}}
\subfigure[$P_1(x, t)$ at $t=9$a.u.]
{\includegraphics[width=0.32\textwidth,height=0.22\textwidth]{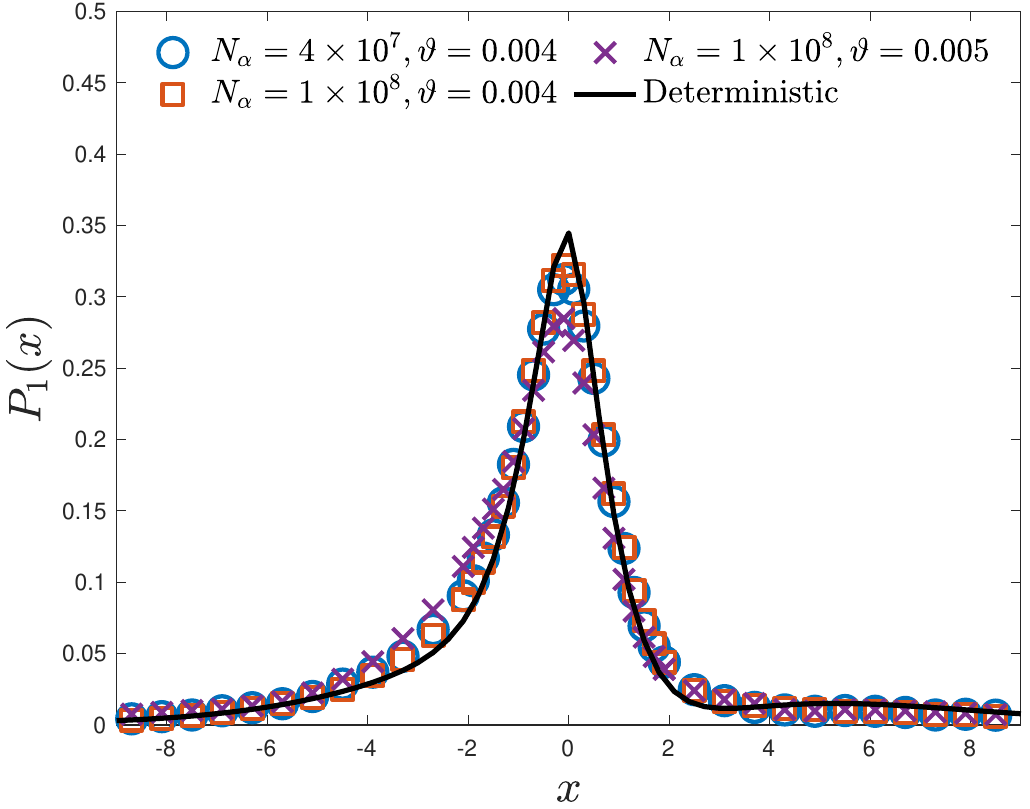}}
\caption{\small   The 12-D proton-electron coupling: Visualization of the marginal spatial distribution $P_{xy}(x_1, x_2, t)$ produced by the deterministic scheme (left) and particle-based method with $N_0 = 1\times 10^8$, $\vartheta=0.01$ (middle), as well as the one-dimensional projection $P_1(x, t)$ (right). The projection of many-body Wigner function seems to coincide with the single-body counterpart under the asymptotic approximation. The stochastic noises can be suppressed by either increasing sample size $N_0$ or refining the partition (choosing smaller $\vartheta$).} 
\label{supp_12d_wigner_snapshots_xdist}
\end{figure}

The snapshots of the spatial marginal density of electron 
\begin{equation}
P(x_1, x_2, t) = \iiiint_{\mathbb{R}\times \mathbb{R}^3 \times \mathbb{R}^3 \times \mathbb{R}^3} f(x_1, x_2, x_3, \bx_p, \bk_e, \bk_p, t) \D x_3 \D \bx_p \D \bk_e \D \bk_p,
\end{equation}
the reduced electron Wigner function
\begin{equation}
W_1(x, k, t) = \iiiint_{\mathbb{R}^2\times \mathbb{R}^3 \times \mathbb{R}^2 \times \mathbb{R}^3} f(\bx_e, \bx_p, \bk_e, \bk_p, t) \D x_{e,2} \D x_{e,3}\D \bx_p \D k_{e,2} \D k_{e,3}\D \bk_p,
\end{equation}
and the reduced proton Wigner function
\begin{equation}
W_4(x, k, t) = \iiiint_{\mathbb{R}^3\times \mathbb{R}^2 \times \mathbb{R}^3 \times \mathbb{R}^2} f(\bx_e, \bx_p, \bk_e, \bk_p, t) \D \bx_{e} \D x_{p, 2} \D x_{p, 3} \D \bk_{e} \D k_{p, 2}\D k_{p,3}
\end{equation}
are visualized in Figures  \ref{supp_12d_wigner_snapshots_reduced_1} and \ref{supp_12d_wigner_snapshots_xdist}, respectively. The following observations are made from the results.

\begin{description}

\item[(1)] From Figure \ref{supp_12d_wigner_snapshots_reduced_1}, the electron almost obeys the single-body Wigner dynamics, which coincides with the prediction of the asymptotic approximation. The main features of the electron Wigner function, including the double-peak structure induced by the Coulomb collisions and the negative valley that manifests the uncertainty, can be captured by the  particle-based stochastic algorithm, albeit with slight stochastic noises.

\item[(2)] The particle-based stochastic algorithm can capture the pattern of spatial unharmonic oscillation of electron. The difference mainly lies at the peak of the wavepacket, which may be smoothed out by the piecewise constant reconstruction. Fortunately, from  the comparison of $P_1(x) = \int_{\mathbb{R}}P_{xy}(x, x_2, t) \D x_2$ on the right column of Figure \ref{supp_12d_wigner_snapshots_xdist}, the difference can be compensated by either increasing $N_0$ or refining the partition (choosing smaller $\vartheta$). 

\end{description}

\clearpage
\section{Calculation of the star discrepancy of a sequence}

In principle, the star discrepancy can be attained by searching all the critical boxes with the upper coordinate $\by$, with its dimension drawn from that of all possibles coordinates in $\mathcal{X}$. However, the calculation of the star discrepancy is in general a NP-hard problem, say, there might not exist an algorithm that can attain the maximal value with polynomial complexity \cite{WinkerFang1997}. 

Several algorithms based on integer optimization are proposed, including the threshold accepting \cite{WinkerFang1997}, the genetic algorithm \cite{Shah2010} and the improved threshold accepting (TA-improved) method \cite{GnewuchWahlstromWinzen2012}. In particular, the heuristic threshold accepting with improved sampling strategy is shown to be efficient in moderate large dimensional problem (especially $d=20$-$60$).

The threshold accepting algorithm, often referred to a refined local search algorithm, is an integer optimization heuristic. For a set $\mathcal{X}$ of $N$ points, the collection of critical boxes is denoted by $\bar{\Gamma}(\mathcal{X})$, we set a total number of iterations $I$ and the number of independent trials $N_{tr}$ and a non-positive threshold value $T$. During the iterations, $T$ will increase until it reaches zero. This procedure helps to get rid of being trapped in local minima. More details, e.g., searching a neighbor by sampling and setting the threshold $T$, can be found in \cite{GnewuchWahlstromWinzen2012}.  

A series of benchmarks has been made for the dimensionality ranging from $d=6$ to $d=360$. Here $N$, $I$ and $N_{tr}$ denote the number of points, number of iterations in threshold accepting and count of independent trials, respectively. The points are generated by random sampling from uniform distribution on $[0,1]^d$. Since the algorithms intends to solve an optimization problem, the maximal value attained in the experiments can be regarded as the references.

All the results are listed in Tables.~\ref{supp_tab_disc_6d} to \ref{supp_tab_disc_360d} and uncover several facts.

\begin{description}

\item[(1)] For low-dimensional points, under a small iteration time and a few trial times, TA-improved algorithm can yields satisfactory results.

\item[(2)] For high-dimensional points, the number of iterations $I$ must be sufficiently large to ensure of the efficiency of searching. Too small $I$ (such as $I =16,32$), even under a large trial time $N_{tr}$ (such as $N_{tr} = 100$), fails to give reasonable approximations.

\item[(3)] When $I$ fixed, the computational time for searching points increases moderately as $d$ increases. However, when $d$ fixed, the computational time grows almost exponentially as $I$ becomes larger.

\item[(4)] For randomly distributed points, a boundary effect is observed. When $N$ is small and $d$ is large, the discrepancy turns out to be near $1$.

\end{description}

\begin{table}[!h]
  \centering
  \caption{\small Discrepancies of $6$-D points up to $N=10^4$ are calculated by TA-improved algorithm. The points are drawn from uniform distribution on $[0,1]^d$.
}\label{supp_tab_disc_6d}
 \begin{lrbox}{\tablebox}
  \begin{tabular}{cccccc}
\hline\hline
\multicolumn{6}{c}{$N= 10^2$, Maximal value is 2.845E-01.}\\
\hline
$N_{tr}$	&	$I=16$	&	$I=32$	&	$I=64$	&	$I=128$	&	$I=256$	\\
\hline
5	&	2.802E-01	&	2.826E-01	&	2.845E-01	&	2.845E-01	&	2.845E-01	\\
10	&	2.826E-01	&	2.845E-01	&	2.845E-01	&	2.845E-01	&	2.845E-01	\\
15	&	2.826E-01	&	2.845E-01	&	2.845E-01	&	2.845E-01	&	2.845E-01	\\
20	&	2.826E-01	&	2.845E-01	&	2.845E-01	&	2.845E-01	&	2.845E-01	\\
50	&	2.845E-01	&	2.845E-01	&	2.845E-01	&	2.845E-01	&	2.845E-01	\\
100	&	2.845E-01	&	2.845E-01	&	2.845E-01	&	2.845E-01	&	2.845E-01	\\
\hline
time per trial (s)&	5.257E-04	&	1.701E-03	&	5.374E-03	&	2.047E-02	&	7.848E-02	\\
\hline\hline
\multicolumn{6}{c}{$N= 10^3$, Maximal value is 6.453E-02.}\\
\hline
$N_{tr}$	&	$I=16$	&	$I=32$	&	$I=64$	&	$I=128$	&	$I=256$	\\
\hline
5	&	5.909E-02	&	6.109E-02	&	6.453E-02	&	6.453E-02	&	6.453E-02	\\
10	&	5.909E-02	&	6.352E-02	&	6.453E-02	&	6.453E-02	&	6.453E-02	\\
15	&	5.909E-02	&	6.352E-02	&	6.453E-02	&	6.453E-02	&	6.453E-02	\\
20	&	6.020E-02	&	6.352E-02	&	6.453E-02	&	6.453E-02	&	6.453E-02	\\
50	&	6.020E-02	&	6.453E-02	&	6.453E-02	&	6.453E-02	&	6.453E-02	\\
100	&	6.126E-02	&	6.453E-02	&	6.453E-02	&	6.453E-02	&	6.453E-02	\\
\hline
time per trial (s)	&	3.029E-03	&	9.893E-03	&	3.714E-02	&	1.442E-01	&	5.775E-01	\\
\hline\hline
\multicolumn{6}{c}{$N= 10^4$, Maximal value is 1.971E-02.}\\
\hline
$N_{tr}$	&	$I=16$	&	$I=32$	&	$I=64$	&	$I=128$	&	$I=256$	\\
\hline
5	&	1.724E-02	&	1.847E-02	&	1.884E-02	&	1.917E-02	&	1.956E-02	\\
10	&	1.768E-02	&	1.847E-02	&	1.888E-02	&	1.917E-02	&	1.956E-02	\\
15	&	1.768E-02	&	1.847E-02	&	1.888E-02	&	1.917E-02	&	1.956E-02	\\
20	&	1.768E-02	&	1.847E-02	&	1.888E-02	&	1.953E-02	&	1.956E-02	\\
50	&	1.768E-02	&	1.895E-02	&	1.933E-02	&	1.953E-02	&	1.957E-02	\\
100	&	1.799E-02	&	1.895E-02	&	1.937E-02	&	1.959E-02	&	1.971E-02	\\
\hline
time per trial (s)&	2.713E-02	&	9.407E-02	&	3.542E-01	&	1.379E+00	&	5.548E+00	\\
\hline\hline
 \end{tabular}
\end{lrbox}
\scalebox{1}{\usebox{\tablebox}}
\end{table}

\begin{table}[!h]
  \centering
  \caption{\small Discrepancies of $12$-D points up to $N=10^4$ are calculated by TA-improved algorithm. The points are drawn from uniform distribution on $[0,1]^d$.
}\label{supp_tab_disc_12d}
 \begin{lrbox}{\tablebox}
  \begin{tabular}{cccccc}
\hline\hline
\multicolumn{6}{c}{$N= 10^2$, Maximal value is 2.925E-01.}\\
\hline
$N_{tr}$	&	$I=16$	&	$I=32$	&	$I=64$	&	$I=128$	&	$I=256$	\\
\hline
5	&	2.843E-01	&	2.764E-01	&	2.870E-01	&	2.925E-01	&	2.925E-01	\\
10	&	2.843E-01	&	2.834E-01	&	2.870E-01	&	2.925E-01	&	2.925E-01	\\
15	&	2.843E-01	&	2.834E-01	&	2.925E-01	&	2.925E-01	&	2.925E-01	\\
20	&	2.843E-01	&	2.854E-01	&	2.925E-01	&	2.925E-01	&	2.925E-01	\\
50	&	2.843E-01	&	2.854E-01	&	2.925E-01	&	2.925E-01	&	2.925E-01	\\
100	&	2.843E-01	&	2.870E-01	&	2.925E-01	&	2.925E-01	&	2.925E-01	\\
\hline
time per trial (s)&	9.089E-04	&	2.691E-03	&	8.675E-03	&	3.272E-02	&	1.279E-01	\\
\hline\hline
\multicolumn{6}{c}{$N= 10^3$, Maximal value is 1.024E-01.}\\
\hline
$N_{tr}$	&	$I=16$	&	$I=32$	&	$I=64$	&	$I=128$	&	$I=256$	\\
\hline
5	&	9.225E-02	&	9.495E-02	&	9.914E-02	&	1.017E-01	&	1.024E-01	\\
10	&	9.225E-02	&	9.673E-02	&	9.914E-02	&	1.018E-01	&	1.024E-01	\\
15	&	9.225E-02	&	9.673E-02	&	1.012E-01	&	1.024E-01	&	1.024E-01	\\
20	&	9.225E-02	&	9.903E-02	&	1.022E-01	&	1.024E-01	&	1.024E-01	\\
50	&	9.225E-02	&	9.903E-02	&	1.022E-01	&	1.024E-01	&	1.024E-01	\\
100	&	9.225E-02	&	9.906E-02	&	1.022E-01	&	1.024E-01	&	1.024E-01	\\
\hline
time per trial (s)&	5.071E-03	&	1.708E-02	&	6.294E-02	&	2.412E-01	&	9.639E-01	\\
\hline\hline
\multicolumn{6}{c}{$N= 10^4$, Maximal value is 2.911E-02.}\\
\hline
$N_{tr}$	&	$I=16$	&	$I=32$	&	$I=64$	&	$I=128$	&	$I=256$	\\
\hline
5	&	2.116E-02	&	2.370E-02	&	2.835E-02	&	2.902E-02	&	2.911E-02	\\
10	&	2.262E-02	&	2.402E-02	&	2.835E-02	&	2.902E-02	&	2.911E-02	\\
15	&	2.560E-02	&	2.402E-02	&	2.835E-02	&	2.903E-02	&	2.911E-02	\\
20	&	2.560E-02	&	2.402E-02	&	2.835E-02	&	2.903E-02	&	2.911E-02	\\
50	&	2.560E-02	&	2.722E-02	&	2.835E-02	&	2.903E-02	&	2.911E-02	\\
100	&	2.560E-02	&	2.779E-02	&	2.836E-02	&	2.909E-02	&	2.911E-02	\\
\hline
time per trial (s)&	4.928E-02	&	1.599E-01	&	5.995E-01	&	2.335E+00	&	9.231E+00	\\
\hline\hline
 \end{tabular}
\end{lrbox}
\scalebox{1}{\usebox{\tablebox}}
\end{table}

\begin{table}[!h]
  \centering
  \caption{\small Discrepancies of $24$-D points up to $N=10^4$ are calculated by TA-improved algorithm. The points are drawn from uniform distribution on $[0,1]^d$.
}\label{supp_tab_disc_24d}
 \begin{lrbox}{\tablebox}
  \begin{tabular}{cccccc}
\hline\hline
\multicolumn{6}{c}{$N= 10^2$, Maximal value is 3.650E-01.}\\
\hline
$N_{tr}$	&	$I=16$	&	$I=32$	&	$I=64$	&	$I=128$	&	$I=256$	\\
\hline
5	&	3.290E-01	&	3.489E-01	&	3.626E-01	&	3.621E-01	&	3.621E-01	\\
10	&	3.302E-01	&	3.581E-01	&	3.626E-01	&	3.682E-01	&	3.621E-01	\\
15	&	3.302E-01	&	3.581E-01	&	3.626E-01	&	3.682E-01	&	3.650E-01	\\
20	&	3.404E-01	&	3.581E-01	&	3.626E-01	&	3.682E-01	&	3.650E-01	\\
50	&	3.515E-01	&	3.581E-01	&	3.650E-01	&	3.682E-01	&	3.650E-01	\\
100	&	3.515E-01	&	3.600E-01	&	3.709E-01	&	3.682E-01	&	3.650E-01	\\
\hline
time per trial (s)&	1.343E-03	&	4.296E-03	&	1.486E-02	&	5.777E-02	&	2.260E-01	\\
\hline\hline
\multicolumn{6}{c}{$N= 10^3$, Maximal value is 1.465E-01.}\\
\hline
$N_{tr}$	&	$I=16$	&	$I=32$	&	$I=64$	&	$I=128$	&	$I=256$	\\
\hline
5	&	1.027E-01	&	1.278E-01	&	1.431E-01	&	1.456E-01	&	1.456E-01	\\
10	&	1.099E-01	&	1.278E-01	&	1.431E-01	&	1.456E-01	&	1.456E-01	\\
15	&	1.185E-01	&	1.354E-01	&	1.431E-01	&	1.461E-01	&	1.465E-01	\\
20	&	1.185E-01	&	1.354E-01	&	1.436E-01	&	1.461E-01	&	1.465E-01	\\
50	&	1.185E-01	&	1.354E-01	&	1.450E-01	&	1.461E-01	&	1.465E-01	\\
100	&	1.197E-01	&	1.354E-01	&	1.450E-01	&	1.465E-01	&	1.465E-01	\\
\hline
time per trial (s)&	9.104E-03	&	3.034E-02	&	1.143E-01	&	4.476E-01	&	1.737E+00	\\
\hline\hline
\multicolumn{6}{c}{$N= 10^4$, Maximal value is 5.254E-02.}\\
\hline
$N_{tr}$	&	$I=16$	&	$I=32$	&	$I=64$	&	$I=128$	&	$I=256$	\\
\hline
5	&	3.889E-02	&	4.663E-02	&	4.937E-02	&	5.198E-02	&	5.254E-02	\\
10	&	4.127E-02	&	4.663E-02	&	4.967E-02	&	5.198E-02	&	5.254E-02	\\
15	&	4.127E-02	&	4.674E-02	&	5.041E-02	&	5.198E-02	&	5.254E-02	\\
20	&	4.322E-02	&	4.674E-02	&	5.041E-02	&	5.198E-02	&	5.254E-02	\\
50	&	4.362E-02	&	4.704E-02	&	5.041E-02	&	5.198E-02	&	5.254E-02	\\
100	&	4.362E-02	&	4.762E-02	&	5.041E-02	&	5.201E-02	&	5.254E-02	\\
\hline
time per trial (s)&	9.229E-02	&	3.404E-01	&	1.273E+00	&	5.054E+00	&	1.829E+01	\\
\hline\hline
 \end{tabular}
\end{lrbox}
\scalebox{0.95}{\usebox{\tablebox}}
\end{table}

\begin{table}[!h]
  \centering
  \caption{\small Discrepancies of $36$-D points up to $N=10^4$ are calculated by TA-improved algorithm. The points are drawn from uniform distribution on $[0,1]^d$.
}\label{supp_tab_disc_36d}
 \begin{lrbox}{\tablebox}
  \begin{tabular}{cccccc}
\hline\hline
\multicolumn{6}{c}{$N= 10^2$, Maximal value is 4.314E-01.}\\
\hline
$N_{tr}$	&	$I=16$	&	$I=32$	&	$I=64$	&	$I=128$	&	$I=256$	\\
\hline
5	&	3.647E-01	&	3.820E-01	&	4.314E-01	&	4.314E-01	&	4.314E-01	\\
10	&	3.647E-01	&	4.257E-01	&	4.314E-01	&	4.314E-01	&	4.314E-01	\\
15	&	3.647E-01	&	4.257E-01	&	4.314E-01	&	4.314E-01	&	4.314E-01	\\
20	&	3.647E-01	&	4.257E-01	&	4.314E-01	&	4.314E-01	&	4.314E-01	\\
50	&	3.802E-01	&	4.257E-01	&	4.314E-01	&	4.314E-01	&	4.314E-01	\\
100	&	3.837E-01	&	4.257E-01	&	4.314E-01	&	4.314E-01	&	4.314E-01	\\
\hline
time per trial (s)&	2.016E-03	&	5.818E-03	&	2.178E-02	&	8.334E-02	&	3.306E-01	\\
\hline\hline
\multicolumn{6}{c}{$N= 10^3$, Maximal value is 1.468E-01.}\\
\hline
$N_{tr}$	&	$I=16$	&	$I=32$	&	$I=64$	&	$I=128$	&	$I=256$	\\
\hline
5	&	1.021E-01	&	1.226E-01	&	1.387E-01	&	1.423E-01	&	1.468E-01	\\
10	&	1.070E-01	&	1.268E-01	&	1.387E-01	&	1.436E-01	&	1.492E-01	\\
15	&	1.098E-01	&	1.268E-01	&	1.387E-01	&	1.445E-01	&	1.492E-01	\\
20	&	1.098E-01	&	1.271E-01	&	1.387E-01	&	1.445E-01	&	1.492E-01	\\
50	&	1.098E-01	&	1.271E-01	&	1.403E-01	&	1.480E-01	&	1.492E-01	\\
100	&	1.120E-01	&	1.276E-01	&	1.403E-01	&	1.480E-01	&	1.492E-01	\\
\hline
time per trial (s)&	1.334E-02	&	4.506E-02	&	1.715E-01	&	6.723E-01	&	2.643E+00	\\
\hline\hline
\multicolumn{6}{c}{$N= 10^4$, Maximal value is 4.800E-02.}\\
\hline
$N_{tr}$	&	$I=16$	&	$I=32$	&	$I=64$	&	$I=128$	&	$I=256$	\\
\hline
5	&	2.822E-02	&	3.726E-02	&	4.267E-02	&	4.654E-02	&	4.800E-02	\\
10	&	2.822E-02	&	3.726E-02	&	4.502E-02	&	4.696E-02	&	4.824E-02	\\
15	&	2.848E-02	&	3.726E-02	&	4.502E-02	&	4.715E-02	&	4.838E-02	\\
20	&	2.848E-02	&	3.726E-02	&	4.502E-02	&	4.715E-02	&	4.838E-02	\\
50	&	3.040E-02	&	3.902E-02	&	4.502E-02	&	4.715E-02	&	4.842E-02	\\
100	&	3.159E-02	&	3.902E-02	&	4.502E-02	&	4.736E-02	&	4.847E-02	\\
\hline
time per trial (s)&	1.625E-01	&	5.829E-01	&	2.121E+00	&	9.191E+00	&	3.071E+01	\\
\hline\hline
 \end{tabular}
\end{lrbox}
\scalebox{0.95}{\usebox{\tablebox}}
\end{table}

\begin{table}[!h]
  \centering
  \caption{\small Discrepancies of $60$-D points up to $N=10^4$ are calculated by TA-improved algorithm. The points are drawn from uniform distribution on $[0,1]^d$. 
}\label{supp_tab_disc_60d}
 \begin{lrbox}{\tablebox}
  \begin{tabular}{cccccc}
\hline\hline
\multicolumn{6}{c}{$N= 10^2$, Maximal value is 5.286E-01.}\\
\hline
$N_{tr}$	&	$I=16$	&	$I=32$	&	$I=64$	&	$I=128$	&	$I=256$	\\
\hline
5	&	4.607E-01	&	5.013E-01	&	5.196E-01	&	5.286E-01	&	5.286E-01	\\
10	&	4.628E-01	&	5.086E-01	&	5.196E-01	&	5.286E-01	&	5.286E-01	\\
15	&	4.841E-01	&	5.086E-01	&	5.263E-01	&	5.286E-01	&	5.286E-01	\\
20	&	4.841E-01	&	5.086E-01	&	5.286E-01	&	5.286E-01	&	5.286E-01	\\
50	&	4.841E-01	&	5.181E-01	&	5.286E-01	&	5.286E-01	&	5.286E-01	\\
100	&	4.841E-01	&	5.181E-01	&	5.286E-01	&	5.286E-01	&	5.286E-01	\\
\hline
time per trial (s)&	2.383E-03	&	7.852E-03	&	2.996E-02	&	1.177E-01	&	4.656E-01	\\
\hline\hline
\multicolumn{6}{c}{$N= 10^3$, Maximal value is 2.027E-01.}\\
\hline
$N_{tr}$	&	$I=16$	&	$I=32$	&	$I=64$	&	$I=128$	&	$I=256$	\\
\hline
5	&	1.398E-01	&	1.679E-01	&	1.903E-01	&	1.939E-01	&	2.014E-01	\\
10	&	1.398E-01	&	1.679E-01	&	1.903E-01	&	1.947E-01	&	2.027E-01	\\
15	&	1.398E-01	&	1.730E-01	&	1.903E-01	&	1.963E-01	&	2.027E-01	\\
20	&	1.429E-01	&	1.730E-01	&	1.904E-01	&	1.963E-01	&	2.027E-01	\\
50	&	1.452E-01	&	1.765E-01	&	1.920E-01	&	1.994E-01	&	2.027E-01	\\
100	&	1.512E-01	&	1.765E-01	&	1.920E-01	&	2.004E-01	&	2.027E-01	\\
\hline
time per trial (s)&	1.981E-02	&	6.779E-02	&	2.552E-01	&	9.924E-01	&	3.914E+00	\\
\hline\hline
\multicolumn{6}{c}{$N= 10^4$, Maximal value is 6.161E-02.}\\
\hline
$N_{tr}$	&	$I=16$	&	$I=32$	&	$I=64$	&	$I=128$	&	$I=256$	\\
\hline
5	&	3.642E-02	&	4.217E-02	&	5.336E-02	&	5.833E-02	&	6.130E-02	\\
10	&	3.642E-02	&	4.602E-02	&	5.349E-02	&	5.833E-02	&	6.130E-02	\\
15	&	3.642E-02	&	4.602E-02	&	5.349E-02	&	5.833E-02	&	6.130E-02	\\
20	&	3.642E-02	&	4.602E-02	&	5.349E-02	&	5.885E-02	&	6.130E-02	\\
50	&	3.642E-02	&	4.602E-02	&	5.349E-02	&	5.885E-02	&	6.154E-02	\\
100	&	3.660E-02	&	4.749E-02	&	5.349E-02	&	5.988E-02	&	6.161E-02	\\
\hline
time per trial (s)&	3.313E-01	&	1.017E+00	&	3.519E+00	&	1.250E+01	&	4.624E+01	\\
\hline\hline
 \end{tabular}
\end{lrbox}
\scalebox{0.95}{\usebox{\tablebox}}
\end{table}

\begin{table}[!h]
  \centering
  \caption{\small Discrepancies of $120$-D points up to $N=10^4$ are calculated by TA-improved algorithm. The points are drawn from uniform distribution on $[0,1]^d$.
}\label{supp_tab_disc_120d}
 \begin{lrbox}{\tablebox}
  \begin{tabular}{cccccc}
\hline\hline
\multicolumn{6}{c}{$N= 10^2$, Maximal value is 7.673E-01.}\\
\hline
$N_{tr}$	&	$I=16$	&	$I=32$	&	$I=64$	&	$I=128$	&	$I=256$	\\
\hline
5	&	6.309E-01	&	6.841E-01	&	7.361E-01	&	7.488E-01	&	7.673E-01	\\
10	&	6.309E-01	&	6.954E-01	&	7.486E-01	&	7.673E-01	&	7.673E-01	\\
15	&	6.309E-01	&	6.973E-01	&	7.486E-01	&	7.673E-01	&	7.673E-01	\\
20	&	6.309E-01	&	7.115E-01	&	7.486E-01	&	7.673E-01	&	7.673E-01	\\
50	&	6.452E-01	&	7.196E-01	&	7.519E-01	&	7.673E-01	&	7.673E-01	\\
100	&	6.669E-01	&	7.196E-01	&	7.527E-01	&	7.673E-01	&	7.673E-01	\\
\hline
time per trial (s)&	4.495E-03	&	1.501E-02	&	6.066E-02	&	2.320E-01	&	9.136E-01	\\
\hline\hline
\multicolumn{6}{c}{$N= 10^3$, Maximal value is 2.576E-01.}\\
\hline
$N_{tr}$	&	$I=16$	&	$I=32$	&	$I=64$	&	$I=128$	&	$I=256$	\\
\hline
5	&	1.649E-01	&	2.122E-01	&	2.313E-01	&	2.477E-01	&	2.534E-01	\\
10	&	1.649E-01	&	2.196E-01	&	2.372E-01	&	2.477E-01	&	2.534E-01	\\
15	&	1.717E-01	&	2.196E-01	&	2.372E-01	&	2.477E-01	&	2.562E-01	\\
20	&	1.717E-01	&	2.196E-01	&	2.373E-01	&	2.477E-01	&	2.562E-01	\\
50	&	1.739E-01	&	2.196E-01	&	2.377E-01	&	2.524E-01	&	2.576E-01	\\
100	&	1.770E-01	&	2.196E-01	&	2.377E-01	&	2.536E-01	&	2.576E-01	\\
\hline
time per trial (s)&	3.880E-02	&	1.316E-01	&	5.017E-01	&	2.020E+00	&	9.670E+00	\\
\hline\hline
\multicolumn{6}{c}{$N= 10^4$, Maximal value is 8.736E-02.}\\
\hline
$N_{tr}$	&	$I=16$	&	$I=32$	&	$I=64$	&	$I=128$	&	$I=256$	\\
\hline
5	&	4.035E-02	&	5.696E-02	&	6.912E-02	&	8.254E-02	&	8.736E-02	\\
10	&	4.175E-02	&	5.760E-02	&	6.912E-02	&	8.254E-02	&	8.736E-02	\\
15	&	4.175E-02	&	6.008E-02	&	7.158E-02	&	8.254E-02	&	8.736E-02	\\
20	&	4.175E-02	&	6.008E-02	&	7.158E-02	&	8.277E-02	&	8.736E-02	\\
50	&	4.182E-02	&	6.008E-02	&	7.158E-02	&	8.277E-02	&	8.736E-02	\\
100	&	4.427E-02	&	6.126E-02	&	7.297E-02	&	8.300E-02	&	8.736E-02	\\
\hline
time per trial (s)&	9.001E-01	&	2.616E+00	&	1.028E+01	&	3.550E+01	&	1.153E+02	\\
\hline\hline
 \end{tabular}
\end{lrbox}
\scalebox{0.95}{\usebox{\tablebox}}
\end{table}

\begin{table}[!h]
  \centering
  \caption{\small Discrepancies of $240$-D points up to $N=10^4$ are calculated by TA-improved algorithm. The points are drawn from uniform distribution on $[0,1]^d$.
}\label{supp_tab_disc_240d}
 \begin{lrbox}{\tablebox}
  \begin{tabular}{cccccc}
\hline\hline
\multicolumn{6}{c}{$N= 10^2$, Maximal value is 9.077E-01.}\\
\hline
$N_{tr}$	&	$I=16$	&	$I=32$	&	$I=64$	&	$I=128$	&	$I=256$	\\
\hline
5	&	7.000E-01	&	8.062E-01	&	8.657E-01	&	9.025E-01	&	9.077E-01	\\
10	&	7.000E-01	&	8.062E-01	&	8.657E-01	&	9.025E-01	&	9.077E-01	\\
15	&	7.000E-01	&	8.062E-01	&	8.657E-01	&	9.025E-01	&	9.077E-01	\\
20	&	7.000E-01	&	8.062E-01	&	8.775E-01	&	9.025E-01	&	9.077E-01	\\
50	&	7.159E-01	&	8.169E-01	&	8.856E-01	&	9.070E-01	&	9.077E-01	\\
100	&	7.225E-01	&	8.265E-01	&	8.947E-01	&	9.070E-01	&	9.077E-01	\\
\hline
time per trial (s)&	9.186E-03	&	3.157E-02	&	1.213E-01	&	4.870E-01	&	1.914E+00	\\
\hline\hline
\multicolumn{6}{c}{$N= 10^3$, Maximal value is 3.792E-01.}\\
\hline
$N_{tr}$	&	$I=16$	&	$I=32$	&	$I=64$	&	$I=128$	&	$I=256$	\\
\hline
5	&	2.462E-01	&	3.076E-01	&	3.521E-01	&	3.689E-01	&	3.773E-01	\\
10	&	2.462E-01	&	3.076E-01	&	3.521E-01	&	3.701E-01	&	3.788E-01	\\
15	&	2.662E-01	&	3.086E-01	&	3.569E-01	&	3.701E-01	&	3.788E-01	\\
20	&	2.662E-01	&	3.086E-01	&	3.569E-01	&	3.701E-01	&	3.791E-01	\\
50	&	2.702E-01	&	3.091E-01	&	3.569E-01	&	3.715E-01	&	3.791E-01	\\
100	&	2.702E-01	&	3.121E-01	&	3.569E-01	&	3.732E-01	&	3.792E-01	\\
\hline
time per trial (s)&	9.448E-02	&	3.128E-01	&	1.149E+00	&	4.562E+00	&	1.789E+01	\\
\hline\hline
\multicolumn{6}{c}{$N= 10^4$, Maximal value is 1.183E-01.}\\
\hline
$N_{tr}$	&	$I=16$	&	$I=32$	&	$I=64$	&	$I=128$	&	$I=256$	\\
\hline
5	&	4.859E-02	&	6.636E-02	&	8.850E-02	&	1.059E-01	&	1.142E-01	\\
10	&	4.865E-02	&	6.745E-02	&	9.172E-02	&	1.059E-01	&	1.183E-01	\\
15	&	4.865E-02	&	7.116E-02	&	9.172E-02	&	1.074E-01	&	1.183E-01	\\
20	&	4.905E-02	&	7.209E-02	&	9.178E-02	&	1.074E-01	&	1.183E-01	\\
50	&	5.081E-02	&	7.491E-02	&	9.254E-02	&	1.074E-01	&	1.183E-01	\\
100	&	5.081E-02	&	7.491E-02	&	9.254E-02	&	1.083E-01	&	1.183E-01	\\
\hline
time per trial (s)&	2.794E+00	&	8.392E+00	&	2.613E+01	&	9.355E+01	&	3.030E+02	\\
\hline\hline
 \end{tabular}
\end{lrbox}
\scalebox{0.95}{\usebox{\tablebox}}
\end{table}

\begin{table}[!h]
  \centering
  \caption{\small Discrepancies of $360$-D points up to $N=10^4$ are calculated by TA-improved algorithm. The points are drawn from uniform distribution on $[0,1]^d$.
}\label{supp_tab_disc_360d}
 \begin{lrbox}{\tablebox}
  \begin{tabular}{cccccc}
\hline\hline
\multicolumn{6}{c}{$N= 10^2$, Maximal value is 9.777E-01.}\\
\hline
$N_{tr}$	&	$I=16$	&	$I=32$	&	$I=64$	&	$I=128$	&	$I=256$	\\
\hline
5	&	7.161E-01	&	8.791E-01	&	9.608E-01	&	9.606E-01	&	9.777E-01	\\
10	&	7.161E-01	&	8.791E-01	&	9.608E-01	&	9.614E-01	&	9.777E-01	\\
15	&	7.268E-01	&	8.791E-01	&	9.608E-01	&	9.614E-01	&	9.777E-01	\\
20	&	7.351E-01	&	8.791E-01	&	9.608E-01	&	9.614E-01	&	9.777E-01	\\
50	&	7.356E-01	&	8.791E-01	&	9.608E-01	&	9.614E-01	&	9.777E-01	\\
100	&	7.356E-01	&	8.791E-01	&	9.608E-01	&	9.693E-01	&	9.777E-01	\\
\hline
time per trial (s)&	1.471E-02	&	5.742E-02	&	2.128E-01	&	8.339E-01	&	3.235E+00	\\
\hline\hline
\multicolumn{6}{c}{$N= 10^3$, Maximal value is 4.473E-01.}\\
\hline
$N_{tr}$	&	$I=16$	&	$I=32$	&	$I=64$	&	$I=128$	&	$I=256$	\\
\hline
5	&	3.033E-01	&	3.520E-01	&	4.138E-01	&	4.355E-01	&	4.456E-01	\\
10	&	3.033E-01	&	3.584E-01	&	4.138E-01	&	4.384E-01	&	4.456E-01	\\
15	&	3.033E-01	&	3.634E-01	&	4.138E-01	&	4.384E-01	&	4.456E-01	\\
20	&	3.033E-01	&	3.634E-01	&	4.138E-01	&	4.384E-01	&	4.456E-01	\\
50	&	3.058E-01	&	3.688E-01	&	4.170E-01	&	4.384E-01	&	4.463E-01	\\
100	&	3.141E-01	&	3.688E-01	&	4.170E-01	&	4.384E-01	&	4.473E-01	\\
\hline
time per trial (s)&	2.144E-01	&	6.508E-01	&	2.146E+00	&	7.514E+00	&	3.104E+01	\\
\hline\hline
\multicolumn{6}{c}{$N= 10^4$, Maximal value is 1.451E-01.}\\
\hline
$N_{tr}$	&	$I=16$	&	$I=32$	&	$I=64$	&	$I=128$	&	$I=256$	\\
\hline
5	&	6.816E-02	&	8.787E-02	&	1.148E-01	&	1.324E-01	&	1.444E-01	\\
10	&	6.816E-02	&	8.787E-02	&	1.148E-01	&	1.332E-01	&	1.448E-01	\\
15	&	6.816E-02	&	8.787E-02	&	1.148E-01	&	1.332E-01	&	1.448E-01	\\
20	&	6.816E-02	&	9.158E-02	&	1.148E-01	&	1.332E-01	&	1.448E-01	\\
50	&	7.070E-02	&	9.158E-02	&	1.148E-01	&	1.332E-01	&	1.449E-01	\\
100	&	7.070E-02	&	9.381E-02	&	1.148E-01	&	1.342E-01	&	1.451E-01	\\
\hline
time per trial (s)&	2.772E+00	&	1.004E+01	&	4.226E+01	&	1.348E+02	&	4.997E+02	\\
\hline\hline
 \end{tabular}
\end{lrbox}
\scalebox{0.95}{\usebox{\tablebox}}
\end{table}

\end{document}